\RequirePackage{fix-cm}
\documentclass[natbib,smallextended]{svjour3}       	
\smartqed  								
\usepackage{graphicx}
\usepackage{amsmath}
\usepackage{tcolorbox}
\usepackage{amssymb}
\usepackage{amsfonts}
\usepackage{booktabs}
\usepackage{pdflscape}
\usepackage[colorlinks=true,citecolor=blue,urlcolor=blue,breaklinks]{hyperref}
\newcommand{\mat}{{\mathrm{mat}}}
\newcommand{\baryon}{{\mathrm{b}}}
\newcommand{\dd}{\mathrm{d}}
\newcommand{\ppn}{\mathrm{PPN}}
\newcommand{\bx}{\mathbf{x}}
\newcommand{\br}{\mathbf{r}}
\newcommand{\bk}{\mathbf{k}}
\newcommand{\bn}{\mathbf{n}}
\newcommand{\aem}{\alpha_{\mathrm{EM}}}
\newcommand{\aw}{\alpha_{\mathrm{W}}}
\newcommand{\ag}{\alpha_{\mathrm{G}}}
\newcommand{\as}{\alpha_{\mathrm{S}}}
\newcommand{\gfermi}{G_{\mathrm{F}}}
\newcommand{\dslash}{D\!\!\!\!/}
\newcommand{\hub}{h_0}
\newcommand{\unit}[1]{\mathrm{\ #1}}

\begin{document}
\title{Fundamental constants: from measurement to the universe, a window on gravitation and cosmology%
  \thanks{This article is a revised version of \url{https://doi.org/10.12942/lrr-2011-2}.\\
    \textbf{Change summary} Major revision, updated and expanded.\\
    \textbf{Change details} Updated and restructured the article, added new subsections. Previous Sect.~5 on ``Theories with varying constants'' has been split into new Sects.~3, 4, and 7. due to the increase of results and also to simplify the discussion of the experimental results. Appendices C and D have been added. The number of references has increased from 554 to 1203.}
}

\titlerunning{Fundamental constants}

\author{Jean-Philippe Uzan}

\institute{J.-P. Uzan \at 
Institut d'Astrophysique de Paris, \\
UMR-7095 du CNRS, Universit\'e Pierre et Marie Curie,\\
98 bis bd Arago, 75014 Paris\\
France\\
and\\
Center for Gravitational Physics and Quantum Information,\\
Yukawa Institute for Theoretical Physics,\\ Kyoto University, 606-8502, Kyoto\\
Japan\\
\email{uzan@iap.fr}
}
\date{Received: date / Accepted: date}

\maketitle

\begin{abstract}
Fundamental constants are a cornerstone of our physical laws. Any constant varying in space and/or time would signal a violation of local position invariance and be associated with a violation of the universality of free fall, and hence of the weak equivalence principle at the heart of the geometrisation of gravity. It will also reflect the existence of new degrees of freedom that couple to standard matter fields. Thus, testing for the stability of fundamental constants is of utmost importance for our understanding of gravity and for characterizing the domain of validity of General Relativity. Besides, it opens an independent window on the dark matter and dark energy components. 

As a consequence, thanks to the active developments of experiments, fundamental constants have become a key player in our search for physics beyond the standard model of particle physics and General Relativity. 

This review details the various roles of the fundamental constants in the laws of physics and in the construction of the International System of units, which now depends strongly on them. This requires to distinguish the concepts of \emph{fundamental units} and \emph{fundamental parameters}. Then, the relations between constants, the tests of the local position invariance and of the universality of free fall are presented, as well as the construction of field theories that account for ``varying constants'' and the motivations arising from high-energy physics and string theory. From a theoretical perspective any varying fundamental parameter is related to a dynamical field, the dynamics of which is dictated from the whole theory so that it remains fully consistent: no arbitrary law of variation has to be postulated. Then, the main experimental and observational constraints that have been obtained from atomic clocks, the Oklo phenomenon, Solar system observations, meteorite dating, quasar absorption spectra, stellar physics, pulsar timing, the cosmic microwave background and big bang nucleosynthesis are described. It details the basics of each system, its dependence with respect to the primary parameters the variation of which can be constrained from observations, the known systematic effects and the most recent constraints. It also describes how these primary parameters can be related to the fundamental constants and the model-dependencies that is involved. Both time and space variations are considered. To finish, it contains a short discussion on the more speculative possibility of understanding the numerical values of the fundamental parameters in view of the apparent fine-tuning that they confront us with, by invoking anthropic arguments. Given the huge increase of data and constraints and the difficulty to standardize them, a general scheme to present experimental and observational results and to construct a collaborative data base that will be more efficient for the community and allow us for better traceability, is proposed.

\end{abstract}
\setcounter{tocdepth}{3}
\tableofcontents

\clearpage
\section{Introduction}\label{section:intro}

Fundamental constants appear everywhere in the mathematical formulation of the laws we use to describe the phenomena of nature. They seem to contain some truth about the properties of the physical world while their real nature seem to evade us. 

While initiated about one century ago, the interest of fundamental constants attracted little attention until the surprising claim by \cite{q-webprl99} that quasar absorption spectra indicated that the fine structure constant was smaller in the past. This relaunched the interest on fundamental constants on two different but complementary fronts: (\textit{1}) \emph{theoretical}, as a way to constrain deviations from General Relativity, the validity of which on astrophysical scales is questioned due to the need of \emph{dark energy} and (\textit{2}) \emph{experimental} and \emph{observational}, motivated by both the necessity to confirm or infirm the claim by \cite{q-webprl99} and by the developments of many complementary methods to test the constancy of fundamental constants on a large band of redshifts \citep{jpu-revue}. This is now recognized as a key issue in views of the current anomalies of our cosmological models \citep{NEW_Peebles:2022akh}. To finish, the new definition of the International System of units adopted in 2018 has radically changed the way we shall conceive the relations between constants and units, which give a third motivation for a better understanding of their nature.

The question of the constancy of the constants of physics was probably first addressed by \cite{dirac37,dirac38} who expressed, in his ``\emph{Large Numbers hypothesis}'', the opinion that very large (or small) dimensionless universal constants cannot be pure mathematical numbers and must not occur in the basic laws of physics. He suggested, on the basis of this purely numerological principle, that these large numbers should rather be considered as variable parameters characterizing the state of the universe.  Dirac formed five dimensionless ratios  among which\footnote{$H_0$ is the Hubble constant today, $\rho_0$ the mean matter energy density today. $h_0$ is the reduced Hubble parameter, a dimensionless number defined in Eq.~(\ref{e.hubble}).The other constants are defined in Tables~\ref{tab-list} and \ref{tab-list2} and the definitions of the cosmological quantities are summarized in Appendix~\ref{app3}. } $\delta\equiv H_0\hbar/m_{\mathrm{p}}c^2\sim 2\hub\times10^{-42}$ and $\epsilon\equiv G\rho_0/H_0^2\sim 5\hub^{-2}\times 10^{-4}$ and asked the question of which of these ratios is constant as the universe evolves. Usually, $\delta$ varies as the inverse of the cosmic time while $\epsilon$ varies also with time if the universe is not described by an Einstein--de Sitter solution -- i.e., when a cosmological constant, curvature or radiation are included in the cosmological model; see Eq.~(\ref{e.FLeq}). Dirac then noticed\footnote{See Eq.~(\ref{edef-mu}) below for the definitions of $\aem$, $\ag$ and $\mu$.} that $\ag/\mu\aem$, representing the relative magnitude of electrostatic and gravitational forces between a proton and an electron, was of the same order as $H_0e^2/m_{\mathrm{e}}c^2=\delta\aem\mu$ representing the age of the universe in atomic units. Hence, this five numbers can be ``harmonized'' if one assumes that $\ag$ vary as the inverse of the cosmic time, as $\delta$ does.

This numerological argument by Dirac is indeed not a physical theory but it opened many doors in the investigation on physical constants, both on questioning whether they are actually constant and on trying to understand the numerical values we measure. Both questions have initiated two fields of research, the first on the use of constants to test General Relativity on cosmological scales and, the second, more speculative, on ways to account for fine tunings, and in particular the one for life.

First, the implementation of Dirac's phenomenological idea into a field-theory framework was proposed by \cite{jordan37}, who realized that the constants have to become dynamical fields. He proposed a theory in which both the gravitational and fine-structure constants can vary -- \cite{unzicker} and Section~II of \cite{jpu-revue} provides a summary of some earlier attempts to quantify the cosmological implications of Dirac's argument and to formulate alternatives.  \cite{fierz56} then realized that in such a case, atomic spectra shall become spacetime-dependent so that these theories can be observationally tested. The sub-case in which only $G$ can vary led to the definition of scalar-tensor theories of gravity, further explored by \cite{brans61}. These theories were generalized to obtain various functional dependencies for $G$ in the formalization of universal scalar-tensor theories (see, e.g., \citealt{damour92}). Today many extensions of General Relativity are, as we shall see, accompanied by the variation of some constants.

Second, \cite{dicke61} pointed out that in fact the density of the universe is determined by its age, the later being related to the time needed to form galaxies, stars, heavy nuclei\dots. This led him to formulate that the presence of an observer in the universe places constraints on the physical laws that can be observed. In fact, what is meant by observer is the existence of (highly?) organized systems and this principle can be seen as a rephrasing of the question ``why is the universe the way it is?'' (see e.g., \citealt{hogan00}). \cite{carter74,carter83}, who actually coined the term ``anthropic principle'', showed that the numerological coincidences found by Dirac can be derived from physical models of stars and the competition between the weakness of gravity with respect to nuclear fusion. \cite{carr79} then showed how one can scale up from atomic to cosmological scales only by using combinations of $\aem$, $\ag$ and $m_{\mathrm{e}}/m_{\mathrm{p}}$.

To summarize, Dirac's insight was to question whether some numerical coincidences between very large numbers, that cannot be themselves explained by the theory in which they appear, was a mere coincidence or whether it can reveal the existence of some new underlying physical laws. This paved the way to three main roads of investigation:
\begin{itemize}
 \item how do we construct theories in which what were thought to be constants are in fact dynamical fields?
 \item how can we constrain, experimentally or  observationally, the spacetime dependencies of the constants that appear in our physical laws?
 \item how can we explain the values of the fundamental constants and the fine-tuning that seems to exist between their numerical values?  This latter issue led to the notion of \emph{naturalness}, that is the not-so-well-defined idea that a model of physics is expected to work without requiring ad-hoc coincidences or ``fine-tunings” of its parameters and that instead of such ``conspiracies'' one needs to find a cause.\footnote{The concept has been widely used in particle physics in the context of renormalisation. According to \cite{Hooft1980}, `it is unlikely that the microscopic equations contain various free parameters that are carefully adjusted by Nature to give cancelling effects such that the macroscopic systems have some special properties. This is a philosophy which we would like to apply to the unified gauge theories: the effective interactions at [...] a low energy scale $\mu_1$ should follow from the properties at a much higher energy scale $\mu_2$ without the requirement that various different parameters at the energy scale $\mu_2$  match with an accuracy of the order of $\mu_1/\mu_2$ . That would be unnatural." or, according to  \cite{gellmann83} ``The situation of having numerous arbitrary dimensional parameters is even more humiliating when some of them are very small. [\dots] First of all, we'd like to interpret small or nearly symmetric quantities as coming from a slightly broken symmetry; otherwise they don’t make sense to us. [\dots] When we can't avoid dialing some renormalized quantity to a small value [\dots] that situation has recently been described as a problem of `naturalness'.''  We refer to \cite{Nelson85,Giudice:2008bi,Craig:2022eqo} for further discussions on this concept, as well as Sect.~\ref{section5}.}
 \end{itemize}

While ``varying constants'' may seem, at first glance, to be a cheap oxymoron, it has to be considered merely as jargon to be understood as ``\emph{revealing new degrees of freedom and their coupling to the known fields of our theory}'', that is a powerful method to test, or eventually reveal, the existence of new fields beyond those of the standard model of particle physics and General Relativity. It has the advantage to be applicable on astrophysical scales and to complement the tests of General Relativity that rely on the large scale structure (see \citealt{Uzan:2003zq,jpu-revu3} for an early discussion of this complementarity). As such, the tests on the constancy of the fundamental constants are indeed very important tests of fundamental physics and of the laws of nature we are currently using. Detecting any such variation will indicate the need for new physical degrees of freedom in our theories, that is \emph{new physics}.

Besides, given the rising number of anomalies in cosmology and the growing importance of the dark sector, constraints on the stability of fundamental constants have an important role to play in particular in their possibility to reveal a new long range composition dependent force and the need for an extension of General Relativity. This was discussed and highlighted for the first time in \cite{Uzan:2003zq}; see also \cite{jpu-revu3,ugrg}. This point was recently raised by \cite{NEW_Peebles:2022akh} who argued for ``a well- supported Fine-Structure Survey", which is indeed what the \emph{constantology} \citep{jpu-revue,uzanleclercqbook} and \emph{$\alpha$-cosmography} \citep{NEW_Martins:2022unf} are supposed to be. We hope that the expansion of this field and the connection with fundamental physics and cosmology and the power of constants to reveal scalar fields will be fully recognized, as well as their power to bridge local, astrophysical and cosmological scales.

In order to use fundamental constants to set constraints on theories beyond the standard model, one needs to go through a series of complementary steps, which are all very technical, namely
\begin{enumerate}
 \item it is necessary to understand and to model the physical systems used to set the constraints. In particular one needs to relate the effective parameters that can be observationally constrained to a set of fundamental constants. One also needs to list and understand all systematics a bad modelization of which could mimic the variation of a constant;
 \item it is necessary to relate and compare different constraints that are obtained at different spacetime positions. This requires a spacetime dynamics and thus to specify a model as well as a cosmology. It follows that these constraints combining several systems at different redshifts are usually more model-dependent;
 \item it is necessary to relate the variations of different fundamental constants in order to discuss degeneracies or to highlight amplifications.
\end{enumerate}
Therefore, we shall start in Sect.~\ref{subsec11} by recalling the links between the constants of physics and the theories in which they appear, as well as with metrology.  This will offer a clear \emph{definition of fundamental constants} and of their splitting as either \emph{fundamental units} or \emph{fundamental parameters} -- see Sect.~\ref{subsubsec213} for detailed definitions --, each playing a deep but different role in the laws of physics. From a theoretical point of view, the constancy of the fundamental constants is deeply linked with the equivalence principle and General Relativity. In Sect.~\ref{subsec12} we recall this relation and in particular the link with the universality of free fall and then highlight the connection with cosmology and the dark sector in Sect.~\ref{subseccosmo}. We then consider two theoretical topics. First, Sect.~\ref{section-theories} describes the \emph{dynamical aspects}, that is the way to make constants dynamical, hence introducing gravity theories beyond General Relativity. Then, Sect.~\ref{subsec5.3} addresses the \emph{structural issue} of the connection between the variations of several constants as the consequence of unification. It must also describe the computation of the effective parameters required to interpret experiments and observations in terms of fundamental constants, focusing on nuclear and atomic physics. Then, we describe and summarize the various constraints that exist on the variation of non-gravitational constants and of the gravitational constant respectively in Sections.~\ref{section3} and~\ref{section4}. Sect.~\ref{subsec81} describes the most popular models used to analyze the different constraints in a consistent way. We finish by a discussion on their spatial variations in Sect.~\ref{section-spatial} and then on the possibility to understand their numerical values in Sect.~\ref{section5}.

Various reviews have been written on this topic over the past years. I will refer to the first review by \cite{jpu-revue} as FVC03 and mention the following later reviews by \cite{jpu-revu2,barrow05,bronni,karshen-metro,karshen-rev, gbisern, olive-rio, damour-issi, jpu-rio, jpu-issi,NEW_Uzan:2015uba,flamrev,NEW_Martins:2017yxk} and I refer to \cite{NEW_Tiesinga:2021myr} for the numerical values of the constants adopted in this review. 

The present review builds on its previous version \citep{NEW_Uzan:2010pm} of which it is an update. First, it contains a detailed description of the role of the fundamental constant in the new SI unit system. While it keeps mostly the same organisation and notations, it updates all the constraints of the former version and includes new techniques that have been proposed since then. In particular, the debate on the results of quasar absorption spectra have witnessed a series of dedicated works and a deeper understanding. We shall mention huge progresses on the constraints from atomic clocks and their possible applications in space and the improvements on the constraints of the violation of the universality of free fall. On the cosmological side, the cosmic microwave background and primordial nucleosynthesis predictions and observations have been improved as well as the constraints on spatial variations. New methods include molecular spectroscopy, the Sunyaev--Zel'dovich effect, and the use of gravitational waves. On the theory sides, many works have tried to clarify the relations between constants.

\section{Constants and fundamental physics}\label{section1}

\subsection{Toward a definitions of fundamental constants}\label{subsec11}

Any physical theories introduce various mathematical structures to describe the phenomena of nature. They involve various fields, symmetries and constants. These structures are postulated in order to construct a mathematically-consistent description of the known physical phenomena in the most unified and simple way.  As such they reflect empirical choices guided by some principles and intuition. Hence, they need to be well-defined and confirmed by experiment.

 \begin{tcolorbox}
We define  the fundamental constants of a physical theory as the minimal set of \emph{any parameters that cannot be explained by this theory}.
 \end{tcolorbox} 
Indeed, we are often dealing with other constants that in principle can be expressed in terms of these fundamental constants.  The existence of these two sets of constants is important and arises from two different considerations. From a theoretical point of view we would like to extract the minimal set of fundamental constants, but often these constants are not directly measurable. From a more practical point of view, we need to measure constants, or combinations of them, which allow us to  reach the highest accuracy.

Therefore, these fundamental constants are \emph{contingent quantities} that can only be measured. Such parameters have to be assumed constant in this theoretical framework for two reasons:

\begin{itemize}
  \item from a \emph{theoretical point of view}, the postulated theoretical framework  does not -- and cannot -- provide any way to compute these parameters, i.e., it does not have any equation of evolution for them since otherwise  it would be considered as a dynamical field,

  \item from an \emph{experimental point of view}, these parameters can only be measured. If the theories in which they appear have been validated experimentally, it means that, at the precisions of these experiments, these parameters have indeed been checked to be constant, as required by the necessity of the reproducibility of the experimental results.
\end{itemize}

This means that testing for the constancy of these parameters is a test of the theories in which they appear and allows us to extend our knowledge of their domain of validity. This also resonates with the definition by \cite{weinberg83} who stated that they cannot be calculated in terms of other constants ``\dots not just because the calculation is too complicated (as for the viscosity of water) but because we do not know of anything more fundamental''. We could say that the more constants a theory involves the higher its contingency, which may link with the discussion we shall have on fine tuning in Sec.~\ref{section5}.

This has a series of implications. \emph{First}, the list of fundamental constants  to consider depends on our theories of physics and, thus, on time. Indeed, when introducing new, more unified or more fundamental, theories the number of constants may change so that this list reflects both our knowledge of physics and, more important, our ignorance. \emph{Second}, it also implies that some of these fundamental constants can become dynamical quantities in a more general theoretical framework so that the tests of the constancy of the fundamental constants are tests of fundamental physics, which can reveal that what was thought to be a fundamental constant is actually a field whose dynamics cannot be neglected. If such fundamental constants are actually dynamical fields it also means that the equations we are using are only approximations of other and more fundamental equations, in a kind of adiabatic limit, and that an equation for the evolution of this new field has to be obtained.

The reflections on the nature of the constants and their role in physics are numerous. We refer to the books by \cite{barrow-book, ul-book,fritzsch-book,  uzanleclercqbook,NEW_uzanleclercqbook} and to \cite{weinberg83,  okun96, bjorken1,  duff01,wilczek07, hfcte,  volovik09, NEW_Langacker:2017uah} for various discussions of this issue that we cannot develop at length here. This paragraph summarizes some of the properties of the fundamental constants that need to be kept in mind.

\subsubsection{Characterizing the fundamental constants}\label{subsec12fc}

Physical constants seem to play a central role in our physical theories since, in particular, they determine the magnitudes of the physical processes. Let us sketch briefly some of their main properties.

\paragraph{How many fundamental constants shall we consider?} \

The set of constants, which are conventionally considered as fundamental \citep{flower01}, consists of the electron charge $e$, the electron mass $m_{\mathrm{e}}$, the proton mass $m_{\mathrm{p}}$, the reduced Planck constant $\hbar$, the velocity of light in vacuum $c$, the Avogadro constant $N_{\mathrm{A}}$, the Boltzmann constant $k_{\mathrm{B}}$, the Newton constant $G$, the permeability and permittivity of vacuum, $\varepsilon_0$ and $\mu_0$. The latter had a fixed exact value in the SI system of unit before 2018 ($\mu_0=4\pi\times10^{-7} \unit{H} \unit{m}^{-1}$), which was implicit in the definition of the Ampere; $\varepsilon_0$ is then fixed by the relation $\varepsilon_0\mu_0=c^{-2}$ (See Sect.~\ref{subsec12} and Appendix~\ref{App0} for a discussion on the {\emph{new SI}).

However, it is clear that this cannot correspond to the list of the fundamental constants, as defined earlier as the free parameters of the theoretical framework at hand. To identify such a list we must specify this framework since it is clear that one cannot discuss fundamental constants without first defining the theoretical framework assumed to give the most fundamental description of nature. As discussed in the previous section, fundamental constants are the unknown constant parameters of this fundamental theory so that this list has evolved with our understanding of physics (see \cite{ul-book} for a documented history of the evolution of the fundamental constants).

Today, gravitation is described by General Relativity, and the three other interactions and the matter fields are described by the standard model of particle physics.  It follows that one has to consider 22 unknown constants (i.e., 19 unknown dimensionless parameters): the Newton constant $G$, 6 Yukawa couplings for the quarks ($h_{\mathrm{u}},h_{\mathrm{d}},h_{\mathrm{c}},h_{\mathrm{s}},h_{\mathrm{t}},h_{\mathrm{b}}$) and 3 for the leptons ($h_{\mathrm{e}},h_\mu,h_\tau$), 2 parameters of the Higgs field potential ($\hat\mu,\lambda$), 4 parameters for the Cabibbo--Kobayashi--Maskawa matrix (3 angles $\theta_{ij}$ and a phase $\delta_{\mathrm{CKM}}$), 3 coupling constants for the gauge groups $SU(3)_c\times SU(2)_L\times U(1)_Y$ ($g_1,g_2,g_3$ or equivalently $g_2,g_3$ and the Weinberg angle $\theta_{\mathrm{W}}$), and a phase for the QCD vacuum ($\theta_{\mathrm{QCD}}$), to which one must add the speed of light $c$  and the Planck constant $h$. See Table~\ref{tab-list} for a summary and their numerical values.

\begin{table}[t]
\caption[Fundamental constants of the standard model of particle physics]{List of the 22 fundamental constants of our standard model.  See \cite{NEW_Tiesinga:2021myr} for further details on the measurements. It is important to stress that, since the detection of the Higgs boson, all free parameters of the standard model have been measured, which makes it a fully predictive theory.  Note that this list does not include the charge of the electron $e=1.602 176 634\,\times 10^{-19}\unit{C}$ that is related to $g_2$ and $\theta_{\mathrm{W}}$ thanks to Eq.~(\ref{e.eg2}).}
\label{tab-list}
{\small
\centering
\begin{tabular}{p{5cm}ll}
 \toprule
 Constant & Symbol & Value \\
 \midrule
 Speed of light & $c$ & 299\,792\,458~m~s$^{-1}$ \\
 Planck constant & $h$ &$6.626\,070\,15\,\times 10^{-34}$~J~s \\
 Newton constant & $G$ & 6.674\,30(15)\,$\times 10^{-11}\unit{m^{2}\ kg^{-1}\ s^{-2}}$\\
 \midrule
 Weak coupling constant (at $m_Z$) & $g_2(m_Z)$ & 0.6520\,$\pm$\,0.0001\\
 Strong coupling constant (at $m_Z$) & $g_3(m_Z)$ & 1.221\,$\pm$\,0.022\\ 
 Weinberg angle & $ \sin^2\theta_{\mathrm{W}}$(91.2~GeV)$_{\overline{\mathrm{MS}}}$ & $0.231\,21\pm 0.00004$ \\
 \midrule
 Electron Yukawa coupling & $h_{\mathrm{e}}$ & $2.94 \times 10^{-6}$\\
 Muon Yukawa coupling & $h_\mu$ & 0.000607\\
 Tauon Yukawa coupling & $h_\tau$ & 0.0102156\\
 Up Yukawa coupling & $h_{\mathrm{u}}$ & 0.000016\,$\pm$\,0.000007\\
 Down Yukawa coupling & $h_{\mathrm{d}}$ & 0.00003\,$\pm$\,0.00002\\
 Charm Yukawa coupling & $h_{\mathrm{c}}$ & 0.0072\,$\pm$\,0.0006\\
 Strange Yukawa coupling & $h_{\mathrm{s}}$ & 0.0006\,$\pm$\,0.0002\\
 Top Yukawa coupling & $h_{\mathrm{t}}$ & 1.002\,$\pm$\,0.029\\
 Bottom Yukawa coupling & $h_{\mathrm{b}}$ & 0.026\,$\pm$\,0.003\\
 \midrule
 Quark CKM matrix angle & $\sin\theta_{12}$ & 0.22650$\pm$0.00048\\
                                          & $\sin\theta_{23}$ & $0.04053^{+0.00083}_{-0.00061}$\\\
                                           & $\sin\theta_{13}$ & $0.00361^{+0.00011}_{-0.00009}$\\
 Quark CKM matrix phase & $\delta_{\mathrm{CKM}}$ & $1.196^{+0.045}_{-0.043}$\\
 \midrule
 Higgs potential quadratic coefficient & $\hat\mu$ & $\simeq 88.4$~GeV \\
 Higgs potential quartic coefficient & $\lambda$ & $\simeq  0.13$\\
 QCD vacuum phase & $\theta_{{\mathrm{QCD}}}$ & $<10^{-10}$\\
\bottomrule 
\end{tabular}
}
\end{table}

Again, this list relies on what we accept as a fundamental theory. Today we have many hints that the standard model of particle physics has to be extended, in particular to include the existence of massive neutrinos. Such an extension comes with at least seven new constants (3 Yukawa couplings and 4 Maki--Nakagawa--Sakata (MNS) parameters, similar to the CKM parameters). On the other hand, the number of constants can decrease if some unifications between various interaction exist (see Sect.~\ref{subsecGUT} for more details) since the various coupling constants may be related to a unique coupling constant $\alpha_U$ and an energy scale of unification $m_{\rm u}$ through
$$
 \alpha_i^{-1}(E) = \alpha_U^{-1} + \frac{b_i}{2\pi}\ln \frac{m_{\rm u}}{E},
$$
where the $b_i$ are numbers, that depend on the explicit model of unification. Note that this would also imply that the variations, if any, of various constants shall be correlated.

\paragraph{Relation to other usual constants} \

The free parameters of the standard model are related to various constants that appear in this review (see Table~\ref{tab-list2}) and actually to all the other constants of physics but also of other fields.

First, the quartic and quadratic coefficients of the Higgs field potential are related to the Higgs mass and vev, 
$$
m_H=\sqrt{-\hat\mu^2/2} \qquad\hbox{and}\qquad v=\sqrt{-\hat\mu^2/\lambda}\,.
$$
The latter is related to the Fermi constant is derived from the $\mu$-lifetime as (see Section~10 of \citealt{NEW_pdg})
$$
\gfermi=(v^2\sqrt{2})^{-1}=  1.166~378~8(6)\,\times 10^{-5}\unit{GeV}^{-2}\,,
$$
which imposes that $v=(246.7\pm0.2) \unit{GeV}$ . Since its discovery,  the Higgs mass has been measured with increasing precision to reach, thanks to the ATLAS experiment at CERN, the unprecedented precision of 0.09\%, $m_H=(125.20\pm0.11)$~ GeV (ATLAS 2023).

Then, the masses of the quarks and leptons are related to their Yukawa couplings and the Higgs vev by $m=hv/\sqrt{2}$. Note however that, due to the long-distance confining property of the strong QCD interaction,  free quarks have never been observed. All quarks but the top  hadronize, i.e., become part of a meson or baryon, on a timescale of order $1/\Lambda_{\mathrm{QCD}}$, the top decaying before it can hadronize.  This implies that the question of what is meant by the quark mass is a complex one, which requires one to adopt a specific prescription; see Section 9 of \cite{NEW_pdg}. Perturbatively,  one can use the prescription of the pole mass $m_{\rm q}$ of the divergence of the quark propagator as a definition which, while close to the physical picture of mass, however suffers from ambiguities of order $\Lambda_{\mathrm{QCD}}$ when relating it to observable quantities. This anticipates the difficult discussion on the relations between the free parameters of the standard model and the observables and nuclear properties.

The values of the gauge couplings depend on energy via the renormalization group so that they are given at a chosen energy scale, here the mass of the $Z$-boson, $m_Z$.  $g_1$ and $g_2$ are related by the Weinberg angle as 
$$
g_1=g_2\tan \theta_{\mathrm{W}} \, .
$$
The electromagnetic coupling constant is not $g_1$ since $SU(2)_L\times U(1)_Y$ is broken to $U(1)_{\mathrm{elec}}$ so that
\begin{equation}\label{e.eg2}
 g_{\mathrm{EM}}(m_Z)=e=g_2(m_Z) \sin\theta_{\mathrm{W}}.
\end{equation}
Defining the fine-structure constant as $\aem= g_{\mathrm{EM}}^2/\hbar c$, the (usual) zero energy electromagnetic fine structure constant can be extracted from the anomalous magnetic moment of the electron, $a_e= (1159652180.59\pm0.13)\times10^{-12}$ applying QED corrections up to five loops allows one to extract to get $\aem= 1/137.035999166(15)$ \citep{NEW_Fan:2022eto}. The combination of the measurements of the Rydberg constant and atomic masses with interferometry of atomic recoil kinematics of $^{87}$Rb and $^{133}$Cs leads to $\aem=1/ 137.035999206(11)$ \citep{Morel:2020dww} and $\aem=1/137.035999046(27)$ \citep{Parker:2018vye} respectively. It is to be noted that these two latter values differ by 5.5$\sigma$! Combining all the latest data yields the world average $\aem = 1/137.035999178(8)$, which differs from the CODATA recommended value at $Q^2 = 0$ $\aem= 1/137.035\,999\,084(21)$, which is the value we report in Table~\ref{tab-list2}, even though it does not use the latest experimental inputs. We refer to Section 10 of \cite{NEW_pdg} for a discussion of these values and their differences. 

\begin{table}[t]
\caption[Related constants of the standard model of particle physics]{List of some related constants that appear in our  discussions. See \cite{NEW_pdg}.}
\label{tab-list2}
\centering
{\small
\begin{tabular}{p{6cm}ll}
 \toprule
 Constant & Symbol & Value \\
 \midrule
 Electromagnetic coupling constant & $g_{\mathrm{EM}}=e=g_2 \sin\theta_{\mathrm{W}}$ & 0.313429\,$\pm$\,0.000022\\
 Higgs mass & $m_H$ & $(125.20\pm0.11)$~ GeV\\
 Higgs vev & $v$ & (246.7\,$\pm$\,0.2)~GeV\\
  Fermi constant & $\gfermi=1/\sqrt{2}v^2$ & 1.166~378~8(6)\,$\times 10^{-5}\unit{GeV}^{-2}$\\
  Mass of the $W^\pm$ & $m_W$ & 80.369~2(133) ~GeV\\
  Mass of the $Z^0$ & $m_Z$ & 91.188~0(20)~GeV\\
  Fine structure constant & $\aem$ & 1/137.035\,999\,084(21)\\
  Fine structure constant at $m_Z$ & $\aem(m_Z)$ & $1/(1127.930\pm0.008)$\\
 Weak structure constant at $m_Z$ & $\aw(m_Z)$ & 0.03383\,$\pm$\,0.00001\\
   Strong structure constant at $m_Z$ & $\as(m_Z)$ & 0.118~0(9)\\
 Gravitational structure constant &$ \ag= Gm_{\mathrm{p}}^2/\hbar c$ & $\sim 5.905 \times 10^{-39}$ \\
 \hline
 Electron mass & $m_{\mathrm{e}}= h_{\mathrm{e}}v/\sqrt{2}$ & 510.998~950~00(15)~keV \\
 & &  $9.109 383 7015(28) \times 10^{-31}$~kg\\
  Mu mass & $m_\mu= h_\mu v/\sqrt{2}$ & (105.658~375~5$\pm$0.000~002~3)~MeV   \\
  Tau mass & $m_\tau= h_\tau v/\sqrt{2}$ & 1776.93\,$\pm$\,0.09~MeV \\
 Up quark mass & $m_{\mathrm{u}}= h_{\mathrm{u}}v/\sqrt{2}$ & $(2.16\pm0.07 )$~MeV\\
 Down quark mass & $m_{\mathrm{d}}= h_{\mathrm{d}}v/\sqrt{2}$ & ($4.70\pm0.07$)~MeV\\
 Strange quark mass & $m_{\mathrm{s}}= h_{\mathrm{s}}v/\sqrt{2}$ &  ($93.5\pm0.8$)~MeV \\
 Charm quark mass & $m_{\mathrm{c}}= h_{\mathrm{c}}v/\sqrt{2}$ &  $(1.2730\pm0.0046)$~GeV\\
 Bottom quark mass & $m_{\mathrm{b}}= h_{\mathrm{b}}v/\sqrt{2}$ & $(4.183\pm0.007)$~GeV\\
 Top quark mass & $m_{\mathrm{t}}= h_{\mathrm{t}}v/\sqrt{2}$ & $(172.57\pm0.29)$~GeV\\
 QCD energy scale & $\Lambda_{{\mathrm{QCD}}}$ & (190\,--\,240)~MeV\\
 \hline
  Mass of the proton & $m_{\mathrm{p}}$ & 938.272~088 16(29)~MeV \\
 &&1.672~621~923~69(51)$\times10^{-27}$~kg\\
  Mass of the neutron & $m_{\mathrm{n}}$ & 939.565~420~52(54)~MeV \\
  proton-neutron mass difference & $Q_{\mathrm{np}}$ & $(1.293~332~4\pm0.000~000~5)$~MeV \\
 \hline
  Proton-to-electron mass ratio & $\mu=m_{\mathrm{p}}/m_{\mathrm{e}}$ &1836.152~673~43(11)\\
 Electron-to-proton mass ratio & $\bar\mu=m_{\mathrm{e}}/m_{\mathrm{p}}$ & 1/1836.152~673~43(11)\\
 $d-u$ quark mean mass & $m_{\mathrm{q}}=(m_{\mathrm{u}}+m_{\mathrm{d}})/2$ & (2.5\,--\,5.0)~MeV\\
 $d-u$ quark mass difference & $\delta m_{\mathrm{q}} = m_{\mathrm{d}}-m_{\mathrm{u}}$ & (0.2\,--\,4.5)~MeV \\
 Proton gyromagnetic factor &$g_{\mathrm{p}}$ & 5.586 \\
 Neutron gyromagnetic factor &$g_{\mathrm{n}}$ & $-3.826$ \\
 Rydberg constant & $R_\infty$ & 10\,973\,731.568\,527(73)~m$^{-1}$\\
 \hline
  Photon mass & $m_\gamma$ &  $<10^{-18}$~eV \\
  Photon charge & $q_\gamma$ &  $<10^{-35}~e$ \\
   Graviton mass & $m_g$ &  $<1.76\times10^{-23}$~eV \\
\bottomrule
\end{tabular}
}
\end{table}

From electroweak renormalization schemes, one can characterize the running of $\aem$ on the energy scale of the process with, given the discussion of the previous paragraph, $\aem\simeq1/137.036$ at very low energy, i.e., close to the Thomson limit.  The running has been observed in small and large angle Bhabha scattering \citep{OPAL:2005xqs,L3:2005tsb}. At energies larger than a few MeV, the hadronic contribution to vacuum polarization introduces a theoretical uncertainty in the determination of $\aem$. Using the modified minimal subtraction (MS) scheme and $\alpha_s(m_Z)  =  0.1187\pm0.0017$, one gets
$$
\aem(m_Z)=1/(1127.930\pm0.008),
$$
that  corresponds to the quark contribution (without the top) to the conventional (on-shell) QED coupling
$$
\aem(m_Z) = \frac{\aem}{1-\Delta\aem(m_Z)}
$$
with $\Delta\aem(m_Z) =  0.02783\pm0.00006$. We refer to Table 10.1 of \cite{NEW_pdg} for the various evaluations of $\Delta\aem(m_Z)$. This already illustrates the model-dependence that will enter the relation between the QCD parameters and the low energy quantities, even when one sticks to the standard model.

We define the QCD energy scale, $\Lambda_{\mathrm{QCD}}$, as the energy at which the strong coupling constant diverges. Note that it implies that $\Lambda_{\mathrm{QCD}}$ also depends on the Higgs and fermion masses through threshold effects.

More familiar constants, such as the masses of the proton and the neutron are, as we shall discuss in more details below (see Sect.~\ref{subsubmass}), more difficult to relate to the fundamental parameters because they depend not only on the masses of the quarks but also on the electromagnetic and strong binding energies.

\paragraph{Are some constants more fundamental than others?} \

As pointed-out by \cite{levy79}, all constants of physics do not play the same role, and some have a much deeper role than others. Following his analysis, we can define three classes of fundamental constants,
\begin{itemize}
\item  \textit{class A} being the class of the constants characteristic of a particular system,
\item \textit{class B} being the class of constants characteristic of a class of physical phenomena, 
\item \textit{class C} being the class of universal constants. 
\end{itemize}
Indeed, the status of a constant can change with time and our understanding of the laws of nature; see Fig.~\ref{fig1-classes}. For instance, the velocity of light was initially a class A constant (describing a property of light alone), which then became a class B constant when it was realized within Maxwell theory  that light electromagnetic phenomena so that its velocity was a general property of all electromagnetic waves and as such related to the vacuum permeability and permittivity.  To finish, it ended as a type C constant: it enters special relativity and is related to the notion of causality, whatever the physical phenomena at stakes. It has even become a much more fundamental constant: since 1983 it enters in the definition of the meter \citep{petley83}; see \cite{ul-book} for a more detailed discussion. We thus define
\begin{itemize}
\item  \textit{class D} being the class of the dimensionful constants that enter the definition of units.
\end{itemize}
This classification has to be contrasted with the proposition of \cite{wilczek07} to distinguish the standard model free parameters as the gauge and gravitational couplings (which are associated to internal and spacetime curvatures) and the other parameters entering the accommodation of inertia in the Higgs sector.

\begin{figure}[htbp]
  \centerline{\includegraphics[scale=0.45]{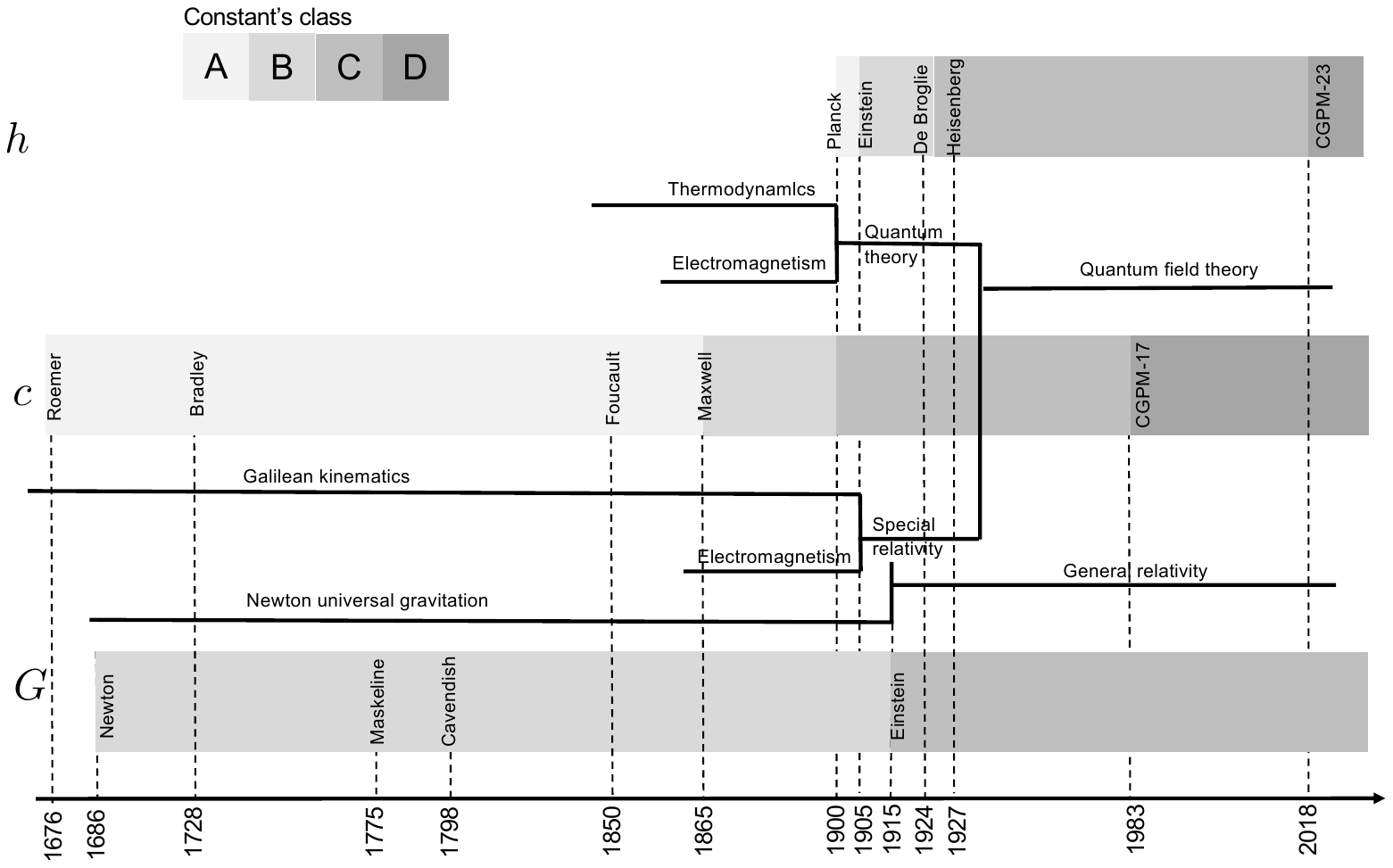}}
  \vskip-1.5cm
  \caption[Evolution of the status of $c$, $G$ and $h$]{(Over-simplified) timeline of the evolution of the status of the three fundamental constants $c,G$ and $h$ that are used to define the Planck units. Only the gravitational constant does not enter the definition of the International System of units. The changes of status of these three constants is related to unification between different physical theories and the emergence of new concepts such as \emph{spacetime} and \emph{wave function} that unified previous non-commensurable concepts (space and times, wave and particle) thanks to one constant. From \cite{ul-book}.}
  \label{fig1-classes}
\end{figure}

\paragraph{Relation with physical laws} \

As explained in the previous paragraph, \cite{levy79}  ranked the constants in terms of their universality and he proposed that only three constants be considered to be of class C, namely $G$, $\hbar$ and $c$. He pointed out two important roles of these constants in the laws of physics.

First, they act as \emph{concept synthesizer} during the process of our understanding of the laws of nature: contradictions between existing theories have
often been resolved by introducing new concepts that are more general or more synthetic than older ones. Constants build bridges between quantities that were thought to be incommensurable and thus allow for new concepts to emerge. For example the speed of light $c$ underpins the synthesis of space and time while the Planck constant $h$ allowed one to relate the concept of energy and frequency and the gravitational constant $G$ creates a link between matter and spacetime. 

Second, it follows that these constants are related  to the \emph{domains of validity} of these theories. For instance, as soon as velocity approaches $c$, relativistic effects become important and cannot be neglected. On the other hand, for speed much below $c$, Galilean kinematics is sufficient. Similarly the Planck constant signals when quantum theory has to be used since if the action $S$ of a system greatly exceeds the value $\hbar$ , classical mechanics will be appropriate to describe this system. 

Third, these two constants are also \emph{related to two universal principles}, i.e., that are independent on the exact form of the laws of nature: $c$ is related to causality and $\hbar$ to the quantum uncertainty principle. These two principles lie above the specific formulation of a precise field theory. No such principle associated to $G$ has been exhibited even though, within General Relativity, one can identify it to a measure of the universal elasticity of spacetime to any matter form. Fig~\ref{fig1-classes} summarizes the (simplified) evolution of the status of these 3 constants in regards of the evolution of theoretical physics theories.

Hence, while the place of $c$ (related to the notion of causality) and $\hbar$ (related to the quantum) in this list are well argued, the place of $G$ remains debated since it is thought that it will have to be replaced by some energy scale of a quantum theory of gravity.
\vfil

\paragraph{Evolutions} \

There are many ways the list of constants can change with our understanding of physics. 

First, \emph{new constants may appear} when new systems or new physical laws are discovered; this is, for instance, the case of the charge of the electron or more recently the gauge couplings of the nuclear interactions. 

A constant can also \emph{move from one class to a more universal} class. An example is that of the electric  charge, initially of class A (characteristic of the electron),  which then became class B when it was understood that it characterizes the strength of the electromagnetic interaction. 

A constant can also \emph{disappear} from the list, because it is either replaced by more fundamental constants (e.g., the Earth acceleration due to gravity and the proportionality constant entering Kepler law both disappeared because they were ``explained'' in terms of the Newton constant and the mass of the Earth or the Sun) or because it can happen that a better understanding of physics teaches us that two hitherto distinct quantities have to be considered as a single phenomenon (e.g., the understanding by Joule that heat and work were two forms of energy led to the fact that the Joule constant, expressing the proportionality between work and heat, lost any physical meaning and became a simple conversion factor between units used in the measurement of heat (calories) and work (Joule). Nowadays the calorie has fallen in disuse. 

Indeed demonstrating that a constant is varying will have direct implications on our list of constants since it would mean that it has to be replaced by a field, the dynamics of which will introduce new constant parameters that would need (\textit{1}) to be measurable and measured for the theory to be predictive and (\textit{2}) related to the former ``constants" it replaces is some limiting regime. This also means that the definitions of most dictionaries shall be modified since a constant cannot be defined by the property that ``it does not change with time'', that is by an empirical statement that shall be tested and as a consequence cannot be elevated to a category.

\paragraph{Conclusion} \

The evolution of the number, status of the constants teaches us a lot about the evolution of the ideas and theories in physics since they reflect the birth of new concepts, their  evolution and unification with other ones. Fundamental constants are both a door on new physics and a pedagogical way to illustrate the evolution of ideas in physics. The discussion of this paragraph highlights that they cannot be defined without a set of reference theories of which they are the free parameters that, indeed, cannot be determined by these theories and hence must be measurable and measured for it to be fully predictive. This discussion also illustrates that an inadequate definition may lead to the conclusion that the question ``Are the constants varying?" is a dull oxymoron while it is actually a deep question on the domain of validity of the theories that are used to describe nature. Constants stand at the crossroad between theoretical physics and history of science, offering a guideline to read the status and evolution of our physical theories. As we shall see, being among the most fundamental concepts of a theory, they have also modified the way we think physical units.

\subsubsection{Constants and metrology}\label{subsec12met}

Since we cannot compute them in the theoretical framework in which they appear, it is a crucial property of the fundamental constants (but in fact of all the constants) that their value can be measured. The relation between constants and metrology is a huge subject to which we just draw the attention on some selected aspects, specially in view of the redefinition of the International System of units that was adopted on November 16$^{\rm th}$ 2018, placing constants at the heart of metrology. For more discussions, see \cite{birge29} and \cite{karshen-metro,qed-karshen1}.

The introduction of constants in physical laws is closely related to the existence of systems of units. For instance, Newton's law states that the gravitational force between two masses is proportional to each mass and inversely proportional to the square of their separation. To transform the proportionality to an equality one requires the use of a quantity with dimension of $\mathrm{m^{3}\ kg^{-1}\ s^{-2}}$ independent of the separation between the two bodies, of their masses, of their composition (equivalence principle) and on the position (local position invariance). With an other system of units the numerical value of this constant could have simply been anything. Indeed, the numerical value of any constant crucially depends on the definition of the system of units.

\paragraph{The need for constants} \

For centuries, physical laws were expressed only with dimensional numbers. For instant, considering falling bodies from different heights, Galileo established a scaling relation between the height $h$ and duration $t$ of the fall,
$$
\frac{h_1}{h_2}=\left(\frac{t_1}{t_2}\right)^2\,,
$$
which implies no constant at all. This is a pure comparison between the evolution of two physical systems. Hence no need for units and constants. The drawback is that the comparison has to be performed locally. Hence, one can think of the introduction of a units system as the invention of some abstract reference physical system to which one can compare local experimental measurements. It follows that we would rewrite the previous law of falling bodies as
 $$
\frac{h}{1\,\unit{m}}\propto\left(\frac{t}{1\,\unit{s}}\right)^2\,,
$$
which is only a proportionality because in general a body does not fall from 1~m in 1~s, and then as
$$
h=\frac{1}{2} g t^2 
$$
with $g$ a constant of proportionality which shall be independent of $h$, $t$ and the mass, with dimension of an acceleration. It is clear that the units used to report the value of $g$ and to measure $h$ and $t$ have to be consistent so that the numerical value of $g$ depends in the choice of the abstract reference physical system (i.e., the units system) to which the experiment is compared.

Hence, the introduction of constants and the one of units are related and they allow for a globalisation of the comparison of experiments performed at different epoch and different laboratories: one shifted from a ``local equality" to a ``global proportionality" using arbitrary units and then a "global equality" requiring a proportionality constant. Indeed the system of units is completely arbitrary, and thus the numerical values of the dimensionful constants. We shall now describe in more details the various connections between constants and units.

\paragraph{Measuring constants} \

The determination of the laboratory value of constants relies mainly on the measurements of lengths, frequencies, times, \dots (see \citealt{petley85} for a treatise on the measurement of constants, and \cite{flower01} for a review). Hence, any question on the variation of constants is linked to the definition of the system of units and to the theory of measurement. The behavior of atomic matter is determined by the value of many constants. As a consequence, if, e.g., the fine-structure constant is spacetime dependent, the comparison between several devices such as clocks and rulers will also be spacetime dependent. This dependence will differ from one clock to another so that metrology becomes both \emph{device and spacetime dependent}, a property that will actually be used to construct tests of the constancy of the constants since it questions the reproductibility of experiments.

Indeed \emph{a measurement is always a comparison between two physical systems of the same dimensions}. This is thus a relative measurement, which gives as result a pure number. This trivial statement is oversimplifying since in order to compare two similar quantities measured separately, one actually needs to perform a number of comparisons. In order to reduce their number -- and in particular to avoid creating every time a chain of comparisons --, a certain set of them has been included in the definitions of units. Each unit can then be seen as an abstract physical system, which has to be realized effectively in the laboratory, and to which another physical system is compared. A measurement in terms of these units is usually called an absolute measurement. Most fundamental constants are related to microscopic physics and their numerical values can be obtained either from a pure microscopic comparison (as is, e.g., the case for $m_{\mathrm{e}}/m_{\mathrm{p}}$) or from a comparison between microscopic and macroscopic values (for instance to deduce the value of the mass of the electron in kilogram). This shows that the choice of the units has an impact on the accuracy of the measurement since the pure microscopic comparisons are in general more accurate than those involving macroscopic physics. \emph{This implies that only the variation of dimensionless constants can be measured and in case such a variation is detected, it is impossible to determine, which dimensional constant is varying} \citep{ellisu}.

It is also important to stress that in order to deduce the value of constants from an experiment, one usually needs to use theories and models. An example \citep{karshen-metro} is provided by the Rydberg constant. It can easily be expressed in terms of some fundamental constants as $R_\infty = {\aem^2m_{\mathrm{e}}c}/{2 h}$. It can be measured from, e.g., the triplet $1s-2s$ transition in hydrogen, the frequency of which is related to the Rydberg constant and other constants by assuming QED so that the accuracy of $R_\infty$  is much lower than that of the measurement of the transition. This could be solved  by defining $R_\infty$ as $4\nu_{\mathrm{H}}(1s-2s)/3c$ but then the relation with more fundamental constants would be much more complicated and actually not exactly known. This illustrates the relation between a practical and a fundamental approach and the limitation arising from the fact that we often cannot both exactly calculate and directly measure some quantity. Note also that some theoretical properties are plugged in the determination of the constants.

The laboratory determination of the fine structure constant $\aem$ arises from the comparison of theory and experiment for anomaly $a_{\rm e}=(|g_{\rm e}|-2)/2$  of the electron magnetic-moment  $\mu_{\rm e}=g_{\rm e}e s/2m_{\rm e}$, $s$ being its spin. It inherits contributions from the pure QED, predominantly electroweak, and predominantly hadronic  sectors \citep{NEW_Mohr:2015ccw}. The most accurate measurement of $a_{\rm e}$ was obtained from a single electron that was suspended for months at a time in a cylindrical Penning trap \citep{NEW_Hanneke:2008tm} leading to $\aem^{-1} = 137.035 999 084(51)$. It can also be obtained from the expression $\aem^2=2R_\infty(M/m_{\rm e})(h/M)/c$ \citep{NEW_Wicht_2002}. Using Bloch oscillations of $^{87}$Rb atoms in an optical lattice, \cite{NEW_Bouchendira:2013mpa} determined $h/M_{\rm Rb}$ and deduced $\aem^{-1} = 137.035 999 037(91)$. The agreement of these determinations of the fine structure constant is a strong validation of QED and of the standard model.

As a conclusion, let us recall that (\textit{i}) in general, the values of the constants are not determined by a direct measurement but by a chain involving both theoretical and experimental steps, (\textit{ii}) they depend on our theoretical understanding, (\textit{iii}) the determination of a self-consistent set of values of the fundamental constants results from an adjustment to achieve the best match between theory and a defined set of experiments (which is important because we actually know that the theories are only good approximation and have a domain of validity) (\textit{iv}) that the system of units plays a crucial role in the measurement chain, since for instance in atomic units, the mass of the electron could have been obtained directly from a mass ratio measurement (even more precise!) and (\textit{v}) fortunately the test of the variability of the constants does not require \textit{a priori} to have a high-precision value of the considered constants.

\paragraph{System of units}\ 

Thus, one needs to define a coherent system of units. This has a long, complex and fascinating history that was driven by  simplicity and universality but also by increasing stability and accuracy \citep{barrow-book, uzanleclercqbook,NEW_bipm,NEW_mesure,NEW_Tzalenchuk:2022yxv}. 

The international system of units (SI) has evolved to slowly replace physical artefacts by fundamental constants. This important story is summarized in Appendix~\ref{App0}. First, Sect.~\ref{app0-1} highlights the process in which it was understood that one shall look for the most fundamental physical objects to define units, opening a debate on the number of basic units; see Sect.~\ref{app0-2}. It follows that the definitions were driven by the will to use more stable and more fundamental quantities so that they closely follow the progress of physics and eventually fundamental constants. While the role of fundamental constants was clear in the construction of natural units described in Sect.~\ref{app0-3}, it increased to lead to the definition of the new (2018) SI system that is described in Sect.~\ref{app0-4}. 

The important point is that a, in principle arbitrary, choice of at least 3 units -- see however the discussion of  Sect.~\ref{app0-5} -- needs to be made in order to allow ease of comparison of global experimental results. But that choice has no impact on the more important question at the heart of this review that is on whether dimensionless constants de keep the same value over space and time.

\subsubsection{Definitions: \emph{fundamental parameters} and \emph{fundamental units}}\label{subsubsec213}

Once a set of three independent constants has been chosen as natural units, then all other constants are dimensionless quantities. The values of these combinations of constants does not depend on the way they are measured \citep{sansdim1, sansdim3, sansdim2,ellisu}, on the definition of the units etc.\dots. It follows that any variation of constants that will leave these numbers unaffected is actually just a redefinition of units.

 \begin{tcolorbox}
Hence, we shall split the constants into two sets respectively defined as (see Fig.\ref{fig2-constantes})
\begin{itemize}
\item \underline{\emph{fundamental units}}, the arbitrary set of dimensionful constants used to define a system of units. Generally, there are 3 fundamental units.
\item \underline{\emph{fundamental parameters}}, the set of dimensionless constants that remains. Changing their value will change the physics; the actual list depends on the laws chosen to describe nature and they can only be measured.
\end{itemize}
 \end{tcolorbox}
These latter dimensionless numbers represent, e.g., the mass ratio, relative magnitude of strength etc.\dots.  Changing their values will indeed have an impact on the intensity  of various physical phenomena, so that they encode some properties of our world. They have specific values (e.g., $\aem\sim1/137$, $m_{\mathrm{p}}/m_{\mathrm{e}} \sim1836$, etc.) that we may hope to understand. Are all these numbers completely contingent, or are some (why not all?) of them related by relations arising from some yet unknown and more fundamental theories. In such theories, some of these parameters may actually be dynamical quantities and, thus, vary in space and time. These are our potential varying constants.

In this review several fundamental parameters will be considered, namely the fine structure constant $\aem$ and the gravitational structure constant $\ag$, the electron-to proton mass ratio,\footnote{The introduction of $\mu$ and $\bar\mu$, while redundant, is due to the use of two conventions in the literature. In so doing, we stay close to the published results without having to change signs in the constraints.}
\begin{equation}\label{edef-mu}
\aem\equiv\frac{e^2}{4\pi\varepsilon_0\hbar x}, \qquad \ag\equiv\frac{G m_{\rm p}^2}{\hbar c},\qquad
\bar\mu\equiv \frac{m_{\rm e}}{m_{\rm p}}, \qquad \mu\equiv \frac{m_{\rm p}}{m_{\rm e}},
\end{equation}
the elementary masses to QCD scale ratio,
\begin{equation}\label{edef-X}
X_{\rm q}\equiv \frac{m_{\rm q}}{\Lambda_{\rm QCD}}, 
\end{equation}
for the light quark q of mass $m_{\rm q}=(m_{\rm u}+m_{\rm d})/2$, the quark s and the electron.

\begin{figure}[htbp]
  \centerline{\includegraphics[scale=0.35]{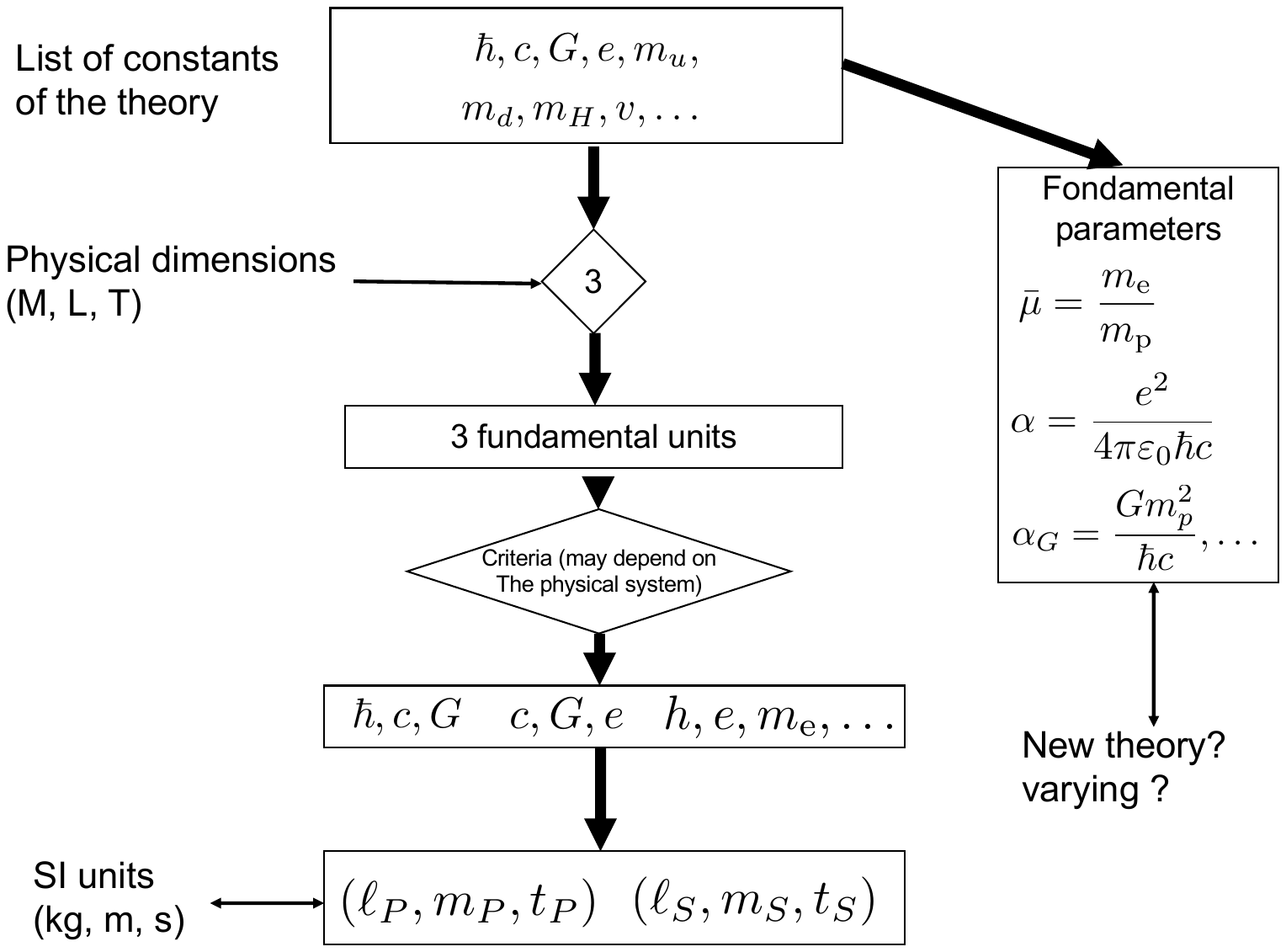}}
  \caption[Fundamental parameters and fundamental units]{Once a set of theories is assumed to describe nature, one can identify its free parameters to get a list of fundamental constants that can then be split in dimensionful \emph{fundamental units}, freely chosen to define a system of units, that can indeed be mapped to the SI units, and a list of dimensionless \emph{fundamental parameters} the value of which are independent of any units system. The latter define the physical properties of the world. One can question their being fine-tuned and their constancy. Translated from \cite{ul-book}.}
  \label{fig2-constantes}
\end{figure}

\subsection{The constancy of constants as a test of General Relativity}\label{subsec12}

The previous paragraphs have emphasize why testing for the consistency of the constants is a test of fundamental physics since it can reveal the need for new physical degrees of freedom in our theory. We now need to stress the relation with other tests of General Relativity and with cosmology.

\subsubsection{General Relativity}

The tests of the constancy of fundamental constants take all their importance in the realm of the tests of the equivalence principle \citep{will-book,NEW_Damour:2012rc}. Einstein General Relativity is based on two independent hypotheses, which can conveniently be described by decomposing the action of the  theory as $S = S_{\text{grav}}+S_{\text{matter}}$. We shall detail both aspects and highlight the connection between variation of the fundamental constants and the existence of a fifth force.

\paragraph{Equivalence principle}\ 

The equivalence principle has strong implication for the functional form of $S_{\text{matter}}$. This principle includes three hypotheses: 
\begin{itemize}
 \item the \emph{universality of free fall}  (UFF), that is the fact that the trajectory of any freely falling test-body in an external gravitational field does not depend on its internal structure, mass, and chemical composition. In Newtonian gravity, it leads to the requirement that the weight of a body is proportional to its inertial mass, i.e., that the inertial and gravitational masses are equal;
 \item the \emph{local position invariance} (LPI), that is the statement that any non-gravitational experiment is independent when and where it is performed;
 \item the \emph{local Lorentz invariance} (LLI). that is the statement that any non-gravitational experiment in a freely falling laboratory is independent of the velocity of its rest frame.
\end{itemize} 
In its weak form (that is for all interactions but gravity), it is satisfied by any metric theory of gravity and General Relativity is conjectured to satisfy it in its strong form (that is for all
interactions gravity included). We refer to \cite{will-book} for a detailed description of these principles. The weak equivalence principle (WEP) can be mathematically implemented by assuming that all matter fields are minimally coupled to a single metric tensor $g_{\mu\nu}$. This metric defines the length and times measured by laboratory clocks and rods so that it can be called the \emph{physical metric}. This implies that the action for any matter field, $\psi$ say, can be written as 
\begin{equation}\label{e.mincoup}
 S_{\text{matter}}(\psi,g_{\mu\nu}).
\end{equation}
This universal \emph{metric coupling} ensures in particular the validity of the universality of free-fall. Since locally, in the neighborhood of the worldline, there always exists a change of coordinates so that the metric takes a Minkowskian form at lowest order, the gravitational field can be locally``effaced''  (up to tidal effects). If we identify this neighborhood to a small lab, this means that any physical properties that can be measured in this lab must be independent of where and when the experiments are carried out. This is indeed the assumption of \emph{local position invariance}, which implies that the constants must take the same value independent of the spacetime point where they are measured. 

Thus, testing the constancy of fundamental constants is a direct test of this principle and therefore of the metric coupling. Interestingly, the tests we are discussing in this review allow one to extend them much further than Solar system scales and even in the early universe, an important information to check the validity of relativity in cosmology.

\paragraph{Test-body in free fall}\ 

The action of a point-body subjected to no other force than gravity reads
\begin{equation}
\label{Spp}
  S_{\text{matter}} = -\int mc\sqrt{-g_{\mu\nu}(x^\alpha) v^\mu v^\nu}\dd t,
\end{equation}
with $v^\mu\equiv\dd x^\mu/\dd t$. The equation of motion that one derives from this action is the usual geodesic equation
\begin{equation}
\label{geo1}
 a^\mu \equiv u^\nu\nabla_\nu u^\mu = 0,
\end{equation}
where $u^\mu= \dd x^\mu/c\dd\tau$, $\tau$ being the proper time; $\nabla_\mu$ is the covariant derivative associated with the metric $g_{\mu\nu}$ and $a^\nu$ is the 4-acceleration. Hence, any test particle in free fall follows a geodesic independent of its mass and chemical composition, as anticipated from the property~(\ref{e.mincoup}) that translates the UFF. Again, this property relies only on the universal minimal coupling in the matter action, independently of the field equations for the spacetime metric.

Any metric theory of gravity enjoys such a matter Lagrangian and the worldline of any test particle shall be a  geodesic of the spacetime with metric $g_{\mu\nu}$, as long as there is no other long range force acting on it (see \cite{gefmotion} for a detailed review of motion in alternative theories of gravity). In the weak field limit $g_{00}=-1-2U/c^2$ where $U$ is the gravitational potential. Note that $U$ may differ from the Newtonian potential $\Phi_{\rm N}$. It follows from Eq.~(\ref{geo1}) that, in the slow velocity limit, the geodesic equation reduces to
\begin{equation}\label{eq.inertie}
 \dot{\mathbf{v}} = \mathbf{a} = -\nabla U \equiv \mathbf{g}, 
\end{equation}
hence defining the acceleration $\mathbf{g}$. If the the metric theory reduces to General Relativity then $U=\Phi_{\rm N}$ and $\mathbf{g}_N=-\nabla\Phi_{\rm N}$. Recall that the proper time of a clock is related to the coordinate time by $\dd\tau = \sqrt{-g_{00}}\dd t$. Thus, if one exchanges electromagnetic signals between two identical clocks in a stationary situation, the apparent difference between their frequencies is
\begin{equation}\label{e.redshift}
 \frac{\nu_1}{\nu_2} = 1 + \frac{U(2)-U(1)}{c^2},
 \end{equation}
at lowest order. This is the \emph{universality} of gravitational redshift, showing a structural relation between the redshift and the free-fall motion~(\ref{eq.inertie}) of massive particles.

\paragraph{Tests of the metric coupling}\ 

The assumption of a metric coupling is actually well tested in the Solar system through 4 types of experiments; see \ \cite{will-book,will-llr}  for a detailed discussion and \cite{NEW_Peik:2020cwm,NEW_Herrmann:2023rzw} for recent developments on quantum tests of relativity.
\begin{itemize}
\item First, the local position invariance implies that all non-gravitational constants are spacetime independent, which have been tested to a very high accuracy in many physical systems and for various fundamental constants; this is the subject of the present review. 
\item Second, the isotropy of space has been tested from the constraint on the possible quadrupolar shift of nuclear energy levels \citep{isotes3,isotest2,isotest} proving that different matter fields couple to a unique metric tensor at the $10^{-27}$ level. 
\item Third, the Einstein effect (or gravitational redshift) given by Eq.~(\ref{e.redshift}) shall be compatible with the potential $U$ given by the equation of motion~(\ref{eq.inertie}). This has been  tested at the $2 \times 10^{-4}$ level \citep{clock1}. It is standard to express the redshift $z_{12}$ between two spacetime positions as
\begin{equation}\label{e.zU}
z_{12}= (1+\gamma)\Delta_{12} U
\end{equation}
where, following \cite{will-book}, $U$ is the potential the gradient of which determines  in the Newtonian limit the acceleration, i.e.,  ${\bf a}=-\nabla U$, of a free particle. Testing the consistency of $\gamma$ with 0 is a
test of LPI. Indeed in General Relativity $U=\Phi_{\rm N}$ and  $\gamma=0$; see Eq.~(\ref{e.redshift}). In alternative theories, one shall identify $U$ and the redshift independently and how they can be measured. As an example, scalar-tensor theories discussed in Sect.~\ref{subsecST} enjoy $\gamma=0$. The parameter is constrained at the $10^{-5}$ level \citep{NEW_Delva:2018ilu,PhysRevLett.121.231102}.
\item Fourth, the universality of free fall can be tested by comparing the accelerations of two test bodies in an external gravitational field.  We shall review its status in the next paragraph.
\end{itemize}
This set of experiments lead us to conclude that the hypothesis of metric coupling is extremely well-tested in the Solar system.

\paragraph{Tests of the universality of free fall}\ 

The test of the universality as a long history, briefly summarized on Fig.~\ref{fig3-uffhisto}. The deviations from UFF are characterized by the E\"otv\"os parameter $\eta_{12}$ defined as
\begin{equation}
 \eta_{12}\equiv 2 \frac{|\mathbf{a}_1-\mathbf{a}_2|}{|\mathbf{a}_1+\mathbf{a}_2|},
\end{equation}
that shall be compatible with 0 as long as UFF holds.

 Indeed, in the Newtonian limit, the Einstein equations and the equation of free fall reduce to the Newton second law and the expression for the weight with the same mass. Nevertheless, contrary to General Relativity, in which UFF is hardwired, one can introduce two notions of masses, the inertial mass $m_{\rm I}$ and the gravitational mass\footnote{We recall that actually one shall distinguish between the \emph{passive gravitational masse} that characterizes the response of a body to the gravitational action of external objects, i.e., defining its weight, that is, the gravitational force $m_{\rm G}{\bf g}$ acting on the body and the \emph{active gravitational mass} that characterizes the object which creates a gravitational field, i.e., its ‘gravitational charge’. Thanks to Newton’s third law of equal action and reaction, one can always choose these two gravitational masses to be equal.} $m_{\rm G}$ so that the free fall in an external gravitational field is dictated by
$$
m_{\rm I} {\bf a} = {\bf P}, \qquad {\bf P}=m_{\rm G} {\bf g}.
$$
Clearly if the UFF holds, then $\left(m_{\rm I}/m_{\rm G}\right)_1 = \left( m_{\rm I}/m_{\rm G}\right)_2 $ so that the inertial and gravitational masses can be chosen equal, from which it follows trivially that
$$
\eta_{12}=2\frac{\left\vert m_{\rm G}^{(1)}/m_{\rm I}^{(1)} - m_{\rm G}^{(2)}/m_{\rm I}^{(2)} \right\vert}{\left\vert m_{\rm G}^{(1)}/m_{\rm I}^{(1)} + m_{\rm G}^{(2)}/m_{\rm I}^{(2)} \right\vert}\,.
$$
It is an \emph{empirical fact}, established by Galileo (in his experiments on inclined planes rather than the probably apocryphal experiment in the Leaning Tower of Pisa), that in the absence of friction, all objects, no matter what their inertial mass, or the nature of their constituents, or the internal energy or cohesive forces of their constituents, fall in the same way in an external gravitational field (in contrast to, for example, the behavior in an electric field of two individual charges of opposite sign and of their neutral ensemble). This principle was stated in the introduction of its \textit{Principia} by Newton and was then one of the cornerstone of General Relativity. Newton manages to test it using pendulums, using the fact that the equation of motion actually becomes
$$
\ddot\theta + \omega^2 \theta=0 \quad\hbox{with}\quad  \omega=\omega_0\sqrt{\frac{m_{\rm G}}{m_{\rm I}}}\quad \hbox{and}\quad  \omega_0=\sqrt{\frac{g}{L}}
$$
so that comparing the oscillations of two different bodies suspended by cables of the same length, one gets $\eta_{12}=2\vert \omega_1-\omega_2\vert/\omega_0$. This method was invented by Galileo, who estimated that $\eta_{12} < 1\%$. Newton repeated the experiment taking into account air resistance to reach an accuracy of $10^{-3}$. In 1827 Bessel reached an accuracy of $10^{-5}$. Then, the tests used torsion balances, as initiated in 1922 by E\"otv\"os to get to a constraint of  $10^{-8}$. Finally, Lunar laser ranging, orbitography and dedicated free fall experiments in orbit manage to reach the higher constraints described below.

\begin{figure}[htbp]
   \vskip-1cm
  \centerline{\includegraphics[scale=0.5]{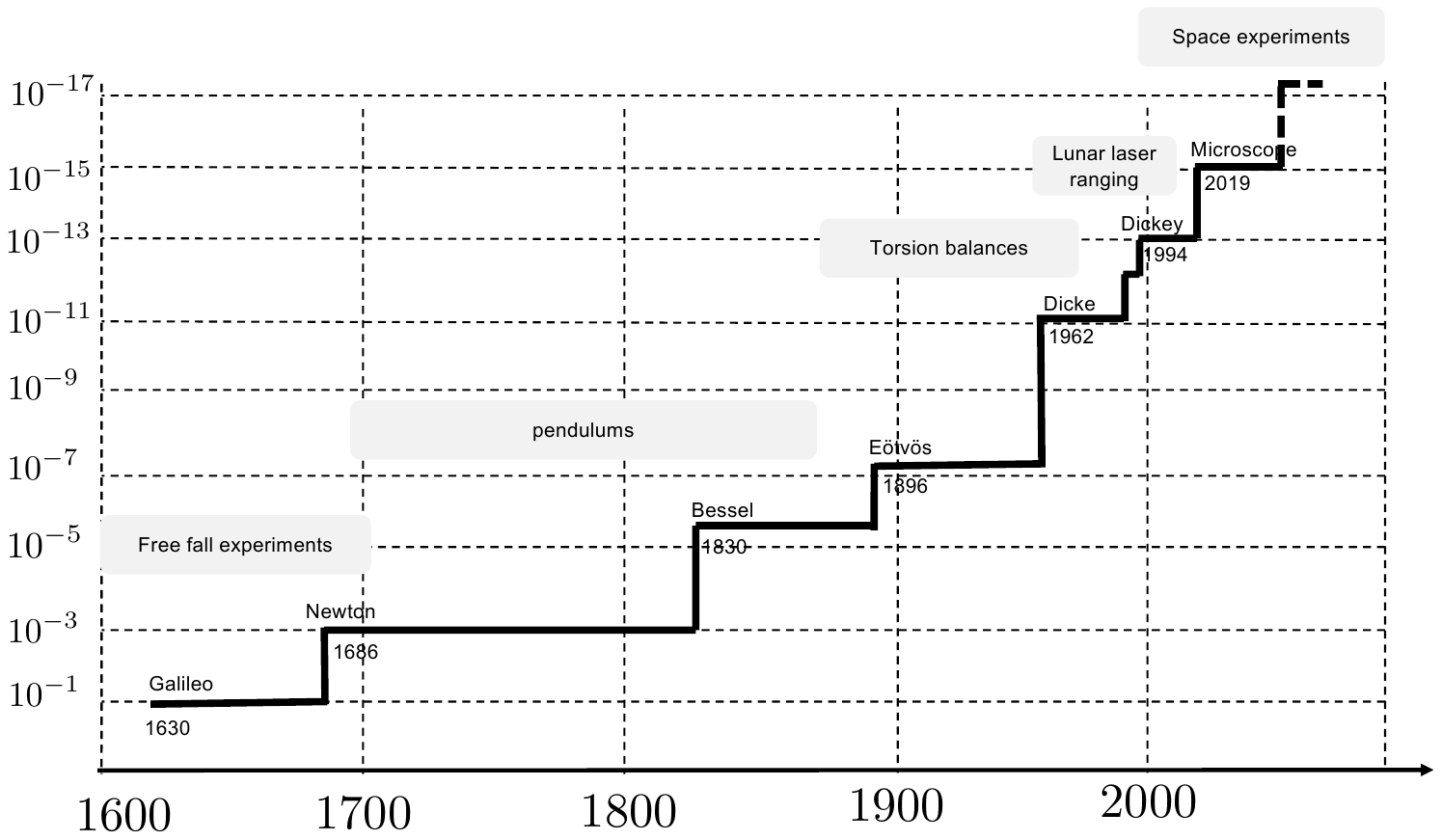}}
    \vskip-1cm
  \caption[Timeline of the tests of the universality of free fall]{The universality of free fall has been a red thread in the understanding of gravitation since Galileo. This explains that is has been testes with higher and higher accuracy with different methods, starting from the study of the universality of free fall, comparison of pendulums with different chemical compositions, torsion balances, Lunar laser ranging and space-designed experiments of free fall in Earth orbit. Adapted from \citealt{ul-book}.}
  \label{fig3-uffhisto}
\end{figure}

As seen from Fig.~\ref{fig3-uffhisto}, the universality of free fall has been confirmed by all experiments so far, with increasing accuracy. Still it remains an \emph{empirical principle} that needs to be better understood. As we shall see later, many theories beyond General Relativity describing varying constant lead to the existence of a fifth force and thus to a violation of the UFF. Note also  that if UFF did not hold, one would have to take extra-care in the measurement of masses and definition of the kilogram.

\begin{table}[t]
\caption[Bounds on the E\"otv\"os parameter $\eta$]{Summary of the constraints on the violation of the universality of free fall. The bounds on the E\"otv\"os parameter are given with a 1$\sigma$ uncertainty. When no mention is added, the two test has been performed in Earth gravitational field unless $[\odot]$ is indicated to specify that it refers to the Sun gravitational field. The top-half of the table concerns macroscopic bodies while the bottom-half summarizes the main tests with atom interferometry.}
\label{tab-eta12}
\centering
{\small
\begin{tabular}{p{3.9cm}ccr}
 \toprule
 Constraint  & Body 1 & Body 2  & Ref. \\
 \midrule
   $(-0.2 \pm 2.8) \times 10^{-12}$ & Be      & Al & \cite{suetal}\\
  $(-1.9 \pm 2.5) \times 10^{-12}$ & Be      & Cu & \cite{suetal}\\
  $(5.1 \pm 6.7) \times 10^{-12}$   & Si/Al   & Cu & \cite{suetal}\\
    $(0.1 \pm 2.7 \pm 1.7) \times 10^{-13}$ & Earth-like rock & Moon-like rock & \cite{uff1} \\
      $(1.1\pm3.0)\times10^{-9}$ & Cu-Pt & U-238 & \cite{Smith:1999cr}\\
   $(-1.9 \pm 2.5) \times 10^{-12}$ & Be      & Cu & \cite{uff2}\\
  $(-1.0 \pm 1.4) \times 10^{-13} \, \, [\odot]$& Earth   & Moon &\cite{uff3} \\
  $(0.3 \pm 1.8) \times 10^{-13}$   & Be      & Ti & \cite{tebi}\\
    $(-3.1 \pm 4.7) \times 10^{-13}\, \, [\odot]$   & Be      & Ti & \cite{tebi}\\
  $(-0.7 \pm1.3)\times 10^{-13}$ & Be & Al &\cite{NEW_Wagner:2012ui}\\
    $(-5.3 \pm4.0)\times 10^{-13}\,\,  [\odot]$ & Be & Al &\cite{NEW_Wagner:2012ui}\\
    $(-0.8 \pm 1.3) \times 10^{-13}\,\,  [\odot]$ & Earth   & Moon &\cite{NEW_Williams:2012nc} \\
  $(-1.5\pm2.3\pm1.5_{\rm syst})\times 10^{-15}$ &  Ti & Pt & \cite{NEW_MICROSCOPE:2022doy} \\
\midrule
 $(0.4\pm1.2)\times10^{-7} $ & Rb-85 (F=2),  & Rb-85 (F=3)& \cite{NEW_PhysRevLett.93.240404}\\
$(0.3\pm 5.4)\times10^{-7}$  & Rb-87  &K-39  &\cite{NEW_PhysRevLett.112.203002}  \\
$(0.2\pm1.2)\times10^{-7} $ & Rb-85 ($m_F=+1$),  & Rb-85 ($m_F=-1$)  & \cite{NEW_PhysRevLett.117.023001}\\
$(1.9\pm 3.2)\times10^{-7}$  & Rb-87  &K-39  &\cite{NEW_Albers2020}\\
 $(1.6\pm1.8_{\rm stat} \pm 3.4_{\rm syst})\times10^{-12}$ &  Rb-87 & Rb-85 & \cite{NEW_PhysRevLett.125.191101} \\
 $(0.8\pm1.4)\times10^{-10} $ & Rb-87 & Rb-85 &  \cite{PhysRevA.104.022822}\\
$(0.9\pm2.9)\times10^{-7} $ & Rb-85 (F=1),  & Rb-85 (F=2)  &\cite{NEW_Xu:2022xzs}\\
  \bottomrule
\end{tabular}
}
\end{table}

First, the UFF has been constrained experimentally, e.g., in the laboratory by comparing the acceleration of a beryllium and a copper masses in the Earth gravitational field \citep{uff2} to get $\eta_{\mathrm{Be,Cu}} = (-1.9\pm2.5)\times10^{-12}$. Similarly the comparison of Earth-core-like and Moon-mantle-like bodies gave \citep{uff1}  $\eta_{\text{Earth,Moon}} = (0.1\pm2.7\pm1.7)\times10^{-13}$, and experiments with torsion balance using beryllium-aluminum and beryllium-titanium test body pairs  set the constraints \citep{tebi,NEW_Wagner:2012ui} $\eta_{\mathrm{Be,Ti}} = (0.3 \pm1.8)\times 10^{-13}$ and $\eta_{\mathrm{Be,Al}} = (-0.7 \pm1.3)\times 10^{-13}$.

While free fall experiments can be performed on Earth, they are limited by the duration of the fall and are limited by errors in initial conditions at release coupling to the gravity gradient of the Earth, they allow one to reach a precision of order $10^{-10}$. High drop towers such as the ZARM tower of the University of Bremen allows for free-fall times up to $9.3$~s. Indeed, space allows  for much longer free falls with a similar acceleration in low-Earth orbits. This led to the MICROSCOPE experiment \citep{NEW_Touboul:2017grn,NEW_MICROSCOPE:2019jix,NEW_MICROSCOPE:2022doy}  that, after two-years mission from 2016 to 2018, provided a sharp constraint on the E\"otv\"os parameter between one proof mass made of titanium and another made of platinum,
\begin{equation}\label{eta-microscope}
 \eta_{\mathrm{Ti, Pt}} = (-1.5\pm2.3\pm1.5_{\rm syst})\times 10^{-15},
\end{equation}
where the statistical error is 1$\sigma$. Besides, the reference instrument provided a null result, $\eta_{\mathrm{Pt, Pt}} = [0.0\pm 1.1 \pm 2.3_{\rm syst})]\times 10^{-15}$  showing no sign of unaccounted systematics. We refer to \cite{NEW_Berge:2023sqt} for an up-to-date review of the MICROSCOPE experiment, its results and implications for theories of gravity beyond General Relativity.

The Lunar Laser ranging experiment \citep{uff3}, compares the relative acceleration of the Earth and Moon in the gravitational field of the Sun. It sets the constraint $\eta_{\text{Earth,Moon}} = (-1.0\pm1.4)\times10^{-13}$,
slightly reevaluated to $\eta_{\text{Earth,Moon}} = (-0.8\pm1.3)\times10^{-13}$ \citep{NEW_Williams:2012nc}. Note that since the core represents only 1/3 of the mass of the Earth, and since the Earth's mantle has the same composition as that of the Moon (and thus shall fall in the same way), one loses a factor of three, so that this constraint is actually similar to the one obtained in the lab. Further constraints are summarized in Table~\ref{tab-eta12}. The latter constraint also contains some contribution from the gravitational binding energy and thus includes the strong equivalence principle. When the laboratory result of \cite{uff1} is combined with the LLR results of \cite{G-llr0} and \cite{muller91}, one gets a constraints on the strong equivalence principle parameter, respectively $\eta_{\text{SEP}} = (3\pm6)\times10^{-13}$ and $\eta_{\text{SEP}} = (-4\pm5)\times10^{-13}$ -- see \cite{NEW_Merkowitz:2010kka} for a review. While similar analysis with satellite laser ranging have reached $\eta\sim 2\times10^{-9}$ for LAGEOS and LAGEOS~II, they remain not competitive with LLR \citep{NEW_nobili2008}.

Thanks to the advent of atom interferometry \citep{NEW_RevModPhys.81.1051}, one can now use atoms, neutrons or charged particles as test masses , allowing one for a better control of systematic effects in free-fall experiments. Atoms have well-known and reproducible properties and it is possible to make very small atomic probes, the position of which can be very precisely controlled, and the possibility to use different atomic states and isotopes allow for the rejection of systematic errors.  Using cold rubidium atoms in an atomic fountain interferometer, \cite{NEW_PhysRevLett.93.240404} compared the free fall of Rb-85 and Rb-87 to reach $\eta=(1.2\pm1.7)\times10^{-7}$ while the comparison between the  two different hyperfine ground states $F = 2, m_F = 0$ and $F = 3, m_F = 0$ of Rb-85 concluded that $\eta_{85 F=2, F=3}=(0.4\pm1.2)\times10^{-7}$  while the comparison of $m_F=+1$ and $m_F=-1$ of Rb-87 \citep{NEW_PhysRevLett.117.023001} gave $\eta=(0.2\pm1.2)\times10^{-7}$. \cite{NEW_PhysRevA.88.043615} reached $\eta=(1.2\pm3.2)\times10^{-7}$ for Rb-87 vs Rb-88 while the comparison of Rb-87 to K-39 concluded that  $\eta=(0.3\pm 5.4)\times10^{-7}$ \citep{NEW_PhysRevLett.112.203002} and $\eta=(0.2\pm 1.6)\times10^{-7}$ for the comparison of Sr-88 (no spin) and Sr-87  (half integer spin) \citep{NEW_PhysRevLett.113.023005}. Rb-85/87 reached the level $10^{-9}$ \citep{NEW_PhysRevA.92.023626} and then \cite{NEW_PhysRevLett.125.191101} obtained
\begin{equation}
\eta=[1.6\pm1.8_{\rm stat} \pm 3.4_{\rm syst}]\times10^{-12}\,.
\end{equation}
We refer to \cite{NEW_Yuan:2023evh} for a dedicated review on the quantum tests of the WEP with cold atom interferometry. These experiments are expected to reach higher precision in space and could test the WEP with ani-matter \citep{NEW_Charlton:2020xxc,NEW_Rousselle_2022}.

These constraints are summarized in Table~\ref{tab-eta12}. While they show no sign of violation of the UFF, they require physical models to be interpreted and related to the variation of the fundamental constants, as we shall describe.

\paragraph{Dynamics}\ 

The second building block of General Relativity  concerns the dynamics of the gravitational sector, assumed to be described by the Einstein--Hilbert action
\begin{equation}
\label{Sgrav}
 S_{\text{grav}} = \frac{c^3}{16\pi G}\int\sqrt{-g_*}R_*\dd^4x.
\end{equation}
This defines the dynamics of a massless spin-2 field $g^*_{\mu\nu}$, called the Einstein metric. General Relativity then assumes that both  metrics coincide, 
\begin{equation}\label{e.SEP-RG}
g_{\mu\nu}=g^*_{\mu\nu}
\end{equation}
(which is related to the strong equivalence principle), but it is possible to design theories in which this is indeed not the case (see the example of scalar-tensor theories below; Sect.~\ref{subsecST}) so that General Relativity is one out of a large family of metric theories.

The variation of the total action with respect to the metric yields the Einstein field equations
\begin{equation}
 R_{\mu\nu}-\frac{1}{2}R g_{\mu\nu} = \frac{8\pi G}{c^4}T_{\mu\nu},
\end{equation}
where $T^{\mu\nu}\equiv (2/\sqrt{-g})\delta S_{\text{matter}}/\delta g_{\mu\nu}$ is the matter stress-energy tensor. The coefficient 
\begin{equation}\label{def.kappa}
\kappa=\frac{8\pi G}{c^4}
\end{equation} 
is determined by the weak-field limit of the theory that should reproduce the Newtonian predictions. In this regime, the Einstein equations reduce to the Poisson equation
$$
\Delta\Phi_{\rm N} = 4\pi G \rho
$$
defining the Newtonian potential. 

To grasp the lowest order relativistic effects in the weak field limit, and in order to test the dynamics of General Relativity in the Solar system, one can rely on the parameterized post-Newtonian formalism (PPN). Its  is a general formalism that introduces 10 phenomenological parameters to describe any possible deviation from General Relativity at the first post-Newtonian order \citep{will-book,will-llr}; see also \citealt{blanchet-lrr} for a review on higher orders. The formalism assumes that gravity is described by a metric and that it does not involve any characteristic scale. In the simplest case and very special case of a single gravitating body, it reduces to the two Eddington parameters introduced in the Schwarzschild metric in isotropic coordinates as
$$
 g_{00} = - 1 + 2 \frac{Gm}{rc^2} -
 2\beta^\ppn\left(\frac{Gm}{rc^2}\right)^2,
 \qquad
 g_{ij} = \left(1+2\gamma^\ppn\frac{Gm}{rc^2}\right)\delta_{ij}.
$$
Indeed, General Relativity predicts $\beta^\ppn=\gamma^\ppn=1$. These two phenomenological parameters are constrained (1) by the shift of the Mercury perihelion \citep{shapiro90}, which implies that $|2\gamma^\ppn-\beta^\ppn-1|<3\times10^{-3}$, (2) the Lunar laser ranging experiments \citep{uff3}, which implies that $|4\beta^\ppn-\gamma^\ppn-3|=(4.4\pm4.5)\times10^{-4}$ and (3) by the deflection of electromagnetic signals, which are all controlled by $\gamma^\ppn$. For instance the very long baseline interferometry \citep{vlbi} implies that $|\gamma^\ppn-1|=4\times10^{-4}$, while the measurement of the time delay variation to the Cassini spacecraft \citep{cassini} sets $\gamma^\ppn-1=(2.1\pm2.3)\times10^{-5}$.

\paragraph{Phenomenological modifications of Newton gravity}\ 

The PPN formalism does not allow one to test finite range effects that could be caused, e.g., by a massive degree of freedom. In that case one expects a Yukawa-type deviation from the Newton potential. The total gravitational potential created by a point-mass of mass $m$ at distance $r$ is parameterized as
\begin{equation}\label{e.yukPot}
 V=-\frac{Gm}{r}\left(1+\alpha\hbox{e}^{-r/\lambda}\right),
\end{equation}
a form that will be physically justified later, where $\alpha$ is the strength of the deviation compared to Newtonian gravity (which may depend on the composition of the interacting masses) and $\lambda$ is the range of the corresponding. For macroscopic extended bodies, the potential is modified by a shape factor arising from the integration of the point-particle potential. As we shall see later, such a potential corresponds to the static limit of an interaction mediated by virtual bosons of mass $m_\phi=\hbar/\lambda c$.

The constraints on $\eta$ from Table~\ref{tab-eta12} can be translated on constraints in the $(\lambda,\alpha)$-plane. For each experiment, one first needs to compute the shape factor to take into account that all element of the body do not contribute  similarly to the  macroscopic Yukawa interaction. For instance, a homogenous sphere of radius $R$ has a shape factor
\begin{equation}\label{e.shapePhi}
\Phi(x) \equiv 3(x \cosh x - \sinh x)/x^3
\end{equation} 
with $x=R/\lambda$, which indeed tends to 1 when $\lambda\rightarrow+\infty$ since in this limit the Gauss theorem is recovered. Then, the new interaction may be composition dependent so that one shall replace $\alpha$ in Eq.~(\ref{e.yukPot}) by $\alpha_{ij}$ that depends on the dimensionless Yukawa charge $q$ characteristic of each material
\begin{equation}
\alpha_{ij}=\alpha\left(\frac{q}{\mu}\right)_i\left(\frac{q}{\mu}\right)_j
\end{equation}
with $\mu$ the atomic mass in atomic units (e.g., $\mu=12$ for carbon-12, or $\mu=47.948$ for titanium).  The definition of $q$ depends on the microphysics of the coupling of the new massive degree of freedom to standard matter fields; see Sect.~\ref{subsecUFF} below for explicit examples. Taking into account the electromagnetic and nuclear binding energies, the charge usually reduces to the baryon or lepton numbers ($B$ and $L$) of the test mass materials; see e.g., \cite{NEW_Fayet:1990wx} and Sect.~\ref{subsecUFF} below. While purely phenomenological, this description is useful since it describes the macroscopic fifth force induced by a massive scalar field in the Newtonian regime. It follows that the E\"otv\"os parameter due to a Yukawa potential between 2 bodies in free-fall around the Earth is
\begin{equation} \label{eq_eta}
\eta = \alpha \left[ \left(\frac{q}{\mu}\right)_i - \left(\frac{q}{\mu}\right)_j\right]\left(\frac{q}{\mu}\right)_E \Phi\left(\frac{R_E}{\lambda} \right)\left( 1 + \frac{r}{\lambda} \right) \hbox{e}^{-\frac{r}{\lambda}}
\end{equation}
($\Phi=1$ for the test masses since their sizes are much smaller than $\lambda$) with $r=R_E+h$, $h$ being the altitude of the satellite.

Figure~\ref{fig-Yuk} summarizes the constraints obtained thanks to the MICROSCOPE experiments \citep{NEW_Berge:2017ovy}. The constraints on $(\lambda,\alpha)$ are summarized in \cite{5force,NEW_Berge:2023sqt}, which typically shows that $\alpha<10^{-2}$ on scales ranging from the millimeter to the Solar system size. One shall however be careful with these constraints since geodesy experiments shall also take into account the shape of the Earth while many analysis assume a spherical Earth; see \cite{NEW_Berge:2018htm} for a discussion of the impact of the shape of the Earth when constraining gravity and the geoid simultaneously.
 
\begin{figure}[htbp]
  \centerline{\includegraphics[scale=0.3]{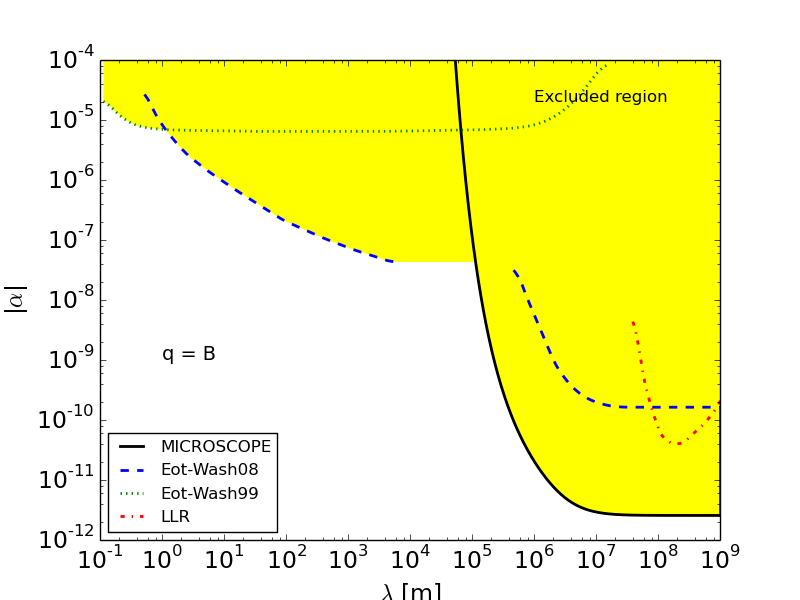} \includegraphics[scale=0.3]{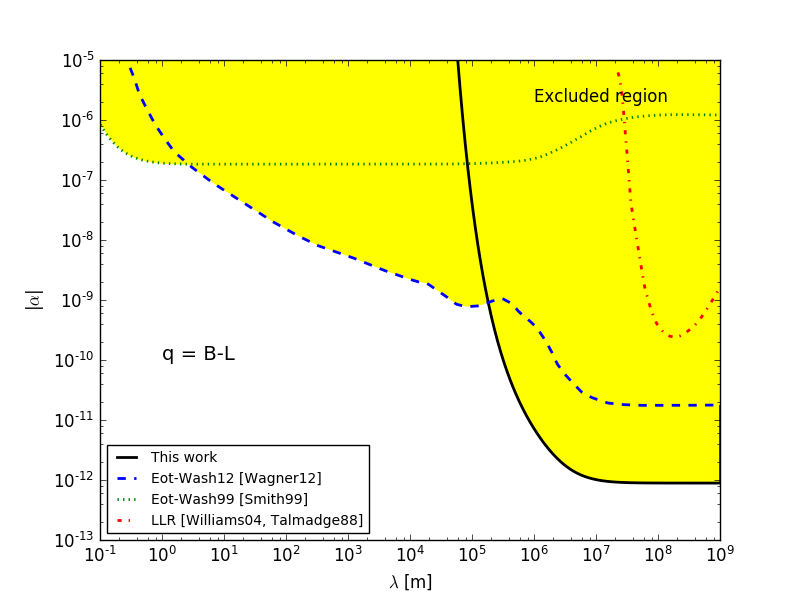}}
  \caption[Constraints on the Yukawa potential parameters $(\alpha,\lambda)$ obtained by MICROSCOPE ]{Constraints on the Yukawa potential parameters $(\alpha,\lambda)$ assuming $q=B$ (left) or $q=B-L$ (right) obtained by MICROSCOPE and compared to previous constraints (dotted: \cite{Smith:1999cr}, Dashed: \cite{Schlamminger:2007ht}, Dot-dahed: \cite{Talmadge:1988qz}). From  \cite{NEW_Berge:2017ovy}.}
  \label{fig-Yuk}
\end{figure}

\paragraph{Conclusions}\ 

General Relativity is our reference theory to describe gravitation, that is the long range non-screened interaction. As we have seen the UFF is extremely well-tested in the Solar system and complemented by the test of the LPI described in this review. Besides, it is also tested  with pulsars \citep{gefpul2,gefpul} in the strong field regime \citep{psaltis} and, more recently thanks to the detection of gravitational waves. For more details we refer to \cite{lilley,tury,will-book,will-llr}. Needless to say that any extension of General Relativity has to pass these constraints. However, deviations from General Relativity can be larger in the past, as we shall see, which makes cosmology an interesting physical system to extend these constraints in spacetime but also to other matter component such as the dark matter that cannot be used, for now, in laboratory experiments.

\subsubsection{Varying constants and free fall: the link between LPI and UFF}\label{subsecUFF}

As the previous discussion shows, the constancy of the fundamental constants and the universality are two pillars of the equivalence principle. \cite{dicke64} realized that they are actually not independent and that if the coupling constants are spatially dependent then this will induce a violation of the universality of free fall.

\paragraph{Fifth force}\ 

The connection lies in the fact that the mass of any composite body, starting, e.g., from neutrons and protons and then nuclei,  depends on the mass of the elementary particles that constitute it (this means that it will depend on the various Yukawa couplings and on the Higgs sector parameters) but also a contribution, $E_{\text{binding}}/c^2$, arising from the binding energies of the different interactions (i.e., strong, weak and electromagnetic) but also gravitational for massive bodies. Thus, the mass of any body is a function of all the constants, $m[\alpha_i]$.

It follows that the action for a point particle is no more given by Eq.~(\ref{Spp}) but by
\begin{equation}
\label{Sppvar}
  S_{\text{matter}} = -\int m_A[\alpha_j]c\sqrt{-g_{\mu\nu}(x^\alpha) v^\mu v^\nu}\dd t,
\end{equation}
where $\alpha_j$ stands for the list of constants on which $m_A$ depends. It includes $\aem$ but also many others constants. The index $A$ in $m_A$ recalls that the dependence in these constants is a priori different for bodies of different chemical composition. The variation of this action gives the equation of motion
\begin{equation}\label{e.fifthgeo}
 u^\nu\nabla_\nu u^\mu = - \left(\sum_i\frac{\partial \ln m_A}{\partial \alpha_i} 
 \frac{\partial \alpha_i}{\partial x^\beta} \right)
 \left( g^{\beta\mu}+u^\beta u^\mu\right).
\end{equation}
It departs from a pure geodesic motion due to the appearance of a fifth force
\begin{equation}
 F_5^\mu \equiv - \left(\sum_i\frac{\partial \ln m_A}{\partial \alpha_i} 
 \frac{\partial \alpha_i}{\partial x^\beta} \right)
 \left( g^{\beta\mu}+u^\beta u^\mu\right)
\end{equation}
that is, thanks to the projection operator $ \left( g^{\beta\mu}+u^\beta u^\mu\right)$, purely spatial, i.e., $ F_5^\mu u_\mu=0$. The first parenthesis contains the sensitivities of the mass on a variation of the various constants $\alpha_i$,
\begin{equation}
\label{deffai0}
 f_{A,i} \equiv \frac{\partial\ln m_A}{\partial\alpha_i}.
\end{equation}
To finish, the force is non-vanishing only if the constants enjoy a 4-gradient, $\partial \alpha_i/\partial x^\beta\not=0$, that is if at least one of them varies in space and/or time.

The expression~(\ref{e.fifthgeo}) provides the link between variation of fundamental constants and the universality of free fall.  This simple argument shows that if the constants depend on time then there must exist an anomalous acceleration that will depend on the chemical composition of the body $A$. It implies that, given a microscopic physical model to compute the sensitivities $f_{A,i}$, any constraint on the UFF can be translated on a constraint on  $\partial \alpha_i / \partial x^\beta$. Similarly, any model with varying constants e.g., on cosmological scales, will inevitably induce a violation of UFF in the Solar system, which creates a strong link between cosmological and local scales. Concrete examples of such fifth force will be  discussed in Sect.~\ref{section-theories}.

\paragraph{Newtonian limit}\ 

In the Newtonian limit $g_{00}=-1+2\Phi_{\rm N}/c^2$, and at first order in $v/c$, the equation of motion of a test particle reduces to
\begin{equation} \label{eq.inertie2}
 \mathbf{a} =  \mathbf{g}_N  +  \delta \mathbf{a}_A, \qquad
  \delta \mathbf{a}_A = -c^2\sum_if_{A,i}\left(\nabla\alpha_i+\dot\alpha_i\frac{\mathbf{v}_A}{c^2}\right)
\end{equation}
so that in the slow velocity (and slow variation) limit it reduces
to
$$
\delta \mathbf{a}_A = -c^2\sum_i f_{A,i}\nabla\alpha_i\,.
$$
This allows one to get the E\"otv\"os parameter as
\begin{equation}\label{e.etaGrad}
\eta_{AB}= \frac{c^2}{g_N}\sum_i \left\vert \left(f_{A,i} -f_{B,i}\right) \nabla\alpha_i\right\vert.
\end{equation}
so that the constraints summarized in Table~\ref{tab-eta12} can be translated on a constraint on the gradient of the fundamental constants on Earth length scale.

The anomalous acceleration~(\ref{eq.inertie2}) is generated by the change in the (electromagnetic, gravitational, \dots) binding energies \citep{dicke64,haugan76,nordtvedt90} but also in the Yukawa couplings and in the Higgs sector parameters so that the $\alpha_i$-dependencies are a priori composition-dependent.  As a consequence, any variation of a fundamental constant will entail a violation of the universality of free fall: the total mass of the body being space dependent, an anomalous force appears if energy is to be conserved. The variation of the constants, deviation from General Relativity and violation of the weak equivalence principle are in general expected together, with quantitative relationships depending on the particular theory and model used; see e.g., Sec.~\ref{subsec81} for concrete examples.

\paragraph{Sensivitiy parameters}\ 

The composition dependence of $\delta\mathbf{a}_A$ and thus of $\eta_{AB}$ can be used to optimize the choice of materials for the experiments testing the equivalence principle \citep{damourmat,damourdono,dado1} but also to distinguish between several models if data from the universality of free fall and atomic clocks are combined \citep{dentuff}.

From a theoretical point of view, the computation of $\eta_{AB}$ requires the determination of the coefficients $f_{Ai}$. This can be achieved in two steps by first relating the new degrees of freedom of the theory to the variation of the fundamental constants  and then relating them to the variation of the masses. As we shall detail in Sect.~\ref{section-theories}, the first issue is very model dependent while the second is especially difficult (see Sect.~\ref{subsubmass}), particularly when one wants to understand the effect of the quark masses, since it is related to the intricate structure of QCD and its role in low energy nuclear reactions.

As an example, the mass of a nuclei of charge $Z$ and atomic number $A$ can be expressed as
\begin{equation}\label{e.mass0}
m(A,Z)=Zm_{\mathrm{p}}+(A-Z)m_{\mathrm{n}}+Zm_{\mathrm{e}}+E_{\mathrm{S}}+E_{\mathrm{EM}},
\end{equation}
where $E_{\mathrm{S}}$ and $E_{\mathrm{EM}}$ are respectively the strong and electromagnetic contributions to the binding energy. The Bethe--Weiz\"acker formula allows one to estimate the latter as
\begin{equation}\label{bethe}
E_{\mathrm{EM}}=98.25\frac{Z(Z-1)}{A^{1/3}}\aem \unit{MeV}.
\end{equation}
If we decompose the proton and neutron masses as \citep{gasser82}
\begin{equation}\label{e.massP}
m_{(\mathrm{p,n})}=u_3 +b_{(\mathrm{u,d})}m_{\mathrm{u}}+b_{(\mathrm{d,u})}m_{\mathrm{d}}+B_{(\mathrm{p,n})}\aem
\end{equation}
where $u_3$ is the pure QCD approximation of the nucleon mass ($b_{\mathrm{u}}$, $b_{\mathrm{d}}$ and $B_{(\mathrm{n,p})}/u_3$ being pure numbers), it reduces to
\begin{align}
\label{mass}
m(A,Z)=&\left(Au_3+E_{\mathrm{S}}\right)+
(Zb_{\mathrm{u}}+Nb_{\mathrm{d}})m_{\mathrm{u}}+(Zb_{\mathrm{d}}+Nb_{\mathrm{u}})m_{\mathrm{d}}\nonumber\\
&+\left(ZB_{\mathrm{p}}+NB_{\mathrm{n}}+98.25\frac{Z(Z-1)}{A^{1/3}} \unit{MeV}\right)\aem \,,
\end{align}
with $N=A-Z$, the neutron number. For an atom, one would have to add the contribution of the electrons, $Zm_{\mathrm{e}}$. This expression clearly depends on strong, weak and electromagnetic quantities. The numerical coefficients $B_{(\mathrm{n,p})}$ are given explicitly by \citep{gasser82}
\begin{equation}
\label{gl}
B_{\mathrm{p}}\aem=0.63 \unit{MeV} \quad B_{\mathrm{n}}\aem=-0.13 \unit{MeV}.
\end{equation}
Such estimations were used in the first analysis of the relation between variation of the constant and the universality of free fall \citep{damour94a,dvaliZ,jpu-revue} but the dependence on the quark mass is
still not well understood and we refer to \cite{damourdono,dado1,bbn-dimi2,donomass,oklo-14} for some attempts to refine this description. It is also important to stress that a microscopic model with only a varying fine structure constant will generate varying proton and neutron masses due from their binding energies, an effect that cannot be neglected.

For macroscopic bodies, the  total mass also inherits a negative contribution,
\begin{equation}
\label{llr8}
  \Delta m(G)=-\frac{G}{2c^2}\int\frac{\rho(\vec r)\rho(\vec r')}{|\vec r-\vec r'|}
\dd^3\vec r\dd^3\vec r'\,,
\end{equation}
from the gravitational binding energy. As a conclusion, from~(\ref{mass}) and (\ref{llr8}), we expect the masses to depend on all the coupling constants. We shall discuss this issue in more detail in Sect.~\ref{subsubmass}.

To finish, note that varying coupling constants can also be associated with violations of local Lorentz invariance and CPT symmetry \citep{lolo1,lolo2, gurz}. 

\subsection{Relations with cosmology}\label{subseccosmo}

Most constraints on the time variation of the fundamental constants will not be local and related to physical systems at various epochs of the evolution of the universe. It follows that the comparison of
different constraints requires a full cosmological model. Besides, many varying fundamental constant models have been related to the study of dark energy models involving a light field. Thus, we need to recall the basics of the current cosmological model.

\subsubsection{$\Lambda$CDM standard cosmological model}\ \label{secLCDM}

Our current cosmological model is known as the $\Lambda$CDM (see \citealt{peteruzanbook,NEW_Uzan:2016wji} for a detailed description, Appendix~\ref{app3} for the main definitions and Table~\ref{tab-cosmo} for the typical value of the cosmological parameters). 

It is important to recall that its construction relies on 4 main hypotheses, as discussed in \cite{jpu-revu3}, 
\begin{itemize}
\item[](H1) a theory of gravity; 
\item[](H2) a description of the matter components contained in the Universe and their non-gravitational interactions; 
\item[](H3) symmetry hypothesis; and 
\item[](H4) a hypothesis on the global structure, i.e., the topology, of the Universe. 
\end{itemize}
These hypotheses are not on the same footing since H1 and H2 refer to the physical theories. However, these hypotheses are not sufficient to solve the field equations and we must make an assumption on the symmetries (H3) of the solutions describing our Universe on large scales while H4 is an assumption on some global properties of hese cosmological solutions, with same local geometry. But the last two hypothesis are unavoidable because the knowledge of the fundamental theories is not sufficient to construct a cosmological model \citep{jpu-model}.

The $\Lambda$CDM model assumes that gravity is described by General Relativity (H1), that the Universe contains the fields of the standard model of particle physics plus some dark matter and a cosmological constant, the latter two having no physical explanation at the moment. It also deeply involves the Copernican principle as a symmetry hypothesis (H3), without which the Einstein equations usually cannot been solved, and assumes most often that the spatial sections are simply connected (H4). H2 and H3 imply that the description of the standard matter reduces to a mixture of a pressureless and a radiation perfect fluids. This model is compatible with all astronomical data despite the  fact that anomalies have been showing up in the past years \citep{NEW_Peebles:2022akh}. It roughly indicates that $ \Omega_{\Lambda0} \simeq 0.73$, $\Omega_{\mathrm{mat0}} \simeq 0.27,$ and $\Omega_{K0} \simeq 0$. Thus, cosmology roughly imposes that $|\Lambda_0|\leq H_0^2$, that is $\ell_\Lambda \leq H_0^{-1} \sim 10^{26} \unit{m} \sim 10^{41} \unit{GeV}^{-1}$.

\subsubsection{Dark energy and its implications}\ 

The analysis of the cosmological dynamics of the universe and of its large scale structures requires the introduction of a new constant, the \emph{cosmological constant}, associated with a recent acceleration of the cosmic expansion, that can be added to the Einstein--Hilbert action as
$$
 S_{\text{grav}} = \frac{c^3}{16\pi G}\int\sqrt{-g}(R-2\Lambda)\dd^4x.
$$
This constant can equivalently be introduced in the matter action. Note, however, that it is disproportionately small compared  to the natural scale fixed by the Planck length
\begin{equation}
 \rho_{\Lambda_0}\sim10^{-120}M_{\mathrm{Pl}}^4\sim10^{-47} \unit{GeV}^4\,.
\end{equation}

Classically, this value is no problem but it was pointed out that at the quantum level, the vacuum energy should scale as $M^4$, where $M$ is some energy scale of high-energy physics.  In such a case, there is a discrepancy of 60\,--\,120 orders of magnitude between the cosmological conclusions and the theoretical expectation. This is the \emph{cosmological constant problem} \citep{weinberg89}.

Two approaches to solve this problem have been considered. Either one accepts such a constant and such a fine-tuning and tries to explain it on anthropic ground.  Or, in the same spirit as Dirac, one interprets it as an indication that our set of cosmological hypotheses have to be extended, by either abandoning the Copernican principle \citep{uce} or by modifying the local physical laws (either gravity or the  matter sector). The way to introduce such new physical degrees of freedom were classified in \cite{jpu-revu3}. In that latter approach, the tests of the constancy of the fundamental constants are central, since they can reveal the coupling of this new degree of freedom to the standard matter fields. Note, however, that the cosmological data still favor a pure cosmological constant.

\subsubsection{Light fields phenomenology}\label{subsecphico}

Among all the proposals, quintessence \citep{PhysRevD.37.3406} involves a scalar  field rolling down a runaway potential hence acting as a fluid with an effective equation of state in the range $-1 \leq w \leq 1$ if the field is minimally coupled. It was proposed that the quintessence field is also  the dilaton \citep{riazuelo2001,gasperini02}. The same scalar field then drives the time variation of the cosmological constant and of the gravitational constant and it has the property to also enjoy tracking solutions \citep{uzan99}. Such models do not solve the cosmological constant problem but only relieve the coincidence problem. One of the underlying motivation to replace the cosmological constant by a scalar field comes from superstring models in which any dimensionful parameter is expressed in terms of the string mass scale and the vacuum expectation value of a scalar field. However, the requirement of slow roll (mandatory to have a negative pressure) and the fact that the quintessence field dominates today imply, if the minimum of the potential is zero, that it is very light, roughly of order $m\sim 10^{-33} \unit{eV}$ \citep{carroll98}. 

Such a light field will inevitably lead to violations of the universality of free fall and variation of the fundamental constants if it is non-universally coupled to the matter fields. \cite{carroll98} considered the effect of the coupling of this very light quintessence field to ordinary matter via a coupling to the electromagnetic field as $\phi F^{\mu\nu}\widetilde F_{\mu\nu}$. \cite{chiba2001} also argued that an ultra-light quintessence field induces a time variation of the coupling constant if it is coupled to ordinary matter and studied a coupling of the form $d_e \phi F^{\mu\nu}F_{\mu\nu}/M_{\mathrm{P}}$, as, e.g., expected from Kaluza--Klein theories (see Sect.~\ref{subsub5.3}  below). Such a coupling with $d_e\ll 1$  generically arise if quantum gravity effects weakly break an underlying global symmetry of $\phi$ near the Planck scale \citep{Kallosh:1995hi}.

As will be described in Sect.~\ref{subsec81}, many models link the acceleration of the cosmic expansion and the variation of fundamental constants. While a deviation from $w=-1$ is difficult to detect, quintessence may reveal itself through the variation of fundamental constants.

\subsubsection{Ultra-light Dark Matter models}\label{secULDM}

As we shall see in more details below, in a Friedmann-Lema\^{\i}tre spacetime, a massive scalar field with linear coupling enjoys a Klein--Gordon  equation of the form $\ddot\varphi+3H\varphi+m^2\varphi= 4\pi G \sigma$ where $H=\dot a/a$ is the Hubble parameter. The second term sources a Hubble damping that can be neglected as long as $m_\varphi\gg H_0\sim 1.5\times 10^{-33}$~eV. $\sigma$ is a source term that arise from the coupling of $\varphi$ to the standard model field, see Eq.~(\ref{e.KGgen2}) with $\alpha(\varphi)=cst$ below. Being related to the baryon density, it will evolve with a characteristic time scale of $H^{-1}$ so that for periods much shorter, it can be considered {\rm constant}. In this regime, the solution of the Klein--Gordon  as a superposition of a term sourced by the standard matter and an oscillating component,
\begin{equation}\label{e.phiDM}
\varphi(t) = \frac{4\pi G\sigma}{m_\varphi^2} + \varphi_0\cos(m_\varphi t + \delta)\,.
\end{equation}
Averaging its energy and pressure, $4\pi G\rho_\varphi=\dot\varphi^2/2 + V$ and $4\pi GP_\varphi=\dot\varphi^2/2 - V$ -- using the notation and normalisation of Sect.~\ref{subsecST} and Eq.~(\ref{einstein*}) -- over a period much smaller than $H^{-1}$, the scalar field behaves as a pressureless component that can be identified with dark matter (DM) with density
\begin{equation}\label{e.phiDM_rho}
4\pi G\rho_\varphi = \frac{1}{2} m_\varphi^2\varphi_0^2 \,.
\end{equation}
Note however that light scalar fields do not behave exactly as perfect cold dark matter on short length scales, where their density perturbations have a nonzero sound speed. For $m_\varphi<10^{-22}$~eV, they would have inhibited cosmological structure growth \citep{Marsh:2013ywa} in conflict with observations. In this mass range, the field would behave classically. Hence, a light dilaton with mass in the range 
\begin{equation}
m_\varphi \in (10^{-22}-1)~\unit{eV}
\end{equation}
can act as a dark matter component. such an ultralight bosonic field as dark matter has large occupation numbers per mode and behaves as a classical wave with a frequency proportional to $m_\varphi$. 

It was realized, as already discussed in the previous section on quintes\-sence, that the light field shall couple to standard model fields so that these dark matter models can be detected in laboratory experiments, in particular through the variation of fundamental constants \citep{NEW_Arvanitaki:2014faa,NEW_Stadnik:2014tta,NEW_VanTilburg:2015oza,NEW_Stadnik:2015kia,Derevianko:2016sgw}. \cite{NEW_Stadnik:2015xbn} proposed to search for this dark component with  laser interferometer measurements, such as large-scale gravitational-wave detectors (e.g., LIGO, Virgo, GEO600,\ldots) and investigate the design of a strontium optical lattice clock – silicon single-crystal cavity interferometer as a novel small-scale platform. \cite{NEW_Savalle:2019jsb} investigate the ddetection of light DM thanks to spacetime separated clocks that are shown to probe different parameter combinations than the ``usual"" co-located clock experiments. We refer to  \cite{NEW_Stadnik:2015upa} for an overview of recent developments in the detection of light bosonic dark matter, including axion, pseudoscalar axion-like and scalar dark matter, to \cite{NEW_Safronova:2017xyt,NEW_Safronova:2019lex} for extensive reviews on the search for light dark matter, and to  \cite{NEW_Budker:2024bzj} for a study of spacecraft missions to detect such halos with instruments such as quantum clocks, atomic and molecular spectrometers designed to search for fast (tens of hertz to gigahertz) oscillations of fundamental constants. \cite{NEW_Gue:2024onx} presented the theoretical investigation of the expected experimental signals produced by freely falling atoms with time oscillating mass and transition frequency that may be induced by the coupling of scalar DM to standard matter showing.

From a theoretical point of view, most analysis rely of the light dilaton model described in details in Sect.~\ref{subsub1b} with linear or quadratic couplings. Note that \cite{NEW_Gan:2023wnp} proposed to consider a dark photon coupled to electromagnetism with a $\varphi$ dependent coupling, which allows one to open the early time parameter space to produce dark matter while being compatible with actual laboratory constraints, making the dark photon a still viable candidate. Note the claim by \cite{Bauer:2024yow} that  the quadratic interactions of axion dark matter lead to a divergence of the axion field value for decay constant smaller than $10^{15}$~GeV so that the field value for the axion close to a massive body like Earth is divergent in a similar way as the spontaneous scalarization \citep{PhysRevLett.70.2220,PhysRevD.54.1474} arising in scalar-tensor theories for light scalars in the presence of massive bodies.

\subsubsection{Dark Matter topological defects}\label{secTopDef}

Note also that such a scalar field can also form topological defects \citep{NEW_topdefbook} that can result in the existence of domains with different values of the fundamental constants \citep{wall1}, as described in Sect.~\ref{secwallalpha}. Besides, the dark matter density would be concentrated into many distinct, compact spatial regions. \cite{Essig:2013lka} proposed a classification of the way  the fields forming the defect interact with the standard matter, using the so-called ``portals'', the collection of gauge-invariant operators of the standard matter coupled with the operators from the dark sector. \cite{NEW_Derevianko:2013oaa} described many field set-ups that could lead to monopole, strings or wall assuming linear and quadratic coupling.  Besides the linear, $\varphi m_\psi \bar\psi\psi$, and quadratic, $\varphi^2 m_\psi \bar\psi\psi$, couplings they considered an axionic portal $\partial_\mu\varphi \bar\psi\gamma_\mu\psi$ and a current-current portal $\varphi^*\partial_\mu\varphi \bar\psi\gamma_\mu\psi$. They emphasized that networks of correlated atomic clocks, such as the Global Positioning System, are a powerful tool to search for this topological defect dark matter, hence providing a new fundamental physics application to the ever-improving accuracy of atomic clocks. Indeed, during the encounter with an extended dark matter object, as it sweeps through the network, initially synchronized clocks will become desynchronized. Time discrepancies between spatially-separated clocks are expected to exhibit a distinct signature, encoding defect’s space structure and its interaction strength with atoms. \cite{NEW_Stadnik:2014tta} detailed the detection of topological defect dark matter through transient-in-time effects and for a relic, coherently oscillating condensate. \cite{Roberts:2017hla}  found no evidence for DM in the form of domain walls  from the mining 16 years of archival GPS data. A Global Network of Optical Magnetometers for Exotic physics searches (GNOME) was proposed to reach higher sensitivities \citep{Afach:2021pfd}.

\subsubsection{Varying constants and cosmology}\label{secosmocte}

We have three means of investigation and types of relations between varying constants and the cosmological model within the scalar field extension of the standard $\Lambda$CDM.

\paragraph{Test field}\ 

The field driving the time variation of the fundamental  constants is not taylored for explaining the acceleration of the universe -- and it never dominates the matter content today or its equation of  state is not negative enough. In such a case, the variation of the  constants is disconnected from the dark energy problem. Cosmology  then allows us to determine the dynamics of this field during the whole history of the universe and thus to compare local and   cosmological constraints and interpret all the data consistently but the new field does not modify the cosmic history. An example is provided by scalar-tensor  theories (see Sect.~\ref{subsecST}) for which one can compare,   e.g., primordial nucleosynthesis to local  constraints \citep{bbn-Gpichon}. 

Note however that in such a situation one should  take into account the effect of the variation of the  constants on the astrophysical observations since it can affect   local physical processes and bias, e.g., the luminosity of  supernovae and indirectly modify the distance luminosity-redshift  relation derived from these observations \citep{barrow01,cmb-G1}.

\paragraph{Dark sector candidate}\ 

The field driving the time variation of the fundamental constants is also responsible for the ``dark sector''.  It can account for dak matter or cause the late time acceleration of the cosmic expansion (see Sect.~\ref{subsub0} for concrete examples and the discussion of Sect.~\ref{subsecphico} above). 

It follows that the local value of the scalar field and time derivatives are related either to the dark matter abundance or to the equation of state of dark energy. Hence, the dynamics of the universe, the level of variation of the constants and the other deviations from general  relativity are connected \citep{damour94a,damourrunaway,jpu-revu2,msu,jpu-revu3,lilley} so that the study of the  variation of the constants can improve, within the model, the reconstruction of the equation state of the dark  energy \citep{avelino06,doran,nunes09,parkinson}.

\paragraph{Dark-visible matter equivalence principle}\ 

Note that the question of the validity of the equivalence principle in the dark sector is also crucial since it cannot be tested in the laboratory or in Solar system experiments. To that purpose, \cite{NEW_Coc:2008yu} derived the constraints from BBN on a scalar-tensor theory with non-universal coupling with dark sector so that no violation of the EEP can be witnessed in the visible sector, following earlier proposals by \cite{damour90,NEW_Fuzfa:2007sv}. \cite{NEW_Mohapi:2015gua} constrain such a coupling from galactic observations of strong lensing and of velocity dispersion, showing that data favor violations through coupling strengths that are of opposite signs for ordinary and dark matter even though no deviation from GR.

From a theoretical perspective,  this opens the question of whether one can modify the dark matter gravity sector independently of the standard matter sector, in particular for since such models have become candidates to a solution of the Hubble tension  \citep{NEW_Pitrou:2023swx,NEW_Uzan:2023dsk}. In some models, a scalar fifth force may potentially lead to a violation of the WEP that could be probed \citep{Carroll:2009dw,Mantry:2009ay}, in particular if dark and standard matter fields interact. Such models are however strongly constrained \citep{Carroll:2008ub}. It has also been suggested that the anomalies in the positron/electron spectra may arise from a long-range dark force mediating the dark matter annihilation and that it could be detected at the LHC \citep{Bai:2009it}. Then, one shall investigate the effects of the scalar force in astrophysical environments and how it modifies DM haloes, see e.g., \cite{Frieman:1991zxc,Gradwohl:1992ue,Nusser:2004qu,Kesden:2006vz,Bean:2008ac,Keselman:2009nx,NEW_Mohapi:2015gua} for tests of the WEP between the visible and DM sectors.  \cite{Peebles:2012sm} suggested that a DM fifth force seems to have beneficial effects for DM distribution on small scales and that an extra evanescent component of matter with evolving mass and a fifth force large enough may be needed to reach a better understanding of early assembly of more nearly isolated protogalaxies.

\paragraph{The Hubble tension}\ 

A new connection has appeared with the \emph{Hubble tension}. In a simplified way, it arises from the fact that the value of the Hubble constant $H_0$ derived from the analysis of CMB+BAO data, which point toward $H_0=(67.4\pm0.5)$~km/s/Mpc,  differs from the one obtained at low redshifts from the Hubble diagram, estimated to be $H_0=(73.04\pm1.04)$~km/s/Mpc; hence a $5\sigma$ discrepancy \citep{NEW_Schoneberg:2021qvd}.

The problem is usually formulated as a low/high-redshift tension since in first approximation, the key physical parameters at the background level are the comoving sound horizon
\begin{equation}\label{e.rs}
r_s=\frac{1}{H_0}\int_{z_*}^\infty \frac{c_s \dd z}{H(z)/H_0} 
\end{equation}
with $z_*\sim1088$ the redshift at recombination, and the comoving angular diameter distance
\begin{equation}\label{e.rs+1}
 R_{\rm ang}(z)=\frac{1}{H_0}f_K\left[\int_0^z \frac{\dd z}{H(z)/H_0} \right],
\end{equation}
related to the angular (or luminosity) distance, $D_A(z)=R_{\rm ang}(z)/(1+z)$ or $D_L=(1+z)R_{\rm ang}(z)$; see Appendix~\ref{app3} for definitions. Their ratio fixes the physical angular scales of the acoustic peaks of the CMB. 

The precision of CMB observations set a strong constraint on this quantity from which one can get the early time estimations of $H_0$ given the knowledge of the baryon and dark matter (DM) energy densities. It follows that most of the arguments on the $H_0$ tension circle around the sound horizon with two main categories of models \citep{DiValentino:2021izs,NEW_Schoneberg:2021qvd,Abdalla:2022yfr}: (\textit{1}) ``\emph{Late time models}'' modify the expansion history after recombination, increasing $H_0$ but keeping $r_s$  unchanged; (\textit{2}) ``\emph{Early time models}'' modify the expansion history before recombination, e.g., through energy injection around the recombination, changing both $H_0$ and $r_s$ so as to have a lower sound horizon to allow for a larger $H_0$ to which one can add (\textit{3}) ``\emph{earlier recombination models}'' \citep{Jedamzik:2020zmd} and (\textit{4}) ``\emph{modified DM gravity models}'' \citep{NEW_Pitrou:2023swx,NEW_Uzan:2023dsk} in which the gravitation in the DM sector is assumed to enjoy a scalar-tensor gravity while the visible sector  remains  unaffected. Many of these models assume a form of dark energy or an interaction of dark matter with a new matter component and in particular a scalar field.  The dynamics of these 4 categories have been compared on Fig.~6 of \cite{NEW_Uzan:2023dsk}.
 
It was realized that the Einstein-Boltzmann equations used to predict CMB+BAO is invariant \citep{NEW_Rich:2015jla,Cyr-Racine:2021oal,NEW_Ge:2022qws}  under the rescaling transformations\footnote{\cite{NEW_Rich:2015jla} derives  the dependence of the CMB temperature angular power spectrum to the 4 dimensionless quantities  where $m_\chi$ is the mass of the dark matter particle, demonstrating in particular that $\Delta [m_{\rm p}/m_\chi]/(m_{\rm p}/m_\chi)=0.09 \pm 0.15$ while $m_\chi$ is not known.}
$$
H\rightarrow \lambda H\, \qquad G\rho\rightarrow \lambda^2G\rho\, \qquad \sigma_T n_e\rightarrow\lambda\sigma_T n_e, \qquad A_s\rightarrow A_s\lambda^{1-n_s}
$$
of the Hubble parameter, gravitational constant, Thomson scattering and amplitude of the cosmological perturbations, the latter involving a dependence on the spectral index $n_s$ of the initial power spectrum. Hence, it led to the idea that the Hubble tension may be related to the variation of constants.  A varying gravitational constant was proposed by \cite{NEW_Begue:2017lcw} and implemented in different ways \citep{NEW_Knox:2019rjx,DiValentino:2021izs,Abdalla:2022yfr} as a generalisation of old extended quintessence models \citep{uzan99}. Most of these models face difficulties with either primordial nucleosynthesis and/or local constraints on deviations from General Relativity.  \cite{Cyr-Racine:2021oal} implemented this scaling by invoking a mirror world but the model requires primordial helium abundance in contradiction with BBN results. \cite{NEW_Zhang:2022ujw} extended this to a model of a massive field with linear and quadratic coupling able to resolve the tension while keeping the same helium-4 abundance from BBN provided $\Delta\aem/\aem\simeq-2\times10^{-5}$ at recombination.

Indeed, in order for the sound horizon, and thus CMB predictions, to remain unchanged,  one needs to also modify the Thomson cross-section $\sigma_T$, i.e., the non-gravitational physics which is highly constrained.  The idea to ``compensate" the variation of the gravitational constant by a variation of the fine structure constant and/or electron mass and thus the possibility to solve or alleviate the Hubble tension thanks to a variation of the constant has first been noticed by \cite{NEW_Hart:2019dxi}. It has attracted an huge attention since then; see the report of the latest constraints in Sect.~\ref{subsec34}. \cite{NEW_Hart:2021kad} then showed that varying $(\aem,m_{\rm e})$ can alleviate the Hubble tension despite the CMB \citep{NEW_Planck:2014ylh} while being compatible with other constraints on the variation of constants, as discussed in this review, that shall indeed be included together with the cosmological observations. 

From the study of the scaling of the parameters, \cite{Sekiguchi:2020teg} demonstrated that if the Hubble constant and electron mass are scaled  $\Delta h/h\sim 3.23 \Delta m_{\rm e}/m_{\rm e}$ then the CMB is barely unchanged. Hence, one could resolve the Hubble tension at the expanse of a higher $m_{\rm e}$. However this would also require that $\Delta\Omega_{\rm m}/\Omega_{\rm m}=-5.46\Delta m_{\rm e}/m_{\rm e}$ resulting in a lower $\Omega_{\rm m}$ \citep{Lee:2022gzh}, which is potentially a critical issue since this would be in contraction with SNIa and BAO data. As understood by \cite{Sekiguchi:2020teg}, this could be resolved by allowing for a non-vanishing space curvature, which actually turned to be the best-fit model in the extensive model comparisons by \cite{NEW_Schoneberg:2021qvd} or  by including early dark energy  \citep{NEW_Hart:2022agu} highlighting clearly  the connection between $m_{\rm e}$ and $H_0$. Note that a non-vanishing space curvature needs to face model-building issues in inflation and in tension with the analysis by \cite{DiDio:2016ykq}.

Most of the studies \citep{NEW_Hart:2017ndk,NEW_Hart:2019dxi, NEW_Schoneberg:2021qvd,NEW_Khalife:2023qbu,NEW_Baryakhtar:2024rky,NEW_Seto:2024cgo,NEW_Toda:2024ncp,NEW_Schoneberg:2024ynd} considered a \emph{constant shift} of the electron mass between CMB and today; see Sect.~\ref{subsec34} and Table~\ref{tabme}. Following \cite{NEW_Martins:2015dqa}, some models assume a phenomenological evolution of the constants parameterized as $[(1+z)/1000]^p$ for $\aem$ and or $m_{\rm e}$ and include $p$ in the fit; see e.g., \cite{NEW_Hart:2019dxi}. While this accounts for a cosmic time evolution such a parameterisation is actually a choice of  a ``solution'' to a non-specified underlying theory. While this could be sufficient to analyse background data, it shall be taken with a grain of salt for CMB analysis since perturbations cannot be treated consistently. As illustrated by \cite{NEW_Uzan:2023dsk} on a scalar-tensor theory, neglecting the treatment of the perturbations can lead to an error larger than a percent for the CMB predictions compared to the full treatment, with same background evolution. Indeed, many dynamical models can and have been constructed on the basis of a light scalar field; see e.g., \cite{NEW_Barrow:2005qf,NEW_Avelino:2008dc,NEW_Hoshiya:2022ady} as well as all the models discussed in Sect.~\ref{subsec81}. As Eq.~(\ref{edotphi0}) shows, all these models induce a time variation of the constant in the Solar system that is sharply constrained. On the example of $B(\varphi)F^2$ models discussed in Sect.~\ref{subsub0} \cite{NEW_Vacher:2024qiq} demonstrated that this is a critical hurdle to construct a viable model of varying constant that would resolve the Hubble tension since the field would have to relax rapidly enough to avoid the local constraints. In particular they demonstrates that in single-field models one has to extremely fine-tune the shape of the potential and/or the initial conditions. Indeed, for single field models in a potential that is not fine-tuned we can put a generic bound at recombination of $\Delta\aem/\aem<5 \times10^{-4}$ 95~C.L.. To finish, let us also mention that \cite{NEW_Schoneberg:2024ynd} cooked up a model in which $m_{\rm e}\propto m_{\rm p}$ to evade the local constraints on the variation of $\bar\mu$ while allowing $m_{\rm e}$ to vary between recombination and today.

\begin{figure}[htbp]
\vskip-.5cm
 \centerline{\includegraphics[scale=0.4]{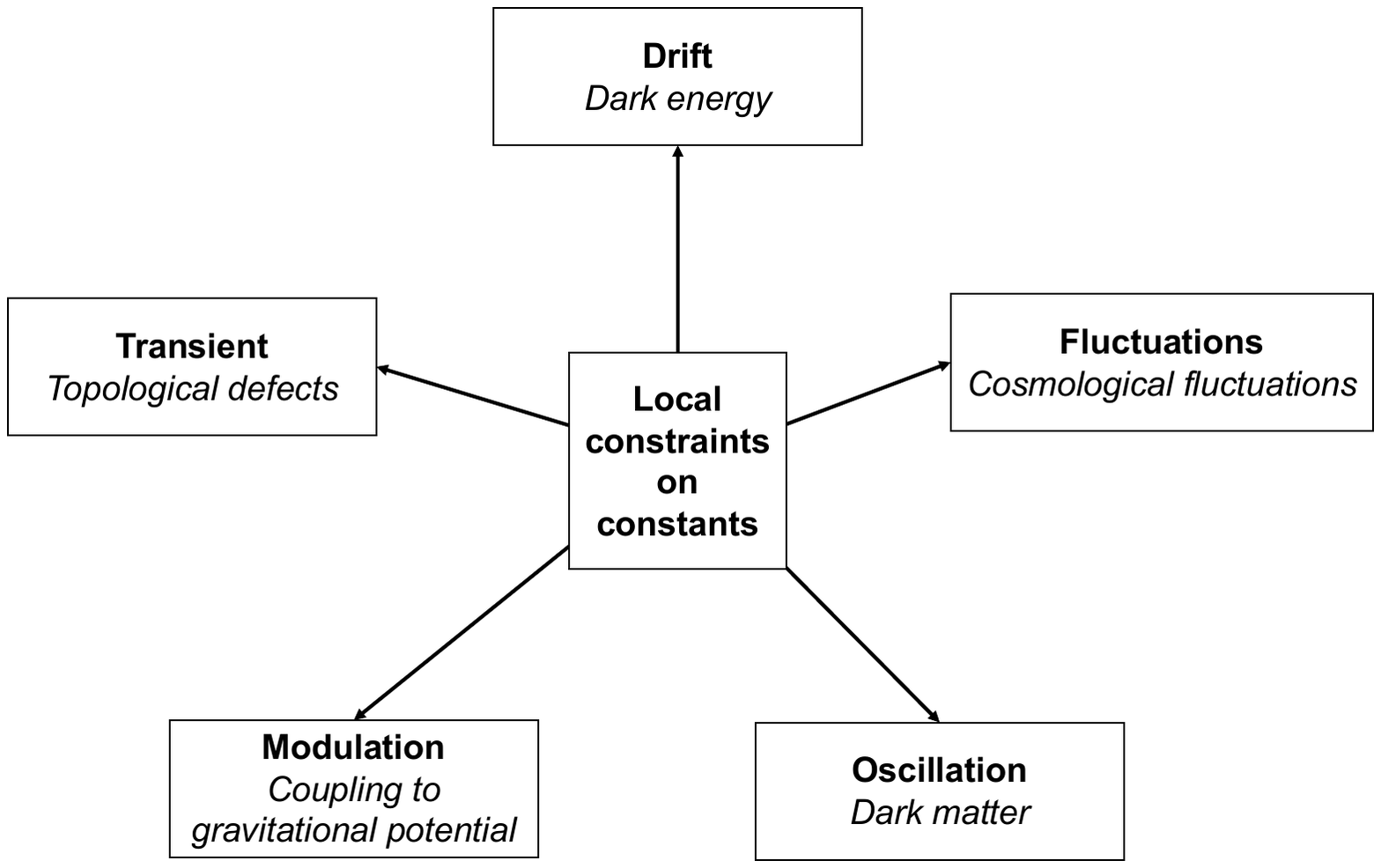}}
\vskip-.5cm
  \caption[Relations between variation of the constants and cosmology]{The local studies of the variation of fundamental constants through either atomic clock experiments or bounds on the universality of free fall offer different connections with cosmology. Constraints on a \emph{linear drift} open to dark energy models (see Sect.~\ref{subsub1}, Sect.~\ref{subsub1b}  or~\ref{subsub0} for concrete examples), while \emph{transients} allow one to study topological defects (see Sect.~\ref{secTopDef} and Sect.~\ref{secwallalpha} for a concrete example). The study of \emph{oscillations} give access to light dark matter models (see Sect.~\ref{secULDM} and Sect.~\ref{subsub1b} for concrete examples) while annual or diurnal \emph{modulations} set bounds on the coupling to the gravitational potential, as described in Sect.~\ref{subsuba}. Finally, the so-far experimentally unexplored \emph{fluctuations} link to cosmological fluctuations; see e.g., Sect.~\ref{subsubb}.}
  \label{fig-clocksum}
\end{figure}

\subsection{Conclusion} 

The cosmological model requires a new constant, the \emph{cosmological constant} $\Lambda$, that needs to be included in our list of fundamental constants. Its value questions naturalness, hence motivating many lines of research. 

Indeed, it may be replaced, as for any other constants, by  a new degree of freedom accounting tor a new component of dark energy. Note that since the cosmological constant is not connected with the standard model of particle physics, this does not imply a violation of the universality of free fall. The cosmological dynamics can be hope to constrain the dark energy equation of state in order to conclude whether it is varying or not. Cosmology also provides a link between the microphysics and macrophysics description of our universe, as foreseen by Dirac. The tests of fundamental constants can discriminate between various explanations of the acceleration of the universe and eventually indicate if the dark sector can be explained by a modification of General Relativity. Once a specific model is specified, cosmology also allows one to set stronger constraints since it relates observables that cannot be compared otherwise; we give several examples in Sect.~\ref{subsec81}. The cosmological model is also needed to get the look-back time/redshift relation as depicted on Fig.~\ref{fig-temps-z} for the standard $\Lambda$CDM model.

To finish, and as summarized on Fig.~\ref{fig-clocksum}, the study of the stability of fundamental constants in term of a drift, modulation, oscillation, transient or fluctuation links to various phenomena of cosmological interest such as dark matter, dark energy, topological defects. This creates a strong bond between local experiments and the global description  of the universe, making the fundamental constants a key system to unveil the dark sector. 

As we shall discuss dimensionless constants, we refer to \cite{NEW_Narimani:2011rb,NEW_Rich:2013oxa} for a rewriting of the cosmological observables in terms of dimensionless parameters, which is a useful insight.

\begin{figure}[htbp]
  \centerline{\includegraphics[scale=0.4]{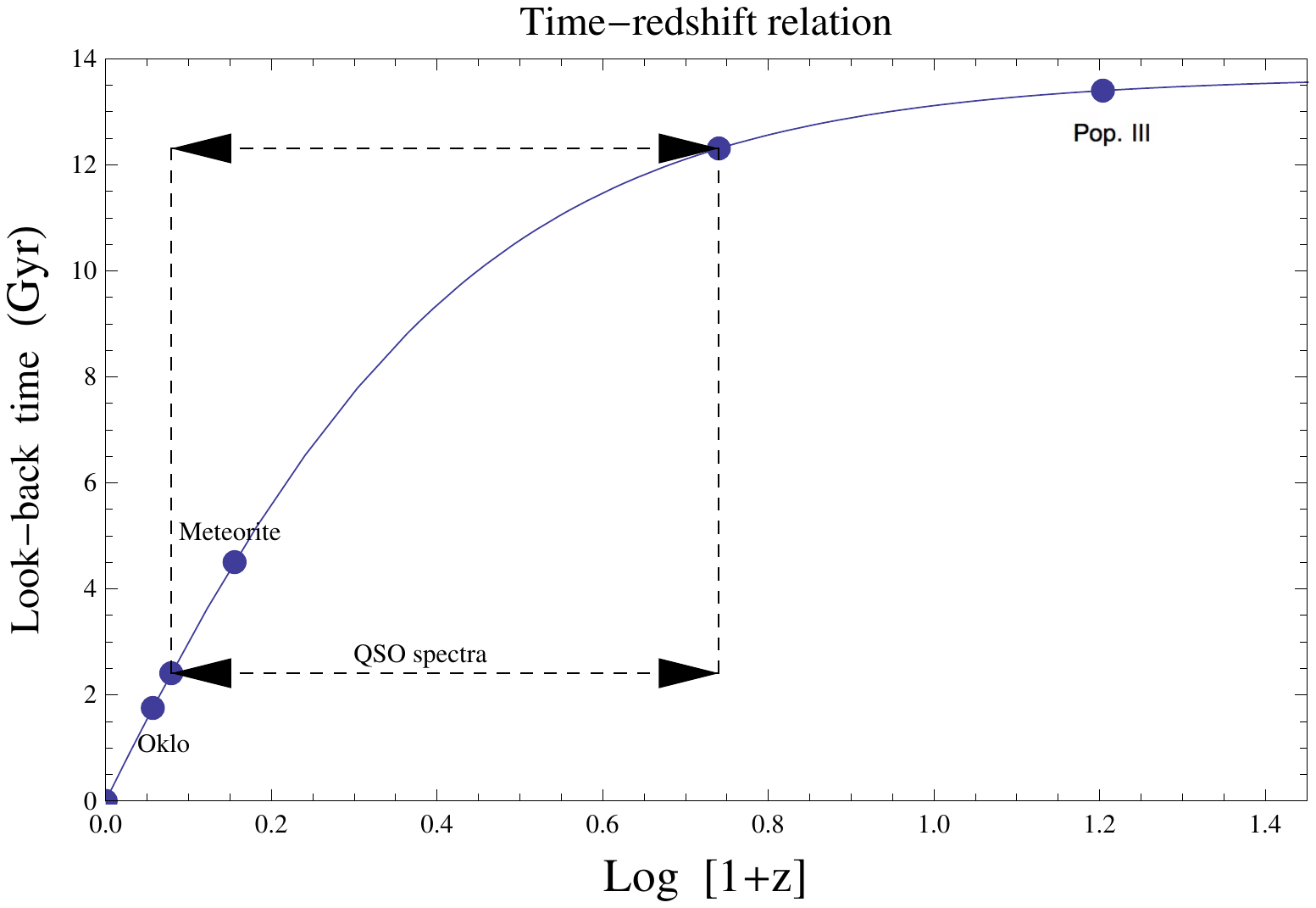}}
  \vskip-0.5cm
  \caption[Look-back time-redshift relation for the $\Lambda$CDM model]{Once a cosmological model is specified, one can deduce the look-back time-redshift relation, here for the standard $\Lambda$CDM model with the cosmological parameters of Table~\ref{tab-cosmo}. We have also indicated some of the systems that will be studied in Sect.~\ref{section3}.}
  \label{fig-temps-z}
\end{figure}

\section{Theories with varying constants}\label{section-theories}

Since the variation of one constant would results on a violation of the LPI and of the UFF, theories with varying constants are strongly related to the study of gravity theories beyond General Relativity. This Section describes this link by first recalling the standard universal scalar-tensor theories in Sect.~\ref{subsecST} mostly to clarify the connection between dynamical fields and constants as well as the role of the choice of a frame. This will be extended in Sect.~\ref{subsecSTmulti} to discuss the simplest varying fine structure constant theory and discuss general properties of non universal coupling while giving a concrete example of the connection between the amplitude of the variation of $\aem$ and the E\"otv\"os parameter. Since both are constrained to a high level in the Solar system, a new dynamical field shall be hidden which has led to the formulation of screening mechanisms described in Sect.~\ref{subsub2}. Section~\ref{subsec314} will then highlight the role of the constants in the general search for any new physical degrees of freedom beyond the standard model, a standard prediction of many high-energy physics theories, summarized in Sect.~\ref{subsub5.3}, that are generically predicting dynamical constants and hence set to the challenge of explaining why the latter are actually observed to vary so little, i.e., they have to provide an efficient stabilisation mechanism to be viable.

\paragraph{Early insights}\ 

As explained in the introduction, Dirac postulated that $G$ varies  as the inverse of the cosmic time. Such an hypothesis is indeed not a theory since the evolution of $G$ with time is postulated and not derived from an equation of  evolution\footnote{Note that the Dirac hypothesis can also be achieved by assuming that $e$ varies as $t^{1/2}$, as pointed out by \cite{gamow67a}. Indeed this reflects a choice of units, either atomic or Planck units. However, there is a difference: assuming that only $G$ varies violates the strong equivalence principle while assuming a varying $e$ results in a theory violating the weak equivalence principle. It does not mean that we are detecting the variation of a dimensionful constant but simply that either $e^2/\hbar c$ or $Gm_{\mathrm{e}}^2/\hbar c$ is varying. This shows that many implementations of the same idea are a priori possible.}  consistent with the other field equations, that indeed shall take into account that $G$ is no more a constant. In particular, in a Lagrangian formulation one needs to take into account that $G$ is no more constant and thus to be considered as a field when the action is varied. This will indeed have consequences in all the field equations and provide a new equation of motion.

The first implementation of Dirac's phenomenological idea into a field-theory framework (i.e., modifying Einstein's gravity and incorporating non-gravitational forces and matter) was proposed by \cite{jordan37}. He realized that the constants have to become dynamical fields and proposed the action
\begin{equation}
S=\int\sqrt{-g}\dd^4\bx\phi^\eta\left[R-\xi\left(\frac{\nabla\phi}{\phi}
\right)^2-\frac{\phi}{2}F^2 \right],
\end{equation}
$\eta$ and $\xi$ being two parameters. It follows that both $G$ and the fine-structure constant have been promoted to the status of dynamical fields.

\cite{fierz56} realized that with such a Lagrangian, atomic spectra will be spacetime-dependent, and he proposed to fix $\eta$ to the value $-1$ to prevent such a spacetime dependence.  This led to the definition of a one-parameter ($\xi$) class of scalar-tensor theories in which only $G$ is assumed to be a dynamical field. This was then further explored by \cite{brans61} (with the change of notation $\xi\rightarrow \omega_{\rm BD}$).   In this Jordan--Fierz--Brans--Dicke theory the gravitational constant is replaced by a scalar field, which can vary both in space and time.  It follows that, for cosmological solutions, $G\propto t^{-n}$ with $n^{-1}=2+3\omega_{\text{BD}}/2$. Thus, Einstein's gravity is recovered when $\omega_{\text{BD}}\rightarrow\infty$.  This kind of theory was further generalized to obtain various functional dependencies for $G$ in the formalisation of scalar-tensor theories of gravitation (see, e.g., \citealt{damour92} or \citealt{will-book}).

The number of theoretical frameworks describing varying constants and gravity beyond General Relativity is too large to be exhaustively reviewed. Hence, we first  present generalities on scalar-tensor theories and their extensions to include non universal couplings. This will bridge with the presentation of Sect.~\ref{subsec12}. Then, the connection with high-energy physics will be considered.

\subsection{Introducing new fields: generalities}\label{subsec_newfields}

\subsubsection{The example of scalar-tensor theories}\label{subsecST}

Let us start to remind how the standard general relativistic framework can be extended to make $G$ dynamical on the example of scalar-tensor theories, in which gravity is mediated not only by a massless spin-2 graviton but also by a spin-0 scalar field that couples universally to matter fields (this ensures the universality of free fall). 

\paragraph{General action}\ 

In the Jordan frame, using the notations by \cite{gefpolar}, the action of the theory takes the form
\begin{equation}
\label{actionJF}
  S =\int \frac{\dd^4 x }{16\pi G_*}\sqrt{-g}
     \left[F(\varphi)R-g^{\mu\nu}Z(\varphi)\varphi_{,\mu}\varphi_{,\nu}
        - 2U(\varphi)\right]+ S_{\text{matter}}[\psi;g_{\mu\nu}],
\end{equation}
where $G_*$ is the bare gravitational constant. This action involves three arbitrary functions ($F$, $Z$ and $U$) but only two are physical since there is still the possibility to redefine the scalar field. The function $F$ needs to be positive to ensure that the graviton carries positive energy. $S_{\text{matter}}$ is the action of the matter fields that are minimally coupled to the metric $g_{\mu\nu}$.  It follows that the lengths and times as measured by laboratory apparatus are defined in this frame. Besides, the UFF and the WEP hold but not the SEP.

Note that the normalisation of the scalar fields $\varphi$, as well as $\varphi_*$ in the Einstein frame formulation below, has been chosen to be dimensionless and we set the mass scale
$$
M_*= \frac{1}{\sqrt{8\pi G_*}}
$$
which would reduce to the Planck mass in General Relativity. We set
\begin{equation}\label{phivarphi}
\phi =M_*\varphi, \qquad
\phi_*= M_*\varphi_*.
\end{equation}
Among all the scalar-tensor theories, the Brans-Dicke theory is defined by the single free dimensionless parameter $\omega_{\rm BD}$ as
\begin{equation}\label{def_BD}
F=\varphi, \qquad Z=\frac{\omega_{\rm BD}}{\varphi}, \qquad U=0\,.
\end{equation}

\paragraph{Field equations in the Jordan frame}\ 

The variation of the action~(\ref{actionJF}) gives the following field equations
\begin{eqnarray}
F(\varphi) \left(R_{\mu\nu}-\frac{1}{2}g_{\mu\nu}R\right)
&=& 8\pi G_* T_{\mu\nu}
+ Z(\varphi) \left[\partial_\mu\varphi\partial_\nu\varphi
- \frac{1}{2}g_{\mu\nu}
(\partial_\alpha\varphi)^2\right]
\nonumber\\
&&+\nabla_\mu\partial_\nu F(\varphi) - g_{\mu\nu}\Box F(\varphi)
- g_{\mu\nu} U(\varphi)\ ,
\label{einstein}\\
2Z(\varphi)~\Box\varphi &=&
-\frac{dF}{d\varphi}\,R - \frac{dZ}{d\varphi}\,(\partial_\alpha\varphi )^2
+ 2 \frac{dU}{d\varphi}\ ,
\label{BoxPhi}\\
\nabla_\mu T^\mu_\nu &=& 0\ ,
\label{matter}
\end{eqnarray}
\label{2.2}
where $T \equiv T^\mu_\mu$ is the trace of the matter energy-momentum tensor $T^{\mu\nu} \equiv (2/\sqrt{-g})\times \delta S_m/\delta g_{\mu\nu}$.  As expected \citep{ellisu}, we have one equation, (\ref{einstein}), which reduces to the standard Einstein equation when $\varphi$ is constant and a new equation, (\ref{BoxPhi}), to describe the dynamics of the new degree of freedom while the conservation equation~(\ref{matter}) of the matter fields is unchanged, as expected from the weak equivalence principle so that any test-particle will follow geodesic of the metric $g_{\mu\nu}$.

\paragraph{Field equations in Einstein frame}\ 

It is useful to define an Einstein frame action through a conformal transformation of the metric 
\begin{equation}
g_{\mu\nu}^* = F(\varphi)g_{\mu\nu} \, \Longleftrightarrow \,
g_{\mu\nu}  = A^2(\varphi_*)g_{\mu\nu}^* \,.
\end{equation}
In the following all quantities labeled by a star (*) refer to Einstein frame. Defining the field $\varphi_*$ and the two functions $A(\varphi_*)$ and $V(\varphi_*)$ (see, e.g., \citealt{gefpolar}) by
\begin{align*}
 \left(\frac{\dd\varphi_*}{\dd\varphi}\right)^2
              &= \frac{3}{4}\left(\frac{\dd\ln F(\varphi)}{\dd\varphi}\right)^2
                  +\frac{1}{2F(\varphi)},\\
 A(\varphi_*) &= F^{-1/2}(\varphi),\\
 2V(\varphi_*)&= U(\varphi) F^{-2}(\varphi),
\end{align*}
the action~(\ref{actionJF}) takes the form
\begin{equation}\label{actionEF}
 S = \frac{1}{16\pi G_*}\int \dd^4x\sqrt{-g_*}\left[ R_*
        -2g_*^{\mu\nu} \partial_\mu\varphi_*\partial_\nu\varphi_*
        - 4V\right]+ S_{\text{matter}}[A^2g^*_{\mu\nu};\psi].
\end{equation}
The kinetic terms have been diagonalized so that the spin-2 and spin-0 degrees of freedom of the theory are perturbations of $g^*_{\mu\nu}$ and $\varphi_*$ respectively. In this frame the field equations are
\begin{align}
R^*_{\mu\nu} - \frac{1}{2} R^* g^*_{\mu\nu} &= 8\pi G_* T^*_{\mu\nu}
+ 2 \partial_\mu\varphi_*\partial_\nu\varphi_* -
g^*_{\mu\nu}(g_*^{\alpha\beta}
\partial_\alpha\varphi_*\partial_\beta\varphi_*) \nonumber\\
&\quad - 2 V(\varphi) g^*_{\mu\nu}\ ,
\label{einstein*}\\
\Box_*\varphi_* &= -4\pi G_*\alpha(\varphi_*)~T_*
+dV(\varphi)/d\varphi_*\ ,
\label{Boxvarphi}\\
\nabla^*_\mu T^\mu_{*\nu} &= \alpha(\varphi_*)~T_*
\partial_\nu\varphi_*\ ,
\label{matter*}
\end{align}
with 
\begin{equation}\label{def_ab_ST}
\alpha\equiv \dd\ln A/\dd\varphi_*\qquad\hbox{and}\qquad \beta\equiv\dd\alpha/\dd\varphi_*\,.
\end{equation}
In this version, the Einstein equations are not modified. But, since the theory can now be seen as the theory in which all the mass are varying in the same way, there is a source term to the conservation equation. It follows that the motion of a test- body will enjoy a fifth force and will not be a geodesic of the metric $g^*_{\mu\nu}$, except for photons.

This shows that the same theory can be interpreted as a varying $G$ theory or a universally varying mass theory, but remember that whatever its form the important parameter is the dimensionless quantity $Gm^2/\hbar c$. Still, all other dimensionless parameters will remain constant, and in particular mass ratios and the fine structure constant, simply as a consequence of the universal coupling.

\paragraph{Gravitational constant}\ 

The action~(\ref{actionJF}) defines an effective gravitational constant $G_{\mathrm{eff}} = G_*/F = G_*A^2$. This constant does not correspond to the gravitational constant effectively measured in a Cavendish experiment, 
\begin{equation}\label{gcav}
 G_{\mathrm{cav}} = G_*A_0^2(1+\alpha_0^2)
                    = \frac{G_*}{F}\left(1 + \frac{F_\phi^2}{2ZF + 3 F_\phi^2} \right)
\end{equation}
where the first term, $G_*A_0^2$, corresponds to the exchange of a graviton while the second term $G_*A_0^2\alpha_0^2$, is related to the long range scalar force, a subscript $0$ referring to the quantity evaluated today. The gravitational constant depends on the scalar field and is thus dynamical. In the case of the Brans-Dicke parameterisation~(\ref{def_BD}) reduces to
\begin{equation}\label{gcavBD}
G_{\mathrm{cav}}  = \frac{G_*}{\varphi} \frac{2\omega_{\rm BD}+4}{2\omega_{\rm BD} + 3}\,.
\end{equation}

\paragraph{Local constraints}\ 

This illustrates the main features common to all models: (\textit{i}) new dynamical fields appear (here a scalar field), (\textit{ii}) some constant will depend on the value of this scalar field (here $G$ is a function of the scalar field). It follows that the Einstein equations will be modified and that there will exist a new equation dictating the propagation of the new degree of freedom.

In this particular example, the coupling of the scalar field is universal so that no violation of the universality of free fall is expected. The deviation  from General Relativity can be quantified in terms of the post-Newtonian parameters, which can be expressed in terms of the values of $\alpha$ and $\beta$ today as
\begin{equation}
 \gamma^{\mathrm{PPN}} - 1 = -\frac{2\alpha_0^2}{1+\alpha^2_0},\qquad
 \beta^{\mathrm{PPN}} - 1 =\frac{1}{2}
 \frac{\beta_0\alpha_0^2}{(1+\alpha_0^2)^2}\,.
\end{equation}
These expressions are valid only if the field is light on the Solar system scales. If this is not the case then these conclusions may be changed \citep{cameleon}. The Solar system constraints imply $\alpha_0$ to be very small, typically $\alpha_0^2<10^{-5}$ while $\beta_0$ can still be large. Binary pulsar observations \citep{gefpul2,gefpul} impose that ${\beta_0>-4.5}$. The time variation of $G$ is then related to $\alpha_0$, $\beta_0$ and the time variation of the scalar field today as
\begin{equation}
\label{Gdotst}
 \frac{\dot{G}_{\mathrm{cav}}}{G_{\mathrm{cav}}} =2\alpha_0\left(1+\frac{\beta_0}{1+\alpha_0^2} 
 \right)\dot\varphi_{*0}\,.
\end{equation}
This example shows that the variation of the constant and the deviation from General Relativity quantified in terms of the PPN parameters are of the same magnitude, because they are all driven by the same new scalar field.

The example of scalar-tensor theories is also very illustrative to show how deviation from General Relativity can be fairly large in the early universe while still being compatible with Solar system constraints.  It relies on the attraction mechanism toward General Relativity \citep{dn1,dn2}.
 
\paragraph{Cosmological evolution}\ 

Consider the simplest model  of a massless dilaton with quadratic coupling ($\ln A =\frac{1}{2}\beta\varphi_*^2$). Note that the linear case correspond to a Brans-Dicke theory with a fixed deviation from General Relativity. It follows that $\alpha_0 = \beta\varphi_{*0}$ and $\beta_0  = \beta$. As long as $V=0$, the Klein--Gordon equation can be rewritten in terms of the variable $p=\ln a$ as
\begin{eqnarray}
\label{kgqq}
 \frac{2}{3-\varphi_*^{'2}}\varphi_*''
 +(1-w)\varphi_*' =-\alpha(\varphi_*)(1-3w),
\end{eqnarray}
with $w$ the equation of state of the fluid dominating the cosmological dynamics. As emphasized by \cite{dn1}, this is the equation of motion of a point particle with a velocity dependent inertial mass, $m(\varphi_*)=2/(3-\varphi_*^{'2})$ evolving in a potential $\alpha(\varphi_*)(1-3w)$ and subject to a damping force, $-(1-w)\varphi_*'$. During the cosmological evolution the field is driven toward the minimum of the coupling function. If $\beta>0$, it drives $\varphi_*$ toward 0, that is $\alpha\rightarrow0$, so that the scalar-tensor theory becomes closer and closer to General Relativity. When $\beta<0$, the theory is driven away from General Relativity and is likely to be incompatible with local tests unless $\varphi_*$ was initially unrealistically close from 0.

\begin{figure}[htbp]
\centerline{\includegraphics[scale=0.5]{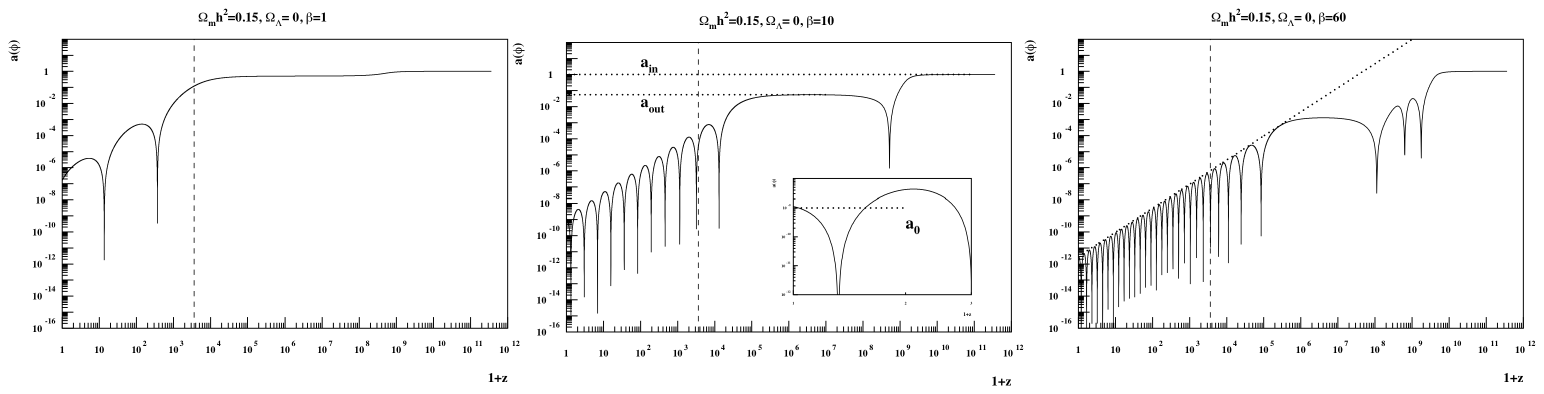}}
\vskip-6cm
  \caption[Dynamics of the attraction mechanism]{Dynamics of a non-minimmally scalar field arising from the Klein--Gordon  equation~(\ref{kgqq}). Since it couples to the trace of the stress-energy tensor, it generate an effective potential during the matter dominated era while it freezes to a constant value during the radiation era. The attraction toward the minimum of the coupling function $a(\varphi)=\ln A(\varphi)=\frac{1}{2}\beta\varphi^2$ is more efficient for large $\beta$. The oscillations during the radiation era are caused by the electron-antielectron, muon-antimuon etc. annihilations. From \cite{couv}.}
  \label{fig-DNmec}
\end{figure}

It follows that the deviation from General Relativity remains constant during the radiation era -- up to threshold effects in the early universe \citep{bbn-Gpichon,couv} and quantum effects \citep{copu} -- and the theory is then attracted toward General Relativity during the matter era. Note that it implies that postulating a linear or inverse variation of $G$ with cosmic time is actually not realistic in this class of models. Since the theory is fully defined, one can easily compute various cosmological observables: late time dynamics \citep{msu}, CMB anisotropy \citep{cmb-G1}, weak lensing \citep{carlowl},  BBN \citep{bbn-Gpichon,couv,coc4a} in a consistent way to confront them with data.

\paragraph{Further WEP preserving theories}\ 

Scalar-tensor theories are a simple example of gravity theories beyond General Relativity. Many such theories have been formulated today and are of interest in particular to study the dark energy problem in cosmology (see e.g., \cite{CLIFTON20121} for a review). The \emph{Lovelock theorem} \citep{10.1063/1.1665613} states that the theory of gravity derived from an action principle. Provided (\textit{1}) the theory is diffeomorphism-invariant, (\textit{2}) the spacetime geometry is described by a Lorentzian $D$-dimensional manifold and the connection compatible with the metric is torsion and metricity free, i.e., it reduces to the Levi-Civita connection, (\textit{3})  the action depends solely on the metric manifold, and (\textit{4}) the field equations are of second order in derivatives of the metric then  the action functional is of the form
$$
S[g]=\int \dd^D x \sqrt{-g}\sum_{i=0}^{(d-1)/2}\beta_i{\cal R}^{(i)}
$$
with $\beta_i$ coupling constants and ${\cal R}^{(i)}$ the curvature invariants of order $i$ of the Riemann tensor,
$$
{\cal R}^{(i)} = \frac{k!}{2^k} \delta^{[\mu_1\nu_1\ldots \mu_k\nu_k}_{[\alpha_1\beta_1\ldots\alpha_k\beta_k]}\prod_{j=1}^i {R^{\alpha_j\beta_j}}_{\mu_j\nu_j}\,.
$$
In $D=4$ dimensions, it reduces to General Relativity plus a cosmological constant together with the Gauss-Bonnet topological term, that does not modify the equations of motion,
$$
S[g]=\int \dd^D x \sqrt{-g}\left[\beta_0 + \beta_1 R + \beta_2\left(R^2-4R_{\mu\nu} R^{\mu\nu} + R_{\mu\nu\alpha\beta } R^{\mu\nu\alpha\beta} \right)    \right]
$$
This tells us that there are 5 ways  in which GR can be modified, by relaxing these assumptions.
respectively,
\begin{enumerate} 
\item Add new field involved in mediating the gravitational force, i.e., new fields coupled to the metric tensor in the Einstein–Hilbert action; this is the situation we have considered here.
\item Work in more than $D=4$ for dimensions, such as, e.g., string theory, Kaluza-Klein theory, or the inclusion of a Gauss-Bonnet term in the action which only becomes relevant in higher dimensions; see Sect.~\ref{subsub5.3} below.
\item Build a higher-order theory whose field equations contain greater than second-order derivatives, see e.g.,the class of dhost theories \citep{NEW_Langlois_2016}.
\item Give up locality, add torsion or break diffeomorphism invariance.
\item Give up on the action principle; which we do not consider in this review.
\end{enumerate}
While this provides a convenient guideline to classify gravity theories beyond General Relativity, this remains an idealized picture since indeed, one can relax several hypothesis simultaneously and that the same theory can be ``read" differently given the way it is written mathematically. Figure~\ref{fig-theories} summarizes the different avenues to extend General Relativity and their relations. Scalar-tensor theories can be considered for many reasons as the simplest extensions of General Relativity. They are part of the larger class of Horndeski theories \citep{NEW_horndenski}, which are the most general 4-dimensional scalar-tensor theory whose Lagrangian leads to second-order equations of motion.  To finish, note that a similar theorem relying on the Hamiltonian formulation was derived by \cite{refHKT}. It suggests another route for extensions beyond General Relativity by considering Dirac’s hypersurface deformation algebra, leading to the idea of emergent spacetime equipped with a deformed general covariance so that modifications of gravity appear even if one does not break the gauge symmetry but instead deform it.

\begin{figure}[htbp]
\vskip-.75cm
 \centerline{\includegraphics[scale=0.4]{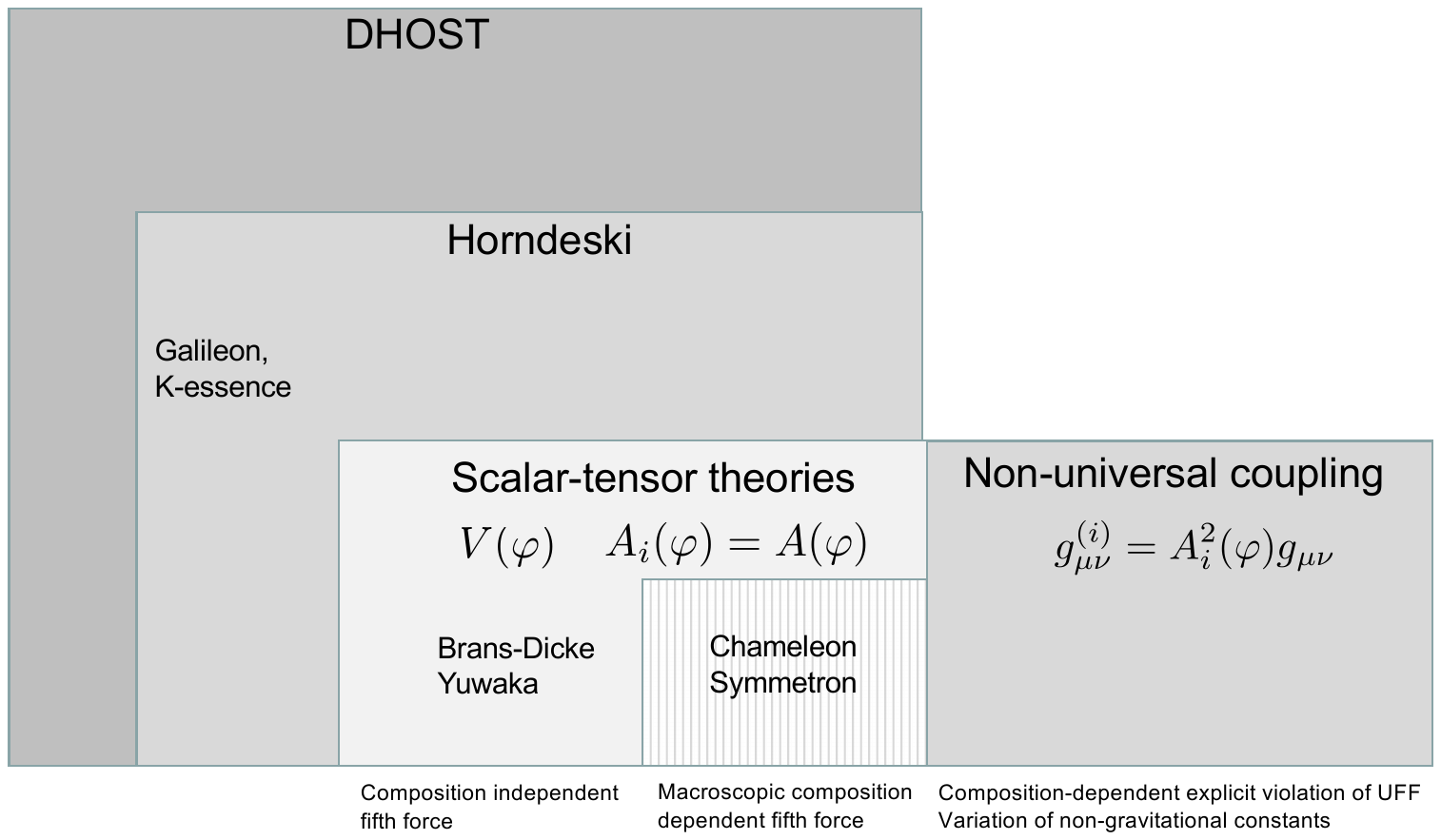}}
 \vskip-1cm
  \caption[Classification of gravity theories beyond General Relativity]{DHOST theories satisfies the WEP at the Lagrangian level (left). They include as subcases the Horndenski theories (among which Galileon and K-essence) and scalar-tensor theories with many models, depending on the choice of the potential and coupling function. They are also a subcase of all non-universal theories (right). While the DHOST theories satisfies the WEP at the microscopic level, note that they may violate if at the macroscopic level, as described in Sect.~\ref{subsub2}.}
  \label{fig-theories} 
\end{figure}

\subsubsection{Making non-gravitational constants dynamical}\label{subsecSTmulti}

\paragraph{Simplistic varying-$\aem$ model}\ 

Given the previous example, it seems a priori simple to cook up a theory that describes a varying fine-structure constant by coupling a scalar field to the electromagnetic Faraday tensor as
\begin{equation}\label{e.BFphi}
 S = \frac{1}{16\pi G} \int\left[R - 2(\partial_\mu\varphi)^2  -\frac{1}{4}B(\varphi)F_{\mu\nu}^2 \right]\sqrt{-g}\dd^4 x+
  S_{\text{matter}}[\psi;g_{\mu\nu}]
\end{equation}
so that the fine-structure is given by $\aem=B^{-1}$. However, such a simple implementation may have dramatic implications on the universality of free fall because the contribution of the electromagnetic binding energy to the mass of any nucleus implies that it become $\varphi$-dependent. One can easily reproduce the derivation of the fifth force in Eqs.~(\ref{e.fifthgeo}) to get
\begin{equation}\label{e.fifthgeo2}
 u^\nu\nabla_\nu u^\mu = F_5^\mu  
 \equiv - \frac{\partial \ln m_A}{\partial \varphi} \left( g^{\beta\mu}+u^\beta u^\mu\right)  \partial_\beta \varphi\,.
\end{equation}
Hence the sensitivity to $\varphi$ is
\begin{equation} \label{deffai}
\alpha_A(\varphi) \equiv \frac{\partial\ln m_A}{\partial\varphi} =  \frac{\partial\ln m_A}{\partial\alpha_i}  \frac{\partial\alpha_i} {\partial \varphi}= \sum_i \alpha_i f_{A,i} s_i(\varphi)
\end{equation}
where the different sensitivity coefficients are $f_{A,i}$ and $s_i$ are defined in Eqs.~(\ref{e.fA}) and~(\ref{def_sphi}) in Appendix~\ref{app2}. Indeed with the choice of a scalar field with dimension of mass, one would have got
$$
\alpha_A(\phi) \equiv M_*\frac{\partial\ln m_A}{\partial\phi} 
$$
making use of the relation~(\ref{phivarphi}).

Let us note before proceeding that in $D=4$ dimensions such a simple coupling cannot be eliminated by a conformal rescaling $g_{\mu\nu}=A^2(\varphi)g_{\mu\nu}^*$ since
$$
 \int B(\varphi)g^{\mu\rho}g^{\mu\nu}F_{\nu\sigma}F_{\rho\sigma}\sqrt{-g}\dd^D x
 \longrightarrow
 \int B(\varphi)A^{D-4}(\varphi)g_*^{\mu\rho}g_*^{\mu\nu}F_{\nu\sigma}F_{\rho\sigma}\sqrt{-g_*}\dd^4 x
$$
so that the action is invariant in $D=4$ dimensions, i.e., if included in a general scalar-tensor theories as discussed in Sect.~\ref{subsecST} it will be present both in Einstein and Jordan frames so that the violation of UFF appears in both frames.

\paragraph{Amplitude of the violation of the UFF}\ 

In the case at hand, the $\varphi$-dependent part of the mass arises from the electromagnetic binding energy that we can estimate thanks to the Bethe--Weiz\"acker formula~(\ref{bethe}). Using the expression~(\ref{e.mass0}) and assuming that the protons and neutrons all have the same mass $m_{\rm N}$ and neglecting the $\varphi$--dependent contributions of the binding energies of the proton and neutron, it follows that
\begin{eqnarray}
 \alpha_A = \frac{\partial\ln m_A}{\partial\varphi}& \simeq& -98.25  \aem  \frac{Z(Z-1)}{A^{1/3}}   \times    \frac{1\unit{MeV} }{A m_{\rm N}} \times(\partial_\varphi\ln\aem)_0 \nonumber\\
  &\sim& 10^{-3}  \frac{Z(Z-1)}{A^{4/3}} (\partial_\varphi\ln B)_0\nonumber
\end{eqnarray}
since $\aem=1/B$. Then, the level of the violation of the universality of free fall is expected to be of the level of $\eta_{12}\sim \alpha_\oplus| \alpha_{1}- \alpha_{2} |\sim 10^{-6}X(A_1,Z_1;A_2,Z_2)\times$ $(\partial_\phi\ln B)^2_0$. Since the factor $X(A_1,Z_1;A_2,Z_2)$ typically ranges as $\mathcal{O}(0.1-10)$, we deduce that $(\partial_\varphi\ln B)_0$ has to be very small then ${\cal O}(10^{-5}-10^{-4})$ for the Solar system constraints to be satisfied -- see Fig.~\ref{fig-deseul0} for a concrete detailed example. It follows that today the scalar field has to be very close to the minimum of the coupling function $\ln B$. This led to the idea of the \emph{least coupling mechanism} \citep{damour94a,damour94b} discussed in Sect.~\ref{subsub1}.  The constraints on such models are described in Sect.~\ref{subsub0}.

\subsubsection{Summary}\ 

This example is indeed very simplistic because it only takes into account the effect of the electromagnetic binding energy (see Sect.~\ref{subsec22}) but it clearly illustrates that the amplitude of variation of the constants and of the violation of the universality free fall are linked and both determined by the $\varphi$-dependence of the mass, $m_A(\varphi)$ of the extended objects starting from the nuclei. It explicitly shows that one cannot couple a light field blindly to, e.g., the Faraday tensor, to make the fine-structure constant dynamics and that some mechanism for reconciling this variation with local constraints, and in particular the university of free fall, will be needed. Hence, the use of both local and cosmological constraints will be an asset to constrain all physical models. 
\paragraph{Generic action}\ 

To finish, let us note that this hints that generic varying-constants theories will take the form
\begin{eqnarray}\label{e.genericS}
 S &=& \frac{1}{16\pi G} \int\left[R - 2(\partial_\mu\varphi)^2 -4V(\varphi) \right] \sqrt{-g}\dd^4 x  \nonumber\\
 && \qquad\qquad\qquad\qquad\qquad+  \!\!\!\! \!\!\!\! \!\!\!\!\sum_{\hbox{matter fields}}\!\!\!\! \!\!\!\! \!\!\!\! S_{\text{matter}}[\psi_i;A^2_i(\varphi)g_{\mu\nu}],
\end{eqnarray}
that would depend on a set of coupling functions $A_i(\varphi)$ that will determine the strength of the fifth force acting on each fundamental fields and from which one can compute the fifth force on a test-body once its chemical composition and mass are specified. It is naturally written in Einstein frame and contrary to the universal scalar-tensor theories, one cannot define a unique and universal Jordan frame. Hence, one shall be careful when referring to densities etc. since one would need to precise in which frame it is defined and how it is measured. Picking-up a Jordan frame  for a component $i$, then all the other components $j\not= i$ will have coupling $A_i/A_j$ and then experience a fifth force. In particular models, see e.g., Sect.~\ref{subsec81}, one can construct an approximate ``hadronic'' frame to disentangle between composition-independent and composition-dependent effect.

\paragraph{Generic weak and cosmological solutions of the Klein--Gordon  equation}\ 

Before we turn to specific models, the action~(\ref{e.genericS}) has some generic properties, in particular concerning the profile et evolution of the scalar field since the Klein--Gordon  equation~(\ref{Boxvarphi}) generalises to
\begin{equation}\label{e.KGgen2}
\Box\varphi = dV(\varphi)/d\varphi + \sum_i 4\pi G\alpha_i(\varphi)(\rho_i-3P_i)
\end{equation}
using that $T_i=3P_i-\rho_i$ so that one identifies the effective potential
\begin{equation}\label{e.Veff2}
V_{\rm eff}=  V +4\pi G \sum_i (\rho_i-3P_i) \ln A_i(\varphi) 
\end{equation}
The sum may be on different species that the sum in Eq.~(\ref{e.genericS}) since the latter concerns fundamental fields while the former is on macroscopic components, obtained as averaging, hence it could be baryonic matter, dark matter, radiation in cosmology, gas of different chemical compositions, etc.

First, let us consider this equation in a Minkowski spacetime. Assuming that $V$ has a minimum so that locally it reduces to a mass term, and that the couplings $\ln A_i$ are linear, it reduces to
$$
-\ddot\varphi+\Delta\varphi= \frac{\dd V_{\rm eff}}{\dd\varphi} =  m_\varphi^2\varphi + 4\pi G\sum_i \alpha_i (\rho_i-3P_i).
$$
This equation can be solved by Fourier decomposing $\varphi$ to get a solution of the homogeneous equation and then use the Yukawa Green function to determine a particular solution, hence
\begin{equation}
\varphi = \int \frac{\dd^3{\bf k}}{(2\pi)^{3/2}} \varphi_{\bf k}\hbox{e}^{i ({\bf k}.{\bf x}-\omega_k t)}
-\int \dd^3{\bf x}' G\alpha_i\left[\rho_i({\bf x}')-3P_i({\bf x}')\right]  \frac{\hbox{e}^{-m_\varphi |{\bf x}-{\bf x}'|}}{|{\bf x}-{\bf x}'|}\nonumber
\end{equation}
with the condition $\omega^2_k={\bf k}^2 + m_\varphi^2$ and $\varphi_{\bf k}^*=\varphi_{-\bf k}$ for a real-valued scalar field. In the case of a point mass, $\rho=M_A\delta({\bf x})$, the static solution is indeed the usual Yukawa potential for a point particle
$$
\varphi_{\rm pp}({\bf x}) = -\frac{G\alpha_AM_A}{r}\hbox{e}^{-m_\varphi r}
$$
while for a homogeneous sphere of density $\rho_A$ and radius $R_A$,
$$
\varphi_{\rm sphere}({\bf x}) = \varphi_{\rm pp}({\bf x}) \Phi(m_\varphi R_A)
$$
where the shape factor $\Phi$ is given in Eq.~(\ref{e.shapePhi}). This confirms the standard result of Eq.~(\ref{eq_eta}). The case of quadratic coupling  was demonstrated by \cite{Hees:2018fpg} to ``naturally'' lead to screening or scalarization, which makes their study in the sSlar system particularly interesting. Such a solution will be useful to study the spatial variation of the constants in the Solar system.

Then, on cosmological scales, as long as the spacetime can be described by a Friedmann--Lema\^{\i}tre spacetime with scale factor $a$ -- see Appendix~\ref{app3} for definitions --, the conservation equation~(\ref{matter*}) that holds for each component $i$ independently reduces to
$$
\dot\rho_i +3H(\rho_i+P_i)= (\rho_i-3P_i) \alpha_i(\varphi) \dot \varphi.
$$
Assuming a constant equation of state $w_i=P_i/\rho_i$ and rewriting the r.h.s. as $\rho_i (1-3w_i) (\ln A_i)^.$, it and can be integrated as
$$
\rho_i(a) = \rho_{i0} a^{-3(1+w_i)}\left[\frac{A_i(\varphi)}{A_i(0)} \right]^{4-3(1+w_i)}.
$$
This shows that the cosmic dilution will be modified in a composition-dependent way. While radiation is not affected, baryonic matter and eventually dark matter will not redshift as $a^{-3}$, which could be interpreted effectively as a non-vanishing equation of state.

\subsection{Hiding light degrees of freedom}\label{subsub2}

While the introduction of a new light degree of freedom is an easy recipe to extend General Relativity and to make some constants dynamical, they face the problem to generate a too large violation of the UFF or of the tests of General Relativity in the Solar system, in particular when the new field is assumed to be light in order to be a dark energy candidate, typically $m\sim10^{-33}$~eV for a scalar field \citep{PhysRevD.37.3406}. Hence, the scalar field shall decouple from the matter sector, at least on Solar system scales. Several mechanisms have been considered to implement such a screening; see e.g.,\cite{NEW_Khoury:2010xi}.

\subsubsection{Decoupling mechanisms} \label{subsecdecoupling}

Two mechanisms fall in this category in which the coupling of the new degree of freedom becomes small in the Solar system.

\paragraph{Least coupling principle}\ 

First, as we have already foreseen in the previous paragraph~\ref{subsecSTmulti}, if the field potential and coupling function enjoy the same minima, then the scalar field is attracted toward its minimum during the cosmological evolution. Hence its coupling to standard matter becomes weak. This idea of the \emph{least coupling mechanism} was first proposed by \cite{dn1,dn2} for scalar-tensor theories and adapted by \cite{gasperini02} to quintessence models and then developed by \cite{damour94a,damour94b} for non-universal couplings. It derives from the attraction of the scalar-tensor theory toward General Relativity during the cosmological evolution since the Klein--Gordon  equation~(\ref{Boxvarphi}) is dictated by the effective potential
\begin{equation}\label{e.Veff}
V_{\rm eff}(\varphi_*)=V(\varphi_*)-4\pi G_* T_*\ln A(\varphi_*).
\end{equation}
As later detailed in Sect.~\ref{subsub1} can be implemented by
\begin{equation}
A= \exp\left(\frac{1}{2}\beta\varphi^2\right)
\,\qquad
V= \frac{1}{2}m_\varphi \varphi^2
\end{equation}
as originally considered by  \cite{dn1,dn2} and then for the light-dilaton model by \cite{damour94a,damour94b},  or as
\begin{equation}
A=\exp\left(-\lambda\varphi\right)
\,\qquad
V\propto \exp\left(-\varphi\right)
\end{equation}
as for the runaway dilaton \citep{gasperini02,damourrunaway}.

\paragraph{Symmetron mechanism}\ 

The \emph{symmetron mechanism} \citep{PhysRevLett.104.231301,PhysRevD.84.103521} is a mechanism that makes the strength of the coupling environmentally dependent so that it becomes sufficiently small in regions of high density, or equivalently in the Newtonian limit of high Newtonian potential. It relies on the properties that  the effective potential vacuum has an expectation value of the scalar field is nonzero in low-density environment and the fact that the $Z_2$-symmetry (i.e., $\varphi_*\leftrightarrow-\varphi_*$) is restored in high-density regions, so that the field have a zero vev in such regions and does not couple to matter. It can be implemented with the choice
\begin{equation}
A(\phi_*)=1+\frac{\phi_*^2}{2M^2}
\,\qquad
V(\phi_*)=-\frac{1}{2}\mu^2\phi_*^2+\frac{1}{4}\lambda\phi_*^4\,.
\end{equation}

While the two mechanisms lead to a suppression of the fifth force in the Solar system today, they differ in the sense that the least coupling principle depends on the actual cosmological value of the scalar field, $\varphi_0$ and thus on the strength of the attraction toward General Relativity and its initial value at the end of inflation while the symmetron mechanism depends on the local matter density. It follows that they have different phenomenologies for cosmology and the variation of the constants. Let us also stress that for non-universal couplings,  the effective potential is given by Eq.~(\ref{e.Veff2}) and  the mechanism requires the potential and the coupling functions have the same minima.

\subsubsection{Chameleon mechanism}\  \label{sec_cham}

\paragraph{Mechanism}\ 

Since the ``local" mass of the scalar field is related to the range of the fifth force it mediates,  another possibility to screen its effect is to make it heavy in high-density environment and light in low-density regions. The chameleon mechanism is an interesting model of this kind \citep{cameleon,cameleon2}. It is defined as a scalar-tensor theory as described in Sect.~\ref{subsecST} with the special feature that the potential $V$ and the coupling function $A$ do not share the same minimum. Hence, the scalar field acquires a density-dependent mass that controls the range of the associated fifth force, from small-range in high-density regions to long-range in low-density regions. This makes it not only a promising candidate to explain the accelerated expansion of the Universe, but also a potentially easily testable model. It is defined by the action~(\ref{actionEF}) in the Einstein frame with a coupling constant associated to the matter fields,
\begin{equation}
A_i(\phi_*) = \hbox{e}^{\beta_i\phi_*/M_*}
\end{equation}
and a  potential monotonically decreasing, tending to 0 at infinity with a null derivative. A prototypical potential is given by the Ratra–Peebles inverse power-law potential \citep{PhysRevD.37.3406} of energy scale $\Lambda$ and exponent $n$
\begin{equation}
V(\phi_*) = \Lambda^4\left(1+\frac{\Lambda^n}{\phi_*^n}\right),
\end{equation}
in the case of a universal coupling $\beta$. Otherwise the minimum will depend on both the density and chemical composition. Clearly, the minimum of the effective potential depends on the local density of matter as
$$
\phi_{\rm min}(\rho) = \left(\frac{n M_*\Lambda^{n+4}}{\beta\rho}  \right)^{1/n+1}.
$$
In a medium of constant density, the field is expected to relax exponentially to this minimum  on a typical distance of the order of its local Compton wavelength $\lambda_c(\rho)=1/\sqrt{V''_{\rm eff}([\phi_{\rm min}(\rho)]}$.

\paragraph{Macroscopic violation of the UFF in a theory satisfying the WEP}\ 

To characterize the effect of the screening on the scalar  fifth force, consider a spherical body with density $\rho$, radius $R$ and mass $M$ in a outside medium of density $\rho_{\rm out}$. The scalar field will relax to the minimum of its potential $\phi_{\rm in}$ inside if the sphere only if the object is dense enough for the field’s Compton wavelength to be much smaller than $R$, and to $\phi_{\rm out}$ outside. One can define the skin thickness $\Delta R$ for $\phi$ to relax to $\phi_{\rm in}$ and the thin-shell parameter
\begin{equation}
Q=3\frac{\Delta R}{R}.
\end{equation}
One shall then consider two effects. First, outside the spherical body,  the environmental screening induces a Yukawa suppression on a typical length $\lambda_c(\rho_{\rm out})$ so that the fifth force will cancel for any  test mass placed further than this distance. Hence the interaction between two bodies is Yukawa-suppressed in a high-density environment but may still be significant in low-density regions. Then if $Q\ll 1$, the field will settle to  $\phi_{\rm in}$  so that $\nabla\phi$ will vanish but on a thin portion of the body. It follows that only a thin shell contributes to the fifth force once integrated on the whole body so that it becomes marginal compared to that of the Newtonian gravity and is effectively hidden to experiments and the source is \emph{screened}. This is in  particular the reason for which planetary orbits do not show any deviation from pure Newtonian gravity: even if the chameleon’s Compton wavelength in space becomes larger than typical distances between planets, the chameleon is hidden as soon as planets are screened. On the opposite, if $Q\sim1$, the chameleon field does not reach its minimum within objects that just appears as a perturbation for $\phi$ and then subject to a chameleon-induced fifth force. As a consequence, the amplitude of the fifth force between two spherical masses can be shown to be
$$
F_{12}(r)= 2\beta^2Q_1Q_2 \frac{G M_1 M_2}{r^2}\,.
$$
Hence two macroscopic bodies with different thin-shell parameters undergo different fifth forces. Even if the fifth force is universal for point particles, it will depend on the density and shapes of macroscopic bodies. This gives an example of violation of the UFF at the macroscopic level while it still holds at the microscopic level. The screening analysis of the MICROSCOPE satellite was a key issue to demonstrate that, contrary to what was expected, it was not designed to test chameleon models, even if it were in space; see Fig.~\ref{fig-cham}.

\begin{figure}[htbp]
\vskip-.75cm
 \centerline{\includegraphics[scale=0.4]{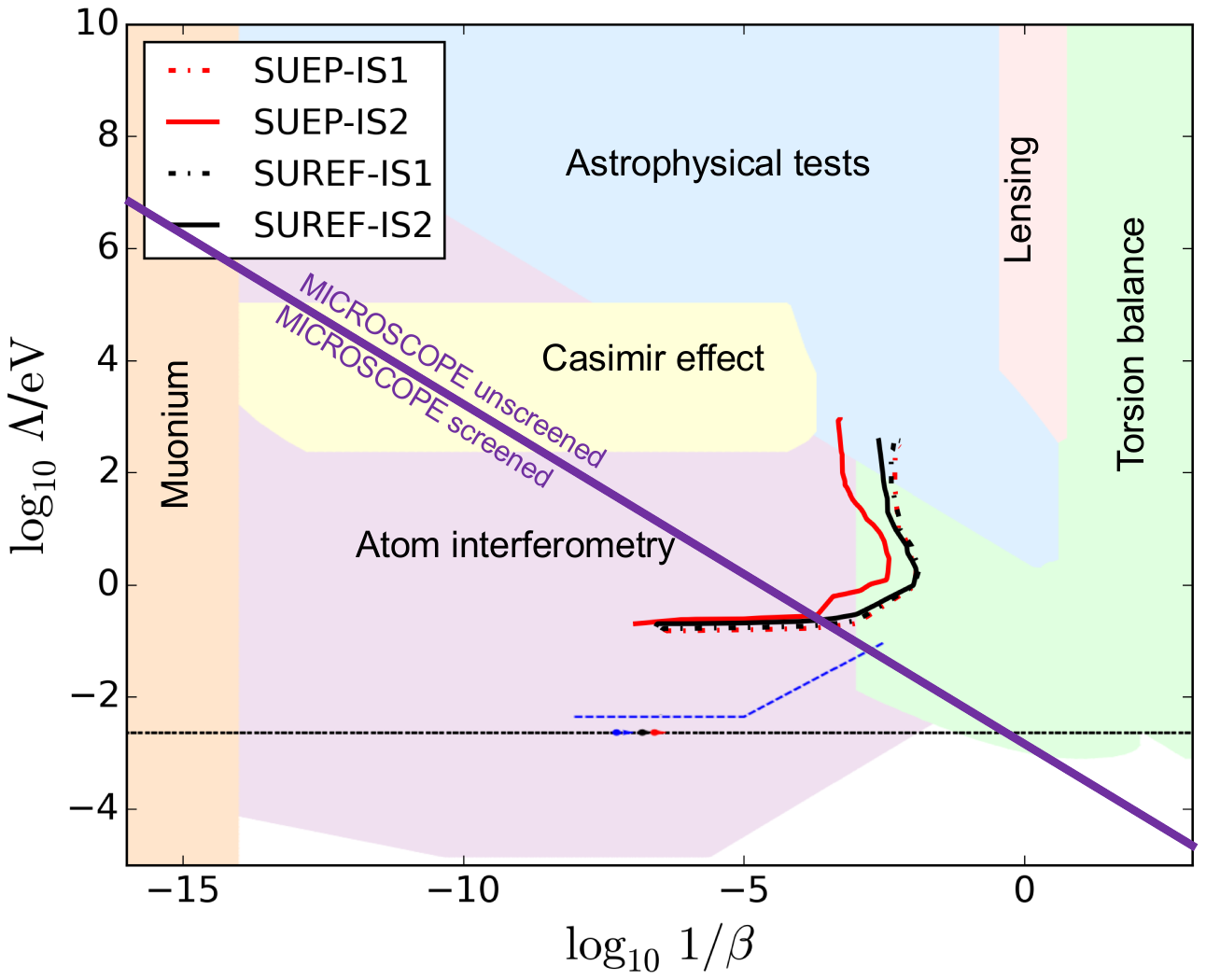}}
 \vskip-.5cm
  \caption[Parameter space and constraints of the chameleon mechanism]{ Chameleon’s parameter space for $n=1$ adapted from \cite{Burrage:2017qrf}.  It includes a purple line indicating the regimes in which the MICROSCOPE satellite is screened/unscreened \citep{NEW_Pernot-Borras:2019gqs} as well as the constraints derived from the stiffness measurement sessions \citep{NEW_Pernot-Borras:2021edr} -- the region excluded at 2$\sigma$ is above the four lines described in the legend. The horizontal doted line denotes the energy scale of dark energy. The experiments include atomic interferometry \citep{hamilton_atom-interferometry_2015,jaffe_testing_2017}, E\"ot-Wash group's torsion balance experiments \citep{upadhye_dark_2012,PhysRevLett.98.021101}, Casimir effect measurements \citep{brax_detecting_2007,PhysRevD.75.077101}, astrophysics tests \citep{Jain_2013,Cabr__2012,PhysRevD.97.104055} and lensing \citep{10.1093/mnras/stw1617}, precision atomic tests \citep{PhysRevD.83.035020, PhysRevD.82.125020} while the dashed-blue line corresponds to microsphere  experiments \citep{PhysRevLett.117.101101} and the blue and red points to neutron interferometry \citep{LEMMEL2015310,PhysRevD.93.062001}.}
 \label{fig-cham}
\end{figure}

\paragraph{Constraints}\ 

We refer to \cite{NEW_Joyce:2014kja,Burrage:2017qrf,NEW_Brax:2018iyo} for reviews on the experimental tests of the chameleon, see Fig.~\ref{fig-cham}. While most of them rely on fifth force effects, screened scalar-tensor models can be constrained by means of gravitational redshift measurements performed with atomic clocks \citep{Brax:2022olf,Levy:2024vyd} and  unconstrained regions of the chameleon parameter space could be ruled out by future redshift experiments with some technical issues to be resolved such as the influence of the nucleus on the chameleon field at the electron cloud. The analysis of the chameleon in space experiments such as MICROSCOPE requires to propagate the field inside the instrument to take into account the screening. Two numerical tools \citep{Briddon:2021etm,Levy:2022xni}  are now available to compute the chamelon field profile and the fifth force it generates for realistic geometries. All this limits the constraints that can be obtained \citep{NEW_Pernot-Borras:2020jev}. Note that at low orbits, one has to model the shape of the Earth \citep{Burrage:2014daa}.  Still, it was demonstrated that a constraint can be derived from the modelization of the electrostatic stiffness \citep{NEW_Pernot-Borras:2021edr,NEW_Berge:2021yye}.

\paragraph{Dependence of the constants on the mass density of the environment}\ 

Indeed, the chameleon mechanism can easily be implemented with non-universal couplings and for any of the model described in Sect.~\ref{subsec81}. They will induce a dependence of the fundamental constant on the energy density $\rho$ of the environment due to the shift of the field  \citep{op2}. \cite{Brax:2013doa} studied the effect of screening to the detectability of the variation of the constants in particular in astrophysical environment showing that if the dipolar variation of $\aem$ -- see Sect.~\ref{subsec-dipalpha} below, -- were to be confirmed this would rule out screened modified gravity models.  The cosmological variation of $\aem$ in chameleon models was investigated in \cite{brax,brax2}.  Models based on the Lagrangian~(\ref{olive}) and exhibiting the chameleon mechanism were investigated by \cite{op2}. Chameleon models coupled to $F^2$ -- see Sect.~\ref{subsub0} -- were discussed by \cite{NEW_Brax:2010uq} and constrained with Solar physics \citep{NEW_OShea:2024jjw}. The axion-homeopathy model was extended to an  axio-ahameleon mechanism by \cite{NEW_Brax:2023qyp} involving  two light scalar fields, an axion and a Brans-Dicke dilaton.

In particular, interstellar absorbing clouds have typical densities $10^{-10}$ smaller than in terrestrial environments \citep{levmusp2} so that the scalar field driving the dynamics of the fundamental constant will be much lighter in space will screened for Earth measurements. Besides, for all models enjoying screening, two absorbing regions at the same redshift would show different values of the fundamental constants if they lie in screened and unscreened regions respectively. This suggests new ways of analyzing data where a tomographic description of the Universe mapping screened and unscreened regions would be correlated to the measured variations of constants \citep{NEW_Cabre:2012tq}.  

The study of molecular CO, CH, NH$_3$ and CH$_3$OH absorption spectra, discussed in details in Sect.~\ref{secmolqso}, have actually allowed to study fundamental constants in the different regions of the Milky Way, opening the way to set limits on matter-density dependence of fundamental constants in different regions of our Galaxy with different densities. The possible shift in the value of $\mu$ in the Milky Way  described in Sect.~\ref{subsec23mw} was related  by \cite{mu-lev,levmusp2,levmusp2b} to the model of \cite{op2} to conclude that such a shift was compatible with a chameleon-like mechanism. Similarly, \cite{NEW_Truppe:2013ypa}
constrained the variation of $\aem$ and $\mu$ between high and low-density environments of the Earth and the interstellar medium thanks to CH and OH, in the Milky way.

A proper and in depth analysis of the dependence of the constant on the local energy density still remains to be performed.

\subsubsection{Environment-dependent dilaton model} \label{sec_rhodil}

Following \cite{NEW_Brax:2018iyo}, the quadratic dilaton model was combined \citep{NEW_Brax:2010gi} with a runaway potential,
\begin{equation}
A(\varphi)=\hbox{e}^{\frac{1}{2}\beta\varphi^2}\sim 1+\frac{1}{2}\beta\varphi^2\,
\quad
V(\varphi)=V_0\hbox{e}^{-\lambda \varphi},
\end{equation}
to make the dilaton dependent on the trace of the matter stress-energy tensor, i.e., on the local energy density $\rho$ for non-relativistic matter. The effective potential~(\ref{e.Veff2}), $V_{\rm eff}=V_0\hbox{e}^{-\lambda \varphi}+2\pi G\rho \beta \varphi^2$ has a minimum explicitly given by
$$
\varphi_{\rm min} = \lambda^{-1} W\left(\frac{\lambda^2V_0}{2\pi G \beta \rho} \right)
$$
where $W$ stands for the Lambert function, i.e., the inverse function of $x\hbox{e}^x$. As for the chameleon, the field is screened in high-density environments. The cosmological dynamics was studied in \citep{NEW_Brax:2022uyh} and the effect on Lunar laser ranging and laboratory experiments detailed in \cite{NEW_Fischer:2023koa,NEW_Fischer:2024eic}. The mechanism is similar to the chameleon and can be combined with composition-dependent couplings, even though it has not been investigated in the literature so far.

\subsubsection{Conclusion}

The interaction between light scalar fields and matter are generically highly constrained by Solar system tests unless, as discussed their couplings to ordinary matter are much suppressed relative to gravitational strength. This is a major hindrance to construct realistic models of light dilatons coupled to matter. We have described four  mechanism of decoupling or screening that allow these models to evade local constraints. They are summarized in Table~\ref{tab-screen}. Even though they have been discussed within composition-independent scalar-tensor theories, they are easily adapted for all the models described in Sect.~\ref{subsec81}.

\begin{table}[htbp]
\caption[Decoupling and screening mechanism]{Main decoupling and screening mechanisms in scalar-tensor gravty.}
\label{tab-screen}
\centering
{\footnotesize
\begin{tabular}{lcccc}
\toprule
Model  & $V$  & $A$ & Environment  \\
 &  & &dependence &         \\
\hline
Light dilaton & $\frac{1}{2}m\varphi^2$ & $ \frac{1}{2}\beta\varphi^2$ & No  \\
Runaway dilaton & $\hbox{e}^{-\varphi}$ & $\hbox{e}^{-\lambda \varphi}$ & No  \\
Environment-dep. dilaton & $\hbox{e}^{-\lambda \varphi}$ & $ 1+\frac{1}{2}\beta\varphi^2$ & Yes&  \\
Chameleon & $\hbox{e}^{\beta\varphi}$ & $\Lambda^4\left(1+\frac{\Lambda^n}{\varphi^n}\right)$ & Yes  \\
Symmetron & $-\frac{1}{2}\mu^2\varphi^2 + \frac{1}{4}\lambda\varphi^4$ & $1+\frac{1}{2}\beta\varphi^2$ & Yes  \\
\bottomrule
\end{tabular}}
\end{table}

\subsection{Searching for new degrees of freedom}\label{subsec314}

This section has illustrated the variety of the extensions of standard framework. As soon as a new degree of freedom is introduced it modifies the equation of motion and consistently a new equation of motion for this degree of freedom follows. Hence the set of equations remains consistent and let no freedom to specify an ad hoc law of variation, i.e., an arbitrary function of time or redshift. This also means that the constants have to be promoted to a dynamical field at the level of the action and indeed not in field equations derived assuming these parameters were constants.

The new degree of freedom can act as \emph{matter} if it is not coupled to the standard matter fields. As such, it only modifies the solution of the Einstein field equations through its stress-energy tensor. This implies that the relative abundance of this new species is arbitrary. When coupled to standard matter, it mediates a \emph{fifth force} that can be long-range if the field is light enough. Among those theories, a large class of models involve universal coupling so that the WEP still holds at the microscopic level, but not necessarily for macroscopic objects; see e.g., chameleon models. When the field is non-universally coupled, the phenomenology exhibits a violation of the UFF and a variation of the fundamental parameters. Both can be tested and are well-constrained, as we shall see, in the Solar system. For this reason, viable theories usually incorporate a screening mechanism. The new degree of freedom can remain a negligible part of the total energy budget but act through the force it generates.

The existence of these new degrees of freedom are strongly motivated by the hypothesis of the existence of a dark matter and dark energy components required by the cosmological model. It is important to remind that their interpretation as \emph{matter} or \emph{new force} depends on the fact that they modify the Friedmann equations through their energy, which is then non-negligible, or by modifying the gravitational interaction, in which case their self-energy may remain negligible. The question is then to prove their existence, characterize them and their interactions, see e.g., Fig.~\ref{fig-newdof}. This is a large research program that bridges particle physics, gravity and cosmology. 

Many tools can be conjointly used: the study of the expansion of the universe and the growth of its large-scale structure to constrain the dark energy equation of state and the properties of dark matter, the use of the large-scale structure  as first proposed in \cite{Uzan:2001} to test the Einstein field equation, in particular thanks to weak lensing, the test of the distance duality relation and many laboratory to constrain a dark matter component. Fundamental constants play a key role in this program and complement the long list of tests of General Relativity in the Solar system \citep{will-llr} and on astrophysical scales \citep{Uzan:2003zq,ugrg}.

\begin{figure}[htbp]
\vskip-.5cm
 \centerline{\includegraphics[scale=0.35]{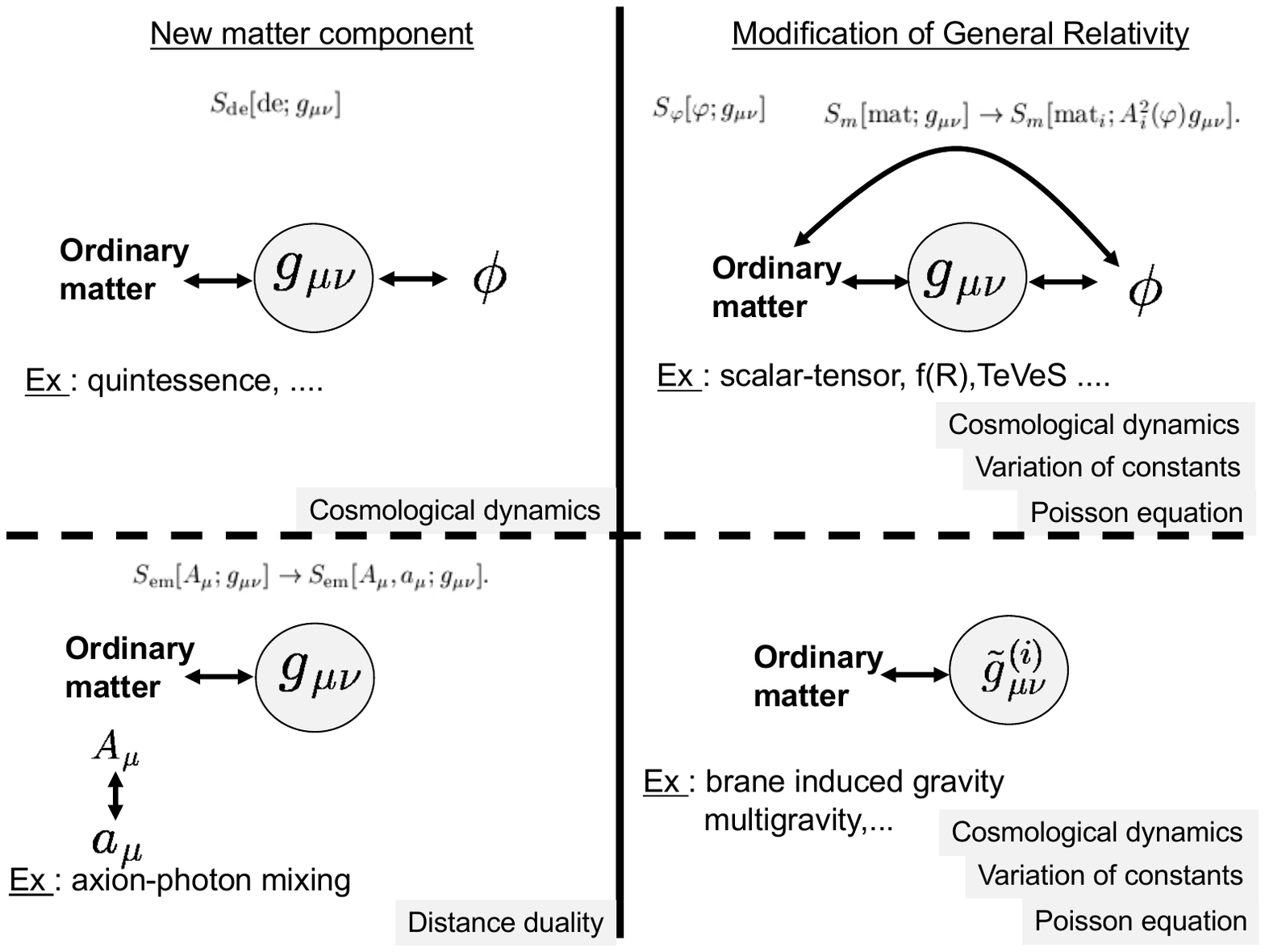}}
 \vskip-.25cm
  \caption[Nature and coupling of new degrees of freedom]{The discussion on physics the standard model can be addressed by trying to unveil the existence of a new dynamical degrees of freedom, their nature and couplings to the standard model fields. While it can be considered as a matter component if it remains universally coupled, it mediates an interaction otherwise. This is in particular the case for non-universal coupling that can be revealed by the variation of the fundamental constants. From \cite{NEW_Uzan:2004my}.}
  \label{fig-newdof}
\end{figure}

\subsection{High-energy theories and varying constants}\label{subsub5.3}

Many high-energy theories involve scalar fields that couple non-universally do the standard model. It seems a generic feature of higher-dimensional gravity theories and of string theories to lead to low-energy effective theories with dynamical constants. Hence, such theories have to face the question: ``why are the low-energy constants so stable?'' which sets a challenge for them to exhibit a stabilisation mechanism.

\subsubsection{Higher-dimension and Kaluza--Klein theory}

Extra-scalar field with non-minimal couplings naturally appear when compactifying a higher-dimensional theory. As an example, let us consider the 5-dimensional Einstein--Hilbert
action (\cite{peteruzanbook}, Chapt.~13)
$$
 S=\frac{1}{12\pi^2 G_5}\int\bar R\sqrt{-\bar g}\dd^5 x.
$$
Decomposing the 5-dimensional metric $\bar g_{AB}$ as
$$
 \bar g_{AB} = \left(
\begin{array}{cc}
  g_{\mu\nu}+\frac{A_\mu A_\nu}{M^2}\phi^2 & \frac{A_\mu}{M}\phi^2\\
  \frac{A_\nu}{M}\phi^2 & \phi^2 \\
\end{array}
 \right),
$$
where $M$ is a mass scale, we obtain
\begin{equation}\label{KKaction}
 S=\frac{1}{16\pi G_*}\int\left(R - \frac{\phi^2}{4M^2}F^2\right)\phi\sqrt{-g}\dd^4 x\,,
\end{equation}
where the 4-dimensional gravitational constant is $G_*=3\pi G_5/4\int\dd y$. The scalar field couples explicitly to the kinetic term of the vector field and cannot be eliminated by a redefinition of the metric: again, this is the well-known conformal invariance of electromagnetism in four dimensions discussed in Sect.~\ref{subsecSTmulti} in which we explicitly showed that such a term induces a variation of the fine-structure constant as well as a violation of the universality of free-fall. Such dependencies of the masses and couplings are generic for higher-dimensional theories and in particular string theory. It is actually one of the definitive predictions for string theory that there exists a dilaton, that couples directly to matter \citep{taylor88} and whose vacuum expectation value determines the string coupling constants \citep{witten84}.

In the models by \cite{kaluza21} and \cite{klein1926} the 5-dimensional spacetime was compactified assuming a single spatial extra-dimension   with topology $S^1$ and radius $R_{\text{KK}}$. It follows that any field $\chi(x^\mu,y$) can be Fourier transformed along the compact dimension (with coordinate $y$), so that, from a 4-dimensional point of view, it gives rise to a tower of of fields $\chi^{(n)}(x^\mu)$ of mass $m_{\mathrm{n}}=n R_{KK}$. At energies small compared to $R_{KK}^{-1}$ only the $y$-independent part of the field remains and the physics looks 4-dimensional.

Assuming that the action~(\ref{KKaction}) corresponds to the Jordan frame action, as the coupling $\phi R$ may suggest, the gravitational constant and the Yang--Mills coupling associated with the vector field $A^\mu$ must scale as 
\begin{equation}
 G\propto \phi^{-1}, \qquad
 g_{YM}^{-2} \propto \phi^2/G \propto \phi^3.
\end{equation}
As explained in Eq.~(\ref{gcav}),  $G$ is not the gravitational constant that would be measured in a Cavendish experiment that is given by $G_{\mathrm{cav}}\propto G_*\phi^{-1}\left(1+\frac{1}{2\phi+3}\right)$. This was generalized to the case of $D$ extra-dimensions \citep{cremmer77} to get
\begin{equation}
\label{kkDdim}
G\propto \phi^{-D},\quad
\alpha_i(m_{\text{KK}})=K_i(D)G\phi^{-2}
\end{equation}
where the constants $K_i$ depends only on the dimension and topology of the compact space \citep{weinberg83b} so that the only fundamental constant of the theory is the mass scale $M_{4+D}$ entering the $4+D$-dimensional theory. A theory on ${\cal M}_4\times {\cal M}_D$ where ${\cal M}_D$ is a $D$-dimensional compact space generates a low-energy quantum field theory of the Yang--Mills type related to the isometries of ${\cal M}_D$. For instance \cite{witten81} showed that for $D=7$, it can accommodate the Yang--Mills group $SU(3)\times SU(2)\times U(1)$. The two main problems of these theories are that (\textit{1}) one cannot construct chiral fermions in four dimensions by compactification on a smooth manifold with such a procedure and (\textit{2}) that gauge theories in five dimensions or more are not renormalizable.

In such a framework, the variation of the gauge couplings and of the gravitational constant arises from the variation of the size of the extra dimensions so that one can derives stronger constraints that by assuming independent variation, but at the expense of being more model-dependent. Let us mention the works by \cite{marciano84} and \cite{wu86}  in which the structure constants at lower energy are obtained by the renormalization group, and the work by \cite{vene02} for a toy model in $D \ge 4$ dimensions, endowed  with an invariant UV cut-off $\Lambda$, and containing a large number $N$  of non-self-interacting matter species. 

\cite{bbnkolb} used the scalings~(\ref{kkDdim}) to constrain the time variation of the radius of the extra dimensions during primordial nucleosynthesis to conclude that$|\Delta R_{\text{KK}}/R_{\text{KK}}|<1\%$. \cite{barrow87} took the effects of the variation of $\as\propto R_{\text{KK}}^{-2}$   and deduced from the helium-4 abundance that $|\Delta R_{\text{KK}}/R_{\text{KK}}|<0.7\%$ and $|\Delta R_{\text{KK}}/R_{\text{KK}}|<1.1\%$ respectively for $D=2$ and $D=7$ Kaluza--Klein theory and that $|\Delta R_{\text{KK}}/R_{\text{KK}}|<3.4\times10^{-10}$ from the Oklo data. An analysis of most cosmological data (BBN, CMB, quasar etc.) assuming that the extra dimension scales as $R_0(1+\Delta t^{-3/4})$ and $R_0[1+\Delta](1-\cos\omega(t-t_0))$ concluded that $\Delta$ has to be smaller than $10^{-16}$ and $10^{-8}$ respectively \citep{landauKK}, while \cite{lichu}  assumed that gauge fields and matter fields can propagate in the bulk, that is in the extra dimensions. \cite{aguilar} evaluated the effect of such a couple variation of $G$ and the structures constants on distant supernova data, concluding that a variation similar to the one reported by \cite{q-webprl01} would make the distant supernovae brighter, hence having an effect opposite to the one of the cosmological constant. We shall stress that \cite{NEW_Tahara:2020rxq} proposed a model to freeze the extra-dimensions thanks to hight-order curvature terms in the Lovelock theory.

\subsubsection{String theory}

There exist five anomaly-free, supersymmetric perturbative string theories respectively known as type I, type IIA, type IIB, SO(32) heterotic and $E_8\times E_8$ heterotic theories; see, e.g., \cite{polchinski98}. 

One of the \emph{definitive predictions} of these theories is the existence of a scalar field, the \emph{dilaton}, that couples directly to matter \citep{taylor88} and whose vacuum expectation value determines the string coupling constant \citep{witten84}. There are two other excitations that are common to all perturbative string theories, a rank two symmetric tensor (the graviton) $g_{\mu\nu}$ and a rank two antisymmetric tensor $B_{\mu\nu}$. The field content then differs from one theory to another. 

It follows that the 4-dimensional couplings are determined in terms of a string scale and various dynamical fields (dilaton, volume of compact space, \dots). When the dilaton is massless, we expect \emph{three} effects:
\begin{enumerate}
\item  a scalar admixture of a scalar component inducing deviations from General Relativity in gravitational effects, 
\item  a variation of the couplings, and 
\item a violation of the weak equivalence principle. 
\end{enumerate}
Our purpose is to show how the 4-dimensional couplings are related to the string mass scale, to the dilaton and the structure of the extra-dimensions mainly on the example of heterotic theories.

To be more specific, let us consider an example. The two \emph{heterotic theories} originate from the fact that left- and right-moving modes of a closed string are independent. This reduces the number of supersymmetry to $N=1$ and the quantization of the left-moving modes imposes that the gauge group is either $SO(32)$ or $E_8\times E_8$ depending on the fermionic boundary conditions. The effective tree-level action is (see, e.g., \citealt{Gross})
\begin{eqnarray}
\label{het}
S_{H}&=&\int\dd^{10}\mathbf{x}\sqrt{-g_{10}}\hbox{e}^{-2\Phi}
         \left[M_{_{H}}^8\left\lbrace R_{10}+4\Box\Phi-4(\nabla\Phi)^2
         \right\rbrace-\frac{M_{_{H}}^6}{4}F_{AB}F^{AB}
         +\ldots\right].
\end{eqnarray}
When compactified on a 6-dimensional Calabi--Yau space, the effective 4-dimensional action takes the form
\begin{eqnarray}\label{het4}
S_{H}&=&\int\dd^{4}\mathbf{x}\sqrt{-g_{4}}\phi
\left[M_{_{H}}^8\left\lbrace R_{4}+\left(\frac{\nabla\phi}{\phi}\right)^2
-\frac{1}{6}\left(\frac{\nabla V_6}{V_6}\right)^2\right\rbrace-\frac{M_{_{H}}^6}{4}F^2\right]+\ldots
\end{eqnarray}
where $\phi\equiv V_6\hbox{e}^{-2\Phi}$ couples identically to the Einstein and Yang--Mills terms. It follows that
\begin{equation}
M_4^2=M_{_{H}}^8\phi,\qquad
g^{-2}_{\text{YM}}=M_{_{H}}^6\phi
\end{equation}
at tree-level. Note that to reach this conclusion, one has to assume that the matter fields -- in the `dots' of Eq.~(\ref{het4}) -- are minimally coupled to $g_4$; see, e.g., \cite{maeda88}.

The strongly coupled $SO(32)$ heterotic string theory is equivalent to the weakly coupled type I string theory. \emph{Type~I superstring} admits open strings, the boundary conditions of which divide the number of supersymmetries by two. It follows that the tree-level effective bosonic action is $N=1$, $D=10$ supergravity, which takes the form, in the string frame,
\begin{eqnarray}
S_{I}&=&\int\dd^{10}\mathbf{x}\sqrt{-g_{10}}M_{_{I}}^6\hbox{e}^{-\Phi}
        \left[\hbox{e}^{-\Phi}
        M_{_{I}}^2R_{10}-\frac{F^2}{4}+\ldots\right]
\end{eqnarray}
where the dots contains terms describing the dynamics of the dilaton, fermions and other form fields. At variance with~(\ref{het}), the field $\Phi$ couples differently to the gravitational and Yang--Mills terms because the graviton and Yang--Mills fields are respectively excitation of close and open strings. It follows that $M_I$ can be lowered even to the weak scale by simply having $\exp\Phi$ small enough. Type I theories require $D9$-branes for consistency. When $V_6$ is small, one can use T-duality (to render $V_6$ large, which allows one to use a quantum field theory approach) and turn the $D9$-brane into a $D3$-brane so that
\begin{eqnarray}
S_{I}&=&\int\dd^{10}\mathbf{x}\sqrt{-g_{10}}
\hbox{e}^{-2\Phi}M_{_{I}}^8R_{10}-\int\dd^{4}\mathbf{x}\sqrt{-g_{4}}\hbox{e}^{-\Phi}
\frac{1}{4}F^2+\ldots
\end{eqnarray}
where the second term describes the Yang--Mills fields localized on the $D3$-brane. It follows that
\begin{equation}
M_4^2=\hbox{e}^{-2\Phi}V_6M_{_{I}}^8,\qquad
g^{-2}_{\text{YM}}=\hbox{e}^{-\Phi}
\end{equation}
at tree-level. If one compactifies the $D9$-brane on a 6-dimensional orbifold instead of a 6-torus, and if the brane is localized at an orbifold fixed point, then gauge fields couple to fields $M_i$ living only at these orbifold fixed points with a (calculable) tree-level coupling $c_i$ so that
\begin{equation}
M_4^2=\hbox{e}^{-2\Phi}V_6M_{_{I}}^8,\qquad
g^{-2}_{\text{YM}}=\hbox{e}^{-\Phi}+c_iM_i.
\end{equation}
The coupling to the field $c_i$ is a priori non universal.  At strong coupling, the 10-dimensional $E_8\times E_8$ heterotic theory becomes M-theory on $R^{10}\times S^1/Z_2$ \citep{horava96}.  The gravitational field propagates in the 11-dimensional space while the gauge fields are localized on two 10-dimensional branes. At one-loop, one can derive the couplings by including Kaluza--Klein excitations to get \citep{dudas00}
\begin{equation}
g^{-2}_{\text{YM}}=M_{_{H}}^6\phi-\frac{b_a}{2}(RM_{_H})^2+\ldots
\end{equation}
when the volume is large compared to the mass scale and in that case the coupling is no more universal. Otherwise, one would get a more complicated function. Obviously, the 4-dimensional effective gravitational and Yang--Mills couplings depend on the considered superstring theory, on the compactification scheme but in any case they depend on the dilaton.

As an example, \cite{maeda88} considered the ($N=1, D=10$)-supergravity model derived from the heterotic superstring theory in the low energy limit and assumed that the 10-dimensional spacetime is compactified on a 6-torus of radius $R(x^\mu)$ so that the effective 4-dimensional theory described by (\ref{het4}) is of the Brans--Dicke type with $\omega=-1$.  Assuming that $\phi$ has a mass $\mu$,  and couples to the matter fluid in the universe as $S_{\text{matter}}=\int\dd^{10}\mathbf{x}\sqrt{-g_{10}}\exp(-2\Phi){\cal L}_{\text{matter}}(g_{10})$, the reduced 4-dimensional matter action is
\begin{equation}
S_{\text{matter}}=\int\dd^{4} \mathbf{x}\sqrt{-g}\phi{\cal L}_{\text{matter}}(g).
\end{equation}
The cosmological evolution of $\phi$ and $R$ can then be computed to deduce that $\dot\aem/\aem\simeq10^{10}$ $(\mu/1 \unit{eV})^{-2} \unit{yr}^{-1}$.  \cite{Vayonakis:1988aq} considered the same model but assumed that supersymmetry is broken by non-perturbative effects such as gaugino condensation. Hence, and contrary to \cite{maeda88}, $\phi$ is stabilized and the variation of the constants arises mainly from the variation of $R$ in a runaway potential.

\cite{kiritsis} considered a probe D3-brane in the context of  AdS/CFT correspondence at finite temperature and provides the predictions for the running electric and magnetic effective couplings, beyond perturbation theory. It allows to construct a varying speed of light model.

To conclude, superstring theories offer a natural theoretical framework to discuss the value of the fundamental constants since they become expectation values of some fields. This is a first step towards their understanding but yet, no complete and satisfactory mechanism for the stabilization of the extra-dimensions and dilaton is known.  It has paved the way for various models that we  detail in Sect.~\ref{subsec81}. 

\section{Relations between constants}\label{subsec5.3}

As soon as one consider unification schemes, various constants can exhibit correlated variations, which indeed are expected to be model-dependent. The second theoretical aspect one shall consider is the possible correlation between the variations of several fundamental constants and the dependencies of primary parameters on fondamental constants. This requires to have a deeper look on the matter sector, i.e., on the standard model of particle physics and its extensions.

First, in quantum field theory, one needs to take into account the running of coupling constants with energy and the possibilities of grand unification to bind them so that their variations will derive from the one of a single high-energy coupling. It will give a link between the QCD scale, the coupling constants and the masses of the fundamental particles (i.e., the Yukawa couplings and the Higgs vev), as described in Sect.~\ref{subsecGUT}. Second, in order to relate fundamental constant to the nuclear primary parameters, one compute the binding energies and the masses of the proton, neutron and different nuclei in terms of the gauge couplings and the quark masses. This step involves QCD and nuclear physics. Similarly, one can express the gyromagnetic factors in terms of the quark masses; see Sect.~\ref{subsec-gyro}. This step, described in Sect.~\ref{subsubmass}, is particularly important to interpret the constraints from the atomic clocks and the QSO spectra. In particular, it allows one to set stronger constraints on the varying parameters at the expense of a model-dependence. To finish, we present in Sect.~\ref{secRS} the phenomenological $(R,S)$-parameterisation.

\subsection{Implication of gauge coupling unification}\label{subsecGUT}

The first theoretical implication of high-energy physics arises  from the unification of the non-gravitational interactions. In these unification schemes,  the three standard model coupling constants derive from one unified coupling constant so that their variation shall be correlated.

\paragraph{Renormalisation and running of the couplings constants}\ 

In quantum field theory, the calculation of scattering processes include higher order corrections of the coupling constants related to loop corrections that introduce integrals over internal 4-momenta. Depending on the theory, these integrals may be either finite or diverging as the logarithm or power law of a UV cut-off. In a class of theories, called \emph{renormalizable}, among which the standard model of particle physics, the physical quantities calculated at any order do not depend on the choice of the cut-off scale. But the result may depend on $\ln E/m$ where $E$ is the typical energy scale of the process. It follows that the values of the coupling constants of the standard model depend on the energy at which they are measured (or of the process in which they are involved). This running arises from the screening due to the existence of virtual particles, which are polarized by the presence of a charge. The renormalization group allows one to compute the dependence of a coupling constants on the energy $E$ as
$$
 \frac{\dd g_i(E)}{\dd\ln E}=\beta_i(E),
$$
where the beta functions, $\beta_i$, depend on the gauge group and on the matter content of the theory and may be expended in powers of $g_i$. For the SU(2) and U(1) gauge couplings of the standard model, they are given by
$$
 \beta_2(g_2)=-\frac{g_2^3}{4\pi^2}\left(\frac{11}{6} - \frac{n_g}{3}\right),\qquad
  \beta_1(g_1)=+\frac{g_1^3}{4\pi^2} \frac{5n_g}{9} 
$$
where $n_g$ is the number of generations for the fermions.  We remind that \emph{the fine-structure constant is defined in the limit of zero momentum transfer so that cosmological variation of $\aem$ are independent of the issue of the renormalization group dependence}. For the SU(3) sector, with fundamental Dirac fermion representations, 
$$
 \beta_3(g_3)=-\frac{g_3^3}{4\pi^2}\left(\frac{11}{4} - \frac{n_f}{6}\right),
$$
$n_f$ being the number of quark flavors with mass smaller than $E$. The negative sign implies that (\textit{1}) at large momentum transfer the coupling decreases and loop corrections become less and less significant: QCD is said to be asymptotically free; (\textit{2}) integrating the renormalization group equation for $\alpha_3$ gives
$$
 \alpha_3(E)= \frac{6\pi}{(33-n_f)\ln(E/\Lambda_c)}
$$ 
so that it diverges as the energy scale approaches $\Lambda_c$ from above, that we decided to call $\Lambda_{\mathrm{QCD}}$. This scale characterizes all QCD properties and in particular the masses of the hadrons are expected to be proportional to $\Lambda_{\mathrm{QCD}}$ up to corrections of order $m_{\mathrm{q}}/\Lambda_{\mathrm{QCD}}$.

\paragraph{Unification}\ 

It was noticed quite early that these relations imply that the weaker gauge coupling becomes stronger at high energy, while the strong coupling becomes weaker so that one can thought the three non-gravitational interactions may have a single common coupling strength above a given energy. This is the driving idea of Grand Unified Theories (GUT) in which one introduces a mechanism of symmetry-breaking from a higher symmetry group, such, e.g., as SO(10) or SU(5),  at high energies. 

It has two important consequences for our present considerations. First there may exist algebraic relations between the Yukawa couplings of the standard model. Second, the  structure constants of the standard model unify at an energy scale $m_{\rm u}$
\begin{equation}
 \alpha_1(m_{\rm u})= \alpha_2(m_{\rm u})= \alpha_3(m_{\rm u})\equiv  \alpha_U(m_{\rm u}).
\end{equation}
We note that the electroweak mixing angle can also be a time dependent parameter, but only for $E\not=m_{\rm u}$ since at  $E=m_{\rm u}$, it is fixed by the symmetry to have the value $\sin^2\theta=3/8$, from which we deduce
$$
\aem^{-1}(M_Z)=\frac{5}{3}\alpha_1^{-1}(M_Z) + \alpha_2^{-1}(M_Z).
$$
It follows from the renormalization group relations that
\begin{equation}
 \alpha_i^{-1}(E)=\alpha_i^{-1}(m_{\rm u}) - \frac{b_i}{2\pi}\ln\frac{E}{m_{\rm u}},
\end{equation}
where the beta-function coefficients are given by $b_i=(41/10,-19/6,7)$ for the standard model (or below the SUSY scale $\Lambda_{\mathrm{SUSY}}$) and by $b_i=(33/5,1,-3)$ for $N=1$ supersymmetric theory. Given a field decoupling at $m_{\mathrm{th}}$, one has 
$$
\alpha_i^{-1}(E_-)=\alpha_i^{-1}(E_+) - \frac{b^{(-)}_i}{2\pi}\ln\frac{E_-}{E_+}
 - \frac{b^{({\mathrm{th}})}_i}{2\pi}\ln\frac{m_{\mathrm{th}}}{E_+}
$$
where $b^{({\mathrm{th}})}_i=b^{(+)}-b^{(-)}$ with $b^{(+/-)}$ the beta-function coefficients respectively above and below the mass threshold, with tree-level matching at $m_{\mathrm{th}}$. In the case of multiple thresholds, one must sum the different contributions. The existence of these thresholds implies that the running of $\alpha_3$ is complicated since it depends on the masses of heavy quarks and colored superpartner in the case of supersymmetry. For non-supersymmetric theories,  the low-energy expression of the QCD scale is 
\begin{equation}
\label{QCDscale}
 \Lambda_{\mathrm{QCD}} = E\left(\frac{m_{\mathrm{c}}m_{\mathrm{b}}m_{\mathrm{t}}}{E}\right)^{2/27}
 \exp\left(-\frac{2\pi}{9\alpha_3(E)} \right)
\end{equation}
for $E>m_{\mathrm{t}}$. This implies that the variation of Yukawa couplings, gauge couplings, Higgs vev and $\Lambda_{\mathrm{QCD}}/ M_{\mathrm{P}}$ are correlated. 

A second set of relations arises in models  in which the weak scale is determined by dimensional transmutation \citep{transmut,transmut2}.  In such cases, the Higgs vev is related to the Yukawa constant of the top quark by \citep{bbnco}
\begin{equation}
 v = M_{\mathrm{P}}\exp\left(-\frac{8\pi^2 c}{h_{\mathrm{t}}^2} \right),
\end{equation}
where $c$ is a constant of order unity. This would imply that
\begin{equation}\label{e.defS}
\delta\ln v=S\delta\ln h
\end{equation} 
with $S\sim160$ \citep{cnouv}.

\paragraph{Consequences for the coupled variation of fundamental constants}\ 

The first consequences of unification were investigated in  \cite{transmut2,bbnco,cf1,cf2,damour94a,damour94b,langacker} where the variation of the 3 coupling constants was reduced to the one of $\alpha_U$ and $m_{\rm u}/ M_{\mathrm{P}}$. It was concluded that, setting 
\begin{equation}\label{e.dfR}
 R\equiv\delta\ln\Lambda_{\mathrm{QCD}}/\delta\ln\aem,
\end{equation}
$R\sim 34$ with a stated accuracy of about 20\% \citep{langacker,langacker0} (assuming only $\alpha_U$ can vary),  $R\sim40.82$ in the string dilaton model assuming Grand Unification \citep{damour94a,damour94b} (see Sect.~\ref{subsub1}), $R=38\pm6$ \citep{cf1}  and then $R=46$ \citep{cf2,cf3}, the difference arising from the quark masses and their associated thresholds. However, these results implicitly assume that the electroweak symmetry breaking and supersymmetry breaking mechanisms, as well as the fermion mass generation, are not affected by the variation of the unified coupling. It was also mentioned in \cite{cf2} that $R$ can reach $-235$ in unification based on SU(5) and SO(10). The large value of $R$ arises from the exponential dependence of $\Lambda_{\mathrm{QCD}}$ on $\alpha_3$. In the limit in which the quark masses are set to zero, the proton mass, as well as all other hadronic masses are proportional to $\Lambda_{\mathrm{QCD}}$, i.e., $m_{\mathrm{p}}\propto\Lambda_{\mathrm{QCD}}[1+{\cal O}(m_{\mathrm{q}}/\Lambda_{\mathrm{QCD}})]$. \cite{langacker} further related the Higgs vev to $\aem$ by $\dd\ln v/\dd\ln\aem\equiv\kappa$ and estimated that $\kappa\sim70$ so that, assuming that the variation of the Yukawa couplings is negligible, it could be concluded that 
$$
\delta \ln \frac{m}{\Lambda_{\mathrm{QCD}}}\, \sim 35 \delta\ln\aem,
$$
for the quark and electron masses. This would also imply that the variations of $\mu$ and $\aem$ are correlated, still in a very model-dependent way, typically one can conclude \citep{cnouv} that
$$
 \frac{\delta\mu}{\mu} =-0.8R \frac{\delta\aem}{\aem} +0.6(S+1) \frac{\delta h}{h},
$$
with $S\sim160$. The running of $\alpha_U$ can be extrapolated to the Planck mass, $ M_{\mathrm{P}}$. Assuming $\alpha_U( M_{\mathrm{P}})$ fixed and letting $m_{\rm u}/ M_{\mathrm{P}}$ vary, it was concluded \citep{dine} that $R=2\pi(b_U+3)/[9\aem(8b_U/3-12)]$ where $b_U$ is the beta-function coefficient describing the running of $\alpha_U$. This shows that a variation of $\aem$ and $\mu$ can open a windows on GUT theories. A similar analysis \citep{dent03} assuming that electroweak symmetry breaking was triggered by non-perturbative effects in such a way that $v$ and $\alpha_U$ are related, concludes that ${\delta\mu}/{\mu} =(13\pm7){\delta\aem}/{\aem}$ in a theory with soft SUSY breaking and ${\delta\mu}/{\mu} =(-4\pm5){\delta\aem}/{\aem}$ otherwise.

From a phenomenological point of view, \cite{dent1}  assumed a proportionality with fixed ``unification coefficients'' and that the variations of the constants at a given redshift $z$ depend on a unique evolution factor $\ell(z)$ and can be derived from those of the unification mass scale (in Planck units), $m_{\rm u}$, the unified gauge coupling $\alpha_U$, the Higgs vev, $v$ and in the case of supersymmetric theories the soft supersymmetry breaking mass, $\tilde m$. Introducing the coefficients $d_i$ by
$$
\Delta\ln\frac{m_{\rm u}}{ M_{\mathrm{P}}} = d_M\ell,\quad
\Delta\ln\alpha_U = d_U\ell,\quad
\Delta\ln\frac{v}{m_{\rm u}} = d_H\ell,\quad
\Delta\ln\frac{\tilde m}{ M_{\mathrm{P}}} = d_S\ell,
$$
($d_S=0$ for non-supersymmetric theories) and assuming that the masses of the standard model fermions all vary with $v$ so that the Yukawa couplings are assumed constant, it was shown that the variations of all constants can be related to $(d_M,d_U,d_H,d_S)$ and $\ell(z)$, using the renormalization group equations (neglecting the effects induced by the variation of $\alpha_U$ on the RG running of fermion masses). This decomposition is a good approximation provided that the time variation is slow, which is actually backed up by the existing constraints, and homogeneous in space (so that it may not be applied as such in the case a chameleon mechanism is at work \citep{braxm}.

This allowed to be defined six classes of scenarios: (\textit{1}) varying gravitational constant ($d_H=d_S=d_U=0$) in which only $m_{\rm u}/ M_{\mathrm{P}}$ or equivalently $G\Lambda^2_{\mathrm{QCD}}$ is varying; (\textit{2}) varying unified coupling $(d_U=1,d_H=d_S=d_M=0)$; (\textit{3}) varying Fermi scale defined by $(d_H=1,d_U=d_S=d_M=0)$ in which one has $\dd\ln\mu/\dd\ln\aem=-325$; (\textit{4}) varying Fermi scale and SUSY-breaking scale $(d_S=d_H=1,d_U=d_M=0)$ and for which $\dd\ln\mu/\dd\ln\aem=-21.5$; (\textit{5}) varying unified coupling and Fermi scale $(d_U=1, d_H=\tilde\gamma d_U, d_S=d_M=0)$ and for which $\dd\ln\mu/\dd\ln\aem=(23.2-0.65\tilde\gamma)/(0.865+0.02\tilde\gamma)$;  (\textit{6}) varying unified coupling and Fermi scale with SUSY $(d_U=1, d_S\simeq d_H=\tilde\gamma d_U, d_M=0)$ and for which $\dd\ln\mu/\dd\ln\aem=(14-0.28\tilde\gamma)/(0.52+0.013\tilde\gamma)$. 

Using the dependence of $m_{\rm p}$ on the Higgs vev  \citep{gasser82}, \cite{NEW_Mohamadnejad:2018vst,NEW_Chakrabarti:2021sgs} deduced that $\Delta \mu/\mu =-0.91\Delta v/v$ in order to constrain the cosmological variation of $v$ assuming its dynamics is incorporated from a Brans-Dicke action.

Each scenario can be compared to the existing constraints to get sharper bounds on them \citep{bbn-dent,dent1,dent2,bbn-muller} and emphasize that the correlated variation between different constants (here $\mu$ and $\aem$) depends strongly on the theoretical hypothesis that are made.

\paragraph{Effect of the QCD vacuum angle $\theta$}\ 

Motivated by the naturalness problem of the QCD vacuum angle $\theta$, whose natural order of magnitude would be of order unity while it is experimentally observed to be smaller than $10^{-10}$ \citep{Abel:2020pzs}. The fact that, contrary to the cosmological constant, there seem to be no anthropic limitations on $\theta$ motivated  \cite{NEW_Lee:2020tmi} to investigate the impact of the QCD vacuum on  light nuclei starting. The question of the $\theta$-dependence of nuclear physics was addressed by \cite{NEW_Ubaldi:2008nf} considering the effect of the $\theta$ of the QCD Lagrangian
\begin{equation}\label{e.thetaQCD}
{\cal L}_\theta = - \frac{g^2 \theta}{32 \pi^2} F\tilde F
\end{equation}
which leads to CP-violation in the strong interaction. It concluded that the binding energies of deuteron and diproton would change by 10\% for $\theta\sim 130^{\rm o}$, which could affect BBN and  that for $\theta\sim 2^{\rm o}$  the effects on   the $3\alpha$-process would imply that the abundances of carbon and oxygen would be ten times greater. This led \cite{NEW_Lee:2020tmi} to investigate the impact of the QCD vacuum on BBN and the $3\alpha$ process. They started from the nucleon mass dependence \citep{Brower:2003yx}
\begin{equation}\label{emNtheta}
m_{\rm n}= m_0 - 4c_1M_\pi^2(\theta) - \frac{3 g_A^2}{32\pi F^2_\pi}M_\pi^3(\theta)
\end{equation}
where $m_0\simeq 865 \unit{MeV}$ is the nucleon mass in the chiral limit, $g_A=1.27$ the axial-vector coupling constant, $F_\pi$ the pion decay constant and $c_1$ a low-energy constantfrom the second order chiral pion-nucleon and where the pion mass behaves as
$$
M_\pi^2(\theta) =M_\pi^2\cos\frac{\theta}{2}\sqrt{1+\varepsilon^2\tan^2\frac{\theta}{2}}
$$
with $M_\pi=139.57 \unit{MeV}$ the charged pion mass and $\varepsilon= (m_{\rm u}-m_{\rm d})/(m_{\rm u}+m_{\rm d})$ and the $\theta$-dependence of the pion-nucleon coupling $g_{\pi NN}(\theta)$ to  compute the neutron-proton mass difference
\begin{equation}
(m_{\rm n}-m_{\rm p})^{\rm QCD}(\theta) = 4 c_5B_0\frac{M^2_\pi}{M^2_\pi(\theta)}(m_{\rm u}-m_{\rm d})
\end{equation}
and show that its magnitude increases with $\theta$ reaching 1\% at $\theta\sim1$. Similarly they showed that the neutron lifetime drops off very quickly when $\theta$ deviates from 0 so that effects on BBN are expected to be important. This allowed them to investigate the effects on primordial and stellar nucleosynthesis to conclude that the dineutron and diproton are bound for $\theta\gtrsim0.2$ and $\theta\gtrsim0.7$ and that $\theta$ must be smaller than $0.1$ to  recover the real nuclear reaction rates, that can then be used safely since experimentally, $\theta<10^{-10}$.

This line of analysis was followed by  \cite{Kim:2022ype}  who considered the coupling of the gluon to an axion $a$ so that Eq.~(\ref{e.thetaQCD}) is modified to ${\cal L}\propto a F\tilde F$. For an ultralight dark matter field, its oscillations will induce fluctuations of all nuclear quantities. Using the dependence~(\ref{emNtheta}) for the nucleon mass, and
$$ 
g_{\rm p}=g_{\rm p}^{(0)}- \frac{g_A^2}{4\pi F_\pi^2} m_{\rm n}M_\pi(\theta)\,
\qquad
g_{\rm n}=g_{\rm n}^{(0)} +  \frac{g_A^2}{4\pi F_\pi^2} m_{\rm n}M_\pi(\theta)
$$
for the proton and nucleon gyromagnetic factors. Following \cite{kappa-tedesco} they computed the $\theta$-dependence for thorium and the standard atomic clocks; see  \cite{Kim:2022ype} for a similar analysis.  \cite{NEW_Davoudiasl:2022kdq} built on this analysis to relate a variation of $\theta$ to a variation of $\bar\mu$ as $\Delta(\theta^2)\sim-1.4\Delta\bar\mu/\bar\mu$ to conclude from the constraints on the variation of $\bar\mu$ thanks to the Yb-clocks experiment \citep{Lange:2020cul} that
\begin{equation}
\frac{\dd \theta^2}{\dd t}<6\times10^{-15} \unit{yr^{-1}}.
\end{equation}
\cite{Flambaum:2023bnw} considered an ultralight dark matter axion $\varphi=\theta/f_a$. Using the results from \cite{kappa-tedesco,Kim:2022ype} they concluded from the analysis of Rb-Cs experiment by \cite{NEW_Hees:2016gop} that the axion decay constant shall satisfy $f_a>1.8\times 10^9 \unit{GeV}(10^{-15}\unit{eV}/m_\varphi)$. Then, from the Rb-Cs clock experiment,  \cite{NEW_Dzuba:2024src} deduced that
\begin{equation}
\frac{\dd \theta^2}{\dd t}=(8.2\pm16)\times10^{-13} \unit{yr^{-1}}.
\end{equation}
Similar analysis for Al/Hg gave $(1.8\pm2.8)\times10^{-13} \unit{yr^{-1}}$ while those for Yb-clocks \citep{Banerjee:2023bjc} and Rb-quartz and Dy \citep{PhysRevLett.130.251002} allowed to constrain linear and quadratic ultralight  scalar and axion dark matter models.

\subsection{Masses and binding energies in terms of the fundamental constants}\label{subsubmass}

For ``composite'' systems such as proton, neutron, nuclei or even planets and stars, we need to  compute their mass, which requires to determine their binding energies. As already seen, the electromagnetic binding energy induces a direct dependence on $\aem$ and can be evaluated using, e.g., the Bethe--Weizs\"acker formula~(\ref{bethe}). The dependence of the masses on the quark masses, via nuclear interactions, and the determination of the nuclear binding energy are especially difficult to estimate.

\paragraph{Nucleon mass}\ 

In the chiral limit of QCD, in which all quark masses are negligible compared to $\Lambda_{\mathrm{QCD}}$, all dimensionful quantities scale as some power of $\Lambda_{\mathrm{QCD}}$. For instance, concerning the nucleon mass, $m_{\mathrm{N}}=c\Lambda_{\mathrm{QCD}}$ with $c\sim3.9$ being computed from lattice QCD. This predicts a mass of order 860~MeV, smaller than the observed value of 940~MeV. The nucleon mass can be computed in chiral perturbation theory and expressed in terms of the pion mass as \citep{leinweber} $ m_{\mathrm{N}} = a_0 + a_2m_\pi^2 + a_4m_\pi^4 + a_6 m_\pi^6 +\sigma_{N\pi} + \sigma_{\Delta\pi} +\sigma_{\mathrm{tad}}$ (where all coefficients of this expansion are defined in \citealt{leinweber}), which can be used to show \citep{clock-muq} that the nucleon mass scales as
\begin{equation}\label{e.MassN}
 m_{\mathrm{N}} \propto\Lambda_{\mathrm{QCD}}X_{\mathrm{q}}^{0.037}X_{\mathrm{s}}^{0.011}
\end{equation}
with the definition~(\ref{edef-X}) for the quantities $X$. It follows that
\begin{equation}\label{e.mubarq}
 \bar\mu \propto  X_{\mathrm{q}}^{-0.037}X_{\mathrm{s}}^{-0.011} X_{\rm e}
\end{equation}
so that when one assumes $X_{\mathrm{s}}\propto X_{\mathrm{q}}$ it reduces to
\begin{equation}\label{e.MassN2}
 \bar\mu \propto  X_{\mathrm{q}}^{-0.048}X_{\rm e}\,.
\end{equation}
Note, however, that such a notation is dangerous since it would imply that $m_{\mathrm{N}}$ vanishes in the chiral limit but it is a compact way to give $\delta m_{\mathrm{N}}/\delta X_{\mathrm{q}}$ etc. It was further extended \citep{oklo-14} by using a sigma model to infer that $ m_{\mathrm{N}} \propto\Lambda_{\mathrm{QCD}}X_{\mathrm{q}}^{0.045}X_{\mathrm{s}}^{0.19}$. These two examples explicitly show the strong dependence in nuclear modeling.

\paragraph{Deuterium binding energy, $B_{\rm D}$}\ 

Before discussing the general case, let us consider the expression of the  deuterium binding energy $B_{\rm D}$  has been discussed in different ways (see Sect.~\ref{bbn2cste}). Many models have been created. 

 A first route relies on the use of the dependence of $B_{\rm D}$ on the pion mass \citep{bbnpi1,bbnpi2,bbn-pudliner,nnbyoo}, which can then be related to $m_{\mathrm{u}}$, $m_{\mathrm{d}}$ and $\Lambda_{\mathrm{QCD}}$. 

A second avenue is to use a sigma model in the framework of the Walecka model \citep{waleka} in which the potential for the nuclear forces keeps only the  $\sigma$, $\rho$ and $\omega$ meson exchanges \citep{oklo-14}.  We also emphasize that the deuterium is only produced during BBN, as it is too weakly bound to survive in the regions of stars where nuclear processes take place. 

The fact that we do observe deuterium today sets a non-trivial constraint on the constants by imposing that the deuterium remains stable from BBN time to today. Since it is weakly bound, it is also more sensitive to a variation of the nuclear force compared to the electromagnetic force. This was used in \cite{dentfair} to constrain the variation of the nuclear strength in a sigma-model.

\paragraph{Atomic nuclei masses}\ 

To go further and determine the sensitivity of the mass of a nucleus to the various constant,
$$
m(A,Z)=Zm_{\mathrm{p}}+(A-Z)m_{\mathrm{n}}+Zm_{\mathrm{e}}+E_{\mathrm{S}}+E_{\mathrm{EM}}
$$
one should determine the strong binding energy [see related discussion below Eq.~(\ref{mass})] in function of the atomic number $Z$ and the mass number $A$.

While the particular case of deuterium has been discussed in the previous paragraph, the situation is more complicated for larger nuclei since there is no simple modeling. For large mass number $A$, the strong binding energy can be approximated by the liquid drop model
\begin{equation}
\label{eq.li-drop}
 \frac{E_{\mathrm{S}}}{A} = a_V -\frac{a_S}{A^{1/3}} - a_A\frac{(A-2Z)^2}{A^2} + a_P
 \frac{(-1)^A+(-1)^Z}{A^{3/2}}
\end{equation}
with $(a_V,a_S,a_A,a_P)=(15.7,17.8,23.7,11.2) \unit{MeV}$ \citep{drop}. It has also  been suggested \citep{lilley} that the nuclear binding energy can be expressed as
\begin{equation}
 E_{\mathrm{S}} \simeq A a_3 + A^{2/3} b_3\qquad \hbox{with}\qquad
 a_3 = a_3^{\text{chiral limit}} + m^2_\pi\frac{\partial a_3}{\partial m_\pi^2}.
\end{equation}
In the chiral limit, $a_3$ has a non-vanishing limit to which we need to add a contribution scaling like $m^2_\pi\propto\Lambda_{\mathrm{QCD}}m_{\mathrm{q}}$.  \cite{lilley} also pointed out that the delicate balance between attractive and repulsive nuclear interactions \citep{waleka} implies that the binding energy of nuclei is expected to depend strongly on the quark masses \citep{donomass}. A fitting
formula derived from effective field theory and based of the semi-empirical formula derived in \cite{furnstahl} was also proposed \citep{damourdono} as
\begin{equation}
 \frac{E_{\mathrm{S}}}{A} = -\left(120-\frac{97}{A^{1/3}}\right)\eta_S +\left(67-\frac{57}{A^{1/3}}\right)\eta_V + \ldots
\end{equation}
where $\eta_S$ and $\eta_V$ are the strength of respectively the scalar (attractive) and vector (repulsive) nuclear contact interactions normalized to their actual value. These two parameters need to be related to the QCD parameters \citep{donomass}. 

We refer to \cite{fw2} for the study of the dependence of the binding of light ($A\leq 8$) nuclei on possible variations of hadronic masses, including meson, nucleon, and nucleon-resonance masses. The following approximate relations have been widely used in BBN analysis \citep{fw2,bbn-dent}
\begin{equation}\label{e.BX}
\Delta B_X/B_X = c_\alpha\Delta\aem/\aem + c_q \Delta m_{\rm q}/m_{\rm q}
\end{equation}
with the sensitivity coefficients summarized in Table~\ref{tab-Bx}.

\begin{table}[htbp]
\caption[Sensitivity coefficients of the binding energies of light elements on $(\aem,m_{\rm q})$]{Sensitivity coefficients of a variation of the binding energies of light elements on a variation of $(\aem,m_{\rm q})$; see Eq.~(\ref{e.BX}).}
\label{tab-Bx}
\centering
{\footnotesize
\begin{tabular}{lcc}
\toprule
$X$  & $c_\alpha$  & $c_q$   \\
\hline
 Tritium & $-0.047$ & $-2.1$  \\
 helium-3 &  $-0.093$& $-2.3$  \\
 helium-4   &$-0.030$  & $-0.94$  \\
  lithium-7    & $-0.046$ &  $-1.4$ \\
   Berrilium-7     &  $-0.089$&  $-1.4$ \\
\bottomrule
\end{tabular}
}
\end{table}

These expressions allow one to compute the sensitivity coefficients that enter in the decomposition of the mass; see Eq.~(\ref{mdephi}). They stress one of the most difficult issue related to the intricate structure of QCD and its role in low energy nuclear physics, which is central to determine the masses of nuclei and the binding energies, quantities that are particularly important for BBN, the universality of free fall and stellar physics.

\subsection{Gyromagnetic factors in terms of the fundamental constants}\label{subsec-gyro}

Since they enter in the constraints arising from the comparison of atomic clocks and QSO, it important to relate the gyromagnetic factors to fundamental constants.

\paragraph{Proton and neutron gyromagnetic factors}\ 

The proton and neutron gyromagnetic factors are respectively given by $g_{\mathrm{p}}=5.586$ and $g_{\mathrm{n}}=-3.826$ and are expected to depend on $X_{\mathrm{q}}=m_{\mathrm{q}}/\Lambda_{\mathrm{QCD}}$ \citep{flambq}. In the chiral limit in which $m_{\mathrm{u}}=m_{\mathrm{d}}=0$, the nucleon magnetic moments remain finite so that one could have thought that the finite quark mass effects should be small. However, it is  enhanced by $\pi$-meson loop corrections, which are proportional to $m_\pi\propto\sqrt{m_{\mathrm{q}}\Lambda_{\mathrm{QCD}}}$. Following \cite{leinweber}, this dependence can be described by the approximate formula
$$
 g(m_\pi) = \frac{g(0)}{1+ a m_\pi + b m_\pi^2}.
$$
The coefficients are given by $a=(1.37,1.85)/\unit{GeV}$ and $b=(0.452,0.271)/\unit{GeV}^2$ respectively for the proton an neutron. This lead \cite{flambq} to $g_{\mathrm{p}}\propto m_\pi^{-0.174}\propto X_{\mathrm{q}}^{-0.087}$ and $g_{\mathrm{n}}\propto m_\pi^{-0.213}\propto X_{\mathrm{q}}^{-0.107}$. This was further extended in \cite{clock-muq} to take into account the dependence with the strange quark mass $m_{\mathrm{s}}$ to obtain
\begin{equation}\label{e.gdeX}
 g_{\mathrm{p}}\propto X_{\mathrm{q}}^{-0.087}X_{\mathrm{s}}^{-0.013},\qquad
 g_{\mathrm{n}}\propto X_{\mathrm{q}}^{-0.118}X_{\mathrm{s}}^{0.0013}
\end{equation}
with the definitions~(\ref{edef-X}). All these expressions assume $\Lambda_{\mathrm{QCD}}$ constant in their derivations \citep{kappa-flamb,kappa-tedesco,dzubapolo2,dinh,NEW_Luo:2011cf,NEW_JacksonKimball:2014vsz}. In the approximation $X_{\rm s} \propto X_{\rm q}$ which is well-motivated by the Higgs mechanism of mass genration, the dependence in the quark mass reduces to
\begin{equation}\label{e.gdeXq}
 g_{\mathrm{p}}\propto X_{\mathrm{q}}^{-0.10},\qquad
 g_{\mathrm{n}}\propto X_{\mathrm{q}}^{-0.1167}.
\end{equation}

\paragraph{Caesium and rubidium gyromagnetic factors}\ 

Concerning caesium-133 and Rubidium-87, the computation of the sensitivity has evolved over time. Using a chiral perturbation theory, it was first deduced \citep{clock-muq} that
\begin{equation}
 g_{\mathrm{Cs}}\propto X_{\mathrm{q}}^{0.110}X_{\mathrm{s}}^{0.017} \propto X_{\mathrm{q}}^{0.127},\qquad
 g_{\mathrm{Rb}}\propto X_{\mathrm{q}}^{-0.064}X_{\mathrm{s}}^{-0.010}\propto X_{\mathrm{q}}^{-0.074}\,. \nonumber
\end{equation}
where the second equality assumes   $X_{\rm s} \propto X_{\rm q}$. \cite{kappa-tedesco} refined the computation thanks to three methods (see their Table~III). Considering only the valence nucleon, they first  got the same values  listed just above as previously obtained by \cite{clock-muq} while including the non-valence nucleons lowered the coefficients respectively to $0.044$ and $-0.056$.  Further, including the effect of quark mass on the spin-spin interaction, they concluded
\begin{equation}\label{e.gdeXCsRb}
 g_{\mathrm{Cs}}  \propto X_{\mathrm{q}}^{0.009},\qquad
  g_{\mathrm{Rb}}  \propto X_{\mathrm{q}}^{-0.016}\, .
\end{equation}
Taking into account the effect of the variation of the nuclear radius, it was corrected \citep{dinh} by a  factor $X_{\rm q}^{-0.007}$ and  $X_{\rm q}^{-0.003}$ respectively so that
\begin{equation}\label{e.gdeXCsRb2}
 g_{\mathrm{Cs}}  \propto X_{\mathrm{q}}^{0.002},\qquad
 g_{\mathrm{Rb}}  \propto X_{\mathrm{q}}^{-0.019}\, .
\end{equation}

One shall take these sensitivities with care. \cite{NEW_Luo:2011cf} compared the computation of the dependencies of the gyromagnetic factor and proton mass to the light quark masses and shown that is strongly model-dependent. They first related the nuclear gyromagnetic factor to the ones of the proton and neutron as well as the spin-spin interaction $b$ taking into account the effects of the polarization of the non-valence nucleons and spin-spin interaction
\begin{eqnarray} \label{new eq27}
\frac{\delta g_{\rm Rb}}{g_{\rm Rb}} &=& 0.764 \frac{\delta g_{\rm p}}{g_{\rm p}} - 0.172 \frac{\delta g_{\rm n}}{g_{\rm n}} - 0.379 \frac{\delta b}{b} \, , \\
\label{new eq28}
\frac{\delta g_{\rm Cs}}{g_{\rm Cs}} &=& -0.619 \frac{\delta g_{\rm p}}{g_{\rm p}} + 0.152 \frac{\delta g_{\rm n}}{g_{\rm n}} + 0.335 \frac{\delta b}{b} \, .
\end{eqnarray}
In comparison, the shell model gives 
\begin{equation}\label{e.shellmodel}
\frac{\delta g_{\rm Rb}}{g_{\rm Rb}} = 0.736 \frac{\delta g_{\rm p}}{g_{\rm p}},\qquad
\frac{\delta g_{\rm Cs}}{g_{\rm Cs}} = -1.266 \frac{\delta g_{\rm p}}{g_{\rm p}}.
\end{equation}
The main difference arises from the dependence in $g_{\rm n}$ and $b$ but the order of magnitude is similar. The comparaison of (\textit{1}) the non-relativistic constituent quark model approach (\textit{2}) chiral perturbation theory approach with and without combining with lattice QCD exhibited an important model-dependence in the computation of the gyromagnetic factors in terms of the quark masses and QCD scale  \citep{NEW_Luo:2011cf}. 

\paragraph{Influence of the $s$ and $c$ quark}\ 

\cite{NEW_Flambaum:2022amo} computed the sensitivities of several secondary parameters to the $s$ and $c$ quark masses. It allowed them to conclude that $\delta B_{\rm D}/B_{\rm D}\propto X_{\rm s}^{-13}X_{\rm c}^{-0.95} $ so that focusing on this single parameter, BBN implies that $|\delta m_{\rm s}/m_{\rm s}|<1.8\times10^{-3}$ and $|\delta m_{\rm c}/m_{\rm c}|<2.5\times10^{-2}$. Similarly, from the behavior of the resonance energy $E_r$ of $^{149}$Sm, they concluded from Oklo data that  $|\delta m_{\rm s}/m_{\rm s}|<1.7\times10^{-10}$ and $|\delta m_{\rm c}/m_{\rm c}|<1.2\times10^{-9}$. Finally, from the scaling $\delta\bar\mu/\bar\mu\propto X_{\rm s}^{0.056}X_{\rm c}^{0.083}$, they concluded that atomic clock experiments set $|\dot m_{\rm s}/m_{\rm s}|<7.9\times10^{-16}$~yr$^{-1}$ and $|\dot m_{\rm c}/m_{\rm c}|<5.3\times10^{-16}$~yr$^{-1}$. 

\paragraph{Conclusions}\ 

Most of the data in the literature are interpreted in terms of the sensitivities computed by \cite{kappa-tedesco} and \cite{dinh} that we adopt for the following numerical applications.
\begin{tcolorbox}
The sensitivities in $X_{\rm q}$ for the hyperfine tandisyions of $^{1}$H, $^{133}$Cs and $^{87}$Rb arising from their gyromagnetic factors are taken to be 
\begin{equation}\label{e.dinhK}
  K_{\rm H}^q =-0.1, \qquad 
 K_{\rm Cs}^q = 0.002, \qquad 
 K_{\rm Rb}^q = -0.019\,.
\end{equation}
When the dependence on $X_{'\rm q}$ arising from $m_{\rm e}/m_{\rm p}$ is included, Eq.~(\ref{e.MassN2}) gives that
\begin{equation}\label{e.dinhK2}
  K^q \rightarrow K^q-0.048.
\end{equation}
 \end{tcolorbox}
 
\subsection{Expression of the atomic transitions}\label{sec-clock-phy}

As detailed in Sect.~\ref{subsec31-e.h}, the combination of clocks allows one to set constraints on the variation of ($\aem,\bar\mu$) and the various gyromagnetic factors, more particularly (see Table~\ref{tab1}) $g_{\rm p}$,  $g_{\rm Cs}$ and $g_{\rm Rb}$. The first step is a standard and straightforward analysis of the data to set constraints on the independent variations of the \emph{five primary QED parameters} $(\aem,\bar\mu, g_{\rm p}, g_{\rm Cs},g_{\rm Rb})$. 

Indeed, these parameters are not independent and extracting constraints on more fundamental parameters requires further theoretical insight. The first step, staying within QED, is to express the gyromagnetic factors in terms of $g_{\rm p}$ and $g_{\rm n}$, as described in Sect.~\ref{subsec-gyro}. This  can be performed in several ways to get either the sets of parameters  ($\aem,\bar\mu, g_{\rm p},g_{\rm n},b$) thanks to Eqs.~(\ref{new eq27}-\ref{new eq28}) or simply $(\aem,\bar\mu,g_{\rm p})$ in the shell model~(\ref{e.shellmodel}). An example of such an analysis can be found in \cite{NEW_Luo:2011cf,NEW_Ferreira:2014fba}. The next steps require the use of quantum chromodynamics, in particular to  relate the nucleon $g$-factors in terms of the quark mass and the QCD scale as described in Eq.~(\ref{e.gdeX}, \ref{e.gdeXq}) to express all the transition frequencies in terms of $(\aem,\bar\mu, X_{\rm q},X_{\rm s})$, recalling that the $X_{\rm q,s}$ are defined in Eq.~(\ref{edef-X}) with $m_{\rm q}=(m_{\rm u}+m_{\rm d})/2$. Indeed $\bar\mu$ includes the proton mass that depends of $X_{\rm q,s}$ and $\aem$ so that one can shift to the parameters $(\aem,X_{\rm q},X_{\rm s},X_{\rm e})$ thanks to Eq.~(\ref{e.MassN},~\ref{e.MassN2}).
 
\paragraph{Definition of the physical interpretation schemes}\ 

Several interpretation schemes have been used in the literature with different model-dependent layers:
\begin{enumerate}
\item sharp and model independent constraints on the variation of the \emph{primary QED parameters} $(\aem,\bar\mu, g_{\rm p}, g_{\rm Cs},g_{\rm Cs})$, using the sensitivity $\lambda$ summarized in Table~\ref{tab0} and defined as
\begin{equation}\label{e.31QED}
\nu_A \propto  g_A^{\lambda_g}\aem^{\lambda_\alpha}\bar\mu^{\lambda_\mu}
\end{equation}
The sensitivity parameters are gathered in Table~\ref{tab0}.

\item Expressing the gyromagnetic factors in terms of the quark masses using the expressions~(\ref{e.gdeX}) and~(\ref{e.gdeXCsRb2}), one gets the frequencies in terms of $(\aem, \bar\mu,  X_{\mathrm{q}}, X_{\mathrm{s}})$, as
\begin{equation}\label{e.QCD1}
 \nu_A \propto \aem^{\lambda_\alpha} \bar\mu^{\lambda_\mu} X_{\mathrm{q}}^{K_{\mathrm{q}}}X_{\mathrm{s}}^{K_{\mathrm{s}}}.
\end{equation}
Most of the constraints on atomic clocks assume $X_{\rm s}\propto X_{\rm q}$ so that is reduces to
\begin{equation}\label{e.QCD2}
 \nu_A \propto \aem^{K_\alpha} \bar\mu^{K_\mu} X_{\mathrm{q}}^{K_{\mathrm{q}}},
\end{equation}
with $K_\alpha=\lambda_\alpha$ and $K_\mu=\lambda_\mu$.

\item The last step is to express the proton mass in $\bar\mu$. Neglecting the $\aem$-dependence of $m_{\rm p}$, \cite{clock-muq} concluded the nucleon masses scale as Eq.~(\ref{e.MassN}). Hence, all transitions are expressed in terms of the fundamental parameters $(\aem,  X_{\mathrm{q}}, X_{\mathrm{s}}, X_{\mathrm{e}})$ as
\begin{equation}\label{e.QCD3}
 \nu_A \propto \aem^{K_\alpha}X_{\mathrm{q}}^{K_{\mathrm{q}}-0.037 }X_{\mathrm{s}}^{K_{\mathrm{s}}-0.011}X_{\rm e}
\end{equation}
and, in the approximation $X_{\mathrm{q}}\propto X_{\mathrm{s}}$,
\begin{equation}\label{e.QCD3b}
\nu_A \propto \aem^{K_\alpha} X_{\mathrm{q}}^{K_{\rm q} -0.048K_{\rm e}} X_{\mathrm{e}}^{K_{\rm e}}
\end{equation}
and $K_{\rm e}=1$ for the transition with $K_\mu\not=0$ and 0 otherwise. Note that the relation (\ref{e.MassN}) assumes that $m_{\rm p}\propto\Lambda_{\rm QCD}$ so that $K_\alpha=\kappa_\alpha$, as assumed here. The values of the sensitivity coefficients are gathered in Table~\ref{tab0}.

\item Sharper constraints under the hypothesis of correlated variations in unification schemes can also be set, e.g., in the $(R,S)$-formalism; see Sect.~\ref{secRS}.
\end{enumerate}

\begin{table}[t]
\caption[ Sensitivities $\lambda$ of atomic transitions to a variation of $(\aem,\bar\mu,g_A)$]{Sensitivities of various transitions on a variation of the fine-structure constant,$\kappa_\alpha$ as defined in Eq.~(\ref{e.31QED}) and~(\ref{clock-sensitivity}).  The coefficients $\lambda$ are defined in Eq.~(\ref{e.31QED}). The coefficients $K$ are defined in Eq.~(\ref{e.QCD2}) with $K^\alpha=\lambda_\alpha$ and $K^\mu =\lambda_\mu$. $K^e$ is defined in Eq.~(\ref{e.QCD3b}) in which the dependence of $m_{\rm p}$ in $\aem$ is neglected so that $K^\alpha=\lambda_\alpha$. Notes. $(\ddagger):$ the early value of this sensitivity was estimated as $-3.2$ \citep{kappa-dzuba} and $-3.19$ \citep{clock-peik06} before reevaluated to $-2.9$ \citep{clock-fortier07} $-2.94$ \citep{dzubapolo2,NEW_Safronova:2019lex}. $(\dag):$ the early value was $0.88$ \citep{kappa-dzuba3,clock-peik04} then reevaluated to $1.03$ \citep{NEW_Safronova:2019lex} and $1.0$ \citep{NEW_Tamm:2013}.}
\label{tab0}
\centering
{\small
\begin{tabular}{p{1.0cm}l|c|ccc | c c c c}
 \toprule
 Atom  & Transition & $\kappa_\alpha$ & $\lambda_\alpha$ & $\lambda_\mu$ & $\lambda_{g_A}$ 	 &   $K^\alpha$ &  $K^\mu$  &  $K^q$  & $K^e$   \\
 \midrule
 $^{1}$H       	 & $1s-2s$	&  0		& 0		 & 0 & 0			 &  0& 0&0 & 0\\
  \midrule
 $^{1}$H 		& hfs			&  0		&2 		&1 &  1  &   2& 1 & $-0.1$  & 1   \\
  $^{133}$Cs	& hfs 		& 0.83   	& 2.83	 & 1 &1  & 2.83&1 & 0.002 &1  \\
 $^{87}$Rb	& hfs 		& 0.34   	& 2.34 	 & 1  &1  & 2.34  & 1 & $-0.019$ & 1 \\
  \midrule
 $^{87}$Sr           & ${}^1$S$_0-{}^3$P$_0$        & 0.06     & 0.06& 0 & 0   & 0.06       & 0&0 & 0 \\
  $^{27}$Al$^{+}$  & ${}^1$S$_0-{}^3$P$_0$        & 0.008    & 0.008 & 0& 0  & 0.008 & 0& 0  &0\\
 $^{199}$Hg$^{+}$ & ${}^2S_{1/2}-{}^2D_{5/2}$  & $-2.94^\ddagger$ &$-2.94$ & 0& 0  &$-2.94$ & 0& 0 &0  \\
$^{171}$Yb$^{+}$ & E2  & 1.0$^\dag$  & 1.0&0 & 0 & 1.0&0 & 0 & 0  \\
  $^{171}$Yb$^{+}$ & E3  &  -5.95 & -5.95& 0&  0  & $-5.95$& 0&  0  & 0 \\
 $^{162}$Dy  & 235 MHz &  $1.72\times10^7$  & $1.72\times10^7$ & 0&  0 & $1.72\times10^7$ & 0&  0 & 0 \\
 $^{163}$Dy  & 3.1 MHz & $8.5\times10^6$&  $8.5\times10^6$ & 0& 0 &  $8.5\times10^6$ & 0& 0 & 0 \\
 $^{164}$Dy  & 754 MHz& $-2.6\times10^6$ & $-2.6\times10^6$  & 0& 0  & $-2.6\times10^6$  & 0& 0 &  0\\
  \midrule
  $^{171}$Yb & ${}^1$S$_{0}-{}^3$P$_{0 }$ & 0.31  &0.31 &0 &  0 &0.31 &0 &  0 & 0 \\
   SF$_6$  &  P(4)E$_0$ & 0  &0 &$1/2$ & 0  &0 &$0.5$ & 0 & 0.5 \\
   Krb &    & 0  &0 & $14890\pm60$ & 0  &0 &$0.5$ & 0 & $14890\pm60$ \\
\bottomrule
\end{tabular}
}
\end{table}

Indeed, these constraints are model-dependent and will depend on the physical modelizations required to compute the sensitivity parameters, as an example, Table~III of \cite{kappa-tedesco} compares their values when different nuclear effects are considered. For instance, $K_{\rm q}$ can vary from 0.127, 0.044 to 0.009 for the caesium according to whether one includes only valence nucleon, non-valence non-nucleon or effect of the quark mass on the spin-spin interaction; see also  \cite{NEW_Luo:2011cf} for a discussion of the model-dependence.

\paragraph{Atomic clock data interpretations}\ 

\begin{tcolorbox}
The comparison of atomic clocks gives access to
\begin{equation}\label{def.yAB}
\delta \ln y_{AB} \equiv \delta \ln\frac{\nu_A}{\nu_B}
\end{equation}
that can be expressed as
\begin{equation}\label{e.dataQED}
\delta \ln y_{AB} = \Delta\lambda_\alpha \frac{\delta\aem}{\aem} + \Delta\lambda_\mu\frac{\delta\bar\mu}{\bar\mu} +  \Delta\lambda_{g_i} \frac{\delta g_i}{g_i} 
\end{equation}
with the sensitivity coefficients summarized in Table~\ref{tab0b} or as
\begin{equation}\label{e.dataQCD}
\delta \ln y_{AB} = \Delta K_\alpha \frac{\delta\aem}{\aem} + \Delta K_\mu\frac{\delta\bar\mu}{\bar\mu} +  \Delta K_q \frac{\delta X_{\rm q}}{X_{\rm q}} 
\end{equation}
with the coefficients
\begin{equation}\label{e.K2}
\Delta_{AB} K^i= K_A^i - K_B^i
\end{equation}
gathered in Table~\ref{tab0b}.
\end{tcolorbox}

\begin{table*}
\begin{center}
\caption[Sensitivity parameters for clock comparison as defined in Eq.~(\ref{e.dataQED})]{Summary of the sensitivity parameters for clock comparison. The first column indicates the clock comparisons, the second gives the expression of $y_{AB}$ as defined in Eq.~(\ref{def.yAB}) in terms of $(\aem,\bar\mu,g_A)$ thanks to Eq.~(\ref{e.31QED}). The columns 3 to 6 give the coefficients $\Delta K_{AB}$ defined in Eq.~(\ref{e.K2}) and derived from the values listed in Table~\ref{tab0}. Note that the value of $\Delta K^q$ in column 5 corresponds to the decomposition~(\ref{e.QCD2}) in terms of  ($\aem, \bar\mu, X_{\rm q})$ and that one needs to take into account the shift~(\ref{e.dinhK2}) for $\Delta K^q$ when using the parameters ($\aem, X_{\rm q}, X_{\rm e})$; see Eq.~(\ref{e.QCD3b}).}
\label{tab0b}
\begin{tabular}{l | c | cccc}
\hline
Clocks $(A, B)$ &  $y_{AB}$ &  $\Delta K^\alpha$ & $\Delta K^\mu$  & $\Delta K^q$ &  $\Delta K^e$   \\
\midrule
Rb-Cs 		&  $(g_{\rm Rb}/g_{\rm Cs})\aem^{-0.49}$ & $-0.49$ & 0 & $-0.021$ & 0 \\
H(hfs)-Cs  	&  $(g_{\rm p}/g_{\rm Cs})\aem^{-0.83}$ &$-0.83$ & 0 & -0.102 & 0\\
\midrule
H(1s-2s)-Cs  		& $(g_{\rm Cs}\bar\mu)^{-1}\aem^{-2.83}$ & $-2.83$ & $-1$ & $-0.002$ & $- 1$ \\
Hg$^+$-Cs   		& $(g_{\rm Cs}\bar\mu)^{-1}\aem^{-5.77}$ & $-5.77$ & $-1$ &  $-0.002$ & $- 1$\\
Sr--Cs   			& $(g_{\rm Cs}\bar\mu)^{-1}\aem^{-2.77}$ & $-2.77$ & $-1$ & $-0.002$ & - 1 \\
Yb($^3$P$_0$)-Cs   & $(g_{\rm Cs}\bar\mu)^{-1}\aem^{-2.52}$ & $-2.52$ & $-1$ & $-0.002$ & - 1\\
Yb(E2)-Cs    		&  $(g_{\rm Cs}\bar\mu)^{-1}\aem^{-1.83}$ & $-1.83$ & $-1$ & $-0.002$ & - 1  \\
Yb(E3)-Cs    		& $(g_{\rm Cs}\bar\mu)^{-1}\aem^{-8.78}$ & $-8.78$ & $-1$ &  $-0.002$ & - 1 \\
CSO -H(hfs)  		&  $ (g_p\bar\mu)^{-1} \aem^3 $   & 3 & $-1$ & 0.1 & $-1$ \\
\midrule
Hg$^+$-Al$^+$  	& $\aem^{-2.948}$ & $-2.948$ & 0 & 0 & 0  \\
Yb(E3)-Yb(E2)  	&  $\aem^{6.95}$  &  6.95 & 0 & 0 & 0 \\
Dy   				&  $\aem$		& 1 & 0 & 0 & 0  \\
\midrule
SF$_6$-Cs  		& $g_{\rm Cs}^{-1}\bar\mu^{-1/2}\aem^{-2.83}$  &$-2.83$ & $-0.5$ & $-0.002$ & $-0.5$ \\
KRb-Cs  			&  $g_{\rm Cs}^{-1}\bar\mu^{14890\pm60}\aem^{-2.83}$     &$-2.83$ & $14890\pm60$ & $-0.002$ & $14890\pm60$ \\
\bottomrule
\end{tabular}
\end{center}
\end{table*}

\paragraph{Astrophysical data interpretation}\ 

Similarly, as the previous analysis for the constraints obtained on QSO observables defined and discussed in Sect.~\ref{secQSOx} to Sect.~\ref{secmolqso} can be expressed as
\begin{eqnarray}
 y		&\equiv g_{\rm p}\aem^2	&\propto \aem^2X_{\mathrm{q}}^{-0.087}X_{\mathrm{s}}^{-0.013},\nonumber\\		
 \bar\mu	&\equiv\frac{m_{\rm e}}{m_{\rm p}}		&\propto X_{\mathrm{q}}^{-0.037}X_{\mathrm{s}}^{-0.011}X_{\mathrm{e}},\nonumber\\
 x 		&\equiv g_{\rm p}\aem^2\mu 		 &\propto\aem^2X_{\mathrm{q}}^{-0.05}X_{\mathrm{s}}^{-0.002}X_{\mathrm{e}}^{-1},\nonumber\\
 F		&\equiv g_{\rm p}(\aem^2\mu)^{1.57}		&\propto\aem^{3.14}X_{\mathrm{q}}^{-0.0289}X_{\mathrm{s}}^{0.0043}X_{\mathrm{e}}^{-1.57},\nonumber\\
 F'		&\equiv  \aem^2\mu					&\propto\aem^2X_{\mathrm{q}}^{0.037}X_{\mathrm{s}}^{0.011}X_{\mathrm{e}}^{-1},\nonumber\\
 G		&\equiv g_{\rm p}(\aem^2\mu)^{1.85}		&\propto\aem^{3.7}X_{\mathrm{q}}^{-0.0186}X_{\mathrm{s}}^{0.0073}X_{\mathrm{e}}^{-1.85},
\end{eqnarray}
once the scaling of the nucleon mass~(\ref{e.MassN}) is used, so that the seven observable quantities can be reduced to only 4 parameters.

\subsection{Phenomenological $(R,S)$-parameterisation}\label{secRS}

As seen from the previous paragraph, the simplest self-consistent way to phenomenologically describe models which correlated variations is to relate them to $\aem$. A broad class  of grand unification models have been encompassed in the $(R,S)$ parameterisation. 

\paragraph{Definition}\ 

Following \cite{bbnco}, these two parameters are defined by Eqs.~(\ref{e.dfR}) and ~(\ref{e.defS}) respectively which implies
$$
\Delta v/v=S\Delta h/h, \qquad
\Delta\Lambda_{\rm QCD}/\Lambda_{\rm QCD}=R\Delta\aem/\aem.
$$
Assuming all Yukawa couplings enjoy a universal variation, one deduces that
\begin{equation}
\Delta m_{\rm e}/m_{\rm e}=\frac{1}{2}(1+S)\Delta\aem/\aem,\qquad
\end{equation}
and, within a dilaton model, that allows a relation between $\Delta h/h$ and $\Delta\aem/\aem$, one has
\begin{equation}
\Delta m_{\rm n}/m_{\rm n}=\Delta m_{\rm p}/m_{\rm p}=[0.8R+0.2(1+S)]\Delta\aem/\aem.
\end{equation}
The relevant BBN parameters can then be shown to behave as \citep{cnouv}
\begin{equation}
\Delta Q_{\rm np}/Q_{\rm np}=[0.1+0.7S -0.6R]\Delta\aem/\aem,
\end{equation}
\begin{equation}
\Delta \tau_{\rm n}/\tau_{\rm n}=[-0.2-2.0S +3.8R]\Delta\aem/\aem,
\end{equation}
\begin{equation}
\Delta B_{\rm D}/B_{\rm D}=[-6.5(1+S)+18R]\Delta\aem/\aem,
\end{equation}
while for clock studies, one has \citep{NEW_Martins:2017qxd}
\begin{equation}
\Delta\bar\mu/\bar\mu = P \Delta\aem/\aem,\qquad
\Delta g_{\rm p}/g_{\rm p}= Q\Delta\aem/\aem
\end{equation}
with $R=10(2P-15Q)$ and $(1+S)=50(P-8Q)$. 

$R$ and $S$ can be taken as pure phenomenological parameters. One expects $S>0$ while $R$ can be any sign. Their typical value in unification scenarios are thought to be $(R,S)=(26,160)$ \citep{langacker,cnouv} and (109.4,0) in dilaton type models \citep{NEW_Nakashima:2009cs}. 

\paragraph{Constraints on $(R,S)$}\ 

These two parameters can be constrained by observation. Many works have interpreted data within this framework (see discussion in Sect.~\ref{subsecGUT}). \cite{NEW_Luo:2011cf} analyzed atomic clocks constraints (see Sect.~\ref{subsec31-e.h}) in this framework as well as \cite{NEW_Juliao:2013xia} that showed that the data are degenerated according to $(S+1)-2.7 R=-5\pm15$ and that $P=1.5\pm4.5$.  \cite{NEW_Ferreira:2013vxa,NEW_Ferreira:2014fba,NEW_Martins:2017qxd} stressed that the various observation of PKS1413+135 (see Table~\ref{tab03}) allows one to set independent constraints on $(\aem,\mu, g_{\rm p})$ and then extracted from the QSO constraints the one-dimensional confidence intervals for $(R,S)=(277\pm24,742\pm65)$ and from molecular data within our Galaxy \citep{NEW_Joao:2015qva}, while \cite{NEW_Thompson:2016qez,NEW_Thompson:2017wvp} interpreted molecular spectra. \cite{NEW_Martins:2019qxe} combined local and astrophysical data. From a model of polytropic white dwarfs with varying constants, \cite{NEW_Magano:2017mqk} determined their mass-radius relation dependence on $(R,S)$. From G191-B2B (see Sect.~\ref{secstrongs}) it was concluded that it constraints a different combination than atomic clocks.

\section{Experimental and observational constraints on non-gravitational constants}\label{section3}

This section focuses on the experimental and observational constraints on the non-gravitational constants, that is assuming $\ag$ remains constant. We use the convention that $\Delta\alpha = \alpha-\alpha_0$ for any constant $\alpha$, so that $\Delta\alpha<0$ refers to a value smaller in the past than today.

\begin{figure}[htbp]
 \centerline{\includegraphics[scale=0.5]{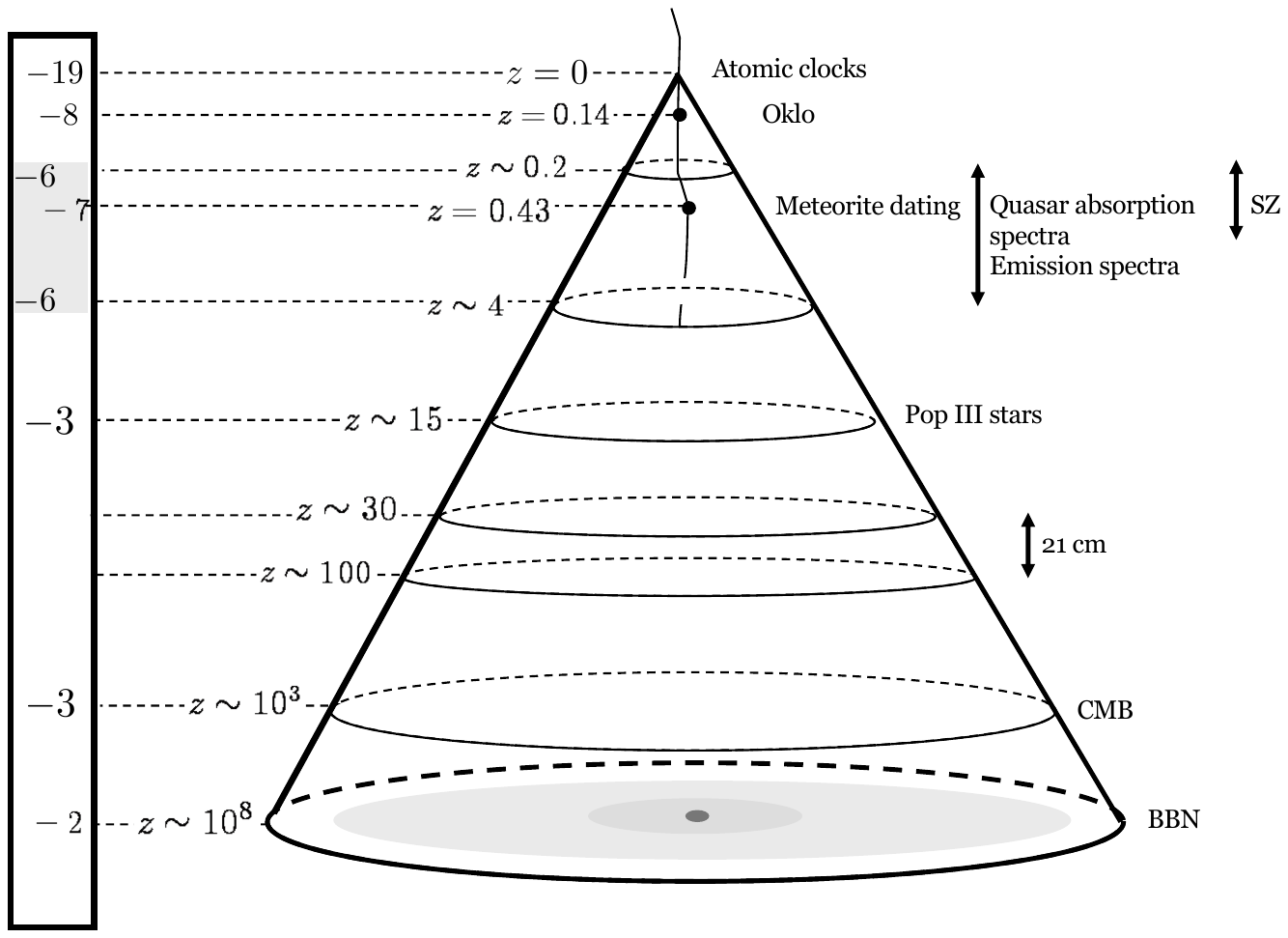}}
  \caption[Systems used to test the constancy of the constants]{Summary of the systems that have been used to probe the constancy of   the fundamental constants and their position in a spacetime diagram in which the cone   represents our past light cone. The shaded areas represents the comoving space probed  by different tests. The scale of the left indicates the typical order of magnitude of the constraints on $\Delta\aem/\aem$; e.g., $-6$ means $\Delta\aem/\aem\lesssim10^{-6}$ at the exception of the first one which refers to $\dot\aem/\aem\lesssim10^{-19}$/yr. We also indicate the typical redshift $z$ of the system.}
  \label{fig-systems}
\end{figure}

\paragraph{Physical systems}\ 

The various physical systems that have been considered to constrain the variation of the fundamental constants can be classified in many ways. We can classify them according to chronologically thanks to their look-back time and more precisely by their spacetime position relative to our actual position. This is summarized in Fig.~\ref{fig-systems}. Indeed higher redshift systems offer the possibility to set constraints on larger time scales, but this is at the expense of usually involving other parameters such as the cosmological parameters. This is, in particular, the case of the cosmic microwave background or of primordial nucleosynthesis. The systems can also be classified in terms of the physics they involve. For instance, atomics clocks, quasar absorption spectra and the cosmic microwave background require only to use quantum electrodynamics to draw the primary constraints while the Oklo phenomenon, meteorites dating and nucleosynthesis require nuclear physics. This is summarized on Table~\ref{tab-sum}.

\begin{table}[htbp]
\caption[Systems considered to set constraints on the variation of the fundamental constants]{Summary of the systems considered to set constraints on the variation of the fundamental constants. We summarize the observable quantities, the primary constants used to interpret the data and the other hypothesis required for their interpretation. All the quantities appearing in this table are defined in the text.}
\label{tab-sum}
\centering
{\small
\begin{tabular}{llllc}
 \toprule
 System  & Observable & Primary constraints  & Other hypothesis & Section  \\
 \midrule
 Atomic clock      & $\delta\ln\nu$           &  $g_i,\aem,\mu$     & --  & \ref{subsec31} \\
 Oklo phenomenon   & isotopic ratio           &  $E_r$              & geophysical model  & \ref{sec:oklo}\\
 Meteorite dating  & isotopic ratio           &  $\lambda$          &  --  & \ref{sec:meteorite} \\
 Quasar spectra    & atomic spectra           &  $g_{\mathrm{p}},\mu,\aem$ & cloud physical properties  & \ref{subsec33}\\
 Stellar physics   & element abundances       &  $B_{\rm D}$               & stellar model   & \ref{secstellar}\\
  CMB               & $\Delta T/T$             &  $\mu,\aem$          & cosmological model & \ref{subsec34} \\
 21~cm             & $T_b/T_{\mathrm{CMB}}$          &  $g_{\mathrm{p}},\mu,\aem$ & cosmological model  & \ref{subsec55} \\
  Galaxy clusters &SZ effect & $y$ & galaxy+cosmological model  & \ref{subsecSZ}\\
 BBN               & light element abundances &  $Q_{\mathrm{np}},\tau_{\mathrm{n}},m_{\mathrm{e}},m_{\mathrm{N}},\aem,B_{\rm D}$
          & cosmological model  & \ref{secbbn}    \\
 \bottomrule
\end{tabular}
}
\end{table}

\paragraph{Strategy}\ 

For any system, setting constraints goes through several steps, summarized on Fig.~\ref{fig-strategie}.
\begin{itemize} 
\item First, we have some observable quantities $O$ from which we can draw constraints on primary parameters $G_k$, which may or may not be fundamental constants (e.g., the BBN parameters, the lifetime of $\beta$-decayers, \dots).  This allows us to state which constants dominate the system and which can be neglected in the analysis. Given data and a study of the effect of the external physical parameters of the system, one can draw constraints on the independent variations of the $G_k$.
\item These primary parameters must then be related to a set of fundamental constants. This requires e.g., to connect nuclear physics to QCD or atomic physics to QED. One can then get constraints on the independent variations of the fundamental constants $\alpha_i$ as described in Sect.~\ref{subsubmass}.
\item In a last step, the number of constants can be reduced by relating them in some unification schemes to get stronger constraints at the expense of being more model-depedent, as described in Sect.~\ref{subsecGUT}.
\end{itemize}
Indeed each step requires a specific modelization and hypothesis and has its own limitations.

\begin{figure}[htbp]
 \centerline{\includegraphics[scale=0.45]{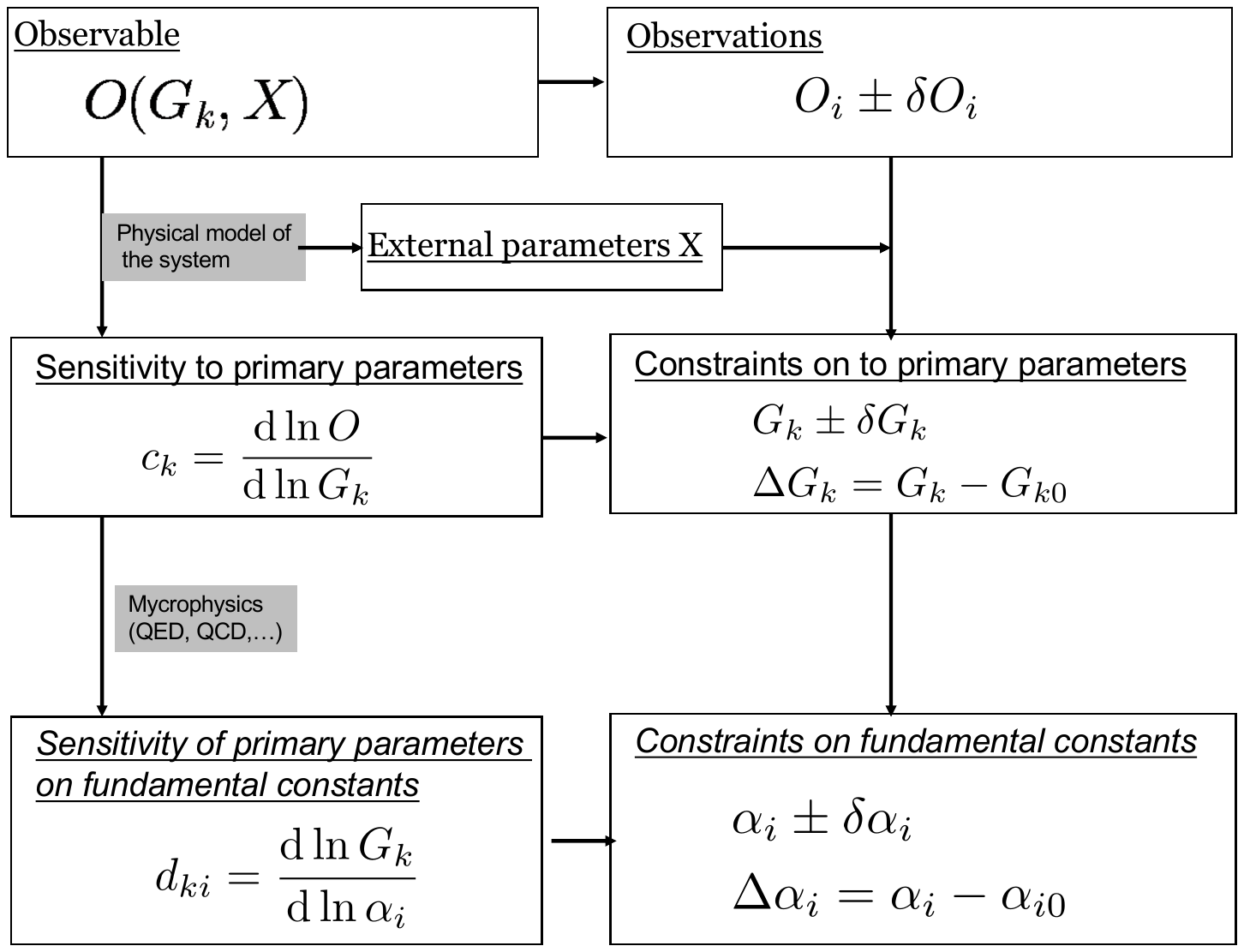}}
  \caption[Relations between observables, primary parameters and fundamental constants]{Given a physical system, its modelization will allow one to (\textit{1}) compute the sensitivity of the observables $O$ to the relevant list of primary parameters $G_k$ by computing the sensitivity coefficients~(\ref{e.ck}) and (\textit{2}) identifying all the external parameters $X$ to be controlled such as e.g., the temperature, magnetic fields etc... This allows one to draw constraints on the independent variation of the primary parameters. Then those parameters need to be expressed in terms of the fundamental constants $\alpha_i$  thanks to the sensitivity coefficients~(\ref{e.dki}). One can then investigate models with correlated or uncorrelated variations to draw final conclusions on the correlated or independent variations of the fundamental constants. The definitions of the sensitivity coefficients are summarized in Appendix~\ref{app2}.}
  \label{fig-strategie}
\end{figure}

\subsection{Atomic and molecular Clocks}\label{subsec31}

The search of the variation of fundamental constants with clocks has followed the developments and progresses of metrology and in particular the tremendous developments of atomic clock precision in the past 20 years; see e.g., the reviews by \cite{NEW_Safronova:2017xyt,NEW_Safronova:2019lex}. While most of the constraints so far have been established with atomic clocks, many new techniques including highly charged ion, molecular and nuclear clocks are now considered.

\subsubsection{Atomic spectra and constants}\label{subsec31-e.h}

The laboratory constraints on the time variation of fundamental constants are obtained by comparing the long-term behavior of several oscillators and rely on frequency measurements. They have witnessed tremendous progresses over the past years, in particular due to the central role of time and frequency measurements in metrology.

\paragraph{Generalities}\ 

The atomic transitions have various dependencies in the fundamental constants that can be used in clocks comparison to constrain their variation. As a textbook example, for the hydrogen atom, the gross, fine and hyperfine-structures are roughly given by
$$
 2p-1s:\,\, \nu\propto cR_\infty,\qquad
 2p_{3/2}-2p_{1/2}:\,\, \nu\propto cR_\infty\aem^2,\qquad
 1s:\,\,\propto cR_\infty\aem^2 g_{\mathrm{p}}\bar\mu,
$$
respectively, where the Rydberg constant set the dimension. $g_{\mathrm{p}}$ is the proton gyromagnetic factor and $\bar\mu=m_{\mathrm{e}}/m_{\mathrm{p}}$. 

In the non-relativistic approximation, the transitions of all atoms have similar dependencies but two effects have to be taken into account.
\begin{itemize}
\item First, the hyperfine-structures involve a gyromagnetic factor $g_i$ (related to the nuclear magnetic moment by $\mu_i=g_i\mu_{\mathrm{N}}$, with $\mu_{\mathrm{N}}= e\hbar/2m_{\mathrm{p}}c$), which are different for each nuclei. 
\item Second, relativistic corrections (including the Casimir contribution), which also depend on each atom (but also on the type of the transition) can be included through a multiplicative function $F_{\mathrm{rel}}(\aem)$. It has a strong dependence on the atomic number $Z$, which can be illustrated on the case of alkali atoms, for which
$$
 F_{\mathrm{rel}}(\aem) =\left[1-(Z\aem)^2\right]^{-1/2}\left[1-\frac43(Z\aem)^2\right]^{-1}
 \simeq 1 +\frac{11}{6}(Z\aem)^2.
$$
\end{itemize}
The developments of highly accurate atomic clocks using different transitions in different atoms offer the possibility to test various combinations of the fundamental constants.

\paragraph{Dependencies on the fundamental constants}\ 

It follows that at the lowest level of description, we can interpret all atomic clocks results in terms of the g-factors of each atoms, $g_i$, the electron to proton mass ration $\bar\mu$ and the fine-structure constant $\aem$. We shall parameterize the hyperfine and fine-structures frequencies as follows.
\begin{itemize}
\item The hyperfine frequency in a given electronic state of an alkali-like atom, such as $^{133}$Cs, $^{87}$Rb, $^{199}$Hg$^{+}$, is
\begin{equation}
\label{e.hf}
 \nu_{\mathrm{hfs}} \simeq R_\infty c \times A_{\mathrm{hfs}} \times g_i
 \times \aem^2\times\bar\mu \times F_{\mathrm{hfs}}(\aem)
\end{equation}
where $g_i=\mu_i/\mu_{\mathrm{N}}$ is the nuclear $g$-factor. $A_{\mathrm{hfs}}$ is a numerical factor depending on each particular atom and we have set $F_{\mathrm{rel}}=F_{\mathrm{hfs}}(\alpha)$. 
\item The frequency of an electronic transition is well-approximated by
\begin{equation}
\label{e.elec}
 \nu_{\text{elec}} \simeq R_\infty c \times A_{\text{elec}} \times F_{\text{elec}}(Z,\alpha),
\end{equation}
where, as above, $A_{\text{elec}}$ is a numerical factor depending on each particular atom and $F_{\text{elec}}$ is the function accounting for relativistic effects, spin-orbit couplings and many-body effects. Even though an electronic transition should also include a contribution from the hyperfine interaction, it is generally only a small fraction of the transition energy and thus should not carry any significant sensitivity to a variation of the fundamental constants.
\end{itemize}

\paragraph{Sensitivity parameters}\ 

The importance of the relativistic corrections was probably first emphasized in \cite{clock-prestage} and their computation through relativistic $N$-body calculations was carried out for many transitions in \cite{kappa-dzuba2, kappa-dzuba,kappa-dzuba3, kappa-flamb}. They can be characterized by introducing the sensitivity of the relativistic factors to a variation of $\aem$,
\begin{equation}\label{clock-sensitivity}
 \kappa_\alpha \equiv \frac{\partial \ln F}{\partial \ln\aem}.
\end{equation}
Table~\ref{tab0} summarizes the values of some of them, as computed in \cite{kappa-dzuba, kappa-tedesco}. Indeed a reliable knowledge of these coefficients at the 1\% to 10\% level is required. The interpretation of the spectra in this context relies, from a theoretical point of view, only on quantum electrodynamics (QED), a theory, which is well tested experimentally \citep{qed-karshen1} so that we can safely obtain constraints on $(\aem,\bar\mu,g_i)$, still keeping in mind that the computation of the sensitivity factors required numerical $N$-body simulations. In terms of the primary parameters  ($\aem,\bar\mu,g_A$), any transitions takes the general form~(\ref{e.31QED}), $\nu_A \propto cR_\infty  g_A^{\lambda_g}\aem^{\lambda_\alpha}\bar\mu^{\lambda_\mu}$ where the sensitivity coefficients $\lambda_i$ are gathered in Table~\ref{tab0}.  

This implies that the comparison of two clocks allows one to constrain different combinations of constants since (\textit{1}) the comparison of two microwave clock frequencies depends on the gyromagnetic factors and $\aem$ but not on $\bar\mu$, (\textit{2}) the comparison of two optical clocks depends on $\aem$ only and (\textit{3}) the comparison of an optical to microwave clocks depends on $(\aem, \bar\mu, g_A)$.

\subsubsection{Experimental constraints from atomic clocks}\label{sec-clock-data}

We present the latest results that have been obtained from the comparison of atomic clocks and refer to Sect.~III.B.2 of FCV \citep{jpu-revue} for earlier studies. They all rely on the developments of new atomic clocks, with the primarily goal to define better frequency standards; see \cite{NEW_Dimarcq:2023lpj}. From an experimental point of view, various combinations of clocks have been performed. It is important to analyze as many species as possible in order to rule-out species-dependent systematic effects. Most experiments are based on a frequency comparison to caesium clocks. The hyperfine splitting frequency between the $F=3$ and $F=4$ levels of its ${}^{2}S_{1/2}$ ground state at 9.192~GHz has been used for the definition of the second since 1967. One limiting effect, that contributes mostly to the systematic uncertainty, is the frequency shift due to cold collisions between the atoms. On this particular point, clocks based on the hyperfine frequency of the ground state of the rubidium at 6.835~GHz, are more favorable.

In order to be as independent as possible from theoretical hypothesis, this section presents only ``raw" constraints on clock comparisons. 

\paragraph{\underline{Atomic hydrogen vs caesium-133}}\
 
The $1s-2s$ transition in atomic hydrogen was compared to the ground state hyperfine  splitting of caesium-133 \citep{clock-fisher04} in 1999 and 2003, setting an upper limit on the variation of $\nu_{\mathrm{H}}$ of  $(-29\pm57) \unit{Hz}$ within 44~months. This can be translated in a
 relative drift
 \begin{equation}
\label{clock-H}
  \frac{\dd}{\dd t}\ln\left(\frac{\nu_{\mathrm{H}}}{\nu_{\mathrm{Cs}}}\right)
    = (-32\pm63)\times10^{-16} \unit{yr}^{-1}.
 \end{equation}
 Since the relativistic correction for the atomic hydrogen transition nearly vanishes, we have $\nu_{\mathrm{H}}\sim R_\infty$ so that
$$
 \frac{\nu_{\mathrm{H}}}{ \nu_{\mathrm{Cs}} }     \propto \left( g_{\mathrm{Cs}}\bar\mu\right)^{-1}  \aem^{-2.83}.
$$
 
\paragraph{\underline{Rubidium-87 vs caesium-133}}\

The comparison of the  ground state hyperfine frequencies of the rubidium-87 and caesium-133 in laser-cooled atomic fountain clocks in their electronic  ground state between 1998 and 2003, with an accuracy of order $10^{-15}$ allowed \cite{clock-marion03}  to get $ \frac{\dd}{\dd t}\ln\left(\frac{\nu_{\mathrm{Rb}}}{\nu_{\mathrm{Cs}}}\right)= (0.2\pm7.0)\times10^{-16} \unit{yr}^{-1}$.  With one more year of experiment, the constraint dropped $\frac{\dd}{\dd t}\ln\left(\frac{\nu_{\mathrm{Rb}}}{\nu_{\mathrm{Cs}}}\right) = (-0.5\pm5.3)\times10^{-16} \unit{yr}^{-1}$. \cite{NEW_Guena:2012zz} then reported the constraint from measurements spanning 14 years of experiment $\frac{\dd}{\dd t}\ln\left(\frac{\nu_{\mathrm{Rb}}}{\nu_{\mathrm{Cs}}}\right) = (-1.36\pm0.91)\times10^{-16} \unit{yr}^{-1}$. It was further improved by taking into account 3 more years of data by \cite{NEW_CRPHYS_2015__16_5_461_0} to
   \begin{equation}
  \frac{\dd}{\dd t}\ln\left(\frac{\nu_{\mathrm{Rb}}}{\nu_{\mathrm{Cs}}}\right)
    = (-1.16\pm0.61)\times10^{-16} \unit{yr}^{-1}.
 \end{equation}
 From Eq.~(\ref{e.hf}), and using the values of the sensitivities  $\kappa_\alpha$, listed in Table~\ref{tab0},  one gets
 $$
   \frac{  \nu_{\mathrm{Rb}}   }{  \nu_{\mathrm{Cs}}  }  \propto\frac{g_{\mathrm{Rb}}}{g_{\mathrm{Cs}}}\aem^{-0.49}\,,
 $$
 that has the advantage to be independent of $\bar\mu$.

\paragraph{\underline{Aluminium and mercury single-ion optical clocks}}\

The comparison of the ${}^1S_{0}- {}^3P_{0}$ transition in   $^{27}\text{Al}^{+}$ and ${}^2S_{1/2}-{}^2D_{5/2}$ in $^{199}\text{Hg}^{+}$  over a year allowed \cite{clock-rosen08} to set the constraint 
  \begin{equation}\label{e.alHg}
    \frac{\dd}{\dd t}\ln\left(\frac{\nu_{\mathrm{Hg}}}{\nu_{\mathrm{Al}}}\right)=
    (5.3\pm7.9)\times10^{-17} \unit{yr}^{-1}.
  \end{equation}
  Proceeding as previously, this tests the stability of\footnote{Note that this sensitivity was estimated as  $-3.02$/$-3.208$ and then reevaluated to $-2.948$; see Table~\ref{tab0}.}
  $$
  \frac{\nu_{\mathrm{Hg}}}{\nu_{\mathrm{Al}}}\propto \aem^{-2.948},
  $$
 hence giving a sharp constraint on $\aem$ alone.  Improvement of one order of magnitude are foreseen from the development of $^{27}{\mathrm{Al}}^{+}$ quantum-logic clock with a systematic uncertainty below ${10}^{\ensuremath{-}18}$ by \cite{NEW_PhysRevLett.123.033201}.

\paragraph{\underline{Mercury-199 vs caesium-133}}\

The $^{199}$Hg$^{+}$ ${}^2S_{1/2}-{}^2D_{5/2}$ optical transition has a high sensitivity to $\aem$ so that it is well suited to test its variation. While $\nu_{\mathrm{Cs}}$ is still given by Eq.~(\ref{e.hf}), $\nu_{\mathrm{Hg}}$ is given by Eq.~(\ref{e.elec}). Using the sensitivities of Table~\ref{tab0}, we conclude that
  $$
 \frac{\nu_{\mathrm{Hg}}}{\nu_{\mathrm{Cs}}}\propto \left(g_{\mathrm{Cs}}\bar\mu\right)^{-1}\aem^{-5.77}.
  $$
  The frequency of the $^{199}$Hg$^{+}$ electric quadrupole transition at 282~nm  was compared to the ground state hyperfine transition of caesium  during a two year period, which lead to \citep{clock-bize03} $\frac{\dd}{\dd t}\ln\left(\frac{\nu_{\mathrm{Hg}}}{\nu_{\mathrm{Cs}}}\right)=    (0.2\pm7)\times10^{-15} \unit{yr}^{-1}$. This was improved by a comparison over a 6 year period \citep{clock-fortier07}  to get
\begin{equation}
   \frac{\dd}{\dd t}\ln\left(\frac{\nu_{\mathrm{Hg}}}{\nu_{\mathrm{Cs}}}\right)=
    (3.7\pm3.9)\times10^{-16} \unit{yr}^{-1}.
  \end{equation}
  
\paragraph{\underline{Strontium-87 vs caesium-133}}\

The comparison of the ${}^1S_0-{}^3P_0$ transition in neutral $^{87}$Sr with a caesium clock was performed in three independent laboratories. The combination of  these three experiments \citep{clock-blatt} led to the  constraint $\frac{\dd}{\dd t}\ln\left(\frac{\nu_{\mathrm{Sr}}}{\nu_{\mathrm{Cs}}}\right)=(-1.0\pm1.8)\times10^{-15} \unit{yr}^{-1}$.

Using a system of 2 Sr-clocks and 3-Cs clocks \cite{New_targat} obtained $\frac{\dd}{\dd t}\ln\left(\frac{\nu_{\mathrm{Sr}}}{\nu_{\mathrm{Cs}}}\right)= (-3.3 \pm 3.0)\times 10^{-16} \unit{yr}^{-1}$, later improved by \cite{NEW_CRPHYS_2015__16_5_461_0} to $\frac{\dd}{\dd t}\ln\left(\frac{\nu_{\mathrm{Sr}}}{\nu_{\mathrm{Cs}}}\right)= (-2.3 \pm 1.8)\times 10^{-16} \unit{yr}^{-1}$/ From a series of 42 measurements of the transition frequency of the $^1$S$_0$-$^3$P$_0$ line in $^{87}$Sr against 2 caesium clocks at PTB between 2017 and 2019, \cite{NEW_Schwarz:2020upg} concluded that
\begin{equation}
   \frac{\dd}{\dd t}\ln\left(\frac{\nu_{\mathrm{Sr}}}{\nu_{\mathrm{Cs}}}\right)=
    (-4.2 \pm 3.3)\times 10^{-17} \unit{yr}^{-1}.
  \end{equation}
Proceeding as previously, this tests the stability of
$$
 \frac{\nu_{\mathrm{Sr}}}{\nu_{\mathrm{Cs}}}\propto \left( g_{\mathrm{Cs}}\bar\mu\right)^{-1} \aem^{-2.77}.
$$

\paragraph{\underline{Ytterbium}}\

It is the only case among the clocks presently under development for which there is more than one clock transition. Yb$^+$ has two ultranarrow optical clock transitions, an   electric octupole ($E3$: ${}^2$S$_{1/2}$-${}^{2}$F$_{7/2}$) at 467 nm and an electric quadrupole ($E2$:  ${}^2S_{1/2}- {}^2D_{3/2}$) at 436 nm (688~THz). This leads to various constraints.
  
\begin{itemize}
\item{\bf Yb$^+$-E2 vs caesium-133.} The $E2$ electric quadrupole transition was compared to  the ground state hyperfine transition of caesium. The constraint of \cite{clock-peik06} was updated,  after comparison over a six year period, which led to \citep{clock-peik04,NEW_Peik:2010zz} $\frac{\dd}{\dd t}\ln\left(\frac{\nu_{\mathrm{Yb-E2}}}{\nu_{\mathrm{Cs}}}\right)= (-0.78\pm1.40)\times10^{-15} \unit{yr}^{-1}$. This was improved by \cite{NEW_Tamm:2013} to get
\begin{equation}
   \frac{\dd}{\dd t}\ln\left(\frac{\nu_{\mathrm{Yb-E2}}}{\nu_{\mathrm{Cs}}}\right)=
     (-0.5 \pm 1.9)\times10^{-16} \unit{yr}^{-1}.
  \end{equation}
 Proceeding as previously, this tests the stability of
  $$
 \frac{\nu_{\mathrm{Yb-E2}}}{\nu_{\mathrm{Cs}}}\propto \left(g_{\mathrm{Cs}}\bar\mu\right)^{-1} \aem^{-1.83}.
  $$

\item{\bf Yb$^+$-E3 vs caesium-133.} A similar analysis for the transition $E3$  at 467~nm gave \citep{NEW_Huntemann:2014dya} $\frac{\dd}{\dd t}\ln\left(\frac{\nu_{\mathrm{Yb-E3}}}{\nu_{\mathrm{Cs}}}\right)=(0.2\pm4.1)\times10^{-16} \unit{yr}^{-1}$.  Besides, the comparison with a caesium clock allow  \cite{Lange:2020cul} to conclude that
\begin{equation}
   \frac{\dd}{\dd t}\ln\left(\frac{\nu_{\mathrm{Yb-E3}}}{\nu_{\mathrm{Cs}}}\right)=
    (-3.1\pm3.4)\times10^{-17} \unit{yr}^{-1}.
  \end{equation}
 Table~\ref{tab0} gives that  
 $$
 \frac{\nu_{\mathrm{Yb-E3}}}{\nu_{\mathrm{Cs}}}\propto \left(g_{\mathrm{Cs}}\bar\mu\right)^{-1} \aem^{-8.78}.
  $$

\item{\bf Yb$^+$-E3 vs Yb$^+$-E2.} \cite{NEW_Godun:2014naa} performed the first measurement of the ratio between  $E2$ and $E3$  without reference to the Cs primary standard, and using the same single ion of $^{171}$Yb. Such a direct measurement of the ratio of the two optical frequencies are free from the additional uncertainties introduced by the primary Cs frequency standard. This was then  used by \cite{Lange:2020cul} who got
\begin{equation}\label{e.Ybe3e2}
   \frac{\dd}{\dd t}\ln\left(\frac{\nu_{\mathrm{Yb-E3}}}{   \nu_{\mathrm{Yb-E2}}}  \right)=
    (-6.8\pm7.6)\times10^{-18} \unit{yr}^{-1}.
  \end{equation}
The analysis of  about 235 days of measurement data accumulated over a period of about 26 months, led \cite{ NEW_Filzinger:2023zrs} to conclude that 
\begin{equation}\label{e.Ybe3e2b}
   \frac{\dd}{\dd t}\ln\left(\frac{\nu_{\mathrm{Yb-E3}}}{   \nu_{\mathrm{Yb-E2}}}  \right)=
    (-1.2\pm1.8)\times10^{-18} \unit{yr}^{-1}
  \end{equation}
once combined with the previous data by  \cite{Lange:2020cul}.  For this transition the sensitivity coefficient \citep{dzubapolo2} is $K^\alpha_{E2,E3}=-6.95$ which implies
\begin{equation}
 \frac{\nu_{\mathrm{Yb-E3}}}{   \nu_{\mathrm{Yb-E2}}} = \aem^{-6.95}\,.
\end{equation} 

\item{\bf Yb vs caesium-133.} The  $^1$S$_0$ - $^3$P$_0$ transition in neutral  $^{171}$Yb serves as a basis for frequency standards \citep{NEW_Brown:2017lvs}. Its sensitivity to $\aem$ was computed by \cite{Yb_Safronova:2018quw}. \cite{NEW_McGrew2017} compared the $^1$S$_0$ - $^3$P$_0$ transition of $^{171}$Yb to an international collection of national primary and secondary frequency standards using satellite time and frequency transfer to get
\begin{equation}
   \frac{\dd}{\dd t}\ln\left(\frac{\nu_{\mathrm{Yb}}}{\nu_{\mathrm{Cs}}}\right)=
   (-4.9 \pm 3.6)\times 10^{-17} \unit{yr}^{-1}
\end{equation}
and Table~\ref{tab0} gives that it behaves
 $$
 \frac{\nu_{\mathrm{Rb}}}{\nu_{\mathrm{Cs}}}\propto \left( g_{\mathrm{Cs}}\bar\mu \right)^{-1} \aem^{-2.52}\,.
 $$
\end{itemize}  
  
\paragraph{\underline{Atomic dyprosium}}\

It was suggested in \cite{kappa-dzuba,kappa-dzuba3}  (see also \cite{Dy-df} for a computation of the transition amplitudes of the low  states of dyprosium) that the electric dipole (E1) transition between two nearly degenerate opposite-parity  states in atomic dyprosium should be highly sensitive to the variation of $\aem$.  It was then demonstrated \citep{nguyen04} that a constraint of the order of  $10^{-18}$/yr can be reached. The frequencies of nearly of two isotopes of dyprosium  were monitored over a 8 months period \citep{clock-cingoz} showing  that the frequency variation of the 3.1-MHz transition in $^{163}$Dy and the 235-MHz transition in  $^{162}$Dy are $9.0 \pm 6.7$\,Hz/yr and $-0.6 \pm 6.5$\,Hz/yr, respectively. As seen from Table~\ref{tab0}, the Dy/Cs frequency comparison is over six orders of magnitude more sensitive to variation of $\aem$ than to variation of $(\mu,X_{\rm q})$ so that it provides a constraints on $\aem$ alone
  \begin{equation}
    \label{clock-dypro}
    \frac{\dot\aem}{\aem} = (-2.7\pm2.6)\times 10^{-15} \unit{yr}^{-1},
  \end{equation}
  at 1$\sigma$ level, without any assumptions on the constancy of other constants.  \cite{NEW_Leefer:2013waa} used the 6735.5-MHz transition in $^{164}$Dy instead of the 3.1-MHz transition in $^{163}$Dy to reach
\begin{equation}
    \label{clock-dypro2}
    \frac{\dot\aem}{\aem} = (-5.8\pm6.9)\times 10^{-17} \unit{yr}^{-1}.
\end{equation}

\paragraph{Summary}\

The experimental constraints on laboratory clock comparison are summarized in Table~\ref{tab1} in which we have included their dependencies in the QED primary parameters using the data from Table~\ref{tab0}. In the end all frequency comparisons depend only on the 4 parameters $(\aem,\bar\mu, g_{\rm Cs},g_{\rm Rb})$. Indeed, they are not independent but the set of data can be used to derived constraints on their independent time variation.

\begin{table}[htbp]
\caption[Experimental constraints obtained from the comparisons of  atomic clocks.]{Summary of the experimental constraints obtained from the comparisons of atomic clocks. For each constraint on the relative drift of the frequency ratio of the two clocks, we provide the dependence in the various constants, using the numbers of Table~\ref{tab0}. The experiments on SF$_6$ and KRb molecular clocks are described in Sect.~\ref{secmol} below. Note that the constraints obtained from the same clock comparison are not independent (since they report results from the same experiment, running over different time spans); see text for discussion.}
\label{tab1}
\centering
{\small
\begin{tabular}{lcrcr}
\toprule 
 Clock 1            & Clock 2       & Constraint  (yr$^{-1}$) & Constants  &  Reference \\
 ~                  & & $\frac{\dd}{\dd t}\ln \left(\frac{\nu_{\text{clock 1}}}{\nu_{\text{clock 2}}} \right)\quad$ &  dependence & ~  \\
\midrule
$^{87}$Rb            & $^{133}$Cs & $(0.2 \pm 7.0) \times 10^{-16}$    & $\frac{g_{\mathrm{Rb}}}{g_{\mathrm{Cs}}}\aem^{-0.49}$ & \cite{clock-marion03} \\
& & $(-0.5 \pm 5.3) \times 10^{-16}$   & & \cite{clock-bize05}\\
& & $(-1.36\pm0.91) \times 10^{-16}$ & ~  & \cite{NEW_Guena:2012zz}\\
&  & $(-1.16\pm0.61)\times10^{-16}$ & ~  & \cite{NEW_CRPHYS_2015__16_5_461_0}\\
\midrule\midrule
$^{1}$H              & $^{133}$Cs & $(-32 \pm 63) \times 10^{-16}$     & $(g_{\mathrm{Cs}}\bar\mu)^{-1} \aem^{-2.83}$ & \cite{clock-fisher04}\\
\midrule
$^{199}$Hg$^{+}$     & $^{133}$Cs & $(0.2 \pm 7) \times 10^{-15}$      & $(g_{\mathrm{Cs}}\bar\mu)^{-1} \aem^{-5.77}$& \cite{clock-bize03}\\
&  & $(3.7 \pm 3.9) \times 10^{-16}$    &  & \cite{clock-fortier07}\\
\midrule
$^{171}$Yb$^{+}$ (E2)      & $^{133}$Cs & $(-0.78\pm1.40)\times10^{-15}$   & $(g_{\mathrm{Cs}}\bar\mu)^{-1} \aem^{-1.83}$ & \cite{clock-peik04}\\
&      & $(-0.5 \pm 1.9) \times 10^{-16}$ &    & \cite{NEW_Tamm:2013}\\
\midrule
$^{171}$Yb$^{+}$  (E3)   & $^{133}${Cs}$$ & $(-0.2 \pm 4.1) \times 10^{-16}$ & $ (g_{\mathrm{Cs}}\bar\mu)^{-1} \aem^{-8.78}$  & \cite{NEW_Huntemann:2014dya}\\
 && $(-3.1\pm3.4)\times10^{-17}$ & &\cite{Lange:2020cul} \\
\midrule
$^{171}$Yb  $^1$S$_0-^3$P$_0$    & $^{133}${Cs}$$ & $ (-4.9 \pm 3.6)\times 10^{-17}$ &  $(g_{\mathrm{Cs}}\bar\mu)^{-1} \aem^{-2.52}$ & \cite{NEW_McGrew2017}\\
\midrule
$^{87}$Sr           & $^{133}$Cs & $(-1.0 \pm 1.8) \times 10^{-15}$   &  $(g_{\mathrm{Cs}}\bar\mu)^{-1} \aem^{-2.77}$& \cite{clock-blatt}\\
& & $ (-3.3 \pm 3.0)\times 10^{-16} $   &  & \cite{New_targat}\\
& & $(-2.3 \pm 1.8) \times 10^{-16}$   &  & \cite{NEW_CRPHYS_2015__16_5_461_0}\\
&& $(-4.2 \pm 3.3)\times 10^{-17}$   & & \cite{NEW_Schwarz:2020upg}\\
\midrule\midrule
$^{199}${Hg}$^{+}$ & $^{27}$Al$^{+}$     & $(5.3 \pm 7.9) \times 10^{-17}$ & $\aem^{-2.948}$ & \cite{clock-rosen08}\\
\midrule
$^{171}$Yb$^{+}$  (E3)    & $^{171}$Yb$^{+}$  (E2) & $(-6.8\pm 7.8) \times 10^{-18}$ & $\aem^{-6.95}$  &  \cite{Lange:2020cul} \\
& & $ (-1.2\pm1.8)\times10^{-18}$ & & \cite{ NEW_Filzinger:2023zrs}\\
\midrule
$^{162}$Dy            & $^{163}$Dy  & $(-2.7 \pm 2.6) \times 10^{-15}$   & $\aem$ & \cite{clock-cingoz}\\
\midrule
$^{162}$Dy     & $^{164}${Dy}  & $(-5.8 \pm 6.9) \times 10^{-17}$ &$\aem$   & \cite{NEW_Leefer:2013waa}\\
\midrule\midrule
 {SF}$^{6}$ &     $^{133}$Cs          &   $(1.9 \pm 2.7) \times 10^{-14}$ & $g_{\mathrm{Cs}}^{-1} \bar\mu^{-1/2} \aem^{-2.83}$   & \cite{clock-mu}\\
 KRb           &     $^{133}$Cs          &   $(-0.44 \pm 1.47_{\rm stat}\pm 0.24_{\rm syst}) \times 10^{-10}$   & $g_{\mathrm{Cs}}^{-1} \bar\mu^{15\,000} \aem^{-2}$  & \cite{NEW_Kobayashi:2019xdt}. \\
\bottomrule
\end{tabular}
}
\end{table}

\subsubsection{Constraints on primary parameters}

\paragraph{Constraint on $\aem$ alone}\
 
 Since four clock comparisons listed in Table~\ref{tab1} depend on $\aem$ only, it can be constrained independently of the 3 others primary parameters. 

The aluminium-mercury clocks comparison \citep{clock-rosen08} depends only on $\aem$ so that the constraint~(\ref{e.alHg}) translates to ${\dot\aem}/{\aem} = (-1.6\pm2.3)\times10^{-17} \unit{yr}^{-1}$ on $\aem$ alone without any use of complementary data. Since its sensitivity to $\aem$ is 6 order of magnitude higher than other sensitivities, see Eq.~(\ref{clock-dypro2}), the dyprosim comparison \citep{NEW_Leefer:2013waa} gives in practice ${\dot\aem}/{\aem} = (-5.8\pm6.9)\times 10^{-17} \unit{yr}^{-1}$. Then, the Yb:E2-E3 comparison by \cite{Lange:2020cul} implies the bound $\dot\aem/\aem= (1\pm1.1)\times10^{-18} \unit{yr}^{-1}$. The improvements by \cite{ NEW_Filzinger:2023zrs} translate to
\begin{equation}\label{e.clock-alpha}
  \frac{\dot\aem}{\aem} = (1.8\pm2.5)\times 10^{-19} \unit{yr}^{-1}.
\end{equation}
Combining the independent measurements does not improve this latter constraint, which can be considered as the sharpest constraint on the independent variation of $\aem$ today.

\paragraph{Constraints on other constants}\

Several analysis have been performed to extract constraints on $\bar\mu$ and $X_{\rm q}$. Their results have evolved with the improvement of experimental data that have been combined in different ways. We refer to Fig.~\ref{fig-clock2} for a up-to-date status on the constraints on $(\aem,g_{\rm Cs}\bar\mu)$.

From their result on Rb/Cs, \cite{NEW_Guena:2012zz} derived $\dd\ln[\aem^{-0.49}X_{\rm q}^{-0.021}]$  $= (-1.36\pm0.91)\times10^{-16} \unit{yr}^{-1}$. Combined with  the contraints on $\aem$ by \cite{clock-rosen08}, they concluded 
\begin{equation}\label{e-XQ0}
K_{\rm Cs}^q \dot X_{\rm q}/X_{\rm q}= (1.4\pm0.9)\times10^{-17} \unit{yr}^{-1}.
\end{equation}
This constraint was then used  by \cite{NEW_Huntemann:2014dya} to analyse their Yb(E3)-Cs results fitted together with the constraints  by \cite{clock-rosen08} and \cite{NEW_Leefer:2013waa} to conclude that $\dot\aem/\aem= (-0.20\pm0.20)\times10^{-16} \unit{yr}^{-1}$ and $\dot{\bar \mu}/\bar\mu=(-0.5\pm1.6)\times10^{-16} \unit{yr}^{-1}$.  \cite{NEW_Godun:2014naa} follows the same route but assumes the sensitivity to $X_{\rm q}$, was zero to get the similar results, $\dot\aem/\aem= (-0.7\pm2.1)\times10^{-17} \unit{yr}^{-1}$ and $\dot{\bar \mu}/\bar\mu=(0.2\pm1.1)\times10^{-16} \unit{yr}^{-1}$.  This series of analysis were summarized in Fig.~3 of \cite{NEW_Huntemann:2014dya}. Combining with the constraint on $\aem$ by \cite{clock-rosen08} and on $X_{\rm q}$ by \cite{NEW_Ashby:2018jdl}, \cite{NEW_McGrew2017b}  concluded that $\dot{\bar \mu}/\bar\mu=(5.3\pm6.5)\times10^{-17}~\unit{yr}^{-1}$.  To finish, \cite{NEW_Schwarz:2020upg}  combined~(\ref{e-XQ0}) with the constraint on $\aem$ by \cite{NEW_Godun:2014naa} and \cite{NEW_Huntemann:2014dya}, to conclude that $\dot{\bar \mu}/\bar\mu=(-6.9\pm6.5)\times 10^{-17} \unit{yr}^{-1}$. 

\paragraph{Clock constraints on time variation of $\aem, \bar\mu$ and $X_{\rm q}$}\
 
Figure~\ref{fig-clock2} reproduces and updates Fig.~3 of \cite{NEW_Huntemann:2014dya} including the new constraints by \cite{Lange:2020cul,NEW_McGrew2017,NEW_Schwarz:2020upg} and \cite{ NEW_Filzinger:2023zrs}. An efficient way to disentangle the constraints is through 3 steps.
 \begin{enumerate}
 \item First, the comparisons of optical clocks set the constraint~(\ref{e.clock-alpha})  on $\aem$ alone.
 \item Then, the Rb/Cs experiments constrain $g_{\rm Rb}/g_{\rm Cs}$ to get
  \begin{align}													
 \frac{\dd}{\dd t}\ln\left(\frac{g_{\mathrm{Rb}}}{g_{\mathrm{Cs}}}\right)&=  \frac{\dd}{\dd t}\ln\left(  X_{\rm q}^{-0.021} \right) \nonumber \\
 													&= (-1.16\pm 0.61)\times10^{-16} \unit{yr}^{-1} \,,\label{clock-gtest2}
\end{align} 
using \cite{NEW_CRPHYS_2015__16_5_461_0}.
\item To finish, the  comparisons of optical clocks to Cs-133 provide constraints on $g_{\rm Cs}\bar\mu$. Given the bound~(\ref{e.clock-alpha}) , the combined analysis of the data by \cite{clock-fortier07,NEW_Tamm:2013,Lange:2020cul,NEW_McGrew2017,NEW_Schwarz:2020upg} gives
\begin{align}\label{clock-gtest}
\frac{\dd}{\dd t}\ln\left(g_{\mathrm{Cs}}\bar\mu\right)&= \frac{\dd}{\dd t}\ln\left( X_{\rm q}^{0.002}\bar\mu  \right) =  \frac{\dd}{\dd t}\ln\left( X_{\rm q}^{-0.046} X_{\rm e} \right) \nonumber\\
 													&=   ( 0.42\pm 0.13)\times 10^{-16} \unit{yr}^{-1} \,.
\end{align} 	
\item These three latter bounds can be combined to get a constraints on $\dot X_{\rm q}$ and   $\dot{\bar\mu} $ or $\dot X_{\rm e}$. 
 \end{enumerate}
 
 \begin{tcolorbox}
 As a conclusion, the global fit of the data discussed above allows us to set the following constraints
 \begin{eqnarray}\label{e.clockfinalbound}
  \left.\frac{\dot\aem}{\aem}\right\vert_0 &=& (1.8\pm2.5)\times 10^{-19} \unit{yr}^{-1},\\
  \left.\frac{\dot{\bar\mu}}{\bar\mu}\right\vert_0 &=& (3.09\pm1.42)\times 10^{-17} \unit{yr}^{-1}, \\
  \left.\frac{\dot X_{\rm q}}{X_{\rm q}}\right\vert_0 &=& (5.5\pm2.9)\times 10^{-15} \unit{yr}^{-1},\\
   K^q_{\rm Cs} \left.\frac{\dot X_{\rm q}}{X_{\rm q}}\right\vert_0 &=& (1.10\pm0.58)\times 10^{-17} \unit{yr}^{-1},\\
  \left.\frac{\dot X_{\rm e}}{X_{\rm e}}\right\vert_0 &=& (2.96\pm1.34)\times 10^{-16} \unit{yr}^{-1}\, ,
  \end{eqnarray}
 where we recall that $\bar\mu=m_{\rm e}/m_{\rm p}$,  $X_{\rm e,q}=m_{\rm e,q}/\Lambda_{\rm QCD}$ and the sensitivity coefficient $K^q_{\rm Cs}= 0.002$, see Eq.~(\ref{e.dinhK}).
 \end{tcolorbox}

\begin{figure}[htbp]
 \centerline{\includegraphics[width=1.5\textwidth]{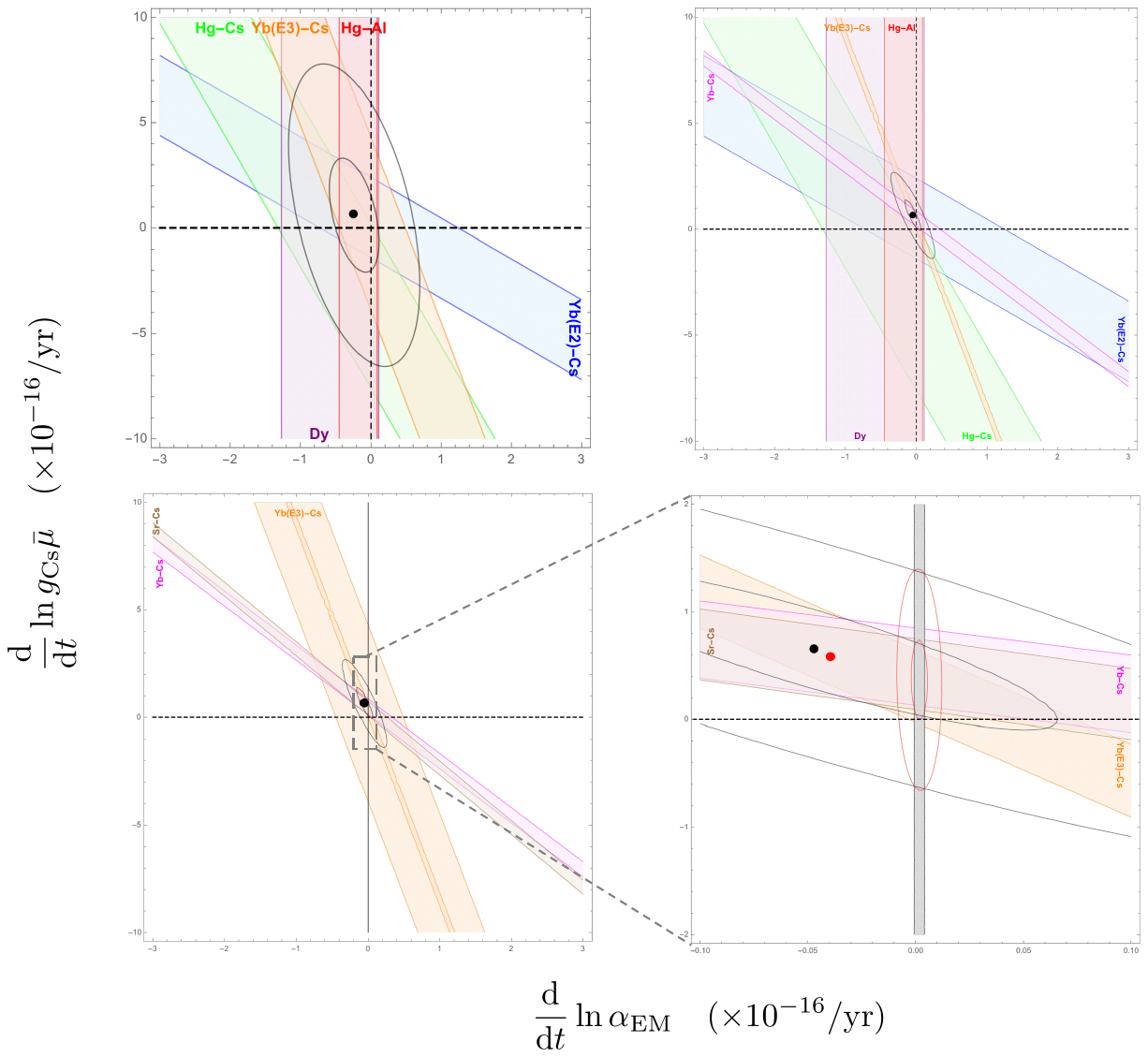}}
 \vskip-0.5cm
  \caption[Atomic clock constraints on $(\aem,g_{\rm Cs}\bar\mu)$]{Synthetic constraints on $(\aem, g_{\rm Cs}\bar\mu)$ from the atomic clock constraints summarized in Table~\ref{tab1}. The upper left-plot reproduces \cite{NEW_Huntemann:2014dya}, note however that it depicts $g_{\rm Cs}\bar\mu$ instead of $\mu=m_{\rm p}/m_{\rm e}$. The contour plots indicate 1$\sigma$ and 2$\sigma$ constraints and the black point the best-fit. The data used combine comparisons of optical clocks [red: Hg-Al  \citep{clock-rosen08}; Purple: Dy 162-164  \citep{NEW_Leefer:2013waa}] and comparison of optical clocks to caesium [Green: Hg-Cs \citep{clock-fortier07}; Blue: Yb(E2)-Cs  \citep{NEW_Tamm:2013}; Orange: Yb(E3)-Cs \citep{NEW_Huntemann:2014dya}]. The upper-right plot updates the Yb(E3)-Cs data by \cite{Lange:2020cul} and adds the Yb-Cs data [purple] by \cite{NEW_McGrew2017}. The lower-left plot compares the Yb(E3)-Cs data by \cite{NEW_Huntemann:2014dya} and \cite{Lange:2020cul}  and adds the Sr-Cs data [Brown] by  \cite{NEW_Schwarz:2020upg}. The lower-right plot zooms on the dashed region and adds the constraints~(\ref{e.clock-alpha}) on $\aem$. The black dot and contour lines correspond to the best-fit,  1$\sigma$ and 2$\sigma$ constraints obtained from all data expect Sr-Cs and the $\aem$ constraint while the red dot is the  best fit for all data but the $\aem$ constraint. The red ellipses correspond to 1$\sigma$ and 2$\sigma$ constraints obtained from all data. }
  \label{fig-clock2}
\end{figure}

\subsubsection{Fiber-linked networks and clocks in space}\label{secfiber}

To extend the analysis of the previous section, on can compare clocks at different locations  since they can be networked online using either satellites \citep{NEW_Schiller:2006uy,NEW_Magnani:2019buu} -- see \cite{NEW_Belenchia:2021rfb,NEW_Alonso:2022oot} for a review and a roadmap on quantum physics in space --, fibre links \citep{NEW_sRoberts:2019sfo}, or offline via time-stamping of measurements \citep{NEW_Wcislo:2018ojh} -- see \cite{NEW_Savalle:2019jsb} for a discussion of spacetime separated clock experiments to test variations of fundamental constants. It has recently been implemented by \cite{Filzinger:2023qqh} to search for ultralight dark matter.

\paragraph{Fiber-linked comparisons}\

\cite{NEW_sRoberts:2019sfo} searched for transient variations of $\aem$ using data from the European network of fiber-linked optical clocks based on Sr, Hg and Yb$^+$ atoms located in France, Germany and United Kingdom. Thanks to the use of fiber links to perform the comparisons, the stability of the measurements is limited by the one of the clocks themselves \citep{NEW_lisdat}. It was concluded that $\Delta\aem/\aem<5\times 10^{-17}$ on a duration of $10^3$~s. The  QSNET project \citep{NEW_Barontini:2021cfm,NEW_Barontini:2021mvu} is a network of clocks that will include existing Sr, Yb$^+$ and Cs atomic clocks at the National Physical Laboratory (NPL) in London and several new clocks (N$^+_2$ molecular ion clock, CaF molecular optical lattice clock and a Cf highly charged ion clock).

\paragraph{Atomic clocks in space (ACES)}\
 
An improvement of at least an order of magnitude on current constraints can be achieved in space. The PHARAO/ACES project \citep{clock-ACES,reynaud09} (Projet d'Horloge Atomique par Refroidissement d'Atomes en Orbite) of the European Spatial Agency combines laser cooling techniques and a microgravity environment in orbit. It consists of a Cs clock installed on the outside pallet of the Columbus module of the ISS that can be compared to ground clocks using two-way microwave links. It aims at performing redshift test at the level of $2-3\times10^{-6}$ during 1.5 to 3 years. Its performances are evaluated in \cite{NEW_Savalle:2019isy}.

The Galileo gravitational redshift test \citep{NEW_Delva:2018ilu,NEW_Delva:2019kxg}, GREAT, has been comparing H-maser clocks onboard of eccentric satellites to ground clocks to conclude that $\beta_{\rm H-maser}=(0.19\pm2.46)\times10^{-5}$ \citep{NEW_Delva:2019kxg}; see Eq.~(\ref{e.betaK}) below for the definition of the redshift parameter. The proposal FOCOS \citep{NEW_Derevianko:2021kye} aims to improve this constraint to $10^{-9}$.

The SAGAS (Search for anomalous gravitation using atomic sensor) project  \citep{reynaud09, wolf09}  aims at flying highly sensitive optical atomic clocks and cold atom accelerometers on a Solar system trajectory on a time scale of 10 years. It could test the constancy of the fine-structure constant along the satellite worldline, which, in particular, can set a constraint on its spatial variation of the order of $10^{-9}$.

Among these recent developments we shall also mention the proposal by \cite{NEW_Terno:2018uak} to supress the first-order Doppler effect which dominates the weak gravitational signal and the mission \citep{NEW_Schkolnik:2022utn} of optical atomic clocks aboard the space station (OACESS).

\subsubsection{Experimental constraints  from molecular transitions}\label{secmol}

As we have seen, only atomic clocks based on a hyperfine transition involve $\bar \mu$ and they all share the same sensitivity which makes it difficult to extract a model-independent constraints on the variation of $\bar\mu$. Molecules and precision molecular spectroscopy provide pathway toward better constraint on the variation of $\bar\mu$ that are complementary to the constraints from atomic clocks; see e.g., \citep{NEW_Jansen:2013paa} for a dedicated review.

\paragraph{Generalities}\
 
The first to be considered were diatomic molecules since, as first pointed out by \cite{mu-theorie}, molecular lines can provide a test of the variation of $\mu$. The energy difference between two adjacent rotational levels in a diatomic molecule is inversely proportional to $M r^{-2}$, $r$ being the bond length and $M$ the reduced mass, and the vibrational transition of the same molecule has, in first approximation, a $\sqrt{M}$ dependence. For molecular hydrogen $M=m_{\mathrm{p}}/2$ so that the comparison of an observed vibro-rotational spectrum with a laboratory spectrum gives an information on the variation of $m_{\mathrm{p}}$ and $m_{\mathrm{n}}$. Comparing pure rotational transitions with electronic transitions gives a measurement of $\mu$. It follows that the frequency of vibro-rotation transitions is, in the Born--Oppenheimer approximation, of the form
\begin{equation}
\label{mu1}
\nu\simeq E_I\left(c_{\text{elec}} +c_{\text{vib}}\sqrt{\bar\mu}
+c_{\text{rot}}\bar\mu\right)
\end{equation}
where $c_{\text{elec}}$, $c_{\text{vib}}$ and $c_{\text{rot}}$ are some numerical coefficients. Molecules such as H$_2$, HD or NH$_3$ have been extensively used in astrophysics (see Sect.~\ref{secmolqso} below).
 
\paragraph{SF$_6$ transition}\
 
 The comparison of the vibro-rotational transition in the molecule SF6, using the transition is P(4)E$_0$ in the 2$\nu_3$ band, was compared to a caesium clock over a two-year period, leading to the constraint \citep{clock-mu}
\begin{equation}
   \frac{\dd}{\dd t}\ln\left(\frac{\nu_{\text{SF6}}}{\nu_{\mathrm{Cs}}}\right)=
    (1.9\pm0.12\pm2.7)\times10^{-14} \unit{yr}^{-1},
\end{equation}
where the second error takes into account uncontrolled systematics. Now, Table~\ref{tab0} gives
$$
 \frac{\nu_{\mathrm{SF6}}}{\nu_{\mathrm{Cs}}}\propto \bar\mu^{1/2}\aem^{-2.83}(g_{\mathrm{Cs}}\bar\mu)^{-1}.
$$
It can be combined with the constraint~(\ref{clock-H}) which enjoys the same dependence to caesium  and~(\ref{e.clock-alpha}) to infer that
\begin{equation}
\label{clock-bound2}
    \frac{\dot{\bar \mu}}{\bar\mu}=(-3.8\pm5.6)\times10^{-14} \unit{yr}^{-1}.
\end{equation}
Combined with Eq.~(\ref{clock-gtest}), we can obtain independent constraints on the time variation of $g_{\mathrm{Cs}}$, $g_{\mathrm{Rb}}$ and $\bar\mu$.

While this is indeed less stringent that the constraints from atomic clocks discussed in the previous section, molecular clocks are a fast developing field; see e.g. \cite{NEW_Oswald:2021vtc}. The success of techniques such as laser cooling, trapping or coherent manipulations in atomic physics have motivated their extension to molecules and the control of their quantum state offers new direction too test fundamental physics and in particular fundamental constants \citep{NEW_Mitra:2022mua}. This has been extended to the proposal to consider optical clocks based on molecular vibrations  \citep{NEW_Hanneke:2020xsm}.

\paragraph{KRb molecule}\
 
 To date, the best constraint has been obtained from KRb  \citep{NEW_Kobayashi:2019xdt}. Using a transition between nearly a degenerate pair of vibrational levels each associated with a different electronic potential of ultracold diatomic alkali KRb molecules that enjoys  a large sensitivity coefficient, $\Delta\nu/\nu \simeq (14\,890\pm60)\Delta\bar\mu/\bar\mu$, it was concluded that
\begin{equation}
\label{clock-bound3}
   \frac{\dot{\bar \mu}}{\bar\mu}=(0.3\pm1.0)\times10^{-14} \unit{yr}^{-1}.
\end{equation}

\paragraph{Other proposals}\

There is a large activity to increase the stability and sensitivity of molecular clocks, both theoretically and experimentally. We list s series of proposals and latest developments on this rich and active field of research.
\begin{itemize}
\item{\bf Simple molecules}, such as H$^+_2$ and HD$^+$, offer the possibility of accurate theoretical computations while many experiments exist with high precision spectroscopy  \citep{NEW_Ubachs:2015fuf,NEW_muc,NEW_mud} making them good candidates for test of $\mu$ variation \citep{NEW_mub,NEW_mua}. They are also of importance for astrophysics; see Sect.~\ref{secmolqso} below.

\item {\bf Diatomic molecules}. \cite{mu-N2} estimated the enhanced sensitivity of very close narrow levels of different nature that exist in diatomic molecules due to cancelation between the fine structure and vibrational intervals in the electronic ground state, such as  Cl$^+_2$, CuS, IrC, SiBr, HfF$^+$, LaS, LuO, HfF$^+$ and $^7$LiH-$^6$LiH \citep{NEW_Constantin:2015eqv} for which the vibrorotational transition in the $v=0\rightarrow 1$ band has a sensitivity of $-585$. We refer to \cite{NEW_Ubachs:2015fro} for a review on H$_2$, \cite{NEW_PhysRevA.77.012511} for $^24$MgH and $^{40}$CaH, \cite{NEW_Kajita_2009} for CaH$^+$ or $^{207}$Pb$^{19}$F that enjoys a pair of closely spaced levels of opposite levels \citep{NEW_Flambaum:2013rma}. We shall also mention transitions with hyperfine-structures, for which  the sensitivity to $\aem$ can reach 600 for instance $^{139}$La$^{32}$S or silicon monobrid \citep{silicon} that allows one to constrain $\aem\bar\mu^{-1/4}$.

\item {\bf Cold diatomic molecules}: 2-photons Raman transitions in lattice-confined photo-associated Sr$_2$ molecules \citep{NEW_Zelevinsky:2007yn,NEW_Reinaudi_2012} or Cs$_2$  molecules \citep{NEW_PhysRevLett.100.043202,NEW_PhysRevA.84.042117} have been considered.

\item {\bf Non-polar molecular ions of homonuclear diatomics}, such as N$_2^+$ or O$_2^+$. \cite{NEW_Kajita:2016spm,NEW_Kajita:2017rsg} estimates that the Stark and Zeeman shifts in the transition frequencies of the O$_2^+$ molecular ion can allow experiments to reach $\Delta\mu/\mu<{\cal O}(10^{-17}$ since the sensitivities $\Delta\ln\nu/\delta\ln\mu$ for the transitions $X^2\Pi_{1/2}$ $v=0\rightarrow v'$ $(v'=1,4,8)$ are of the order $-0.48$ at frequencies (56.5, 219, 421.9)~THz. It can even reach 140 for the transition $X^2\Pi_{1/2} v=21\rightarrow a^4\Pi_{1/2} v=0$ at 2.7~THz \citep{NEW_Hanneke:2016yot} leading \cite{NEW_Carollo:2018arm}  to estimate one could reach $\Delta\mu/\mu\sim{\cal O}(10^{-18})$.

\item {\bf  Molecular ions}. \cite{NEW_Beloy:2011th} have computed the sensitivities of the  NH$^+$ rotational spectrum to variations of ($\aem,\mu$).

\item {\bf Polar molecules} with deep potential such as TeH$^+$ allow for high sensitivity dipole-allowed vibrational transitions. \cite{NEW_Stollenwerk:2018grb} estimated that with enough vibrational cooling laser power to saturate all the transitions after one can reach $\Delta\mu/\mu\sim 3.6\times10^{-17}$, as confirmed by \cite{NEW_Kokish:2017fyn}. They mention that simulations also support the potential for fluorescence state read-out of TeH$^+$ suggesting  the possibility of searching for $\mu$-variation using multi-ion spectroscopy on laser-coolable polar species.

\item{\bf Dihalogens and hydrogen halides:} \cite{NEW_Pasteka:2015hla} identified strong candidates among HBr$^+$, HI$^+$, Br$^+_2$ , I$^+_2$ , IBr$^+$, ICl$^+$, and IF$^+$. 

\item{\bf Tunneling-Rotational Transitions} have enhanced sensitivities. Ethylene Glycol (C$_2$H$_6$O$_2$) in its ground conformation has such a transition with the frequency about 7 GHz. Since tunneling and rotational energies have different dependencies on $\bar\mu$, the spectrum is highly sensitive to the possible its  variation with sensitivities ranging from $-17$ to $+18$ \citep{NEW_Viatkina:2014jpa}. Ethylene glycol has been detected in the interstellar medium.

\item{\bf Internal rotor molecules} in molecules exhibiting hindered internal rotation, such as methyl mercaptan (CH$_3$), have torsion-rotation transitions with enhanced sensitivities to a variation of $\bar\mu$\citep{NEW_Jansen:2013qaa}. This enhancement occurs due to a cancellation of energies associated with the torsional and rotational degrees of freedom of the molecule. It is also exceptionally large in methanol that is detected in quasar absorption spectra.

\item {\bf Polyatomic molecules}. \cite{NEW_wens:2015hla} computed the sensitivities of $^{14}$NH$_3$, $^{15}$NH$_3$, $^{14}$ND$_3$, and $^{15}$ND$_3$ allowing \cite{NEW_Owens:2016xml} to show that the senssitivity of amonia vibro-rotation lines can reach $-16.738$; see also \cite{NEW_Jansen:2013paa}. \cite{NEW_SANTAMARIA2014116} proposed an experiment to constrain, over a-few-year timescale, the fractional temporal variation of $\mu$ at the level of $10^{-15}$/yr by means of a spectroscopic frequency measurement on a beam of cold CF$_3$H molecules. \cite{NEW_Kozlov:2010zb} stressed that the tunneling transition of H$_3$O$^+$ occurs in the far-infrared and that such transitions are observed within the interstellar medium in the Milky Way as well as nearby galaxies. Hence, they computed its sensitivity to a variation of $\mu$ and argued it can be used as an independent target to test hypothetical changes in $\mu$ in  different ambient conditions of high (terrestrial) and low (interstellar medium) matter densities. They also considered H$_2$O$_2$ \citep{NEW_Kozlov:2011qp} to have exhibited a transition at 14.8~GHz with a sensitivity of 37 that can be observed astrophysically \citep{NEW_Bergman:2011pn}. \cite{NEW_Kozlov:2012au} also mentioned that the rovibrational spectrum  of polyatomic molecules with $\Pi$-electronic ground state can be strongly modified by the Renner-Teller effect, leading increased sensitivities.

\item Precision measurement of the scattering lengths in Bose-Einstein condensate and Feshbach molecular experiments were also discussed \citep{clock-further4} and argued to reach a $10^{-15}-10^{-16}$ level on the variation of $\mu$ \citep{NEW_GACESA2014124}.

\end{itemize}

While there is still a gap of 2 to 3 orders of magnitude on the constraint on $\dot\mu/\mu$ between molecular and atomic clocks, the former is expected to become competitive in a near future. Molecular structure and dynamics offer rich energy scales that are at the heart of new protocols in precision measurement and quantum information science \citep{NEW_Kondov:2019jzq} and many candidates with high sensitivity have been singled out. These clocks will allow one to test both a drift and rapid oscillations and will complete the constraints obtained from atomic clocks. This will complement the use of molecular spectra in astrophysics by providing both theoretical computations of the sensitivity coefficients and high-precision laboratory spectra for comparison.

\subsubsection{Nuclear clocks as a possible new system}\label{subsec_nuclock}

The transition frequency of nuclear energy levels are generally outside the laser accessible range by several orders of magnitude. It has however been noticed that the nuclear transitions of an optical clock based on a very narrow ultraviolet nuclear transition between the low lying $^{229\rm{m}}$Th isomer (i.e.,  a long-lived excited nuclear state ($10^3-10^4$~s) to its ground state with predicted energy of $7.8\pm0.05$~eV and $\Delta E/E\sim10^{-20}$ \citep{NEW_Beck:2007zza,NEW_beeks} is an exception. The existence of the isomere was confirmed by \cite{NEW_vonderWense:2016fto}. The transition has been discovered experimentally \citep{PhysRevLett.132.182501,PhysRevLett.133.013201} in CaF$_2$ thorium-doped crystals at an energy of 8.4~eV, i.e., at about $10\sigma$ of the theoretical predictions. State-resolved laser spectroscopy at the $10^{-12}$ precision level was recently reported by \cite{Zhang:2024ngu} and the first constraints on the ultra-light dark matter models -- see Sect.~\ref{secULDM} for description -- have been obtained by \cite{Fuchs:2024edo}.

On the one hand, it has attracted interest in the design of a clock based on such a nuclear transition \citep{NEW_thclock,NEW_Campbell:2012zzb} who showed that such a clock can reach $\delta\omega/\omega\sim10^{-19}$.  On the other hand, using a Walecka model for the nuclear potential, \cite{clock-further6}  concluded that the sensitivity of the   transition to the fine-structure constant and quark mass was  typically
$$
   \frac{\delta\omega}{\omega}\sim 10^5\left(4\frac{\delta\aem}{\aem}
   + \frac{\delta X_{\mathrm{q}}}{X_{\mathrm{q}}} - 10 \frac{\delta X_{\mathrm{s}}}{X_{\mathrm{s}}}\right)\, .
$$
This roughly provides a five orders of magnitude amplification, which can lead to a constraint at the level of $10^{-23}$/yr   on the time variation of $X_{\mathrm{q}}$. Such a method is promising  and would offer different sensitivities to systematic effects  compared to atomic clocks. However, this sensitivity is not clearly   established since different nuclear calculations do not   agree \citep{berengutTh} and \cite{hayes} questioned the result on the basis of the Feynman-Hellmann theorem -- that relates the derivative of the total energy with respect to a parameter to the expectation value of the derivative of the Hamiltonian with respect to that same parameter. One expects that
$$
\Delta\omega/\omega = K \Delta\aem/\aem \qquad K=\Delta V_c/\omega\,.
$$
$\Delta V_c$, the change in the Coulomb energy, is the central physical quantity to understand the high sensitivity since it derives from an almost perfect cancelation of its change, $\Delta V_C= E_c^{\rm m}-E_c^{\rm ground}\sim-1$~MeV. \cite{hayes} argued that no sensitivity to $\aem$ should arise while \cite{NEW_He:2008zzf} confirmed that it shall be $-9.2\times10^{4}$ pointing out the importance of the correlation between the nuclear and electromagnetic interactions. The small difference between the Coulomb energies of the two states, which both are of the order $10^9$~eV, is the cause of the large amplification of the sensitivity to a $\aem$ variation. \cite{NEW_Litvinova:2009vp,Feldmeier:2017olg} used Hartree-Fock and Hartree-Fock-Bogoliubov calculations  to compute the Coulomb and kinetic energies between the $3/2^+$ and $5/2^+$ states. It concluded to the limited precision of the nuclear model calculations. \cite{NEW_Flambaum:2008ij} then estimated the contribution of the polarization of the Coulomb energy to the spacing between the two states as a function of the nuclear deformation again concluding to a sensitivity of order $10^4$.  \cite{berengutTh} proposed a method to extract the $\aem$ dependence and it was concluded by \cite{NEW_Thielking:2017qet}, based on \cite{NEW_Litvinova:2009vp}, that
$$
\frac{\Delta E_c}{1\,{\rm MeV} }= -485\left[\frac{\left< r^2_{229{\rm m}}\right>}{\left< r^2_{229}\right>} -1\right] +11.6 \left[\frac{Q_0^{\rm m}}{Q_0} -1\right] \simeq -0.29\pm 0.0043\,,
$$
with $\left< r^2_{229}\right>$ the nuclear mean-square charge radius and $Q_0$ the electric quadrupole moment. This value is dominated by the experimental unbcertainty of $Q^{\rm m}_0/Q_0$ of about 4\%.  It follows that the nuclear properties of both states have to be determined precisely to derive the actual sensitivity (see e.g., \cite{NEW_Safronova:2018fue} for $\left< r^2_{229}\right>$). We refer to \cite{NEW_Thirolf:2019deu,NEW_Thirolf:2019ocm,NEW_Peik:2020cwm} for a detailed description of these modelizations and for the prospective of nuclear clocks for fundamental physics. We refer to \cite{NEW_Caputo:2024doz} for further developments. Note that,  assuming a prolate spheroid nucleus, \cite{Beeks:2024xnc} estimated the sensitivity of the nuclear transition frequency to variations of $\aem$ to be $K_\alpha=5~900\pm2~300$, i.e., a three orders of magnitude enhancement over atomic clock schemes based on electron shell transitions. 

To finish let us mention that \cite{NEW_Fadeev:2021odr} studied the sensitivity of M\"ossbauer transitions on $X_{\rm q}$ to conclude that
$$
\frac{\delta\omega}{\omega} 1.45\left( \frac{\delta\aem}{\aem}-1\right)\frac{\delta m_{\rm q}}{m_{\rm q}}
$$
pointing to a sensitivity of $10^4$ for $^{229\rm{m}}$Th and $10^3$ for $^{235\rm{m}}$U.

\subsubsection{Future evolutions}\label{futureClo}

The constraints from clock comparisons of is mostly determined by the uncertainties of the clocks, their sensitivities to the variation of the constants, and the time interval over which the comparisons are made. All these factors have been continuously improved in the past years. Hence one has 3 strategies:
\begin{enumerate}
\item repeat experiments on a longer time scale.
\item improve existing clocks. The accuracy of atomic clocks has improved by a factor $10^3$ in the past 15 years to reach a fractional frequency uncertainty of two parts in $10^{18}$. The most rapid improvement is expected to come from optical to optical clock comparison -- see e.g., \cite{PhysRevLett.133.023401} who reported an optical lattice clock with a total systematic uncertainty of $8.1\times10^{-19}$ -- while optical to microwave comparison being limited by the ultimate accuracy of microwave clocks. Comparison of different clocks beyond 10$^{-18}$ will become more challenging due to the sensitivity to the environment, including temperature and gravitational potential; for a review see e.g.,  \cite{NEW_RevModPhys.87.637}.
\item develop new clocks with higher sensitivities. First, a good candidate are \emph{highly charged ions} \citep{th4,hci00,NEW_Kozlov:2018mbp,NEW_Rehbehn:2021zlr} and in particular optical of heavy actinides from uranium to einsteinium \citep{NEW_Allehabi:2024sym,NEW_Dzuba:2024idl}. Many systems, offering transitions are between ground and excited metastable states of the ions, are promising candidates for optical clocks. These clocks are less sensitive to external perturbations and their sensitivity to variation of $\aem$ is enhanced due to larger relativistic effects. \emph{Molecular transitions}, described in Sect.~\ref{secmol} have significantly improved and many candidates are being studied. Then we have discussed in Sect.~\ref{subsec_nuclock} the developments of \emph{nuclear clocks}.
 \end{enumerate}
We refer to the reviews \citep{NEW_Safronova:2017xyt,NEW_Safronova:2019lex} on these three issues. Indeed this goes together with the developments of quantum sensors \citep{NEW_Tsai:2021lly} and algorithm \citep{NEW_Zaheer:2023ulf}.  Besides, there is a strong developments to compare clocks at different spacetime positions, as discussed in Sect.~\ref{secfiber}, either thanks to clock network or in space; see e.g., \cite{NEW_Derevianko:2021kye,NEW_Barontini:2021mvu}.  This allows one to probe spacetime correlations and is the only possibility of detecting transient events linked to macroscopic dark objects. These developments are motivated by the search of light matter candidate and the tests of General Relativity. As explained on Fig.~\ref{fig-clocksum} various modes of operation target different physical sectors.

\subsection{Macroscopic bodies and chemistry}\label{sec:chimie}

Let us briefly  mention the idea that a variation of constants would induce the size and shape of material objects. 

\cite{NEW_Stadnik:2015xbn} consider the effect of a time variation of the constants on laser interferometer since it would alter the accumulated phase, $\varphi=\omega L/c$  of the light beam inside. Indeed both the transition frequency $\omega$ and the size of the interferometer $L\propto a_{\rm B}\propto \hbar^2/m_{\rm e}2e^2$ would vary so that $\delta\varphi\simeq\varphi\delta\aem/\aem$ for an atomic transition. 

Similarly, \cite{NEW_Pasteka:2018tho} remarked that  in non-relativistic physics the size of molecules and solids is proportional to the Bohr radius. Since this dependence cancels out in the ratio of the sizes, the individual dependence of different compounds on  $\aem$ arises from the difference in the relativistic effects. It led them to investigate the variation of crystal lattice parameters  and molecular bond lengths due to variation of the fine structure constant and the proton-to-electron mass ratio for selected solid state and molecular systems.  

\cite{NEW_ues} discussed the change in macroscopic objects and the influence on the structure of molecules, including bond lengths and geometry. In the case of the water molecules their dependencies on $(\aem,\bar\mu)$ was determined from heavy numerical simulations \citep{NEW_PhysRevA.81.042523}. This was extrapolated to the structure of organic molecules such as DNA. Following the statement by \cite{NEW_DNA} that ``the distant between adjacent sugars or phosphates in the DNA chain must be between 5.5 and 6.5 Angstroms" for the DNA helix structure to exist. They claim one would need  $\delta(\ell_{A+T}/\ell_{G+C})$ shall be smaller than 10\%, where $ \ell_{A+T}$ and $\ell_{G+C}$ stand for the lengths of the $A-T$ and $G-C$ bonds.

 \cite{NEW_Trachenko:2023pub} investigated the effects of a variation of the constants on condensed matter and liquid physics, focusing on viscosity and diffusion in order to translate the range of bio-friendly viscosity and diffusion on the range of fundamental constants which favor cellular life.
 
 Indeed those systems are not competitive with quantum systems but they may be of interest for discussions circling around the anthropic principle; see Sec.~\ref{section5}.

\subsection{The Oklo phenomenon}\label{sec:oklo}

\subsubsection{A natural nuclear reactor}\label{sec:natural-reactor}

Oklo is the name of a town in the Gabon republic (West Africa) where an open-pit uranium mine is situated. About $1.8 \times 10^{9}$~yr ago (corresponding to a redshift of $\sim$~0.14 with the cosmological concordance model), in one of the rich vein of uranium ore, a natural nuclear reactor went critical, consumed a portion of its fuel and then shut a few millions years later (see, e.g., \citealt{uzanleclercqbook,NEW_Davis:2014nga} for more details). This phenomenon was discovered by the French Commissariat \`a l'\'Energie Atomique (CEA)  in 1972 while monitoring for uranium ores \citep{oklo-1}. Sixteen natural uranium reactors have been identified. Among the 15-17 reaction zones, well studied reactors include the zone RZ2 (about 60 bore-holes, 1800~kg of $^{235}$U fissioned during $8.5 \times 10^{5}\unit{yr})$ and zone RZ10 (about 13 bore-holes, 650~kg of $^{235}$U fissioned during $1.6 \times 10^{5}\unit{yr}$). As reviewed by  \citealt{NEW_Davis:2014nga}, the age of the phenomena is constrained by the facts that (1) the stabilisation of the Oklo geological basement happened not earlier 2.7 Gyr ago, (2) the geological of the local Francevillian sediments are estimated to $2.265\pm0.15$~Gyr and (3) the Great Oxydation, that happened 2.2~Gyr ago, during which cyanobacteria triggered the increase of atmospheric oxygen by a factor $\sim100$ which allowed for uranium to be converted to a soluble form and hence transported and precipitated.

The existence of such a natural reactor was predicted by \cite{oklo-3} who showed that under favorable conditions, a spontaneous chain reaction could take place in rich uranium deposits. Indeed, two billion years ago, uranium was naturally enriched (due to the difference of decay rate between $^{235}$U and $^{238}$U) and $^{235}$U represented about 3.68\% of the total uranium (compared with 0.72\% today and to the 3\,--\,5\% enrichment used in most commercial reactors). Besides, in Oklo the conditions were favorable: (\textit{1}) the concentration of neutron absorbers, which prevent the neutrons from being available for the chain fission, was low; (\textit{2}) water played the role of moderator (the zones RZ2 and RZ10 operated at a depth of several thousand meters, so that the water pressure and temperature was close to the pressurized water reactors of 20~Mpa and 300$^{\circ}$ Cand slowed down fast neutrons so that they can interact with other $^{235}$U and (\textit{3}) the reactor was large enough so that the neutrons did not escape faster than they were produced. Typically, the reaction zones are lens shaped layers about 10~m long, 10~m wide and up to 0.8~m thick. It is estimated that the Oklo reactor powered 10 to 50~kW. This explanation is backed up by the substantial depletion of $^{235}$U as well as a correlated peculiar distribution of some rare-earth isotopes. These rare-earth isotopes are abundantly produced during the fission of uranium and, in particular, the strong neutron absorbers like ${}^{149}_{62}\mathrm{Sm}$, ${}^{151}_{63}\mathrm{Eu}$, ${}^{155}_{64}\mathrm{Gd}$ and ${}^{155}_{64}\mathrm{Gd}$ are found in very small quantities in the reactor.

From the isotopic abundances of the yields, one can extract information about the nuclear reactions at the time the reactor was operational and reconstruct the reaction rates at that time. One of the key quantity measured is the ratio ${}^{149}_{62}\mathrm{Sm}/{}^{147}_{62}\mathrm{Sm}$ of two light isotopes of samarium, which are not fission products. This ratio of order of 0.9 in normal samarium, is about 0.02 in Oklo ores. This low value is interpreted \citep{oklo-2} by the depletion of ${}^{149}_{62}\mathrm{Sm}$ by thermal neutrons produced by the fission process and to which it was exposed while the reactor was active. The capture cross-section of thermal neutron by ${}^{149}_{62}\mathrm{Sm}$
\begin{equation}
\label{oklo1}
n+{}^{149}_{62}\mathrm{Sm}\longrightarrow {}^{150}_{62}\mathrm{Sm}+\gamma
\end{equation}
is dominated by a capture resonance of a neutron of energy of about 0.1 eV ($E_r=97.3 \unit{meV}$ today). The existence of this resonance is a consequence of an almost cancellation between the electromagnetic repulsive force and the strong interaction.

\cite{oklo-2} pointed out that this phenomenon can be used to set a constraint on the time variation of fundamental constants. His argument can be summarized as follows.

\begin{itemize}
\item First, the cross-section $\sigma_{(n,\gamma)}$ strongly depends on the energy of a resonance at $E_{r}=97.3 \unit{meV}$.
\item Geochemical data allow one to determine the isotopic composition of various element, such as uranium, neodynium, gadolinium and samarium. Gadolinium and neodium allow one to determine the fluence (integrated flux over time) of the neutron while both gadolinium and samarium are strong neutron absorbers.
\item From these data, one deduces the value of the averaged value of the cross-section on the neutron flux, $\hat\sigma_{149}$. This value depends on hypothesis on the geometry of the reactor zone.
\item The range of allowed values of $\hat\sigma_{149}$ was  translated into a constraint on $E_r$. This step involves an assumption on the form and temperature of the neutron spectrum.
\item $E_r$ was related to some fundamental constant, which involve a model of the nucleus.
\end{itemize}

In conclusion, we have different steps, which all involve assumptions:

\begin{itemize}
 \item Isotopic compositions and geophysical parameters are  measured in a given set of bore-hold in each zone. A choice has
 to be made on the sample to use, in order, e.g., to ensure that they are not contaminated.
 \item With hypothesis on the geometry of the reactor, on the spectrum and temperature of the neutron flux, one can deduce the  effective value of the cross-sections of neutron absorbers (such as samarium and gadolinium). This requires one to solve a network of nuclear reactions describing the fission.
 \item One can then infer the value of the resonance energy $E_r$, which again depends on the assumptions on the neutron spectrum.
 \item $E_r$ needs to be related to fundamental constant, which involves a model of the nucleus and high energy physics hypothesis.
\end{itemize}

We shall now detail the assumptions used in the various analyses that have been performed since the pioneering work of \cite{oklo-2}.

\subsubsection{Constraining the shift of the resonance energy}

\paragraph{Cross-sections.}\

The cross-section of the neutron capture~(\ref{oklo1}) strongly depends on the energy of a resonance at $E_{r}=97.3 \unit{meV}$ and is well described by a Breit--Wigner formula
\begin{equation}\label{oklo2}
 \sigma_{(n,\gamma)}(E)=\frac{g_0\pi}{2}\frac{\hbar^2}{m_{\mathrm{n}}E}
  \frac{\Gamma_{\mathrm{n}}\Gamma_\gamma}{(E-E_r)^2+\Gamma^2/4}
\end{equation}
where $g_0\equiv(2J+1)(2s+1)^{-1}(2I+1)^{-1}$ is a statistical factor, which depends on the spin of the incident neutron $s=1/2$, of the target nucleus $I$, and of the compound nucleus $J$. For the reaction~(\ref{oklo1}), we have $g_0=9/16$. The total width $\Gamma\equiv\Gamma_{\mathrm{n}}+\Gamma_\gamma$ is the sum of the neutron partial width $\Gamma_{\mathrm{n}}=0.533 \unit{meV}$ (at $E_r=97.3 \unit{meV}$ and it scales as $\sqrt{E}$ in the center of mass) and of the radiative partial width $\Gamma_\gamma=60.5 \unit{meV}$. ${}^{155}_{64}\mathrm{Gd}$ has a resonance at $E_r=26.8 \unit{meV}$ with $\Gamma_{\mathrm{n}}=0.104 \unit{meV}$, $\Gamma_\gamma=108 \unit{meV}$ and $g=5/8$ while ${}^{157}_{64}\mathrm{Gd}$ has a resonance at $E_r=31.4 \unit{meV}$ with $\Gamma_{\mathrm{n}}=0.470 \unit{meV}$, $\Gamma_\gamma=106 \unit{meV}$ and $g=5/8$.

As explained in Sect.~\ref{sec:natural-reactor}, this cross-section cannot be measured from the Oklo data, which allow one to only measure its value averaged on the neutron flux $n(v,T)$, $T$ being the temperature of the moderator. It is conventionally defined as
\begin{equation}
\label{hatsig}
 \hat\sigma =\frac{1}{nv_0}\int\sigma_{(n,\gamma)}n(v,T)
 v\dd v,
\end{equation}
where the velocity $v_0=2200 \unit{m\cdot s}^{-1}$ corresponds to an energy $E_0=25.3 \unit{meV}$ and $v=\sqrt{2E/m_{\mathrm{n}}}$, instead of
$$
  \bar\sigma = \frac{\int\sigma_{(n,\gamma)}n(v,T)v\dd v}{\int n(v,T)
 v\dd v}.
$$
When the cross-section behaves as $\sigma=\sigma_0v_0/v$, which is the case for nuclei known as ``$1/v$-absorbers'', $\hat\sigma=\sigma_0$ and does not depend on the temperature, whatever the distribution $n(v)$. In a similar way, the effective neutron flux,
\begin{equation}
 \hat\phi=v_0\int n(v,T)\dd v\,,
\end{equation}
differs from the true flux
$$
 \phi= \int n(v,T)v\dd v.
$$
However, since $\bar\sigma\phi = \hat\sigma\hat\phi$, the reaction rates are not affected by these definitions.

\paragraph{Extracting the effective cross-section from the data.}\ 
 
 To ``measure'' the value of $\hat\sigma$ from the data, we need to solve the nuclear reactions network that controls the isotopic composition during the fission.

The samples of the Oklo reactors were exposed \citep{oklo-1} to an integrated effective fluence $\int\hat\phi\dd t$ of about $10^{21} \text{neutron} \cdot \unit{cm}^{-2}=1 \unit{kb}^{-1}$. Assuming a steady state mode of operation (as in most of the works) during $3\times10^5$~yr, this leads to a neutron flux density of $10^{8} \text{neutron} \cdot \unit{cm}^{-2}\unit{s}^{-1}$ (typically 5 orders of magnitude smaller than in present day reactors) but one needs to keep in mind that the instantaneous flux can be much higher in the the reactor operated by pulses. It implies that any process with a cross-section smaller than 1~kb can safely be neglected in the computation of the abundances. This includes neutron capture by ${}^{144}_{62}\mathrm{Sm}$ and ${}^{148}_{62}\mathrm{Sm}$, as well as by ${}^{155}_{64}\mathrm{Gd}$ and ${}^{157}_{64}\mathrm{Gd}$. On the other hand, the fission of ${}^{235}_{92}\mathrm{U}$, the capture of neutron by ${}^{143}_{60}\mathrm{Nd}$ and by ${}^{149}_{62}\mathrm{Sm}$ with respective cross-sections $\sigma_{5}\simeq0.6 \unit{kb}$, $\sigma_{143}\sim0.3 \unit{kb}$ and $\sigma_{149}\geq70 \unit{kb}$ are the dominant processes. It follows that the equations of evolution for the number densities $N_{147}$, $N_{148}$, $N_{149}$ and $N_{235}$ of ${}^{147}_{62}\mathrm{Sm}$, ${}^{148}_{62}\mathrm{Sm}$, ${}^{149}_{62}\mathrm{Sm}$ and ${}^{235}_{92}\mathrm{U}$ take the form
\begin{eqnarray}
 \frac{\dd N_{147}}{\hat\phi\dd t}&=&-\hat\sigma_{147} N_{147}+
       \hat\sigma_{f235}y_{147} N_{235}
\label{e1}\\
\frac{\dd N_{148}}{\hat\phi\dd t}&=&\hat\sigma_{147} N_{147}
\label{e2}\\
\frac{\dd N_{149}}{\hat\phi\dd t}&=&-\hat\sigma_{149} N_{149}+
       \hat\sigma_{f235}y_{149} N_{235}
\label{e3}\\
\frac{\dd N_{235}}{\hat\phi\dd t}&=&-\sigma_5 N_{235},\label{e4}
\end{eqnarray}
where $y_i$ denotes the yield of the corresponding element in the fission of ${}^{235}_{92}\mathrm{U}$ and $\hat\sigma_5$ is the fission cross-section. This system can be integrated under the assumption that the cross-sections and the neutron flux are constant and the result compared with the natural abundances of the samarium to extract the value of $\hat\sigma_{149}$ at the time of the reaction. Here, the system has been closed by introducing a modified absorption cross-section \citep{oklo-4} $\sigma_5^*$ to take into account both the fission, capture but also the formation from the $\alpha$-decay of ${}^{239}_{94}\mathrm{Pu}$. One can instead extend the system by considering ${}^{239}_{94}\mathrm{Pu}$, and ${}^{235}_{92}\mathrm{U}$ (see \citealt{oklo-9}). While most studies focus on the samarium, \cite{oklo-5} also includes the gadolinium even though it is not clear whether it can reliably be measured \citep{oklo-4}, still giving similar results.

By comparing the solution of this system with the measured isotopic composition, one can deduce the effective cross-section. At this step, the different analyses \citep{oklo-2,oklo-2bis,oklo-4,oklo-5,oklo-7,oklo-8,oklo-9} differ from the choice of the data. The measured values of $\hat\sigma_{149}$ can be found in these articles. They are given for a given zone (RZ2, RZ10 mainly) with a number that correspond to the number of the bore-hole and the depth (e.g., in Table~2 of \cite{oklo-4}, SC39-1383 means that we are dealing with the bore-hole number 39 at a depth of 13.83~m). Recently, another approach \citep{oklo-8,oklo-9} was proposed in order to take into account of the geometry and details of the reactor. It relies on a full-scale Monte-Carlo simulation \citep{NEW_OkloCode} dealing with realistic geometries (but also operating temperature, amounts of uraninite, gangue and water) and a computer model of the reactor zone RZ2 \citep{oklo-8} and both RZ2 and RZ10 \citep{oklo-9}. It allows one to take into account the spatial distribution of the neutron flux.

\begin{table}[t]
\caption[Analysis of the Oklo data]{Summary of the analysis of the Oklo data. The principal assumptions to infer the value of the resonance energy $E_r$ are the form of the neutron spectrum and its temperature. Concerning the neutron spectrum, ``Max" stands for Maxwellian and ``Max+ept" for Maxwellian+epithermal.}
\label{tab-oklo}
\centering
{\small
\begin{tabular}{lccccr}
\toprule
Ore & neutron spectrum & Temperature ($^{\circ}\mathrm{C}$) & $\hat\sigma_{149}$ (kb)  & $\Delta E_r$ (meV) & Ref. \\
\midrule
 ?        &  Max      &  20          & 55\,$\pm$\,8  & 0\,$\pm$\,20       & \cite{oklo-2}\\
 RZ2 (15) &  Max      & 180\,--\,700 & 75\,$\pm$\,18 & --1.5\,$\pm$\,10.5 & \cite{oklo-4}\\
 RZ10     &  Max      & 200\,--\,400 & 91\,$\pm$\,6  & 4\,$\pm$\,16 & \cite{oklo-5}\\
 RZ10     &               &              &               & --97\,$\pm$\,8      & \cite{oklo-5}\\
 --       &  Max+ept        & 327           & 91\,$\pm$\,6 & $-45^{+7}_{-15}$ & \cite{oklo-7}\\
 RZ2      &  Max+ept      &               & 73.2\,$\pm$\,9.4 & $-5.5 \pm 67.5$ & \cite{oklo-8}\\
 RZ2      &  Max+ept       & 200\,--\,300  & 71.5\,$\pm$\,10.0  & -- & \cite{oklo-9}\\
 RZ10     & Max+ept      & 200\,--\,300  & 85.0\,$\pm$\,6.8 & -- & \cite{oklo-9}\\
 RZ2+RZ10 &  ~            & ~            & ~             & 7.2\,$\pm$\,18.8 & \cite{oklo-9}\\
 RZ2+RZ10 &  ~            & ~            & ~             & 90.75\,$\pm$\,11.15 & \cite{oklo-9}\\
 RZ10 &  ~            & ~            & 85.0\,$\pm$\,6.8            & $[-11.6,26.0]$ & \cite{NEW_Davis:2014nga}\\
 RZ10 &  ~            & ~            & ~             & $[-101.9,-79.6]$ & \cite{NEW_Davis:2014nga}\\ 
\bottomrule
\end{tabular}
}
\end{table}

\paragraph{Determination of \boldmath$E_r$.}\ 
 
 To convert the constraint on the effective cross-section, one needs to specify the neutron spectrum. In the earlier studies \citep{oklo-2,oklo-2bis}, a Maxwell distribution,
$$
 n_{\mathrm{th}}(v,T) = \left(\frac{m_{\mathrm{n}}}{2\pi T}
 \right)^{3/2}\hbox{e}^{-\frac{m v^2}{2 k_{\mathrm{B}}T}},
$$
was assumed for the neutron with a temperature of $20^{\circ}\mathrm{C}$, which is probably too small. Then $v_0$ is the mean velocity at a temperature $T_0=m_{\mathrm{n}}v_0^2/2k_{\mathrm{B}}=20.4^{\circ}\mathrm{C}$. \cite{oklo-4,oklo-5} also assumed a Maxwell distribution but let the moderator temperature vary so that they deduce an effective cross-section $\hat\sigma(R_r,T)$. They respectively restricted the temperature range to $180^{\circ}\mathrm{C}<T<700^{\circ}\mathrm{C}$ and $200^{\circ}\mathrm{C}<T<400^{\circ}\mathrm{C}$, based on geochemical analysis. The issue of the temperature is crucial and strongly debated. Based on lutetium thermometry \cite{NEW_Oklo_temp0} estimated that $T=260^{\circ}\mathrm{C}$ for RZ2 and $T=280^{\circ}\mathrm{C}$ for RZ3 while \cite{NEW_Onegin:2010kq} concluded that $T=182\pm 80^{\circ}\mathrm{C}$ for RZ3. The temperature of RZ10 is estimated to $380^{\circ}\mathrm{C}$ \citep{NEW_Oklo_temp1} while the reanalysis by \cite{NEW_Davis:2014nga} gave $T=(100+30)^{\circ}\mathrm{C}$. The advantage of the Maxwell distribution assumption is that it avoids to rely on a particular model of the Oklo reactor since the spectrum is determined solely by the temperature.

It was then noted \citep{oklo-7,oklo-8} that above an energy of several eV, the neutron spectrum shifted to a $1/E$ tail because of the absorption of neutrons in uranium resonances. Thus, the distribution was adjusted to include an epithermal distribution
$$
 n(v) = (1 - f) n_{\mathrm{th}}(v,T) + f n_{\mathrm{epi}}(v),
$$
with $n_{\mathrm{epi}}=v_c^2/v^2$ for $v>v_c$ and vanishing otherwise. $v_c$ is a cut-off velocity that also needs to be specified. The effective cross-section can then be parameterized \citep{oklo-9} as
\begin{equation}
 \hat\sigma = g(T)\sigma_0 + r_0 I,
\end{equation}
where $g(T)$ is a measure of the departure of $\sigma$ from the $1/v$ behavior, $I$ is related to the resonance integral of the cross-section and $r_0$ is the Oklo reactor spectral index. It characterizes the contribution of the epithermal neutrons to the cross-section. Among the unknown parameters, the most uncertain is probably the amount of water present at the time of the reaction. \cite{oklo-9} chose to adjust it so that $r_0$ matches the experimental values.

These hypothesis on the neutron spectrum and on the temperature, as well as the constraint on the shift of the resonance energy, are summarized in Table~\ref{tab-oklo}. Many analyses \citep{oklo-5,oklo-8,oklo-9} find two branches for $\Delta E_r=E_r - E_{r0}$, with one (the left branch) indicating a variation of $E_r$. Note that these two branches disappear when the temperature is higher since $\hat\sigma(E_r,T)$ is more peaked when $T$ decreases but remain in any analysis at low temperature. This shows the importance of a good determination of the temperature. Note that the analysis of \cite{oklo-8} indicates that the curves $\hat\sigma(T,E_r)$ lie appreciably lower than for a Maxwell distribution and that \cite{oklo-5} argues that the left branch is hardly compatible with the gadolinium data.

\subsubsection{From the resonance energy to fundamental constants}

The energy of the resonance depends a priori on many constants since the existence of such resonance is mainly the consequence of an almost cancellation between the electromagnetic repulsive force and the strong interaction. But, since no full analytical understanding of the energy levels of heavy nuclei is available, the role of each constant is difficult to disentangle.

In his first analysis, \cite{oklo-2} stated that for the neutron, the nucleus appears as a potential well with a depth $V_0\simeq 50 \unit{MeV}$. He attributed the change of the resonance energy to a modification of the strong interaction coupling constant and concluded that $\Delta g_{\mathrm{S}}/g_{\mathrm{S}}\sim \Delta E_r/V_0$. Then, arguing that the Coulomb force increases the average inter-nuclear distance by about 2.5\% for $A\sim150$, he concluded that $\Delta\aem/\aem\sim20\Delta g_{\mathrm{S}}/g_{\mathrm{S}}$, leading to $|\dot\aem/\aem|<10^{-17} \unit{yr}^{-1}$, which can be translated to $|\Delta\aem/\aem|<1.8\times10^{-8}$.

The following analysis focused on the fine-structure constant and ignored the strong interaction. \cite{oklo-4} related the variation of $E_r$ to the fine-structure constant by taking into account that the radiative capture of the neutron by ${}^{149}_{62}\mathrm{Sm}$ corresponds to the existence of an excited quantum state of ${}^{150}_{62}\mathrm{Sm}$ (so that $E_r=E_{150}^*-E_{149}-m_{\mathrm{n}}$) and by assuming that the nuclear energy is independent of $\aem$. It follows that the variation of $\aem$ can be related to the difference of the Coulomb binding energy of these two states. The computation of this latter quantity is difficult and must be related to the mean-square radii of the protons in the isotopes of samarium. In particular this analysis \citep{oklo-4} showed that the Bethe--Weiz\"acker formula overestimates by about a factor the 2 the $\aem$-sensitivity to the resonance energy.  It follows from this analysis that
\begin{equation}
\label{okdamdy}
   \aem\frac{\Delta E_r}{\Delta\aem}\simeq-1.1 \unit{MeV}.
\end{equation}
The analysis by \cite{NEW_Janecke:1972ukb} showed that for the isotopes considered in their analysis, the sensitivity of the ground state energy is accurate at less than 3\%.

The sensitivity (\ref{okdamdy}) implies that the  constraint on $\Delta E_r$ translates to
\begin{equation}
  -0.9\times10^{-7}<\Delta\aem/\aem<1.2\times10^{-7}
\end{equation}
at $2\sigma$ level, corresponding to the range $-6.7\times10^{-17} \unit{yr}^{-1}< \dot\aem/\aem <5.0\times10^{-17}  \unit{yr}^{-1}$ if $\dot\aem$ is assumed constant. This tight constraint arises from the large amplification between the resonance energy ($\sim0.1 \unit{eV}$) and the sensitivity ($\sim1 \unit{MeV}$). The re-analysis of these data together with those of \cite{oklo-5} including gadolinium found the favored result $\dot\aem/\aem=(-0.2\pm0.8)\times10^{-17} \unit{yr}^{-1}$, which corresponds to
\begin{equation}
\Delta\aem/\aem=(-0.36\pm1.44)\times10^{-8}
\end{equation}
and the other branch (indicating a variation; see Table~\ref{tab-oklo}) leads to $\dot\aem/\aem =(4.9\pm0.4) \times10^{-17} \unit{yr}^{-1}$. This non-zero result cannot be eliminated.

The more recent analysis, based on a modification of the neutron spectrum led respectively to \citep{oklo-8}
\begin{equation}
 \Delta\aem/\aem=(3.85\pm5.65)\times10^{-8}
\end{equation}
and \citep{oklo-9}
\begin{equation}
 \Delta\aem/\aem=(-0.65\pm1.75)\times10^{-8},
\end{equation}
at a 95\% C.L., both using the formalism of \cite{oklo-4}.

\cite{oklo-11}, inspired by grand unification models, reconsider the analysis of \cite{oklo-4} by letting all gauge and Yukawa couplings vary.  Working within the Fermi gas model, the over-riding scale dependence of the terms, which determine the binding energy of the heavy nuclei was derived. Parameterizing the mass of the hadrons as $m_i\propto\Lambda_{\mathrm{QCD}}(1+\kappa_im_{\mathrm{q}}/\Lambda_{\mathrm{QCD}}+\ldots)$, they estimated that the nuclear Hamiltonian was proportional to $m_{\mathrm{q}}/\Lambda_{\mathrm{QCD}}$ at lowest order, which allows them to conclude that the energy of the resonance is related to the quark mass by
\begin{equation}\label{e33}
 \frac{\Delta E_r}{E_r}\sim (2.5-10)\times 10^{17}
 \Delta\ln\left(\frac{m_{\mathrm{q}}}{\Lambda_{\mathrm{QCD}}}\right) .
\end{equation}
Using the constraint~(\ref{okdamdy}), they first deduced that
$$
 \left|\Delta\ln\left(\frac{m_{\mathrm{q}}}{\Lambda_{\mathrm{QCD}}}\right)\right|<(1-4)\times10^{-8}.
$$
Then, assuming that $\aem\propto m_{\mathrm{q}}^{50}$ on the basis of grand unification (see Sect.~\ref{subsec22} for details), they concluded that
\begin{equation}
 \left|\Delta\aem/\aem \right|<(2-8)\times10^{-10}.
\end{equation}

Similarly, \cite{oklo-12,oklo-13,oklo-15} related the variation of the resonance energy to the quark mass.  Their first estimate \citep{oklo-12} assumed that it is related to the pion mass, $m_\pi$, and that the main variation arises from the variation of the radius $R\sim 5{\mathrm{fm}}+1/m_\pi$ of the nuclear potential well of depth $V_0$, so that
$$
 \delta E_r\sim -2V_0\frac{\delta R}{R}
 \sim 3\times10^8\frac{\delta m_\pi}{m_\pi},
$$
assuming that $R\simeq 1.2A^{1/3}r_0$, $r_0$ being the inter-nucleon distance. \cite{NEW_Davis:2014nga,NEW_Davis:2014eua} proposed the natural parameterisation
\begin{equation}
\label{e33b}
 \ \Delta E_r = a  \Delta\frac{X_{\rm q}}{X_{\rm q}} + b\frac{Z^2}{A^{4/3}}\frac{\Delta\aem}{\aem}
\end{equation}
with $|b|\sim0.5$~MeV within the factor 2 (see however \cite{NEW_Davis:2015ila}) and the more uncertain parameter $a$ ranging from 10~MeV \citep{oklo-15} to $-40$~MeV \citep{NEW_Davis:2014eua}. This latter result would lead to the constraint on the variation of the quark mass
\begin{equation}
 \left|\Delta m_{\rm q}/m_{\rm q} \right|<10^{-9}.
\end{equation}

Then, \cite{oklo-13} described the nuclear potential by a Walecka model, which keeps only the $\sigma$ (scalar) and $\omega$ (vector) exchanges in the effective nuclear force. Their masses was related to the mass $m_{\mathrm{s}}$ of the strange quark to get $m_\sigma\propto m_{\mathrm{s}}^{0.54}$ and $m_\omega\propto m_{\mathrm{s}}^{0.15}$. It follows that the variation of the potential well can be related to the variation of $m_\sigma$ and $m_\omega$ and thus
on $m_{\mathrm{q}}$ by $V\propto m_{\mathrm{q}}^{-3.5}$. The constraint~(\ref{okdamdy}) then implies
$$
 \left|\Delta\ln\left(\frac{m_{\mathrm{s}}}{\Lambda_{\mathrm{QCD}}}\right)\right|< 1.2\times10^{-10}.
$$
By extrapolating from light nuclei where the $N$-body calculations can be performed more accurately, it was concluded \citep{oklo-14} that the resonance energy scales as $\Delta E_r\simeq10(\Delta\ln X_{\mathrm{q}} -0.1\Delta\ln\aem)$, so that the the constraint from \citep{oklo-8} would imply that $\Delta\ln (X_{\mathrm{q}}/\aem^{0.1})<7\times10^{-9}$.

\subsubsection{Conclusion}

The various constraints obtained from the analysis of the Oklo phenomena are summarized in Table~\ref{tab-oklo2} showing that typically, one shall keep in mind the constraints
$$
\left\vert \Delta\aem/\aem \right\vert \lesssim 10^{-8}.
$$
These last results illustrate that a detailed theoretical analysis and quantitative estimates of the nuclear physics (and QCD) aspects of the resonance shift still remain to be carried out. In particular, the interface between the perturbative QCD description and the description in term of hadron is not fully understand: we do not know the exact dependence of hadronic masses and coupling constant on $\Lambda_{\mathrm{QCD}}$ and quark masses. The second problem concerns modeling nuclear forces in terms of the hadronic parameters.

At present, the Oklo data, while being stringent and consistent with no variation, have to be considered carefully. While a better understanding of nuclear physics is necessary to understand the full constant-dependence, the data themselves require more insight, particularly to understand the existence of the left-branch.

\begin{table}[t]
\caption[Main constraints on the variation of $\aem$ from the Oklo phenomena]{Summary of the main constraints on the variation of $\aem$, assuming all other constants remains fixed, from the Oklo phenomena.}
\label{tab-oklo2}
\centering
{\small
\begin{tabular}{lccr}
\toprule
Ore & neutron spectrum & $\Delta\aem/\aem (10^{-8})$ & Ref. \\
\midrule
 ?        &  Maxwell      &  $(0\pm1.8)$       & \cite{oklo-2}\\
 RZ2 (15) &  Maxwell      &  $(1.5\pm10.5)$ & \cite{oklo-4}\\
 RZ10,13     &  Maxwell      & $(-0.36\pm1.44)$  & \cite{oklo-5}\\
 RZ2      &  Maxwell + epithermal        &     $(3.85\pm5.65)$  & \cite{oklo-8}\\
 RZ2,10      &  Maxwell + epithermal        & $(-0.65\pm1.75)$  & \cite{oklo-9}\\
 RZ3,5 &  Numerical          & $(-0.15\pm 0.85)$ & \cite{NEW_Onegin:2010kq}\\
\bottomrule
\end{tabular}
}
\end{table}

\subsection{Meteorite dating}\label{sec:meteorite}

Long-lived $\alpha$- or $\beta$-decay isotopes may be sensitive probes of the variation of fundamental constants on geological times ranging typically to the age of the Solar system, $t\sim(4-5) \unit{Gyr}$, corresponding to a mean redshift of $z\sim~0.43$. Interestingly, it can be compared with the shallow universe quasar constraints. This method was initially pointed out by \cite{meteo0} and then revived by \cite{revueDyson}. The main idea is to extract the $\aem$-dependence of the decay rate and to use geological samples to bound its time variation.

The sensitivity of the decay rate of a nucleus to a change of the fine-structure constant is defined, in a similar way as for atomic clocks [Eq.~(\ref{clock-sensitivity})], as
\begin{equation}
\label{L-sensitivity}
 s_\alpha \equiv \frac{\partial \ln \lambda}{\partial \ln\aem}.
\end{equation}
$\lambda$ is a function of the decay energy $Q$. When $Q$ is small, mainly due to an accidental cancellation between different contributions to the nuclear binding energy, the sensitivity $s_\alpha$ maybe strongly enhanced. A small variation of the fundamental constants can either stabilize or destabilize certain isotopes so that one can extract bounds on the variation of  their lifetime by comparing laboratory data to geophysical and Solar system probes.

Assume some meteorites containing an isotope $X$ that decays into $Y$ are formed at a time $t_*$. It follows that
\begin{equation}
\label{adecT}
 N_X(t) = N_{X*}\hbox{e}^{-\lambda(t-t_*)},\qquad
 N_Y(t) = N_{X*}\left[1-\hbox{e}^{-\lambda(t-t_*)}\right] + N_{Y*}
\end{equation}
if one assumes the decay rate constant. If it is varying then these relations have to be replaced by
$$
 N_X(t) = N_{X*}\hbox{e}^{\int_{t_*}^{t}\lambda(t')\dd t'}
$$
so that the value of $N_X$ today can be interpreted with Eq.~(\ref{adecT}) but with an effective decay rate of
\begin{equation}
\label{Leff}
 \bar\lambda = \frac{1}{t_0-t_*}\int_{t_*}^{t_0}\lambda(t')\dd t'.
\end{equation}
From a sample of meteorites, we can measure $\{ N_X(t_0),N_Y(t_0)\}$ for each meteorite. These two quantities are related by
$$
N_Y(t_0) = \left[\hbox{e}^{\bar\lambda(t_0-t_*)} - 1 \right]
N_X(t_0) + N_{Y*},
$$
so that the data should lie on a line (since $N_{X*}$ is a priori different for each meteorite), called an ``isochron'', the slope of which determines $\bar\lambda(t_0-t_*)$. It follows that meteorites data only provide an \emph{average} measure of the decay rate, which complicates the interpretation (see \citealt{meteo-fuji2,meteo-fuji1} for explicit examples). To derive a bound on the variation of the constant we also need a good estimation of $t_0-t_*$, which can be obtained from the same analysis for an isotope with a small sensitivity $s_\alpha$, as well as an accurate laboratory measurement of the decay rate.

\subsubsection{Long lived $\alpha$-decays}

The $\alpha$-decay rate, $\lambda$, of a nucleus ${}^A_Z{\mathrm{X}}$
of charge $Z$ and atomic number $A$,
\begin{equation}
{}_{Z+2}^{A+4}{\mathrm{X}}\longrightarrow {}_Z^A{\mathrm{X}}+ {}_2^4{\mathrm{He}},
\end{equation}
is governed by the penetration of the Coulomb barrier that can be described by the Gamow theory. It is well approximated by
\begin{equation}
\lambda\simeq\Lambda(\aem,v)\exp\left(-4\pi Z\aem
\frac{c}{v}\right),
\end{equation}
where $v/c=\sqrt{Q/2m_{\mathrm{p}}c^2}$ is the escape velocity of the
$\alpha$ particle. $\Lambda$ is a function that depends slowly on
$\aem$ and $Q$. It follows that the sensitivity to the
fine-structure constant is
\begin{equation}
 s_\alpha \simeq -4\pi Z \frac{\aem}{\sqrt{Q/2m_{\mathrm{p}}}}
   \left(1- \frac{1}{2}\frac{\dd\ln Q}{\dd\ln\aem}\right).
\end{equation}
The decay energy is related to the nuclear binding energies
$B(A,Z)$ of the different nuclei by
$$
 Q = B(A,Z) + B_\alpha -B(A+4,Z+2)
$$
with $B_\alpha=B(4,2)$. Physically, an increase of $\aem$ induces an increase in the height of the Coulomb barrier at the nuclear surface while the depth of the nuclear potential well below the top remains the same. It follows that $\alpha$-particle escapes with a greater energy but at the same energy below the top of the barrier. Since the barrier becomes thiner at a given energy below its top, the penetrability increases.  This computation indeed neglects the effect of a variation of $\aem$ on the nucleus that can be estimated to be dilated by about 1\% if $\aem$ increases by 1\%.

As a first insight, when focusing on the fine-structure constant, one can estimate $s_\alpha$ by varying only the Coulomb term of the binding energy. Its order of magnitude can be estimated from the Bethe--Weiz\"acker formula~(\ref{bethe}).

\begin{table}[htbp]
\caption[Properties of the nuclei used in $\alpha$-decay studies]{Summary of the main nuclei and their physical properties that have been used in $\alpha$-decay studies.}
\label{tab-alphadec}
\centering
{\small
\begin{tabular}{p{3.0cm} ccccc}
\toprule
 Element   &  $Z$ & $A$ & Lifetime (yr) & $Q$ (MeV)  & $s_\alpha$ \\
 \midrule
 Sm   &  62 & 147 & $1.06 \times 10^{11}$ & 2.310 &  774 \\
 Gd   &  64 & 152 & $1.08 \times 10^{14}$ & 2.204 &  890 \\
 Dy   &  66 & 154 & $3 \times 10^{6}$     & 2.947 &  575 \\
 Pt   &  78 & 190 & $6.5 \times 10^{11}$  & 3.249 &  659 \\
 Th   &  90 & 232 & $1.41 \times 10^{10}$ & 4.082 &  571 \\
 U    &  92 & 235 & $7.04 \times 10^{8}$  & 4.678 &  466 \\
 U    &  92 & 238 & $4.47 \times 10^{9}$  & 4.270 &  548 \\
\bottomrule
\end{tabular}
}
\end{table}

Table~\ref{tab-alphadec} summarizes the most sensitive isotopes, with the sensitivities derived from a semi-empirical analysis for a spherical nucleus \citep{oklo-11}. They are in good agreement with the ones derived from Eq.~(\ref{bethe}) (e.g., for $^{238}$U, one would obtain $s_\alpha=540$ instead of $s_\alpha=548$).

The sensitivities of all the nuclei of Table~\ref{tab-alphadec} are similar, so that the best constraint on the time variation of the fine-structure constant will be given by the nuclei with the smaller $\Delta\lambda/\lambda$.

\cite{meteo0} considered the most favorable case, that is the decay of $^{238}_{92}\mathrm{U}$ for which $s_\alpha=548$ (see Table~\ref{tab-alphadec}). By comparing the geological dating of the Earth by different methods, he concluded that the decay constant $\lambda$ of $^{238}$U, $^{235}$U and  $^{232}$Th have not changed by more than a factor 3 or 4 during the last $3-4\times10^{9}$~years from which it follows
\begin{equation}
  \left|{\Delta\aem}/{\aem}\right|<8\times10^{-3}.
\end{equation}
It was revised by \cite{revueDyson} who claimed that the decay rate has not changed by more than 20\% during the
past $2\times10^9$ years, which implies
\begin{equation}
  \left|{\Delta\aem}/{\aem}\right|<4\times10^{-4}.
\end{equation}
Uranium has a short lifetime so that it cannot be used to set constraints on longer time scales. It is also used to calibrate the age of the meteorites. Therefore, \cite{oklo-11}  suggested to consider $^{147}$Sm. Assuming that $\Delta\lambda_{\mathrm{147}}/\lambda_{\mathrm{147}}$ is smaller than the fractional uncertainty of $7.5\times10^{-3}$ of its half-life
\begin{equation}
  \left|{\Delta\aem}/{\aem}\right|\lesssim\times10^{-5}.
\end{equation}

As for the Oklo phenomena, the effect of other constants has not been investigated in depth. It is clear that at lowest order both $Q$ and $m_{\mathrm{p}}$ scales as $\Lambda_{\mathrm{QCD}}$ so that one needs to go beyond such a simple description to determine the dependence in the quark masses. Taking into account the contribution of the quark masses, in the same way as for Eq.~(\ref{e33}), it was argued that $\lambda\propto X_{\mathrm{q}}^{300-2000}$, which leads to $|\Delta\ln X_{\mathrm{q}}|\lesssim 10^{-5}$. In a GUT framework, that could lead to a constraint of the order of $|\Delta\ln\aem|\lesssim 2\times10^{-7}$.

\subsubsection{Long lived $\beta$-decays}

\cite{meteoDicke} stressed that the comparison of the rubidium-strontium and potassium-argon dating methods to uranium and thorium rates constrain the variation of $\aem$.

As long as long-lived $\beta$-decay isotopes are concerned for which the decay energy $Q$ is small, we can use a non-relativistic approximation for the decay rate
\begin{equation}
\lambda=\Lambda_\pm Q^{p_\pm}
\end{equation}
respectively for $\beta^-$-decay and electron capture. $\Lambda_\pm$ are functions that depend smoothly on $\aem$ and which can thus be considered constant, $p_+=\ell+3$ and $p_-=2\ell+2$ are the degrees of forbiddenness of the transition. For high-$Z$ nuclei with small decay energy $Q$, the exponent $p$ becomes $p=2+\sqrt{1-\aem^2Z^2}$ and is independent of $\ell$. It follows that the sensitivity to a variation of $\aem$ is
\begin{equation}
 s_\alpha=p\frac{\dd\ln Q}{\dd\ln\aem}.
\end{equation}
The second factor can be estimated exactly as for $\alpha$-decay. We note that $\Lambda_\pm$ depends on the Fermi constant and on the mass of the electron as $\Lambda_\pm\propto\gfermi^2m_{\mathrm{e}}^5 Q^p$. This dependence is the same for any $\beta$-decay so that it disappears in the comparison of two dating methods relying on two different $\beta$-decay isotopes, in which case only the dependence on the other constants appear again through the nuclear binding energy. Note, however, that comparing a $\alpha$- to a $\beta$-decay may lead to interesting constraints.

We refer to Sect.~III.A.4 of FVC03 \citep{jpu-revue} for earlier constraints derived from rubidium-strontium, potassium-argon and we focus on the rhenium-osmium case,
\begin{equation}
{}^{187}_{75}{\mathrm{Re}}\longrightarrow{}^{187}_{76}{\mathrm{Os}}+\bar\nu_e+e^-
\end{equation}
first considered by \cite{meteoPD}. They noted that the very small value of its decay energy $Q=2.6 \unit{keV}$ makes it a very sensitive probe of the variation of $\aem$. In that case $p\simeq 2.8$ so that $s_\alpha\simeq-18000$; a change of $10^{-2}\%$ of $\aem$ will induce a change in the decay energy of order of the keV, that is of the order of the decay energy itself. \cite{meteoPD} did not have reliable laboratory determination of the decay rate to put any constraint. \cite{meteoDyson} compared the isotopic analysis of molybdenite ores ($\lambda_{187}=(1.6\pm0.2)\times10^{-11} \unit{yr}^{-1}$), the isotopic analysis of 14 iron meteorites ($\lambda_{187}=(1.4\pm0.3)\times10^{-11} \unit{yr}^{-1}$) and laboratory measurements of the decay rate ($\lambda_{187}=(1.1\pm0.1)\times10^{-11} \unit{yr}^{-1}$). Assuming that the variation of the decay energy comes entirely from the variation of $\aem$, he concluded that $\left|{\Delta\aem}/{\aem}\right| <9\times10^{-4}$ during the past $3\times10^9$ years. Note that the discrepancy between meteorite and lab data could have been interpreted as a time-variation of $\aem$, but the lab measurement were complicated by many technical issues so that Dyson only considered a conservative upper limit.

The modelization and the computation of $s_\alpha$ were improved in \cite{oklo-11}, following the same lines as for $\alpha$-decay.
$$
\frac{\Delta\lambda_{187}}{\lambda_{187}} = p \frac{\Delta Q}{Q}
 \simeq p\left(\frac{20 \unit{MeV}}{Q} \right)\frac{\Delta\aem}{\aem}
 \sim -2.2\times10^4\frac{\Delta\aem}{\aem}
$$
if one considers only the variation of the Coulomb energy in $Q$. A similar analysis \citep{dent1} led to $\Delta\ln \lambda_{187} \simeq10^4\Delta\ln[\aem^{-2.2}X_{\mathrm{q}}^{-1.9}(X_{\mathrm{d}}-X_{\mathrm{u}})^{0.23}X_{\mathrm{e}}^{-0.058}]$.

The dramatic improvement in the meteoric analysis of the Re/Os ratio \citep{meteodata} led to a re-analysis of the constraints on the fundamental constants. The slope of the isochron was determined with a precision of 0.5\%. However, the Re/Os ratio is inferred from iron meteorites the age of which is not determined directly. Models of formation of the Solar system tend to show that iron and angrite meteorites form within the same 5~million years. The age of the latter can be estimated from the $^{207}$Pb-$^{208}$Pb method, which gives 4.558~Gyr \citep{meteodata2} so that $\lambda_{187}= (1.666\pm0.009)\times 10^{-11} \unit{yr}^{-1}$. Thus, we could  adopt \citep{oklo-11}
$$
 \left|\frac{\Delta \lambda_{187}}{\lambda_{187}}\right|<5\times10^{-3}.
$$
However, the meteoritic ages are determined mainly by $^{238}$U dating so that effectively we have a constraint on the variation of $\lambda_{187}/ \lambda_{238}$. Fortunately, since the sensitivity of $^{238}$U is much smaller than the one of the rhenium, it is safe to neglect its effect. Using the recent laboratory measurement \citep{meteodata4} ($\lambda_{187}= (-1.639\pm 0.025)\times 10^{-11} \unit{yr}^{-1}$), the variation of the decay rate is not given by the dispersion of the meteoritic measurement, but by comparing to its value today, so that
\begin{equation}
 \left|\frac{\Delta \lambda_{187}}{\lambda_{187}}\right|=
 -0.016\pm0.016.
\end{equation}
The analysis of \cite{meteo-olive}, following the assumption of \cite{oklo-11}, deduced that
\begin{equation}
 \Delta\aem/\aem = (-8\pm16)\times10^{-7},
\end{equation}
at a 95\% confidence level, on a typical time scale of 5~Gyr (or equivalently a redshift of order $z\sim0.2$).

As pointed out in \cite{meteo-fuji2,meteo-fuji1}, these constraints really represent a bound on the average decay rate $\bar\lambda$ since the formation of the meteorites. This implies in particular that the redshift at which one should consider this constraint depends on the specific functional dependence $\lambda(t)$. It was shown that well-designed time dependence for $\lambda$ can obviate this limit, due to the time average.

The direct test of a possible time variation of the decay rate, or equivalently the half-life, of long lived radioisotopes in the laboratory is difficult mostly because of (\textit{1}) the duration of the experiments and (\textit{2}) the necessity to control background phenomena on the long-term since most of them (seasonal cosmic-ray flux modulation, Solar wind and activity, Lunar cycles, tides, ground radon level, etc.) enjoy time and seasonal periodicities. For the first time, such a test was achieved by  \cite{NEW_NEMO-3:2020mcq} based on 6.0195~yr of NEMO-3 data for the double-$\beta$ decay of Molybdenum to Ruthenium, ${}^{100}{\rm Mo}\longrightarrow {}^{100}{\rm Ru}$. The half-life of $7\times10^{18}$~yr exhibits no time modulation larger than 2.5\% in the frequency range $[0.33225,365.25]~{\rm yr}^{-1}$. So far, this constraint has not been translated to any bound on the time variation of constants.

\subsubsection{Conclusions}

Meteorites data allow to set constraints on the variation of the fundamental constants, which are comparable to the ones set by the Oklo phenomenon. Similar constraints can also bet set from spontaneous fission (see Sect.~III.A.3 of FVC03 \citealt{jpu-revue}) but this process is less well understood and less sensitive than the $\alpha$- and $\beta$- decay processes.

From an experimental point of view, the main difficulty concerns the dating of the meteorites and the interpretation of the effective decay rate. Recently, the possibility to directly constrain the half-life of long lived radioisotopes has been proposed.

As long as we only consider $\aem$, the sensitivities can be computed mainly by considering the contribution of the Coulomb energy to the decay energy, that reduces to its contribution to the nuclear energy. However, as for the Oklo phenomenon, the dependencies in the other constants, $X_{\mathrm{q}}$, $\gfermi$, $\mu$\ldots, require a nuclear model and remain very model-dependent.


\subsection{Quasar absorption spectra}\label{subsec33}

Astrophysics relies heavily on the measurements of redshifts, that is the comparison of the laboratory wavelengths of a particular atomic or molecular transition to is observed value in an astrophysical system, in absorption or emission. The cosmic expansion implies an achromatic redshift of all spectra. There are additional sources of redshift due mostly to peculiar velocity and the Doppler effect they induce and the local gravitational potential of the region from which light is emitted through the Einstein effect. Both are achromatic so that they impact any wavelength in the same way. Indeed, if the values of fundamental constants are spacetime dependent, one expects additional \emph{chromatic} red or blue shift,  depending on the detailed structure of the atom or molecule in question. The situation is actually similar to clock comparisons in the laboratory with the main difference that the physical environment of the observed spectra is not controlled in terms of velocity dispersion of the absorbers/emitters, exact chemical composition, existence of cosmic electric/magnetic fields etc.  Among all the astrophysical systems quasi-stellar objects  (QSO), aka. ``quasars",  absorption  spectra provide a powerful probe of the variation of fundamental constants. Absorption lines in intervening clouds along the line of sight of the QSO give access to the spectra of the atoms present in the cloud, that is to paleo-spectra. As for clocks, the sensitivity of the atomic or molecular transitions that can be observed and the possibility to correlate them will define the level of constraint that can be set. The method was first used by \cite{q-savedo} who constrained the time variation of the fine-structure constant from the doublet separations seen in galaxy emission spectra. For general introduction to these observations, we refer to \cite{dlmu,srintro,radiointro}. Indeed, one cannot use a single transition to tackle down a variation of the fundamental constants, since one should resort on various transitions and look for chromatic effects that can indeed not be reproduced by the expansion of the universe.

\subsubsection{Elements for a $\aem$ measurement}

To achieve such a test, one needs to consider a series of technical practicalities.

\paragraph{Sensitivities of the atomic transitions}\

As for clock comparisons, one needs to understand the dependencies of different types of transitions. An extensive work has been achieved in this direction, still progressing with the discovery of new atomic or molecular lines in astronomical observations.

\cite{kappa-dzuba,quasar-q1} proposed to use the convenient formulation
\begin{equation}
\label{qpara}
 \omega = \omega_0 + q\left[\left(\frac{\aem}{\aem^{(0)}}\right)^2-1\right]
  + q_2\left[\left(\frac{\aem}{\aem^{(0)}}\right)^4-1\right],
\end{equation}
in order to take into account the dependence of the spectra on $\aem$. $\omega$ is the energy in the rest-frame of the cloud, that is at a redshift $z$, $\omega_0$ is the energy measured today in the laboratory. $q$ and $q_2$ are two coefficients that determine the frequency dependence on a variation of $\aem$ and that arise from the relativistic corrections for the transition under consideration. The coefficient $q$ is typically an order of magnitude larger than $q_2$ so that the possibility to constrain a variation of the fine-structure constant is mainly determined by $q$. These coefficients are obtained as
\begin{equation}
q = \frac{\omega(\Delta x) - \omega(-\Delta x )}{2\Delta x}\nonumber
\end{equation}
by varying the value of $\aem$ in numerical codes; see \cite{q-murphyMMlast} for a list of these sensitivities. They were computed for a large set of transitions, first using a relativistic Hartree--Fock method and then  using many-body perturbation theory. We refer to \cite{kappa-dzuba,q-calc1,q-calc2} for an extensive discussion of the computational methods and a list of the $q$-coefficients for various transitions relevant for both quasar spectra and atomic clock experiments. The uncertainty in $q$ are typically smaller than 30~cm$^{-1}$ for Mg, Si, Al and Zn, but much larger for Cr, Fe and Ni due to their more complicated electronic configurations. The accuracy for $\omega_0$ from dedicated laboratory measurements now reach $0.004 \unit{cm}^{-1}$. It is important to stress that the form~(\ref{qpara}) ensures that errors in the $q$-coefficients cannot lead to a non zero detection of $\Delta\aem$.

\begin{figure}[hptb]
 \vskip-.5cm
 \centerline{\includegraphics[scale=0.4]{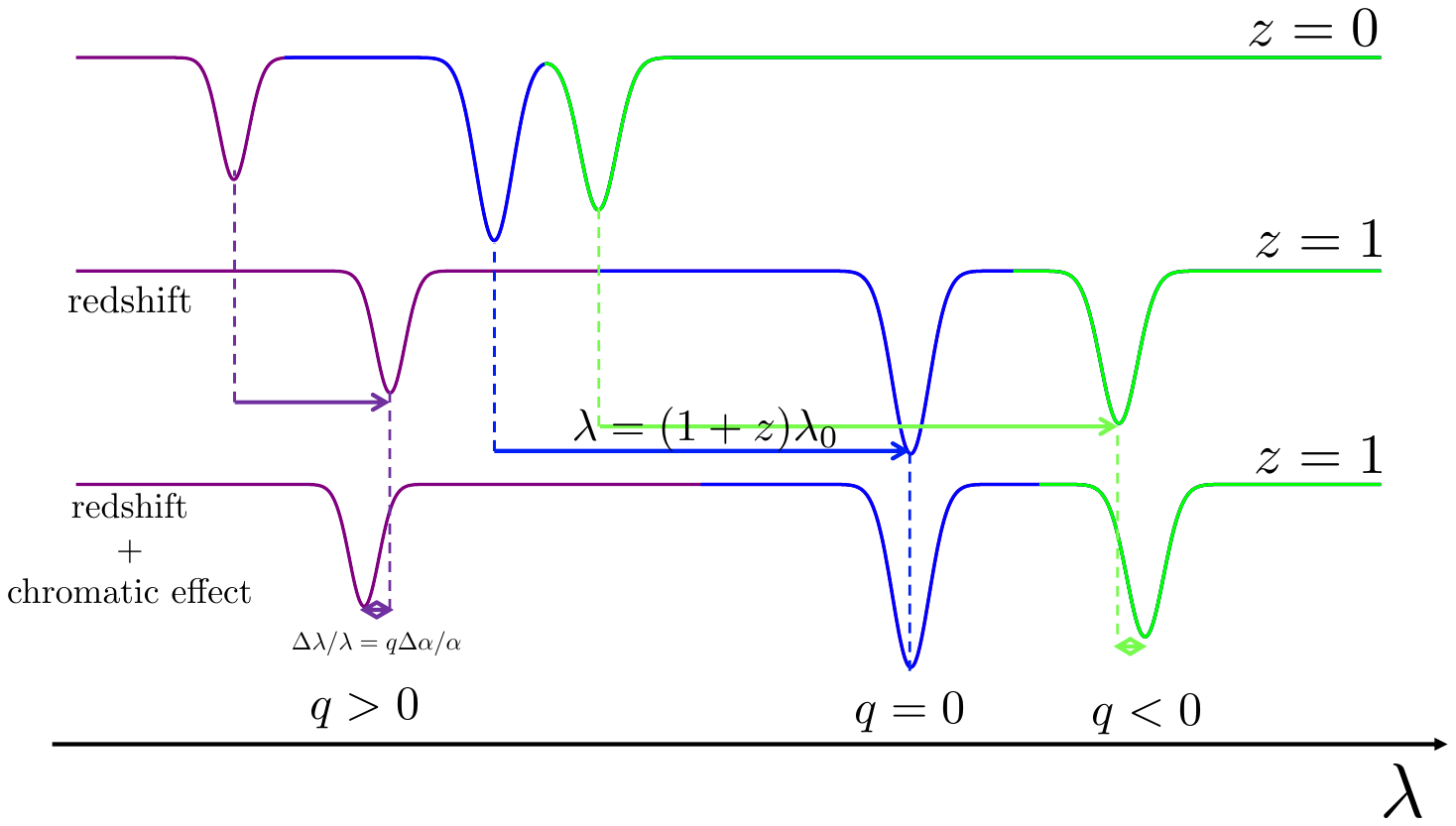}}
 \vskip-2.5cm
  \caption[Comarative effect of the cosmic expansion and the variation of constants on an absorption spectra]{The cosmic expansion induces an achromatic redshift of all wavelengths which distort all emission/absorption spectra. The variation of a fundamental constant $\alpha$ reveals itself through achromatic effects. Depending on the amplitude and sign of the sensitivity $q$, a transition can have a wavelength that is either blue- or red-shifted compared to the cosmological redshift. The idea is then to correlate different lines, in particular with those of very low $q$ that can be used as anchors, which amounts to question if the redshift determined by all transitions is the same or not.}
  \label{fig-defq}
\end{figure}

\paragraph{Accuracy of the measurements}\

The shifts in the absorption lines to be detected are extremely small. For instance a change  of $\aem$ of order $10^{-5}$ corresponds a shift of at most 20~m\AA\ for a redshift of $z\sim2$, which would corresponds to a shift of order $\sim 0.5 \unit{km/s}$, or to about a third of a pixel at a spectral resolution of $R\sim40000$, as achieved with Keck/HIRES or VLT/UVES. Note that since a few years ago uncertainties in laboratory wavelengths provided the dominant part of the error budget of many measurements.

 As we shall discuss later, there are several sources of uncertainties that hamper the  measurement. In particular, the absorption lines have complex profiles (because they result from the propagation of photons through a highly inhomogeneous medium) that are fitted using a combination of Voigt profiles. Each of these components depends on several parameters including the redshift, the column density and the width of the line (Doppler parameter) to which one now needs to add the constants that are assumed to be varying. These parameters are constrained assuming that the profiles are the same for all transitions, which is indeed a non-trivial assumption for transitions from different species (this was one of the driving motivations to use the transition from a single species and of the SIDAM method). More important, the fit is usually not unique. This is not a problem when the lines are not saturated but it can increase the error on $\aem$ by a factor 2 in the case of strongly saturated lines \citep{q-chandAD}. To finish, knowing the relative isotopic abundances of some species is also important; see  \cite{Murphy:2013jpa} for a  recent compilation of the data for $\aem$ measurements.
 
It follows that not all transitions are sufficiently sensitive and  that only  few astrophysical systems are clean enough to provide accurate measurements. Roughly, more than one hundred lines of sight allow for $\aem$ measurements with  only a fraction of them being ideal. Besides atomic absorption spectra, molecular absorption spectra allow for $\mu$-measurements -- see our discussion in Sect.~\ref{secmolqso} -- and emission spectra are in general less  vulnerable to some systematics but less sensitive; see Sect.~\ref{subsec_O3}.

\paragraph{Hunting systematics.}\

In order to claim for a variation of $\aem$, one shall eliminate any systematics that could mimic is effect on an absorption spectra. Many systematics have been listed and studied.

(1) Errors in the determination of laboratory wavelengths to which the observations are compared. 

(2) While comparing wavelengths from different atoms one has to take into account that they may be located in different regions of the cloud with different velocities and hence different Doppler
shifts. 

(3) One has to ensure that there is no transition blended by transitions of another system.

(4) The differential isotopic saturation has to be controlled. Usually quasar absorption systems are expected to have lower heavy element abundances. The spatial inhomogeneity of these abundances may also play a role. 

(5) Hyperfine splitting can induce a saturation similar to isotopic abundances. 

(6) The variation of the velocity of the Earth during the integration of a quasar spectrum can also induce differential Doppler shift.

 (7) Atmospheric dispersion across the spectral direction of the spectrograph slit can stretch the spectrum. It was shown that, on average,  this can, for low redshift observations, mimic a negative
$\Delta\aem/\aem$, while this is no more the case for high redshift observations, hence emphasizing their complementarity.

 (8) The presence of a magnetic field will shift the energy levels by Zeeman effect. 
 
 (9) Temperature variations during the observation will change the air refractive index in the spectrograph. In particular, flexures in the instrument are dealt with by recording a calibration lamp spectrum before and after the science exposure and the signal-to-noise and stability of the lamp is crucial
 
(10) Instrumental effects such as variations of the intrinsic instrument profile have to be controlled.

All these effects have been discussed in details; see e.g.,  \cite{q_murphysyt1,q_murphysyt2} and one shall keep in mind when comparing them that VLT/UVES and Keck/HIRES have, for this particular science case, an irreducible systematic of about 3 parts per million \citep{NEW_Whitmore:2014ina}.

\paragraph{Calibration}\

This was complemented by a study on the crucial issue of calibration since any distortion of the wavelength scale could lead to a non-zero value of $\Delta\aem$.  The wavelengths scale were calibrated by mean of a thorium-argon emission lamp. The quality of the calibration of the Keck/HIRES spectrograph was discussed in \cite{q_murphysyt1,q-thar} and argued to have negligible effects on the measurements.   \cite{q-griest} provided an analysis of is wavelength accuracy of the Keck/HIRES. An absolute uncertainty of  $\Delta z\sim10^{-5}$, corresponding to $\Delta\lambda\sim0.02 \unit{\AA}$ with daily drift of $\Delta z\sim5\times10^{-6}$ and multiday drift of $\Delta z\sim2\times10^{-5}$, arguing that this level of systematic uncertainty makes it difficult to use the Keck/HIRES to constrain the time variation of $\aem$ (at least for a single system or a small sample since the distortion pattern pertains to the echelle orders as they are recorded on the CDD, that is it is similar from exposure to exposure, the effect on $\Delta\aem/\aem$ for an ensemble of absorbers at different redshifts  would be random since the transitions fall in different places with respect to the pattern of the distortion). A similar result has been obtained for the VLT/UVES data \citep{whit}. It was concluded that  the ThAr lamp calibration of the quasar wavelength scale was distorted with respect to that established from the Solar spectrum -- via reflection from asteroids; see \cite{NEW_Rahmani:2013dia} -- following that  spurious velocity shifts were applied to different transitions at different wavelengths, most-likely causing the observed deviations in the determination of $\aem$ \citep{NEW_Whitmore:2014ina}.

\paragraph{Effect of the isotopic abundances}\

As will be detailed below, one concern of the MM method is the isotopic abundances of Mg\,{\sc ii} that can affect the low-$z$ sample since any changes in the isotopic composition will alter the value of effective rest-wavelengths. This isotopic composition is assumed to be close to terrestrial $^{24}$Mg:$^{25}$Mg:$^{26}$Mg = 79:10:11.  While no direct measurement of $r_{\mathrm{Mg}}=({}^{26}{\mathrm{Mg}}+{}^{25}{\mathrm{Mg}})/{}^{24}{\mathrm{Mg}}$ in QSO absorber was feasible for a long time due to the small separation of the isotopic absorption lines, it was shown \citep{q-gaylamber}, on the basis of molecular absorption lines of MgH that $r_{\mathrm{Mg}}$ generally decreases with smaller metallicity. From the absorption spectrum of the quasar HE0001-2340 observed with VLT/UVES \cite{NEW_Agafonova:2011sp} measured the isotopic ratio $^{24}$Mg:$^{25}$Mg:$^{26}$Mg = $(19\pm11):(22\pm13):(59\pm6)$ leading to $r_{\rm Mg}=4$ at $z=0.45$, hence showing over-abundance of heavy Mg isotopes compared to the Solar system value of $r_{\rm Mg}=0.3$. In systems at $z=1.58$ and 1.65 they conclude that $r_{\rm Mg}\lesssim0.7$ (resp. 2.6). While the first system is thought to be the fragment of the outflow caused by a SNIa of a high-metallicity white-dwarf, the two others are enriched by AGB stars. This shows the high variability of $r_{\rm Mg}$ and its dependence on the stellar history of the absorption system. In standard models it should be near 0 at zero metallicity since type~II supernovae are primarily producers of $^{24}$Mg. It was also argued that $^{13}$C is a tracer of $^{25}$Mg and was shown to be low in the case of HE~0515-4414 \citep{levCMg}. However, contrary to this trend, \cite{q-yong}  found that $r_{\mathrm{Mg}}$ can reach high values for some giant stars in the globular cluster NGC~6752 with metallicity [Fe/H]$\sim-1.6$. This led \cite{q-aften} to propose a chemical evolution model with strongly enhanced population of intermediate ($2-8\,M_{\odot}$) stars, which in their asymptotic giant branch phase are the dominant factories for heavy Mg at low metallicities typical of QSO absorption systems, as a possible explanation of the low-$z$ Keck/HIRES observations without any variation of $\aem$. It would require that $r_{\mathrm{Mg}}$ reaches 0.62, compared to 0.27 (but then the UVES/VLT constraints would be converted to a detection).  Care needs to be taken since the star formation history can be different ine each region, even in each absorber, so that one cannot a priori use the best-fit obtained from the Keck data to the UVES/VLT data. However, such modified nucleosynthetic history will lead to an overproduction of elements such as P, Si, Al, P above current constraints \citep{q-fenner}, but this later model is not the same as the one of \cite{q-aften} that was tuned to avoid these problems. The  calculation of isotope shifts in atoms with a few valence electrons was proposed by \cite{NEW_Berengut:2003flq} to determine whether differences in isotope abundances in early universe can contribute to the observed anomalies in quasar absorption spectra, with a negative answer.

\subsubsection{Evolution of the debate}

The claim by \citep{q-webprl99,q-murphy03a,King:2012id} that the Keck/HIRES data indicated that the fine structure constant was smaller in the past relaunched the interest in fundamental constants, both from an experimental and theoretical point of view. For many years, data from the VLT and Keck gave discordant results and were at the heart of a lively debate on the hunt for systematics. It mostly emerged for the result by \cite{webspace} over about two decades and their 4$\sigma$ of a variation of $\aem$ at low redshift $(1<z<4)$. Indeed, despite their obvious successes in other fields, spectrographs such as UVES, HARPS or Keck/HIRES were not built with this science case in mind and were far from optimal for it.  

Trying to confirm these results was the main motivation for an ESO/UVES Large Program \citep{NEW_Molaro:2013saa,NEW_Rahmani:2013dia,NEW_Evans:2014yva}, a dedicated program to test for the variation of fundamental constants with an optimized sample and methodology with about 40 VLT nights of observations over 2010-2013 with a typical resolution of $R\sim60~000$ and a signal-to-noise per pixel of order 100, leading to an expected accuracy of order $10^{-6}$ on $\aem$. The survey was unfortunately not optimized to test the dipole hypothesis by \cite{webspace} since it included no target towards the North pole of the dipole.

This led to a huge amount of data and, today, the spectroscopic QSO are usually split into two samples \citep{NEW_Martins:2017yxk,NEW_Aluri:2022hzs}.
\begin{itemize}
\item[$\bullet$] The {\bf\emph{Archival data set}} of \cite{webspace} contains 293 archival measurements from the  Keck/HIRES and  VLT/UVES spectrographs up to redshift of $4.18$.  These data, with resolution $R\sim50~000$, were  originally not designed for the study of fundamental constants, by a large number of observers, under a broad range of observing conditions, and over a time span of almost a decade. This sample was reanalyzed for the study of the spatial variation of $\aem$ \cite{webspace} -- see Sect.~\ref{section-spatial} -- and shall be then taken with caution. Indeed, it has been well-documented that  $\aem$ measurements require particularly careful wavelength calibration procedures, which rely on additional data coeval with the QSO observations. Such additional data are not ordinarily taken for standard observations and cannot be obtained a posteriori. Moreover, these two spectrographs are now known to suffer from significant intra-order and long-range distortions \citep{NEW_Whitmore:2014ina}. Such limitations may be partially mitigated but cannot be fully eliminated. Its global analysis concludes \citep{webspace}
$$
\Delta\aem/\aem = (-2.16\pm0.86)\times 10^{-6}, \quad z_{\rm eff}=1.50\,.
$$
Note however that concerns do exist concerning the existence of some systematics in the data that  have not been fully modeled or corrected so far \citep{NEW_Whitmore:2014ina,Dumont:2017exu,NEW_Webb:2023jnj}.

\item[$\bullet$] The {\bf\emph{Dedicated data set}}  contains about 30 measurements designed for the specific analysis of $\aem$ variation, where ancillary data enabled a more robust wavelength calibration procedure, or using more modern spectrographs that do not suffer from the limitations of VLT/UVES or Keck/HIRES. This includes measurements listed in Table 1 of \cite{NEW_Martins:2017yxk} and more recent ones from the Subaru telescope \citep{NEW_10.1093/mnras/stx1949} and the X-SHOOTER \citep{NEW_Wilczynska:2020rxx}, HARPS \citep{NEW_Milakovic:2020tvq} and ESPRESSO \citep{NEW_10.1093/mnras/staa807,NEW_Murphy:2021xhb} spectrographs. The original spectra have resolution $R\sim 50~000-150~000$ (with the upper end of the range due to ESPRESSO), with the exception of the X-SHOOTER data which only has $R\sim10000$. The latter are the first direct measurements of $\aem$ in the IR part of the electromagnetic spectrum, extending the redshift range up to  $z\sim7.06$ but their sensitivity is only at the level of tens of ppm, so they do not carry significant weight in the statistical analysis. The weighted mean of all the values in each data set. Similarly, we can identify $\aem$ and effective redshift of Table~\ref{tabSIDAM} gives
$$
\Delta\aem/\aem = (-0.23\pm 0.56)\times 10^{-6}, \quad z_{\rm eff}=1.29\,.
$$
\end{itemize}
Clearly these two data sets are discrepant; see e.g., \cite{NEW_Martins:2022unf,NEW_Martins:2017yxk} for their comparison.The former has a preference for a negative variation at more than two standard deviations, while the latter value is consistent with the null result.  The dedicated database sets more stringent measurements but is, for now, smaller than the archival data set so that their constraining power remains comparable \citep{NEW_Martins:2019qxe}.

The \emph{Dedicated data set} is given in Table~\ref{tabSIDAM} and Fig.~\ref{fig-sqoalpha} and described in the text below.

\begin{table*}
\begin{center}
\begin{tabular}{ccccr}
\toprule
 Object & $z_{\rm abs}$ & $\Delta\aem/\aem$ & Spectrographs & Reference \\
 \midrule
3 sources  &$[0.7-1.5]$ &  $ (4.3\pm3.4)\times10^{-6} $ & HIRES & \cite{NEW_Songaila:2014fza}\\
		 &$z_{\rm min}=1.08$ &  $ $ &  &\\
\midrule
J0120+2133 & 0.576&   $(-9.12\pm39.80\pm3.64_{\rm syst})\times10^{-6}$ & HDS & \cite{NEW_Murphy:2017xaz} \\
          & 0.729&   $(0.73\pm6.17\pm1.77_{\rm syst})\times10^{-6}$ & HDS & \cite{NEW_Murphy:2017xaz} \\
          & 1.048&   $(5.47\pm18.26\pm4.03_{\rm syst})\times10^{-6}$ & HDS & \cite{NEW_Murphy:2017xaz} \\
	 &1.325 & $ (2.60\pm3.45\pm2.38_{\rm syst})\times10^{-6}$   & HDS & \cite{NEW_Murphy:2017xaz} \\
	& 1.343 &   $(8.36\pm11.82\pm2.84_{\rm syst})\times10^{-6}$   & HDS & \cite{NEW_Murphy:2017xaz} \\
\midrule
 J0026-2857& 1.023 &$( 3.54\pm8.54\pm2.38_{\rm syst} )\times10^{-6}$& UVES & \cite{NEW_Murphy:2016yqp} \\
 \midrule
J0058-0041& 1.072 &$( -1.35\pm6.71 \pm2.51_{\rm syst} )\times10^{-6}$& HIRES & \cite{NEW_Murphy:2016yqp} \\
\midrule
HE0515-4414& 1.15& $(-0.12\pm1.79)\times10^{-5}$ &UVES  &\cite{q-sidam1}  \\
     &  & $(0.5\pm2.4)\times10^{-5}$ &HARPS  &  \cite{q-chandSIDAM} \\
\midrule
HS1549+1919& 1.143 &$( -7.49\pm 4.63 \pm3.02_{\rm syst} )\times10^{-6}$& UVES/HIRES/HDS & \cite{NEW_Evans:2014yva} \\
& 1.342 &$( -0.70\pm 6.43 \pm1.55_{\rm syst} )\times10^{-6}$& UVES/HIRES/HDS & \cite{NEW_Evans:2014yva} \\
& 1.802 &$( -6.42\pm 6.52 \pm3.16_{\rm syst} )\times10^{-6}$& UVES/HIRES/HDS & \cite{NEW_Evans:2014yva} \\
\midrule
 HE514-4414 & 1.158 & $ (-1.42\pm0.55\pm0.65_{\rm syst})\times10^{-6} $ & UVES & \cite{Kotus:2016xxb}\\
                       &  & $(-0.27\pm2.41)\times10^{-6}$ & HARPS & \cite{NEW_Milakovic:2020tvq}\\
                       &  & $(1.3\pm1.3\pm0.4_{\rm syst})\times10^{-6}$ & ESPRESSO & \cite{NEW_Murphy:2021xhb}\\
\midrule
J1237+0106& 1.305 &$( -4.54\pm8.08 \pm3.13_{\rm syst} )\times10^{-6}$& HIRES & \cite{NEW_Murphy:2016yqp} \\
 \midrule
J0058-0041& 1.342 &$( 3.05\pm 3.30 \pm2.13_{\rm syst} )\times10^{-6}$& HIRES & \cite{NEW_Murphy:2016yqp} \\
		&  &$( 5.67\pm 4.19 \pm2.16_{\rm syst} )\times10^{-6}$& UVES & \cite{NEW_Murphy:2016yqp} \\
 \midrule
J0108-0037& 1.371 &$(-8.45\pm 5.69 \pm4.64_{\rm syst} )\times10^{-6}$& VLT & \cite{NEW_Murphy:2016yqp} \\
\midrule
HE0001-2340 & 1.58 & $(-1.5\pm2.6)\times10^{-6}$&  UVES & \cite{NEW_Agafonova:2011sp}\\
\midrule
J1029+1039& 1.622 &$(-1.70\pm 9.80 \pm 2.47_{\rm syst} )\times10^{-6}$& HIRES & \cite{NEW_Murphy:2016yqp} \\
\midrule
HE1104-1805 & 1.661 &  $(-4.70\pm5.30)\times10^{-6} $ & HIRES & \cite{NEW_Songaila:2014fza}\\
\midrule
 HE2217-2818 & 1.692 & $(1.3\pm2.4\pm1.0_{\rm syst})\times10^{-6}$ & UVES &  \cite{NEW_Molaro:2013saa} \\
\midrule
HS1946+7658  &1.738 &  $ (-7.90\pm6.20)\times10^{-6} $ & HIRES & \cite{NEW_Songaila:2014fza}\\
\midrule
J1944+7705 & 1.738 &  $(12.70\pm16.30\pm1.47_{\rm syst})\times10^{-6}$  & HDS & \cite{NEW_Murphy:2017xaz} \\
\midrule
Q1101-264 &1.84& $(5.66\pm2.67)\times10^{-6}$ &UVES  & \cite{Q1101} \\
J11032-2645 & 1.839 & $(3.3\pm2.9)\times10^{-6}$ & UVES & \cite{NEW_10.1093/mnras/stx179} \\
\midrule
Q2206-1958& 1.921 &$(-4.65\pm 6.01 \pm 2.24_{\rm syst} )\times10^{-6}$& VLT & \cite{NEW_Murphy:2016yqp} \\
\midrule
Q1755+57& 1.971 &$(4.72\pm 4.18 \pm 2.16_{\rm syst} )\times10^{-6}$& HIRES & \cite{NEW_Murphy:2016yqp} \\
\midrule
PHL957& 2.309 &$(-0.65\pm 6.42 \pm 2.26_{\rm syst} )\times10^{-6}$& HIRES & \cite{NEW_Murphy:2016yqp} \\
	&  &$(-0.20\pm 12.44 \pm 3.51_{\rm syst} )\times10^{-6}$& VLT & \cite{NEW_Murphy:2016yqp} \\
\midrule
J0035-0918& 2.34 & $(-1.2\pm1.1)\times10^{-5}$ & ESPRESSO &\cite{NEW_10.1093/mnras/staa807} \\
\midrule
J1120+0641 	& 5.50726 & $(7.42\pm9.60\pm1.52_{\rm syst})\times10^{-5}$ &X-SHOOTRER &  \cite{NEW_Wilczynska:2020rxx} \\
			& 5.95074& $(-22.85\pm17.11\pm0.32_{\rm syst})\times10^{-5}$ &X-SHOOTRER &  \cite{NEW_Wilczynska:2020rxx} \\
			&6.17097&  $(-10.16\pm14.80\pm0.42_{\rm syst})\times10^{-5}$ &X-SHOOTRER &  \cite{NEW_Wilczynska:2020rxx} \\
			& 7.05852&   $(12.79\pm48.66\pm19.74_{\rm syst}))\times10^{-5}$ &X-SHOOTRER &  \cite{NEW_Wilczynska:2020rxx} \\
\bottomrule
\end{tabular}
\caption[Dedicated $\aem$ dataset]{Summary of the dedicated measurements of $\aem$. Each quasar specifies a line of sight and thus a celestial direction. The redshifts correspond to the various absorption systems on the line of sight leading to independent constraints on $\aem$.}
\label{tabSIDAM} 
\end{center}
\end{table*}

\begin{figure}[hptb]
 \centerline{\includegraphics[scale=0.45]{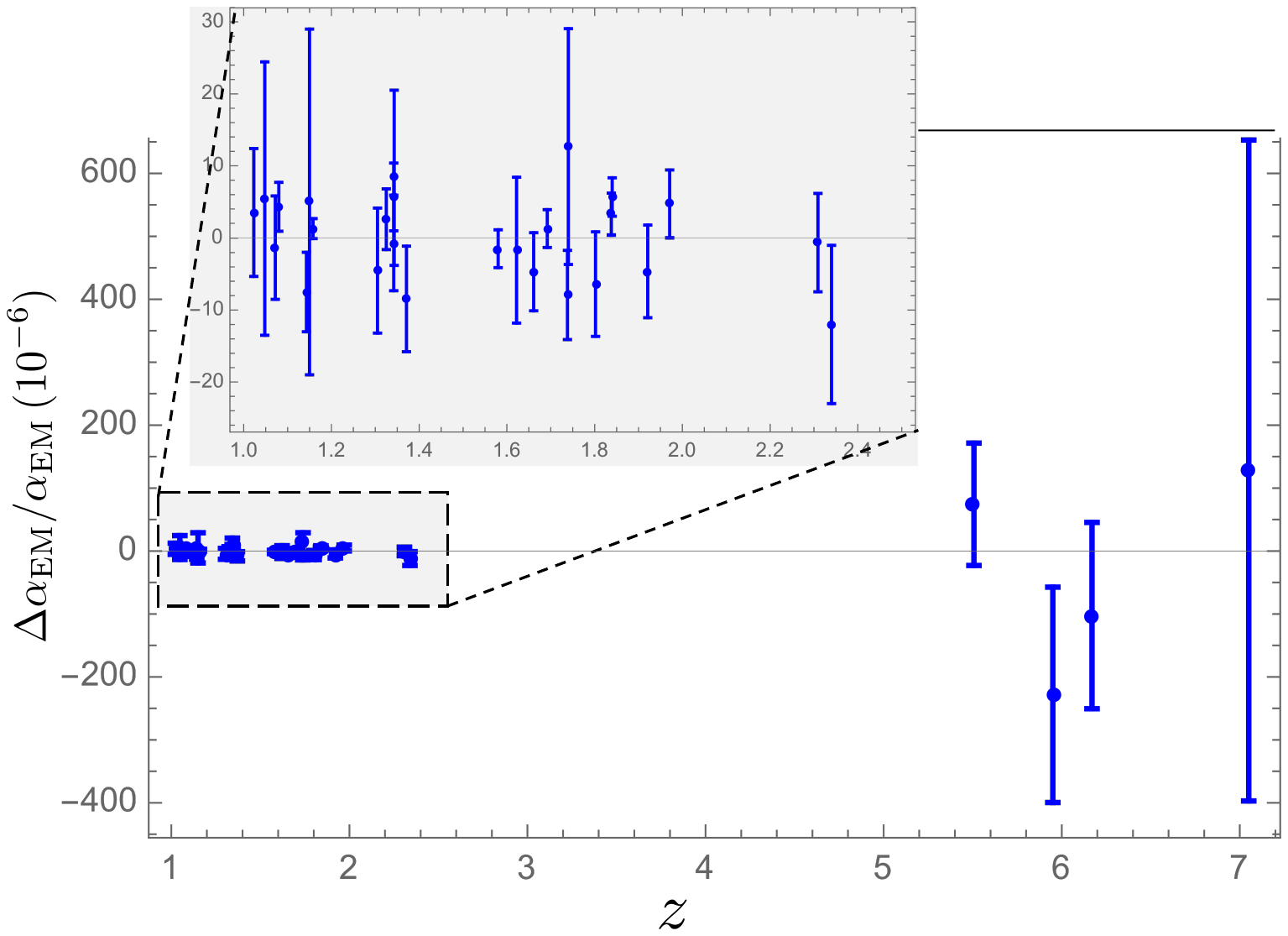}}
 \vskip-0.5cm
  \caption[QSO absorption spectra constraints on $\aem$]{Constraints on the variation of $\aem$ from the analysis of QSO absorption spectra (dedicated data set); see  Table~\ref{tabSIDAM}.}
  \label{fig-sqoalpha}
\end{figure}

\subsubsection{Different methods to track for the variation of constants}\label{subsecAD}

As emphasized, one needs to measure the redshifts from more than one transitions in the same absorption system. The shift between two lines is easier to measure when the difference between the $q$-coefficients is large, which occurs, e.g., for two levels with large $q$ of opposite signs. Many methods were developed to take this into account. 
\begin{itemize}
\item The {\bf\emph{ alkali doublet method}} (AD) focuses on the fine-structure doublet of alkali atoms. This method completely avoids the assumption of homogeneity because, by construction, the two lines of the doublet must have the same profile. Indeed the AD method avoids the implicit assumption of the MM method that chemical and ionization inhomogeneities are negligible. 
\item It was then generalized to the {\bf\emph{many-multiplet method}} (MM), which uses correlations between various transitions in different atoms using the theoretical information that transitions are almost insensitive to a variation of $\aem$. This is the case of Mg\,{\sc ii}, which can be used as an \emph{anchor}, i.e., a reference point. To obtain strong constraints one can either compare transitions of light atoms with those of heavy atoms (because the $\aem$ dependence of the ground state scales as $Z^2$) or compare $s-p$ and $d-p$ transitions in heavy elements (in that case, the relativistic correction will be of opposite signs). This latter effect increases the sensitivity and strengthens the method against systematic errors. However, the results of this method rely on two assumptions: (\textit{i}) ionization and chemical homogeneity and (\textit{ii}) isotopic abundance of Mg\,{\sc ii} close to the terrestrial value. Even though these are reasonable assumptions, one cannot completely rule out systematic biases that they could induce. 
\item The {\bf\emph{single ion differential alpha measurement method}} (SIDAM) avoids the influence of small spectral shift due to ionization inhomogeneities within the absorber and due to possible non-zero offset between different exposures since it relies on different transitions of a single ion in individual exposure. 
\end{itemize}

Most studies are based on \emph{optical techniques} due to the profusion of strong UV transitions  that are redshifted into the optical band (this includes AD, MM, SIDAM and it implies that they can be applied only above a given redshift, e.g., Si\,{\sc iv} at $z>1.3$, Fe\,{\sc ii}$\lambda 1608$ at $z>1$) or on \emph{radio techniques} since radio transitions arise from many different physical effects (hyperfine splitting and in particular H\,{\sc i} 21~cm hyperfine transition, molecular rotation, Lambda-doubling, etc). In the latter case, the line frequencies and their comparisons yield  constraints on different sets of fundamental constants including $\aem$, $g_{\mathrm{p}}$ and $\mu$. Thus, these techniques are complementary since systematic effects are different in optical and radio regimes. Also the radio techniques offer some advantages: (\textit{1}) to reach high spectral resolution ($<1 \unit{km/s}$), alleviating in particular problems with line blending and the use of, e.g., masers allow to reach a frequency calibration better than roughly 10~m/s; (\textit{2}) in general, the sensitivity of the line position to a variation of a constant is higher; (\textit{3}) the isotopic lines are observed separately, while in optical there is a blend with possible differential saturations (see, e.g., \citealt{calma} for a discussion).

\paragraph{Alkali doublet method (AD)}\

The first method used to constraint the time variation of the fine-structure constant relies on fine-structure doublets splitting for which $\Delta\nu \propto \aem^2 Z^4 R_\infty/2 n^3$. Hence,  $\aem$, $\Delta\nu/\bar\nu\propto\aem^2$, so that the variation of the fine structure constant at a redshift $z$ can be obtained as
$$
  \left(\frac{\Delta\aem}{\aem}\right)(z)= \frac{c_r}{2}
 \left[\left(\frac{\Delta\lambda}{\bar\lambda}\right)_z/\left(
 \frac{\Delta\lambda}{\bar\lambda}\right)_0 -1\right],
$$
where $c_r\sim1$ is a number taking into account the relativistic  corrections. This expression is indeed a simple approach since one should, as for atomic clocks, take into
account the relativistic corrections more precisely. Using the formulation~(\ref{qpara}), one can deduce that
$ c_r = (\delta q +\delta q_2)/(\delta q + 2\delta q_2)$, where the $\delta q$ are the differences between the $q$-coefficients for the doublet transitions.

The AD method has been applied to several species such as, e.g., C\,{\sc iv}, N\,{\sc v}, O\,{\sc vi}, Mg\,{\sc ii}, Al\,{\sc iii}, Si\,{\sc ii}, Si\,{\sc iv}. We refer to Sect.~III.3 of FVC03 \citep{jpu-revue} for a summary of
their results (see also \citealt{Levseul}) and focus on the three most recent analysis, based on the Si\,{\sc iv} doublet. In this particular case, $q=766$ (resp.\ 362)~cm$^{-1}$ and $q_2=48$ (resp.\ --8)~cm$^{-1}$ for Si\,{\sc iv} $\lambda1393$ (resp. $\lambda1402$) so that $c_r=0.8914$. The method is based on a $\chi^2$ minimization of multiple components Voigt profile fit to the absorption features in the QSO spectra. In general such a profile depends on three parameters, the column density $N$, the Doppler width and the redshift. It is now extended to include $\Delta\aem/\aem$. The fit is carried out by simultaneously varying these parameters for each component.

\cite{q-murphyAD} analyzed 21 Keck/HIRES  Si\,{\sc iv} absorption systems toward 8 quasars to obtain the weighted  mean
 \begin{equation}
   \label{AD-m}
   \Delta\aem/\aem = (-0.5\pm1.3)\times10^{-5}, \qquad 2.33< z < 3.08,
 \end{equation}
 with a mean redshift of $z=2.6$. The S/N ratio of these data is in the  range 15\,--\,40 per pixel and the spectral resolution is  $R\sim34000$. \cite{q-chandAD} analyzed  15  Si\,{\sc iv} absorption systems selected from a VLT-UVES sample  containing 31 systems (eliminating contaminated, saturated or very broad systems; in particular a lower limit on the  column density was fixed so that both lines of the doublets  are detected at more than $5\sigma$) to get the weighted  mean,
 \begin{equation}
   \label{a-AD1}
   \Delta\aem/\aem = (-0.15\pm0.43)\times10^{-5}, \qquad 1.59< z < 2.92.
 \end{equation}
The improvement of the constraint arises mainly from a better S/N  ratio, of order 60\,--\,80 per pixel, and resolution  $R\sim45000$. Note that combining this result with~(\ref{AD-m}) leads to the weighted mean $\Delta\aem/\aem =  (-0.04\pm0.56)\times10^{-5}$ in the range $1.59< z < 3.02$. The analysis \citep{q-martinezAD} of seven C\,{\sc iv} systems and   two Si\,{\sc iv} systems in the direction of a single quasar,   obtained by the VLT-UVES (during the science verification) led to
 \begin{equation}
   \Delta\aem/\aem = (-3.09\pm8.46)\times10^{-5}, \qquad 1.19< z < 1.84.
 \end{equation}
This is less constraining than the two previous analyses, mainly because the $q$-coefficients are smaller for C\,{\sc iv} (see \citealt{ppc4} for the calibration of the laboratory spectra). One limitation may arise from the isotopic composition. Silicium has three naturally occurring isotopes with terrestrial abundances $^{28}$Si:$^{29}$Si:$^{30}$Si = 92.23:4.68:3.09 so that each absorption line is a composite of absorption lines from the three isotopes. However, it was shown that this effect of isotopic shifts \citep{q-murphyAD} is negligible in the case of Si\,{\sc iv}.

\paragraph{Many multiplet method (MM)}\label{MMQSO}\

A generalisation of the AD method, known as the many-mulptiplet was proposed by \cite{q-MMmethod}. It relies on the combination of transitions from different species, taking into account the theoretical informlation that some transitions are fairly unsensitive to a change of the fine-structure constant (e.g., Mg\,{\sc ii} or Mg\,{\sc  i}, hence providing good anchors) while others such as Fe\,{\sc ii} are more sensitive.\\
 
 \noindent{\bf\em Early Keck/HIRES data analysis.}\ The MM-method was first applied by \cite{q-webprl99} who analyzed one transition of the Mg\,{\sc ii} doublet and five Fe\,{\sc ii} transitions from three multiplets. Using 30 absorption systems toward 17 quasars, they obtained
\begin{eqnarray}
 &&\Delta\aem/\aem=(-0.17\pm0.39)\times 10^{-5},\qquad 0.6<z<1\nonumber\\
 &&\Delta\aem/\aem=(-1.88\pm0.53)\times 10^{-5},\qquad
 1<z<1.6.\nonumber
\end{eqnarray}
This was  indeed the first claim that a constant may have varied during the evolution of the universe. It was later confirmed by \citep{q_murphysyt2,q-webprl01} from a reanalysis of the initial sample and by including new optical QSO data to reach 28 absorption systems with redshift $z=0.5-1.8$ plus 18 damped Lyman-$\alpha$ absorption systems towards 13 QSO plus 21 Si\,{\sc iv} absorption systems toward 13 QSO. The analysis used mainly the multiplets of Ni\,{\sc ii}, Cr\,{\sc ii} and Zn\,{\sc ii} and Mg\,{\sc i}, Mg\,{\sc   i}, Al\,{\sc ii}, Al\,{\sc iii} and Fe\,{\sc ii} was also included. The analysis \citep{q-murphy03a} relies on 128 absorption spectra, later updated \citep{q-murphyMMlast} to include 143 absorption systems. The more robust estimates is the weighted mean
\begin{equation}
\label{a-MMw}
 \Delta\aem/\aem = (-0.57\pm0.11)\times10^{-5},\qquad 0.2<z<4.2.
 \end{equation}
The resolution for most spectra was $R\sim45000$ and the S/N per pixel ranges from 4 to 240, with most spectral regions with S/N$\sim$~30.

The low-$z$ ($z<1.8$) and high-$z$ rely on different ions and transitions with very different $\aem$-dependencies. At low-$z$, the Mg transitions are used as anchors against which the large positive shifts in the Fe\,{\sc ii} can be measured. At high-$z$, different transitions are fitted (Fe\,{\sc ii}, S\,{\sc ii}, Cr\,{\sc ii}, Ni\,{\sc ii}, Zn\,{\sc ii}, Al\,{\sc ii}, Al\,{\sc iii}). The two sub-samples respond differently to simple systematic errors due to
their different arrangement of $q$-coefficients in wavelength space. The analysis for each sample gives the weighted mean
\begin{eqnarray}
 \Delta\aem/\aem = (-0.54\pm0.12)\times10^{-5},\qquad
 0.2<z<1.8\nonumber\\
 \Delta\aem/\aem = (-0.74\pm0.17)\times10^{-5},\qquad
 1.8<z<4.2,
 \end{eqnarray}
with respectively 77 and 66 systems.\\
 
\noindent{\bf\em VLT/UVES data analysis.}\ The previous results led another team to check this detection using the UVES spectrograph operating on the VLT. In order to avoid as much systematics as possible, they apply a series of selection criteria \citep{q-chandMM} on the systems used: (\textit{1}) consider only lines with similar ionization potentials (Mg\,{\sc ii}, Fe\,{\sc ii}, Si\,{\sc ii} and Al\,{\sc ii}) as they are most likely to originate from similar regions in the cloud; (\textit{2}) avoid absorption lines contaminated by atmospheric lines; (\textit{3}) consider only systems with hight enough column density to ensure that all the mutiplets are detected at more than $5\sigma$; (\textit{4}) demand than at least one of the anchor lines is not saturated to have a robust measurement of the redshift; (\textit{5}) reject strongly saturated systems  with large velocity spread; (\textit{6}) keep only systems for which the majority of the components are separated from the neighboring by more than the Doppler shift parameter. The advantage of this choice is to reject most complex or degenerate systems, which could result in uncontrolled systematics effects. The drawback is indeed that the analysis is based on less systems.

\cite{q-chandMM} and \cite{q-chandMM2} analyzed the observations of 23 systems satisfying these criteria in direction of 18 QSO with a S/N ranging between 50 and 80 per pixel and a resolution $R>44000$ to conclude that $\Delta\aem/\aem = (-0.06\pm0.06)\times10^{-5}$ for $0.4<z<2.3$, hence giving a $3\sigma$ constraint on a variation of $\aem$. This analysis was challenged by \cite{q-contr3,q-contr2, q-contr1} who claimed that the same data lead to a weighted mean, $\Delta\aem/\aem = (-0.44\pm0.16)\times10^{-5}$. These arguments were responded in \cite{q-contr3b} that  revised the VLT/UVES constraint, rejecting two more than 4$\sigma$ deviant systems that were claimed to dominate the re-analysis, to conclude that
\begin{equation}
\label{a-MMpp}
 \Delta\aem/\aem = (0.01\pm0.15)\times10^{-5},\qquad 0.4<z<2.3.
\end{equation}
Let us mention that \cite{q-sidam1} reanalyzed some systems of \cite{q-chandMM,q-chandMM2} by means of the SIDAM method (see below) and disagree with some of them, claiming for a problem of calibration. They also claim that the errors quoted in \cite{q-murphyMMlast} are underestimated by a factor 1.5. This debate has greatly motivated the developments of the Large Program and set the stage for better-controlled analysis of $\aem$ systems.\\

\noindent{\bf\em  Regressional MM.}\ The MM method was adapted to use a linear regression method \cite{q-quast}. The idea is to measure the redshift $z_i$ deduced from the transition $i$ and plot $z_i$ as a function of the sensitivity coefficient. If $\Delta\aem\not=0$ then there should exist a linear relation with a slope proportional to $\Delta\aem/\aem$. On a single absorption system (VLT/UVES), on the basis of Fe\,{\sc ii} transition, they concluded that
\begin{equation}\label{qrmm}
 \Delta\aem/\aem = (-0.4\pm1.9\pm2.7_{\text{syst}})\times10^{-6},\qquad z=1.15,
\end{equation}
compared to $\Delta\aem/\aem = (0.1\pm1.7)\times10^{-6}$ that is obtained with the standard MM technique on the same data. This is also consistent with the constraint~(\ref{qch}) obtained on the same system with the HARPS spectrograph.\\

\noindent{\bf\em Analysis of the archival data set.}\  \cite{webspace} split the VLT/UVES data at $z=1.8$ to get $\left(\Delta\aem/\aem\right)_{\text{VLT};\, z<1.8} = (-0.06\pm0.16)\times10^{-5}$, in agreement with the former study \citep{q-contr3b}, while at higher redshifts $\left(\Delta\aem/\aem\right)_{\text{VLT}\, z>1.8} = (+0.61\pm0.20)\times10^{-5}$. This higher-redshift set exhibits a positive a variation of $\aem$ of opposite sign with respect to the previous Keck/HIRES detection \citep{q-murphyMMlast}  that gave $\left(\Delta\aem/\aem\right)_{\text{Keck};\, z<1.8} = (-0.54\pm0.12)\times10^{-5}$ and $\left(\Delta\aem/\aem\right)_{\text{Keck};\, z>1.8} = (-0.74\pm0.17)\times10^{-5}$. It was pointed out that the Keck/HIRES and VLT/UVES observations can be made consistent in the case the fine structure constant is spatially varying \citep{webspace,berenspa,berenspa2}, as discussed in Sect.~\ref{section-spatial} since they do not correspond to the same hemisphere.

\subsubsection{Single ion differential measurement (SIDAM)}\label{subsecSIDAM}

This variation \citep{q-sidam0} on the MM method avoids the influence of small spectral shifts due to ionization inhomogeneities within the absorbers as well as to non-zero offsets between different exposures. It was mainly used with Fe\,{\sc ii}, which provides transitions with positive and negative $q$-coefficients. Since it relies on a single ion, it is less sensitive to isotopic abundances, and in particular not sensitive to the one of Mg.

The first analysis relies on the QSO HE~0515-4414 \citep{q-quast} set the constraint~(\ref{qrmm}). An independent analysis \citep{q-sidam1}  gave a weighted mean $\Delta\aem/\aem =(-0.12\pm1.79)\times10^{-6}$ for this system at $z=1.15$ at $1\sigma$. It was also  independently studied using the HARPS spectrograph -- that has a higher resolution that UVES; $R\sim112~000$ -- mounted on the 3.6~m telescope at La Silla observatory \citep{q-chandSIDAM}. Observations based on Fe\,{\sc ii} with a S/N of about 30\,--\,40 per pixel set the constraint
\begin{equation}\label{qch}
 \Delta\aem/\aem =(0.5\pm2.4)\times10^{-6},\qquad z=1.15.
\end{equation}
The second constraint \citep{Q1101,q-sidam1} was obtained from a system at $z=1.84$ toward Q~1101-264,  $\Delta\aem/\aem =(5.66\pm2.67)\times10^{-6}$.  A potential systematic  uncertainty  is the relative shift of the wavelength calibration in the blue and the red arms of UVES where the distant Fe lines are recorded simultaneously (see, e.g., \citealt{molaro} for a discussion of the systematics).

A long series of studies, mostly motivated by the Large program with the VLT/UVES led to the construction of the dedicated data set. The actual measurements are all gathered in in Table~\ref{tabSIDAM} and we shall only give a short description of these works.

From the study on magnesium isotope abundances with the VLT/UVES \cite{NEW_Agafonova:2011sp} set the strong constraint $\Delta\aem/\aem =(-1.5\pm2.6)\times10^{-6}$  in a system at $z_{\rm abs}=1.5864$ towards HE0001-2340 mainly from Fe{\sc ii}, Si{\sc ii}, Al{\sc ii}, Al{\sc iii}, Mg{\sc i} and Mg{\sc ii}.

\cite{NEW_Evans:2014yva} observed 3 systems at $z_{\rm abs}=1.143, 1.342$ and 1.802 in the direction of HS1549+1919 with 3 telescopes (VLT/UVES, Keck/HIRES and Subaru/HDS). The two low-$z$ systems are constrained thanks to Fe{\sc ii}, Mg{\sc i}, and Mg{\sc ii} while Al{\sc ii}, Al{\sc iii} and Fe{\sc ii} are used for the higher-$z$ system. The constraint for each system is obtained from the combination of 3 measurements Table~\ref{tabSIDAM} and the average on the 3 systems gives $\Delta\aem/\aem =(-5.4\pm3.3\pm1.5_{\rm syst})\times10^{-6}$ at 1$\sigma$. 

\cite{NEW_Songaila:2014fza} used the MM method on 3 quasars observed with Keck/HIRES with Mg{\sc ii} and Fe{\sc ii} lines  at low-$z$ and  Cr{\sc ii}, Zn{\sc ii}, Ni{\sc ii} and Mn{\sc ii} lines at higher-$z$. 8 systems with $z$ between 0.7 and 1.5 along the 3 lines of sight are averaged while individual constraints at $z_{\rm abs}=1.661$ and~1.738 are obtained for two lines of sight. They concluded  that Mn{\sc ii}, Ni{\sc ii} and Cr{\sc ii} give the most robust results. Their constraints are gathered in Table~\ref{tabSIDAM}.

The first result from the Large program with the VLT/UVES \citep{NEW_Molaro:2013saa} is the analysis of an absorption system at $z_{\rm abs}=1.692$ towards H22217-2818 from 6 lines of Al{\sc ii}, {\sc iii} and Fe{\sc ii} that led to $\Delta\aem/\aem =(1.3\pm2.4\pm1.0_{\rm syst})\times10^{-6}$.

\cite{NEW_Murphy:2016yqp} analyzed 9 systems in which Zn and Cr{\sc ii} are strong enough to constrain  $\Delta\aem/\aem$, 3 of which are observed by both Keck/HIRES and VLT/UVES providing 12 independent data point with $z\in[1.0–2.4]$. The 11 independent constraints are gathered in Table~\ref{tabSIDAM} and the a weighted mean $\Delta\aem/\aem =(1.2\pm1.7\pm0.9_{\rm syst})\times10^{-6}$ at 1$\sigma$.
 
\cite{NEW_Murphy:2017xaz} tackled the issue the long-range distortions of the wavelength calibration thanks to a supercalibration procedure at  the Subaru Telescope/HDS. The first line of sight towards J0120+2133 enjoys 6 absorption systems, only three of which give a constraint on $\aem$. Their average gives $\Delta\aem/\aem =(2.53\pm2.87\pm2.10_{\rm syst})\times10^{-6}$. 

From the analysis of O{\sc i}, Al{\sc ii}, Si{\sc ii} and Fe{\sc ii} with VLT/ESPRESSO in a damped Ly$\alpha$ system at $z_{\rm abs}=2.34$ towards  J0035-0918 \cite{NEW_10.1093/mnras/staa807}  got  $\Delta\aem/\aem =(-1.2\pm1.1)\times10^{-6}$.

The system at redshift $z_{\rm abs}=1.158$ towards the brightest Southern quasar HE514-4414 is probably the most studied absorber for measuring possible cosmological variations of $\aem$ since \cite{q-quast} at VLT with UVES \citep{Kotus:2016xxb}  and ESPRESSO \citep{NEW_Murphy:2021xhb} and HARPS at ESO 2.6m telescope \citep{NEW_Milakovic:2020tvq}. The latter analysis made 3 advances on (i) the calibration, using both laser frequency comb and ThAr methods, (ii) the modelization of the spectra with artificial intelligence techniques and (iii) the introduction of additional parameters to measure $\aem$ in each absorption component. They also dispatch their 47 measurements into 37 bins of $\delta z=10^{-4}$ to analyze the effects of the calibration. Their three measurements are summarized in Table~\ref{tabSIDAM}. Combining their ESPRESSO results with 28 measurements from other spectrographs in which wavelength calibration errors have been mitigated, allowed \cite{NEW_Murphy:2021xhb} to conclude  $\Delta\aem/\aem =(-0.5\pm0.5\pm0.4_{\rm syst})\times10^{-6}$ in the redshift band $0.6-2.4$. \cite{NEW_Lee:2022ppu} examined the impact of blinding procedures applied in the recent analysis of the same data thanks to supercomputer Monte Carlo AI calculations to generate a large number of independently constructed models of the absorption complex. They conclude that to avoid bias, all future measurements must include $\aem$ as a free parameter from the beginning of the modeling process.
 
 \cite{NEW_Wilczynska:2020rxx} made the  first measurements of $\aem$ with a near-IR spectrograph, VLT/X-SHOOTER. Among the 11 absorption systems identified towards J1120+0641, 4 allowed for a measurement of $\aem$ at redshift $z_{\rm abs} =7.059$ (C{\sc iv}, Si{\sc iv}, N{\sc v}), 6.171 (Al{\sc ii}, Si{\sc ii}, {\sc iv}, Fe{\sc ii}, Mg{\sc ii}), 5.951 (Fe{\sc ii}), Mg{\sc ii}), Si{\sc ii})), and 5.507 (Al{\sc ii}), Fe{\sc ii}), Mg{\sc ii}), Si{\sc ii})) among a sample with a total of 323 measurements spanning the redshift range 0.2 to 0.71 on which they applied the MM method  with $\aem$ as a free fitting parameter with IA techniques. Data are gathered in Table~\ref{tabSIDAM} and the weighted mean is and is $\Delta\aem/\aem= (-2.18\pm7.27)\times10^{-5}$.
 
\subsubsection{Summary of the observable constraints on $\aem$}\label{subsecalphadata}

The analysis of QSO spectra to infer constraints on the variation of $\aem$ is difficult. It has witnessed a lot of developments, in particular to understand all the systematics that can biais the analysis and lead to a spurtious non-zero detection. The early Keck/HIRES and VLT/UVES observations and the lively debate between their discordant conclusions led to the development of the Large Program at the VLT and the construction of a dedicated data set.

Among the important advances for $\aem$ measurements, we have mentioned -- see e.g., \cite{Levshakov:2016oua,NEW_Milakovic:2022gvb} for prospective discussions  -- (\textit{1}) the control of the calibration, and in particular the joint use of laser frequency comb \citep{steinmetz,Schmidt:2020ywz} and ThAr lamps, the former leading to a wavelength calibration residuals six times smaller than when using the standard ThAr calibration; (\textit{2}) the use of Artificial Intelligence and the use of genetic algorithm to develop automated pipelines \citep{NEW_10.1093/mnras/stx179,Lee:2021kjr,NEW_Webb:2023jnj} and (\textit{3}) the use of additional model parameters to measure $\aem$ in each individual absorption component in order to reduce the statistical uncertainty.

The search for a better resolution is being investigated in many directions. With the a resolution of $R\sim40~000$, the line wavelengths can be determined with an accuracy of $\sigma_\lambda\sim 1 \unit{m\AA}$. This implies an accuracy of the order of $10^{-5}$  on $\Delta\aem/\aem$  for lines with typical $q$-coefficients. This limit can indeed be pushed to $10^{-6}$ when more transitions or systems are used together. Any improvement is then related to the possibility to measure line positions with higher accuracy. This can be done by increasing $R$ up to the point at which the narrowest lines in the absorption systems are resolved. The Bohlin formula \citep{bohlin} gives the estimates
$$
 \sigma_\lambda\sim \Delta\lambda_{\text{pix}}
     \left(\frac{\Delta\lambda_{\text{pix}}}{W_{\text{obs}}}\right)
     \frac{1}{\sqrt{N_e}}\left(\frac{M^{3/2}}{\sqrt{12}}\right),
$$
where $\Delta\lambda_{\text{pix}}$ is the pixel size, $W_{\text{obs}}$ is the observed equivalent width, $N_e$ is the mean number of photoelectron at the continuum level and $M$ is the number of pixel covering the line profile.

The developments of high resolution and ultras-stable spectrographs, such as EXPRESSO (Echelle Spectrograph for PREcision Super Stable Observation)  \citep{Expresso,1750848,Pepe:2020ent,Schmidt:2020ywz} on 4 VLT units  are key developments with a resolution $R\sim 145~000$. Projects such as CODEX (COsmic Dynamics EXplorer) on E-ELT  \citep{CODEX,CODEX2,ubm}, ELET/HIRES \citep{Maiolino:2013bsa} or ANDES \citep{ANDES:2023cif} consider fundamental constants as both science and design drivers.  ANDES shall be able to compare the values of fundamental constants up to 12 Gyr~ago and 15~Gpc and provide a photon-limited uncertainty of about $3\times10^{-7}$, corresponding to 6~m/s relative line shifts, on their measurements.; see Fig.~3 of \cite{ANDES:2023cif} for synthetic spectra with $\aem$ variation. Besides, it will provide measurement of molecular H$_2$ allowing it to combine $\aem$ and $\mu$ measurements. At lower redshifts, ALMA may provide additional information \citep{Fish:2013dqb,Tilanus:2014iha}.

As a concrete example, ESPRESSO was specifically designed to suppress wavelength calibration errors and allows for cm/s (photon-limited) calibration precision; see \cite{NEW_Murphy:2022fac,NEW_Murphy:2021xhb} for a description of the progresses it led to. As an example, the analysis of the single system HE 0515-4414 \citep{NEW_Murphy:2022fac} reached an total uncertainty of 1.4 ppm, i.e., similar to the ensemble precision of the previous large samples of absorbers from HIRES and UVES that indicated variations at about the 5 ppm level \citep{webspace}. Hence one can safely conclude that it arose from long-range distortions in the wavelength scale, as discussed above. Recently, \cite{Schmidt:2024ttr}  improved the calibration accuracy to provide fully consistent measurements with a scatter of the order of a m/s ensuring that the instrument-related systematics can be nearly eliminated over most of the spectral range.

Thanks to the progresses to reach an increasing precision of the laboratory measurements of the rest frame frequencies of molecular transitions, and an increasing sensitivity and spectral resolution of astronomical observations, one may hope to reach constraints at the level of $10^{-8}$ for $\aem$ and $\mu$.

\subsection{Quasar absorption spectra: contraints on combination of $\aem$, $\mu$ and $g_{\rm p}$}\label{subsec33b}

\subsubsection[H\,{\sc i}-21~cm vs.\ UV: $x= g_{\mathrm{p}}\aem^2/\mu$]{H\,{\sc i}-21~cm vs.\ UV: \boldmath$x= g_{\mathrm{p}}\aem^2\mu$}\label{secQSOx}

The comparison of UV heavy element transitions with the hyperfine H\,{\sc i} transition allows to extract \citep{UV-1} a measurement of
$$
 x\equiv  g_{\mathrm{p}}\aem^2 \mu,
$$
since the hyperfine transition is proportional to $\aem^2 g_{\mathrm{p}}\mu^{-1} R_\infty$ while optical transitions are simply proportional to $R_\infty$. Hence constraints on the time variation of $x$ can be obtained from high resolution 21~cm spectra compared to UV lines, e.g., of Si\,{\sc ii}, Fe\,{\sc ii} and/or Mg\,{\sc ii}, as first performed in \cite{x-wolfe} in $z\sim0.524$ absorber.

Using 9 absorption systems, \cite{UV-2} showed no evidence for any variation of $x$ , $ \Delta x/x=(-0.63\pm0.99)\times10^{-5}$ for $0.23<z<2.35$. This constraint was criticized by \cite{x-kanekar} on the basis that the systems have multiple components and that it is not necessary that the strongest absorption arises in the same component in both types of lines. However, the error analysis of \cite{UV-2} tries to estimate the effect of this assumption. \cite{dent1} noticed that the systems lie in two widely-separated ranges and that the two samples have completely different scatter. Therefore it can be split into two samples of respectively 5 and 4 systems to get
\begin{eqnarray}
 &&\Delta x/x=(1.02\pm1.68)\times10^{-5},\qquad 0.23<z<0.53,\\
 &&\Delta x/x=(0.58\pm1.94)\times10^{-5},\qquad 1.7<z<2.35.
\end{eqnarray}

In such an approach two main difficulties arise: (1) the radio and optical source must coincide (in the optical, QSO can be considered pointlike and it must be checked that this is also the case for the radio source), (2) the clouds responsible for the 21~cm and UV absorptions must be localized in the same place. Therefore, the systems must be selected with care and today the number of such systems is small and are still actively looked for \citep{dlasearch}. 

The recent detection of 21~cm and molecular hydrogen absorption lines in the same damped Lyman-$\alpha$ system at $z_{\mathrm{abs}}=3.174$ towards SDSS J1337+3152 constrains \citep{x21cm} the variation $x$ to $\Delta x/x=-(1.7\pm1.7)\times10^{-6}$ at $z=3.174$. This system is unique since it allows for 21~cm, H$_{2}$ and UV observation so that in principle one can measure $\aem$, $x$ and $\mu$ independently. However, as the H$_{2}$ column density was low, only Werner band absorption lines are seen so that the range of sensitivity coefficients is too narrow to provide a stringent constraint, $\Delta\mu/\mu<4\times10^{-4}$. It was also shown that the H$_{2}$ and 21~cm are shifted because of the inhomogeneity of the gas, hence emphasizing this limitation. \cite{dlasearch} also mentioned that 4 systems at $z=1.3$ sets $\Delta x/x = (0.0\pm1.5)\times10^{-6}$ and that another system at $z=3.1$ gives  $\Delta x/x = (0.2\pm0.5)\times10^{-6}$. Note also that the comparison \citep{OH-3b} with C\,{\sc i} at $z \sim 1.4-1.6$ towards Q0458-020 and Q2337-011, yields $\Delta x/x = (6.8 \pm 1.0 \pm 6.7_{\rm syst}) \times 10^{-6}$ over the band o redshift $0 < \langle z\rangle \le 1.46$, but this analysis ignores an important wavelength calibration estimated to be of the order of $6.7\times10^{-6}$. It was argued that, using the existing constraints on $\Delta \mu/\mu$, this measurement is inconsistent with claims of a smaller value of $\aem$ from the many-multiplet method, unless fractional changes in $g_{\rm p}$ are larger than those in $\aem$ and $\mu$. 

From the Green Bank Telescope digital data, \cite{NEW_Darling:2012jf} selected 10 objects with H{\sc i} absorption lines with redshifts spanning from 0.09 to 0.69 with the primary goal to compare literature analog spectra to contemporary digital spectra and then measure the time drift of redshift, $\dot z$. Besides, by comparing UV metal absorption lines (from the data by \cite{UV-2}) to the 21 cm line from GBT allowed them to set constraints on a variation of $x$. Their seven data points are gathered in their Table~1 and summarized in Table~3 of \cite{NEW_Martins:2017yxk}. This leads to the average constraint
\begin{equation}
\Delta x/x =(-1.2\pm1.4)\times10^{-6}
\end{equation}
in the redshift range 0.24-2.04.

\cite{NEW_Rahmani:2012ze} analyzed high resolution optical spectra from VLT/UVES and and 21-cm absorption spectra from  Giant Metrewave Radio Telescope nd the Green Bank Telescope for 5 quasars with redshifts ranging from 1.17 to 1.56 and for which 21~cm has already been detected. The data of 4 out of 5 systems (towards J0501-0159, J1623+0718, J2340-0053 and J2358-1020) are used to constrain $x$ and were confirmed for 2 systems thanks to Keck/HIRES spectra. They constrain  the weighted and the simple means 
\begin{equation}
\Delta x/x =(-0.1\pm1.3)\times10^{-6},\quad\hbox{and}\quad 
\Delta x/x =(0.0\pm1.5)\times10^{-6}
\end{equation}
for $z\in[1.17,1.56]$ with mean redshift $\langle z\rangle=1.36$. We summarized the measurements for these 4 systems in Table~\ref{tab03}.

\subsubsection[H\,{\sc i} vs.\ molecular transitions: $y\equiv  g_{\mathrm{p}}\aem^2$]{H\,{\sc i} vs.\ molecular transitions:  \boldmath$y\equiv g_{\mathrm{p}}\aem^2$}\label{secQSOy}

The H\,{\sc i} 21~cm hyperfine transition frequency is proportional to $g_{\mathrm{p}}\mu^{-1}\aem^2R_\infty$ (see Sect.~\ref{subsec31-e.h}). On the other hand, the rotational transition frequencies of diatomic are inversely proportional to their reduced mass $M$. As on the example of Eq.~(\ref{mu1}) where we compared an electronic transition to a vibro-rotational transition, the comparison of the hyperfine and rotational frequencies is proportional to
$$
 \frac{\nu_{\mathrm{hf}}}{\nu_{\mathrm{rot}}} \propto
 g_{\mathrm{p}}\aem^2\frac{M}{m_{\mathrm{p}}} \simeq
 g_{\mathrm{p}}\aem^2 \equiv y,
$$
where the variation of $M/m_{\mathrm{p}}$ is usually suppressed by a large factor of the order of the ratio between the proton mass and nucleon binding energy in nuclei, so that we can safely neglect it. The constraint on the variation of $y$ is directly determined by comparing the redshift as determined from H\,{\sc i} and molecular absorption lines,
$$
\frac{\Delta y}{y} = \frac{z_{\text{mol}}-z_{\text{H}}}{1+z_{\text{mol}}}.
$$

This method was first applied \citep{y-var} to the CO molecular absorption lines \citep{y-combes} towards PKS~1413+135 to get $ \Delta y/y=(-4\pm6)\times10^{-5}$ at $z=0.247$.  The most recent constraint \citep{y-murph} relies on the comparison of the published redshifts of two absorption systems determined both from H\,{\sc i} and molecular absorption. The first is a system at $z=0.6847$ in the direction of B~0218+357 for which the spectra of CO(1-2), $^{13}$CO(1-2), C$^{18}$O(1-2), CO(2-3), HCO$^{+}$(1-2) and HCN(1-2) are available. They concluded that
\begin{equation}
 \Delta y/y=(-0.16\pm0.54)\times10^{-5}\qquad z=0.6847.
\end{equation}
The second system lies in direction of PKS~1413+135 for which the molecular lines of CO(1-2), HCO$^{+}$(1-2) and HCO$^{+}$(2-3) have been detected. Its analysis led to
\begin{equation}
 \Delta y/y=(-0.2\pm0.44)\times10^{-5},\qquad z=0.247.
\end{equation}
\cite{y-carilli} obtained the constraints $|\Delta y/y|<3.4\times10^{-5}$ at $z\sim 0.25$ and $z\sim0.685$.

The radio domain has the advantage of heterodyne techniques, with a spectral resolution of $10^{6}$ or more, and dealing with cold gas and narrow lines. The main systematics is the kinematical bias, i.e., that the different lines do not come exactly from the same material along the line of sight, with the same velocity. To improve this method one needs to find more sources, which may be possible with the radio telescope ALMA.\footnote{\url{http://www.eso.org/sci/facilities/alma/}}

\subsubsection[OH - 18~cm: $F=g_{\mathrm{p}}(\aem^2\mu)^{1.57}$]{OH -  18~cm: \boldmath$F=g_{\mathrm{p}}(\aem^2\mu)^{1.57}$} \label{secQSOF}

Using transitions originating from a single species, as with SIDAM, allows to reduce the systematic effects. The 18~cm lines of the OH radical offers such a possibility \citep{OH-1,OH-2}.

The ground state, ${}^2\Pi_{3/2} J=3/2$, of OH is split into two levels by $\Lambda$-doubling and each of these doubled level is further split into two hyperfine-structure states. Thus, it has two ``main'' lines ($\Delta F=0$) and two ``satellite'' lines ($\Delta F=1$). Since these four lines arise from two different physical processes ($\Lambda$-doubling and hyperfine splitting), they enjoy different dependencies in $g_{\mathrm{p}}$
and $\aem$. By comparing the four transitions to the H\,{\sc i} hyperfine line, one has access to
\begin{equation}
 F\equiv g_{\mathrm{p}}(\aem^2\mu)^{1.57}
\end{equation}
and it was also proposed to combine them with HCO$^{+}$ transitions to lift the degeneracy.

Using the four 18~cm OH lines from the gravitational lens at $z\sim0.765$ toward J0134-0931 and comparing the H\,{\sc i} 21~cm and OH absorption redshifts of the different components allowed \cite{OH-3} to set $\Delta F/F=(-0.86\pm0.86)\times10^{-5}$ at $z_{\rm abs}=0.765$. Combining with the constraint by \cite{OH-1} in the lens toward B0218+357,
\begin{equation}
 \Delta F/F=(0.35\pm0.40)\times10^{-5},\qquad z=0.685,
\end{equation}
led  o $\Delta F/F=(-0.44\pm0.36\pm1.0_{\text{syst}})\times10^{-5}$ \citep{OH-3} t at  $z_{\rm abs}=0.765$ where the second error is due to velocity offset between OH and H\,{\sc i} velocity dispersion of 3~km/s. Thanks to observations with  deep Green Bank Telescope spectroscopy, \cite{NEW_Kanekar:2012fy} concluded that
\begin{equation}
 \Delta F/F=(-5.2\pm4.3)\times10^{-6},\qquad  z=0.765.
\end{equation}
A similar analysis \citep{darling0} in a system in the direction of PKS1413+135 gave
\begin{equation}
 \Delta F/F=(0.51\pm1.26)\times10^{-5},\qquad z=0.2467.
\end{equation}
From the first detection of OH in a system in a star-forming galaxy at $z_{\rm abs}=0.0519$ toward QO248+430 and its combination with H{\sc i}~21cm, \cite{NEW_Gupta_2018} derived
\begin{equation}
 \Delta F/F=(5.2\pm4.5)\times10^{-6},\qquad  z=0.0519.
\end{equation}

\subsubsection[Far infrared fine-structure lines: $F'=\aem^2\mu$]{Far  infrared fine-structure lines: \boldmath$F'=\aem^2\mu$}\label{subsec358}

Another combination \citep{FIR-2} of constants is obtained from the comparison of far infrared fine-structure spectra with rotational transitions, which respectively behaves as $R_\infty\aem^2$ and $R_\infty/\mu$ so that they give access to
$$
F'=\aem^2\mu .
$$
A good candidate for the rotational lines is CO since it is the second most abundant molecule in the Universe after H$_{2}$.

Using the C\,{\sc ii} fine-structure and CO rotational emission lines from the quasars J1148+5251 and BR~1202-0725, \cite{FIR-1} concluded
that
\begin{eqnarray}
 &&\Delta F'/F'=(0.1\pm1.0)\times10^{-4},\qquad z=6.42,\\
 &&\Delta F'/F'=(1.4\pm1.5)\times10^{-5},\qquad z=4.69,
\end{eqnarray}
which represent the constraints at highest redshift. As usual, when comparing the frequencies of two different species, one must account for random Doppler shifts caused by non-identical spatial
distributions of the two species. Several other candidates for microwave and FIR lines with good sensitivities are discussed in \cite{kozrio}. \cite{NEW_Lentati:2012re}  used two sensitive observations toward these two systems. First, combining [C{\sc ii}] fine structure and CO(2-1) rotational transitions towards J1148+5251 taken with the Plateau de Bure Interferometer (PdBI) and Jansky Very Large Array (JVLA) respectively, led to
\begin{equation}
\Delta F'/F' =(-33\pm23)\times10^{-5} \qquad z=6.42\,.
\end{equation}
Then, combining [C{\sc ii}] fine structure and CO(2-1) transitions  from BR1202-0725 and its sub-millimeter companion galaxy at $z = 4.695$ taken with ALMA and the PdBI, they derived
\begin{equation}
\Delta F'/F' =(-5\pm 15)\times10^{-5},\qquad z=4.69\,.
\end{equation}

\cite{NEW_Curran:2011qu} fit the observed [C{\sc i}] and CO profiles of the redshifted systems known at the time in order to derive constraints on $F'$. They exhibit an anti-correlation between $\Delta F'$ and the  quality of the carbon detection and claim that ``current instruments are incapable of the sensitivities required to measure changes in the constants through the comparison of CO and carbon lines" before studying the case for ALMA. We shall not describe their analysis further and refer to their Table~1. Their results for their 8 systems  with absorption redshift between $2.285$ and $4.11$ are indeed included in our summary table~\ref{tab02}.

The quasar host galaxy RXJ0911.4+0551 at $z=2.796$ is considered very attractive for $F'$-measurements since it emits very strong and narrow CO(7-6) and [C~{\sc i}] lines. \cite{NEW_Weiss:2012cn} derived
\begin{equation}
 \Delta F'/F'=(6.9\pm3.7)\times10^{-6},\qquad z=2.796.
\end{equation}
\cite{NEW_Levshakov:2012kv} used unique observations of the CO(7-6) rotational line and the  [C~{\sc i}] a lensed galaxy  HLSJ091828.6+514223 at redshift $z=5.243$ to derive the upper bound
\begin{equation}
\vert\Delta F'/F' \vert< 2\times 10^{-5}.
\end{equation}
From IRAM/NOEMA spectra of J0439+1634 at $z=6.519$ that exhibit  four rotational transitions of CO(6-5), CO(9-8) and CO(10-9), \cite{NEW_Levshakov:2020ule}  used only the two first since the third is blended, together with [C{\sc i}] and [C{\sc ii}]  lines to get
\begin{equation}
\Delta F'/F' =(0.0\pm 2.7)\times10^{-5}.
\end{equation}
Similarly from new ALMA observations of the quasar J2310+1855 at $z=6.003$ let them conclude that
\begin{equation}
\Delta F'/F' =(2.3\pm 3.4)\times10^{-5}.
\end{equation}

\subsubsection[``Conjugate'' satellite OH lines:  $G=g_{\mathrm{p}}(\aem^2\mu)^{1.85}$]{``Conjugate'' satellite OH  lines: \boldmath$G=g_{\mathrm{p}}(\aem^2\mu)^{1.85}$}\label{secQSOG}

The satellite OH~18~cm lines are conjugate so that the two lines have the same shape, but with one line in emission and the other in absorption. This arises due to an inversion of the level of populations within the ground state of the OH molecule. This behavior has recently been discovered at cosmological distances  and it was shown \citep{OH-1} that a comparison between the sum and difference of satellite line redshifts probe
$$
G=g_{\mathrm{p}}(\aem^2\mu)^{1.85}.
$$

From the analysis of the two conjugate satellite OH systems at $z\sim0.247$ towards PKS~1413+135 and at $z\sim0.765$ towards PMN~J0134-0931, \cite{OH-1} concluded  that
\begin{equation}
 |\Delta G/G| < 7.6\times10^{-5} \qquad z\sim0.247.
\end{equation}
It was also applied to a nearby system, Centaurus~A, to give $ |\Delta G/G| < 1.6\times10^{-5}$ at $z\sim0.0018$. A more recent analysis \citep{OH-2b} claims for a tentative evidence (with 2.6$\sigma$ significance, or at 99.1\% confidence) for a smaller value of $G$
$$
 \Delta G/G = (-1.18\pm0.46)\times10^{-5} \qquad z\sim0.247
$$
for the system  towards PKS~1413+135. Further observation of the same system at the Arecibo telescope, in emission (1720 MHz) and absorption (1612 MHz) by \cite{NEW_Kanekar:2018mxs} led however to the constraint
\begin{equation}
 \Delta G/G=(0.97\pm1.52)\times10^{-6},\qquad z=0.2467.
\end{equation}
Combined with their former data from the Arecibo Telescope and the Westerbork Synthesis Radio Telescope \citep{OH-2b}, they concluded $\Delta G/G=(-1.0\pm1.3)\times10^{-6}$.

One strength of this method is that it guarantees that the satellite lines arise from the same gas, preventing from velocity offset between the lines. Also, the shape of the two lines must agree if they arise from the same gas.

\subsubsection{Summary of the constraints on $x,y,F,F'$ and $G$}\label{subsecxyFF}

The constraints discussed in Sect.~\ref{secQSOx} to~\ref{secQSOG} are summarized in Table~\ref{tab03} and Fig.~\ref{fig-xyFFG} for each of the combination of $g_{\rm p}$, $\aem$ and $\mu$ as a function of the redshift of the absorption system.

\begin{table*}
\begin{center}
\begin{tabular}{llccr }
\hline
 & Object & $z_{\rm abs}$ &  constraint & Reference \\ 
\toprule
$x$ 	& 3 sources &  0.23-2.35 &$(-0.63\pm0.99)\times10^{-5}$ & \cite{UV-2} \\
         & 5 sources & 0.23-0.53  & $(1.02\pm1.68)\times10^{-5} $ & \cite{dent1} \\
         & 4 sources &  1.7-2.35 & $(0.58\pm1.94)\times10^{-5} $ & \cite{dent1} \\
	& J1337+3152 & 3.174 & $(1.7\pm1.7)\times10^{-6}$ &\cite{x21cm} \\
         & 7 sources &  0.24- 2.04 & $(-1.2\pm1.4)\times10^{-6} $ & \cite{NEW_Darling:2012jf} \\	
         &J2358-1020& 1.173 & $(1.8\pm2.7)\times10^{-6}$ & \cite{NEW_Rahmani:2012ze} \\
	& J1623+0718&1.336& $(-3.7\pm3.4)\times10^{-6}$ & \cite{NEW_Rahmani:2012ze} \\
	&J2340-0053 &1.361& $(-1.3\pm2.0)\times10^{-6}$ & \cite{NEW_Rahmani:2012ze} \\
	& J0501-0159&1.5605 & $(3.0\pm3.1)\times10^{-6}$ & \cite{NEW_Rahmani:2012ze} \\
\midrule
$y$ & PKS1413+135 & 0.247 &$(-4\pm6)\times10^{-5}$ &\cite{y-combes} \\
&   & &$(-0.2\pm0.44)\times10^{-5}$  & \cite{y-murph}\\
& B0218+357& 0.6847&$(-0.16\pm0.54)\times10^{-5}$  &  \cite{y-murph}\\
\midrule
$F$ &     Q0248+430 & 0.0519 & $(5.2\pm4.5)\times10^{-6}$ & \cite{NEW_Gupta_2018}\\
 &PKS1413+135  & 0.247&$(-0.51\pm1.26)\times10^{-5}$  & \cite{darling0}\\
& B0218+357 & 0.685 & $(0.35\pm0.4)\times10^{-5}$ & \cite{OH-1}\\
 & J0134-0931 & 0.765 & $(-0.44\pm0.36\pm1_{\rm syst})\times10^{-5}$ &\cite{OH-3} \\
   & & &  $(-5.2\pm4.3)\times10^{-6}$ &\cite{NEW_Kanekar:2012fy}\\
\midrule
$F'$   & J1024+4709 & 2.285 & $(100\pm40)\times10^{-6}$ & \cite{NEW_Curran:2011qu}\\
   &J2135-0102  & 2.326 & $(100\pm100)\times10^{-6}$ & \cite{NEW_Curran:2011qu}\\ 
    & J636+6612 & 2.517 & $(-100\pm120)\times10^{-6}$ & \cite{NEW_Curran:2011qu}\\
     & H1413+117 & 2.558 & $(-40\pm80)\times10^{-6}$ & \cite{NEW_Curran:2011qu}\\
        & J1401+0252 &2.565  & $(-140\pm80)\times10^{-6}$ & \cite{NEW_Curran:2011qu}\\
          &J0911+0551 &2.796 & $(6.9\pm3.7)\times10^{-6}$& \cite{NEW_Weiss:2012cn} \\
   & APM0828+5255 & 3.913  & $(-360\pm90)\times10^{-6}$ & \cite{NEW_Curran:2011qu}\\
      & MM1842+5938&3.930  & $(-180\pm40)\times10^{-6}$ & \cite{NEW_Curran:2011qu}\\
         & PSS2322+1944&4.112  & $(170\pm130)\times10^{-6}$ & \cite{NEW_Curran:2011qu}\\                   
  &BR1202-0725 & 4.695& $(1.4\pm1.5)\times10^{-5}$& \cite{FIR-1} \\
 & &  & $(-5\pm 15)\times10^{-5}$ & \cite{NEW_Lentati:2012re}\\
 &J0918+5142 & 5.243 & $ <2\times 10^{-5}$ & \cite{NEW_Levshakov:2012kv} \\
  &J2310+1855 & 6.003 & $(2.3\pm 3.4)\times10^{-5}$  & \cite{NEW_Levshakov:2020ule} \\
   &J0439+1634&6.519 & $(0.0\pm 2.7)\times10^{-5}$  & \cite{NEW_Levshakov:2020ule} \\
& J1148+5251& 6.420& $(1\pm10)\times10^{-5}$&\cite{FIR-1} \\
 & &  &  $(-33\pm23)\times10^{-5}$ & \cite{NEW_Lentati:2012re} \\
\midrule
$G$& PKS1413+135& 0.247& $(-1.18\pm0.46)\times10^{-5}$& \cite{OH-2b}\\
&&&$(0.97\pm1.52)\times10^{-6}$&\cite{NEW_Kanekar:2018mxs}\\
&&&$(-1.0\pm1.3)\times10^{-6}$&combined\\
\bottomrule
\end{tabular}
\caption[Absorption spectra constraints on $x,y,F,F'$ and $G$]{\label{tab03} Constraints on the combination of $g_{\rm p}$, $\aem$ and $\mu=m_{\rm p}/m_{\rm e}=1/\bar\mu$. The object determines the line of sight and the redshift refers to the absorption system. We recall that $x\equiv\aem^2  g_{\mathrm{p}}/\mu$, $y\equiv g_{\mathrm{p}}\aem^2$,  $F\equiv g_{\mathrm{p}}(\aem^2\mu)^{1.57}$, $F'\equiv\aem^2\mu$ and  $G=g_{\mathrm{p}}(\aem^2\mu)^{1.85}$. The observations are ordered by increasing redshift and then by the date of the constraint for each system.}
\end{center}
\end{table*}

\begin{figure}[hptb]
 \centerline{\includegraphics[scale=0.5]{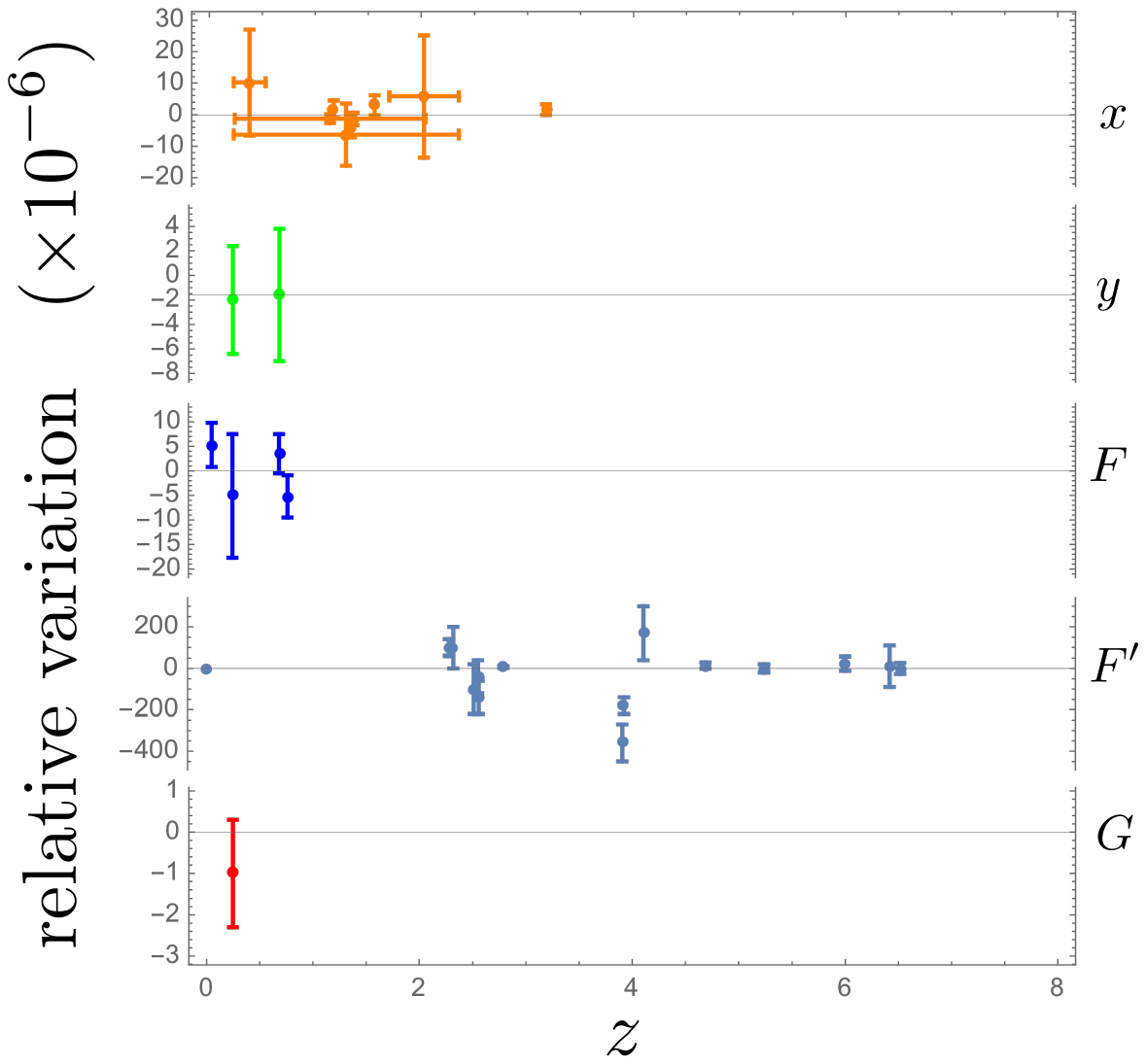}}
 \vskip-0.5cm
  \caption[Absorption spectra constraints on $x,y,F,F'$ and $G$]{Constraints on the variation of -- from the top to bottom panels --  $x\equiv\aem^2  g_{\mathrm{p}}/\mu$, $y\equiv g_{\mathrm{p}}\aem^2$,  $F\equiv g_{\mathrm{p}}(\aem^2\mu)^{1.57}$, $F'\equiv\aem^2\mu$ and  $G=g_{\mathrm{p}}(\aem^2\mu)^{1.85}$ as function of the redshift of the absorption system, as summarized in Table~\ref{tab03}.}
  \label{fig-xyFFG}
\end{figure}

\subsection{Molecular absoprtion spectra and the electron-to-proton mass ratio}\label{secmolqso}

\subsubsection{Introduction}\label{subsecmuqsointro}

As was pointed out in Sect.~\ref{subsec31}, molecular lines provide a test of the variation\footnote{Again, $\mu$ is used either from $m_{\mathrm{e}}/m_{\mathrm{p}}$ or $m_{\mathrm{p}}/m_{\mathrm{e}}$ in the literature. In this review, it has been chosen to use  $\mu=m_{\mathrm{p}}/m_{\mathrm{e}}$ and  $\bar\mu=m_{\mathrm{e}}/m_{\mathrm{p}}$.} \citep{mu-theorie} of $\mu$ since rotational and vibrational transitions are respectively inversely proportional to their reduce mass and its square-root [see  Eq.~(\ref{mu1})].  

While many molecules are observed in our local environment, only 19 diatomic molecules have been observed extragalactically and about 125 polyatomic molecules have been detected in the interstellar medium \citep{NEW_2018ApJS..239...17M}. There are two main issues to select the astrophysical molecules to be used: (\textit{1}) their sensitivity to a variation of $\mu$ and (\textit{2}) how abundant and ``easy'' to observe them and (\textit{3}) the intensity of the transitions. This depends on both the intrinsic strength of the transition (i.e., its Einstein A coefficient) and the population of the lower energy states, i.e., of the temperature of the system. Unfortunately intergalactic absorption systems are usually low temperature. Moreover molecules with different sensitivities, with different signs, allow for a better control of the systematic effects and (\textit{4}) the availability of telescope to measure those molecular spectral lines (while observing in the optical and radio band is quite easy from Earth, IR and far-IR band are important to study nearby systems and require space observations).

For almost 3 decades, vibrorotational lines of H$_2$ were the sole method to access $\mu$ but H$_2$ absorbers were difficult to detect even though it is the most abundant molecule in the universe. Today about 10 absorption systems have been extensively studied with both H$_2$ and HD transitions \citep{NEW_Ubachs:2017zmg}. However so far one has been unable to observe H$_2$ directly in systems below redshift 2. It is now complemented by constraints arising from amonia (2 systems), methanol (Milky Way+1 system), carbon monoxyde (2 systems) and preliminary results from oxygenated water and methylamine.

Intensive works have been performed to compute the sensitivities of the required transitions and to compile data on molecular spectra with the highest accuracy. We refer e.g., to \cite{NEW_Ubachs:2015fro,NEW_Ubachs:2017zmg,NEW_Ubachs:2018wan} for H$_2$ and HD and to \cite{NEW_Kozlov:2013lha} for the computation of the sensitivities of CH, OH, NH$^+$, NH$_3$, ND$_3$, NH$_2$D, NHD$_2$, H$_2$O$_2$, H$_3$O$^+$, CH$_3$OH, and CH$_3$NH$_2$,  \cite{NEW_Yurchenko:2013nym} for CH$_4$. \cite{NEW_Syme:2020rde,NEW_Syme:2020roz} compiled the sensitivities of all diatomic molecules that have been observed astrophysically.

From a theoretical point of view, constraints from molecules are complementary to those from atoms since molecules are usually found in denser environment. Hence comparison of $\mu$ constraints from atoms and molecules allow one  to question the chameleon mechanism (see Sect.~\ref{subsub2}) and independent constraints on $\mu$ and $\aem$ allows one to test unification scenarios; see Sect.~\ref{secRS}.

\subsubsection{Constraints with H$_{2}$}

H$_{2}$ is the most abundant molecule in the universe and there were many attempts to use it to set constraints on the variation of $\mu$ despite the fact that H$_{2}$ is very difficult to detect \citep{noterdame}. As proposed in \cite{mu-vl}, the sensitivity of a vibro-rotational wavelength to a variation of $\mu$ can be parameterized as
$$
 \lambda_i =  \lambda_i^0(1+z_{\mathrm{abs}})\left(1+ K_i\frac{\Delta\mu}{\mu}\right),
$$
where $ \lambda_i^0$ is the laboratory wavelength (in the vacuum) and $ \lambda_i$ is the wavelength of the transition $i$ in the rest-frame of the cloud, that is at a redshift $z_{\mathrm{abs}}$ so that the observed wavelength is $ \lambda_i/(1+z_{\mathrm{abs}})$. $K_i$ is a sensitivity coefficient analogous to the $q$-coefficient introduced in Eq.~(\ref{qpara}), but with different normalization since in the parameterisation we would have $q_i=\omega_i^0K_i/2$,
$$
 K_i \equiv \frac{\dd\ln \lambda_i}{\dd\ln\mu}
$$
corresponding to the Lyman and Werner bands of molecular hydrogen. From this expression, one deduces that the observed redshift measured from the transition $i$ is simply
$$
 z_i = z_{\mathrm{abs}} + bK_i,\qquad b\equiv -(1+z_{\mathrm{abs}})\frac{\Delta\mu}{\mu},
$$
which implies in particular that $z_{\mathrm{abs}}$ is not the mean of the $z_i$ if $\Delta\mu\not=0$ . Indeed $z_i$ is measured with some uncertainty of the astronomical measurements $ \lambda_i$ and by errors of the laboratory measurements $ \lambda_i^0$. But if $\Delta\mu\not=0$ there must exist a correlation between $z_i$ and $K_i$ so that a linear regression of $z_i$ (measurement) as a function of $K_i$ (computed) allows to extract $(z_{\mathrm{abs}},b)$ and their statistical significance. We refer to Sect.~V.C of FVC03 \citep{jpu-revue} for earlier studies.

A first analysis of VLT/UVES of the molecular hydrogen of two damped Lyman-$\alpha$ absorption systems at $z=2.3377$ and $3.0249$ in the direction of two quasars (Q1232+082 and Q0347-383)  showed \citep{mu-ivan1} a slight indication of a variation, $\Delta\mu/\mu=(5.7\pm3.8)\times10^{-5}$ at 1.5$\sigma$ for the combined analysis. The lines were selected so that they are isolated, unsaturated and unblended (using only 12 lines over 50 detected for the first quasar and 18 over 80 for the second) and the two selected spectra had no transition in common. The authors performed their analysis with two laboratory catalogs and got different results, pointing out that the errors on the laboratory wavelengths are comparable to those of the astronomical measurements.

It was further improved with an analysis of two absorption systems at $z=2.5947$ and $z=3.0249$ in the directions  of Q0405-443 and Q~0347-383 observed with the VLT/UVES spectrograph. The same selection criteria where applied, letting respectively 39 (out of 40) and 37 (out of 42) lines for each spectrum and 7 transitions in common. The combined analysis of the two systems led \citep{mu-ivan2} to $\Delta\mu/\mu=(1.65\pm0.74)\times10^{-5}$ or $\Delta\mu/\mu=(3.05\pm0.75)\times10^{-5}$, according to the laboratory measurements that were used. The same data were reanalyzed with new and highly accurate measurements of the Lyman bands of H$_{2}$, which implied a reevaluation of the sensitivity coefficient $K_i$. It leads to the two constraints \citep{mu-rein}
\begin{eqnarray}
 &&\Delta\mu/\mu = (2.78\pm0.88)\times10^{-5},\qquad z= 2.59, \quad[\hbox{Q00405-443}]\\
 &&\Delta\mu/\mu = (2.06\pm0.79)\times10^{-5},\qquad z = 3.02,\quad[\hbox{Q0347-383}]
 \label{muxx0}
\end{eqnarray}
leading to a 3.5$\sigma$ detection for the weighted mean $\Delta\mu/\mu = (2.4\pm0.66)\times10^{-5}$. \cite{mu-rein} did not claim for a detection and are cautious enough to state that systematics dominate the measurements. \cite{mu-king} employed a comprehensive fitting method for these 2 quasars to which they added a new system at $z_{\rm abs}=2.81$ in direction of Q0528-250, to get
\begin{eqnarray}
 &&\Delta\mu/\mu = (10.1\pm6.2)\times10^{-6},\qquad z= 2.59, \quad[\hbox{Q00405-443}]\\
 &&\Delta\mu/\mu = (-1.4\pm3.9)\times10^{-6},\qquad \!\! z= 2.80, \quad[\hbox{Q0528-250}]\\
 &&\Delta\mu/\mu = (5.2\pm7.4)\times10^{-6},\qquad\,\,\, z = 3.02, \quad[\hbox{Q0347-383}]
\end{eqnarray}
with a weighted mean of $(2.6\pm3.0)\times10^{-6}$ at $z\sim2.81$. Concerning Q0347-383, \cite{mu-thom} and \cite{mu-thom2} reanalyzed the data using an advanced line-by-line method to get
\begin{equation}
 \Delta\mu/\mu = (-28\pm16)\times10^{-6}. \qquad z= 3.02,\quad [\hbox{Q0347-383}].
\end{equation}
New observations led to the bound  \citep{mu-wendt} $|\Delta\mu/\mu|<4.9\times10^{-5}$ at a 2$\sigma$, thus contradicting~(\ref{muxx0}). Then   \cite{NEW_Wendt:2010qe} derived $\Delta\mu/\mu = (15\pm9\pm6_{\text{syst}})\times 10^{-6}$ at $z= 3.02$ towards Q0347-383 while \cite{NEW_Wendt:2012ea} improved to
\begin{equation}
 \Delta\mu/\mu = (4.3\pm7.2)\times10^{-6}. \qquad z= 3.02, \quad  [\hbox{Q0347-383}],
\end{equation}
leading to a weighted mean of the four most accurate results of $\Delta\mu/\mu = (5.1\pm4.5)\times10^{-6}$ \citep{NEW_Ubachs:2015fro}.

Concerning Q0528-250, renewed VLT observations \citep{NEW_King:2011km} concluded from the analysis of Using 76 H$_2$ and 7 HD transitions
\begin{equation}
\Delta\mu/\mu = (0.3\pm3.2\pm1.9_{\text{syst}})\times 10^{-6} \qquad z= 2.80, \quad [\hbox{Q0528-250}].
\end{equation}
so that an average value is $\Delta\mu/\mu = (-0.5\pm2.7)\times10^{-6}$ \citep{NEW_Ubachs:2015fro}.

Concerning Q1232+082, from the analysis of VLT/UVES observations of 106 H$_2$ and HD lines with a comprehensive fitting technique, \cite{NEW_Dapra:2016dqh} concluded that
\begin{equation}
\Delta\mu/\mu = (19\pm9\pm5_{\text{syst}})\times 10^{-6} \qquad z= 2.34, \quad[\hbox{Q1232+082}].
\end{equation}

A line-by-line analysis of Q00405-443 \citep{mu-thom} gave
\begin{equation}
 \Delta\mu/\mu = (0.6\pm10)\times10^{-6}. \qquad z= 2.59, \quad [\hbox{Q00405-443}],
\end{equation}
leading to an average value of $\Delta\mu/\mu = (7.5\pm5.3)\times10^{-6}$ \citep{NEW_Ubachs:2015fro}.

The molecular system at $z=2.059$ toward J2123-0050 was observed by the HIRES spectrometer at the Keck telescope. It exhibited  86 H$_{2}$ transitions and 7 HD transitions.  The analysis by \citep{HD-1} concluded that
\begin{equation}
 \Delta\mu/\mu= (5.6\pm5.5\pm2.7_{\text{syst}})\times 10^{-6},\qquad z=2.059, \quad [\hbox{J2123-0050}].
\end{equation}
An independent spectrum was obtained with the VLT/UVES for 90 H$_2$ lines and 6 HD lines from which \cite{NEW_vanWeerdenburg:2011ru} derived
\begin{equation}
 \Delta\mu/\mu= (8.5\pm3.6\pm2.2_{\text{syst}})\times 10^{-6},\qquad z=2.059, \quad  [\hbox{J2123-0050}].
\end{equation}
Averaging over these two independent results \citep{NEW_Ubachs:2015fro} lead to $\Delta\mu/\mu= (7.6\pm3.5)\times 10^{-6}$ for J2123-0050.

\cite{NEW_Rahmani:2013dia} observed a system at $z_{\rm abs}=2.4018$ towards HE0027-1836 with VLT over 3 years to obtain
\begin{equation}
 \Delta\mu/\mu= (-7.6\pm10.2)\times 10^{-6},\qquad z=2.4018, \quad[\hbox{HE0027-1836}].
\end{equation}
The system at $z_{\rm abs}=2.659$ in the direction of J0643-5041 exhibiting a single H$_2$ absorption feature, was analyzed in a line-by-line analysis by \cite{NEW_AlbornozVasquez:2013wfw,NEW_Rahmani:2014opa}  from high-resolution VLT-UVES data from more than 23 hours exposure, to get $\Delta\mu/\mu= (7.4\pm4.3\pm5.1_{\text{syst}})\times 10^{-6}$. Reanalyzed by \cite{NEW_Bagdonaite:2013eia} it gave $\Delta\mu/\mu= (12.7\pm4.5\pm4.2_{\text{syst}})\times 10^{-6}$ and  \cite{NEW_Ubachs:2015fro} adopted the mean value
 \begin{equation}
 \Delta\mu/\mu= (10.3\pm4.6)\times 10^{-6},\qquad z=2.659 \quad[\hbox{J0643-5041}],
\end{equation}
which corresponds to $(7.4\pm6.7)\times 10^{-6}$ after combining the uncertainties.

The absorption system at $z=2.426$ in direction of Q2348-011  has a complex velocity structure featuring at least 7 H$_2$ absorption sub-systems. Analyzed by \cite{NEW_Bagdonaite:2011ab} it led to
 \begin{equation}
 \Delta\mu/\mu= (-6.8\pm27.8)\times 10^{-6},\qquad  z=2.426, \quad  [\hbox{Q2348-011}].
\end{equation}

The analysis of an absorption system at $z_{\rm abs}=2.69$ towards J1237+064 which exhibits 3 clouds with 137  lines of H$_2$ and HD was analyzed by \cite{NEW_Dapra:2015yva} to give
 \begin{equation}
 \Delta\mu/\mu= (-5.4\pm6.3\pm4.0_{\rm syst})\times 10^{-6},\qquad z=2.69, \quad [\hbox{J1237+064}]
\end{equation}
and \cite{NEW_Ubachs:2015fro} adopt $(-5.4\pm7.2)\times 10^{-6}$ after combining the uncertainties.

The high redshift system at $z_{\rm abs}=4.224$ in the direction of J1443+2724 was observed by VLT/UVES. Combining archival data \citep{NEW_Ledoux:2006mi} with 2013-data by \cite{NEW_Bagdonaite:2015kga} gave
 \begin{equation}
 \Delta\mu/\mu= (-9.5\pm5.4\pm5.3_{\rm syst})\times 10^{-6},\qquad z=4.224 \quad[\hbox{J1443+2724}]
\end{equation}
which corresponds to $(-9.5\pm7.5)\times 10^{-6}$ after combining the uncertainties.

Note that the detection of several deuterated molecular hydrogen HD transitions makes it possible to test the variation of $\mu$ in the same way as with H$_{2}$ but in a completely independent way, even though today it has been detected only in 2 places in the universe. The sensitivity coefficients have been published in \cite{mu-ivanov} and HD was first detected by \cite{noterdame}. HD was recently detected \citep{HDdetect} together with CO and H$_{2}$ in a DLA cloud at a redshift of 2.418 toward SDSS1439+11 with 5 lines of HD in 3 components together with several H$_{2}$ lines in 7 components. It allowed to set the 3$\sigma$ limit of $|\Delta\mu/\mu|<9\times10^{-5}$ \citep{dlmu}.
Even though the small number of lines does not allow to reach the level of accuracy of H$_{2}$ it is a very promising system in particular to obtain independent measurements.

This method is subject to important systematic errors among which (\textit{1}) the sensitivity to the laboratory wavelengths -- since the use of two different catalogs yields different results \citep{mu-rein} -- (\textit{2}) the molecular lines are located in the Lyman-$\alpha$ forest where they can be strongly blended with intervening H\,{\sc i} Lyman-$\alpha$ absorption lines, which requires a careful fitting \citep{mu-king} since it is hard to find lines that are not contaminated. From an observational point of view, very few damped Lyman-$\alpha$ systems have a measurable amount of H$_{2}$ so that only a dozen systems is actually known even though more systems will hopefully be obtained soon \citep{dlasearch}. 

To finish, the sensitivity coefficients are usually low, typically of the order of $10^{-2}$. Some advantages of using H$_{2}$ arise from the fact there are several hundred available H$_{2}$ lines so that many lines from the same ground state can be used to eliminate different kinematics between regions of different excitation temperatures. The overlap between Lyman and Werner bands also allows one to reduce the errors of calibration.

\subsubsection{Amonia observations}

Amonia is one of the mot abundant polyatomic molecules observed in the interstellar medium. Unfortunately only two extragalactic sources, B0218+357 and  PKS1830-211, are known and both are in low redshift galaxies. Following the idea by \cite{NEW_ND3} to use inversion transition in ND$_3$ to probe a variation of $\mu$,  \cite{mu-N1,mu-N2} adapted this technique to compute inversion spectrum of NH$_3$ and show that this allows for a better sensitivity to $\mu$ (see \cite{NEW_Owens:2016xml} for extensive computation of the sensitivity coefficients). The inversion vibro-rotational mode is described by a double well with the first two levels below the barrier. The tunneling implies
that these two levels are split in inversion doublets. It was concluded that the inversion transitions scale as
$\nu_{\mathrm{inv}}\sim \bar\mu^{4.46}$, compared with a rotational
transition, which scales as $\nu_{\mathrm{rot}}\sim \bar\mu$. This
implies that the redshifts determined by the two types of transitions
are modified according to $\delta
z_{\mathrm{inv}}=4.46(1+z_{\mathrm{abs}})\Delta\mu/\mu$ and $\delta
z_{\mathrm{rot}}\sim(1+z_{\mathrm{abs}})\Delta\mu/\mu$ so that
$$
 \Delta\mu/\mu = 0.289\frac{z_{\mathrm{inv}} - z_{\mathrm{rot}}}{1+z_{\mathrm{abs}}}.
$$

For the first quasar absorption system displaying NH$_{3}$ at $z=0.68466$ in the direction of B0218+357, \cite{mu-N1} estimated from the published redshift uncertainties that a precision of $\sim2\times10^{-6}$ on $\Delta\mu/\mu$ can be achieved. A detailed measurement \citep{mu-N3} of the ammonia inversion transitions by comparison to HCN and HCO$^{+}$ rotational transitions concluded that
\begin{equation}\label{muweb}
 \Delta\mu/\mu =(0.74\pm.0.47\pm0.76_{\rm syst})\times10^{-6}, \qquad z=0.685,  \quad \hbox{[B0218+357]}
\end{equation}
which corresponds to the 2$\sigma$ bound on the variation of $\mu$, $|\Delta\mu/\mu| < 1.8\times10^{-6}$.  Combining inversion (NH$_3$) and rotational (CS, H$_2$CO) absorption lines detected with the Green Bank Telescope allowed \cite{NEW_Kanekar:2011yb}
to conclude that
\begin{equation}
 \Delta\mu/\mu =(-0.35\pm 0.12)\times10^{-6}, \qquad z=0.685, \quad \hbox{[B0218+357]}\,.
\end{equation}

For the second known system, the analysis of the comparison of NH$_{3}$ to HC$_{3}$N spectra was performed toward the gravitational lens system PKS~1830-211 ($z\simeq0.89$), which is a much more suitable system, with 10 detected NH$_{3}$ inversion lines and a forest of rotational transitions. \cite{mu-henkel} reached the conclusion that
\begin{equation}
\label{muhenkel}
 |\Delta\mu/\mu| < 1.4\times10^{-6}, \qquad z=0.89,  \quad \hbox{[PKS~1830-211]}
\end{equation}
at a 3$\sigma$ level. From a comparison of the ammonia inversion lines with the NH$_{3}$ rotational transitions,  \cite{menten} concluded
\begin{equation}
\label{mumentel}
 |\Delta\mu/\mu| < 3.8\times10^{-6}, \qquad z=0.89, \quad \hbox{[PKS~1830-211]}
\end{equation}
at 95\% C.L. Recent works \citep{Owens:2015hla,Owens:2016xml} took into account that the rotational energy levels of amonia  exhibit a non-negligible centrifugal distortion dependence, which was disregarded. Hence the single sensitivity $K=4.46$ shall be replaced with ensitivities ranging from 4.3 to 4.9. From the  spectroscopic measurements by \cite{Henkel:2008hy}, the reanalysis of this system \citep{NEW_Bagdonaite:2015wju} led to $\Delta\mu/\mu = (1.1\times 7.0)\times10^{-6}$.

One strength of this analysis is to focus on lines arising from only one molecular species but it was mentioned that the frequencies of the inversion lines are about 25 times lower than the rotational ones, which might cause differences in the absorbed background radio continuum. From a survey with the Australia Telescope Compact Array \cite{NEW_Muller:2011yu} detected 28 different molecules toward the south-west absorption region, located about 2 kpc from the center, making this system  the one with the largest number of detected molecular species of any extragalactic object at the time of the work. The analysis of the inversion lines of amonia led to
\begin{equation}
 \Delta\mu/\mu= (-2.04\pm0.74)\times10^{-6}, \qquad z=0.89,\quad \hbox{[PKS~1830-211]}
\end{equation}
assuming no time variation between the different observations.

\subsubsection{Methanol observations}

\cite{NEW_Levshakov:2011su} identified methanol as a tracer of $\mu$. They pointed out that its spectrum  includes purely rotational transitions as well as transitions with contributions of the internal degrees of freedom associated with the hindered rotation of OH and that the latter is highly sensitive to $\mu$. Sensitivity coefficients are calculated and range between $-88$ and 330.  Note that there 2 types of methanol: in E-type, one of the protons in the hydrogen atoms of the CH$_3$ group has an antiparallel nuclear spin with respect to the others while in A-type methanol, the nuclear spins of the three protons in the CH$_3$  methyl group are parallel. The two methanol types have different transition frequencies and may arise in different physical environments and have both been observed in molecular clouds. Recently, \cite{NEW_Agafonova:2024xpe} presented the simultaneous observations of Class~I methanol masers at 25, 36, and 44 GHz towards 22 Galactic targets thanks to the Effelsberg 100-m telescope.

This led \cite{NEW_Jansen:2011zza} to conclude that the 6.7 and 12.2 GHz masers transitions in methanol, corresponding to the transitions $5_1\rightarrow6_0$A$^+$ and $2_0\rightarrow 3_{-1}$, are among the brightest radio objects in the sky and enjoy a sensitivity of $K_\mu=-42$ and $K_\mu=-33$ respectively, making methanol an excellent candidate to constrain the variation of fundamental constants. The first detection of methanol at cosmological distance \citep{NEW_Muller:2011yu} in the molecular system located in the disk of a spiral galaxy at $z_{\rm abs}=0.8859$ in direction of the quasar PKS1830-211.  They concluded that
\begin{equation}
\Delta\mu/\mu =(-1.95\pm0.47)\times10^{-6}\qquad z=0.8859  \quad\hbox{[PKS1830-211]}.
\end{equation}
While a $4\sigma$ detection in agreement with their NH3 results,  \cite{NEW_Muller:2011yu} considered their results as ``tentative". Taking into account the velocity dispersion of a large number of molecular species, they concluded  $\vert\Delta\mu/\mu\vert<4\times10^{-6}$. The first constraint based only on methanol was obtained by \cite{NEW_Ellingsen:2012cd}  from the observation  of the former transitions in the same system to get
\begin{equation}
\Delta\mu/\mu =(0.8\pm 2.1)\times10^{-7} \qquad z=0.8859  \quad\hbox{[PKS1830-211]}
\end{equation}
which corresponds to a $3\sigma$ upper bound of $6.3\times10^{-7}$. \cite{NEW_Bogdanaite} combined the 12.2~GHz line with nine other transitions and performed a detailed analysis to the error budget to conclude that
\begin{equation}
\Delta\mu/\mu=(0.0\pm1.0)\times10^{-7} \qquad z=0.8859   \quad\hbox{[PKS1830-211]}.
\end{equation}
After a study of the systematic effects of chemical segregation, excitation temperature, frequency dependence and time variability of the background source are quantified, \cite{NEW_Bagdonaite:2013sia} concluded that
\begin{equation}
\Delta\mu/\mu=(1.5\pm1.5)\times10^{-7} \qquad z=0.8859   \quad\hbox{[PKS1830-211]}.
\end{equation}
These studies did not test whether the different lines arise from the same region of the gas. \cite{NEW_Kanekar:2014ota} actually concluded from their VLA spectrum that the 12.2~GHz line has different properties than the 3 other transitions they consider ($0_0\rightarrow 1_0$E at 48.382, $0_{0}\rightarrow 1_0$A$^+$ at 48.377 and $2_{-1}\rightarrow 1_0$E at 60.531~GHz) and shall not be included in the analysis since it introduces a biais. From these 3 lines they deduced that 
\begin{equation}
\vert\Delta\mu/\mu\vert< 4\times10^{-7}
\end{equation}
at $2\sigma$ and that including the 12.2~GHz line would biais the result to $\Delta\mu/\mu=(-2.9\pm5.7)\times10^{-8}$. Indeed,  for gravitational lensed systems, additional systematics have to be considered since, even though lensing is achromatic, the structure of the background quasar being lensed does vary both with frequency and time.  Thanks to observations up to a few milliarcseconds  \cite{NEW_Marshall:2016tke} resolved  the northeastern and southwestern as well as the Einstein ring (The quasar jet has a steep spectral index and in PKS B1830-211 it is this which causes an Einstein ring to be present only at lower frequencies). \cite{NEW_Muller:2021ttk} confirmed that methanol is detected in only one, the southwest, of the two lines of sight. The observation of 14 transitions of methanol, five of the A-form and nine of the E-form with ALMA allowed them to set the 1$\sigma$ constraint
\begin{equation}
\Delta\mu/\mu=(-1.8\pm1.2)\times10^{-7} \qquad z=0.8859   \quad\hbox{[PKS1830-211]}\,,
\end{equation}
corresponding to at $\vert\Delta\mu/\mu\vert< 3.6\times10^{-7}$ at $3\sigma$. 

In conclusion, methanol offers the advantage of a test based on a single species, avoiding  the location biais of multi-species techniques. But for now it has been observed in a single absorption system that has been shown to be subject to many disturbing phenomena such as the time variability of the background blazar and the existence of chromatic substructures.

\subsubsection{CO observations}

CO is the second most abundant molecule in gas form in the universe. From the study of transitions from electronic A-X, \cite{NEW_Salumbides:2012qb} suggested to use the electronic A$^1\Pi-$X$^1\Sigma^+$ transition in CO to supplement H$_2$ even though their sensitivities are low \citep{NEW_Dapra:2016jyh} and thus expected to set less stringent constraints.

A combined analysis of CO and H$_2$ in a system at $z_{\rm abs}=2.69$ towards J1237+0647 observed with the VLT/UVES. Thirteen CO vibrational bands in this absorber were detected  set
\begin{equation}
\Delta\mu/\mu =(0.7\pm1.6\pm0.5_{\rm syst})\times10^{-5}\qquad\hbox{[J1237+0647]}
\end{equation}
from the analysis of CO only while it reaches
\begin{equation}
\Delta\mu/\mu =(-5.6\pm5.6\pm3.1_{\rm syst})\times10^{-6}\qquad\hbox{[J1237+0647]}
\end{equation}
once combined with H$_2$ in the same system

From the observations of 8 vibrational singlet-singlet bands and 1 singlet-triplet band of carbon monoxide in the spectrum of  a damped Lyman-$\alpha$ system at $z =2.52$ towards quasar SDSS J000015.16+004833.2 with VLT/UVES and correction for long-range wavelength scale distortions using the supercalibration technique, \cite{NEW_Dapra:2017vlu} concluded that
\begin{equation}
\Delta\mu/\mu =(1.8\pm2.2\pm0.4_{\rm syst})\times10^{-5}\qquad\hbox{[J0000+0048]}\,.
\end{equation}

So far, only two systems have been studied while  optical absorption bands of CO have been detected in 6  absorption systems at redshifts larger than 1,  SDSS J16045+2203 \citep{NEW_Noterdaeme:2008ii}, SDSS J0857+1855, SDSS J1047+2057, SDSS J1705+3543 \citep{NEW_Noterdaeme:2010tm}, SDSS J1439+1117 \citep{HDdetect} and SDSS J1237+0647 \citep{NEW_Noterdaeme:2010gv}.

\subsubsection*{\textit{Other molecules}}

\cite{NEW_Ilyushin:2012vb} identified enhancement of the sensitivity to $\mu$ in  torsion-wagging-rotation transitions in the ground state of methylamine (CH$_3$NH$_2$) due to energy cancellations between internal rotational, overall rotational and inversion energy splitting. This molecule is present in our Galaxy and in a system at $z_{\rm abs}=0.8859$ in direction of the quasar PKS1830-211  \citep{NEW_Muller:2011yu}. Using 3 transitions at  78.135, 79.008 and 89.956~GhZ with $k_\mu=-0.87$ for the first two and $-1.4$ for the third, \cite{NEW_Ilyushin:2012vb} concluded that $\vert\Delta\mu/\mu\vert<9\times10^{-6}$. Combining with the radial velocity of the methanol  line at 60.531~GHz in methanol $(K_\mu=-7.4$, \cite{NEW_Jansen:2011zza}), they conclude that  $\vert\Delta\mu/\mu\vert<10^{-6}$, with the caveat that the two molecules may not be distributed in the same region of the absorption system.

The detection of H$_2$O$_2$ in interstellar clouds have been reported by \cite{NEW_Bergman:2011pn}. It has been identified as a good candidate for testing variation of $\mu$ \citep{NEW_Kozlov:2011qp,NEW_Polyansky:2013cc} since the estimations of the sensitivity coefficients of its microwave transitions lead to the largest coefficient for 14.8~GHz transition: $K_\mu=37$.

\subsubsection*{\textit{Summary and further possibilities}}

All these constraints are summarized in Table~\ref{tab02}, showing the tremendous progresses of the observation of the molecular absorption spectra. Molecular H$_2$ observations allowing one to reach $10^{-8}$ are expected from the E-ELT/ANDES spectrograph \citep{ANDES:2023cif}. We refer to \cite{Chen:2019ltf} for the prospects on methanol observation with the FAST telescope. Let us also mention that FIR lines are expected to be observed by a new generation of telescopes such as HERSCHEL\footnote{\url{http://sci.esa.int/science-e/www/area/index.cfm?fareaid=16}} and that surveys in radio are being carried out so that the number of known redshift OH, HI and HCO$^{+}$ absorption systems will increase. For instance the future Square Kilometer Array (SKA) will be able to detect relative changes of the order of $10^{-7}$ in $\aem$.

\begin{table*}
\begin{center}
\begin{tabular}{lcccr}
\toprule
 Object & $z$ & ${\Delta\mu}/{\mu}$ & Molecules & Reference \\ 
\toprule
B0218+357&  0.685 & $(0.74\pm.0.47\pm0.76_{\rm syst})\times10^{-6}$ & NH$_3$, HCN, HCO$^+$ &  \cite{mu-N3}\\
 & & $(-0.35\pm.012)\times10^{-6}$ & NH$_3$, CS, H$_2$CO &  \cite{NEW_Kanekar:2011yb}\\
 \midrule
PKS1830-211 & 0.886 & $(0\pm 1.4)\times10^{-6}$ & NH$_3$ &  \cite{mu-henkel}\\
			 & & $(-2.04\pm0.74)\times10^{-6}$ & NH$_3$ & \cite{NEW_Muller:2011yu}\\
			 && $(-1.95\pm0.47))\times10^{-6}$ &CH$_3$OH  & \cite{NEW_Muller:2011yu}\\
   			& & $(0.8\pm 2.1)\times10^{-7}$ &CH$_3$OH  & \cite{NEW_Ellingsen:2012cd}\\ 
 			&       & $(0.0\pm 1.0)\times10^{-7}$ &  CH$_3$OH &  \cite{NEW_Bogdanaite}\\
			& & $(1.5\pm 1.5)\times10^{-7}$  &  CH$_3$OH  & \cite{NEW_Bagdonaite:2013sia}\\
			 &       & $(0\pm4)\times10^{-7}$ &  CH$_3$OH &   \cite{NEW_Kanekar:2014ota}\\
			 &       & $(-1.8\pm1.2)\times10^{-7}$ &  CH$_3$OH &  \cite{NEW_Muller:2021ttk}\\
 			&  & $<9\times10^{-6}$ & CH$_3$NH$_2$ & \cite{NEW_Ilyushin:2012vb}\\ 
\midrule
 J2123-0050 & 2.059 & $(5.6\pm5.5\pm2.9_{\text{syst}})\times 10^{-6}$ & H$_2$, HD (HIRES)  &  \cite{HD-1} \\
&  & $(8.5\pm3.6\pm2.2_{\text{syst}})\times 10^{-6}$ & H$_2$, HD (UVES) & \cite{NEW_vanWeerdenburg:2011ru} \\
&  & $(7.6\pm3.5)\times 10^{-6}\, [*]$ & H$_2$, HD & \cite{NEW_Ubachs:2015fro} \\
\midrule
Q1232+082 & 2.34 & $(19\pm9\pm5_{\text{syst}})\times 10^{-6}$ & H$_2$, HD &  \cite{NEW_Dapra:2016dqh}\\
\midrule
 HE0027-1836 & 2.402& $(-7.6\pm8.1\pm6.3_{\rm syst})\times 10^{-6}$ & H$_2$ & \cite{NEW_Rahmani:2013dia} \\
 \midrule
  Q2348-011 & 2.426 &  $(-6.8\pm27.8)\times 10^{-6}$ & H$_2$ &\cite{NEW_Bagdonaite:2011ab}  \\
 \midrule
 J0000+0048 & 2.52 & $(1.8\pm2.2\pm0.4_{\rm syst})\times10^{-5}$ & CO, H$_2$ & \cite{NEW_Dapra:2017vlu} \\
 \midrule
 Q00405-443 & 2.595 & $(27.8\pm 8.8)\times10^{-6}$ &H$_2$ &\cite{mu-rein} \\
 &  & $(5.2\pm7.4)\times10^{-6}$ &H$_2$ &\cite{mu-king} \\
 &  & $(0.6\pm10)\times10^{-6}$ &H$_2$ &\cite{mu-thom}  \\
 & & $(7.5\pm5.3)\times10^{-6} \, [*]$ &H$_2$ & \cite{NEW_Ubachs:2015fro}  \\
 \midrule
 J0643-5041& 2.659 & $(7.4\pm4.3\pm5.1_{\text{syst}})\times 10^{-6}$ & H$_2$ &  \cite{NEW_AlbornozVasquez:2013wfw}  \\
 B0642-5038            & & $(12.7\pm4.5\pm4.2_{\text{syst}})\times 10^{-6}$ & H$_2$ & \cite{NEW_Bagdonaite:2013eia}\\
				&& $ \Delta\mu/\mu= (10.3\pm4.6)\times 10^{-6}\, [*]$ & H$_2$ & \cite{NEW_Ubachs:2015fro}\\
\midrule
 J1237+0647 & 2.69 & $(-5.4\pm6.3\pm4.0_{\rm syst})\times 10^{-6}$ & H$_2$, HD & \cite{NEW_Dapra:2015yva} \\
    		& &   $(0.7\pm 1.6\pm0.5_{\rm syst})\times10^{-5}$& CO & \cite{NEW_Dapra:2016jyh}\\
        &   &   $(5.6\pm 5.6\pm 3.1_{\rm syst})\times10^{-6}$& CO,H$_2$ & \cite{NEW_Dapra:2016jyh}\\
\midrule
Q0528-250 & 2.811 & $(-1.4\pm3.9)\times10^{-6}$ &H$_2$ &\cite{mu-king} \\
        & & $ (0.3\pm3.2\pm1.9_{\text{syst}})\times 10^{-6}$  &H$_2$, HD &\cite{NEW_King:2011km} \\
& & $ (-0.5\pm2.7)\times10^{-6}\, [*]$  &H$_2$ &  \cite{NEW_Ubachs:2015fro}  \\
\midrule
Q0347-383 & 3.025 & $(20.6\pm 7.9)\times10^{-6}$ &H$_2$ &\cite{mu-rein} \\
   &  & $(5.2\pm7.4)\times10^{-6}$ &H$_2$ &\cite{mu-king} \\
   &  & $(-28\pm16)\times10^{-6}$ &H$_2$ &\cite{mu-thom2} \\
    &  & $(15\pm9\pm6_{\text{syst}})\times 10^{-6}$ &H$_2$ &\cite{NEW_Wendt:2010qe} \\
     &  & $(4.3\pm7.2)\times10^{-6}$  &H$_2$ &\cite{NEW_Wendt:2012ea} \\
     &  & $ (5.1\pm4.5)\times10^{-6} \, [*] $ &H$_2$ & \cite{NEW_Ubachs:2015fro}  \\
 \midrule
J1443+2724 & 4.224 & $(-9.5\pm5.4\pm5.3_{\rm syst})\times 10^{-6}$ & H$_2$ & \cite{NEW_Bagdonaite:2015kga} \\
\bottomrule
\end{tabular}
\caption[Molecular absorption spectra constraints on the variation of $\mu$]{\label{tab02}Available measurements of $\mu$ from molecular absorption spectra. Listed are, respectively, the object along each line of sight, the redshift of the measurement, the measurement itself, the molecule(s) used, and the original reference. Low-redshift measurements were obtained with various facilities in the radio/mm band, while high-redshift ones were obtained in the UV/optical with the UVES spectrograph. [*] indicates that this value is the mean obtained from other values and thus is not an independent measurement; see text.}
\label{tab3mu}
\end{center}
\end{table*}

\begin{figure}[hptb]
 \centerline{\includegraphics[scale=0.5]{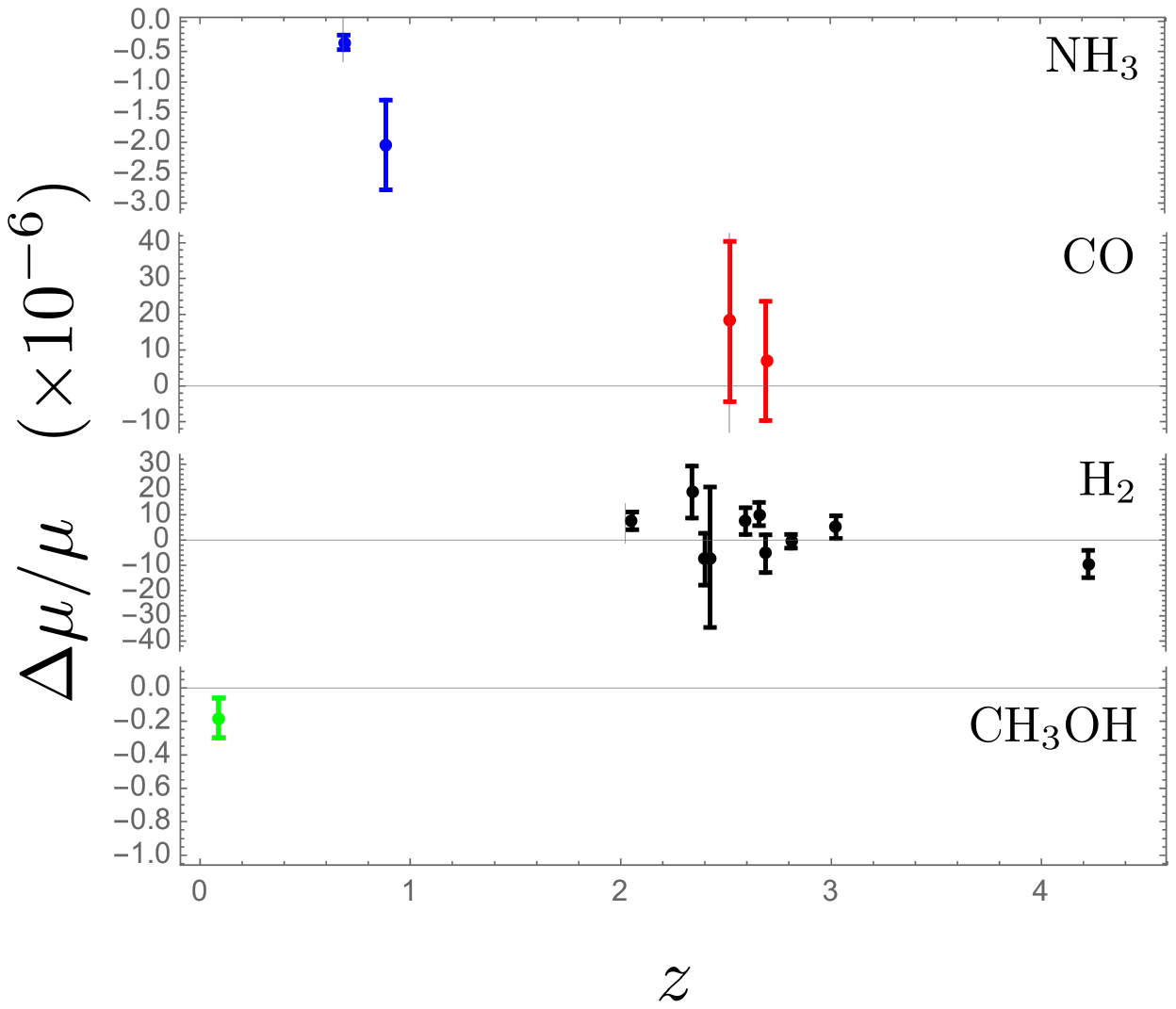}}
 \vskip-0.5cm
  \caption[Molecular bsorption spectra constraints on $\mu$]{Molecular absorption spectra constraints on the variation $\mu=m_{\rm p}/m_{\rm e}$ from H$_2$, NH$_3$, CO and CH$_3$OH as summarized in Table~\ref{tab3mu}.}
  \label{fig-mu}
\end{figure}

\subsection{Emission spectra}\label{subsec_O3}

Similar analysis to constrain the time variation of the fundamental constants were also performed with emission spectra. Very few such estimates have been performed, since it is less sensitive and harder to extend to sources with high redshift. In particular, emission lines are usually broad as compared to absorption lines and the larger individual errors need to be beaten by large statistics.

The O\,{\sc iii} analysis considered the forbidden line doublet of O\,{\sc iii} at $\lambda=4960.295$~\r{A} and $\lambda=5008.239$~\r{A}. These two particular spectral lines are produced by magnetic dipole interactions with a small contribution of electric quadrupole radiation. It can be shown \citep{em-o3}  that
\begin{equation}
R\equiv \frac{\lambda_{5008}-\lambda_{4960}}{\lambda_{5008}+\lambda_{4960}}\propto \aem^2\,.
\end{equation}
The early analysis by \cite{PhysRevLett.19.1294} concluded that $\Delta\aem/\aem = (1\pm2)\times10^{-3}$ in the redshift band $[0.17-0.26]$. 

The [O {\sc iii}] doublet method was sidelined for decades thought to be non-competitive compared to absorption spectra. It was revived in the 2000s thanks to the SDSS project that  produced a large sample of QSO spectra with resolution of order $R\sim2000$. Most the constraints gets an improved precision by averaging on a large number of QSOs so that constraints are obtained for a band of redshifts. As such, the analysis of a sample of 42 quasars from SDSS early data release gave \citep{em-o3}
\begin{equation}
 \Delta\aem/\aem = (7\pm14)\times10^{-5},\qquad 0.16<z<0.8,
\end{equation}
while the DR1 release (165 quasars) gave $\Delta\aem/\aem = (12\pm7)\times10^{-5}$. From the analysis of a sample of 1568 quasar of the SDSS-DR2, \cite{NEW_OIII2} concluded that
\begin{equation}
 \Delta\aem/\aem = (2.4\pm2.5)\times10^{-5},\qquad 0.0<z<0.8.
\end{equation}
{Using 2347 quasars of the SDSS-DR7, \cite{NEW_10.1093/mnrasl/slt183} obtained
\begin{equation}
 \Delta\aem/\aem = (-2.1\pm1.6)\times10^{-5},\qquad 0.02<z<0.74,
\end{equation}
iconsistent with a non-varying $\aem$ at a level of $2\times10^{-5}$. Applying the same method on a much larger data base of 13,175 quasars from SDSS DR12 yielded \citep{NEW_10.1093/mnras/stv1406}
\begin{equation}
 \Delta\aem/\aem = (0.9\pm1.8)\times10^{-5},\qquad 0.04<z<1.0.
\end{equation}
They also provide the measurement of $\Delta\aem/\aem$ in 10 redshift bins from 0.390 to 1.00. \cite{NEW_Laker:2022mik} analyzed 12~000 galaxies from the SDSS-DR8 to get
\begin{equation}
 \Delta\aem/\aem = (1.3\pm1.26)\times10^{-5},\qquad 0.0<z<0.467.
\end{equation}

The method was then extended straightforwardly along the lines of the MM method and applied \citep{em-grup} to the fine-structure transitions in Ne\,{\sc iii}, Ne\,{\sc v}, O\,{\sc iii}, O\,{\sc i} and S\,{\sc ii} multiplets from a sample of 14 Seyfert 1.5 galaxies to derive the constraint
\begin{equation}
 \Delta\aem/\aem = (150\pm70)\times10^{-5},\qquad 0.035<z<0.281.
\end{equation}
Ne\,{\sc iii} and S\,{\sc ii} were also analyzed by \cite{NEW_OIII2} to get $ \Delta\aem/\aem = (36\pm1)\times10^{-4}$ while  \cite{NEW_10.1093/mnras/stv1406} got $ \Delta\aem/\aem = (34\pm1)\times10^{-4}$. These doublets are more difficult to analyze since they are fainter than those of O\,{\sc iii}. These results are significantly different from zero with a clear tendency for a positive variation of $\aem$, a general tendency incompatible with the O\,{\sc iii} constraints and that need to be better understood and it was demonstrated \citep{NEW_10.1093/mnras/stv1406} that the accuracy of Ne\,{\sc iii} is below $10^{-3}$ due to contamination by other emission lines.

This method requires to use large samples (typically $N > 20$) and has been applied mostly with QSO probably because they have higher O\,{\sc iii}  luminosities or simply due to  the observational strategy of SDSS. \cite{NEW_Li:2023fqw} noted that (\textit{1}) starburst galaxies far outnumber QSO  and (\textit{2}) have narrower  O\,{\sc iii} emission lines, which is advantageous for improving the accuracy of the $\aem$ measurement. They claim that it can compensate their disadvantage of lower O\,{\sc iii}  luminosity. Besides, it will allow one to reach higher redshifts. To that goal, \cite{NEW_Li:2023fqw} constructed a sample of 40 spectra of Ly$\alpha$ emitting galaxies and of 46 spectra of QSO with redshift ranging from $1.09$ to  $3.73$ using the VLT/X-Shooter  near-IR spectra that are publicly available \citep{NEW_Vernet:2011fy}. The analysis of the  86 spectra gave
  \begin{equation}
 \Delta\aem/\aem = (-3\pm6)\times10^{-5},\qquad 1.09<z<3.73.
\end{equation}
 
\cite{NEW_DESI:2024yok}  used the DESI (Dark Energy Spectroscopic Instrument) observation of about 110~000  galaxies with strong and narrow [O\,{\sc iii}] emission lines on a redshift band up to $0.95$. The sample was split in 10 bins of $\Delta z=0.1$ and $\Delta\aem/\aem$ was measured for each subsample. They concluded for  an apparent variation, $\Delta\aem/\aem=(2-3)\times10^{-5}$ and argued that this was likely to be due to systematics associated with wavelength calibration. Since it allows for the use of other emission lines, in particular [Ne~{\sc iii}] and a better calibration, it is claimed that DESI may  probe higher redshifts (up to 1.45 in the near future). 

\cite{NEW_Jiang:2024hiv} applied the same method to a galaxy sample consists of 572 spectra with strong and narrow [O\,{\sc iii}] emission lines from 522 galaxies with 267 spectra at $z > 5$. The sample was split into 5 subsamples,  all consistent with zero within $1\sigma$ error of $(1-2)\times 10^{-4}$; see Fig.~5 of \cite{NEW_Jiang:2024hiv} . The whole sample led to the conclusion that
  \begin{equation}
 \Delta\aem/\aem = (4\pm7)\times10^{-5},\qquad 3<z<7.
\end{equation}
This is the first constraint to reach such hight redshifts. From the LAMOST (Large Sky Area Multi-object Fiber Spectroscopic Telescope) Data Release 9 quasar catalog, \cite{Wei:2024trz} exhibited a sample of  209 quasar spectra with strong and narrow [O ~{\sc iii}] emission lines over a redshift range between 0 and 0.8 to conclude that
  \begin{equation}
 \Delta\aem/\aem = (0.5\pm3.7)\times10^{-4},\qquad 0<z<0.8.
\end{equation}
so far not competitive with the previous constraints from the Sloan Digital Sky Survey.

The [O~{\sc iii}]  emission spectra method has regained interest. In particular because it relies on fewer assumptions and suffers from fewer systematic errors compared the MM method: no assumptions on chemical composition, ionization state, and distribution of energy levels have to be made since the doublet lines originate in the transitions from the same upper level of the same ion. The method is also considered more tolerant of the wavelength distortion because of the small wavelength range used in the measurement  \citep{NEW_Li:2023fqw}. Besides, while it is difficult to detect absorption spectra at high redshifts (see e.g., \cite{NEW_Wilczynska:2020rxx} for a discussion of this limitation), many starburst galaxies have been discovered in the early universe, and their doublet emission lines can be used. The data discussed in this paragraph are summarized in  Table~\ref{tab02b} and Fig.~\ref{fig-03alpha}.
 
\begin{table*}
\begin{center}
\begin{tabular}{lccr}
\toprule
 Redshift band & $\Delta\aem/\aem$ & Data & Reference \\ 
\toprule
 $[0.17-0.26]$ & $(1\pm2)\times10^{-3}$  & 5 QSO & \cite{PhysRevLett.19.1294}\\
$[0.16-0.8]$ & $(7\pm14)\times10^{-5}$  & 42 QSO (SDSS EDR)&  \citep{em-o3}\\
$[0.16-0.8]$   &  $ (12\pm7)\times10^{-5}$  &   165 QSO (SDSS-DR1) &   \citep{em-o3}\\
$[0-0.8]$   & $(2.4\pm2.5)\times10^{-5}$ & 1568 QSO (SDSS-DR2)  & \cite{NEW_OIII2} \\
 $[0.02-0.74]$ & $(-2.1\pm1.6)\times10^{-5}$ &  2347 QSO (SDSS-DR7) & \cite{NEW_10.1093/mnrasl/slt183} \\
$[0.04-1.0]$ & $(0.9\pm1.8)\times10^{-5}$ &13~175 QSO (SDSS DR12) & \citep{NEW_10.1093/mnras/stv1406} \\
$[0-0.467]$ & $(1.3\pm1.26)\times10^{-5}$ &  12000 galaxies (SDSS-DR8) & \cite{NEW_Laker:2022mik}\\
 $[ 0.035-0.281]$ & $(150\pm70)\times10^{-5}$ &  14 Seyfert 1.5 galaxies  &\cite{em-grup}\\
$[1.09- 3.73]$ & $(-3\pm6)\times10^{-5}$ & 40 galaxies+46 QSO  (VLT/X-Shooter) &  \cite{NEW_Li:2023fqw} \\
$[3-7]$ & $(4\pm7)\times10^{-5}$ &522 galaxies (JWST)&  \cite{NEW_Jiang:2024hiv} \\
$[0-0.8]$ & $(0.5\pm3.7)\times10^{-4}$ & 209 quasras (LAMOST) & \cite{Wei:2024trz}\\
 \bottomrule
\end{tabular}
\caption[Emission spectra constraints on $\aem$]{Available measurements of $\aem$ from the emission spectra. These constraints are averaged on a redshift bin but do not require any  assumptions on chemical composition, ionization state. Besides they are sensitive to $\aem$ alone.}
\label{tab02b}
\end{center}
\end{table*} 

\begin{figure}[hptb]
  \centerline{\includegraphics[scale=0.4]{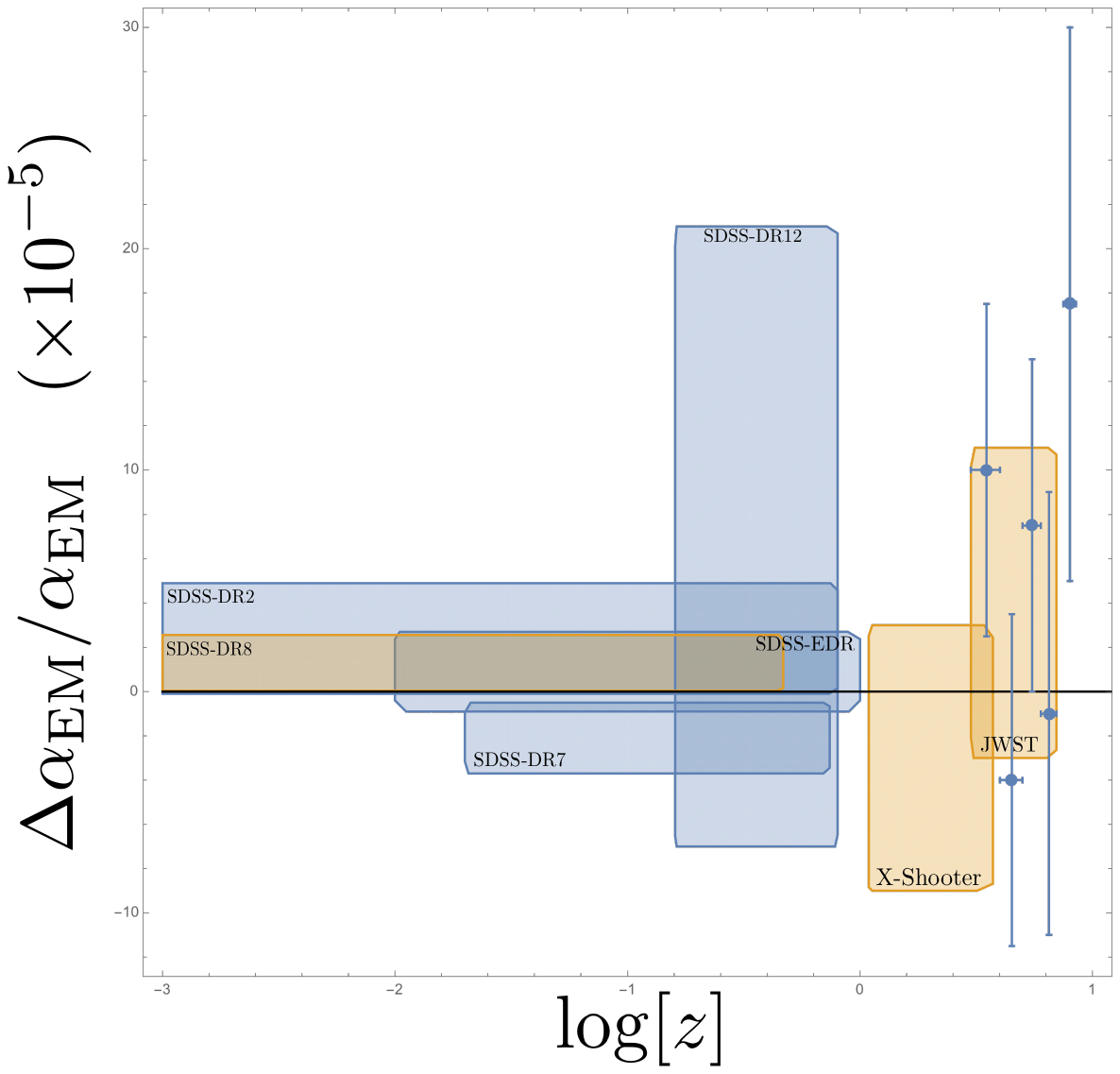}}
  \vskip-.25cm
  \caption[Emission spectra constraints on $\aem$]{Constraints on the variation of $\aem$ as a function of $\log(z)$. The plot gathered the data summarized in Table~\ref{tab02b} together with the 5 data points from Table~5 of  \cite{NEW_Jiang:2024hiv}. The orange and blue zones correspond respectively to QSO sources and galaxy sources. The extension in redshift accounts for the width of the bin on which the data are averaged.}
  \label{fig-03alpha}
\end{figure}

\subsection{Stellar constraints}\label{secstellar}

Stars start to accumulate helium produced by the pp-reaction and the CNO cycle in their core. Furthermore, the products of further nuclear reactions of helium with either helium or hydrogen lead to isotopes with $A=5$ or $A=8$, which are highly unstable. In order to produce elements heavier than $A>7$ by fusion of lighter isotopes, the stars need to reach high temperatures and densities. In these conditions, newly produced $^{12}$C would almost immediately be fused further to form heavier elements so that one expects only a tiny amount of $^{12}$C to be produced, in contradiction with the observed abundances. This led \cite{star-hoyle} to conclude that a then unknown excited state of the $^{12}$C with an energy close to the $3\alpha$-threshold should exist since such a resonance would increase the probability that $^{8}$Be captures an $\alpha$-particle. It follows that the production of $^{12}$C in stars relies on the three conditions:

\begin{itemize}
 \item the decay lifetime of $^{8}$Be, of order $10^{-16}$~s, is   four orders of magnitude longer than the time for two $\alpha$   particles to scatter, so that a macroscopic amount of beryllium can   be produced, which is sufficient to lead to considerable production   of carbon,
 \item an excited state of $^{12}$C lies just above the energy of  $^{8}$Be+$\alpha$, which allows for
   $$
         {}^4{\mathrm{He}} + {}^4{\mathrm{He}} \leftrightarrow {}^8{\mathrm{Be}},\qquad
         {}^8{\mathrm{Be}}+{}^4{\mathrm{He}}\leftrightarrow {}^{12}{\mathrm{C}}^*
         \rightarrow {}^{12}{\mathrm{C}} + 7.367 \unit{MeV},
   $$
 \item the energy level of $^{16}$O at 7.1197~MeV is non resonant   and below the energy of $^{12}\text{C} + \alpha$, of order 7.1616~MeV,   which ensures that most of the carbon synthesized is not destroyed   by the capture of an $\alpha$-particle. The existence of this   resonance, the $J^\pi_l=0_2^+$-state of $^{12}$C was actually    discovered \citep{star-cooke} experimentally later, with an energy   of $372\pm4 \unit{keV}$ [today, $E_{0_2^+}=379.47\pm0.15     \unit{keV}$], above the ground state of three $\alpha$-particles   (see Fig.~\ref{fig-3a1}).
\end{itemize}

\begin{figure}[hptb]
  \centerline{\includegraphics[scale=0.35]{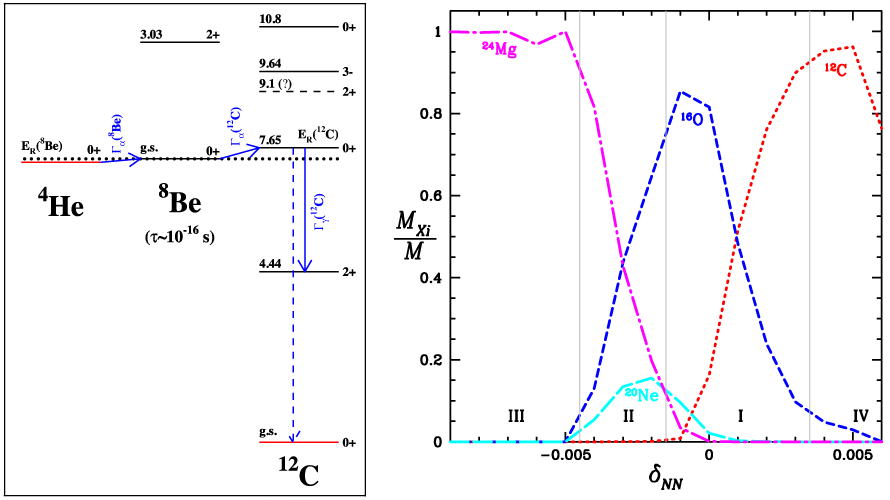}}
  \caption[Stellar constraints on $\delta_{NN}$ from the $3\alpha$-reaction]{\textit{Left:} Level scheme of nuclei participating to the  $^{4}\text{He}(\alpha\alpha,\gamma)^{12}\text{C}$ reaction.  \textit{Right:} Central abundances at the end of the CHe burning as a function of $\delta_{NN}$ for a $60\,M_{\odot}$ star with $Z=0$. From \cite{coc3a}.}
  \label{fig-3a1}
\end{figure}

The variation of any constant that would modify the energy of this resonance would also endanger the stellar nucleosynthesis of carbon, so that the possibility for carbon production has often been used in anthropic arguments. Qualitatively, if $E_{0_2^+}$ is increased then the carbon would be rapidly processed to oxygen since the star would need to be hotter for the triple-$\alpha$ process to start. On the other hand, if $E_{0_2^+}$ is decreased, then all $\alpha$-particles would produce carbon so that no oxygen would be synthesized. It was estimated \citep{star-livio} that the carbon production in intermediate and massive stars is suppressed if the various of the energy of the resonance is outside the range $-250 \unit{keV} \lesssim \Delta E_{0_2^+}\lesssim 60 \unit{keV}$, which was further improved \citep{star-ober1} to, $-5 \unit{keV} \lesssim \Delta E_{0_2^+}\lesssim 50 \unit{keV}$ in order for the C/O ratio to be larger than the error in the standard yields by more than 50\%. Indeed, in such an analysis, the energy of the resonance was changed by hand. However, we expect that if $E_{0_2^+}$ is modified due to the variation of a constant other quantities, such as the resonance of the oxygen, the binding energies and the cross-sections will also be modified in a complex way.

In practice, to draw a constraint on the variation of the fundamental constants from the stellar production of carbon, one needs to go through different steps, each of them involving assumptions,

\begin{enumerate}
 \item to determine the effective parameters, e.g., cross-sections, which affects the stellar evolution. The simplest choice is to modify only the energy of the resonance but it may not be  realistic since all cross-sections and binding energies should also be affected. This requires one to use a stellar evolutionary model;
  \item relate these parameters to nuclear parameters. This involves the whole nuclear physics machinery;
 \item to relate the nuclear parameters to fundamental constants. As for the Oklo phenomenon, it requires to link QCD to nuclear physics.
\end{enumerate}

A first analysis \citep{star-ober3,star-ober4,star-ober1} used a model that treats the carbon nucleus by solving the 12-nucleon Schr\"odinger equation using a three-cluster wavefunction representing the three-body dynamics of the $^{12}$C state. The NN interaction was described by the Minnesota model \citep{star-min,star-thom} and its strength was modified by multiplying the effective NN-potential by an arbitrary number $p$. This allows to relate the energy of the Hoyle level relative to the triple alpha threshold,  $\varepsilon\equiv Q_{\alpha\alpha\alpha}$, and the gamma width, $\Gamma_\gamma$, as a function of the parameter $p$, the latter being almost not affected. The modified $3\alpha$-reaction rate was then given by
\begin{equation}
 r_\alpha = 3^{3/2}N_\alpha^3\left(\frac{2\pi \hbar^2}{M_\alpha k_{\mathrm{B}}T}\right)^3
 \frac{\Gamma}{\hbar} \exp\left[{-\frac{\varepsilon(p)}{k_{\mathrm{B}}T}}\right],
\end{equation}
where $M_\alpha$ and $N_\alpha$ are the mass and number density of the $\alpha$-particle, The resonance width $\Gamma=\Gamma_\alpha\Gamma_\gamma/ (\Gamma_\alpha + \Gamma_\gamma) \sim \Gamma_\gamma$. This was included in a stellar code and ran for red giant stars with $1.3$, $5$ and $20\,M_{\odot}$ with Solar metallicity up to thermally pulsating asymptotic giant branch \citep{star-ober3} and in low, intermediate and high mass ($1.3, 5, 15, 25\,M_{\odot}$) with Solar metallicity also up to TP-AGB \citep{star-ober1} to conclude that outside a window of respectively 0.5\% and 4\% of the values of the strong and electromagnetic forces, the stellar production of carbon or oxygen will be reduced by a factor 30 to 1000.

In order to compute the resonance energy of the $^{8}$Be and $^{12}$C a microscopic cluster model was developed \citep{star-min}. The Hamiltonian of the system is then  of the form $H=\sum_i^A T(\mathbf{r}_i +\sum_{j<i}^A V(\mathbf{r}_{ij}) $, where $A$ is the nucleon number, $T$ the kinetic energy and $V$ the NN interaction potential. In order to implement the variation of the strength of the nuclear interaction with respect to the electromagnetic interaction, it was taken as
$$
V(\mathbf{r}_{ij}) = V_C(\mathbf{r}_{ij}) + (1+\delta_{NN})V_N(\mathbf{r}_{ij}),
$$
where $\delta_{NN}$ is a dimensionless parameter that describes the change of the nuclear interaction, $V_N$ being described in \cite{star-thom}.  When $A>4$ no exact solution can be found and approximate solutions in which the wave function of the $^{8}$Be and $^{12}$C are described by clusters of respectively 2 and 3 $\alpha$-particle is well adapted. 

First, $\delta_{NN}$ can be related to the deuterium binding energy as
\begin{equation}
 \Delta B_{\rm D}/B_{\rm D} = 5.7701\times \delta_{NN},
\end{equation}
which, given the discussion in Sect.~\ref{bbn2cste}, allows one to relate $\delta_{NN}$ to fundamental constants, as, e.g., in \cite{cnouv}. Then, the resonance energy of the $^{8}$Be and $^{12}$C scale as
\begin{align}
 E_R({}^8{\mathrm{Be}}) = (0.09208-12.208\times\delta_{NN}) \unit{Mev}\,, \nonumber\\
 E_R({}^{12}{\mathrm{C}}) = (0.2877-20.412\times\delta_{NN}) \unit{Mev}\,,
\end{align}
so that the energy of the Hoyle level relative to the triple-$\alpha$ threshold is $Q_{\alpha\alpha\alpha}= E_R({}^8{\mathrm{Be}})+E_R({}^{12}{\mathrm{C}})$.

This was implemented in \cite{coc3a,star-new} to population~III stars with typical masses, $15$ and $60\,M_{\odot}$ with zero metallicity, in order to compute the central abundances at the end of the core He burning. From Fig.~\ref{fig-3a1}, one can distinguish 4 regimes (\textit{I}) the star ends the CHe burning phase with a core composed of a mixture of $^{12}$C and $^{16}$O, as in the standard case; (\textit{II}) if the $3\alpha$ rate is weaker, $^{12}$C is produced slower, the reaction $^{12}\text{C}(\alpha,\gamma)^{16}\text{O}$ becomes efficient earlier so that the star ends the CHe burning phase with a core composed mostly of $^{16}$O; (\textit{III}) for weaker rates, the $^{16}$O is further processed to $^{20}$Ne and then  $^{24}$Mg so that the star ends the CHe burning phase with a core composed of $^{24}$Mg and (\textit{IV}) if the $3\alpha$ rate is stronger, the $^{12}$C is produced more rapidly and the star ends the CHe burning phase with a core composed mostly of $^{12}$C. Typically this imposes that
\begin{equation}
 -5\times10^{-4}<\delta_{NN}<1.5\times 10^{-3}, \quad
 -3\times10^{-4}<\Delta B_{\rm D}/B_{\rm D}<9\times10^{-3},
\end{equation}
at a redshift of order $z\sim 15$, to ensure the ratio C/O to be of order unity.

To finish, a recent study \citep{star-adam} focus on the existence of stars themselves, by revisiting the stellar equilibrium when the values of some constants are modified. In some sense, it can be seen as a generalisation of the work by \cite{gamow67a} to constrain the Dirac model of a varying gravitational constant by estimating its effect on the lifetime of the Sun. In this semi-analytical stellar structure model, the effect of the fundamental constants was reduced phenomenologically to 3 parameters, $G$, which enters mainly on the hydrostatic equilibrium, $\aem$, which enters in the Coulomb barrier penetration through the Gamow energy, and a composite parameter $\mathcal{C}$, which describes globally the modification of the nuclear reaction rates. The underlying idea is to assume that the power generated per unit volume, $\varepsilon(r)$, and which determines the luminosity of the star, is proportional to the fudge factor $\mathcal{C}$, which would arise from a modification of the nuclear fusion factor, or equivalently of the cross-section. Thus, it assumes that all cross-sections are affected is a similar way. The parameter space for which stars can form and for which stable nuclear configurations exist was determined, showing that no fine-tuning seems to be required.

This system is very promising and will provide new information on the fundamental constants at redshifts smaller than $z\sim15$ where no constraints exist at the moment, even though drawing a robust constraint seems to be difficult at the moment. In particular, an underlying limitation arises from the fact that the composition of the interstellar media is a mixture of ejecta from stars with different masses and it is not clear which type of stars contribute the most the carbon and oxygen production. Besides, one would need to include rotation and mass loss \citep{meynet}. As for the Oklo phenomenon, another limitation arises from the complexity of nuclear physics.

\subsection{Cosmic Microwave Background}\label{subsec34}

The CMB radiation is composed of photons emitted at the time of the recombination of hydrogen and helium when the universe was about 300000 years old [see, e.g.,\ \cite{peteruzanbook} for details on the physics of the CMB]. This radiation is observed to be a black-body with a temperature $T_0=2.725 \unit{K}$ with small anisotropies of order of the $\mu$K. The temperature fluctuation in a direction $(\vartheta,\varphi)$ is usually decomposed on a basis of spherical harmonics as
\begin{equation}
\label{cmb1}
 \frac{\delta T}{T}(\vartheta,\varphi)=\sum_{\ell}\sum_{m=-\ell}^{m=+\ell}a_{\ell
  m}Y_{\ell m}(\vartheta,\varphi).
\end{equation}
The angular power spectrum multipole $C_\ell=\langle \vert a_{lm}\vert^2 \rangle$ is the coefficient of the decomposition of the angular correlation function on Legendre polynomials. Given a model of structure formation and a set of cosmological parameters, this angular power spectrum can be computed and compared to observational data in order to constrain this set of parameters.

\paragraph{Influence of the fundamental constants}\ 

The CMB temperature anisotropies mainly depend on  $G$, $\aem$ and $m_{\mathrm{e}}$.

The gravitational constant enters in the Friedmann equation and in the evolution of the cosmological perturbations. It has mainly three effects \citep{cmb-G1} that are detailed in Sect.~\ref{subsecGcmb}. $\aem$, $m_{\mathrm{e}}$ affect the dynamics of the recombination. Their influence is complex and must be computed numerically. However, we can trace their main effects since they mainly modify the CMB spectrum through the change in the differential optical depth of photons due to the Thomson scattering
\begin{equation}
\label{cmb2}
 \dot\tau=x_{\mathrm{e}}n_{\mathrm{e}}c\sigma_{\mathrm{T}},
\end{equation}
which enters in the collision term of the Boltzmann equation describing the evolution of the photon distribution function and where $x_{\mathrm{e}}$ is the ionization fraction (i.e., the number density of free electrons with respect to their total number density $n_{\mathrm{e}}$).

The first dependence arises from the Thomson scattering cross-section given by
\begin{equation}
\label{cmb3}
 \sigma_{\mathrm{T}}=\frac{8\pi}{3}\frac{\hbar^2}{m_{\mathrm{e}}^2c^2}\aem^2
\end{equation}
and the scattering by free protons can be neglected since $m_{\mathrm{e}}/m_{\mathrm{p}}\sim5\times10^{-4}$.

The second, and more subtle dependence, comes from the ionization fraction. Recombination proceeds via 2-photon emission from the $2s$ level or via the Ly-$\alpha$ photons, which are redshifted out of the resonance line \citep{cmb-peebles68} because recombination to the ground state can be neglected since it leads to immediate re-ionization of another hydrogen atom by the emission of a Ly-$\alpha$ photons. Following \cite{cmb-peebles68} and \cite{cmb-ma-b} and taking into account, for the sake of simplicity, only the recombination of hydrogen, the equation of evolution of the ionization fraction takes the form
\begin{equation}
\label{eecmb1}
 \frac{\dd x_{\mathrm{e}}}{\dd t}={\cal C}\left[\beta\left(1-x_{\mathrm{e}}\right)
  \hbox{exp} \left(-\frac{B_1-B_2}{k_{\mathrm{B}}T_M}\right)
  -{\cal R}n_{\mathrm{p}}x_{\mathrm{e}}^2\right],
\end{equation}
where $T_M$ is the temperature. At high redshift, $T_M$ is identical to the one of the photons $T_\gamma=T_0(1+z)$ but evolves according to
\begin{equation}
\label{e_cmbT}
 \frac{\dd T_M}{\dd t} = -\frac{8\sigma_{\mathrm{T}} a_R}{3 m_{\mathrm{e}}}
 T_R^4\frac{x_{\mathrm{e}}}{1+x_{\mathrm{e}}}(T_M - T_\gamma) - 2 H T_M
\end{equation}
where the radiation constant $a_R=4\sigma_{\mathrm{SB}}/c$ with $\sigma_{\mathrm{SB}} =k^4_{\mathrm{B}}\pi^2/(60\pi c^2\hbar^3)$ the Stefan--Boltzmann constant. In Eq.~(\ref{eecmb1}), $B_n=-E_I/n^2$ is the energy of the $n^{\rm th}$ hydrogen atomic level, $\beta$ is the ionization coefficient, ${\cal R}$ the recombination coefficient, ${\cal C }$ the correction constant due to the redshift of Ly-$\alpha$ photons and to 2-photon decay and $n_p=n_e$ is the number density of protons. $\beta$ is related to ${\cal R}$ by the principle of detailed balance so that
\begin{equation}
\label{cmb5}
\beta={\cal R}\left(\frac{2\pi m_{\mathrm{e}}
  k_{\mathrm{B}}T_M}{h^2}\right)^{3/2}\hbox{exp}\left(-\frac{B_2}{k_{\mathrm{B}}T_M}\right).
\end{equation}
The recombination rate to all other excited levels is
$$
{\cal R}=\frac{8\pi}{c^2}\left(\frac{k_{\mathrm{B}}T}{2\pi
  m_{\mathrm{e}}}\right)^{3/2}
\sum_{n,l}^*(2l+1)\hbox{e}^{B_n/k_{\mathrm{B}}T}\int_{B_n/k_{\mathrm{B}}T}^\infty
\sigma_{nl}\frac{y^2\dd y}{\hbox{e}^y-1}
$$
where $\sigma_{nl}$ is the ionization cross-section for the $(n,l)$ excited level of hydrogen. The star indicates that the sum needs to be regularized and the $\aem$-, $m_{\mathrm{e}}$-dependence of the ionization cross-section is complicated to extract. However, it can be shown to behave as $\sigma_{nl}\propto\aem^{-1}m_{\mathrm{e}}^{-2}f(h\nu/B_1)$. Finally, the factor ${\cal C}$ is given by
\begin{equation}
\label{cmb7}
{\cal
C}=\frac{1+K\Lambda_{2s}(1-x_{\rm e})}{1+K(\beta+\Lambda_{2s})(1-x_{\rm e})}
\end{equation}
where $\Lambda_{2s}$ is the rate of decay of the $2s$ excited level to the ground state via 2 photons; it scales as $m_{\mathrm{e}}\aem^8$. The constant $K$ is given in terms of the Ly-$\alpha$ photon $\lambda_{\alpha}=16\pi\hbar/(3m_{\mathrm{e}}\aem^2c)$ by $K=n_p\lambda_\alpha^3/(8\pi H)$ and scales as $m_{\mathrm{e}}^{-3}\aem^{-6}$.

In summary, both the temperature of the decoupling and the residual ionization after recombination are modified by a variation of $\aem$ or $m_{\mathrm{e}}$. This was first discussed in \cite{cmb-bat} and \cite{cmb-kap}. The last scattering surface can roughly be determined by the maximum of the visibility function $g=\dot\tau\exp(-\tau)$, which measures the differential probability for a photon to be scattered at a given redshift. Increasing $\aem$ shifts $g$ to a higher redshift at which the expansion rate is faster so that the temperature and $x_{\rm e}$ decrease more rapidly, resulting in a narrower $g$. This induces a shift of the $C_\ell$ spectrum to higher multipoles and an increase of the values of the $C_\ell$. The first effect can be understood by the fact that pushing the last scattering surface to a higher redshift leads to a smaller sound horizon at decoupling. The second effect results from a smaller Silk damping.

\paragraph{Degeneracy}\ 

\begin{figure*}[htb]
\centering
\vskip1cm
\includegraphics[width=5.5cm]{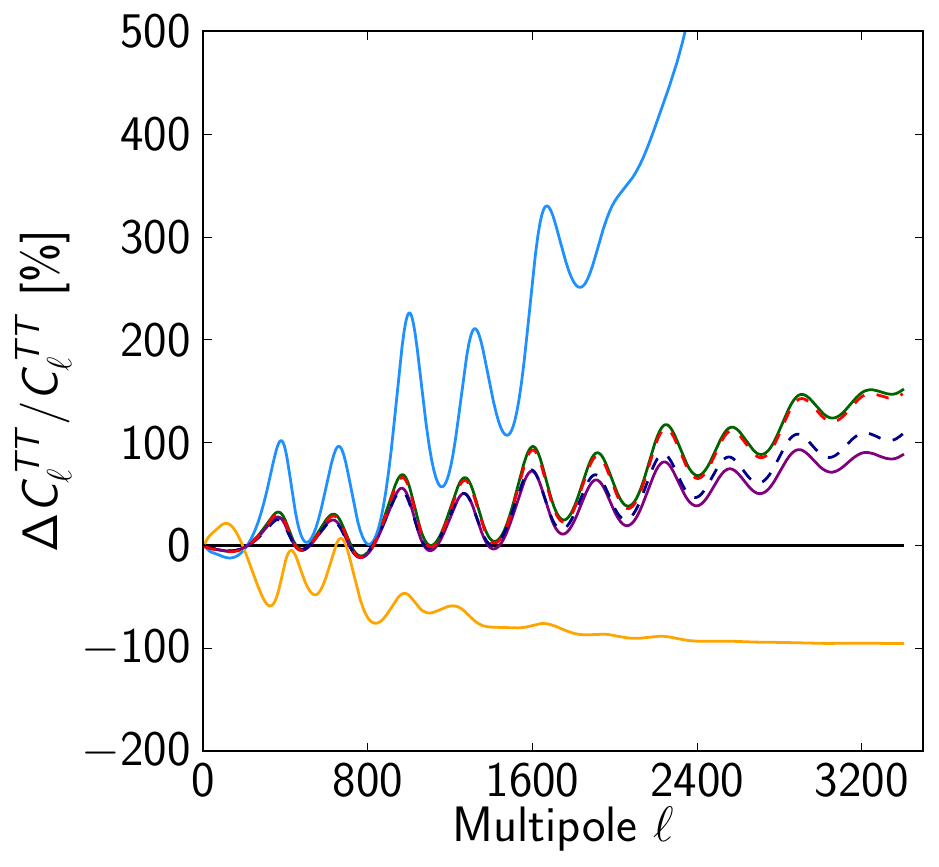}$\qquad$
\includegraphics[width=5.5cm]{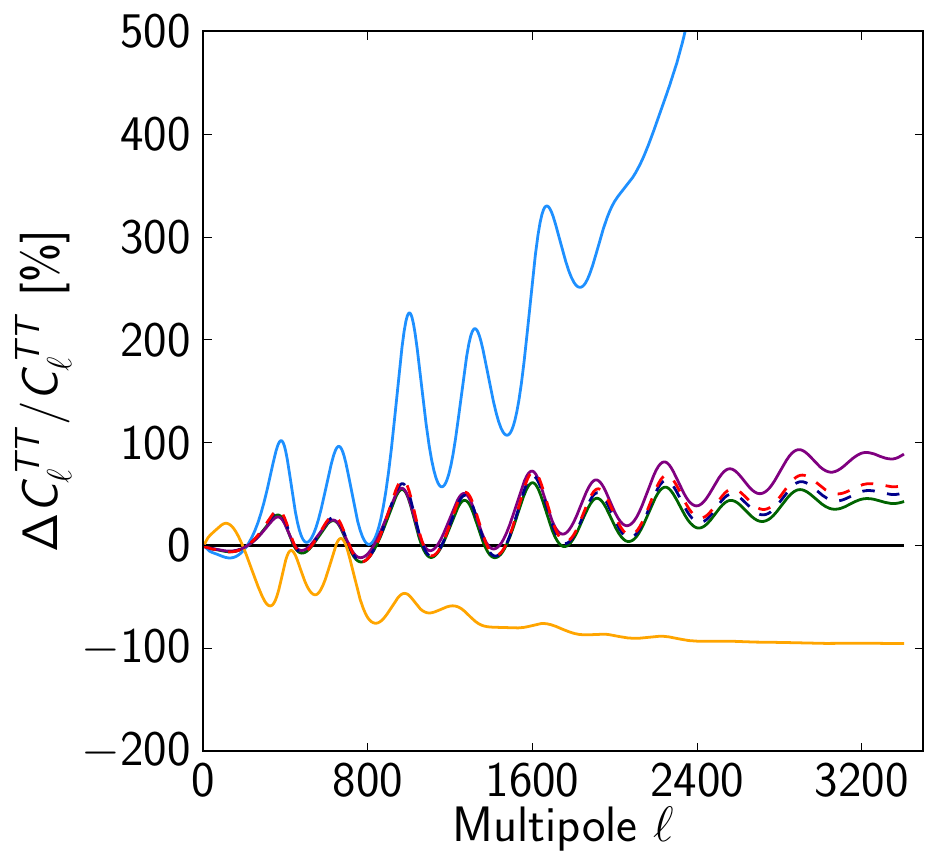}
\caption[Sensitivity of CMB spectra to a change of $\aem$ or $m_{\rm e}$]{\textit{Left}: relative difference between a CMB TT power spectrum calculated using (i) a value of $\alpha/\alpha_0$ different from $1$ in one, few or all the terms where it appears, and a power spectrum calculated using ({\rm ii}) a standard value of $\alpha/\alpha_0=1$. We thus plot $\Delta C_\ell/C_\ell=(C_\ell-C_{\ell,{\rm st}})/C_{\ell,{\rm st}}$[\%]. The cases considered are: $\alpha$ varying only in the hydrogen binding energy (solid light blue), only in the Ly$\alpha$ energy (solid yellow), in both the previous two terms (solid purple), in both the previous terms and in the Thomson scattering cross-section (dashed dark blue), in the previous three terms and in the $2-$photon decay rate (dashed red), in all terms (solid green). In each case, we assume that $\alpha$ varies of $+5\%$ ($\alpha/\alpha_0=1.05$) only in the terms considered, while it is $\alpha/\alpha_0=1$ in all the others. \textit{Right}: same cases as left, but for a variation of $m_{\rm e}$ of $10.025\%$ ($m_{\rm e}/m_{\rm e0}=1.1025$). From \cite{NEW_Planck:2014ylh}.}
\label{fig:effects_alpha_me}
\end{figure*}

Since the atomic energy levels scale as $B\propto\aem^2m_{\rm e}$, this ensures that the effect of a 5\% change in $\aem$ or $10.025\%$ in $m_{\rm e}$ in the hydrogen binding energy and in the Ly$\alpha$  energy level produces identical effects on the angular power spectra. This degeneracy is in Fig.~\ref{fig:effects_alpha_me}: the blue line (relative to the change of hydrogen binding energy only), the yellow line (relative to a change in the Ly$\alpha$ energy level only) and the purple line (sum of the previous two effects are identical for $\aem$ and for $m_{\rm e}$.

It is evident from the figures that the major contribution to the change in the angular power spectrum induced by a variation of $\aem$ or $m_{\rm e}$ comes from the change in the hydrogen binding energy  and Ly$\alpha$ energy which leads to a degeneracy at small angular scales. A difference in the effects of $\aem$ or $m_{\rm e}$ is however introduced when one considers the impact on the Thomson scattering cross-section $\sigma_{\rm T}$. Adding the effect of the constants on the Thomson cross-section, shown in the dark-blue dashed lines in Fig.~\ref{fig:effects_alpha_me}, increases the amplitude of the peaks for a larger value of $\alpha$, while it decreases it for a larger value of $m_{\rm e}$. Alternatively, this is the reason why $\aem$ and $m_{\rm e}$ have different effects on the amplitude of the peaks.

\paragraph{CMB code modifications}\ 

Most studies have introduced those modifications in the RECFAST code \citep{cmb-seager} including similar equations for the recombination of helium. Our previous analysis shows that the dependencies in the fundamental constants have various origins, since the binding energies $B_i$ scale has $m_{\mathrm{e}}\aem^2$, $\sigma_T$ as $\aem^2m_{\mathrm{e}}^{-2}$,  $K$ as $m_{\mathrm{e}}^{-3}\aem^{-6}$, the ionisation coefficients $\beta$ as $\aem^3$, the transition frequencies as $m_{\mathrm{e}}\aem^{2}$, the Einstein's coefficients as $m_{\mathrm{e}}\aem^{5}$, the decay rates $\Lambda$ as $m_{\mathrm{e}}\aem^{8}$ and $\cal{R}$ has complicated dependence, which roughly reduces to $\aem^{-1}m_{\mathrm{e}}^{-2}$. In earlier works \citep{cmb-han,cmb-kap} it was approximated by the scaling ${\cal R}\propto\aem^{2(1+\xi)}$ with $\xi\sim0.7$. Note that a change in the fine-structure constant and in the mass of the electron are degenerate according to $\Delta\aem\approx0.39\Delta m_{\mathrm{e}}$ but this degeneracy is broken for multipoles higher than 1500 \citep{cmb-bat}. This is illustrated in the analysis using the first Planck data release \citep{NEW_Planck:2014ylh} that implemented a variation of $(\aem,m_{\rm e})$ in RECFAST. Fig.~\ref{fig-cmb01} illustrates the evolution of the free electron fraction $x_{\rm e}(z)$ under a relative variation of either $\aem$ or $m_{\rm e}$. One can clearly witness the shift of the recombination epoch to earlier time/higher redshift when the constants increase, and the effect on the $TT$, $TE$ and $EE$ angular power spectra as depicted in Fig.~\ref{fig-cmb02}. An earlier recombination has 3 main effects: (\textit{1}) a smaller sound horizon et recombination and hence a larger diameter distance resulting in a shift of the Doppler peaks to higher multipoles; (\textit{2}) a decrease of the Silk damping since at lowest order the Thomson cross-section scales as $\sigma_{\rm T}\propto\aem^2/m_{\rm e}^2$; (\textit{3}) an increase of the spectra at large scales due to the shorter time interval between matter-radiation equality and recombination (see \cite{NEW_Planck:2014ylh} for a detailed discussion).

\begin{figure}[hptb]
  \centerline{\includegraphics[scale=0.4]{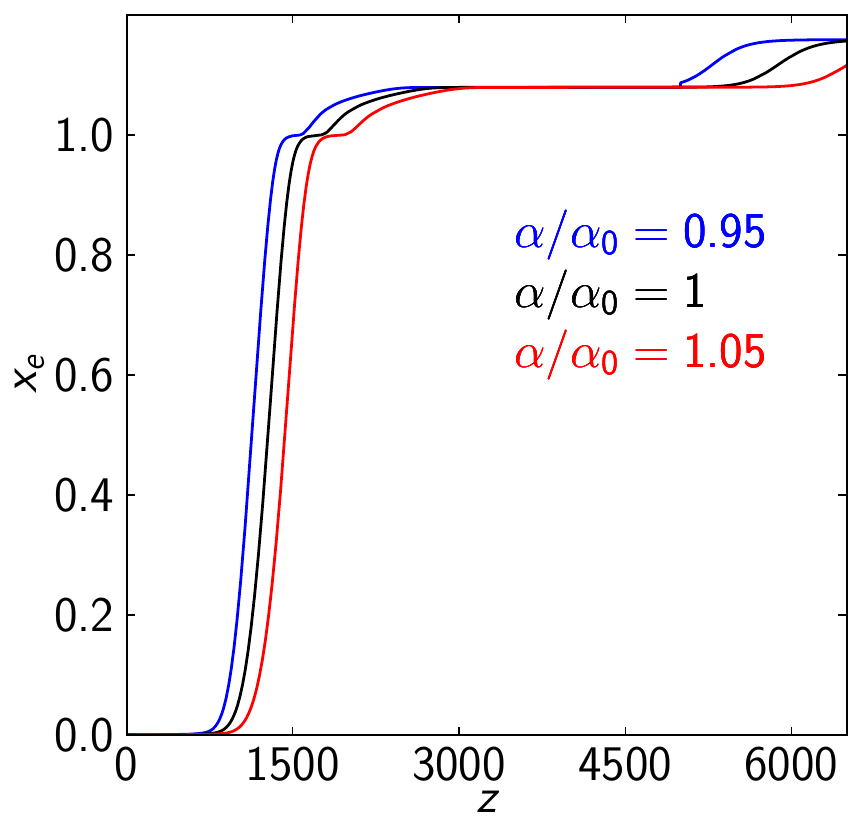}\qquad \includegraphics[scale=0.4]{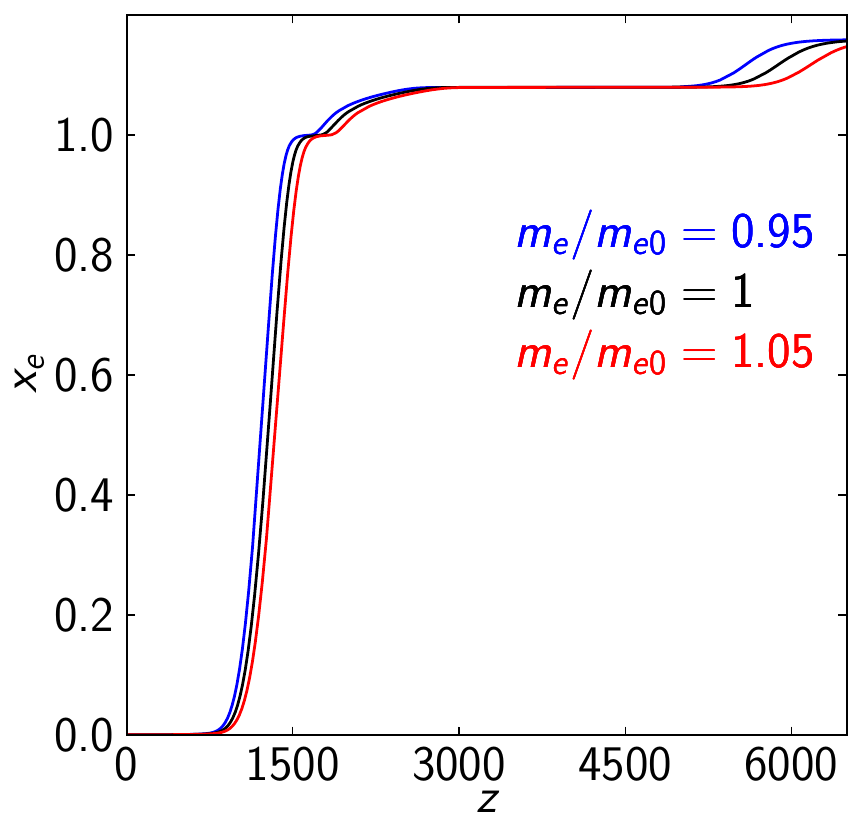}}
  \caption[Sensitivity of the electron fraction to $\aem$ and $m_{\rm e}$]{Evolution of the free electron fraction with redshift for various values of $\aem$ (left) or $m_{\rm e}$ (right) of $-5\%$ (blue), 0\% (black) and $+5\%$ (red).  The decrease in the plateaus at $z\sim 6~000$ and $2~000$ correspond to the first and second recombination of helium while the larger one at $z\sim1300$ is due to the recombination of hydrogen. From \cite{NEW_Planck:2014ylh}.}
  \label{fig-cmb01}
\end{figure}

\paragraph{Observational constraints}\ 

The first studies \citep{cmb-han,cmb-kap} focused on the sensitivity that can be reached by WMAP\footnote{\url{http://map.gsfc.nasa.gov/}} and Planck\footnote{\url{http://astro.estec.esa.nl/SA-general/Projects/Planck/}}. They concluded that they should provide a constraint on $\aem$ at recombination, i.e., at a redshift of about $z\sim1$, with a typical precision $\vert\Delta\aem/\aem\vert\sim10^{-2}-10^{-3}$.

\begin{figure}[hptb]
  \centerline{\includegraphics[scale=0.25]{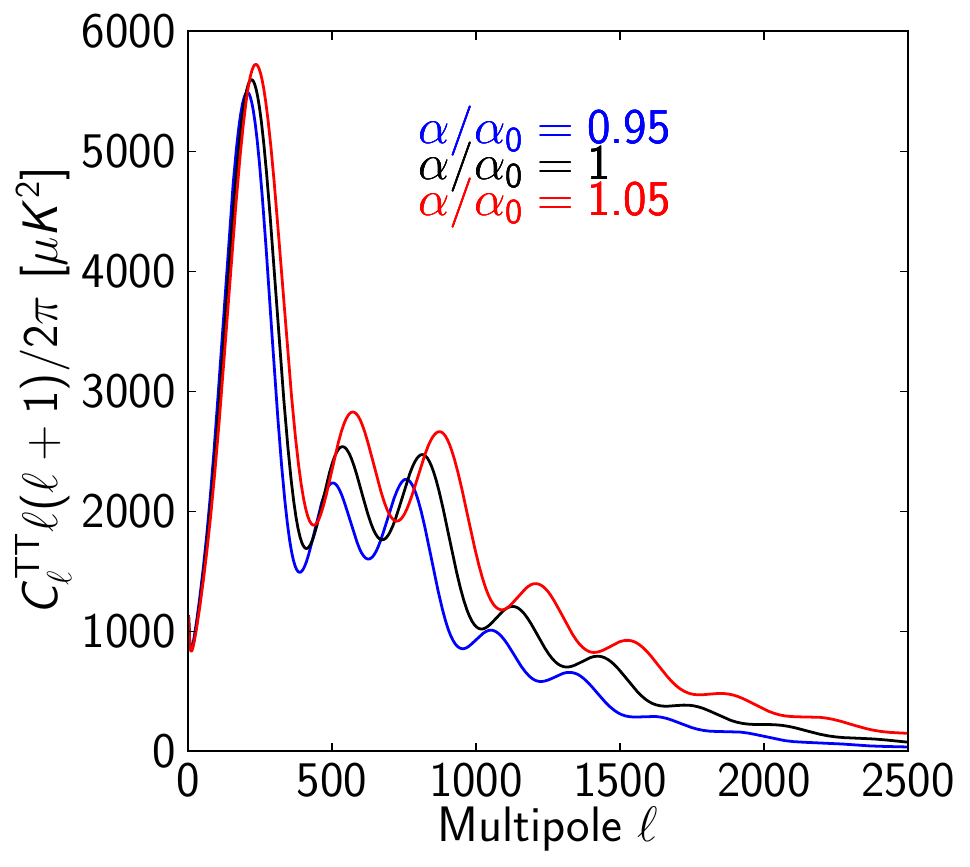}\quad \includegraphics[scale=0.25]{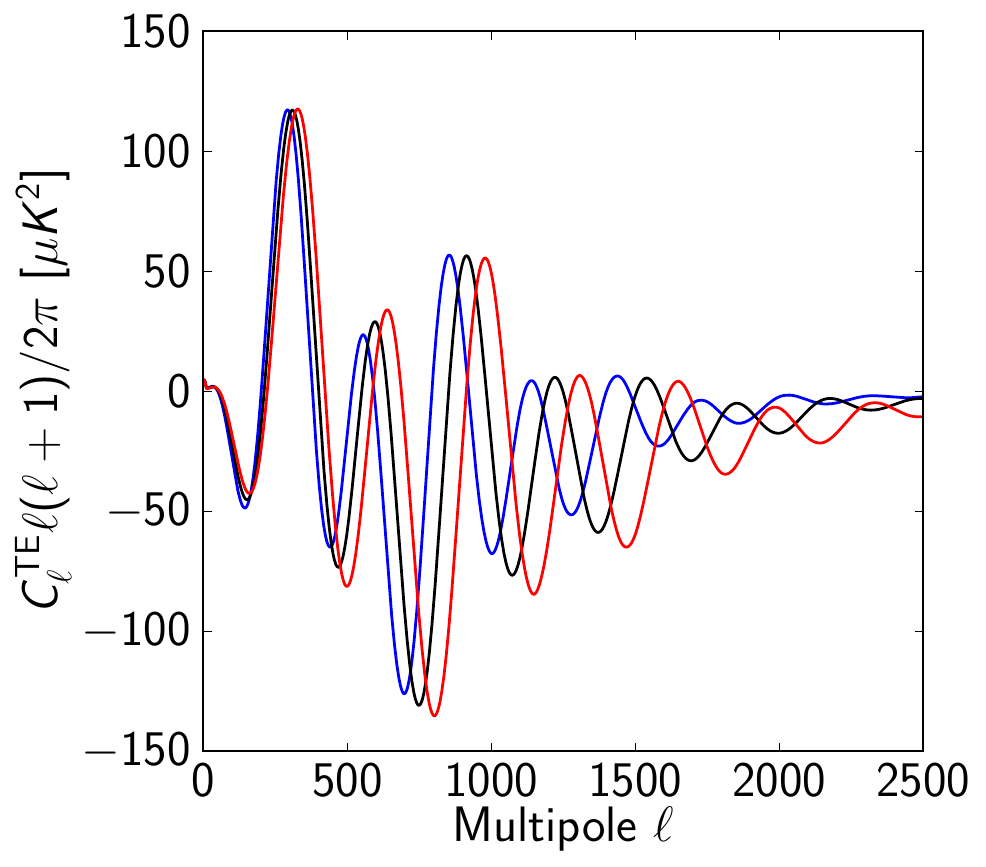} \quad \includegraphics[scale=0.25]{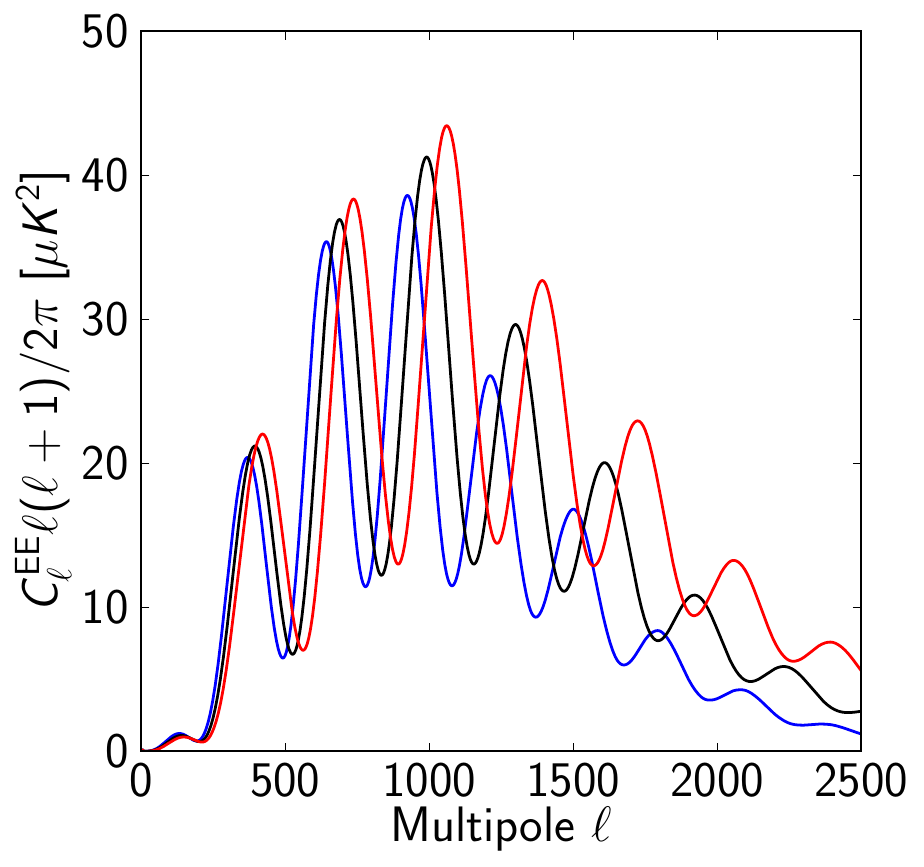}}
    \centerline{\includegraphics[scale=0.25]{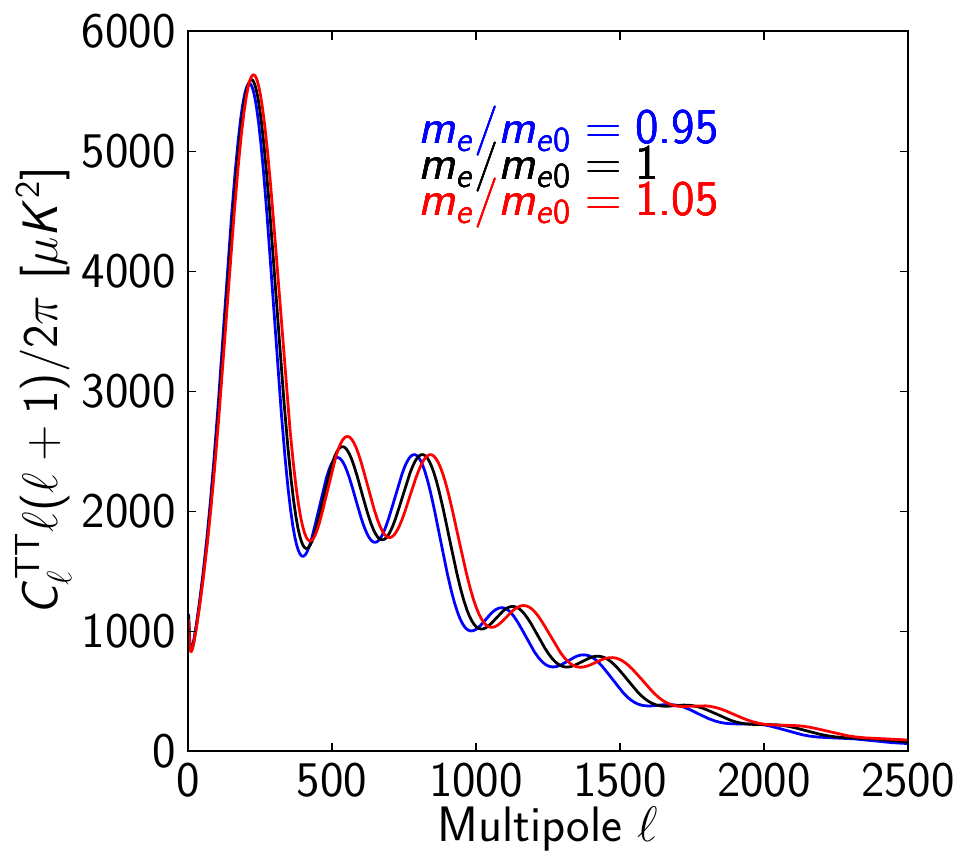}\quad \includegraphics[scale=0.25]{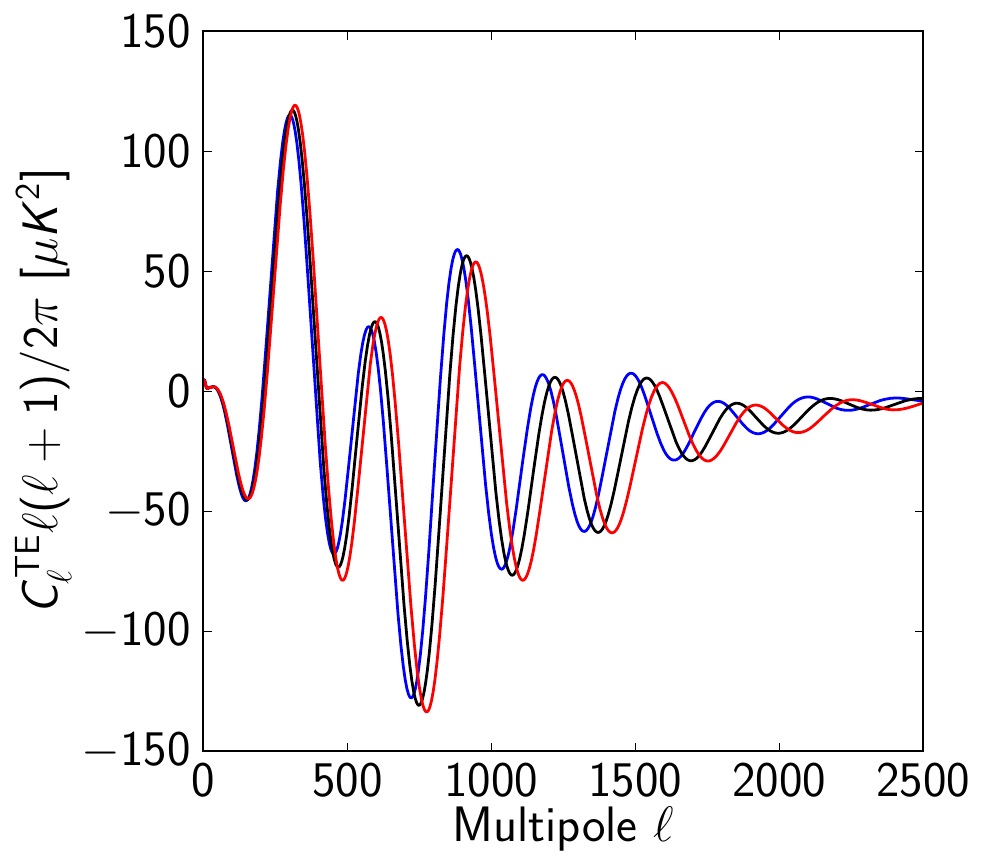} \quad \includegraphics[scale=0.25]{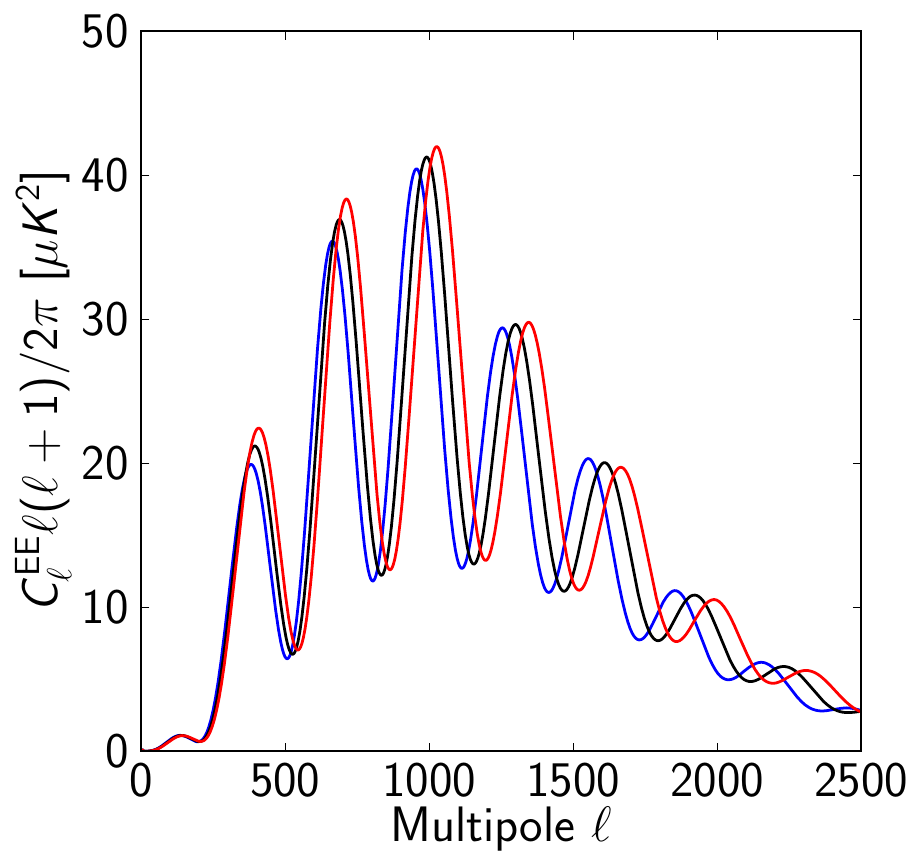}}
  \caption[Sensitivity of CMB spectra to a change of $\aem$ or $m_{\rm e}$]{ CMB $TT$, $TE$ and $EE$ angular power spectrum from left to right for various values of $\aem$ (left) or $m_{\rm e}$ (right) of $-5\%$ (blue), 0\% (black) and $+5\%$ (red). From \cite{NEW_Planck:2014ylh}.}
  \label{fig-cmb02}
\end{figure}

The first attempt \citep{cmb-avelino00} to actually set a constraint was performed on the first release of the data by BOOMERanG and MAXIMA. It concluded that a value of $\aem$ smaller by a few percents in the past was favored but no definite bound was obtained, mainly due to the degeneracies with other cosmological parameters. It was later improved \citep{cmb-avelino01} by a joint analysis of BBN and CMB data that assumes that only $\aem$ varies and that included 4 cosmological parameters ($\Omega_\mat,\Omega_\baryon,h,n_s)$ assuming a universe with Euclidean spatial section, leading to $-0.09<\Delta\aem<0.02$ at 68\% confidence level. A similar analysis \citep{cmb-landau01}, describing the dependence of a variation of the fine-structure constant as an effect on recombination the redshift of which was modeled to scale as $z_* = 1080[1+2\Delta\aem/\aem]$, set the constraint $-0.14<\Delta\aem<0.02$, at a $2\sigma$ level, assuming a spatially flat cosmological models with adiabatic primordial fluctuations that. The effect of re-ionisation was discussed in \cite{cmb-mart}. These works assume that only $\aem$ is varying but, as can been seen from Eqs.~(\ref{cmb1}--\ref{cmb7}), assuming the electron mass constant.

With the WMAP first year data, the bound on the variation of $\aem$ was sharpened \citep{cmb1-2003} to $-0.05<\Delta\aem/\aem<0.02$,  after marginalizing over the remaining cosmological parameters  ($\Omega_\mat h^2,\Omega_\baryon h^2,\Omega h^2,n_s,\alpha_s,\tau)$ assuming a universe with Euclidean spatial sections. Restricting to a model with a vanishing running of the spectral index ($\alpha_s\equiv\dd n_s/\dd\ln k=0$), it gives $-0.06<\Delta\aem/\aem<0.01$, at a 95\% confidence level. In particular it shows that a lower value of $\aem$ makes $\alpha_s=0$ more compatible with the data. These bounds were obtained without using other cosmological data sets. This constraint was confirmed by the analysis of \cite{cmb2-2006}, which got $-0.097<\Delta\aem\aem<0.034$, with the WMAP-1yr data alone and $-0.042<\Delta\aem/\aem<0.026$,  at a 95\% confidence level, when combined with constraints on the Hubble parameter from the HST Hubble Key project. 

The analysis of the WMAP-3yr data allows  \cite{cmb3-2007} to improve this bound to $-0.039<\Delta\aem/\aem <0.010$, at a 95\% confidence level, assuming ($\Omega_\mat,\Omega_\baryon,h, n_s,z_{\mathrm{re}},A_s$) 
for the cosmological parameters ($\Omega_\Lambda$ being derived from the assumption $\Omega_K=0$, as well as $\tau$ from the re-ionisation redshift, $z_{\mathrm{re}}$)  and using both temperature and polarization data ($TT$, $TE$, $EE$). 

The WMAP 5-year data were analyzed, in combination with the 2dF galaxy redshift survey, assuming that both $\aem$ and $m_{\mathrm{e}}$ can vary and that the universe was spatially Euclidean. Letting 6 cosmological parameters [($\Omega_\mat h^2,\Omega_\baryon   h^2,$ $\Theta,\tau, n_s,A_s$), $\Theta$ being the ratio between the   sound horizon and the angular distance at decoupling] and 2 constants vary, \cite{scoccola, scoccola2}  concluded $-0.012<\Delta\aem/\aem<0.018$ and $-0.068<\Delta m_{\mathrm{e}}/m_{\mathrm{e}}<0.044$, the bounds fluctuating slightly with the choice of the recombination scenario. A similar analysis \citep{wmap-alpha} not including $m_{\mathrm{e}}$ gave $-0.050<\Delta\aem/\aem<0.042$, which can be reduced by taking into account some further prior from the HST data. Including polarisation data data from ACBAR, QUAD and BICEP, it was also obtained \citep{menegoni} $-0.043<\Delta\aem/\aem<0.038$ at 95\% C.L.\ and $-0.013<\Delta\aem/\aem<0.015$ including HST data, also at 95\% C.L. Let us also mention the work by \cite{martins37} trying to include the variation of the Newton constant by assuming that $\Delta\aem/\aem=Q\Delta G/G$, $Q$ being a constant and the investigation of \cite{naka00} taking into account $\aem$, $m_{\mathrm{e}}$ and $\mu$, $G$ being kept fixed. Considering ($\Omega_\mat,\Omega_\baryon,h, n_s,\tau$) for the cosmological parameters they concluded from WMAP-5 data ($TT$, $TE$, $EE$) that $-8.28\times10^{-3}<\Delta\aem/\aem<1.81\times10^{-3}$ and $-0.52<\Delta\mu/\mu<0.17$

The analysis of \cite{scoccola,scoccola2} was updated by \cite{scoccola3} to the WMAP-7yr data \citep{NEW_WMAP:2010qai}, including polarisation and SDSS data. It leads to $-0.025<\Delta\aem/\aem<-0.003$ and $0.009<\Delta m_{\mathrm{e}}/m_{\mathrm{e}}<0.079$ at a 1$\sigma$ level. \cite{NEW_Menegoni:2012tq} focused on the small angular scale up to $\ell_{\rm max}=3000$ by adding data from ACT \citep{NEW_Dunkley:2010ge}, ACBAR \citep{NEW_Reichardt:2008ay} and SPT \citep{NEW_Keisler:2011aw} in order to study te relation between $\aem$ and the number of relativistic degrees of freedom $N_{\rm eff}$ that is reported to be higher than its nominal value $N_{\rm eff}=3.046$ -- the deviation from $3$ is related to the non-spontaneous neutrino decoupling, see e.g., \cite{NEW_Pitrou:2018cgg}. Assuming a 6 parameters cosmological model ($\Omega_\mat h^2,\Omega_\baryon h^2, H_0, \tau, n_s,A_s$) they consider 3 models with (1) $\aem$ free ($N_{\rm eff},Y_{\rm p}$ fixed to (3.046,0.24), (2) ($\aem,N_{\rm eff}$) free and $Y_{\rm p}$ fixed and (3) ($\aem,N_{\rm eff},Y_{\rm p}$) free to conclude $\aem/\aem(0) = 0.984\pm 0.005$ (case 1), $\aem/\aem(0) = 0.99\pm 0.006$ (case 2), $\aem/\aem(0) = 0.987\pm 0.014$ (case 3). The freedom of a free $\aem$ increases the error bar on $N_{\rm eff}$ by about 30\%. \citep{NEW_Planck:2014ylh} analyzed the first Planck data release combined with several other data as described in \cite{NEW_Planck:2013nga} and fit a model with 6 cosmological parameters ($\Omega_\mat h^2,\Omega_\baryon h^2, H_0, \tau, n_s,A_s$). Assuming $m_{\rm e}$ constant, they get for the Planck data + WMAP polarization data + high $\ell$ data
\begin{equation}
 \Delta\aem/\aem =(3.6\pm3.7)\times10^{-3}\,.
\end{equation}
From CMB alone, Planck data the constraint on the variation of $\aem$ by a factor of about 5 compared to WMAP-9 and also better that WMAP combined with small scale experiment \citep{NEW_AtacamaCosmologyTelescope:2013swu}, mainly because Planck data is able to break the strong $\aem-H_0$ degeneracy from the observation of the damping tail. The study of the correlation shows that the constraints on the cosmological parameters change very little, exceptions being $n_s$ and $H_0$ and the degeneracy with $Y_{\rm p}$ is confirmed. Finally, they also conclude that $\aem$ is weakly degenerate with foreground, beam and calibration parameters. Assuming $\aem$ constant, \cite{NEW_Planck:2013nga} notes that Planck cannot break the strong degeneracy between $H_0$ and $m_{\rm e}$ since the latter does not affect much the damping tail. But adding external data such as BAO decreases the uncertainty by a factor of order 5 to get
\begin{equation}
\Delta  m_{\rm e}/m_{\rm e} = (4\pm11)\times10^{-3}\,.
\end{equation}
When considering the simultaneous variation of $\aem$ and $m_{\rm e}$ it is shown that thanks to small scale data, the degeneracy is lifted and that the constraints are not substantially changed.

\cite{NEW_Hart:2017ndk} revisited this result with the recombination codes {\tt CosmoRec} \citep{NEW_Chluba:2010ca} and {\tt Recfast++}. From Planck-2015 data combined with  BAO data, they concluded that $\Delta\aem/\aem= (0.7\pm 2.5)\times10^{-3}$ and $\Delta m_{\rm e}/m_{\rm e} = (3.9\pm7.4)\times10^{-3}$ for independent variations,  while for a joint variation they get
\begin{equation}
\Delta\aem/\aem= (1.1\pm 2.6)\times10^{-3}
\quad 
\Delta m_{\rm e}/m_{\rm e} = (5.6\pm8)\times10^{-3}
\end{equation}
including BAO data. \cite{NEW_Hart:2019dxi} included Planck-2018 polarisation to get, with either $m_{\rm e}$ or $\aem$  fixed,
\begin{equation}
\Delta\aem/\aem= (0.05\pm 0.24)\times10^{-2}
\quad 
\Delta m_{\rm e}/m_{\rm e} = (-11.2\pm5.9)\times10^{-2}
\end{equation}
from CMB alone and
\begin{equation}
\Delta\aem/\aem= (0.19\pm 0.22)\times10^{-2}
\quad 
\Delta m_{\rm e}/m_{\rm e} = (0.78\pm0.67)\times10^{-2}
\end{equation}
once BAO are included, while the joint analysis on $(\aem,m_{\rm e})$ gives
\begin{equation}
\Delta\aem/\aem= (0.10\pm 0.24)\times10^{-2}
\quad 
\Delta m_{\rm e}/m_{\rm e} = (0.54\pm0.80)\times10^{-2}.
\end{equation}
while adding SNIa data pushes the electron mass to $\Delta\aem/\aem = (1.9\pm0.55)\times10^{-2}$. These two works also more model-dependent constraints that assume a redshift dependence of the form $(1+z)^p$. To go beyond the hypothesis of a time-independent changes of $(\aem,m_{\rm e})$, \cite{NEW_Hart:2021kad} proposed a model-independent principal component analysis using an eigenmode decomposition of the varying constant during recombination.  From Planck-2018 data, they show that for each constant, three independent modes can be constrained at present. The analysis was extended by \cite{NEW_Hart:2022agu} to include early dark energy. This results highlight the connection between $m_{\rm e}$ and $H_0$ and the dependence of the constraints on the time-variation around the decoupling as well as the importance to extend them up to reionisation.  An attempt to deal with the whole ionization history (i.e., recombination + reionization) was addressed in \cite{NEW_Wang:2022wug} by adding SFR density from UV and IR measurements to the Planck-2018+BAO+SNIa data. They concluded that at 68\% C.L. $(\Delta\aem/\aem)_{\rm CMB}= (1.49 \pm 2)\times10^{-3}$ and $(\Delta\aem/\aem)_{\rm Rei} = -1.46^{+0.31}_{-0.27} \times10^{-1}$.  These constraints suffer a 4.64$\sigma$ discrepancy so that the work mostly illustrate the importance not to restrict to CMB.

\paragraph{Links with the Hubble tension}\ 

The possibility to solve or alleviate the Hubble tension thanks to a variation of the constant has first been stressed by \cite{NEW_Hart:2019dxi} while it was noticed by \cite{NEW_Planck:2013nga} that ``for WMAP there is a strong degeneracy between $H_0$ and $m_{\rm e}$, which is why the uncertainty on $m_{\rm e}/m_{\rm e0}$ is much larger than for Planck''. This hints  that a combined modification of recombination and reionisation physics could be at work.  This has motivated a renewed attention to constrain a \emph{constant shift} of the electron mass between CMB and today with many studies \citep{NEW_Hart:2017ndk,NEW_Hart:2019dxi, NEW_Schoneberg:2021qvd,NEW_Khalife:2023qbu,NEW_Baryakhtar:2024rky,NEW_Seto:2024cgo,NEW_Toda:2024ncp,NEW_Schoneberg:2024ynd,NEW_Toda:2024ncp} considering different data sets. Most of them consider the Planck-2015 \citep{NEW_Planck:2013nga} or Planck-2018 \citep{NEW_Planck:2018vyg} for the CMB, BAO data mostly from the SDSS-DR12 LRG \citep{BOSS:2016wmc}, DR14-QSO/Ly$\alpha$ \citep{eBOSS:2019qwo} and DR7 MGS \citep{Ross:2014qpa} or DESI \citep{DESI:2024lzq} SNIa data from either Pantheon \citep{Pan-STARRS1:2017jku} or PantheonPlus \citep{Scolnic:2021amr}.  Then, from a theoretical point of view they may or may  not consider a non-vanishing curvature, $\Omega_K$, massive neutrinos, $m_\nu$, a varying $\aem$, an equation of state for the dark energy component or a sign shift of the cosmological constant in \cite{NEW_Toda:2024ncp}. The results of their analysis are summarized in Table~\ref{tabme}.  We refer to \cite{NEW_Chluba:2023xqj} and \cite{NEW_Schoneberg:2024ynd} for summaries of the current status of this question as well as to Sect.~\ref{secosmocte}.

\begin{table}[htbp]
\caption[CMB constraints on $(m_{\rm e},H_0)$ in view of the Hubble tension.]{Summary of the main results on a constant shift on $m_{\rm e}$ and their consequences for the Hubble constant. The data use and the extra parameters than have been varied are also indicated; see text for definitions.}
\label{tabme}
\centering
{\footnotesize
\begin{tabular}{cc|c|ccc|r}
\toprule
 $\Delta m_{e}/m_{\rm e}\,  (\times 10^2)$ &  $H_0 \unit{[km/s/Mpc]}$ & Hypothesis &  CMB & BAO & SNIa & Ref. \\
 \midrule
 $1.1^{+7.7}_{-5.7}$ & $72\pm10$  &   & W9 &  &  & \cite{NEW_Planck:2013nga}\\     
  $4\pm9.1$ & $73\pm10$  &  $\aem$  & W9 &  &  & \cite{NEW_Planck:2013nga}\\     
  $2.7\pm1.2$ & $73.5\pm2.4$  &   & P13&  SDSS+BOSS& HST  & \cite{NEW_Planck:2013nga}\\    
 $0.39\pm0.74$ & $68.1\pm1.3$  &   & P15 & SDSS &  & \cite{NEW_Hart:2017ndk}\\          
  $0.56\pm0.80$  & $68.1\pm1.3$  & $\aem$  & P15 & SDSS &  & \cite{NEW_Hart:2017ndk}\\    
   $0.47\pm0.66$ & $68.46\pm1.26$  &   & P18 & SDSS &  & \cite{NEW_Hart:2019dxi}\\          
  $1.5\pm1.8$  & $69.29\pm2.11$  & $\Omega_K$  & P18 & SDSS &  & \cite{NEW_Hart:2019dxi}\\    
    $0.3\pm0.6$  & $68.0\pm1.1$  &  & P18+SPT & SDSS &  Pan & \cite{NEW_Khalife:2023qbu}\\    
   $0.35\pm1.64$ & $68.2\pm1.6$  &  $\Omega_K$ & P18+SPT & SDSS & Pan & \cite{NEW_Khalife:2023qbu}\\          
  $3\pm3$  & $69.8^{+1.8}_{-2.9}$  & $\Omega_K, m_\nu$  & P18+SPT & SDSS & Pan  & \cite{NEW_Khalife:2023qbu}\\     
   $0.92\pm0.55$ & $69.44\pm0.84$  &  & P18 & SDSS +DESI& Pan & \cite{NEW_Seto:2024cgo}\\          
  $1.3\pm1.4$  & $69.7\pm 1.4$  & $\Omega_K$  & P18& SDSS+DESI & Pan  & \cite{NEW_Seto:2024cgo}\\    
     $-1.00^{+1.09}_{-1.04}$ & $65.1^{+2.2}_{-2.0}$ &  & P18 &  & Pan+ & \cite{NEW_Baryakhtar:2024rky}\\       
       $1.21\pm0.63$  & $70.03\pm1.06$  & & P18 & DESI &   & \cite{NEW_Schoneberg:2024ynd}\\     
   $1.88\pm0.52$ & $71.61\pm1.00$  &  $\Omega_K$   & P18 & DESI& Pan+ & \cite{NEW_Schoneberg:2024ynd}\\          
  $0.72\pm0.84$  & $69.38\pm 2.17$  & $w$  & P18& DESI & Pan+  & \cite{NEW_Schoneberg:2024ynd}\\         
  $0.61\pm0.65$  & $68.8\pm 1.1$  &  & P18& SDSS & Pan  & \cite{NEW_Toda:2024ncp}\\      
    $1.9\pm 1.8$  & $69.7\pm 1.7$  &$\Omega_K$   & P18& SDSS & Pan  & \cite{NEW_Toda:2024ncp}\\      
\bottomrule
\end{tabular}
}
\end{table}

\begin{figure}[hptb]
  \vskip-.25cm
  \centerline{\includegraphics[scale=0.4]{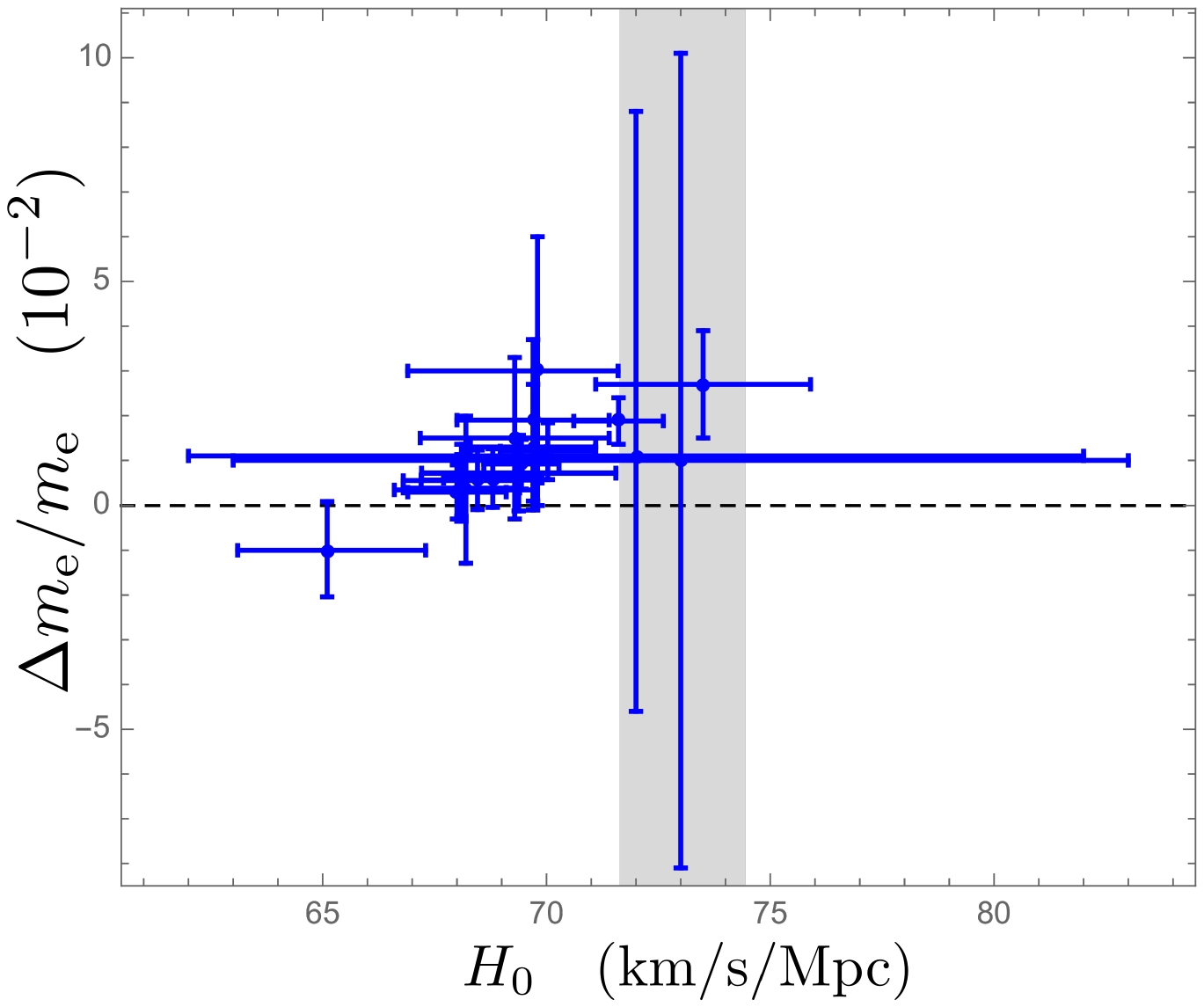}}
  \vskip-.5cm
  \caption[CMB constraints on $(m_{\rm e},H_0)$ in view of the Hubble tension]{Summary of the main results on a constant shift on $m_{\rm e}$ and their consequences for the Hubble constant; see Table~\ref{tabme}. The vertical gray zone indicates the  Cepheid-calibrated supernovae determination of the Hubble constant, $H_0 = (73.04\pm1.04)$~km/s/Mpc by the SH0ES experiment \citep{Riess_2022}.}
  \label{fig-home}
\end{figure}

\paragraph{Summary}\ 

The main limitation of these analyses lies in the fact that the CMB angular power spectrum depends on the evolution of both the background spacetime and the cosmological perturbations. It follows that it depends on the whole set of cosmological parameters as well as on initial conditions, that is on the shape of the initial power spectrum. Hence the results will always be conditional to the model of structure formation. The constraints on $\aem$ or $m_{\mathrm{e}}$ can then be seen mostly as constraints on a delayed recombination. A strong constraint on the variation of $\aem$ can be obtained from the CMB only if the cosmological parameters are independently known. \cite{cmb1-2003} forecasts that CMB alone can determine $\aem$ to a maximum accuracy of 0.1\% and \cite{NEW_Martinelli:2012zf} showed that including lensing data from Euclid experiment will also to reach an accuracy of $8\times10^{-4}$. These tests push our understanding of the recombination and turn to be important in view of the Hubble tension.

\begin{table}[htbp]
\caption[Constraints on the variation of $\aem$ from CMB data.]{Summary of the constraints on the variation of $\aem$ obtained from  the analysis of CMB  data. All, unless specified, assume $\Omega_K=0$. Concerning the data \emph{W-(1,3,5,7)} refers to the WMAP 1,..,7 years data, \emph{P15/P18} to the Planck 2015 and 2018 releases and \emph{pW} to pre-WMAP data such as BOOMERanG, MAXIMA or COBE -- see text -- specifying the use of temperature only (\emph{T}) or temperature +polarisation (\emph{T+Pol}).  \emph{BBN} refers to the use of BBN data. The cosmological parameters involved on the statistical analysis are specified in the text.}
\label{tab-cmb}
\centering
{\footnotesize
\begin{tabular}{lllr}
\toprule
 Constraint  					&  Data  	& Comment   	& Ref. \\
($\Delta\aem/\aem \times 10^{2}$)  	& 		&  		& \\
 \midrule
  $(-3.5\times5.5)$ 		& pW, BBN & $\aem$ only & \cite{cmb-avelino01}\\
  $(-6\pm8)$	   		 & pW    &  $\aem$ only	& \cite{cmb-landau01} \\
  $(-1.5\pm3.5)$		& W1 & include $\alpha_s$ & \cite{cmb1-2003}\\
   $(-2.5\pm3.56$		 & W1 & $\alpha_s=0$ &  \cite{cmb1-2003}\\
     $(-3.15\pm6.55)$   & W1 &  $m_{\mathrm{e}}$ free & \cite{cmb2-2006}\\
  $(10.9\pm15.1)$   & W1, $H_0$ & same &  \cite{cmb2-2006}\\
    $(-1.45\pm2.45)$   & W3 (T+Pol), $H_0$ & & \cite{cmb3-2007}\\
  $(0.3\pm1.5)$   & W5, 2df & $m_{\mathrm{e}}$ free & \cite{scoccola}\\ 
  $(-0.4\pm4.6)$   & W5, $H_0$  &  & \cite{wmap-alpha}\\     
    $(-0.25\pm4.05)$   & W5 + Pol &  & \cite{menegoni}\\ 
  $(0.1\pm 1.4 )$   & W5 + Pol,  $H_0$ & & \cite{menegoni}\\  
    $(-0.324\pm0.505)$ & W5 (T+Pol) & $m_{\rm e}$ free & \cite{naka00}\\
      $(-1.4\pm1.1)$ & W7 (T+Pol), SDSS & $m_{\mathrm{e}}$ free & \cite{scoccola3}\\      
     $(1.6\pm 0.5)$ & W7+Pol &  & \cite{NEW_Menegoni:2012tq}\\ 
   $(-1\pm 0.6)$ &  W7+Pol& $Y_{\rm p}$ fixed & \cite{NEW_Menegoni:2012tq}\\ 
    $(1.3\pm 1.4)$  & W7+Pol & $(N_{\rm eff},Y_{\rm p})$ free & \cite{NEW_Menegoni:2012tq}\\     
     $ (0.36\pm0.37)$ & P15  + W7(Pol) & $m_{\mathrm{e}}$ free  & \cite{NEW_Planck:2014ylh}  \\
$(0.07\pm0.25)$ & P15 + BAO  &  & \cite{NEW_Hart:2017ndk}\\      
$(0.11\pm0.26)$ & P15 + BAO  & $m_{\mathrm{e}}$ free & \cite{NEW_Hart:2017ndk}\\      
 $(0.05\pm0.24)$ & P18 (T+Pol)&  & \cite{NEW_Hart:2019dxi}\\          
 $(0.19\pm0.22)$ & P18 (T+Pol), BAO  & & \cite{NEW_Hart:2019dxi}\\       
 $(0.10\pm 0.24)$ & P18 (T+Pol), BAO & $m_{\mathrm{e}}$ free & \cite{NEW_Hart:2019dxi}\\      
$(0.149\pm0.2)$& P18, BAO, SNI  & &  \cite{NEW_Wang:2022wug}\\         
\bottomrule
\end{tabular}
}
\end{table}

\subsection{21 cm} \label{subsec55}

After recombination, the CMB photons are redshifted and their temperature drops as $(1+z)$. However, the baryons are prevented from cooling adiabatically since the residual amount of free electrons, that can couple the gas to the radiation through Compton scattering, is too small. It follows that the matter decouples thermally from the radiation at a redshift of order $z\sim200$.

The intergalactic hydrogen atoms after recombination are in their ground state, which hyperfine-structure splits into a singlet and a triple states ($1s_{1/2}$ with $F=0$ and $F=1$ respectively, see Sect.~III.B.1 of FCV [\citealt{jpu-revue}]). \cite{21cm-1} proposed  that the observation of the 21~cm emission can provide a test on the fundamental constants. We refer to \cite{21cm-2} for a detailed review on 21~cm.

The fraction of atoms in the excited (triplet) state versus the ground (singlet) state is conventionally related by the spin temperature $T_{\mathrm{s}}$ defined by the relation
\begin{equation}
 \frac{n_t}{n_s} = 3 \exp\left(-\frac{T_*}{T_{\mathrm{s}}}\right)
\end{equation}
where $T_*\equiv hc/(\lambda_{21}k_{\mathrm{B}})=68.2 \unit{mK}$ is the temperature corresponding to the 21~cm transition and the factor 3 accounts for the degeneracy of the triplet state (note that this is a very simplified description since the assumption of a unique spin temperature is probably not correct \citep{21cm-2}. The population of the two states is determined by two processes, the radiative interaction with CMB photons with a wavelength of $\lambda_{21}=21.1 \unit{cm}$ (i.e., $\nu_{21}=1420 \unit{MHz}$) and spin-changing atomic collision. Thus, the evolution of the spin temperature is dictated by \citep{21cm-2}.
\begin{equation}\label{e_21Ts}
 \frac{\dd T_{\mathrm{s}}}{\dd t} = 4C_{10}\left(\frac{1}{T_{\mathrm{s}}}-
                                    \frac{1}{T_{\mathrm{g}}}\right)T_{\mathrm{s}}^2
                                +(1+z)HA_{10}\left(\frac{1}{T_{\mathrm{s}}}-
                                    \frac{1}{T_\gamma}\right)\frac{T_\gamma}{T_*}\,.
\end{equation}
The first term corresponds to the collision de-excitation rate from triplet to singlet and the coefficient $C_{10}$ is decomposed
as
$$
 C_{10} = \kappa_{10}^{HH}n_p + \kappa_{10}^{eH}x_{\mathrm{e}}n_p
$$
with the respective contribution of H-H and $e$-H collisions. The second term corresponds to spontaneous transition and $A_{10}$ is the Einstein coefficient. The equation of evolution for the gas temperature $T_{\mathrm{g}}$ is given by Eq.~(\ref{e_cmbT}) with $T_M=T_{\mathrm{g}}$ (we recall that we have neglected the contribution of helium) and the electronic density satisfies Eq.~(\ref{eecmb1}).

\begin{table}[htbp]
\caption[Main dependencies of the 21~cm parameters in $(\aem,m_{\rm e})$]{Summary of the main dependencies of the 21~cm relevant parameters as listed by \cite{NEW_Lopez-Honorez:2020lno} assuming $m_{\rm p}$ and $g_{\rm p}$ constant. The expression of $S_\alpha$ if given in Eq.~(2.17) of \cite{NEW_Lopez-Honorez:2020lno}, as obtained from \cite{NEW_Furlanetto:2006fs} and ${\cal G}_i$ a function detailed in \cite{NEW_Madau:2016jbv}.}
\label{tab-21cm}
\centering
{\footnotesize
\begin{tabular}{lllc}
\toprule
 Constraint  &   & dependence  \\
 \midrule
 hyperfine transition & $\nu_{21}$   & $\aem^4m_{\rm e}^2$  \\
 Spontaneous emission coefficient of the 21~cm transition & $A_{10}$   &  $\aem^{13}m_{\mathrm{e}}$ \\
 Brightness temeperature factor & $\tau_{\nu_{21}} T_{\rm S}$  & $\aem^5$  \\
 Lyman-$\alpha$ frequency   &   $\nu_\alpha$ &  $\aem^2m_{\rm e}^2$ \\
  Proper Lyman-$\alpha$ intensity   & $\tilde J_{\alpha,\star}$  & $\aem^{-2}m_{\rm e}^{-1}$  \\
 Spontaneous emission coefficient of the Lyman-$\alpha$ transition & $A_{\alpha}$   &  $\aem^{5}m_{\mathrm{e}}$ \\
  Gunn-Peterson optical depth&   $\tau_{\rm GP}$&  $\aem^{-1}m_{\rm e}^{-2}$ \\
   Lyman-$\alpha$ coupling     & $ x_\alpha$  &  $S_\alpha\aem^{-10}m_{\rm e}^{-4}$  \\
    Recombination case B-coefficient&  $\alpha_{\rm B}$ &  $\aem^{3.4}m_{\rm e}^{-3/2}$ \\
     Ground state energy of specie $i$&   $h\nu_i$& $\aem^{2}m_{\rm e}$  \\
      Ionization cross-section of specie $i$ &  $\sigma_i$ & ${\cal G}_i\aem^{-1}m_{\rm e}^{-2}$  \\
\bottomrule
\end{tabular}
}
\end{table}

It follows \citep{21cm-1,21cm-2b} that the change in the brightness temperature of the CMB at the corresponding wavelength scales as $T_{\mathrm{b}}\propto A_{12}/\nu_{21}^2$, where the Einstein coefficient $A_{12}$ is defined below. Observationally, we can deduce the brightness temperature from the brightness $I_\nu$, that is the energy received in a given direction per unit area, solid angle and time, defined as the temperature of the black-body radiation with spectrum $I_\nu$. Thus, $k_{\mathrm{B}}T_{\mathrm{b}} \simeq I_\nu c^2/2\nu^2$. It has a mean value, $\bar T_{\mathrm{b}}(z_{\text{obs}})$ at various redshifts where $1+z_{\text{obs}} = \nu_{21}^{\text{today}}/\nu_{\text{obs}}$. Besides, as for the CMB, there will also be fluctuation in $T_{\mathrm{b}}$ due to imprints of the cosmological perturbations on $n_p$ and $T_{\mathrm{g}}$. It follows that we also have access to an angular power spectrum $C_\ell(z_{\text{obs}})$ at various redshifts (see \citealt{21cm-3} for details on this computation).

Both quantities depend on the value of the fundamental constants. Beside the same dependencies of the CMB that arise from the Thomson scattering cross-section, we have to consider those arising from the collision terms. In natural units, the Einstein coefficient scaling is given by $A_{12}=\frac23\pi\aem\nu_{21}^3m_{\mathrm{e}}^{-2}\sim 2.869\times10^{-15} \unit{s}^{-1}$. It follows that it scales as $A_{10}\propto g_{\mathrm{p}}^3\mu^3\aem^{13}m_{\mathrm{e}}$. The brightness temperature depends on the fundamental constant as $T_{\mathrm{b}}\propto g_{\mathrm{p}}\mu\aem^{5}/m_{\mathrm{e}}$. Note that the signal can also be affected by a time variation of the gravitational constant through the expansion history of the universe. \cite{21cm-1} (see also \citealt{21cm-2} for further discussions), focusing only on $\aem$, showed that this was the dominant effect on a variation of the fundamental constant (the effect on $C_{10}$ is much complicated to determine but was argued to be much smaller). It was estimated that a single station telescope like LWA\footnote{\url{http://lwa.unm.edu}} or LOFAR\footnote{\url{http://www.lofar.org}} can lead to a constraint of the order of $\Delta\aem/\aem\sim0.85\%$, improving to 0.3\% for the full LWA. The fundamental challenge for such a measurement is the subtraction of the foreground.

The 21~cm absorption signal in a available on a band of redshift typically ranging from $z\lesssim1000$ to $z\sim20$, which is between the CMB observation and the formation of the first stars, offering a unique window on  the ``dark age''. Thus, it offers an interesting possibility to trace the constraints on the evolution of the
fundamental constants between the CMB epoch and the quasar absorption spectra.

As for CMB, the knowledge of the cosmological parameters is a limitation since a change of 1\% in the baryon density or the Hubble parameter implies a 2\% (3\% respectively) on the mean bolometric temperature. The effects on the angular power spectrum have been estimated but still require an in depth analysis along the lines of, e.g., \cite{21cm-3}. It is motivating since $C_{\ell}(z_{\text{obs}})$ is expected to depend on the correlators of the fundamental constants, e.g., $\langle\aem(\bx,z_{\text{obs}})\aem(\bx',z_{\text{obs}})\rangle$ and thus in principle allows to study their fluctuation, even though it will also depend on the initial condition, e.g., power spectrum, of the cosmological perturbations.

Using the public tool {\tt 21vmFASTv2} \citep{NEW_Mesinger:2010ne} and the {\tt Recfast++} code for reionisation (in the version by \cite{NEW_Hart:2017ndk} implementing varying constant to get the initial conditions on $x_{\rm e}$ and the temperature of the gas at $z=30$, \cite{NEW_Lopez-Honorez:2020lno} discussed the dependence of the 21-cm signal on $(\aem,m_{\rm e})$ and their degeneracies with astrophysical parameters (see Table~\ref{tab-21cm}) to conclude that their variations can be hope to be constrained at a level of ${\cal O}(10^{-3})$ with future SKA data \citep{NEW_Weltman:2018zrl}.

Thanks to its high resolution in radio spectral lines, SKA1-Low has good prospects to use e.g., lines from HI and the OH radical) to constrain $\aem$  \citep{NEW_Weltman:2018zrl}.  The covered redshifts for SKA1-Low will be, e.g., $z<13$ for the HI-21~cm absorption and $z<16$ for the ground-state OH-18~cm absorption \citep{NEW_Curran:2004mg}.  \cite{21cm-1}  proposed another method to measure $\aem$ through the 21 cm absorption of CMB photons. They estimated that the change of 1\% of its value shall modify the mean brightness temperature decrement of the CMB due to 21~cm absorption by 5\% over the redshift range $z\in[30,50]$.

In conclusion, the 21~cm observation will open a observational window on the fundamental at redshifts ranging typically from 30 to 100, but full in-depth analysis is still required (see \citealt{21com,21reply} for a critical discussion of this probe). 

\subsection{Galaxy clusters}\label{subsecSZ}

\cite{NEW_Galli:2012bf} proposed a new method based on the idea that hot clusters radiate  in the X-ray mainly due to bremsstrahlung, while they leave an imprint on the CMB frequency spectrum through the Sunyaev-Zel’dovich effect. The SZ Compton parameter $y$ depends on the Thomson cross-section as
\begin{equation}
 y = \sigma_{\rm T} \int n_{\rm e} \frac{k_{\rm B}T}{m_{\rm e}c^2} \dd l =\frac{ \sigma_{\rm T}}{m_{\rm e}c^2} \int  P \dd l 
\end{equation}
with $P=n_ek_{\rm B}T$ the pressure of the intracluster medium. The integrated Compton parameter, $Y_{{\rm SZ}}$, is defined as
\begin{equation}
Y_{{\rm SZ}}=\int y({\bf n})\dd^2\Omega = \frac{ \sigma_{\rm T}}{m_{\rm e}c^2} \int  P \dd l \dd^2\Omega = \frac{ \sigma_{\rm T}}{m_{\rm e}c^2} \int  D_A^{-2} P \dd^3V 
\end{equation}
with $D_A$ the angular distance. Using Eq.~(\ref{cmb3}) for $\sigma_{\mathrm{T}}$ and since the integral on the line of sight peaks at the redshift on the cluster one gets
\begin{equation}
Y_{{\rm SZ}}  D_A^{2}(z) = \frac{8\pi}{3}\frac{\hbar^2}{m_{\rm e}^3c^4}\aem^2 \int  P \dd^3V.
\end{equation}
Concerning the X-ray emission, one observes the surface brightness derived from the thermal bremsstrahlung emissivity \citep{NEW_Galli:2012bf}. It is common to define the parameter $Y_X=M T_X$ in term of the mass  and temperature of the cluster. The former is shown to scale \citep{NEW_Sasaki:1996jr} as 
$$
M\sim \aem^{-3/2}m_{\rm e}^{3/4} m_{\rm H}D_LD_A^{3/2},
$$
with $m_{\rm H}$ the mass of hydrogen, that we identify to $m_{\rm p}$ for simplicity here. The ratio $Y_{\rm SZ}D_A^2/Y_X$ is then expected to be constant from numerical simulations and current observations, and it scales as $\aem^{7/2}$. More precisely we get that
\begin{eqnarray}
\frac{Y_{{\rm SZ}}  D_A^{2}}{Y_X} &=& \left(\frac{\sigma_{\rm T}}{m_{\rm e}m_{\rm p}\mu_{\rm e}c^2}\right)\left(\frac{\aem}{\aem(0)}\right)^{3/2}
\left(\frac{m_{\rm e}}{m_{\rm e}(0)}\right)^{-3/4}\left(\frac{m_{\rm H}}{m_{\rm H}(0)}\right)^{-1} \nonumber\\
&& \qquad \times \eta(z) 
\frac{\int n_{\rm e}T\dd^3V}{T_X\int n_{\rm e}\dd^3V}
\end{eqnarray}
with $\eta(z)\equiv D_{\rm L}/(1+z)^2D_{\rm A}$ a parameter that is equal to 1 as long as distance duality holds \citep{NEW_Uzan:2004my}. Hence
\begin{equation}
\left(\frac{Y_{{\rm SZ}}  D_A^{2}}{Y_X}\right)_i = \left(\frac{\aem}{\aem(0)}\right)^{7/2} \! \left(\frac{m_{\rm e}}{m_{\rm e}(0)}\right)^{-15/4}\! \left(\frac{m_{\rm p}}{m_{\rm p}(0)}\right)^{-2}\! \eta(z) \left(\frac{Y_{{\rm SZ}}  D_A^{2}}{Y_X}\right)_0\! .
\end{equation}
Using a catalog of 61 selected clusters  with $z\in[0.2,1.5]$ with SZ detected by SPT and X-ray measurement from XMM-Newton, and assuming no variation in $m_{\rm e}$ and $m_{\rm p}$ and that $\eta(z)=1$, \cite{NEW_Galli:2012bf} concluded that
\begin{equation}
\left|\Delta \aem/\aem\right|<8\times10^{-3}
\end{equation}
at 66\% C.L., not including uncertainties on the cosmological parameters used to determine the angular diameter distance. As pointed out by the author, the data  are neither a complete nor a representative sample of clusters and used to provide a first estimate of the accuracy that can be reached by this new method. This constraint was later confirmed by \cite{NEW_Bora:2020sws} with a catalog of 58 SPT clusters with $Y_X$ determined by XMM-Newton. Assuming a redshift evolution of the form $\Delta\aem/\aem =-\gamma\ln(1+z)$ they concluded that $\gamma=0.95^{+0.12}_{-0.11}$ hence showing no evolution with $z$. \cite{NEW_Colaco:2019fvl} got $\gamma=-0.15\pm0.17$ and \cite{NEW_DeMartino:2016hxb} used the Planck 2013 data to measure the thermal SZ effect at the location of 618 X-ray selected clusters but focused on a dipole variation (see Sect.~\ref{subsec-dipalpha}). \cite{NEW_Albuquerque:2024upw} applied this test by combining the catalog of 44 X-ray data by \cite{SPT:2021vsu} and SNIa data to infer that $\Delta\bar\mu<1\%$ on a redshift band $[0.018,1.160]$. See also \cite{NEW_Goncalves:2019xtc} for the influence of $\eta$ on this test.

This method is promizing given the growing activity of SZ surveys (such as SPTpol, SPT-3G, ACTpol) that shall detect an order of magnitude more clusters up to $z\sim2$ and X-ray surveys (such as eROSITA that shall discover approx. $10^5$ clusters). It allows one for a test of $\aem$ variation and a possible dipolar modulations. It still assumes the validity of the distance duality that is intertwined with a variation of $\aem$ and $m_{\rm e}$ (see e.g., \cite{NEW_Holanda:2015oda,NEW_Holanda:2019vmh})
which is today bounded to $|\Delta\eta|<10^{-4}$ \citep{NEW_Ellis:2013cu} between decoupling and today.

\subsection{Fast Radio Bursts}\label{subsecFRB}

Fast Radio Bursts are millisecond transient events in radio frequency; see \cite{Platts:2018hiy,Petroff:2021wug} for reviews. While many models have been proposed to explain the burst’s origin, the emission mechanism remains unknown  \citep{Platts:2018hiy}. One of the key parameter is the FRB’s dispersion measure (DM) related to the density of free electrons along the line of sight. Starting from the time for a pulse to reach Earth from a source at distance $d$ with frequency $\omega$ and groupe velocity $v_g$, 
$$
t_p=\int_0^d \frac{\dd s}{v_g}\simeq d+\int_0^d \frac{\omega_p^2}{2\omega^2} \dd s
$$
where $\omega_p=4\pi e^2 n_e/m_{\rm e}$ is the plasma frequency, one defines
\begin{equation}
DM(d)=\int_0^d n_e \dd s \qquad\hbox{so that} \qquad \frac{\dd t_p}{\dd \omega} = -\frac{4\pi e^2}{m_{\rm e}\omega^3} DM(d).
\end{equation}
This expression neglects the expansion of the universe and assumes the constancy of the fundamental constants. In a cosmological framework, this rewrites as \citep{Lemos:2022kdh}
$$
 \frac{\dd t_p}{\dd \omega_0} = -\frac{4\pi \aem(0)}{m_{\rm e}(0)\omega_0^3} DM(z)
$$
with
\begin{equation}
  DM=\int_0^z \frac{n_e(z')}{(1+z')^2H(z')}\frac{\aem(z')}{\aem(0)} \frac{m_{\rm e}(0)}{m_{\rm e}(z')} \dd z'\,,
\end{equation}
where $\omega_0$ is the observed frequency. The main difficulty is to evaluate the contribution of the interstellar medium, Milky Way halo, host galaxy and intergalactic medium. This requires a difficult modelization to extract the contribution of the intergalactic medium. \cite{NEW_Lemos:2024jbl} applied the method to a catalog of 17 FRB upto a redshift of $0.47$ assuming that $\Delta\aem/\aem =-\gamma \ln(1+z)$ and $\mu$ constant to conclude that $\gamma<1\%$. While it still has to face many systematics and difficult modelizations, the growth of FRB catalogues still makes it an idea to be developed further.

\subsection{Big Bang nucleosynthesis}\label{secbbn}

\subsubsection{Overview}\label{secbbnoverview}

The amount of $^{4}$He produced during the big bang nucleosynthesis (BBN) is mainly determined by the neutron to proton ratio at the freeze-out of the weak interactions that interconvert neutrons and protons. The results thus depend on $G$, $\aw$, $\aem$ and $\as$ respectively through the expansion rate, the neutron to proton ratio, the neutron-proton mass difference and the nuclear reaction rates, besides the standard parameters such as, e.g., the number of neutrino families.

The standard BBN scenario \citep{bbn-cyburt2,peteruzanbook,NEW_Pitrou:2018cgg} proceeds in three main steps:

\begin{enumerate}
\item for $T>1 \unit{MeV}$, ($t<1 \unit{s}$) a first stage during which the neutrons, protons, electrons, positrons an neutrinos are kept in statistical equilibrium by the (rapid) weak interaction
\begin{equation} \label{bbn0}
n\longleftrightarrow p+e^-+\bar\nu_e, \quad
n+\nu_e\longleftrightarrow  p+e^-, \quad
n+e^+\longleftrightarrow p+\bar\nu_e.
\end{equation}
As long as statistical equilibrium holds, the neutron to proton ratio is
\begin{equation}
(n/p)=\hbox{e}^{-Q_{\mathrm{np}}/k_{\mathrm{B}}T}
\end{equation}
where $Q_{\mathrm{np}}\equiv (m_{\mathrm{n}}-m_{\mathrm{p}})c^2=1.29 \unit{MeV}$. The abundance of the other light elements is given by \citep{peteruzanbook}
\begin{align}
Y_A=&g_A\left(\frac{\zeta(3)}{\sqrt{\pi}}\right)^{A-1}2^{(3A-5)/2}A^{5/2}
     \left[\frac{k_{\mathrm{B}}T}{m_{\mathrm{N}}c^2}\right]^{3(A-1)/2} \nonumber\\
     & \eta^{A-1}Y_{\mathrm{p}}^ZY_{\mathrm{n}}^{A-Z}\hbox{e}^{B_A/k_{\mathrm{B}}T},
\end{align}
where $g_A$ is the number of degrees of freedom of the nucleus $_Z^A{\mathrm{X}}$, $m_{\mathrm{N}}$ is the nucleon mass, $\eta$ the baryon-photon ratio and $B_A\equiv(Zm_{\mathrm{p}}+(A-Z)m_{\mathrm{n}}-m_A)c^2$ the binding energy.
\item Around $T\sim0.8 \unit{MeV}$ ($t\sim2$~s), the weak interactions freeze out at a temperature $T_{\mathrm{f}}$ determined by the competition between the weak interaction rates and the expansion rate of the universe and thus roughly determined by $\Gamma_{\mathrm{w}}(T_{\mathrm{f}})\sim H(T_{\mathrm{f}})$ that is
\begin{equation}
\gfermi^2(k_{\mathrm{B}}T_{\mathrm{f}})^5\sim\sqrt{GN_*}(k_{\mathrm{B}}T_{\mathrm{f}})^2
\end{equation}
where $\gfermi$ is the Fermi constant and $N_*$ the number of relativistic degrees of freedom at $T_{\mathrm{f}}$. Below $T_{\mathrm{f}}$, the number of neutrons and protons change only from the neutron $\beta$-decay between $T_{\mathrm{f}}$ to $T_{\mathrm{N}}\sim0.1 \unit{MeV}$ when $p+n$ reactions proceed faster than their inverse dissociation.

\item For $0.05 \unit{MeV}<T<0.6 \unit{MeV}$ ($3 \unit{s}<t<6 \unit{min}$), the synthesis of light elements occurs only by two-body reactions. This requires the deuteron to be synthesized ($p+n\rightarrow D$) and the photon density must be low enough for the photo-dissociation to be negligible. This happens roughly when
\begin{equation} \label{n0}
\frac{n_{\mathrm{d}}}{n_\gamma}\sim\eta^2\exp(-B_{\rm D}/T_{\mathrm{N}})\sim 1
\end{equation}
with $\eta\sim3\times10^{-10}$. The abundance of $^{4}$He by mass, $Y_{\mathrm{p}}$, is then well estimated by
\begin{equation} \label{n1}
Y_{\mathrm{p}}\simeq2\frac{(n/p)_{\mathrm{N}}}{1+(n/p)_{\mathrm{N}}}
\end{equation}
with
\begin{equation}
\label{n2}
(n/p)_{\mathrm{N}}=(n/p)_{\mathrm{f}}\exp(-t_{\mathrm{N}}/\tau_{\mathrm{n}})
\end{equation}
with $t_{\mathrm{N}}\propto G^{-1/2}T_{\mathrm{N}}^{-2}$ and $\tau_{\mathrm{n}}^{-1}=1.636\,\gfermi^2(1+3g_A^2)m_{\mathrm{e}}^5/(2\pi^3)$, with $g_A\simeq1.26$ being the axial/vector coupling of the nucleon. Assuming that $B_{\rm D}\propto\as^2$, this gives a dependence $t_{\mathrm{N}}/\tau_{\mathrm{p}}\propto G^{-1/2} \as^2 \gfermi^2$.
\item The light element abundances, $Y_i$, are then obtained by solving a series of nuclear reactions
$$
 \dot{Y}_i = J_i - \Gamma_{ij} Y_j,
$$
where $J_{i}$ and $\Gamma_{ij}$ are time-dependent source and sink terms.
\end{enumerate}

\subsubsection{Status of standard BBN}

Many public codes implement the standard BBN model, {\tt PArthENoPE} \citep{NEW_Consiglio:2017pot}, {\tt PRIMAT} \citep{NEW_Pitrou:2018cgg}, {\tt PRyMordial} \citep{NEW_Burns:2023sgx}. All the reaction rates required for the BBN predictions are measured in accelerators so that, assuming the Copernican principle, BBN has only two free parameters, namely $\Omega_{\rm b}h_0^2$ and $N_{\rm eff}$. The predictions of  the light elements abundances can be computed as a function of $\eta$ and compared to their observed abundance (see Fig.~\ref{fig-bbn} that summarizes the observational constraints obtained on helium-4, helium-3, deuterium and lithium-7). $\eta$ is related to the baryon density, a parameter measured by other cosmological probes such as the CMB by the relation \citep{NEW_Pitrou:2018cgg,NEW_Fields:2019pfx,NEW_Yeh:2020mgl}
\begin{equation}
\frac{\Omega_{\rm b}h_0^2}{0.0224} =\left(\frac{\eta}{6.13197 \times10^{-10}}\right)
\left(\frac{T_{\rm CMB}}{2.7255~\unit{K}}\right)
\left( \frac{1-1.759\times10^{-3} \frac{Y_{\rm_{\rm p}}}{0.2471}}{1-1.759\times10^{-3} }\right)
\end{equation}
Prior to WMAP, these parameters were adjustable but are now determined with high accuracy from the CMB analysis. WMAP data \citep{wmap} have led to the conclusion that ${\eta = \eta_{\text{WMAP}} = (6.19\pm0.15)\times10^{-10}}$.  This value has been reevaluated by Planck-2018 \citep{NEW_Planck:2018vyg} to
\begin{eqnarray}
  \Omega_{\rm b}h_0^2 &=& 0.02237\pm0.00015\qquad\hbox{(CMB)}\nonumber\\
  \Omega_{\rm b}h_0^2& =& 0.02242\pm0.00014\qquad\hbox{(CMB +BAO)}.
\end{eqnarray}
Then, the effective number of neutrinos has been reevaluated to $N_{\rm eff}=3.044$ for 3 families, taking into account their non-instantaneous decoupling \citep{NEW_Froustey:2020mcq} and the neutron decay constant is now measured to ${\tau_{\rm n}=879.4\pm0.6}$~s \citep{NEW_ParticleDataGroup:2020ssz}.

Concerning  the observations, the spectroscopic abundances. \emph{helium-3} is not very constraining because it is both produced and destroyed in stars so that the evolution of its abundance in time is not very precise. \emph{Lithium-7} exhibits a factor of order 3 discrepancy, which is usually discarded quietly, the consensus being that it cannot arise from the nuclear sector. Finally, the most most recent recommended observed value for  \emph{Deuterium} \citep{NEW_Cooke:2017cwo} is D/H$=(2.527\pm0.030)\times10^{-5}$ at a redshift $z\sim2.5-3.1$, making it the most constraining BBN observable because both its observational measurement and its theoretical prediction reach 1\% accuracy. It has recently been pointed out that differences between BBN codes are attributed to different choices made when modelling the nuclear cross-sections, and not on weak rates. It is important to keep in mind that one would need to control their accuracy at least at the percent level and to take into account the latest data; see \cite{NEW_Fields:2019pfx,NEW_Pisanti:2020efz,NEW_Pitrou:2020etk,NEW_Pitrou:2021vqr}.

There exists a long-time discrepancy between the predicted abundance of lithium-7 based on the CMB results \citep{cocbbn,cocnew} for $\eta$, ${}^7{\mathrm{Li}}/{\mathrm{H}}=(5.14\pm0.50)\times10^{-10}$ and its values measured in metal-poor halo stars in our galaxy \citep{bbnbonif}, ${{}^7{\mathrm{Li}}/{\mathrm{H}}=(1.26\pm0.26)\times10^{-10}}$, which is a factor of three lower, at least \citep{bbn-cyburt} (see also \citealt{spite2}), than the predicted value. No solution to this \emph{lithium-7} problem is known. A back of the envelope estimate shows that we can mimic a lower $\eta$ parameter, just by modifying the deuterium binding energy, letting $T_{\rm N}$ unchanged, since from Eq.~(\ref{n0}), one just need $\Delta B_{\rm D}/T_{\mathrm{N}}\sim -\ln 9$ so that the effective $\eta$ parameter, assuming no variation of constant, is three times smaller than $\eta_{\text{WMAP}}$. This rough rule of thumb explains that the solution of the lithium-7 problem may lie in a possible variation of the fundamental constants.

\subsubsection{Constants everywhere}

In complete generality, the effect of varying constants on BBN is difficult to model because of the intricate structure of QCD and its role in low energy nuclear reactions. Thus, a solution is to proceed in \emph{two steps}, first by determining the dependencies of the light element abundances on the BBN parameters and then by relating those parameters to the fundamental constants.

The analysis of the previous Sect.~\ref{secbbnoverview}, that was restricted to the helium-4 case, clearly shows that the abundances will depend on: (\textit{1}) $\ag$, which will affect the Hubble expansion rate at the time of nucleosynthesis in the same way as extra-relativistic degrees of freedom do, so that it modifies the freeze-out time $T_{\mathrm{f}}$. This is the only gravitational sector parameter. (\textit{2}) $\tau_{\mathrm{n}}$, the neutron lifetime dictates the free neutron decay and appears in the normalization of the proton-neutron reaction rates. It is the only weak interaction parameter and it is related to the Fermi constant $\gfermi$, or equivalently the Higgs vev. (\textit{3}) $\aem$, the fine-structure constant. It enters in the Coulomb barriers of the reaction rates through the Gamow factor, in all the binding energies. (\textit{4}) $Q_{\mathrm{np}}$, the neutron-proton mass difference enters in the neutron-proton ratio and we also have a dependence in (\textit{5}) $m_{\mathrm{N}}$ and $m_{\mathrm{e}}$ and (\textit{6}) the binding energies.

Clearly all these parameters are not independent but their relation is often model-dependent. If we focus on helium-4, its abundance mainly depends on $Q_{\mathrm{np}}$, $T_{\mathrm{f}}$ and $T_{\mathrm{N}}$ (and hence mainly on the neutron lifetime, $\tau_{\mathrm{n}}$). Early studies (see Sect.~III.C.2 of FVC03 [\citealt{jpu-revue}]) generally focused on one of these parameters. For instance, \cite{bbnkolb} calculated the dependence of primordial $^{4}$He on $G$, $\gfermi$ and $Q_{\mathrm{np}}$ to deduce that the helium-4 abundance was mostly sensitive in the change in $Q_{\mathrm{np}}$ and that other abundances were less sensitive to the value of $Q_{\mathrm{np}}$, mainly because $^{4}$He has a larger binding energy; its abundances is less sensitive to the weak reaction rate and more to the parameters fixing the value of $(n/p)$. To extract the constraint on the fine-structure constant, they decomposed $Q_{\mathrm{np}}$ as $Q_{\mathrm{np}}=\aem Q_\alpha+\beta Q_\beta$ where the first term represents the electromagnetic contribution and the second part corresponds to all non-electromagnetic contributions. Assuming that $Q_\alpha$ and $Q_\beta$ are constant and that the electromagnetic contribution is the dominant part of $Q$, they deduced that $|\Delta\aem/\aem|<10^{-2}$. \cite{bbnco} kept track of the changes in $T_{\mathrm{f}}$ and $Q_{\mathrm{np}}$ separately and deduced that ${\frac{\Delta Y_{\mathrm{p}}}{Y_{\mathrm{p}}}\simeq\frac{\Delta   T_{\mathrm{f}}}{T_{\mathrm{f}}}-\frac{\Delta Q_{\mathrm{np}}}{Q_{\mathrm{np}}}}$ while more recently the analysis \citep{landaubbn} focused on $\aem$ and $v$.\\

Let us now see how these parameters are now accounted for in BBN codes.

\cite{bbnberg} started to focus on the $\aem$-dependence of the thermonuclear rates (see also \citealt{bbnichi}). In the non-relativistic limit, it is obtained as the thermal average of the product of the cross, the relative velocity and the the number densities. Charged particles must tunnel through a Coulomb barrier to react. Changing $\aem$ modifies these barriers and thus the reaction rates. Separating the Coulomb part, the low-energy cross-section can be written as
\begin{equation}
\label{bir}
 \sigma(E)=\frac{S(E)}{E}\hbox{e}^{-2\pi\eta(E)}
\end{equation}
where $\eta(E)$ arises from the Coulomb barrier and is given in terms of the charges and the reduced mass $M_r$ of the two interacting particles as
\begin{equation}
 \eta(E)=\aem Z_1Z_2\sqrt{\frac{M_r c^2}{2E}}.
\end{equation}
The form factor $S(E)$ has to be extrapolated from experimental nuclear data but its $\aem$-dependence as well as the one of the reduced mass were neglected.  Keeping all other constants fixed, assuming no exotic effects and taking a lifetime of 886.7~s for the neutron, it was deduced that ${\left|{\Delta\aem}/{\aem}\right| <2\times10^{-2}}$. This analysis was then extended \citep{bbnnollet} to take into account the $\aem$-dependence of the form factor to conclude that
$$
 \sigma(E) = \frac{2\pi\eta(E)}{\exp^{2\pi\eta(E)}-1}
   \simeq 2\pi\aem
   Z_1Z_2\sqrt{\frac{M_rc^2}{c^2}}\exp^{-2\pi\eta(E)}.
$$
\cite{bbnnollet} also took into account (\textit{1}) the effect that when two charged particles are produced they must escape the Coulomb barrier. This effect is generally weak because the $Q_i$-values (energy release) of the different reactions are generally larger than the Coulomb barrier at the exception of two cases, ${}^3{\mathrm{He}}(n,p){}^3{\mathrm{H}}$ and ${}^7{\mathrm{Be}}(n,p){}^7\mathrm{Li}$. The rate of these reactions must be multiplied by a factor $(1+a_i\Delta\aem/\aem)$. (\textit{2}) The radiative capture (photon emitting processes) are proportional to $\aem$ since it is the strength of the coupling of the photon and nuclear currents. All these rates need to be multiplied by $(1+\Delta\aem/\aem)$. (\textit{3}) The electromagnetic contribution to all masses was taken into account, which modify the $Q_i$-values as $Q_i\rightarrow Q_i+ q_i\Delta\aem/\aem$. For helium-4 abundance these effects are negligible since the main $\aem$-dependence arises from $Q_{\mathrm{np}}$. Equipped with these modifications, it was concluded that ${\Delta\aem}/{\aem}=-0.007^{+0.010}_{-0.017}$ using only deuterium and helium-4 since the lithium-7 problem was still present. This was updated by \cite{NEW_Meissner:2023voo} that study the $\aem$-dependence on $Q_{\rm np}$, $\beta$-decays and nuclear cross-sections to conclude that $|\Delta\aem/\aem|<0.2$ assuming all other constants fixed.

Then the focus fell on the deuterium binding energy, $B_{\rm D}$. \cite{oklo-12,oklo-14,bbn-dimi,bbn-dimi2} illustrated the sensitivity of the light element abundances on $B_{\rm D}$. Its value mainly sets the beginning of the nucleosynthesis, that is of $T_{\mathrm{N}}$ since the temperature must low-enough in order for the photo-dissociation of the deuterium to be negligible (this is at the origin of the deuterium bottleneck). The importance of $B_{\rm D}$ is easily understood by the fact that the equilibrium abundance of deuterium and the reaction rate $p(n,\gamma){\mathrm{D}}$ depends exponentially on $B_{\rm D}$ and on the fact that the deuterium is in a shallow bound state. Focusing on the $T_{\mathrm{N}}$-dependence, it was concluded \citep{oklo-12} that $\Delta B_{\rm D}/B_{\rm D}<0.075$.

This shows that the situation is more complex and that one cannot reduce the analysis to a single varying parameter. Many studies then tried to determinate the sensitivity to the variation of many independent parameters.

\begin{figure}[hptb]
  \vskip-1cm
  \centerline{\includegraphics[scale=0.5]{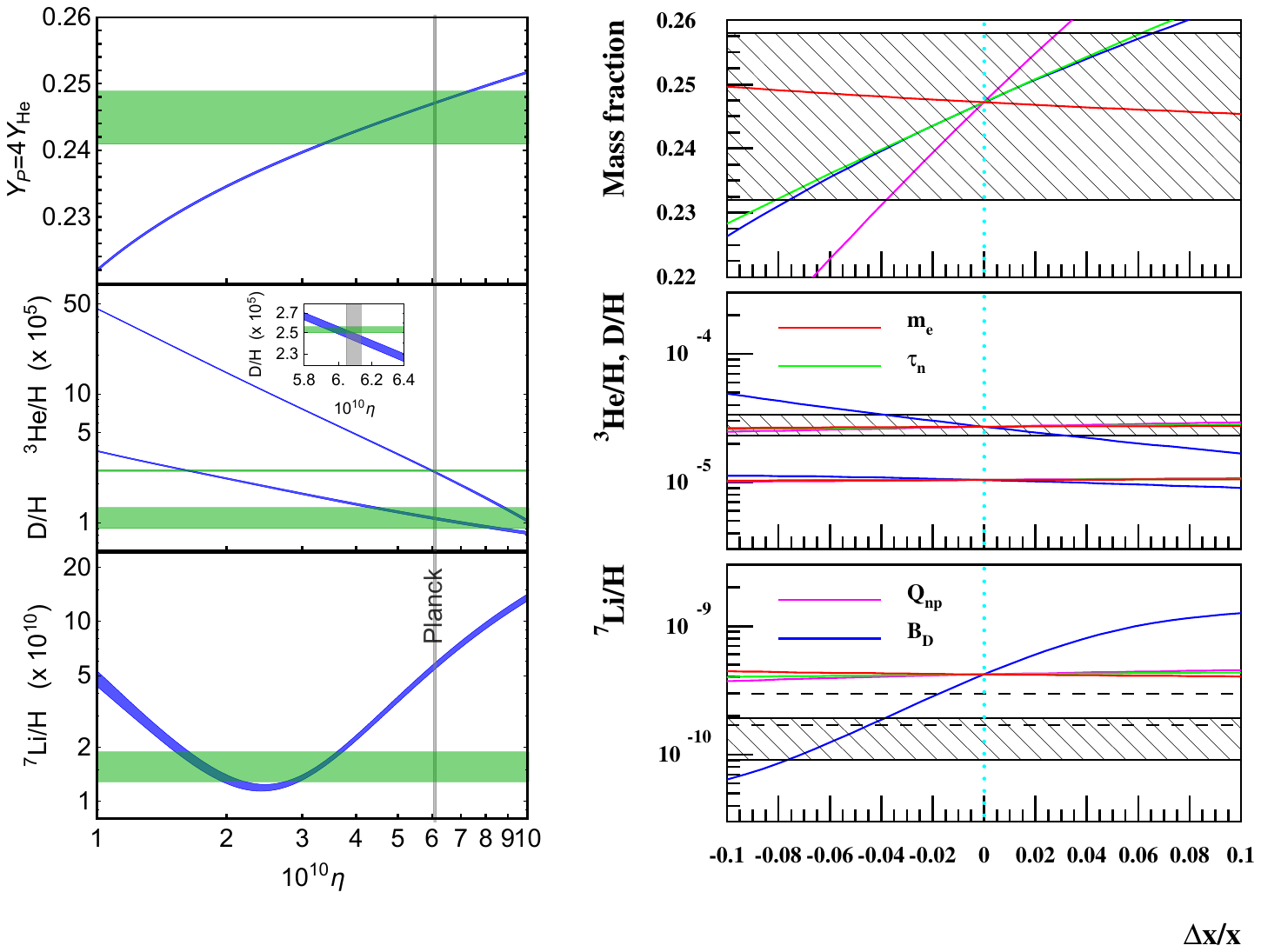}}
  \vskip-.5cm
  \caption[BBN light elements abundances and constraints on the variation of the BBN primary parameters]{(Left): variation of the light element abundances  in function of $\eta$ compared to the spectroscopic abundances.  The vertical line depicts the constraint obtained on $\eta$ from  the study of the cosmic microwave background. The lithium-7  problem lies in the fact that  $\eta_{\text{spectro}}<\eta_{\text{WMAP}}$. From \cite{NEW_Pitrou:2018cgg}. (right):  dependence of the light elements abundances on the independent  variation of the BBN parameters, assuming  $\eta=\eta_{\text{WMAP}}$. From \cite{couv}.}
  \label{fig-bbn}
\end{figure}

The sensitivity of the helium-4 abundance to the variation of 7 parameters was first investigated by  \cite{bbn-muller} considering the dependence on the parameters $\{X_i\}\equiv\{G,\aem,v,m_{\mathrm{e}},\tau_{\mathrm{n}},Q_{\mathrm{np}},$ $B_{\rm D}\}$ independently,
$$
\Delta \ln Y_{\mathrm{He}} = \sum_i c_i^{(X)}\Delta \ln X_i
$$
and assuming $\Lambda_{\mathrm{QCD}}$ fixed (so that the seven parameters are in fact dimensionless quantities). The $c_i^{(X)}$ are the sensitivities to the BBN parameters, assuming the six others are fixed. It was concluded that 
$$
Y_{\mathrm{He}}\propto \aem^{-0.043}v^{2.4}m_{\mathrm{e}}^{0.024}\tau_{\mathrm{n}}^{0.24}Q_{\mathrm{np}}^{-1.8} B_{\rm D}^{0.53} G^{0.405}
$$ 
for independent variations. They further related $(\tau_{\mathrm{n}},Q_{\mathrm{np}},B_{\rm D})$ to $(\aem,v,m_{\mathrm{e}},$ $m_{\mathrm{N}},m_{\mathrm{d}}-m_{\mathrm{u}})$, as we shall discuss in the next Sect.~\ref{bbn2cste}.

This was generalized by \cite{bbn-landau} up to lithium-7 considering the parameters $\{\aem,\gfermi,$ $\Lambda_{\mathrm{QCD}},\Omega_{\rm b} h_0^2\}$, assuming $G$ constant where the variation of $\tau_{\mathrm{n}}$ and the variation of the masses where tied to these parameters but the effect on the binding energies were not considered.

\cite{cnouv} considered the effect of $(Q_{\mathrm{np}}, B_{\rm D}, \tau_{\mathrm{n}}, m_{\mathrm{e}})$ on the abundances of the light elements up to lithium-7 neglecting the effect of $\aem$ on cross-sections. Their dependencies on the independent variation of each of them are depicted on Fig.~\ref{fig-bbn}. It confirmed the result of \cite{oklo-12,olive2-rio} that the deuterium binding energy is the most sensitive parameter. From the helium-4 data alone, the bounds
\begin{equation}
 -8.2\times10^{-2}\lesssim\frac{\Delta\tau_{\mathrm{n}}}{\tau_{\mathrm{n}}}\lesssim 6\times10^{-2},
 \quad
 -4\times10^{-2}\lesssim\frac{\Delta Q_{\mathrm{np}}}{Q_{\mathrm{np}}}\lesssim 2.7\times10^{-2},
\end{equation}
and
\begin{equation}
-7.5\times10^{-2}\lesssim\frac{\Delta B_{\rm D}}{B_{\rm D}}\lesssim 6.5\times10^{-2},
\end{equation}
at a 2$\sigma$ level, were set (assuming $\eta_{\text{WMAP}}$). The deuterium data set the tighter constraint $-4\times10^{-2}\lesssim\Delta\ln B_{\rm D}\lesssim 3\times10^{-2}$. Note also on Fig.~\ref{fig-bbn} that the lithium-7 abundance can be brought in concordance with the spectroscopic observations provided that
$B_{\rm D}$ was smaller during BBN
$$
-7.5\times10^{-2}\lesssim\frac{\Delta B_{\rm D}}{B_{\rm D}}\lesssim -4\times10^{-2},
$$
so that $B_{\rm D}$ may be the most important parameter to resolve the lithium-7 problem. The effect of the quark mass on the binding energies was described in \cite{BBNberengut}. They concluded that a variation of $\Delta m_{\mathrm{q}}/m_{\mathrm{q}}=0.013\pm0.002$ could reconcile the abundance of lithium-7 and the value of $\eta$  deduced from WMAP.

This analysis was extended \citep{bbn-dent} to incorporate the effect of 13 independent BBN parameters including those considered before plus the binding energies of deuterium, tritium, helium-3, helium-4, lithium-6, lithium-7 and beryllium-7. The sensitivity of the light element abundances to the independent variation of these parameters is summarized in Table~I of \cite{bbn-dent}. These BBN parameters were then related to the same 6 ``fundamental'' parameters used in \cite{bbn-muller}.

All these analyses demonstrate that the effects of the BBN parameters on the light element abundances are now under control. They have been implemented in BBN codes and most results agree, as well as with semi-analytical estimates. As long as these parameters are assume to vary independently, no constraints sharper than $10^{-2}$ can be set. One should also not forget to take into account standard parameters of the BBN computation such as $\eta$ and the effective number of relativistic particles.

\subsubsection{From BBN parameters to fundamental constants}\label{bbn2cste}

To reduce the number parameters, we need to relate the BBN parameters to more fundamental ones, keeping in mind that this can usually be done only in a model-dependent way. We shall describe some of the relations that have been used in many studies. They mainly concern $Q_{\mathrm{np}}$, $\tau_{\mathrm{n}}$ and $B_{\rm D}$.

At lowest order, all dimensional parameters of QCD, e.g., masses, nuclear energies etc., are to a good approximation simply proportional to some powers of $\Lambda_{\mathrm{QCD}}$. One needs to go beyond such a description and takes the effects of the masses of the quarks into account.

$\bullet$\emph{Proton-neutron mass difference}. $Q_{\mathrm{np}}$ can be expressed in terms of the mass on the quarks u and d and the fine-structure constant as
$$
 Q_{\mathrm{np}} = a\aem\Lambda_{\mathrm{QCD}} +(m_{\mathrm{d}}-m_{\mathrm{u}}),
$$
where the electromagnetic contribution today is $(a\aem\Lambda_{\mathrm{QCD}})_0=-0.76 \unit{MeV}$ and therefore the quark mass contribution today is $(m_{\mathrm{d}}-m_{\mathrm{u}})=2.05$ \citep{gasser82} so that
\begin{equation}
 \frac{\Delta Q_{\mathrm{np}}}{Q_{\mathrm{np}}} = -0.59\frac{\Delta\aem}{\aem} +
 1.59\frac{\Delta(m_{\mathrm{d}} - m_{\mathrm{u}})}{(m_{\mathrm{d}} - m_{\mathrm{u}})}.
\end{equation}
All the analyses cited above agree on this dependence.

$\bullet$\emph{Neutron lifetime}. It can be well approximated by
$$
 \tau_{\mathrm{n}}^{-1} = \frac{1+3g_A^2}{120\pi^3}\gfermi^2
 m_{\mathrm{e}}^5\left[\sqrt{q^2-1}(2q^4-9q^2-8) + 15\ln\left(q+\sqrt{q^2-1} \right)
 \right],
$$
with $q\equiv Q_{\mathrm{np}}/m_{\mathrm{e}}$ and $\gfermi=1/\sqrt{2}v^2$. Using the former expression for $Q_{\mathrm{np}}$ we can express $\tau_{\mathrm{n}}$ in terms of $\aem$, $v$ and the u, d and electron masses. It follows
\begin{equation}
 \frac{\Delta\tau_{\mathrm{n}}}{\tau_{\mathrm{n}}} = 3.86\frac{\Delta\aem}{\aem} +
   4\frac{\Delta v}{v} + 1.52\frac{\Delta m_{\mathrm{e}}}{m_{\mathrm{e}}}
   -10.4\frac{\Delta(m_{\mathrm{d}} - m_{\mathrm{u}})}{(m_{\mathrm{d}} - m_{\mathrm{u}})}.
\end{equation}
Again, all the analyses cited above agree on this dependence.

$\bullet$\emph{Binding energies}. Let us focus on  $B_{\rm D}$ that, as we have seen, is a crucial parameter. This is one the better known quantities in the nuclear domain and it is experimentally measured to a precision better than $10^{-6}$ \citep{bbn-bdmes}. Two approaches have been followed.
\begin{itemize}
 \item \emph{Pion mass}. A first route is to use the dependence of the binding energy on the pion mass \citep{bbnpi1,bbnpi2}, which is related to the u and d  quark masses by
 $$
 m_\pi^2 =  m_{\mathrm{q}} \langle \bar uu+\bar dd\rangle f_\pi^{-2}\simeq\hat m\Lambda_{\mathrm{QCD}},
$$
where $ m_{\mathrm{q}} \equiv\frac12(m_{\mathrm{u}}+m_{\mathrm{d}})$ and assuming that the  leading order of $\langle \bar uu+\bar dd\rangle f_\pi^{-2}$  depends only on $\Lambda_{\mathrm{QCD}}$, $f_\pi$ being the pion decay constant.  This dependence was parameterized \citep{nnbyoo} as
 $$
 \frac{\Delta B_{\rm D}}{B_{\rm D}} = -r\frac{\Delta m_\pi}{m_\pi},
 $$
 where $r$ is a fitting parameter found to be between 6 \citep{bbnpi1} and 10 \citep{bbnpi2}. Prior to this result, the analysis of \cite{oklo-12} provided two computations, which respectively  lead to $r=-3$ and $r=18$ while, following the same lines, \cite{bbn-landau2} got $r=0.082$. \cite{bbn-muller}, following \cite{bbn-pudliner}, added an electromagnetic contribution to get
\begin{equation}
 \frac{\Delta B_{\rm D}}{B_{\rm D}} = -\frac{r}{2}\frac{\Delta
 m_{\mathrm{q}}}{m_{\mathrm{q}}}-0.0081\frac{\Delta\aem}{\aem},
 \end{equation}
which has not been included in other work.  
\item\emph{Sigma model}. In the framework of the Walecka model,  where the potential for the nuclear forces keeps only the  $\sigma$ and $\omega$ meson exchanges,
 $$
 V=-\frac{g_s^2}{4\pi r}\exp(-m_\sigma r) + \frac{g_v^2}{4\pi r}\exp(-m_\omega
 r),
 $$
 with $g_s$ and $g_v$ two coupling constants. Describing  $\sigma$ as a SU(3) singlet, its mass was related to the one of the strange quark. In this way, one can hope to take into  account the effect of the strange quark, both on the nucleon mass and  the binding energy. In a second step $B_{\rm D}$ is related to  the meson and nucleon masses by
 $$
  \frac{\Delta B_{\rm D}}{B_{\rm D}} = -48\frac{\Delta m_\sigma}{m_\sigma}
     +50\frac{\Delta m_\omega}{m_\omega}+ 6\frac{\Delta m_{\mathrm{N}}}{m_{\mathrm{N}}}
 $$
so that ${\Delta B_{\rm D}}/{B_{\rm D}}\simeq-17{\Delta  m_{\mathrm{s}}}/{m_{\mathrm{s}}}$ \citep{oklo-14}. Unfortunately, a complete treatment of all the nuclear quantities on $m_{\mathrm{s}}$ has not been performed yet.

\end{itemize}

The binding energies of the other elements have been less studied. \cite{bbn-dent} follows a route similar than for $B_{\rm D}$ and relates them to pion mass and assumes that
$$
 \frac{\partial B_i}{\partial m_\pi}=
 f_i(A_i-1)\frac{B_{\rm D}}{m_\pi}r\simeq-0.13f_i(A_i-1),
$$
where $f_i$ are unknown coefficients assumed to be of order unity and $A_i$ is  the number of nucleons. No other estimates has been performed. Other nuclear potentials (such as Reid 93 potential, Nijmegen potential, Argonne $v18$ potential and Bonn potential) have been used in \cite{civitarese} to determine the dependence of $B_{\rm D}$ on $v$ and agree with previous studies. Note also that the effect of a possible change of the nucleon-nucleon interaction on the reaction rates involving $A=5$ and $A=8$ unstable nuclei was investigated in \cite{NEW_Coc:2012xk,NEW_Coc:2012uyk} to conclude that no significant effect on BBN appear seven if beryllium-8 were stable.

These analyses allow one to reduce all the BBN parameters to the physical constants ($\aem$, $v$, $m_{\mathrm{e}}$, $m_{\mathrm{d}}-m_{\mathrm{u}}$, $m_{\mathrm{q}}$) and $G$ that is not affected by this discussion. This set can be further reduced, since all the masses can be expressed in terms of $v$ as $m_i=h_iv$, where $h_i$ are Yukawa couplings. We also refer to the extensive study \citep{Birrell:2014uka,Rafelski:2024fej} for a discussion of the effects of fundamental constants of the standard model of particle physics on the neutrino freeze-out process showing that it mostly depends on the Weinberg angle $\theta_{\rm W}$ and the combination $m_{\rm e}^3G_{\rm F}^2/G_{\rm N}^{1/2}$.

Focusing on the resonant nuclear process $^{3}$He$(d,p)$$^{4}$He dependence on $m_{\rm q}$, \cite{NEW_Cheoun:2011yn} concluded that $-0.5\times10^{-2}<\Delta m_{\rm q}/m_{\rm q}<2.5 \times10^{-2}$ from D/H alone and $\Delta m_{\rm q}/m_{\rm q}<10^{-2}$ from helium-4 alone assuming that $\eta$ is fixed to its WMAP-7 value $\eta=6.23\times10^{-10}$, hence concluding that $-0.5\times10^{-2}<\Delta m_{\rm q}/m_{\rm q}<10^{-2}$. \cite{NEW_Bedaque:2010hr} considered several effective field theories as well as lattice QCD results, including a nuclear pionless effective theory, to compute the sensitivities of the binding energies on $m_{\rm q}$. They concluded that BBN implies $-10^{-2}<\Delta m_{\rm q}/m_{\rm q}<7 \times10^{-3}$. \cite{NEW_Berengut:2013nh} presented a derivation of the sensitivities of the light elements of the binding energies ans nuclear scattering lengths in the chiral perturbation theory in combination with non-perturbative methods. From deuterium and helium-4 data, they concluded $\Delta m_{\rm q}/m_{\rm q}=(2\pm4) \times10^{-2}$, a constraints that drop to $|\Delta m_{\rm q}/m_{\rm q}|<9 \times10^{-3}$ if the variation of $m_{\rm q}$ is assumed to be related to a variation of $v$ that translated to a variation of $m_{\rm e}$ that impacts the neutron lifetime. Following \cite{fw2} for the sensitivities of the binding energies of nucleus masses, \cite{NEW_Mori:2019cfo} investigated the sensitivity of the ${}^7$Be$(n,p){}^7$Li reaction to a change of $m_{\rm q}$. They concluded that if the variation of the excitation energies of the compound nucleus ${}^8$Be$^*$ is the same as that of the ground state, the shift of the resonance at $E_r=0.33$~MeV can decrease the ${}^7$Be abundance significantly for $\Delta m_{\rm q}/m_{\rm q}<-5 \times10^{-3}$.

To go further, one needs more assumptions, such as grand unification, or by relating the Yukawa coupling of the top to $v$ by assuming that weak scale is determined by dimensional transmutation \citep{cnouv}, or that the variation of the constant is induced by a string dilaton \citep{bbnco}. At each step, one gets more stringent constraints, which can reach the $10^{-4}$ \citep{bbn-dent} to $10^{-5}$ \citep{cnouv} level but indeed more model-dependent!

\subsubsection{Constants and the lithium problem}\label{secBBNLi}

The success of the BBN model is limited by the long-standing lithium problem, i.e., thet fact that the theoretically expected abundance of lithium-7 (given our present knowledge of astrophysics, nuclear and particle physics) exceeds the observed one by a factor of about 3.5.

As seen on Fig~\ref{fig-bbn}, \cite{cocnew} pointed out that a lower value of the deuterium binding energy at BBN can solve this  problem. This was later marginally ruled out by the improvement of the D/H measurements by \cite{NEW_Cooke:2017cwo}. This fact nevertheless rose attention of the importance on correlated variations of constants. \cite{NEW_Bedaque:2010hr} stressed that since the values fot lithium-7 binding energies are not reliable, the nuclear pionless effective theory may change this and allow us to interpret the lithium problem as a signal of quark mass variation. Then, \cite{NEW_Mori:2019cfo} proposed a solution with a finite value of $\Delta m_{\rm q}/m_{\rm q}=(4-8)\times 10^{-3}$ using the sensitivities of the binding energies  and nucleus masses from \cite{fw2} while assuming that the resonance energies of excited states do not vary.  \cite{NEW_Mosquera:2017rjp} concluded that the lithium abundance is reduced for $3\sigma$ level variation of ($\aem$, $v$) with introduction of dark energy. \cite{NEW_Martins:2020syb,NEW_Clara:2020efx}  reconsidered the problem within the $(R,S)$-parameterisation (see \ref{secRS}) that allowed them to express the variation of all the primordial abundances as a linear combinations of $R$ and $S$, $\Delta Y_i/Y_i=(x_i +y_iS+z_iR)\Delta\aem/\aem$ to conclude that constants remain an alternative for a solution to the lithium problem.  \cite{NEW_Deal:2021kjs} extended this to include $\Delta Y_i/Y_i=(x_i +y_iS+z_iR)\Delta\aem/\aem+t_i\Delta\tau_{\rm n}/\tau_{\rm n}+v_i\Delta N_{\rm eff}/N_{\rm eff}+w_i\Delta\eta_{10}/\eta_{10}$  to conclude  that the lithium problem most likely has an astrophysical solution while the deuterium discrepancy provides a possible hint for $\Delta\aem/\aem>0$. Similarly, \cite{NEW_Franchino-Vinas:2021nsf} claimed that a $(G,N_{\rm eff},\aem,v)$ variation can ease both the lithium and Hubble tensions and, to finish, \cite{NEW_Seto:2023yal}  stressed a variation of $\aem$ to explain the helium-4 data by \cite{NEW_Matsumoto:2022tlr}.

\subsubsection{Conclusion}

Primordial nucleosynthesis offers a possibility to test almost all fundamental constants of physics at a redshift of $z\sim10^8$. This makes it very rich but indeed the effect of each constant is more difficult to disentangle. The effect of the BBN parameters has been quantified with precision and they can be constrained typically at a $10^{-2}$ level, and in particular it seems that the most sensitive parameter is the deuterium binding energy.

The link with more fundamental parameters is better understood but the dependence of the deuterium binding energy still left some uncertainties and a good description of the effect of the strange quark mass is missing.

We have not considered the variation of $G$ in this section. Its effect is disconnected from the other parameters. Let us just stress that assuming the BBN sensitivity on $G$ by just modifying its value may be misleading. In particular $G$ can vary a lot during the electron-positron annihilation so that the BBN    constraints can in general not be described by an effective speed-up factor \citep{bbn-Gpichon,couv}.

\section{The gravitational sector and $G$}\label{section4}

The gravitational constant was the first constant whose constancy was questioned \citep{dirac37}. From a theoretical point of view, theories with a varying gravitational constant can be designed to satisfy the equivalence principle in its weak form but not in its strong form \citep{will-book} (see Sect.~\ref{subsecST} for details).  Most theories of gravity that violate the strong equivalence principle predict that the locally measured gravitational constant may vary with time.

The value of the gravitational constant is 
$$
G = 6.674~30(15)  \times 10^{-11}\unit{m^{3}\ kg^{-1}\ s^{-2}}
$$  
so that its relative standard uncertainty fixed by the CODATA\footnote{The CODATA is the COmmittee on Data for Science and Technology, see \url{http://www.codata.org/}.} is 0.01\%  \citep{NEW_Tiesinga:2021myr}; see \cite{NEW_Wu:2019pbm} for a review on the latest measurement of $G$. Interestingly, the disparity between different experiments led, in 1998, to a temporary increase of this uncertainty to 0.15\% \citep{G-marko}, which demonstrates the difficulty in measuring the value of this constant. This explains partly why the constraints on the time variation are less stringent than for the other constants. 

A variation of the gravitational constant, being a pure gravitational phenomenon, does not affect the local physics, such as, e.g., the atomic transitions or the nuclear physics. In particular, it is equivalent to stating that the masses of all particles are varying in the same way so that their ratios remain constant. Similarly all absorption lines will be shifted in the same way. It follows that most constraints are obtained from systems in which gravity is non-negligible, such as the motion of the bodies of the Solar system, astrophysical and cosmological systems. They are mostly related in the comparison of a gravitational time scale, e.g., period of orbits, to a non-gravitational time scale. It follows that in general the constraints assume that the values of the other constants are fixed. Taking their variation into account would add degeneracies and make the constraints cited below less stringent.

Indeed since $G$ has dimensions, any constraint cited here has to assume that some mass $m$ is kept constant so that it refer to a constraint on $G m^2/\hbar c$, see Eq.~(\ref{edef-mu}). Most studies  assume implicitly universal scalar-tensor theories so that any mass can be chosen since all mass ratios, $m_i/m_j$, remain constant. Otherwise most cases considered the proton mass $m_{\rm p}$ to be constant. Hence, constraints on the variation of $G$ are meant to be constraints on the dimensionless parameter $\ag$; see Eq.~(\ref{edef-mu}). The main constraints are summarized in Table~\ref{tab-Gccm}.

We refer to Sect.~IV of FVC03 \citep{jpu-revue} for earlier constraints based, e.g., on the determination of the Earth surface temperature, which roughly scales as $G^{2.25}\,M_{\odot}^{1.75}$ and gives a constraint of the order of $|\Delta G/G|<0.1$ \citep{gamow67a}, or on the estimation of the Earth radius at different geological
epochs. 

\begin{table}[htbp]
\caption[Constraints on a secular change of $G$]{Summary of the main constraints on a secular change of $G$. All details in the text.}
\label{tab-Gccm}
\centering
{\footnotesize
\begin{tabular}{lccr}
\toprule
System  & $\Delta G/G$  & $\dot G/G\quad ( \unit{yr}^{-1})$ & Ref. $\qquad\qquad$ \\
\midrule
Earth-Moon  & x  &$(-0.7\pm3.8)\times10^{-13}$ &  \cite{NEW_LLR2010} \\
Earth-Moon  & x& $(-5.0\pm9.6)\times10^{-15} $ &   \cite{NEw_Biskupek:2020fem} \\
Planet ranging & x & $(-0.01\pm 0.91)\times10^{-13}$ &  \cite{NEW_Pinto:2011en} \\
Planet ranging     & x & $-2.9\times 10^{-14} <\dot G/G < 4.6\times 10^{-14}$ & \cite{NEW_Pitjeva:2011ts}  \\
Large Magelanic cloud      & $ -0.07_{-0.04}^{+0.05}$  & x & \cite{NEW_Desmond:2020nde}  \\
 PSR~B1913+16       & x & $(4\pm5)\times10^{-12} $  & \cite{kaspi}  \\
 PSR~B1855+09         & x & $(-9\pm18)\times10^{-12} $  &   \cite{kaspi}\\  
  PSR J1738+0333,    & x & $(-0.7\pm7.6)\times10^{-13}$ & \cite{NEW_Freire:2012mg} \\
PSR J1713+0747   & x &$(-0.6\pm1.1)\times10^{-13}$ &   \cite{NEW_Zhu:2015mdo} \\    
PSR~J0437-4715   &x  & $<2.3\times10^{-11}$ &  \cite{G-pulsar2}\\
Globular clusters        & x & $(-1.4\pm2.1)\times10^{-11} $ & \cite{G-globular} \\            
Helioseismology         & x &$<1.6\times10^{-12} $ &   \cite{guenther98} \\
Asteroseismology          & x  & $1.2\pm2.6)\times10^{-12} $ &  \cite{NEW_Bellinger:2019lnl}  \\           
 NGC 6791        & x &$(- 0.9\pm 0.9) \times10^{-12}$ &\cite{NEW_Garcia-Berro:2011kvq}  \\         
 G117-B15A      & x & $(- 0.9\pm 0.9)\times10^{-10}$& \cite{NEW_Corsico:2001be} \\         
 R548       &x  & $(- 0.65\pm 0.65) \times10^{-10}$ & \cite{NEW_Corsico:2001be} \\         
 G117-B15A         & x & $< 4.1\times10^{-11}$ & \cite{G-puls1,G-white2} \\            
SNIa          & x & $-3\times 10^{-11} <\dot G/G < 7.3\times 10^{-11}$ & \cite{NEW_Mould:2014iga}  \\         
SNII          &x  & $-0.6\pm4.2)\times10^{-12}$ &  \cite{thorsett96}  \\        
 BBN         &  $-0.01^{+0.06}_{-0.05},$ & x &  \cite{NEW_Alvey:2019ctk} \\     
\bottomrule
\end{tabular}
}
\end{table}

Note an important difference between $G$ and other fundamental constants. A variation of $G$ would affect all observables achromatically, hence the effect on  cosmic spectra would be undistinguishable from a global redshift effect that can e.g., arise from the cosmic expansion. If rewritten in terms of  varying masses then all mass ratios remain unchanged.  To finish, note that while one can measure $GM$ with hight accuracy, its is more difficult to measure $G$ and $M$ independently. This explains why the constraints on the time variation of $G$ are several order of magnitude worse than those on non-gravitational constants.

\subsection{Solar system constraints}

Monitoring the orbits of various bodies of the Solar system offers a possibility to constrain deviations from General Relativity, and in particular the time variation of $G$. This accounts for comparing a gravitational time scale (related to the orbital motion) and an atomic time scale and it is thus assumed that the variation of atomic constants is negligible over the time of the experiment.

\paragraph{Earth--Moon system}\ 

A time variation of $G$ can be related to a variation of the mean motion ($n=2\pi/P$) of the orbit of the Moon around the Earth.  A decrease in $G$ would induce both the Lunar mean distance and
period to increase. As long as the gravitational binding energy is negligible, one has
\begin{equation}
 \frac{\dot{P}}{P} = -2\frac{\dot{G}}{G}.\nonumber
\end{equation}
Earlier constraints rely on paleontological data and ancient eclipses observations (see Sect.~IV.B.1 of FVC03 [\citealt{jpu-revue}]) and none of them are very reliable. A main difficulty arises from tidal dissipation that also causes the mean distance and orbital period to increase (for tidal changes $2\dot{n}/n+3\dot{a}/a=0$), but not as in the same ratio as for $\dot{G}$.

The Lunar Laser Ranging (LLR) experiment has measured the Earth-Moon distance with an accuracy of the order of 1~cm over 4 decades. An early analysis of this data \citep{williams76} assuming a Brans--Dicke theory concluded that $|\dot{G}/G| \leq 3\times10^{-11} \unit{yr}^{-1}$, later improved \citep{muller91},  by using 20 years of observation, to $|\dot{G}/G| \leq 1.04\times10^{-11} \unit{yr}^{-1}$, the main uncertainty arising from Lunar tidal acceleration. With, 24 years of data, one reached \citep{G-llr0} $|\dot{G}/G| \leq 6\times10^{-12} \unit{yr}^{-1}$ and then \citep{uff3} reached $|\dot{G}/G| \leq (4\pm9)\times10^{-13} \unit{yr}^{-1}$. This was improved \citep{NEW_LLR2010} to
\begin{equation}
 \left.\frac{\dot{G}}{G}\right|_0 = (-0.7\pm3.8)\times10^{-13} \unit{yr}^{-1}
\end{equation}
by using  data from 1970 to 2009  and including  the effect of a fluid Lunar core in the model. A better analysis of of the Lunar orbit, a better distribution of measurements over the Lunar retro-reflectors, as well as higher accuracy of the data, allowed \cite{NEw_Biskupek:2020fem} to claim that LLR constraints are improved to
\begin{equation}
 \left.\frac{\dot{G}}{G}\right|_0 = (-5.0\pm9.6)\times10^{-15} \unit{yr}^{-1}.
\end{equation}

\paragraph{Planet ranging}\ 

Similarly, \cite{shapiro} compared radar-echo time delays between Earth, Venus and Mercury with a caesium atomic clock between 1964 and 1969. The data were fitted to the theoretical equation of motion for
the bodies in a Schwarzschild spacetime, taking into account the perturbations from the Moon and other planets. They concluded that $|\dot{G}/G|<4\times10^{-10} \unit{yr}^{-1}$. The data concerning Venus cannot be used due to imprecision in the determination of the portion of the planet reflecting the radar. This was improved to $|\dot{G}/G|<1.5\times10^{-10} \unit{yr}^{-1}$ by including Mariner~9 and Mars orbiter data \citep{reasenberg78}.   The analysis was further extended \citep{shapiro90}  to give $\dot{G}/G=(-2\pm10)\times10^{-12} \unit{yr}^{-1}$. The combination of Mariner 10 an Mercury and Venus ranging data gives \citep{anderson91} gave $\dot{G}/G=(0.0\pm2.0)\times10^{-12}  \unit{yr}^{-1}$. \cite{reasenberg79} considered the 14 months data obtained from the ranging of the Viking spacecraft and deduced, assuming a Brans--Dicke theory, $|\dot{G}/G|<10^{-12} \unit{yr}^{-1}$. \cite{hellings83} using all available astrometric data and in particular the ranging data from Viking landers on Mars deduced that
$|\dot{G}/G|=(2\pm4)\times10^{-12} \unit{yr}^{-1}$. The major contribution to the uncertainty is due to the modeling of the dynamics of the asteroids on the Earth-Mars range.   \cite{hellings83} also tried to attribute their results to a variation of the atomic constants.  Using the same data but a different modeling of the asteroids, \cite{reasenberg83} got $|\dot{G}/G|<3\times10^{-11} \unit{yr}^{-1}$, later improved by \cite{chandler93} to $|\dot{G}/G|<10^{-11} \unit{yr}^{-1}$.

From the planetary ephemerides (INPOP13c) \cite{NEW_Fienga:2014njl}  deduced bounds on the variation of $GM_\odot$. Assuming that $M_\odot(t)=(t-J2000)\times M_\odot(J2000)$ and $G(t)=(t-J2000)\times G(J2000)$  and fixing the Sun total mass loss (including radiation and Solar winds) to $\dot M_\odot/M_\odot=(-0.55\pm0.15)\times10^{-13}$ \citep{NEW_Pinto:2011en}, they reached
\begin{equation}
 \left.\frac{\dot{G}}{G}\right|_0 =(-0.01\pm0.91)\times10^{-13} \unit{yr}^{-1}.
\end{equation}
\cite{NEW_Pitjeva:2021hnc} estimated the effect of a secular change of $GM_\odot$ on  positional observations of planets and spacecrafts using the ephemeris EPM2019. From the determination of $\dot M_\odot$ from outgoing Solar radiation and wind and infalling material, they concluded that
\begin{equation}
-2.9\times10^{-14}\,{\rm yr}^{-1}<\dot G/G < 4.6\times10^{-14}\,{\rm yr}^{-1}
\end{equation}
at $3\sigma$, improving their former result \citep{NEW_Pitjeva:2011ts}, $-4.2\times10^{-14}\,{\rm yr}^{-1}<\dot G/G < 7.5\times10^{-14}\,{\rm yr}^{-1}$.

\paragraph{Extension  beyond the Solar system}\ 

\cite{NEW_Desmond:2020nde} proposed to extend these tests to the  Large Magellanic Cloud, (LMC) that contains 6 well-studied Cepheid variable stars in detached eclipsing binaries. Radial velocity and photometric observations enable a complete orbital solution, and precise measurements of the Cepheids’ periods permit a detailed stellar modelling. Both are sensitive to $G$, the former via Kepler’s third law and the latter through the gravitational free-fall time. They concluded that the gravitational constant in the LMC is
\begin{equation}
G_{\rm LMC}/G = 0.93_{-0.04}^{+0.05}\,.
\end{equation}

\subsection{Pulsar timing}

Contrary to the Solar system, the dependence of the gravitational binding energy cannot be neglected while computing the time variation of the period of denser objects such as pulsars. Here, two approaches can be followed. Either one sticks to a model (e.g., scalar-tensor gravity) to compute all the effects in this model or one relies on a more phenomenological approach to set model-independent bounds.

\paragraph{Binary pulsars}\ 

\cite{G-eardley75} followed the first route to discuss the effects of a time variation of the gravitational constant on binary pulsar in the framework of the Brans--Dicke theory. In that case, both a dipole gravitational radiation and the variation of $G$ induce a periodic variation in the pulse period. \cite{nordtvedt90} showed that the orbital period changes as
\begin{equation}
\frac{\dot{P}}{P}=-\left[2+\frac{2(m_1c_1+m_2c_2)+3(m_1c_2+m_2c_1)}{m_1+m_2}
\right]\frac{\dot{G}}{G}
\end{equation}
where $c_i\equiv\delta\ln m_i/\delta\ln G$. He concluded that for the pulsar
PSR~1913+16 ($m_1\simeq m_2$ and $c_1\simeq c_2$) one gets
\begin{equation}
\label{pn}
\frac{\dot{P}}{P}=-\left[2+5c\right]\frac{\dot{G}}{G},
\end{equation}
the coefficient $c$ being model dependent. As another application, he estimated that $c_{\text{Earth}}\sim-5\times10^{-10}$, $c_{\text{Moon}}\sim-10^{-8}$ and $c_{\text{Sun}}\sim-4\times10^{-6}$ justifying the formula used in the Solar system.

\cite{damour88} used the timing data of the binary pulsar PSR~1913+16. They implemented the effect of the time variation of $G$ by considering the effect on $\dot{P}/P$. They defined, in a phenomenological way, that $\dot{G}/G=-0.5\delta\dot{P}/P$, where $\delta\dot{P}$ is the part of the orbital period derivative that is not explained otherwise (by gravitational waves radiation damping). This theory-independent definition has to be contrasted with the theory-dependent result~(\ref{pn}) by
\cite{nordtvedt90}. They got $\dot{G}/G=(1.0\pm2.3)\times10^{-11} \unit{yr}^{-1}$. \cite{damour91} reexamined the data of PSR~1913+16 to establish the bound $\dot{G}/G<(1.10\pm1.07)\times10^{-11} \unit{yr}^{-1}$. Using data from PSR~B1913+16 and PSR~B1855+09,  \cite{kaspi} got respectively
\begin{equation}
\dot{G}/G=(4\pm5)\times10^{-12} \unit{yr}^{-1}
\qquad\hbox{and}
\qquad
\dot{G}/G=(-9\pm18)\times10^{-12} \unit{yr}^{-1},
\end{equation}
the latter case being more ``secure'' since the orbiting companion is not a neutron star. A 10-year timing campaign of PSR J1738+0333, a 5.85-ms pulsar in a low-eccentricity 8.5-hour orbit with a low-mass white dwarf companion, allowed  \cite{NEW_Freire:2012mg} to get
\begin{equation}
\label{freire}
  \vert\dot{G}/G\vert=(-0.7\pm7.6)\times10^{-13} \unit{yr}^{-1}.
\end{equation}
From 21-year timing of observation of the 1.3 $M_\odot$ millisecond pulsar PSR J1713+0747, one of the most precise pulsars known, orbiting a 2.9~$M_\odot$ white dwarf,  \cite{NEW_Zhu:2015mdo} set the constraint
\begin{equation}
\label{zhu}
  \vert\dot{G}/G\vert=(-0.6\pm1.1)\times10^{-13} \unit{yr}^{-1}.
\end{equation}

\paragraph{Single pusars}\ 

All the previous results concern binary pulsars but isolated ones can also be used.  \cite{heitzmann75} related the spin-down of the pulsar JP1953 to a time variation of $G$. The spin-down is a combined effect of electromagnetic losses, emission of gravitational waves, possible spin-up due to matter accretion. Assuming that the angular momentum is conserved so that $I/P$~=~constant, one deduces that
\begin{equation}
 \left.\frac{\dot{P}}{P}\right\vert_G=\left(\frac{\dd\ln I}{\dd\ln G}\right)
\frac{\dot{G}}{G}.
\end{equation}
The observational spin-down can be decomposed as
\begin{equation}
 \left.\frac{\dot{P}}{P}\right\vert_{\text{obs}}= \left.\frac{\dot{P}}{P}\right\vert_{\text{mag}}
+ \left.\frac{\dot{P}}{P}\right\vert_{\text{GW}}+ \left.\frac{\dot{P}}{P}\right\vert_G.
\end{equation}
Since ${\dot{P}}/{P}_{\text{mag}}$ and ${\dot{P}}/{P}_{\text{GW}}$ are positive definite, it follows that ${\dot{P}}/{P}_{\text{obs}}\geq{\dot P}/{P}_G$ so that a bound on $\dot{G}$ can be inferred if the main pulse period is the period of rotation. \cite{heitzmann75} then modeled the pulsar by a polytropic $(P\propto\rho^n$) white dwarf and deduced that ${\dd\ln I}/{\dd\ln G}=2-3n/2$ so that $\vert\dot{G}/G\vert<10^{-10} \unit{yr}^{-1}$. \cite{mansfield76} assumed a relativistic degenerate, zero temperature polytropic star and got that, when $\dot{G}<0$, $0\leq-{\dot{G}}/{G}<6.8\times10^{-11} \unit{yr}^{-1}$ at a $2\sigma$ level. He also noted that a positive $\dot{G}$ induces a spin-up counteracting the electromagnetic spin-down, which can provide another bound if an independent estimate of the pulsar magnetic field can be obtained.  \cite{goldman90}, following \cite{G-eardley75}, used the scaling relations $N\propto G^{-3/2}$ and $M\propto G^{-5/2}$ to deduce that $2{\dd\ln I}/{\dd\ln G}=-5+3{\dd\ln I}/{\dd\ln N}$. He used the data from the pulsar PSR~0655+64 to deduce that
\begin{equation}
\label{goldman}
  0\leq-\dot{G}/G<5.5\times10^{-11} \unit{yr}^{-1}.
\end{equation}
The analysis \citep{G-pulsar2} of 10 years high precision timing data on the millisecond pulsar PSR~J0437-4715 reached
\begin{equation}
\label{goldman2}
  \vert\dot{G}/G\vert<2.3\times10^{-11} \unit{yr}^{-1}.
\end{equation}

It was also argued \citep{jofre,G-neutronstar3} that a variation of $G$ would induce a departure of the neutron star matter from $\beta$-equilibrium, due to the changing hydrostatic equilibrium. This would force non-equilibrium $\beta$-processes to occur, which releases energy that is invested partly in neutrino emission and partly in heating the stellar interior. Eventually, the star arrives at a stationary state in which the temperature remains nearly constant, as the forcing through the change of $G$ is balanced by the ongoing reactions. Comparing the surface temperature of the nearest millisecond pulsar, PSR~J0437-4715, inferred from ultraviolet observations, two upper limits for  variation were obtained, $|\dot{G}/G| < 2 \times 10^{-10} \unit{yr}^{-1}$, direct Urca reactions operating in the neutron star core are allowed, and $|\dot{G}/G| < 4 \times 10^{-12} \unit{yr}^{-1}$, considering only modified Urca reactions. This was extended in \cite{G-neutronstar} in order to take into account the correlation between the surface temperatures and the radii of some old neutron stars to get $\vert\dot{G}/G\vert < 2.1
\times 10^{-11} \unit{yr}^{-1}$.

\subsection{Stellar constraints}\label{secGstar}

Early works, reviewed in Sect.~IV.C of FVC03 \citep{jpu-revue}, studied the Solar evolution in presence of a time varying gravitational constant, concluding that under the Dirac hypothesis, the original nuclear resources of the Sun would have been burned by now. This results from the fact that an increase of the gravitational constant is equivalent to an increase of the star density (because of the Poisson equation).

The idea of using stellar evolution to constrain the possible value of $G$ was originally proposed by \cite{teller48}. The luminosity of a main sequence star can be expressed as a function of Newton's gravitational constant and its mass by using homology relations \citep{teller48,gamow67a}. In the particular case in which the opacity is dominated by free-free transitions, \cite{gamow67a} found that the luminosity of the star approximately scales as $L\propto  G^{7.8} M^{5.5}$. In the case of the Sun, this would mean that for higher values of $G$, the burning of hydrogen will be more efficient and the star evolves more rapidly, therefore we need to increase the initial content of hydrogen to obtain the present observed Sun. In a numerical test of the previous expression, \cite{G-globular}  found that low-mass stars evolving from the Zero Age Main Sequence to the red giant branch satisfy $L\propto G^{5.6}M^{4.7}$, which agrees to within 10\% of the numerical results, following the idea that Thomson scattering contributes significantly to the opacity inside such stars. Indeed, in the case of the opacity being dominated by pure Thomson scattering, the luminosity of the star is given by $L\propto G^4M^3$. This shows how the evolution of the main sequence stars is sensitive to $G$.

The driving idea behind the stellar constraints is that a secular variation of $G$ would affect the hydrostatic equilibrium of the star and in particular its pressure profile. In the case of non-degenerate stars, the temperature, being the only control parameter, will adjust to compensate the modification of the intensity of the gravity. It will then affect the nuclear reaction rates, which are very sensitive to the temperature, and thus the nuclear time scales associated to the various processes. It follows that the main stage of the stellar evolution, and in particular the lifetimes of the various stars, will be modified. As we shall see, basically two types of methods have been used, the first in which on relates the variation of $G$ to some physical characteristics of a star (luminosity, effective temperature, radius), and a second in which only a statistical measurement of the change of $G$ can be inferred. Indeed, the first class of methods are more reliable and robust but is usually restricted to nearby stars. Note also that they usually require to have a precise distance determination of the star, which may depend on $G$.

Let us mention that \cite{NEW_Sahni:2014cra} suggested a time variation of $G$ as a possible explanation of the \emph{faint young sun} problem \citep{NEW_Sunpb}, that is to the fact that numerical models of the Sun could indicate its luminosity was 75\% of its present value 4 Gyr ago which is not supported by geophysical and climatological data. One usually assumes an early greenhouse effect, but one can estimate that an increase  $\Delta G/G\sim0.02$ on a timescale of 4~Gyr offers a solution compatible with current constraints on $\dot G/G$.

\subsubsection{Ages of globular clusters}

These ideas were first applied with globular clusters. Their ages, determined for instance from the luminosity of the main-sequence turn-off, have to be compatible with the estimation of the age of the galaxy. This gave the constraint \citep{G-globular}
\begin{equation}
\dot{G}/G=(-1.4\pm2.1)\times10^{-11} \unit{yr}^{-1}.
\end{equation}
The effect of a possible time dependence of $G$ on luminosity has been studied in the case of globular cluster H-R diagrams but has not yielded any stronger constraints than those relying on celestial mechanics

\subsubsection{Solar and stellar seismology}

A side effect of the change of luminosity is a change in the depth of the convection zone. This induces a modification of the vibration modes of the star and particularly to the acoustic waves, i.e., $p$-modes \citep{demarque94}. 

\paragraph{Helioseismology}\ 

These waves are observed for our star, the Sun, and helioseismology allows one to determine the sound speed in the core of the Sun and, together with an equation of state, the central densities and abundances of helium and hydrogen. \cite{demarque94} considered an ansatz in which $G\propto t^{-\beta}$ and showed that $|\beta|<0.1$ over the last $4.5\times 10^9$ years, which corresponds to $\vert\dot{G}/G\vert<2\times10^{-11} \unit{yr}^{-1}$. \cite{guenther95} also showed that $g$-modes could provide even much tighter constraints but these modes are up to now very difficult to observe. Nevertheless, they concluded, using the claim of detection by \cite{hill90}, that $\vert\dot{G}/G\vert<4.5\times10^{-12} \unit{yr}^{-1}$. \cite{guenther98} then compared the $p$-mode spectra predicted by different theories with varying gravitational constant to the observed spectrum obtained by a network of six telescopes and deduced that
\begin{equation}
\label{gsun3}
  \left\vert\dot{G}/G\right\vert<1.6\times10^{-12} \unit{yr}^{-1}.
\end{equation}
The standard Solar model depends on few parameters and $G$ plays a important role since stellar evolution is dictated by the balance between gravitation and other interactions. Astronomical observations determine $GM_{\odot}$ with an accuracy better than $10^{-7}$ and a variation of $G$ with $GM_{\odot}$ fixed induces a change of the pressure ($P=GM_{\odot}^2/R_{\odot}^2$) and density ($\rho=M_{\odot}/R_{\odot}^3$).  The experimental uncertainties in $G$ between different experiments have important implications for helioseismology. In particular the uncertainties for the standard Solar model lead to a range in the value of the sound speed in the nuclear region that is as much as 0.15\% higher than the inverted helioseismic sound speed \citep{lopesilk}. While a lower value of $G$ is preferred for the standard model, any definite prediction is masked by the uncertainties in the Solar models available in the literature.  \cite{rv02} studied the effect of a variation of $G$ on the density and pressure profile of the Sun and concluded that present data cannot constrain $G$ better than $10^{-2}\%$. It was also shown \citep{lopesilk}  that the information provided by the neutrino experiments is quite significant because it constitutes an independent test of $G$ complementary tohelioseismology.

To finish, let us mention a recent claim on a possible evolution of $G$, that requires confirmation. Using 8640 days of low-l BiSON data, corrected for the Solar cycle variation, \cite{NEW_Bonanno:2017dcx} obtained 
\begin{equation}
\label{gsunnew1}
  \dot{G}/G =(1.25\pm0.3)\times10^{-13} \unit{yr}^{-1}
\end{equation}
including possible systematic effects such as uncertainties on the efficiency of the proton-proton fusion cross-section. This $4\sigma$ effect is argued to significantly outclass models with no secular variation of $G$.

\paragraph{Asteroseismolog}\ 

This method can be extended to other stars. The analysis of the stellar oscillation of the low mass Solar-like star on the main sequence KIC~7970740 for which one can determine dozens of oscillation modes (3 years of observation with the Kepler mission) allowed \cite{NEW_Bellinger:2019lnl} to conclude that
\begin{equation}
\label{gsunnew2}
  \dot{G}/G =(1.2\pm2.6)\times10^{-12} \unit{yr}^{-1}.
\end{equation}

\subsubsection{White dwarfs pulsation}

The observation of the period of non-radial pulsations of white dwarfs allows one to set similar constraints.  White dwarfs represent the final stage of the stellar evolution for stars with a mass smaller to about $10\,M_{\odot}$. Their structure is supported against gravitational collapse by the pressure of degenerate electrons. It was discovered that some white dwarfs are variable stars and in fact non-radial pulsator. This opens the way to use seismological techniques to investigate their internal properties. In particular, their non-radial oscillations are mostly determined by the Brunt--V\"ais\"al\"a frequency
$$
 N^2 = g \frac{\dd\ln P^{1/\gamma_1}/\rho}{\dd r}
$$
where $g$ is the gravitational acceleration, $\Gamma_1$ the first adiabatic exponent and $P$ and $\rho$ the pressure and density (see, e.g., \citealt{G-white1} for a white dwarf model taking into account a varying $G$). A variation of $G$ induces a modification of the degree of degeneracy of the white dwarf, hence on the frequency $N$ as well as the cooling rate of the star, even though this is thought to be negligible at the luminosities where white dwarfs are pulsationally unstable \citep{G-white2}.  \cite{NEW_Althaus:2011ca}  stressed that the impact of a varying $G$ is more notorious in the case of more massive white dwarfs so that the very accurate white dwarf cooling age derived for the old and metal-rich open cluster NGC 6791 makes it a excellent candidate to constrain $\dot G/G$.

Using the observation of G117-B15A that has been monitored during 20 years, \cite{G-puls1} was concluded  that $-2.5\times10^{-10} \unit{yr}^{-1}<\dot{G}/G< 4.0\times10^{-11} \unit{yr}^{-1}$ at a 2$\sigma$-level. The same observations were reanalyzed in \cite{G-white2} to obtain
\begin{equation}
 \vert\dot{G}/G\vert< 4.1\times10^{-11} \unit{yr}^{-1}.
\end{equation}
\cite{NEW_Corsico:2001be} pointed out that among white dwarfs there is a specific class of stars, known as ZZ-Ceti objects, which have a hydrogen-rich envelope and show periodic variations in their light curves. From the study secularly varying $G$ on the pulsational properties of such variable white dwarfs G117-B15A and R548, they concluded
\begin{align}
 \dot{G}/G&=(- 0.9\pm 0.9)\times10^{-10} \unit{yr}^{-1} &\hbox{(G117-B15A)}
\\
  \dot{G}/G&=(- 0.65\pm 0.65) \times10^{-10} \unit{yr}^{-1} & \hbox{(R548)}.
\end{align}

\subsubsection{Late stages of stellar evolution and supernovae}

A variation of $G$ can influence the white dwarf cooling and the light curves ot Type~Ia supernovae.

\paragraph{White dwarf cooling}\ 

\cite{gbwd} considered the effect of a variation of the gravitational constant on the cooling of white dwarfs and on their luminosity function. As first pointed out by \cite{vila76}, the energy of white dwarfs, when they are cool enough, is entirely of gravitational and thermal origins so that a variation of $G$ will induce a modification of their energy balance and thus of their luminosity. Restricting to cold white dwarfs with luminosity smaller than ten Solar luminosity, their luminosity can be related to the star binding energy $B$ and gravitational energy, $E_{\text{grav}}$, as
\begin{equation}
L=-\frac{\dd B}{\dd t}+\frac{\dot{G}}{G}E_{\text{grav}},
\end{equation}
which simply results from the hydrostatic equilibrium. Again, the variation of the gravitational constant intervenes via the Poisson equation and the gravitational potential. The cooling process is accelerated if $\dot{G}/G<0$, which then induces a shift in the position of the cut-off in the luminosity function. \cite{gbwd} concluded that $0\leq-\dot{G}/G<(1\pm1)\times10^{-11} \unit{yr}^{-1}$. The result depends on the details of the cooling theory, on whether the C/O white dwarfs are stratified or not and on hypothesis on the age of the galactic disk. For instance, with no stratification of the C/O binary mixture, one would require $\dot{G}/G=-(2.5\pm0.5)\times10^{-11} \unit{yr}^{-1}$ if the Solar neighborhood has a value of 8~Gyr (i.e., one would require a variation of $G$ to explain the data). In the case of the standard hypothesis of an age of 11~Gyr, one obtains $0\leq-\dot{G}/G<3\times10^{-11} \unit{yr}^{-1}$. More recently, from the effect of a secularly varying $G$ on the main sequence ages and on white dwarf cooling ages, \cite{NEW_Garcia-Berro:2011kvq} concluded
\begin{equation}
 \dot{G}/G=(- 0.9\pm 0.9) \times10^{-12} \unit{yr}^{-1}
\end{equation}
using the white dwarf luminosity function and the distance of the open Galactic cluster NGC 6791. 

\paragraph{Supernovae}\ 

The late stages of stellar evolution are governed by the Chandrasekhar mass $(\hbar c/G)^{3/2}m_{\mathrm{n}}^{-2}$ mainly determined by the balance between the Fermi pressure of a degenerate electron gas and gravity. 

Simple analytical models of the light curves of Type~Ia supernovae predict that the peak of luminosity is proportional to the mass of nickel synthesized. In a good approximation, it is a fixed fraction of the Chandrasekhar mass. In models with a varying $G$, this induces a modification of the luminosity distance-redshift relation \citep{G-sn1,gaztanaga02,cmb-G1}. \cite{NEW_Mould:2014iga} concluded from the analysis of the Hubble diagram of SNIa \citep{NEW_SupernovaCosmologyProject:2011ycw} that
\begin{equation}
-3\times 10^{-11}~{\rm yr}^{-1} <\dot G/G < 7.3\times 10^{-11}~{\rm yr}^{-1} 
\end{equation}
over the past 9 Gyr. Note that this constraint is degenerate with the cosmological parameters. In particular, the Hubble diagram is sensitive to the whole history of $G(z)$ between the highest redshift observed and today so that one needs to rely on a better defined model, such as, e.g., scalar-tensor theory \citep{cmb-G1}. The effect of the Fermi constant was also considered in \citealt{sn1GF}.

In the case of Type~II supernovae, the Chandrasekhar mass also governs the late evolutionary stages of massive stars, including the formation of neutron stars. Assuming that the mean neutron star mass is given by the Chandrasekhar mass, one expects that $\dot{G}/G=-2\dot{M}_{\text{NS}}/3\,M_{\text{NS}}$. \cite{thorsett96} used the observations of five neutron star binaries for which five Keplerian parameters can be determined (the binary period $P_b$, the projection of the orbital semi-major axis $a_1\sin i$, the eccentricity $e$, the time and longitude of the periastron $T_0$ and $\omega$) as well as the relativistic advance of the angle of the periastron $\dot{\omega}$. Assuming that the neutron star masses vary slowly as $M_{\text{NS}}=M_{\text{NS}}^{(0)}-\dot{M}_{\text{NS}} t_{\text{NS}}$ -- so that their age is  determined by the rate at which $P_b$ is increasing, i.e., $t_{NS}\simeq2P_b/\dot{P}_b$ --  and that the mass follows a normal distribution, \cite{thorsett96} deduced that, at $2\sigma$,
\begin{equation}
\label{gsun4}
  \dot{G}/G=(-0.6\pm4.2)\times10^{-12} \unit{yr}^{-1}.
\end{equation}

\subsubsection{New developments from gravitational wave astronomy}

\cite{NEW_Zhao:2018gwk} pointed out that the luminosity distance can be independently determined from GW standard siren caused by the coalescence of binary neutron stars, which could leas to a constraint of the order of 1.5\% on $\Delta G/G$ for redshifts up to 1.3. Note however that the effect of the variation of $G$ on the standard siren was not taken into account in this analysis. Indeed, a variation of $G$ induces a modification of the binary's binding energy so that it should affect the gravitational wave luminosity, hence leading to corrections in the chirping frequency \citep{yunes}. It was estimated that a LISA observation of an equal-mass inspiral event with total redshifted mass of $10^{5}\,M_{\odot}$ for three years should be able to measure $\dot{G}/G$ at the time of merger to better than $10^{-11}$/yr. This method paves the way to constructing constraints in a large band of redshifts as well as in different directions in the sky, which would be an invaluable constraint for many models. 

The development of GW astronomy opens new possibilities. \cite{NEW_Barbieri:2022zge} considered the possibility for spaceborne GW detectors at milli-Hz and deci-Hz frequencies, such as LISA or DECIGO, to measure the first and second time derivatives of the GW frequency and how this can set bounds of the time evolution of $G$. They reached the conclusion that the most favorable sources  among a simulated population of LISA galactic binaries could yield $\dot G/G\sim10^{-6}$/yr and that chirping stellar-mass compact binaries detected by DECIGO-like detectors at Mpc scales can lead to $\dot G/G\sim10^{-11}$/yr. From he observation of the binary neutron stars  GW170817, \cite{NEW_Vijaykumar:2020nzc} got $-1\leq\Delta G/G\leq 8$, corresponding to $-7\times10^{-9}{\rm yr}^{-1}\leq\dot G/G\leq5\times10^{-8}{\rm yr}^{-1}$ assuming a monotonic variation. See also \cite{NEW_Amendola:2017ovw,NEW_LISACosmologyWorkingGroup:2019mwx,Sun:2023bvy}.

More speculative is the idea \citep{G-grb} that a variation of $G$ can lead a neutron stars  to enter into the region where strange or hybrid stars are the true ground state. This would be associated with gamma-ray bursts that are claimed to be able to reach the level of $10^{-17}$/yr on $\dot G$.

\cite{NEW_Feldman:2016pws} proposed an experiment in deep space employing the classic gravity train mechanism. With a setup with three bodies (a larger layered solid sphere with a cylindrical hole through its center, a much smaller retroreflector which will undergo harmonic motion within the hole and a host spacecraft with laser ranging capabilities to measure round trip light- times to the retroreflector), measurements of the period of oscillation of the retroreflector in terms of host spacecraft clock time using existing technology could give determinations of $G$ at $6\times 10^{-8}$, i.e., nearly three orders of magnitude more accurate than current measurements here on Earth. 

We refer to \cite{Yunes:2024lzm} for a recent review on the possibilities opened by gravitational wave astronomy.

\subsection{Cosmological constraints}

Cosmological observations are more difficult to use in order to set constraints on the time variation of $G$. In particular, they require to have some ideas about the whole history of $G$ as a function of time but also, as the variation of $G$ reflects an extension of General Relativity, it requires to modify all equations describing the evolutions of the universe and of the large-scale structure in a consistent way. We refer to \cite{jpu-model,jpu-revu3,ugrg} for an early but detailed discussion of the use of cosmological data to test deviations from General Relativity.

\subsubsection{Cosmic microwave background}\label{subsecGcmb}

A time-dependent gravitational constant will have mainly three effects on the CMB angular power spectrum (see \citealt{cmb-G1} for discussions in the framework of scalar-tensor gravity in which $G$ is considered as a field):

\begin{enumerate}
 \item The variation of $G$ modifies the Friedmann equation and therefore the age of the Universe (and, hence, the sound horizon). For instance, if $G$ is larger at earlier time, the age of the Universe is smaller at recombination, so that the peak structure is shifted towards higher angular scales.
 \item The amplitude of the Silk damping is modified.  At small scales, viscosity and heat conduction in the photon-baryon fluid produce a damping of the photon perturbations. The damping scale is determined by the photon diffusion length at recombination, and therefore depends on the size of the horizon at this epoch, and hence, depends on any variation of the Newton constant throughout the history of the Universe.
\item The thickness of the last scattering surface is modified. In the same vein, the duration of recombination is modified by a variation of the Newton constant as the expansion rate is different. It is well known that CMB anisotropies are affected on small scales because the last scattering ``surface'' has a finite thickness. The net effect is to introduce an extra, roughly exponential, damping term, with the cutoff length being determined by the thickness of the last scattering surface. When translating redshift into time (or length), one has to use the Friedmann equations, which are affected by a variation of the Newton constant.  The relevant quantity to consider is the visibility function $g$. In the limit of an infinitely thin last scattering surface, $\tau$ goes from $\infty$ to 0 at recombination epoch. For standard cosmology, it drops from a large value to a much smaller one, and hence, the visibility function still exhibits a peak, but it is much broader.
\end{enumerate}

In full generality, the variation of $G$ on the CMB temperature anisotropies depends on many factors: (\textit{1}) modification of the background equations and the evolution of the universe, (\textit{2}) modification of the  perturbation equations, (\textit{3}) whether the scalar field inducing the time variation of $G$ is negligible or not compared to the other matter components, (\textit{4}) on the time profile of $G$ that has to be determine to be consistent with the other equations of evolution. \emph{This explains why it is difficult to state a definitive and model-independent constraint}. For instance, in the case of scalar-tensor theories, one has two arbitrary functions that dictate the variation of $G$. As can be seen, e.g., from \cite{cmb-G1,G-cmb}, the profiles and effects on the CMB can be very different and difficult to compare since they are entangled with the cosmological parameters.

In the case of Brans--Dicke theory, one just has a single constant parameter $\omega_{\text{BD}}$ characterizing the deviation from General Relativity and the time variation of $G$; see Eq.~(\ref{def_BD}). Thus, it is easier to compare the different constraints. \cite{chen99} showed that CMB experiments such as WMAP will be able to constrain these theories for $\omega_{\text{BD}}<100$ if all parameters are to be determined by the same CMB experiment, $\omega_{\text{BD}}<500$ if all parameters are fixed but the CMB normalization and $\omega_{\text{BD}}<800$ if one uses the polarization. For the Planck mission these numbers are respectively, 800, 2500 and 3200. \cite{acqua} concluded from the analysis of WMAP, ACBAR, VSA and CBI, and galaxy power spectrum data from 2dF, that  $\omega_{\text{BD}}>120$, in agreement with the former analysis of \cite{G-cmb}. \cite{wmapBD} indictated that The `WMAP-5yr data' and the `all CMB data' both favor a slightly non-zero (positive) $\dot{G}/G$ but with the addition of the SDSS power spectrum data, the best-fit value is back to zero, concluding that $-0.083<\Delta G/G<0.095$ between recombination and today, which corresponds to
\begin{equation}
-1.75\times 10^{-12} \unit{yr}^{-1} < \dot{G}/G < 1.05\times 10^{-12} \unit{yr}^{-1}\,.
\end{equation} 
Assuming Brand-Dicke theory, \cite{NEW_Li:2013nwa} concluded from Planck that 
\begin{equation}
\dot G/G = (-1.315\pm2.375)\times 10^{-13}~{\rm yr}^{-1}\,.
\end{equation}
A more recent analysis based on Planck-2018 CMB temperature, polarization and lensing, with a compilation of BAO measurements from the release DR12 of the BOSS Survey has been performed by \cite{NEW_Ballardini:2021evv} in the framework of a Brans-Dicke model.

From a more phenomenological prospect, some works modeled the variation of $G$ with time in a purely ad-hoc way, for instance \citep{cmb-cc} by assuming a linear evolution with time or a step function. The latter hypothesis has however gain some visibility since it is a candidate to resolve the $H_0$ tension; see e.g., \cite{NEW_Ballesteros:2020sik,NEW_Braglia:2020iik,NEW_Marra:2021fvf,NEW_Braglia:2020auw,NEW_Sakr:2021nja,NEW_Alestas:2022xxm,NEW_Ruchika:2023ugh}.

\subsubsection{BBN}

As detailed in Sect.~\ref{secbbnoverview}, the gravitational constant affects the freeze-out temperature $T_{\mathrm{f}}$. A larger value of $G$ corresponds to a higher expansion rate. This rate is determined by the combination $G\rho$ and in the standard case the Friedmann equations imply that $G\rho t^2$ is constant.  The density $\rho$ is determined by the number $N_*$ of relativistic particles at the time of nucleosynthesis so that nucleosynthesis allows to put a bound on the number of neutrinos $N_\nu$. Equivalently, assuming the number of neutrinos to be three, leads to the conclusion that $G$ has not varied from more than 20\% since nucleosynthesis.  But, allowing for a change both in $G$ and $N_\nu$ allows  one for a wider range of variation.  Contrary to the fine structure constant the role of $G$ is less involved.

It can be described, in its simplest but still useful form, by introducing a speed-up factor, $\xi=H/H_{GR}$, that arises from the modification of the value of the gravitational constant during BBN. Other approaches consider the full dynamics of the problem but restricted themselves to the particular class of Jordan--Fierz--Brans--Dicke  theories \citep{barrow78,yang79,rothman82,arai87,accetta90,dg91,casas92,clifton05}. \cite{casas92} concluded from the study of helium and deuterium that $\omega_{\text{BD}}>380$ when $N_\nu=3$ and $\omega_{\text{BD}}>50$ when $N_\nu=2$. Then, other analysis  considered a massless dilaton with a quadratic coupling \citep{couv,coc4a,bbn-Gpichon,santiago97} or a general massless dilaton \citep{serna96}. It should be noted that a combined analysis of BBN and CMB data was investigated in \cite{copi,kneller03}. The former considered $G$ constant during BBN while the latter focused on a non-minimally quadratic coupling and a runaway potential. It was concluded that from the BBN in conjunction with WMAP determination of $\eta$ set that $\Delta G/G$ has to be smaller than 20\%. However, we stress that the dynamics of the field can modify CMB results (see previous Sect.~\ref{subsecGcmb}) so that one needs to be careful while inferring $\Omega_\baryon$ from WMAP unless the scalar-tensor theory has converged close to General Relativity at the time of decoupling.

In early studies, \cite{barrow78} assumed that $G\propto t^{-n}$ and obtained from the helium abundance that $-5.9\times10^{-3}<n<7\times10^{-3}$, which implies $|{\dot{G}}/{G}|<(2\pm9.3)\,h\times 10^{-12} \unit{yr}^{-1}$, assuming a flat universe.  This corresponds in terms of the Brans--Dicke parameter to $\omega_{\text{BD}}>25$.  \cite{yang79} included the deuterium and lithium to improve the constraint to $n<5\times10^{-3}$, which corresponds to $\omega_{\text{BD}}>50$. It was further improved by \cite{rothman82} to $|n|<3\times10^{-3}$ implying
$|{\dot{G}}/{G}|<1.7\times 10^{-13} \unit{yr}^{-1}$. \cite{accetta90} studied the dependence of the abundances of D, $^{3}$He, $^{4}$He and $^{7}$Li upon the variation of $G$ and concluded
that $-0.3<{\Delta G}/{G}<0.4$, which roughly corresponds to $|\dot{G}/G|<9\times10^{-13} \unit{yr}^{-1}$. All these investigations assumed that the other constants are kept
fixed and that physics is unchanged. \cite{bbnkolb} assumed a correlated variation of $G$, $\aem$ and $\gfermi$ and got a bound on the variation of the radius of the extra dimensions.

Although the uncertainty in the helium-4 abundance has been argued to be significantly larger that what was assumed in the past \citep{osli}, interesting bounds can still be  derived \citep{bbn-cyburt2}. In particular translating the bound on extra relativistic degress of freedom ($-0.6<\delta N_\nu<0.82$) to a constraint on the speed-up factor ($0.949<\xi<1.062$), it was concluded \citep{bbn-cyburt2}, since $\Delta G/G=\xi^2-1=7\delta N_\nu/43$, that
\begin{equation}
 -0.10<\frac{\Delta G}{G}<0.13.
\end{equation}
With similar hypothesis, this was improved to \citep{NEW_Yeh:2022heq}
\begin{equation}
 -0.040<\frac{\Delta G}{G}<0.006.
\end{equation}
Using the {\tt PRIMAT} code \citep{NEW_Pitrou:2018cgg} and assuming only a variation of $G$ (i.e., all cosmological parameters are fixed), \cite{NEW_Alvey:2019ctk} concluded that
\begin{equation}
\Delta G/G=-0.01^{+0.06}_{-0.05},
\end{equation}
at $2\sigma$. \cite{NEW_Giri:2023pyy} considered in which the coupling function in Jordan frame behaves as $F^{-1}\propto 1+A\cos(\omega t + \phi)$ to conclude that if $\rho_\varphi$ is negligible during the whole cosmic history, BBN imposes that $A\omega<4.4\times10^{-4}\unit{s^{-1}}$. Note that $G_{\rm eff}$ oscillating does not imply that the Newton constant measured in a Cavendish experiment~(\ref{gcav}) is oscillating. For instance, it is clear that with $A=\cos\varphi$, $G_{\rm cav}=G_*$ while $G_{\rm eff}=G_*\cos\varphi$.

The relation between the speed-up factor, or an extra number of relativistic degrees of freedom, with a variation of $G$ is only approximate since it assumes that the variation of $G$ affects only the Friedmann equation by a renormalization of $G$. This is indeed accurate only when the scalar field is slow-rolling. For instance \citep{couv}, the speed-up factor is given (with the notations of Sect.~\ref{subsecST}) by
$$
 \xi = \frac{A(\varphi_*)}{A_0}\frac{1+\alpha(\varphi_*)\varphi_*'}{\sqrt{1-\varphi_*^{2\prime}/3}}\frac{1}{\sqrt{1+\alpha_0^2}}
$$
so that
\begin{equation}
 \xi^2 = \frac{G}{G_0}\frac{(1+\alpha(\varphi_*)\varphi_*')^2}{(1+\alpha^2)(1-\varphi_*^{2\prime}/3)},
\end{equation}
so that $\Delta G/G_0=\xi^2-1$ only if $\alpha\ll1$ (small deviation from General Relativity) and $\varphi_*'\ll 1$ (slow rolling dilaton). The BBN in scalar-tensor theories was investigated \citep{couv,bbn-Gpichon} in the case of a two-parameter family involving a non-linear scalar field-matter coupling function. Both concluded that even in the cases where before BBN the scalar-tensor theory was far from General Relativity, BBN enables to set quite tight constraints on the deviations from General Relativity today. In particular, neglecting the cosmological constant, BBN imposes $\alpha_0^2 < 10^{-6.5}  \beta^{-1} (\Omega_\mat h^2 / 0.15)^{-3/2}$ when $\beta > 0.5$ -- with the definitions introduced below Eq.~(\ref{matter*}).

\subsubsection{New proposals: Strong lensing}

Lensing is a generic prediction of General Relativity. In the thin lens approximation, the lensing equation relates the angular positions of the image $\theta$ and of the source $\theta_{\rm s}$ by
\begin{equation}\label{e.sl}
\theta=\theta_{\rm s}+\frac{4GM}{\theta}\frac{D_{\rm LS}}{D_{\rm OL}D_{\rm OS}}
\end{equation}
where $D_i$ are angular distances between the source S, the lens L and the observer O. This equation is used to describe the position of multiple images and times delays in strong lensing; see e.g., \cite{peteruzanbook}. 

Then, in a Freidmann-Lema\^{\i}tre spacetime, independently of the gravity theory, the kinematics of the cosmic expansion implies that redshifts are evolving with cosmic time \citep{NEW_Sandage,NEW_Mvzdot} as
\begin{equation}
\frac{\dd z}{\dd t}= H_0(1+z)-H(z).
\end{equation}
The expansion of the universe also induces an aberration of the directions of observations (see \cite{NEW_Marcori:2018cwn}). High precision astrometry now enables us to measure the time drift of astrophysical observables in real time, hence providing new ways to probe different cosmological models. 

\cite{NEW_Piattella:2017uat} remarked that a similar effects the positions and time delays of a strong lensing system. \cite{NEW_Giani:2020fpz} proposed to extend this very general result to the situation in which $G$ is time dependent. From Eq.~(\ref{e.sl}), they concluded that the positions of the multiple images $\theta$ and time delays $\Delta$ for a lens at redshift $z_{\rm L}$ evolve as
\begin{equation}
2\frac{\dot\theta}{\theta}= H_0 - \frac{H(z_{\rm L}}{1+z_{\rm L}}+\frac{\dot G}{G}, \qquad
\frac{\dot\Delta}{\Delta}= 2\left(H_0 - \frac{H(z_{\rm L})}{1+z_{\rm L}}+\frac{\dot G}{G} \right)\frac{\theta_{\rm E}^2}{(\theta_+-\theta_-)^2}
\end{equation}
where $\theta_{\rm E}$ is the Einstein radius and $\theta_\pm$ the angular positions of the two images. As an example, the quasar such as QSO0957+561 at $z=1.41$ lensed by a cluster at $z_{\rm L}=0.36$ enjoys 2 images separated by $6.1''$ with a time delay of $417\pm3$~days. It follows that $\dot\theta_{\rm E}\sim 10^{-10}$~arcsec/yr and $\Delta\sim10^{-3}$~s/yr.  \cite{NEW_Covone:2022mlt} argued that this lensing drifts may be accessible by Gaia or VLBI in the future. On that basis and using the light curves of the lens quasar system DES J0408-5354,  \cite{NEW_Giani:2020fpz} concluded that a constraints of $10^{-1}-10^{-2}$/yr on $\dot G/G$ could be reached. Applying this technique to the apparent drift of the angular size of the shadow of a black hole \cite{NEW_Frion:2021jse} argued that the amplitude of this effect is of order $10^{-16}$~day$^{-1}$ for M87* so that the observation by the Event Horizon Telescope set a constraint on either the maximum accretion rate or on  a time variation of $|\dot G/G |<10^{-3}$~yr$^{-1}$.

This is indeed not competitive with existing bounds and one also needs to be aware that this computation assumes  comoving observer, lens, and sources, and lens with static mass distribution, which is an ideal situation. Effects of transverse proper motions, mass accretion, growth of density perturbations can modify a lens system over time and shall be estimated. Besides, the use of the standard lensing equation implicitly assumes that the geodesic equation is not modified when $G$ becomes dynamical. This is the case in scalar-tensor theories since the Maxwell Lagrangian is conformally invariant in 4 dimensions. But one shall keep in mind these theoretical limitations.

\subsubsection{New proposals: Gravitational waves}

Among the recent developments in physics and astrophysics, the detection of gravitational waves \citep{Abbott_2016} has opened a whole new field of investigation. Some applications for fundamental constants are starting to be explored.

In General relativity, structurally the speed of gravitational waves is equal to the speed of light in vacuum. Gravitational waves detection offers the possibility to measure GW speed $c_T$. Setting $c_T^2=(1+\alpha_T)c^2$ the lack of gravi-Cerenkov effect from cosmic rays sets $\alpha_T>-10^{-15}$ \citep{Moore:2001bv}. Then, a upper bound from the travel time of GW detected by LIGO imply $\alpha_T<0.42$ \citep{Blas:2016qmn,Cornish:2017jml}. Thanks to the electromagnetic counterpart  \cite{NEW_Baker:2017hug} concluded $\vert\alpha_T\vert\lesssim10^{-15}$ and \cite{NEW_Mastrogiovanni:2020gua} $\vert\alpha_T\vert\lesssim10^{-17}$ .

Then GW observations allow for the measurement of the luminosity distance of standard sirens, that can be compared to the one derived from electromagnetic measurement, such as SNIa. This has opened a new and widely studied field to constrain theories of gravitation beyond General Relativity. An application can be done thanks to the determination of the gravitational luminosity distance \citep{NEW_LIGOScientific:2017adf} and the electromagnetic  luminosity distance \citep{NEW_Coughlin:2019vtv}  for the system  GW170817 \citep{NEW_LIGOScientific:2017vwq}; see also \cite{NEW_Colaco:2023iel}.

\cite{NEW_Calmet:2019nfj} pointed out that LISA could probe the cosmological evolution of the Higgs vev $v$ through the GW produced during the electroweak phase transition. Indeed that would require the Higgs mass to be below 72~GeV at temperature above 100~GeV, much below the actual mass of $125$~GeV. Nevertheless any sign of GW at frequencies marger than $10^{-5}$~Hz could be a sign of a smaller Higgs self-coupling constant in the past.

Then, \cite{NEW_Vijaykumar:2020nzc} argues that observations of gravitational waves emitted during the merging of binary neutron stars can be a window on the time variation of $G$. This relies on the determination of the masses of the neutron stars and on them being consistent with the theoretically allowed range since a varying-$G$ would induce a shift of all the masses by $(1+\Delta G/G)^{-3/2}$. From the data of GW170817, they concluded that $-1\lesssim \Delta G/G\lesssim 8$. The idea can also be applied to black hole-neutron star mergers.

\section{Models with varying constants}\label{subsec81}

The models that can be constructed are numerous and cannot all be reviewed. Several frameworks have however become widely used to interpret experimental and observational data. We shall thus discuss (\textit{1}) the string dilaton model in Sect.~\ref{subsub1},  (\textit{2}) Damour-Donoghue model in Sect.~\ref{subsub1b} and (\textit{3}) the Bekenstein framework in Sect.~\ref{subsub3}. To finish, we shall reconsider in Sect.~\ref{subsub0} the case of a scalar field coupled to electromagnetism, already introduced in Sect.~\ref{subsecSTmulti} in the discussion on dark energy, and even though it is formally a subcase of the three other models. These studies are of importance to investigate how well new data can improve constraints on a set of parameters. They also allow one to evaluate to which extent some conclusions are dependent on the model. They also allow one to connect different constraints and compare the efficiency of each data set constraints.

\subsection{String dilaton and Runaway dilaton models}\label{subsub1}

\paragraph{Model definition}\ 

\cite{damour94a,damour94b} considered the effective action for the massless modes of string theory taking into account the full string loop expansion. They argued it should be of the form
\begin{eqnarray}
S&=&\int\dd^4\mathbf{x}\sqrt{-\hat g}\left[m_{\rm s}^2
\left\lbrace B_g(\Phi)\hat R+4B_\Phi(\Phi)\left[\hat {\Box}\Phi
-(\hat\nabla\Phi)^2\right]
\right\rbrace-B_F(\Phi)\frac{k}{4}\hat F^2\right.\nonumber\\
&&\left.-B_\psi(\Phi)\bar{\hat
\psi}\hat\dslash\hat\psi+\ldots\right]
\end{eqnarray}
in the string frame, $m_{\rm s}$ being the string mass scale. The functions $B_i$ are not known but can be expanded in the limit $\Phi\rightarrow-\infty$ as
\begin{equation}
\label{ans}
B_i(\Phi)=\hbox{e}^{-2\Phi}+c^{(i)}_0+c^{(i)}_1\hbox{e}^{2\Phi}+
c^{(i)}_2\hbox{e}^{4\Phi}
+\ldots
\end{equation}
where the first term is the tree level term. It follows that these functions can exhibit a local maximum. After a conformal transformation ($g_{\mu\nu}=CB_g\hat g_{\mu\nu}, \psi=(CB_g)^{-3/4}B_\psi^{1/2}\hat\psi$), the action in the Einstein frame takes the form
\begin{eqnarray}
S&=&\int\frac{\dd^4\mathbf{x}}{16\pi G}\sqrt{-g}\left[ R-2(\nabla\varphi)^2-\frac{k}{4}B_F(\varphi)F^2-\bar \psi\dslash\psi+\ldots\right]
\end{eqnarray}
where the field $\varphi$ is defined as
$$
 \varphi\equiv\int \left[\frac{3}{4}\left(\frac{B_g'}{B_g}\right)^2+2\frac{B_\Phi'}{B_\Phi} + 2\frac{B_\Phi'}{B_g}\right]\dd\Phi.
$$
It follows that the Yang--Mills couplings all behave as $g^{-2}_{\text{YM}}=kB_F(\varphi)$. This implies that the QCD mass scale is given by
\begin{equation}
\label{qcd}
\Lambda_{\text{QCD}}\sim m_{\rm s}(CB_g)^{-1/2}\hbox{e}^{-8\pi^2kB_F/b}
\end{equation}
where $b$ depends on the matter content. Hence, the mass of any hadron, proportional to $\Lambda_{\text{QCD}}$ in first approximation, depends on the dilaton, $m_A(B_g, B_F,\ldots)$.

\paragraph{Phenomenology}\ 

If, as allowed by the ansatz (\ref{ans}), $m_A(\varphi)$ has a minimum $\varphi_m$ then the scalar field will be driven toward this minimum during the cosmological evolution. However, if the various coupling functions have different minima then the minima of $m_A(\varphi)$ will depend on the particle $A$.  To avoid violation of the equivalence principle at an unacceptable level, it is necessary to assume that all the minima coincide in  $\varphi=\varphi_m$, which can be implemented by setting $B_i=B$. This was shown to be realized when assuming that $\varphi_m$ is a special point in field space, for instance it could be associated to the fixed point of a $Z_2$ symmetry of the $T$- or $S$-duality \citep{lilley}.

Expanding $\ln B$ around its maximum $\phi_m$ as $\ln B\propto-\kappa(\varphi-\varphi_m)^2/2$, \cite{damour94a,damour94b}  constrained the set of parameters $(\kappa,\Delta\varphi_0)$ with
$$
\Delta\varphi_0\equiv\varphi_0-\varphi_m
$$
using the different observational bounds. $\kappa$ is the only model parameter while $\Delta\varphi_0$ is an environmental parameter, the value of which depends on the cosmological evolution.  The model can be extended to include a mass term, $V=m_\varphi(\varphi-\varphi_m)^2/2$, with same minimum has the coupling function $B$ to ensure the attraction mechanism is not spoiled.

This model allows one to address the unsolved problem of the dilaton stabilization, to study all the experimental bounds together and to relate them in a quantitative manner (e.g., by deriving a link between equivalence-principle violations and time-variation of $\aem$). In particular, the Klein--Gordon  equation~(\ref{e.KGgen2}) together with the Friedmann equations~(\ref{e.FLeq}) allows one to determine $\varphi(z)$ and then $\aem(z)$, $\mu(z)$, etc. so that local constraints can be consistently interpreted together with astrophysical and cosmological data. The Klein--Gordon equation involves the effective potential~(\ref{e.Veff}) with $\beta_i$ defined in Eq.~(\ref{deffai}), hence offering a screening mechanisms as discussed in Sect.~\ref{subsub2}.

\paragraph{Lowest order composition-independent effects}\ 

An important feature of this model lies in the fact that at lowest order the masses of all nuclei are proportional to $\Lambda_{\mathrm{QCD}}$ so that at this level of approximation, the coupling is universal and the theory reduces to a scalar-tensor theory, as described in Sect.~\ref{subsecST}. Hence, there will be no violation UFF and the deviation from General Relativity  are characterized by the PPN parameters
$$
 \gamma^\ppn-1\simeq -2\alpha^2_A=-2\beta_s^2\kappa^2\Delta\varphi_0^2,\qquad
 \beta^\ppn-1\simeq\frac{1}{2}\alpha^2_A\frac{\dd \alpha_A}{\dd\varphi}=\frac{1}{2}\beta_s^3\kappa^3\Delta\varphi_0^2
$$
with 
\begin{equation}\label{e.beta-DP}
 \alpha_A=\frac{\partial\ln \Lambda_{\mathrm{QCD}}(\phi)}{\partial\varphi}  = -\left[\ln\frac{m_{\rm s}}{m_A}+\frac{1}{2} \right]\frac{\dd\ln B}{\dd\varphi}
 	\equiv -\beta_s\frac{\dd\ln B}{\dd\varphi}=\beta_s\kappa\Delta\varphi_0
\end{equation}
with $\beta_s\sim40$ \citep{damour94a}.  The variation of the gravitational constant is, from the general expression~(\ref{Gdotst}), simply
$$
\frac{\dot{G}}{G} =2\alpha_A\dot\varphi_0 = -2\left[\ln\frac{m_{\rm s}}{m_A}
        +\frac{1}{2} \right]\frac{\dd\ln B}{\dd\varphi}\dot\varphi_0.
$$
The value of $\dot\varphi_0=H_0\varphi_0'$ is obtained from the Klein--Gordon  equation~(\ref{kgqq}) and is typically given by $\varphi_0'=-Z\beta_s\kappa H_0\Delta\varphi_0$ were $Z$ is a number that depends on the equation of state of the fluid dominating the matter content of the universe in the last $e$-fold  and the cosmological parameters so that
\begin{equation}
 \left.\frac{\dot{G}}{G}\right\vert_0 =2\alpha_A\dot\varphi_0 = -2ZH_0\beta^2_s\kappa^2\Delta\varphi_0^2.
\end{equation}
The factor $Z$ is model-dependent. Another way to estimate $\dot\varphi_0$ is to use the Friedmann equations, which imply that $\dot\varphi_0=H_0 \sqrt{1+q_0-\frac{3}{3}\Omega_{\mathrm{m0}}}$ where $q$ is the deceleration parameter defined in Eq.~(\ref{e.defq}).

\paragraph{Composition-dependent effects}\ 

When one considers the quark masses and binding energies, various composition-dependent effects appear. First, the fine-structure constant scales as $B^{-1}$ so that
\begin{equation}
 \left.\frac{\dot \alpha}{\alpha}\right\vert_0 =\kappa\Delta\varphi_0\dot\varphi_0  = -ZH_0\beta_s\kappa^2\Delta\varphi^2_0.
\end{equation}
Then, as already discussed in Sect.~\ref{section-theories}, one expects a violation of UFF since
\begin{equation}
\label{mdephi}
  m_A(\varphi)= N\Lambda_{\mathrm{QCD}}(\varphi)\left[  1 + \sum_{\mathrm{q}}\epsilon^q_A\frac{m_{\mathrm{q}}}{\Lambda_{\mathrm{QCD}}} + \epsilon^{\mathrm{EM}}_A\aem
  \right].
\end{equation}
Using an expansion of the form~(\ref{mass}), it was concluded that 
\begin{equation}
 \eta_{AB}=\kappa^2\Delta\varphi_0^2\left[  C_B\Delta\left(\frac{B}{M}\right)+ C_D\Delta\left(\frac{D}{M}\right)+ C_E\Delta\left(\frac{E}{M}\right)
 \right]
\end{equation}
with $B=N+Z$, $D=N-Z$ and $E=Z(Z-1)/(N+Z)^{1/3}$ and where the value  of the parameters $C_i$ are model-dependent.

\paragraph{Generic conclusions}\ 

It follows from this model that:
\begin{itemize}
 \item The PPN parameters, the time variation of $\alpha$ and $G$ today and the  violation of the university of free-fall all scale as $\Delta\varphi_0^2$.
 \item The field is driven toward $\varphi_m$ during the cosmological evolution,  a point at which the scalar field decouples from the matter field. The  mechanism is usually called \emph{the least coupling principle}; see Sect.~\ref{subsub2} for the general discussion on screening mechanisms.
 \item Once the dynamics for the scalar field is solved, $\Delta\varphi_0$ can be  related to $\Delta\varphi_i$ at the end of inflation. Interestingly, this quantity can  be expressed in terms of amplitude of the density contrast at the end of inflation,  that is to the energy scale of inflation.
 \item The numerical estimations \citep{damour94a} indicate that $\eta_{U,H}\sim  -5.4\times10^{-5}(\gamma^\ppn-1)$ showing that in such a class  of models, the constraint on $\eta$, once reevaluated to take into account the MICROSCOPE bound~(\ref{eta-microscope}), $\eta\sim10^{-15}$ implies $1-\gamma^\ppn\sim  2\times10^{-11}$, hence $\alpha_A^2\lesssim 10^{-11}$,  which is a better constraint that the one obtained directly.
 \item The model was compared to Oklo, QSO and CMB data by \cite{landaudp} to conclude that  $\vert\Delta\varphi_0\vert<3.4\kappa 10^{-6}$.
\end{itemize}

\paragraph{Runaway dilaton model}\ 

Following \cite{gasperini02}, the light dilaton model was extended \citep{damourrunaway} to the case where the coupling functions have a smooth finite limit for infinite value of the bare string coupling, so that 
$$
B_i=C_i+{\cal O}(\mathrm{e}^{-\varphi}) \,.
$$
Hence, the dilaton runs away toward its attractor at infinity during a stage of inflation and then during the matter dominated era.  The late time dynamics of the scalar field is similar as in quintessence models, so that it can also explain the late time acceleration of the cosmic expansion. The amplitude of residual dilaton interaction is related to the amplitude of the primordial density fluctuations and it induces a variation of the fundamental constants, provided it couples to dark matter or dark energy. It is concluded that, in this framework, the largest allowed variation of $\aem$ is of order $2 \times 10^{-6}$, which is reached for a violation of the universality of free fall of order $10^{-12}$ and it was established that
\begin{equation}
\label{leastuff}
\left.\frac{\dot\aem}{\aem}\right\vert_0\sim\pm10^{-16}\sqrt{1+q_0-\frac{3}{2}\Omega_{\mathrm{m0}}}
\sqrt{10^{12}\eta} \unit{yr}^{-1},
\end{equation}
where the first square-root arises from the computation of $\dot\varphi_0$. The formalism was later used to discuss the time variation of $\aem$ and $\mu$ \citep{chibarun}. Hence, the Klein--Gordon  equation links $\dot\varphi_0$ to the deceleration parameter (if the field is slow-rolling) and hence to the dark energy equation of state and the matter content of the universe. It follows that the model can be constrained on the one side locally (i.e., using the Solar system and laboratory constraints on the fundamental constants) and on the other hand by low-redshift astrophysical data and cosmological data than constrain the cosmological parameters.

As examples, \cite{NEW_Martins:2015dqa} and then \cite{NEW_Martins:2019uxo}  have considered the simplified case in which $\varphi$ has either (1) a constant coupling $\alpha_V$ to the dark matter or (2) $\alpha_A$, as defined in Eq.~(\ref{e.beta-DP}) to the hadronic cosmological matter, or (3) self-coupling $\alpha_\varphi$ to dark energy to set constraint from low-redshift QSO data and local clock and WEP constraints. Assuming that the dilaton behaves as $\varphi\sim\varphi_0 + \varphi'_0\ln a$ at low-redshift, they concluded, taking into account the constraint on the dark matter equation of state \citep{Tutusaus:2016kyl}, that $|\varphi'_0|<0.1$, $|\alpha_A|<1.5\times10^{-5}$ and $|\alpha_V|<0.09$. \cite{NEW_Vacher:2023gnp} followed the same line using local (clocks + MICROSCOPE) data, Oklo and astrophysical (QSO) data to constrain the redshift evolution of $\aem$ and the latest BBN+CMB+BAO+SNIa to constrain the background cosmology. The model assumes a coupling to hadron $\alpha_A=\alpha_{A0}\hbox{e}^{-(\varphi-\varphi_0)}$ and to dark matter $\alpha_m=\alpha_{m0}\hbox{e}^{-(\varphi-\varphi_0)}$ and either (\textit{1}) a massless field and cosmological constant or (\textit{2}) a model with a potential $V\propto \hbox{e}^{c(\varphi-\varphi_0)}$ to conclude that $\alpha_{A0}\lesssim 3\times10^{-6}$, $\alpha_{m0}\lesssim 6\times 10^{-2}$

\paragraph{Conclusions}\ 

In these two string-inspired scenarios, the amplitude of the variations of the constants are related to the one of the density fluctuations during inflation and the cosmological evolution. Both the amplitude of the violation of UFF and the variation of all constants are related to the same scalar field. The models then required an attraction mechanism to avoid the Solar system constraints, either at a finite $\varphi_m$ for the light dilaton or at infinity for the runaway dilaton. The scalar field can also act as a quintessence field so that the time variations of the fundamental constants are also related to the equation of state of the dark energy component and thus to the cosmological expansion.

\subsection{Damour--Donoghue interaction light dilaton model}\label{subsub1b}

\paragraph{Model definition}\

The Damour--Donoghue model \citep{dado2,dado1} considered the standard action of General Relativity and the standard matter field plus a scalar field
$$
S=\int\left\lbrace\frac{1}{16\pi G}\left[R-2g^{\mu\nu}\partial_\mu\varphi\partial_\nu\varphi - V(\varphi)\right]  +{\cal L}_{\rm SM}(\Psi_i,g_{\mu\nu})\right\rbrace\sqrt{-g}\dd^4x
$$
to which one adds an interaction term to the baryonic sector that takes into account a direct coupling to the Faraday tensor of electromagnetism $F_{\mu\nu}$ the gluon tensor $F^A_{\mu\nu}$ and the fermion spinors $\psi_i$ as
\begin{equation}\label{L_dd1}
S_{\rm int}=\int \varphi\left[\frac{d_{e}}{4 e^2}F^2-\frac{\beta_3d_{g}}{2g_3} F^A_{\mu\nu}F_A^{\mu\nu} -
\sum_{i=e,u,d}(d_{{m}_i} +\gamma_{m_i }d_g )m_i \bar\psi_i\psi_i\right]\sqrt{-g}\dd^4x\,. 
\end{equation}
$g_3$ stands for the QCD gauge coupling and $\beta_3$ for the $\beta$-function of the running of $g_3$. One considers 3 fermions (electrons, up and down quarks) of mass $m_i$. Their anomalous dimensions $\gamma_{m_i}$ give the energy running of the quark masses. The interaction is thus characterized by 5 dimensionless couplings ($d_{e}, d_{g}, d_{{m}_i} $) that characterize the strength of the coupling of $\varphi$ to the matter sectors: $d_e$ and $d_g$ parametrise the coupling with the electromagnetic and gluonic fields while $d_{m_{\rm e}} , d_{m_{\rm u}}$ and $d_{m_{\rm d}}$ are the couplings to the electron, $u$ and $d$ quarks mass terms.  It is usual to redefine the latter two as
\begin{equation}\label{def_hatm}
\hat m=\frac{1}{2}(m_{\rm u}+m_{\rm d}), \qquad
\delta m= m_{\rm d}-m_{\rm u}\,.
\end{equation}
While Eq.~(\ref{L_dd1}) is linear in $\varphi$, a similar interaction term with higher powers of $\varphi$ have been considered \citep{Hees:2018fpg}. In such a  case the previous couplings are labelled $d^{(1)}$ while the couplings to  $\varphi^2/2$ are  ($d_{e}^{(2)}, d_{g}^{(2)}, d_{{m}_i}^{(2)} $) respectively. \cite{Stadnik:2016zkf} considered a coupling of the form $d_j(\varphi-\varphi_j)^2$ which then boils down to a combination of the parameters $d^{(1)}$ and $d^{(2)}$. The model can be easily generalized to any coupling functions by replacing in the action~(\ref{L_dd1}) $1+d_i\varphi$ by $B_i(\varphi)$ and then $d_i$ by $\alpha_i=B'_i/B_i$.

This model has been constrained from the explicit violation of the UFF it generates as well as all the others tests on the fundamental constants. It has become a prototypical example for several
ultra-light dark matter scenarios. All the constraints  depend on the choice of the potential, with the two cases of a massless ($V=0$) and light ($V=2m_\varphi^2\varphi^2$).

\paragraph{Field dependencies of the constants}\ 

It follows that the fine structure constant, the quark masses and the QCD mass scale behave as
\begin{eqnarray}\label{cdephi}
&&\aem(\varphi)=(1+d_{e}\varphi)\aem, \nonumber\\
&&m_i (\varphi)= (1+d_{{m}_i}\varphi) m_i,  \nonumber\\
&&\Lambda_3(\varphi)=(1+d_g\varphi) \Lambda_{\rm QCD}
\end{eqnarray}
and Eq.~(\ref{def_hatm}) implies that
\begin{equation}
\hat m (\varphi)= (1+d_{\hat m}\varphi)\hat m\,,\qquad
\delta m (\varphi)= (1+d_{\delta m}\varphi)\delta m
\end{equation}
with
$$
d_{\hat m;\delta m}=\frac{m_{\rm d} d_{m_{\rm d}}\pm m_{\rm u} d_{m_{\rm u}} }{m_{\rm d}\pm m_{\rm u}}.
$$
Hence, the sensitivities~(\ref{def_sphi}) of the constants to the scalar field are simply $s_i(\varphi)=d_i$. In the case of a quadratic coupling \citep{Hees:2018fpg}, the expressions~(\ref{cdephi}) have to be  modified by relabelling the linear coefficients $d_i$ as $d_i^{(1)}$ and then defining the second order coefficients according to the replacement  $d_i\varphi \rightarrow d^{(2)}_i \varphi^2/2$  so that $s_i^{(2)}(\varphi)=d^{(2)}_i\varphi$.

\paragraph{Note on normalisation}\ 

Again, we assume that $\varphi$ is dimensionless so that the action is consistent with the normalisation of e.g., \cite{dn1,dn2,NEW_Stadnik:2015kia,Hees:2018fpg} while other works such as e.g., \cite{NEW_Arvanitaki:2014faa,NEW_Stadnik:2015upa,NEW_Stadnik:2014tta,NEW_Arvanitaki:2015iga,NEW_Berge:2017ovy} consider the coupling to $\phi=\varphi/\sqrt{4\pi G}$. In that case, energy scales $\Lambda_i$, are introduced in the relations~(\ref{cdephi}), i.e., the terms $(1+d_i\varphi)$ are expressed as $(1+ \phi/\Lambda_i)$ so that $d_i=\sqrt{4\pi G}/\Lambda_i$ for the linear couplings. The comparison of normalisations of the quadratic coupling models \citep{NEW_Stadnik:2015kia,NEW_Stadnik:2015upa,NEW_Stadnik:2014tta,NEW_Stadnik:2015xbn,Stadnik:2016zkf,NEW_Kalaydzhyan:2017jtv,NEW_Stadnik:2018sas} are detailed in Appendix~B of \cite{Hees:2018fpg}.

\paragraph{Sensitivities of the atomic masses}\ 

As explained in Sect.~\ref{subsecUFF} and illustrated in Sect.~\ref{subsecSTmulti}, this inplies that all the atomic masses, as well as the masses of the proton and neutron, depend on $\varphi$. Starting from the general expression of the atomic masses~(\ref{e.mass0}-\ref{mass}), the sensitivity $\beta_A$ defined in Eqs.~(\ref{def_betaA}) have been obtained to be  \citep{dado2,dado1}
\begin{eqnarray}\label{e.betaADD}
\beta_A = d_g + (d_{\hat m}-d_g) Q_{\hat m}  + (d_{\delta m}-d_g) Q_{\delta m} + (d_{m_{\rm e}}-d_g) Q_{m_{\rm e}} +d_e Q_e
\end{eqnarray}
ìn terms of the four charges $Q_i$ that depend on the atomic and mass numbers $(Z,A)$ as
\begin{eqnarray}
Q_{\hat m} &=& F_A \left[  93 - \frac{36}{A^{1/3}} - 20\left(\frac{A-2Z}{A}\right)^2 -0.14\frac{Z(Z-1)}{A^{4/3}}\right]\times10^{-3} \\
Q_{\delta m} &=& F_A \left[17\frac{A-2Z}{A}  \right] \times10^{-4} \\
Q_{m_{\rm e}} &=& F_A \left[5.5\frac{Z}{A}  \right]\times10^{-4}\\
Q_e &=& F_A \left[-1.4+8.2\frac{Z}{A}+7.7\frac{Z(Z-1)}{A^{4/3}} \right]\times10^{-4}
\end{eqnarray}
with 
\begin{equation}
 F_A = A \frac{931\unit{MeV}}{m_A}.
\end{equation}

\paragraph{Universality of free fall}\ 

This allows one to translate the experimental tests of the UFF to constraints on the model parameters. 

First, it has been shown \citep{dado2,dado1} (see also \cite{Hees:2018fpg,NEW_Berge:2023sqt} for technical details) that the expression~(\ref{e.betaADD}) for the coupling can be approximated as 
$$
\alpha_A\simeq d_g^* + \left[  (d_{\hat m} - d_g)Q'_{\hat m} +d_e Q_e' +  (d_{m_{\rm e}} - d_g)Q'_{m_{\rm e}} +  (d_{\delta m} - d_g)Q'_{\delta m}
\right]_A,
$$
in which the first term $d_g^*=d_g+0.093(d_{\hat m}-d_g)+2.7\times10^{-4}d_e$ represents the composition independent part and where $Q'_{\hat m}=-0.036A^{1/3}-1.4\times10^{-4}Z(Z-1)/A^{4/3}-0.02(A-2Z)^2/A^2$, $Q'_e=[7.7Z(Z-1)/A^{4/3}-4.1(A-2Z)/Z]\times10^{-4}$, $Q'_{m_{\rm e}}=-2.75\times10^{-4}(A-2Z)/A$ and $Q'_{\delta m}=1.7\times10^{-3}(A-2Z)/A$. Their values for different materials used experimentally are summarized in Table~\ref{tab-Qprimparameters}. Note that the term in $A-2Z$ in these expressions can be neglected for heavy  elements for which $Z/A\sim1/2$ so that, given the experimental precision, one can rely on the value 
$$
\alpha_A\simeq d_g^* + \left[  (d_{\hat m} - d_g)Q'_{\hat m} +d_e Q_e' +  (d_{m_{\rm e}} - d_g)Q'_{m_{\rm e}} \right]_A\,.
$$

\begin{table}[t]
\caption[Dilatonic charges for the materials used to test the UFF]{Dilatonic charges for the materials used to test the UFF. Adapted from \cite{Hees:2018fpg,NEW_Berge:2023sqt}.}
\label{tab-Qprimparameters}
\centering
{\small
\begin{tabular}{p{3cm}lcccc}
 \toprule
 Material &  $-Q'_{\hat m}\, [\times10^{-3}]$ & $Q'_{e}\, [\times10^{-3}]$ & $-Q'_{m_{\rm e}}\, [\times10^{-5}]$ &  $Q'_{\delta m}\, [\times10^{-4}]$  \\
 \midrule
H/He [70:30] & 45.51 & 0.36 & $-18.9$& $-11.7$ \\
Fe & 9.94 & 2.32 & 1.89& 1.17 \\
Be  & 17.64& 0.45& 3.05& 1.91\\
Al   & 12.30& 1.47& 1.00& 0.62 \\
Ti   & 10.42& 2.01& 2.24 & 1.38\\
U-238   &7.63 & 4.28& 6.24& 3.86 \\
Cu   & 6.93&2.46 & 2.18 & 1.35 \\
Pb   & 7.73& 3.92& 5.30& 3.28\\
Pt/Rh [90:10]   & 7.83& 3.92& 5.30& 3.28\\
Ti/Al/V [90:6:4]   & 10.52 & 1.98 & 2.17 & 1.34 \\
\bottomrule
\end{tabular}
}
\end{table}

For a {\emph{massless scalar field} ($V=0$) or any field for which the Compton wavelength, $\lambda\propto m_\varphi^{-1}$ is much larger than any other spatial scales, the E\"otv\"os parameter~(\ref{eq_eta}) reduces to $\eta_{ij}=(\alpha_i-\alpha_j)\alpha_{\rm Earth})$ for two test bodies falling in the gravitational field of the Earth. Hence,
\begin{equation} \label{eq_eta_dilaton}
\eta_{ij}(m_\varphi=0)   =  D_{\tilde{m}} \left([Q'_{\tilde{m}}]_i - [Q'_{\tilde{m}}]_j \right) + D_e \left( [Q'_e]_i - [{Q'_e}]_j\right),
\end{equation}
with the coefficients 
$$
D_{\tilde{m}} = d_g^* (d_{\tilde{m}} - d_g),\qquad\hbox{and}\qquad  D_e = d_g^* d_e.
$$
The constraints are summarized on Fig.~\ref{fig-DaDo} [left]  combining the E\"otv\"os constraints from \cite{Braginskii:1971tn,Wagner:2012ui} and \cite{NEW_MICROSCOPE:2019jix}.

For a \emph{massive scalar field}, the range of the composition-dependent fifth force is finite so that the expression for the E\"otv\"os parameter is modified to
\begin{equation} \label{eq_eta_massive_dilaton}
\eta(m_\varphi)  =  \eta(m_\varphi=0)  \times \Phi\left(\frac{R_E}{\lambda_\varphi}\right) \left( 1+ \frac{r}{\lambda_\varphi}\right) \hbox{e}^{-m_{\varphi}r};
\end{equation}
see Eq.~(\ref{eq_eta}) and $r=R_E+h$ with $h$ the altitude of the satellite. The term $\Phi(R_E/\lambda_\varphi)$ arises from the integration of the Yukawa potential on the source, i.e., the Earth and we recall that $\lambda_\varphi\propto m_\varphi^{-1}$. The constraints for scalar masses ranging from $10^{-12}$~eV to $10^{-14}$~eV are summarized on Fig.~\ref{fig-DaDo} [right].  

\begin{figure}[htbp]
  \centerline{\includegraphics[scale=0.32]{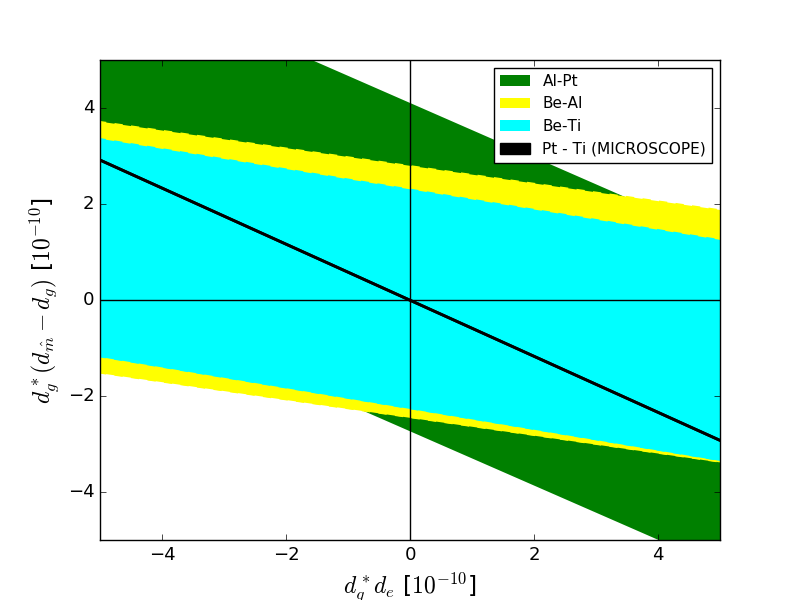}  \includegraphics[scale=0.32]{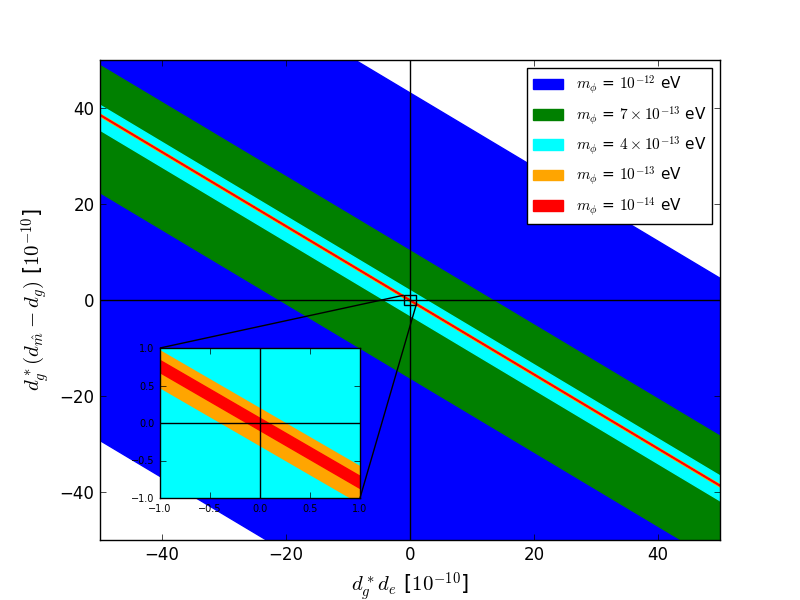}}
  \caption[Constraints on the massless dilaton model]{Constraints on  the couplings of a massless dilaton $(D_{\tilde m},D_e$). The region allowed by the MICROSCOPE \citep{NEW_MICROSCOPE:2019jix} measurement (black band) is compared to earlier constraints by torsion pendulum experiments by \cite{Braginskii:1971tn} (green) and \cite{Wagner:2012ui} (yellow, cyan). The difference of slopes arises from the difference of material used in these 3 experiments. MICROSCOPE allows one to shrink the allowed region by one order of magnitude.  From  \cite{NEW_Berge:2017ovy}.}
  \label{fig-DaDo}
\end{figure}

\paragraph{Atomic clocks}\ 

Using the decomposition of the frequencies of atomic clocks, it is clear, given the Lagrangian~(\ref{L_dd1}) that the best set of parameters to be used is $(\aem,X_{\rm q},X_{\rm e})$ and thus the expression~(\ref{e.QCD3b}) from which one easily deduces that the sensitivity of the frequency $\nu_A$ to the scalar field $\varphi$ defined by
$$
\dd\ln\nu_A = \kappa_A \dd\varphi
$$
is given by
\begin{equation}\label{e.ddclock}
\kappa = K_\alpha d_e + K_{\rm e} (d_{m_{\rm e}}-d_g) + (K_{\rm q}-0.048)(d_{\hat m}-d_g)
\end{equation}
with the coefficients $K$ given in Table~\ref{tab0} for the experiments described in Sect.~\ref{subsec31} and summarized in Table~\ref{tab1}. 

 \begin{tcolorbox}
 The most stringent constraint on the time variation of $\aem$ alone~(\ref{e.clock-alpha}) translates to
\begin{equation}\label{edotphi0}
d_e\dot\varphi_0= (1.8\pm2.5)\times 10^{-19} \unit{yr}^{-1}
\Longleftrightarrow
d_e\left.\frac{\dd \varphi}{\dd\ln a}\right\vert_{a=a_0} = (1.76\pm2.45)\times10^{-9}
\end{equation}
using the value~(\ref{e.hubble}) of the Hubble constant. 
 \end{tcolorbox}
These bounds are valid  for linear couplings. They set strong constraints on the decoupling of the field in the Solar system and hence on all the cosmological dynamics.

\paragraph{Constraints}\ 

The model has been extensively used to interpret local and astrophysical data to the point that it can be considered as a standard lore to that goal, in particular in connection with the search for a ligh scalar dark matter component \citep{NEW_Arvanitaki:2014faa,NEW_Arvanitaki:2015iga,NEW_Stadnik:2015kia,NEW_Stadnik:2014tta,NEW_VanTilburg:2015oza,Derevianko:2016sgw} as described in Sect.~\ref{secULDM}.

The first MICROSCOPE constraints on linear case where published in \citep{NEW_Berge:2017ovy}, see Fig.~\ref{fig-DaDo}. From the modified post-Newtonian equations of motion of an N-body system and the light time travel for a massless dilaton with quadratic coupling, \cite{Bernus:2022zad} predicted the ephemeris of the main planets of the Solar system. The comparison with  the planetary ephemeris INPOP19a, \cite{NEW_Fienga:2014njl} constrained the coupling at the level of ${\cal O}(10^{-5})$; see also \cite{Fienga:2023ocw} for a review on the tests of gravity with ephemeris.

A series of works analyzed clock experiments in this framework. \cite{NEW_Hees:2016gop} searched for oscillations in the Cs-Rb clock data by \cite{NEW_Guena:2012zz} induced by a massive oscillating scalar field with linear coupling. While no sign of a scalar dark matter was found, it constrained the combination $d_{\rm e}+k_{\rm q}(d_{\rm m}-d_{\rm g}/k_\alpha$ of the non universal coupling to standard matter.   \cite{Stadnik:2016zkf} constrained the linear coupling of the light dilaton to  the Higgs boson thanks to atomic clock spectroscopy. \cite{NEW_Stadnik:2014tta} constrained the linear model through variation of fundamental constants with laser and maser interferometry. \cite{Hees:2018fpg} performed an extensive comparison to UFF and clocks and extended their analysis to quadratic couplings. Regarding the linear case, the scalar field being the sum of an oscillating contribution and a Yukawa contribution, the oscillating contribution can be identified as dark matter while the Yukawa interaction leads to a ``standard'' fifth force. Interestingly clocks are more sensitive to the first while UFF are better designed to constrain the Yukawa component. It was concluded that ``natural coupling" of order unity are excluded for mass smaller than $10^{-5}$~eV ($d_e$ only). They pointed out that the quadratic coupling  has a richer phenomenology since no Yukawa interaction is generated and that, Instead, the scalar field exhibits an oscillatory behavior that is perturbed or enhanced by the presence of a massive body. The scalar field oscillations can be amplified (negative $d^{(2)}$) or screened (positive) with the consequence that experiments in space are more interesting to detect or constrain DM with a quadratic coupling - the field is mediating long-distance forces despite being massive.  \cite{PhysRevLett.129.241301}  searched for the oscillating scalar DM fields with frequency comparison between an Yb-171 optical lattice clock and a Cs-133 fountain clock that span 298 days. \cite{NEW_Brzeminski:2022sde} investigated the prospect of increasing the sensitivities of atomic and nuclear clocks by placing them in an eccentric orbit around the Earth. \cite{NEW_Sherrill:2023zah} restricted to two couplings, one to $F^2$ and the second to the electron mass, and the considered scalar field undergoes damped harmonic oscillations with amplitude $\phi_0\exp(-\Gamma t/2)$, pulsation $\omega_d$. The model was constrained thanks to atomic clock through the induced variations of $\aem$ and $\bar\mu$. It was extended by \cite{NEW_Banerjee:2022sqg} to include both a CP odd pseudo-scalar, and a CP-even scalar.

\subsection{Bekenstein and related models}\label{subsub3}

\cite{beken1,beken2} introduced a theoretical framework in which only  the electromagnetic sector was modified by the introduction of a dimensionless scalar field $\epsilon$ so that all electric charges vary in unison $e_i=e_{0i}\epsilon(x^\alpha$) so that only $\aem$ is assumed to possibly vary. 

\paragraph{Early formulation}\ 

To avoid the arbitrariness in the definition of $\epsilon$, which can be rescaled by  a constant factor while $e_{0i}$ is inversely rescaled, it was postulated that the dynamics of $\epsilon$ be invariant under global rescaling so that its action should be of the form
\begin{equation}
 S_\epsilon = -\frac{\hbar c}{2l^2}\int \frac{g^{\mu\nu}\partial_\mu\epsilon\partial_\nu\epsilon}{\epsilon^2}
 \sqrt{-g}\dd^4x,
\end{equation}
$l$ being a constant length scale. Then, $\epsilon$ is assumed to enter all electromagnetic interaction via $e_iA_\mu\rightarrow e_{0i}\epsilon A_\mu$ where $A_\mu$ is the usual electromagnetic potential and the gauge invariance is then preserved only if  $\epsilon A_\mu\rightarrow \epsilon A_\mu + \lambda_{,\mu}$ for any scalar function $\lambda$. It follows that the the action for the electromagnetic sector is the standard Maxwell action
\begin{equation}
 S_\epsilon = -\frac{1}{16\pi}\int F^{\mu\nu}F_{\mu\nu}
 \sqrt{-g}\dd^4x,
\end{equation}
for the generalized Faraday tensor
\begin{equation}
 F_{\mu\nu} =\frac{1}{\epsilon}\left[
 (\epsilon A_\nu)_{,\mu} - (\epsilon A_\mu)_{,\nu}
 \right]\,.
\end{equation}
To finish the gravitational sector is assumed to be described by the standard Einstein--Hilbert action. Finally, the matter action for point particles of mass $m$ takes the form $S_m=\sum\int[-mc^2+(e/c)u^\mu
A_\mu]\gamma^{-1}\delta^3(x^i-x^i(\tau))\dd^4\bx$ where $\gamma$ is the Lorentz factor and $\tau$ the proper time. Note that the Maxwell equation becomes
\begin{equation}
\label{beken4}
\nabla_\mu\left(\epsilon^{-1}F^{\mu\nu}\right)=4\pi j^\nu,
\end{equation}
which is the same as for electromagnetism in a material medium with dielectric constant $\epsilon^{-2}$ and permeability $\epsilon^2$ (this was the original description proposed by \citealt{fierz56} and \citealt{lichne}; see also \citealt{dicke64}). 

As discussed previously, this class of models predict a violation of the  universality of free fall and, from Eq.~(\ref{eq.inertie2}), it is expected that the anomalous acceleration is given by  $\delta\mathbf{a}=-M^{-1}(\partial E_{\mathrm{EM}}/\partial\epsilon)\nabla\epsilon$. From the confrontation of the local and cosmological constraints on the variation of $\epsilon$ \cite{beken1} concluded, given his assumptions on the couplings, that $\aem$ ``\emph{is a parameter, not a dynamical variable}'' (see, however, \citealt{beken2} and then \citealt{kraise}) and concluded the model not to be viable.

\paragraph{Extensions}\ 

It was proposed \citep{sbm} to rewrite this theory by introducing the two fields $a_\mu \equiv \epsilon A_\mu$ and $\psi\equiv \ln\epsilon$ so that the theory takes the form
\begin{equation}
  S= \frac{c^3}{16\pi g}\int R\sqrt{-g}\dd^4x 
        -\frac{1}{16\pi}\int\hbox{e}^{-2\psi} f^{\mu\nu}f_{\mu\nu}\sqrt{-g}\dd^4x 
       -\frac{1}{8\pi\kappa^2} \int(\partial_\mu\psi)^2\sqrt{-g}\dd^4x 
\end{equation}
with $\kappa=l/(4\pi\hbar c)$ and $f_{\mu\nu}$ the Faraday tensor associated with $a_\mu$. The model was further extended to include a potential for $\psi$ \citep{barrowli} and to include the electroweak theory \citep{shaw}.

\cite{olive01} generalized the model to allow for additional couplings of a scalar field $\epsilon^{-2}=B_F(\phi)$ to non-baryonic dark matter (as first proposed in \citealt{damour90}) and cosmological constant. They argue that in supersymmetric dark matter, it is natural to expect that $\phi$ would couple more strongly to dark matter than to baryon. For instance, supersymmetrizing Bekenstein model, $\phi$ will get a coupling to the kinetic term of the gaugino of the form $M_*^{-1}\phi\bar\chi\partial\chi$. Assuming that the gaugino is a large fraction of the stable lightest supersymmetric particle, the coupling to dark matter would then be of order $10^3-10^4$ times larger. Such a factor could almost reconcile the constraint arising from the test of the universality of free fall with the order of magnitude of the cosmological variation. This generalisation of the Bekenstein model is defined as
\begin{align}
\label{olive}
S&=\frac{1}{2}M_4^2\int R\sqrt{-g}\dd^4\bx
-\int\left[\frac{1}{2}M_*^2\partial_\mu\varphi\partial^\mu\varphi+\frac{1}{4}
   B_F(\varphi)
    F_{\mu\nu}F^{\mu\nu}\right]\sqrt{-g}\dd^4\bx
    \nonumber\\
 &\qquad\qquad-\int\left\lbrace\sum\bar
   N_i[i\dslash-m_iB_{N_i}(\varphi)]N_i+\frac{1}{2}\bar\chi\partial\chi
   \right.\nonumber\\
&\left.  \qquad\qquad\qquad\qquad\qquad
   +M_4^2B_\Lambda(\varphi)\Lambda+\frac{1}{2}M_\chi
    B_\chi(\varphi)\chi{}^T\chi
    \right\rbrace \sqrt{-g}\dd^4\bx
\end{align}
where the sum is over proton [$\dslash=\gamma^\mu(\partial_\mu-ie_0A_\mu)$] and neutron [$\dslash=\gamma^\mu\partial_\mu$]. The functions $B$ can be expanded (since one focuses on small variations of the fine-structure constant and thus of $\varphi$) as $B_X=1+\zeta_X\varphi+ \xi_X\varphi^2/2$. It follows that $\aem(\varphi)={e_0^2}/{4\pi B_F(\varphi)}$ so that $\Delta\aem/\aem=\zeta_F\varphi+(\xi_F-2\zeta_F^2)\varphi^2/2$.  Under this form, the effective theory is very similar to the one detailed in Sect.~\ref{subsub2}.

This framework extends the analysis of \cite{beken1} to a 4-dimensional parameter space ($M_*,\zeta_F,\zeta_m,\zeta_\Lambda$). It contains
\begin{itemize}
\item the original Bekenstein model ($\zeta_F=-2$, $\zeta_\Lambda=0$, $\zeta_m\sim10^{-4}\xi_F$), 
\item a Jordan--Brans--Dicke model ($\zeta_F=0$, $\zeta_\Lambda=-2\sqrt{2/2\omega+3}$, $\zeta_m=-1/\sqrt{4\omega+6}$), 
\item a string-like model ($\zeta_F=-\sqrt{2}$, $\zeta_\Lambda=\sqrt{2}$, $\zeta_m=\sqrt{2}/2$) so that $\Delta\aem/\aem=3$) and
\item a  supersymmetrized Bekenstein model ($\zeta_F=-2$, $\zeta_\chi=-2$, $\zeta_m=\zeta_\chi$ so that $\Delta\aem/\aem\sim5/\omega$).
\end{itemize}
In all these models, the universality of free fall sets a strong constraint on $\zeta_F/\sqrt{\omega}$ (with $\omega\equiv M_*/2M_4^2$) and the authors showed that a small set of models may be compatible with a variation of $\aem$ from quasar data while being compatible with  the equivalence principle tests. A similar analysis \citep{marra} concluded that such models can reproduce the variation of $\aem$ from quasars while being compatible with Oklo and meteorite data. \cite{NEW_Onegin:2014hua} argued that the $|\zeta_m(l/l_{\rm P})|<6\times 10^{-7}$ to fullfil Oklo bounds so that the characteristic length of the model shall be smaller than the Planck scale if $\zeta_m\sim10^{-4}$.

\paragraph{Constraints}\ 

A series of works \citep{NEW_Vielzeuf:2013aja,NEW_Martins:2017yxk,NEW_Martins:2022pue,NEW_Vacher:2022sro,NEW_Tohfa:2023zip}  have considered the constraints that arise from cosmological data, the latest astrophysical data on $\aem$ and local constraints . They all point to the conclusion that typically $\zeta<10^{-7}$ and that  the local data dominate the constraints. In particular, the connection between $\eta$ and $\zeta$ \citep{sbm}, $\eta =3\times10^9 \zeta$, implies that MICROSCOPE sets immediately $\eta<10^{-5}$.  \cite{NEW_Leite:2016buh} estimated that astrophysical data ae about 20 times more constraining for $\zeta$ and that the coming E-ELT observations could reach a factor 50 compared to the universality of free fall. \cite{NEW_Li:2010zt} developed $N$-body simulations including the resolution of the scalar field equation. While it  does not cluster significantly, because of its weak coupling to matter, it modifies the structure formation. This was complemented by the study of gravitational collapse \citep{NEW_Chakrabarti:2023zud} in which dark matter is modeled by a Bekenstein scalar field. \cite{NEW_Liu:2021mfk} constrained the model by use of the SZ effect (see Sect.~\ref{subsecSZ}) using two galaxy cluster samples, the 61 clusters provided by the Planck collaboration and the 58 clusters detected by the South Pole Telescope, to conclude that $\zeta>{\cal O}(10^{-2})$.
 
On smaller scales, \cite{NEW_Danielsson:2016nyy} investigated the phenomenology of a heavy Bekenstein scalar for accelerator physics, in particular at LHC since it would be produced through photon-photon fusion, leading to diphoton final state, and by quark-antiquark fusion. This was later generalized to include varying gauge coupling in the electroweak sector \citep{NEW_Danielsson:2019ftq} with either one or two scalars and the signatures were discussed. This allowed them to set constraints on the masses of these scalars and the energy scale of the theory from the LHC $\gamma\gamma, \gamma Z, ZZ, WW$ and $jj$ resonance search results. For instance $\gamma\gamma$ implies that the energy scale for new physics has to be larger than 20~TeV for a scalar of mass $\sim1$~TeV. 

\paragraph{Further studies}\ 

This theory was also used \citep{beken3} to study the spacetime structure around  charged black-hole, which corresponds to an extension of dilatonic charged black hole. It was concluded that a cosmological growth of $\aem$ would decrease the black-hole entropy but with half the rate expected from the earlier analysis \citep{bha1,bah2}.

\cite{NEW_Leszczynska:2018juk} considered  an extension of the Bekenstein theory in which the scalar field enjoys an extra-coupling to the baryon current, i.e., a term $\propto \lambda J^\mu_{\rm baryon} \partial_\mu\varphi$ that spontaneously breaks the baryon symmetry \citep{NEW_Dimopoulos:1978kv}, hence relating baryogenesis to varying constants. It enjoys the interesting feature that the current baryon entropy $\eta\sim 8.6\times10^{-11}$ can then be obtained for a large range of parameters.

\subsection{Light scalar quintessence}\label{subsub0}

\paragraph{Model}\ 

A series of works have investigated the connection between quintessence and the variation of $\aem$ with an action of the form~(\ref{e.BFphi}) assuming $\varphi$ is a quintessence field, i.e., that it enjoys a potential leading to a late time cosmic acceleration. Even though, they are a subcase of the light-dilaton models described in Sect.~\ref{subsub1b} with only $d_e$ non-vanishing, this class of models has been extensively studied since it provides a simple example to study the connection between the dark energy equation of state and the time variation of the fine structure constant. As we have seen in Sect.~\ref{subsecSTmulti}, even if there is no direct coupling to the matter field but the Faraday tensor, the masses of protons, neutrons and all nuclei will be time varying due to their electromagnetic binding energy so that such a coupling leads to a strong violation of the universality of free fall. 

We recall that the action is given by
$$
 S = \frac{1}{16\pi G}  \int\left[R- 2(\partial_\mu\varphi)^2 - V(\varphi)  -\frac{1}{4}B_F(\varphi)F_{\mu\nu}^2 \right]\sqrt{-g}\dd^4 x+
  S_{\text{matter}}[\psi;g_{\mu\nu}].
$$
This model was initially considered by \cite{beken1} with the coupling
\begin{equation}\label{e.356}
B_F=1-\zeta (\varphi-\varphi_0)\,.
\end{equation}
It follows that
\begin{equation}\label{e.BFalpha}
\frac{\Delta\aem}{\aem} = \frac{1-B_F(\varphi)}{B_F(\varphi)}   \simeq \zeta(\varphi-\varphi_0)
\end{equation}
As discussed in Sect.~\ref{subsecphico}, light field models are generically thought to lead to a coupling to $F^2$ so that linear coupling were first investigated. This was then generalized to quintessence models with  a couplings of the form $Z(\phi) F^{\mu\nu}F_{\mu\nu}$ \citep{xyz3,anchor03,parkinson,copQ,doran,lee2,marra,lee} and then to the runaway dilaton \citep{damourrunaway2,damourrunaway}; Sect.~\ref{subsub1}. The evolution of the scalar field drives both the acceleration of the universe at late time and the variation of the constants. 

The dynamics of $\varphi$ depends on its potential. It has two main regimes, either as dark energy if $V$ allows for slow-roll or as dark matter if $\varphi$ oscillates; see Sect.~\ref{secULDM}.  In both case there is a strong connection between either $\dot\varphi_0$ or $\varphi_0$ and the cosmological parameters hence offering a local/global issue. In particular, whatever the model, it has to satisfy the strong bound~(\ref{edotphi0}).

\paragraph{Connection with the universality of free fall}\ 

As detailed in Sect.~\ref{subsecSTmulti} such a coupling induces a strong violation of the universality of free fall, which was one of the motivation of the original work by \cite{beken1}.

This  was revisited in \cite{chiba2001,dvaliZ,wetterich02} in which the dependence of $\aem$ on the scalar field responsible for its variation is expanded as  $\aem=\aem(0)+\lambda\varphi+{\cal O}(\varphi^2)$. The QSO data on $\aem$ were then pointing to $\lambda\Delta\varphi \sim 10^{-7}$ at best during the last Hubble time.  

Concentrating only on the electromagnetic binding energy contribution to the proton and of the neutron masses -- see Eq.~(\ref{gl}) -- it was concluded that a test body composed of $n_{\mathrm{n}}$ neutrons and $n_{\mathrm{p}}$ protons will have a sensitivity ${\lambda}(\nu_{\mathrm{p}}B_{\mathrm{p}}+\nu_{\mathrm{n}}B_{\mathrm{n}})/{m_{\mathrm{N}}}$
where $\nu_{\mathrm{n}}$ (resp.\ $\nu_{\mathrm{p}}$) is the ratio of neutrons (resp.\ protons). Assuming\footnote{For copper $\nu_{\mathrm{p}}=0.456$, for uranium $\nu_{\mathrm{p}}=0.385$ and for lead $\nu_{\mathrm{p}}=0.397$.}  that $\nu_{\mathrm{n,p}}^{\text{Earth}}\sim1/2$ and using that the compactness of the Moon-Earth system $\partial\ln(m_{\text{Earth}}/m_{\text{Moon}})/\partial\ln\aem\sim10^{-3}$, one gets $\eta_{12}\sim10^{-3}\lambda^2$.  \cite{dvaliZ} obtained the same result by considering that $\Delta\nu_{\mathrm{n,p}}\sim6\times10^{-2}-10^{-1}$. This implies that $\lambda<10^{-5}$, which is compatible with a variation of $\aem$ if $\Delta\varphi>10^{-2}$ during the last Hubble period. From the cosmology one can deduce that $(\Delta\varphi)^2\sim (\rho_\varphi+P_\varphi)/\rho_{\text{total}}$. If $\varphi$ dominates the matter content of the universe, $\rho_{\text{total}}$, then $\Delta\varphi\sim 1$ so that $\lambda\sim 10^{-7}$ whereas if it is sub-dominant $\Delta\varphi\ll 1$ and $\lambda\gg 10^{-7}$. In conclusion $10^{-7}<\lambda<10^{-5}$. This makes explicit the tuning of the parameter $\lambda$. This analysis was extended in \cite{dentuff} who included explicitly the electron and related the violation of the universality of free fall to the variation of $\mu$.

Similarly,  \cite{wetterich02} assumed the scalar field $\varphi$ to be responsible for both a variation of $\aem$ and the acceleration of the cosmic expansion. Assuming its equation of state is $w_\varphi\not=1$, one can express its time variation as
\begin{equation}\label{e.phiw}
 \dot\varphi= H\sqrt{3\Omega_\varphi(1+w_\varphi)}.
\end{equation}
Hence, the expected violation of the universality of free fall is related to the time variation of $\aem$ today by
$$
 \eta=-1.75\times10^{-2}\left(\frac{\partial\ln\aem}{\partial z}\right)^2_{z=0}  \frac{(1+\tilde Q)\Delta\frac{Z}{Z+N}}{\Omega_\varphi^{(0)}\left(1+w_\varphi^{(0)}\right)},
$$
where $\tilde Q$ is a parameter taking into account the influence of the mass ratios. Again, this shows that in the worse case in which the Oklo bound is saturated (so that $\partial\ln\aem/\partial z\sim10^{-6}$), one requires $1+w_\varphi^{(0)}\gtrsim10^{-2}$ for $\eta\lesssim10^{-13}$, hence providing a string bond between the dark energy equation of state and the violation of the universality of free fall. \cite{dent2} extended it  to the phenomenological unification model presented in Sect.~\ref{subsecGUT}. In the case of the string dilaton and runaway dilaton models, one reaches a similar conclusion [see Eq.~(\ref{leastuff}) in Sect.~\ref{subsub1}]. Similarly, \cite{msu}  related the equation of state to the post-Newtonian parameters. 

In all these models, the link between the local constraints and the cosmological constraints arise from the fact that local experiments constrain the upper value of $\dot\varphi_0$, which quantify both the deviation of its equation of state from $-1$ and the variation of the constants. It was conjectured that most realistic quintessence models suffer from such a problem \citep{braxm2}.

\paragraph{Generic local constraints: UFF vs $\aem$}\ 

This simplest model allow one to compare the strength of the different local constraints. 

To that goal, we can simply starts from the analysis of Sect.~\ref{subsub1b} in which the only non vanishing coupling is $B_F$ and perform the replacement $1+d_e\varphi\rightarrow B_F(\varphi)$ so that $d_e\rightarrow \alpha_F(\varphi)\equiv\partial_\varphi\ln B_F$. In such a case $d_g^`*$ just gets the contribution for the electromagnetic binding energy so that $d_g^*\rightarrow 2.7\times10^{-4}\alpha_F$ and $D_e\rightarrow d_g^*\alpha_F$. Hence, Eq.~(\ref{eq_eta_dilaton}) for the E\"otv\"os parameter simplifies to 
$$
\eta_{ij}(m_\varphi=0)=2.7\times 10^{-4} ([Q_e']_i - [Q_e']_j)\alpha_F^2(\varphi_0).
$$
Hence from MICROSCOPE~(\ref{eta-microscope}), $\eta_{ij}(m_\varphi=0)=5.2\times10^{-7}\alpha_F^2$ and from  atomic clocks.~(\ref{edotphi0}), one gets  can be rewritten thanks to Eq.~(\ref{e.BFalpha}),
$$
\left.\frac{\dot\aem}{\aem}\right\vert_0 = \alpha_F(\varphi_0)\dot\varphi_0\,.
$$
 \begin{tcolorbox}
 It follows that the combination atomic clocks and UFF provides a constraint on $(\alpha_F(\varphi_0),\dot\varphi_0)$ that is on $(\varphi_0,\dot\varphi_0)$,
\begin{eqnarray}
\alpha_F(\varphi_0)\dot\varphi_0&=& (1.8\pm2.5)\times 10^{-19} \unit{yr}^{-1}\label{edotphi1}\\
\alpha_F^2(\varphi_0)\ &=& (-7.8\pm 11.96 \pm 7.8_{\rm syst})\times 10^{-9},\label{eeta1}
\end{eqnarray}
for a massless scalar field. This sets a strong constraints on $(\varphi_0,\dot\varphi_0)$ whatever $B_F(\varphi)$ and $V(\varphi)$.
 \end{tcolorbox}
One shall also take the constraints arising from the annual modulation (see~Sect.~\ref{subsuba} below for the discussion on those constraints). Since $k_\alpha=\alpha_\odot(\varphi_0)\alpha_F(\varphi_0)$ and since in this restricted model with only $B_F$ non-vanishing, $\alpha_\odot=3.6\times10^{-4}\alpha_F(\varphi_0)$ -- see Table~\ref{tab-Qprimparameters} -- we conclude that the bound~(\ref{e.kalphaLast}) translates to
$$
\alpha_F^2(\varphi_0)=(-6.6\pm8.3)\times10^{-6},
$$
which remains 3 orders of magnitude less constraining than MICROSCOPE. Indeed, in a more realistic model, the two observables are complementary since they exhibit different dependencies on the $\alpha_i$. The MICROSCOPE constraints on the case of a linearly coupled massive scalar field are summarized on Fig.~\ref{fig-deseul1}.

Then for all models in which this scalar fields also account for dark energy, we shall have
$$
\dot\varphi^2=3H_0^2 \Omega_{\varphi} (1+w_\varphi)
$$
so that the constraints on the dark energy equation of state, $1+w=(-0.03\pm0.03)$ \citep{NEW_Planck:2018vyg}, bounds $\dot\varphi_0$ independently.

\begin{figure}[htbp]
 \centerline{\includegraphics[scale=0.4]{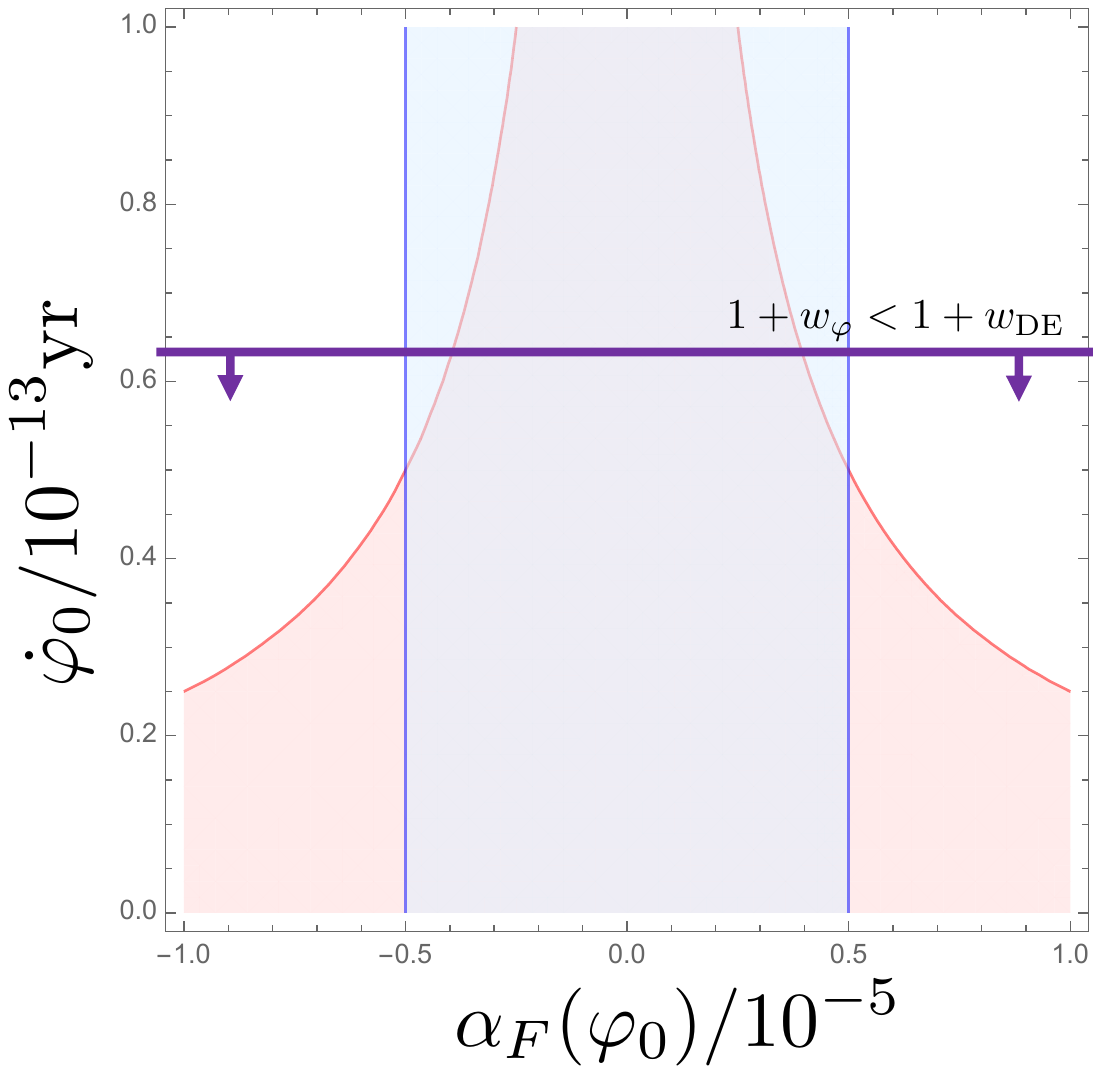}}
\vskip-.35cm
  \caption[Complementarity of local and cosmological data]{Example of the  complementarity of the local and astrophysical data. The blue and red regions represent the allowed values of $\alpha_F(\varphi_0)$ and $\dot\varphi_0$ -- see Eqs.~(\ref{edotphi1}-\ref{eeta1}) --  for a for a massless scalar field  with unique $B_F F^2$ coupling. This shows the complementarity of clock and UFF experiments to constrain $(\varphi_0,\dot\varphi_0)$. Note that the UFF constraint is boosted when other couplings are active since the electromagnetic binding energy is only a small part of the total  mass of the Earth and test masses. Then, if this field accounts for the acceleration of the cosmic expansion,  any constraints on the dark energy equation of state would imply an upper bound on $\dot\varphi_0$. Within a theoretical framework, one can make similar plot at different redshifts, hence providing a constraints on the dynamics of the new degree of freedom.}
  \label{fig-deseul0}
\end{figure}

\paragraph{Constraints}\ 

This simple extension of scalar field cosmology enjoys 2 free functions, the potential that can turn $\varphi$ as either a dark energy or dark matter component, and the coupling function $B_F$. It offers an easy way to compare observations and connect them to the cosmological dynamics, in particular to evaluate the gain of including non-cosmological data when forecasting future experiments. In particular Eq.~(\ref{e.phiw}) implies that
$$
\varphi-\varphi_0=\int_{0}^z\sqrt{3\Omega_{\varphi}(z) [1+\omega_\varphi(z)]}\frac{\dd z}{1+z}.
$$
Besides, the conservation equation, $\dot\rho_\varphi + 3H(1+w_\varphi)\rho_{\varphi}=0$ when one neglects the coupling to matter integrates to
$$
\rho_\varphi = \rho_{\varphi0}\exp\left( 3\int_0^z [1+w_\varphi(z)] \frac{\dd z}{1+z} \right).
$$
Besides Eq.~(\ref{e.356}) implies that
$$
\frac{\Delta\aem}{\aem} = \zeta \int_{0}^z\sqrt{3\Omega_{\varphi}(z) (1+\omega_\varphi(z)}\frac{\dd z}{1+z}.
$$
It is clear that the local, astrophysical and cosmological constraints on $\aem(z)$ provide a direct monitoring on the field dynamics $\varphi(z)$ and thus new bounds on the cosmological dynamics.

As a consequence, \cite{NEW_Avelino:2011dh} addressed the reconstruction problem of the dark energy equation of state in order to discuss its degeneracies with $\aem$ and the limits of the reconstruction from data. This was applied to quasar absorption spectra and clocks data for $\aem$ and SN~Ia data for dark energy \citep{NEW_Martins:2015ama,NEW_Martins:2015jta,NEW_Martins:2016oyv} for various quintessence models; see also \cite{NEW_Leite:2016sff,NEW_Alves:2018mef} for the discussion of the improvement expected from ESPRESSO and later E-ELT. \cite{NEW_Zhai:2012de} constrained a model with inverse power law or exponential potential from QSO data on $\aem$ to get $\zeta\simeq10^{-5}$ but they did not check whether this fit is viable with respect to UFF constraints. 

This is complemented by more phenomenological studied parametrizing the dark energy equation of sates or the scalar field redshift evolution. \cite{NEW_Martins:2022unf} discussed the possibility to constrain a series expansion of $\Delta\aem/\aem$ in powers of $z/(1+z)$ in a similar way as what is done for the dark energy equation of state. \cite{NEW_Leite:2014vka} described the quintessence equation of state by a step function in redshift space to get a parametric form of $\phi(z)$ and thus of $\aem(z)$ in terms of $N$ parameters to investigate the power of astrophysical data on $\aem$ to unveil the equation of state of dark energy. Assuming $\varphi-\varphi_0\propto \lambda\ln a$, so that $\Delta\aem/\aem=-\zeta\lambda\ln(1+z)$, which implies that the quintessence equation of state is $w=-1+\lambda^2/3(1-\Omega_{\rm m})$. \cite{NEW_daFonseca:2022qdf} showed that the constraints $\zeta\lesssim 1.4\times 10^{-4}$ (CMB  + QSO data including the latest ESPRESSO data) increases to $\zeta\lesssim 1.1\times 10^{-4}$ when atomic clocks are added and to  $\zeta\lesssim 1.6\times 10^{-6}$ with MICROSCOPE. The constraints on $\lambda$ do not change much  and remain at the level of $(2-3)\times10^{-2}$. More speculatively, \cite{NEW_Schoneberg:2023lun} showed, using CMB, SNIa, BAO, Òclocks, QSO, Oklo for $\aem$, that a linear coupling to electromagnetism sets a new constraint on the swampland criteria since data favor $\zeta<1.5\times10^{-1}\ll 1-10$ as expected from naturalness \cite{NEW_Euclid:2021cfn} investigated how well the ESO/Euclid data can improve the bounds on $\zeta$ thanks to its sensitivity the dark energy equation of state through weak lensing and large-scale structure growth. It is expected that the Euclid satellite mission will shrink the allowed values for $\zeta$. 

The latest developments are related to the new constraint on $\dot \aem/\aem$ by \cite{NEW_Filzinger:2023zrs} which implies the constraint~(\ref{edotphi0}). First, it allowed \cite{NEW_Filzinger:2023zrs} to improve the bound on $d_e$ by more than one order of magnitude in the mass range $(10^{-24}-10^{-17})$~eV. Then, it allowed \cite{NEW_Vacher:2024qiq} to show that models of single field coupled to $F^2$ with a potential $V$ have either a too large fine-structure constant at recombination while remaining compatible with \cite{NEW_Filzinger:2023zrs} today  or require an extreme fine tuning of the potential or its initial conditions in order to alleviate the Hubble tension. Again, this is a  nice example of the importance of the complementarity between local and cosmological data.

\begin{figure}[htbp]
 \centerline{\includegraphics[scale=0.25]{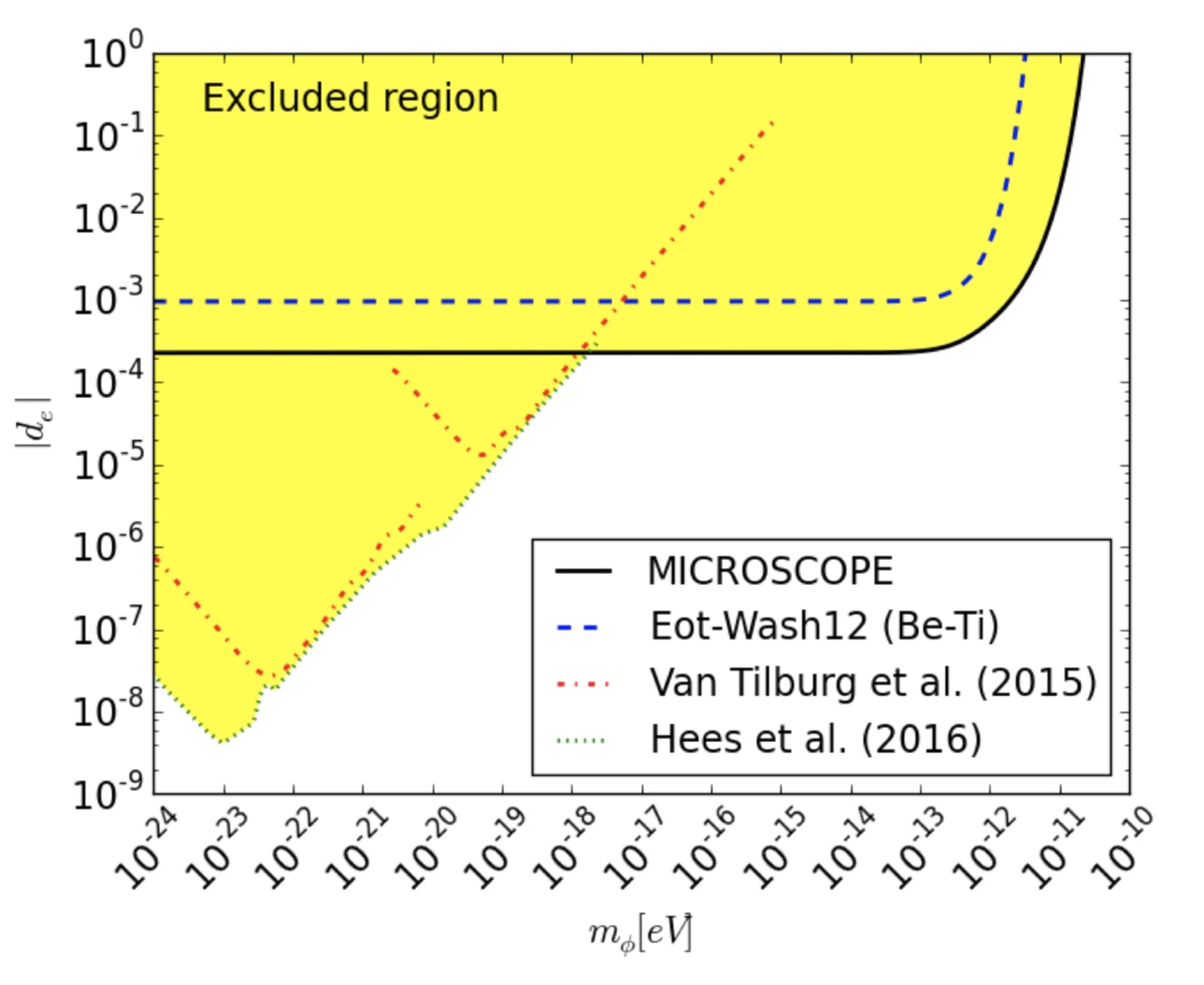}}
  \caption[Constraints on $d_e$ for massive dilaton models]{Constraints on $d_e$ for a massive dilaton coupled linearly only to the electromagnetic sector, compared with constraints (dot-dahed) from atomic spectroscopy \citep{NEW_VanTilburg:2015oza,NEW_Hees:2016gop} and E\"ot-Wash WEP test (dashed) by \cite{Schlamminger:2007ht}. From  \cite{NEW_Berge:2017ovy}.}
  \label{fig-deseul1}
\end{figure}

\subsection{Other ideas}

Since it is impossible to give a exhaustive description of all the models, let us mention without details other theoretical frameworks proposed to accommodate varying constants.

\begin{itemize}

 \item Models involving a late time phase transition in the  electromagnetic sector \citep{chako,anchot};
 \item Braneworld models \citep{aguilar,palma,byrne,librane,bw0} or extra-dimensions \citep{steinh};
 \item Models with pseudo-scalar couplings \citep{flamst};
 \item Growing neutrino models \citep{gnu2,gnu1} in which the neutrino  masses are function of a scalar field, that is also responsible   for the late time acceleration of the universe. In these models the   neutrinos freeze the evolution of the scalar field when they become  non-relativistic while its evolution is similar as in quintessence   when the neutrinos are ultra-relativistic;
 \item Models based on discrete quantum gravity \citep{gambini} or on   loop quantum gravity in which the Barbero--Immirzi parameter   controls the minimum eigenvalue of the area operator and could be   promoted to a field, leading to a classical coupling of Einstein's   gravity with a scalar-field stress-energy tensor \citep{lqg2,lqg1}
 \item ``Varying speed of light'' models for which we refer to the   review by \cite{vsl} and \citep{ellisu} for a   critical view;
 \item Quintessence models with a non-minimal coupling of the   quintessence field \citep{chiba2001,anchor03,xyz3,copQ,parkinson,doran,lee2,marra,avelino06,lee, fujii0, nunes09} [see discussion Sect.~\ref{subseccosmo}];
 \item It has be shown \citep{vonHarling:2016vhf} that, within the Randall-Sundrum mechanism, a; Goldberger-Wise scalar field couples to the fermions,  leading to a model of varying Yukawa couplings;
  \item Holographic dark energy models with non-minimal  couplings. \citep{granda};
 \item \cite{NEW_vandeBruck:2015rma} extended the light field model coupled solely to $F^2$ to a situation in which electromagnetism is disformally coupled to the scalar field, i.e., the electromagnetism Lagrangian is of the form $h(\varphi)\tilde g^{\mu\nu}\tilde g^{\alpha\beta}F_{\mu\alpha}F_{\nu\beta}$ with $\tilde g_{\mu\nu}= C(\varphi) g_{\mu\nu}+ D(\varphi)\partial_\mu\varphi\partial_\nu\varphi$. Similarly matter is coupled to the same metric but with possibly different functions $C$ and $D$;
 \item \cite{NEW_Barros:2022kpo} generalized this same model to include a coupling to the scalar field kinetic term, i.e., an interaction Lagrangian of the form $h[\phi,X]F^2$ with $X\equiv -g^{\mu\nu}\partial_\mu\varphi\partial_\nu\varphi/2$. Assuming $h=h(X)\propto X^{-\eta}$ and that $\varphi$ accounts for quintessence on cosmological scales. CMB, BBN, QSO, atomic clocks and MICROSCOPE imply $\zeta\lesssim 10^{-7}$;
 \item Brans--Dicke theories are generically highly constrained by Solar system tests unless --  see Sect.~\ref{subsecST} --, their couplings to ordinary matter are much suppressed relative to gravitational strength. As discussed, this is a major obstacle to construct realistic models of light dilatons. On the other hand, these scalars appear naturally in UV complete theories like string theory, where they arise as dilatons for the various accidental scaling symmetries that are generic in higher-dimensional supergravities. This led \cite{NEW_Burgess:2021qti} to remark that if matter also couples to a light axion, this would qualitatively change how the fields respond to a gravitating source. As  consequence, matter-dilaton couplings that would be excluded in the absence of an axion can become acceptable given an additional small axion-matter coupling. Fifth forces on matter test bodies are then controlled by the much weaker axion-matter couplings and can easily be small enough to escape detection. This  \emph{axion homeopathy} mechanism is further explored by \cite{NEW_Burgess:2021obw,NEW_Brax:2022vlf}.
 \item \cite{NEW_Brax:2023udt} studied the dynamics of a light scalar field responsible for the $\mu$ term of the Higgs potential and coupled to matter via the Higgs-portal mechanism.  These models could be subject to a screening akin to the symmetron mechanism so that local gravitational tests are evaded thanks to the weakness of the quadratic coupling in the dark matter halo.
\end{itemize}

\section{Spatial variations}\label{section-spatial}

The constraints on the variation of the fundamental constants that we have described so far are mainly related to their cosmological evolution so that, given the Copernican principle, they reduce to constrain on their eventual time variation. Indeed, spatial variations shall also occur, at least in two regimes:
\begin{itemize}
 \item On \emph{cosmological scales}, one shall consider two situations since (1) the fields dictating the variation of  the constants have fluctuations that can let their imprint in some cosmological observables, in particular on large scales if they are light and (2) one can check if the distributions are compatible with the cosmological principle, in particular by constraining their possible dipolar, quadrupolar, etc. deviations from homogenity.
 \item On \emph{local scales} (e.g.,  our Solar system or the Milky Way) the fields at  the origin of the variation of the constants are sourced by the  local matter distribution so that one expect that the constants are  not homogeneous on these scales. The motion of the Earth and the laboratory in which experiments are performed can then reveal this spatial dependence through modulations of the experimental results.
\end{itemize}
Besides, these studies can set constraints on the variation of the constant with either the gravitational potential, as expected in many scalar field models, or the local density of matter, as in chameleon mechanism; see Sect.~\ref{subsub2}. They are thus highly complementary to those in the laboratory and in intergalactic absorption systems.

\subsection{Generalities}

In order to determine the profile of the constants in the Solar system, assume that their values derive from a scalar field. As in Sect.~\ref{subsec_newfields}, we consider that at lowest order the profile of this scalar field is described by the scalar-tensor theory of the type~(\ref{e.genericS}) in which fall most of the models studied in Sect.~\ref{subsec81}, taking into account that all masses scale as  $\Lambda_{\mathrm{QCD}}(\phi_*)$ where $\phi_*$ is the value of the field in the Einstein frame.

\subsection{Local scales}\label{subsuba}

\subsubsection{Solar system scales}

In order to infer the field distribution within the Solar system, we restrict to the weakly self-gravitating ($\Phi_*/c^2\ll 1$) and slow moving ($T^{01}\ll T^{00}$)
localized material systems and follow \cite{damour92}. 

\paragraph{Scalar field profile}\ 

In the Einstein frame and for generic scalar field theories of the form~(\ref{actionEF}) or~(\ref{e.genericS}), the Einstein equation~(\ref{einstein*}) reduces to
\begin{equation}
   \Box_{*}\Phi_* = -4\pi G_*\sigma_*
\end{equation}
where $\Box_*$ is the d'Alembertian in Minkowski spacetime and $\sigma_*$ is given by
$$
 \sigma_*=T^{00}_* + T^{ii}_* = \rho_*+3P_*\,.
$$
The Klein--Gordon  equation~(\ref{Boxvarphi}) reduces to Eq.~(\ref{e.KGgen2}) with the generic effective potential~(\ref{e.Veff2}). Assuming the scalar-field energy density is negligible compared to the matter energy density and restricting to non-relativistic matter $(\rho\gg P/3 )$, we end up with the system
\begin{equation}
   \Box_{*}\Phi_* = -4\pi G_*\rho_*, \qquad
   \Box_*\varphi_* = \frac{\dd V}{\dd\varphi_*} + 4\pi G\alpha(\varphi_*)\rho_*
\end{equation}
where have gathered the sum on $\alpha_i\rho_i$ of Eq.~(\ref{e.Veff2}) in a single term. Restricting to the static case with a single massive point source, the only non-vanishing source terms are $\sigma_*(\br)=M_*\delta^{3}(\br_*)$ so that the set of equations reduces to two Poisson equations
\begin{equation}\label{daleq}
   \Delta_{*}\Phi_* = -4\pi G_*M_*\delta^{3}(\br_*) \qquad
     \Delta_{*}\varphi_* = \frac{\dd V}{\dd\varphi_*}+ 4\pi G_*M_*\alpha(\varphi_*) \delta^{3}(\br_*).
\end{equation}
The equation for $\varphi_*$ can be solved iteratively by expanding the field as $\varphi_*=\varphi_{0*}+ \varphi_{1*}(r_*)$ so that at lowest order $\varphi_{1*}$ is solution of
$$
\left(\Delta_{*}-m_\varphi^2\right) \varphi_{1*} =  4\pi G_*M_* \alpha_0 \delta^{3}(\br_*)
$$
with $\alpha_0\equiv \alpha(\varphi_{0*})$ and $m_\varphi^2=V''(\varphi_{0})$. It follows that we end up with
\begin{equation}\label{daleq1}
   \Delta_{*}\Phi_* = -4\pi G_*M_*\delta^{3}(\br_*) \qquad
   \left(\Delta_{*}-m_\varphi^2\right) \varphi_{1*} =  4\pi G_*M_* \alpha_0 \delta^{3}(\br_*)\,.
\end{equation}
This set of equations are solved by means of the Green function
$$
{\cal G}_{m_\varphi}({\bf x}-{\bf x}') = -\frac{1}{4\pi}  \frac{\hbox{e}^{-m_\varphi |{\bf x}-{\bf x}'|}}{|{\bf x}-{\bf x}'|}
$$
to get
$$
\Phi_*(r_*) = \frac{G_*M_*}{r_*}, \quad
\varphi_{1*}(r)= -\frac{G_*M_*\alpha_0}{r_*} \hbox{e}^{-m_\varphi r_*}.
$$
This can be used to determine the  metric and scalar field in the Jordan frame. It follows that at lowest order, they are given by
\begin{equation}
 \Phi_{\rm N} =  \frac{GM}{r},\qquad
  \varphi_*(r) = \varphi_{*0} - \alpha_0 \Phi_{\rm N}(r)\hbox{e}^{-m_\varphi r}\,,
\end{equation}
where we have neglected the corrections $-\alpha(\varphi)(\varphi-\varphi_0)$ for the gravitational potential, which, given the Solar system constraints on $\alpha_0$, is a good approximation.

\paragraph{Seasonal modulations of the fundamental constant}\ 

Now, let us consider any constant $c_i$ function of $\varphi$. Thus, its profile is given by $c_i(r) =c_i[\varphi(r)] =  c_i(\varphi_0) - \alpha_0 ( \dd c_i/\dd\varphi)_{\phi_0} \Phi_{\rm N}(r)\hbox{e}^{-m_\varphi r}$
so that
\begin{equation}\label{e.272}
 \frac{\Delta c_i}{c_i}(r)=  - s_i(\phi_0)\alpha_0 \Phi_{\rm N}(r)\hbox{e}^{-m_\varphi r}
\end{equation}
where $s_i(\phi_0)$ is the sensitivity of the constant $\alpha_i$ to a variation of the scalar field as defined in Eq.~(\ref{def_sphi}), i.e., $s_i\equiv\dd\ln\alpha_i/\dd\phi$. 

For a laboratory in orbit on an elliptic trajectory,
$$
 r = \frac{a(1-e^2)}{1+e\cos\psi}, \qquad
 \cos\psi = \frac{\cos E - e}{1- e\cos E},\qquad
 t = \sqrt{\frac{a^3}{GM}}(E - e\sin E)
$$
where $a$ is the semi-major axis, $e$ the eccentricity and $\psi$ the true anomaly, which holds if the mass of the scalar field is small enough on the scale of the orbit, $m_\varphi a\ll 1$, which we assume in the following. Hence, expending in $e$, 
$$
  \frac{\Delta c_i}{c_i}(a,\psi) = -s_0\alpha_0\frac{GM}{a}  -s_0\alpha_0\frac{GM}{a}e\cos\psi +{\mathcal O}(e^2) \,.
$$
The first term represents the variation of the mean value of the constant on the orbit compared to its cosmological value. This shows that  local terrestrial and Solar system experiments do measure the effects of the cosmological variation of the constants \citep{damour92,shaw,shaw2,shaw0}. The second term is a seasonal modulation arising from the motion of the Earth within the static profile of $\varphi$.

\paragraph{Parameterisation of the seasonal modulation}\ 

It is usually  parameterized \citep{local1} -- coming back to our original notion $\alpha_i$ for a fundamental constant now that no confusion with $\alpha_0$ is possible -- as
\begin{equation} \label{defki}
  \left.\frac{\Delta\alpha_i}{\alpha_i}\right\vert_{\text{seasonal}} \equiv k_i\frac{\Delta\Phi_{\rm N}}{c^2},
\end{equation}
defining the parameters $k_i$. 

For a clock relying on the transition $A$, one can define the gravitational redshift -- see Eq.~(\ref{e.zU}) --  as
\begin{equation}\label{e.dddbetaA}
\frac{\Delta\nu_A}{\nu_A}\equiv (1+\beta_A) \frac{\Delta\Phi_{\rm N}}{c^2}\,.
\end{equation}
In General Relativity, as a consequence of LPI and as discussed in Eq.~\ref{e.redshift}, $\beta_A=0$. In particular when the WEP holds, the redshift does not depend on the clock. Note that the definition~(\ref{e.dddbetaA}) involves $\Phi_{\rm N}$ which may differ from $U$ used in the definition~(\ref{e.zU}).  If LPI is violated, then the relative frequency drift between two clocks,
\begin{equation}\label{e.gh}
\Delta \ln y_{AB}=\beta_{AB}\frac{\Delta\Phi}{c^2} \qquad \beta_{AB}=\beta_A-\beta_B\,,
\end{equation}
may depend on the clock so that redshifts are no more achromatic. 

From the definition of the sensitivity coefficients~(\ref{e.dataQED}, \ref{e.dataQCD}), $\Delta\ln y_{AB} = \sum_i K_{AB}^i \Delta\ln \alpha_i$, which reduces thanks to~(\ref{defki}) to $\sum_i K_{AB}^i k_i \Delta\Phi_{\rm N}$. Hence,
\begin{equation}\label{e.betaK}
\beta_{AB}= \sum_{i=\aem,\mu,q} K_{AB}^i k_i \quad \Leftrightarrow\quad
\beta_{A}= \sum_{i=\aem,\mu,q} K_{A}^i k_i .
\end{equation}
In this expression the relative redshift, $\beta_{AB}$, is split into a linear combination of the clocks properties encoded in the sensitivity coefficients $K_A^i$ and on the parameter $k_i$ that describes the physics beyond the standard model, e.g., how the light field couples to the standard model fields. It follows that one can extract constraints on the $k_i$ from different experiments.

Similarly, for the universality of free fall, the definition~(\ref{defki}) implies that the general expression~(\ref{e.etaGrad}) for the E\"otv\"os parameter reduces
\begin{equation}
 \eta_{AB}  = \sum_i\left\vert \lambda_{Ai} - \lambda_{Bi}  \right\vert k_i,
\end{equation}
since $\vert\nabla\vert\alpha_i=\alpha_i k_i g_N$ and making use of the relation~(\ref{e.lambdafi}) between the sensitivity coeficients $f_{Ai}$ and $\lambda_{Ai}$.

\subsubsection{Atomic clock constraints}\label{subsec23clock}

The parameters $k_i$ can be constrained from laboratory measurements on Earth and in particular atomic clocks experiments described in Sect.~\ref{sec-clock-data}. by searching seasonal modulations instead of a linear drift.

\paragraph{Methodology}\ 

To that goal, one needs to recall that the Earth has an excentricity $e\simeq0.0167$  so that the Sun gravitational potential on the Terrestrial orbit  should have a peak-to-peak amplitude of $2\,{GMe}/{ac^2}\sim3.3\times10^{-10}$ on a period of 1~year. This shows that the order of magnitude of the constraints will be roughly of $10^{-16}/10^{-10}\sim10^{-6}$ for clocks with an accuracy of $10^{-16}$. The expected magnitude of the constraints on $\beta$ for other systems are summarized in Table~\ref{tab-phi_p}.

\begin{table}[t]
\caption[Characteristics of the Systems to constrain $\beta$]{The typical values of the potential and accuracy of frequency measurements for different systems allow one to estimate the expected magnitude of the constraints they can set on the parameters $\beta$.}
\label{tab-phi_p}
\centering
{\small
\begin{tabular}{lccc}
 \toprule
 Physical system & $\Delta \Phi_{\rm N}/c^2$ & Typical clock accuracy & Expected $\beta$ constraint  \\
 \midrule
Earth orbit & $3.3\times10^{-10}$  & $10^{-16}-10^{-18}$ & $10^{-6}-10^{-8}$ \\
Solar stars &  $2\times10^{-7}$ & $10^{-7}-10^{-8}$ & $10^{-1}$ \\
White dwarfs&  $3\times10^{-4}$  &   $10^{-7}-10^{-8}$ & $10^{-3}$  \\
Neutron stars & 0.5 & $10^{-2}$ & $10^{-1}$ \\
  \bottomrule
\end{tabular}
}
\end{table}

More precisely, the gravitational potential of the Sun on the Earth orbit can be approximated by
$$
\Phi(t) = \Phi_p + \Delta\Phi\cos\left[2\pi\frac{(t-t_p)}{T_0}\right], \qquad \Delta\Phi/c^2\sim 1.65\times10^{-10}
$$
with $\Phi_p$ and $t_p$ the potential and time at perihelion. $\Omega_0=2\pi/T_0$ is the pulsation of the motion so that $T_0$ can be chosen as annual, diurnal, etc. modulation timescale. Then, from the data, one can fit a profile of the form 
$$
y_{AB}(t)= y_0+A_{AB}\cos\left[2\pi\frac{(t-t_p)}{T_0}\right].
$$
The constraint on the amplitude $A_{AB}$ of the modulation of the relative frequencies with periodicity $T_0$ translates in a constraint on the redshift parameter
\begin{equation}\label{e.A-beta}
\beta_{AB} = A_{AB}\left(\Delta\Phi/c^2\right)^{-1}.
\end{equation}
$A_{AB}$ is an experimental constraint, $\Delta\Phi/c^2$ is known from the trajectory. Thanks to the $K_{AB}^i$ listed in Table~\ref{tab0}, this provides a contraint on $k_i$.

\begin{table}[t]
\caption[Expression of the $\beta$ coefficients]{Expressions of the $\beta_{12}=\beta_1-\beta_2$ coefficients derived from Table~(\ref{tab0b}) and Eq.~(\ref{e.betaK}) for the clock experiments summarized in Table~\ref{tab-beta}.}
\label{tab-betaK}
\centering
{\small
\begin{tabular}{cccc}
 \toprule
Clock 1 & Clock 2  & $(\aem,\bar\mu,X_{\rm q})$ & $(\aem,X_{\rm e}, X_{\rm q})$  \\
 \midrule
 H & Cs 			&   $-0.83k_\alpha  -0.102 k_q $  &   id. \\
 H & Rb 			&  $-0.34k_\alpha  -0.081 k_q $  &	id. \\
 Rb & Cs 			&  $-0.49k_\alpha  -0.021 k_q $ & 	id.  \\
\midrule
 Sr & Cs 			&   $-2.77k_\alpha - k_\mu -0.002 k_q $  &  $-2.77k_\alpha - k_e  + 0.046 k_q $   \\
 Hg & Cs 			&  $-5.77k_\alpha - k_\mu -0.002 k_q $ 	&  $-5.77k_\alpha - k_e  + 0.046 k_q $  \\
 Yb& Cs 			& $-2.52k_\alpha - k_\mu -0.002 k_q $  &  $-2.52k_\alpha - k_e  + 0.046 k_q $   \\
 Yb-E3 & Cs 		&  $-8.78 k_\alpha - k_\mu -0.002 k_q $  &  $-8.78k_\alpha - k_e  + 0.046 k_q $   \\
 CSO & H 			&   $3k_\alpha-k_\mu +0.1 k_q$  &  $3k_\alpha - k_e  + 0.148 k_q $  \\
\midrule
 Al$^+$ & Hg$^+$  	&   $-2.948k_\alpha$ & id.  \\
 Dy-162 & Dy-164 	&  $k_\alpha$ & id.  \\
  Yb-E3 & Yb-E2 	&  $6.95 k_\alpha$ & id.   \\
  \bottomrule
\end{tabular}
}
\end{table}

\paragraph{Atomic clock constraints}\ 

All the experiment used here are described in Sect.~\ref{sec-clock-data} and the summary of all the experiments are gathered in Table~\ref{tab1} and Fig.~\ref{fig-deseul}.

\noindent$\bullet${\bf Hg/Cs} The data of \cite{clock-fortier07} gives the constraint
\begin{equation}\label{e.bHgcs1}
\beta_{{\rm Hg, Cs}}=(2\pm3.5)\times10^{-6}.
\end{equation}

\vskip0.5cm 
\noindent$\bullet${{\bf H/Cs} and {\bf H/Rb}} \cite{bauch} compared a caesium atomic fountain frequency standard with a hydrogen maser for about 1~year to conclude that the frequency ratio did not change by more than $7\times10^{-15}$ over 6~months, hence
\begin{equation}\label{e.bHcs1}
|\beta_{{\rm H,Cs}}|< 2.1\times10^{-5}.
\end{equation}
This was improved thanks to  the 7-yr comparison of caesium and hydrogen atomic clocks by \cite{ashby} to
\begin{equation}\label{e.bHcs2}
\beta_{{\rm H,Cs}}=(0.1\pm1.4)\times10^{-6}.
\end{equation}
\cite{NEW_PhysRevA.87.010102} compared 3 atomic clocks based on the hyperfine transitions of  a set of 70 commercial Cs-133 clocks, a  Rb-87 fountain and a set of H-masers  in a continuous operation over 1.5 years. They concluded
\begin{eqnarray}
\beta_{\rm H, Cs}& =& (-0.7\pm1.1)\times10^{-6} \\
\beta_{\rm H, Rb} &=& (2.7\pm4.9)\times10^{-7} \\
\end{eqnarray}
as well as a constraint on $\beta_{{\rm Rb, Cs}}$; see below. The comparison of H-maser to 3 separate Cs fountain clocks and one Rb fountain clock over more than 8~years \citep{NEW_Tobar:2013pwa} concluded that
\begin{eqnarray}
\beta_{{\rm H,Rb}} &=& (6.3\pm10)\times10^{-6}  \nonumber\\
\beta_{{\rm H,Cs}} &=& (3.6\pm4.8)\times10^{-6}\,.
\end{eqnarray}
The latter was improved by the comparison between the long-term fractional frequency variation of 4 H-masers with caesium clocks \citep{NEW_Ashby:2018jdl}
\begin{equation}\label{e.betaHCS2018}
\beta_{\rm H,Cs}=(2.2\pm2.5)\times10^{-7}.
\end{equation}

\vskip0.5cm 
\noindent$\bullet${\bf Rb/Cs} The data by \cite{NEW_Guena:2012zz} translated to
\begin{equation}
\beta_{{\rm Rb, Cs}} = (0.11\pm1.04)\times10^{-6}.
\end{equation}
\cite{NEW_PhysRevA.87.010102} concluded from the experiment described above that
\begin{equation}
\beta_{{\rm Rb, Cs}} = (-1.6\pm1.3)\times10^{-6}.
\end{equation}

\vskip0.5cm 
\noindent$\bullet${\bf Dy-162/164} The atomic dyprosium experiment  by \cite{clock-cingoz}, was shown to imply \citep{ferrel}
\begin{equation}\label{ebetaDy1}
 \beta_{\rm Dy-162, Dy-164} =(-8.7\pm6.6)\times10^{-6}\,.
\end{equation}
Similarly, the local dyprosium-clock constraints~(\ref{clock-dypro2}) by \cite{NEW_Leefer:2013waa} can be translated to
\begin{equation}
\beta_{\rm Dy-162, Dy-164} = (-5.5\pm5.2)\times10^{-7}\,.
\end{equation}

\vskip0.5cm 
\noindent$\bullet${\bf Sr/Cs} The caesium-strontium experiment by \cite{clock-blatt} led to
\begin{equation}\label{eq:ka}
\beta_{{\rm Sr, Cs}}= (-5.7\pm 9.1)\times10^{-6} \,.
\end{equation}
\cite{NEW_Schwarz:2020upg} later used 42 measurements of the transition frequency of $^1$S$_0$-$^3$P$_0$ in $^{87}$Sr over 3~years from 2017 to 2019 compared to caesium to set
\begin{equation}\label{e.jkl}
\beta_{\rm Sr, Cs}=(-1.1\pm5.2)\times10^{-7}.
\end{equation}

\begin{figure}[htbp]
 \centerline{\includegraphics[scale=0.4]{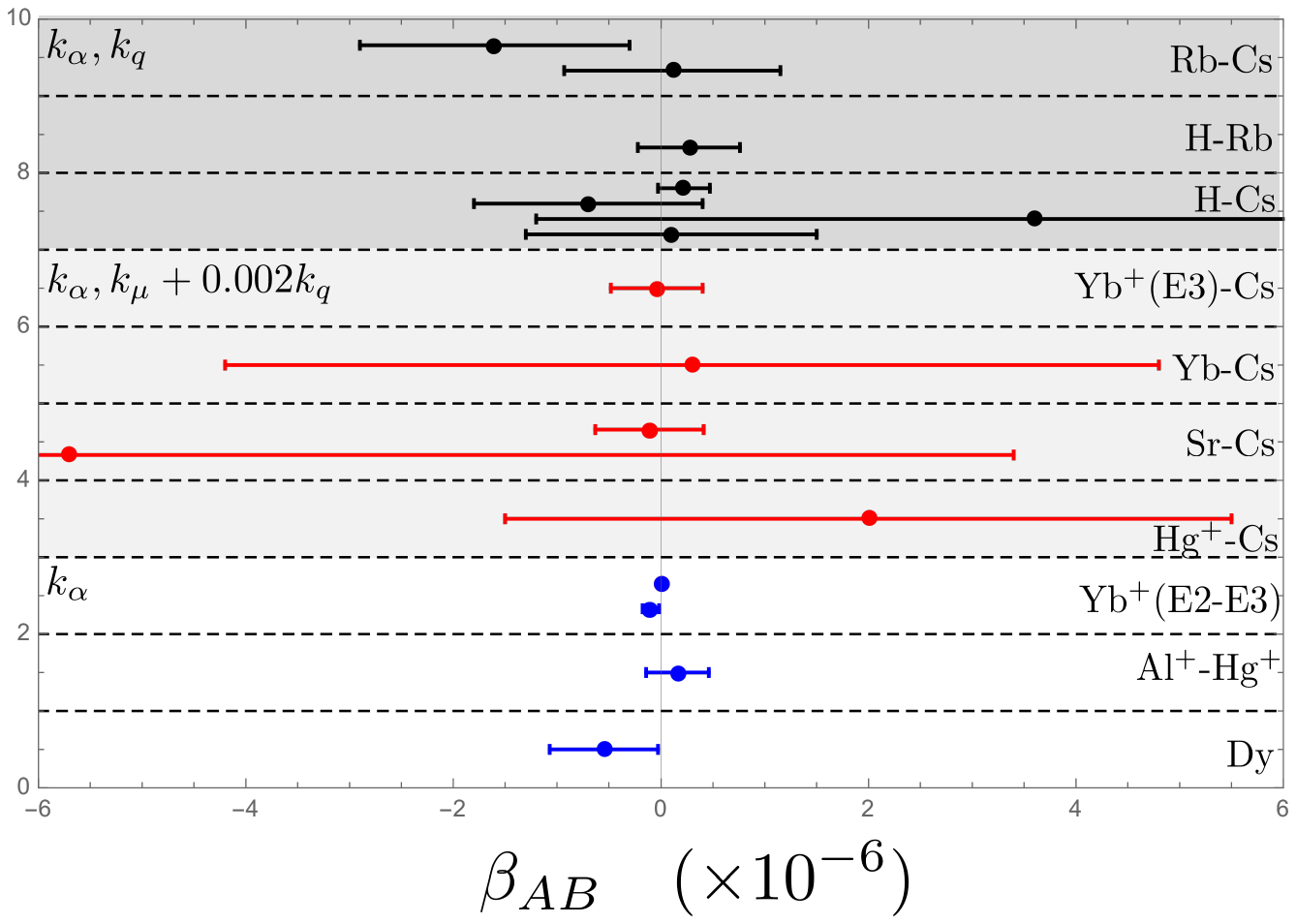}}
  \vskip-.5cm
  \caption[Atomic clock constraints on $\beta_{AB}$]{Atomic clock constraints on $\beta_{AB}$. The data are given in Table~\ref{tab-beta} and are split in 3 families according to the way they depend on $(k_\alpha,k_\mu,k_q)$; see Table~\ref{tab-betaK}.}
  \label{fig-deseul}
\end{figure}

\vskip0.5cm 
\noindent$\bullet${\bf CSO/H-maser} \cite{NEW_Tobar:2009dm} compared a cryogenic sapphire oscillator (CSO)  with various H-masers from 2001 to 2008. They constrained both the annual and diurnal modulations to set the bounds $\beta_{{\rm CSO},{\rm H-maser}} = (-2.7\pm1.4)\times10^{-4}$ and $\beta_{{\rm CSO},{\rm H-maser}} = (-6.9\pm4.0)\times10^{-4}$ respectively with a weighted mean of
\begin{equation}
\beta_{{\rm CSO},{\rm H}} = (-3.2\pm1.3)\times10^{-4}.
\end{equation}
The sensitivity coefficients $K_{{\rm CSO,H}}^i$ are gathered in Table~\ref{tab-betaK}.

\vskip0.5cm 
\noindent$\bullet${\bf Al/Hg} \cite{NEW_Dzuba:2016aiy} reanalyzed the constraint by \cite{clock-rosen08} on  Al$^{+}$-Hg$^{+}$ to conclude that
\begin{equation}\label{ealhg00}
\beta_{{\rm Al, Hg}} = (0.16\pm 0.3)\times10^{-6}\,.
\end{equation}

\vskip0.5cm 
\noindent$\bullet${\bf Yb/Cs} \cite{NEW_McGrew2017} compared their 8 month data of the measured Yb/Cs frequency ratio  to get an amplitude $A=(0.1\pm1.5)\times10^{-16}$ that translates, using Eq.~(\ref{e.A-beta}), to
 \begin{equation}\label{e.ggg}
\beta_{\rm Yb, Cs}=(0.3\pm4.5)\times10^{-7}
\end{equation}

\noindent$\bullet${\bf Yb E2-E3}  The most recent comparisons of these Yb$^+$ E2 and E3 transitions and Cs clocks allowed \cite{Lange:2020cul} to conclude that
\begin{eqnarray}
\beta_{\rm Yb-E3, Cs}&=&(-4\pm44)\times10^{-8}, \label{e.be23-1}\\
\beta_{\rm Yb-E3, Yb-E2}&=&(-9.7\pm7.7)\times10^{-8}. \label{e.be23-2}
\end{eqnarray}
The second constraints was improved by \cite{ NEW_Filzinger:2023zrs} to
\begin{eqnarray}
\beta_{\rm Yb-E3, Yb-E2}&=&(-2.4\pm3.0)\times10^{-9}. \label{e.be23-3}
\end{eqnarray}

\begin{table}[t]
\caption[Constraints on the couplings $\beta$ to the gravitational potential (Solar system)]{Summary of the constraints on the $\beta$ discussed in the text.}
\label{tab-beta}
\centering
{\small
\begin{tabular}{ccccr}
 \toprule
Clock 1 & Clock 2  & $\beta_{12}$ & Note & Ref. \\
 \midrule
H & Cs &  $< 2.1\times10^{-5}$  & $k_\alpha, k_q$&\cite{bauch} \\
 H & Cs & $(0.1\pm1.4)\times10^{-6}$ && \cite{ashby} \\
H & Cs & $ (3.6\pm4.8)\times10^{-6}$ & & \cite{NEW_Tobar:2013pwa}  \\
H & Cs  &  $(-0.7\pm1.1)\times10^{-6}$ & & \cite{NEW_PhysRevA.87.010102}\\
H & Cs  & $(0.22\pm 0.25)\times10^{-6}$ & & \cite{NEW_Ashby:2018jdl} \\ 
\midrule
  H & Rb  &  $(0.27\pm0.49)\times10^{-6}$ & & \cite{NEW_PhysRevA.87.010102}\\
  H & Rb & $(6.3\pm10)\times10^{-6}$ & &  \cite{NEW_Tobar:2013pwa} \\
 \midrule 
 Rb & Cs &$(0.11\pm1.04)\times10^{-6}$ & & \cite{NEW_Guena:2012zz} \\
  Rb & Cs  & $(-1.6\pm1.3)\times10^{-6}$  & & \cite{NEW_PhysRevA.87.010102}\\
 \midrule \midrule
 Hg & Cs & $(2\pm3.5)\times10^{-6}$ &$k_\alpha, k_\mu+0.002k_q $&\cite{clock-fortier07}\\
 \midrule 
Sr & Cs & $(-5.7\pm 9.1)\times10^{-6}$ && \cite{clock-blatt} \\
Sr & Cs & $(-0.11\pm0.52)\times10^{-6}$ && \cite{NEW_Schwarz:2020upg} \\
 \midrule 
 Yb& Cs & $ (0.3\pm4.5)\times10^{-6}$ &&  \cite{NEW_McGrew2017b} \\
 \midrule  
 Yb-E3 & Cs & $ (-4\pm44)\times10^{-8}$ & &  \cite{Lange:2020cul} \\
 \midrule  \midrule
 Dy-162 & Dy-164 & $ (-0.55\pm0.52)\times10^{-6} $ & $k_\alpha$ & \cite{NEW_Leefer:2013waa}\\
 \midrule 
 Al$^+$ & Hg$^+$  &$ (0.16\pm 0.3)\times10^{-6}$  & & \cite{clock-rosen08} \\
 \midrule 
 Yb-E3 & Yb-E2 & $ (-9.7\pm7.7)\times10^{-8}$ & &  \cite{Lange:2020cul} \\
 Yb-E3 & Yb-E2 & $(-2.4\pm3.0)\times10^{-9}$ &  &  \cite{NEW_Filzinger:2023zrs}\\
 \midrule  \midrule 
 CSO & H & $(-2.7\pm1.4)\times10^{-4} $  &[annual]   $k_\alpha,k_\mu,k_q$& \cite{NEW_Tobar:2009dm}  \\
CSO & H & $ (-6.9\pm4.0)\times10^{-4}$ &[diurnal]  & \cite{NEW_Tobar:2009dm}  \\
CSO & H &  $ (-3.2\pm1.3)\times10^{-4}$ &[mean] & \cite{NEW_Tobar:2009dm}  \\
\bottomrule
\end{tabular}
}
\end{table}

\paragraph{Constraints on the parameters $k_i$}\ 

The previous experimental constraints can be combined to get bounds on ($k_\alpha, k_\mu, k_q, k_e$). They have evolved with the accuracy of the experiments and some of the sensifivity coefffcients were reevaluated.

The first bounds~(\ref{e.bHgcs1}) and~(\ref{e.bHcs1})  were translated by \cite{local1}  into $k_\alpha+0.17k_{e}=(-3.5\pm6)\times10^{-7}$ and $\vert k_\alpha+0.13k_{q}\vert<2.5\times10^{-5}$, neglecting the contribution in $k_q$ for Hg/Cs.  Then, it was combined with the dyprosium result~(\ref{ebetaDy1})  by \cite{ferrel} to conclude that $k_{e}=(4.9\pm3.9)\times10^{-5}$ and
$k_{q}=(6.6\pm5.2)\times10^{-5}$ while the \cite{ashby} data on H/Cs improved to $k_\alpha+0.13k_{q}=(-1\pm17)\times10^{-7}$. When combined with H-maser \citep{ashby}, the Sr/Cs bound~(\ref{eq:ka}) allowed \cite{clock-blatt} to get
\begin{equation}\label{eq:blatt}
 k_\alpha=(2.5\pm3.1)\times10^{-6} \quad
  k_{\mu}=(-1.3\pm1.7)\times10^{-5} \quad
   k_{q}=(-1.9\pm2.7)\times10^{-5}.
\end{equation}
\cite{barrowkk,sbkk} reanalyzed the data by \cite{clock-peik04} on Yb$^+$-E2/Cs to conclude that $ k_\alpha+0.51k_{\mu}=(7.1\pm3.4)\times10^{-6}$. Combined with~(\ref{eq:ka}), they got
\begin{equation} \label{eq:barrow}
   k_{\mu}=(3.9\pm3.1)\times10^{-6}, \qquad   k_{q}=(0.1\pm1.4)\times10^{-5}.
\end{equation}
\cite{barrowkk} also  used the data by \cite{clock-rosen08} on Al/Hg clocks to get $k_\alpha=(-5.4\pm5.1)\times10^{-8}$.  Combining with~(\ref{e.ggg}) the former constraints on $k_\alpha$ and $k_q$, \cite{NEW_McGrew2017}  concluded $k_\mu =(0.7\pm1.4)\times 10^{-6}$. Similarly from their result~(\ref{e.jkl}) combined with those by \cite{clock-rosen08,NEW_Dzuba:2016aiy,NEW_Ashby:2018jdl}, \cite{NEW_Schwarz:2020upg} they reached $k_\mu=(3.5\pm5.9)\times10^{-7}$. Then, to tighten the contraints, \cite{NEW_McGrew2017b}  employed a multi-species analysis that can be used to link absolute frequency measurements of different types of atomic clocks with optical ratios that have been measured with sufficient precision to get $k_\mu =(-0.19\pm0.94)\times 10^{-6}$. \cite{NEW_PhysRevA.87.010102} performed a global fit including their data on H/Cs, H/Rb and Rb/Cs together with those by \cite{ashby,clock-fortier07,clock-blatt,NEW_Guena:2012zz} to conclude
\begin{equation}\label{eq:peil}
 k_\alpha=(1.7\pm37.5)\times10^{-7} \quad
 k_{\mu}=(-2.5\pm5.4)\times10^{-6} \quad
 k_{q}=(3.8\pm4.9)\times10^{-6}.
\end{equation}

Then, a series of constraints on $k_\alpha$ independently of $(k_\mu, k_q)$ were obtained (\textit{1}) from the Al/Hg~(\ref{ealhg00}) as $k_\alpha =(5.3\pm10)\times 10^{-8}$ which, combined with (\ref{e.betaHCS2018}) implies $k_q =(-2.6\pm2.6)\times 10^{-6}$; (\textit{2}) from Dy-clocks \citep{NEW_Leefer:2013waa} as $(-5.5\pm5.2)\times10^{-7}$ and (\textit{3}) from Yb:E3-E2, first by \cite{Lange:2020cul} to get $k_\alpha = (1.4\pm1.1)\times10^{-8}$, which combined with their YE3-Cs data implied $k_\mu=(7\pm45)\times10^{-8}$, and finally improved by  \cite{ NEW_Filzinger:2023zrs} to
\begin{equation}\label{e.kalphaLast}
k_\alpha = (-2.4\pm3.0)\times10^{-9}\,.
\end{equation}
\paragraph{Conclusion}\ 
 
 To conclude this analysis, we perform a combined fit of the latest data; see Fig.~\ref{fig-kbeta}. First, the best constraint on $k_\alpha$ alone is given by Eq.~(\ref{e.kalphaLast}).
 \begin{tcolorbox}
The atomic clock constraints on seasonal modulations are summarized in Table~\ref{tab-beta}. It allows us to get the up-to-date constraints on the parameters $k$
\begin{eqnarray}
k_\alpha &=& (-2.4\pm3.0)\times10^{-9} \\
k_\mu&=& (0.28\pm4.2)\times10^{-7}  \\
k_q&=& (-2.31\pm2.27)\times10^{-6},
\end{eqnarray}
from which one extracts, thanks to Eq.~(\ref{e.MassN2}),
\begin{equation}
k_e= (1.2\pm4.3)\times10^{-7}\,.
\end{equation}
All were derived using the sensitivity coefficients computed in \cite{q-calc2,kappa-tedesco} and gathered in Tables~\ref{tab-betaK}. 
\end{tcolorbox}

Still, we shall mention \cite{jenkins} who claimed for an unexplained seasonal variation that demonstrated the difficulty to interpret this phenomena. We also note that \cite{NEW_sRoberts:2019sfo} used data from a European network of fiber-linked optical atomic clocks to constrain transient variation of the fine structure constant. 

\begin{figure}[htbp]
  \vskip-.5cm
 \centerline{\includegraphics[scale=0.45]{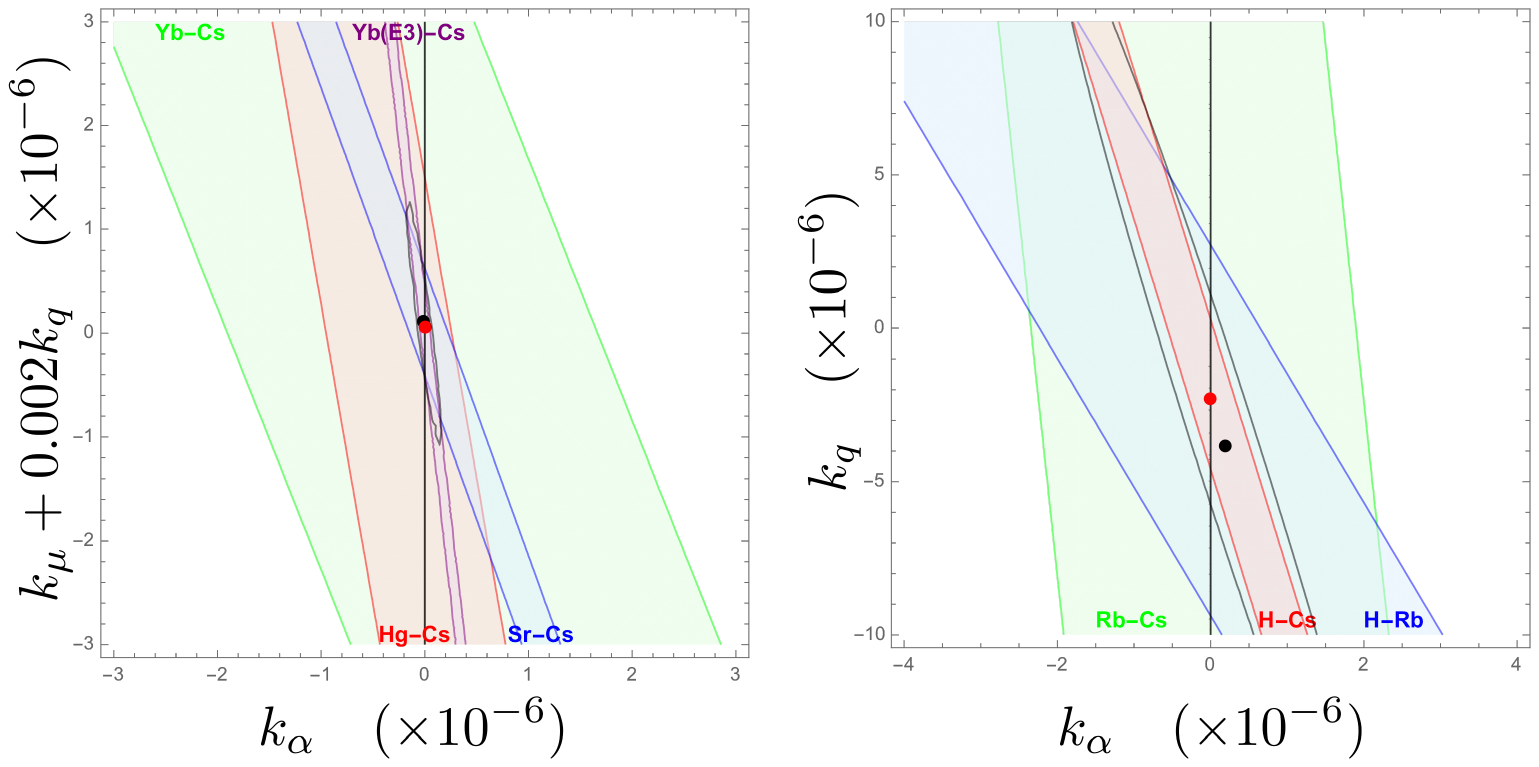}}
  \vskip-1cm
  \caption[Atomic clock constraints on $(k_\alpha,k_\mu,k_q)$]{Atomic clock constraints on $(k_\alpha,k_\mu,k_q)$. The data are given in Table~\ref{tab-beta} and dependence on $(k_\alpha,k_\mu,k_q)$ in Table~\ref{tab-betaK}. The left plot gathers the experiments that depend on $(k_\alpha,k_\mu+0+002k_q)$, Red: Hg-Cs \citep{clock-fortier07}; Blue: Sr-Cs \citep{NEW_Schwarz:2020upg}; Green: Yb-Cs \citep{NEW_McGrew2017b}; and Purple: Yb(E3)-Cs \citep{Lange:2020cul}.  The right plot gathers the experiments that depend on $(k_\alpha,k_q)$, Red: H-Cs \citep{NEW_Ashby:2018jdl}; Blue: H-Rb \citep{NEW_PhysRevA.87.010102} and Green: Rb-Cs \citep{NEW_Guena:2012zz};. On both plot the vertical line stands for the 1$\sigma$ contour on $k_\alpha$ from \cite{NEW_Filzinger:2023zrs}; see Eq.~(\ref{e.kalphaLast}) while the red and black dots are the best fits of all the data and all the data but the conbstraints on $k_\alpha$ respectively.}
  \label{fig-kbeta}
\end{figure}

All these bounds can and will be improved by comparing clocks on Earth and onboard of satellites \citep{clock-ACES,maleki,local1} while the observation of atomic spectra near the Sun can lead to an accuracy of order unity \citep{local1}. A space mission with atomic clocks onboard and sent to the Sun  \citep{maleki}, on highly eccentric terrestrial orbit \citep{NEW_kspace} or on a Solar system escape trajectory \citep{wolf09} could reach an accuracy of $10^{-8}$; see Sect.~\ref{secfiber}.

\subsubsection{Milky Way}\label{subsec23mw}

All the observational methods that have been described in Sect.~\ref{subsec33} can be applied locally within our Galaxy, the Milky Way, provided the detection of local interstellar absorption system.

\paragraph{Amonia inversion method}\ 

The first attempt \citep{mu-lev} to study $\mu$ in the Milky Way was based on 207 individual measurements in the Perseus molecular cloud (PC), the Pipe Nebula (PN), and the infrared dark clouds (IRDCs) and the analysis relied on measurements of the relative radial velocity offsets between NH$_3$ inversion lines and other molecular transitions.  It concluded that $\Delta\mu/\mu=(3.5\pm1.4)\times10^{-8}$, which was at that time the most accurate estimate of this quantity based on the spectral analysis of astronomical objects. The Perseus molecular cloud is located towards the galactic anti-center, whereas the Pipe Nebula and the IRDCs in the direction of the galactic center. Modeling the gravitational potential as 
$$
\Phi = \Phi_{\rm MW} + \Phi_{\rm cloud}
$$
where $\Phi_{\rm MW}$ can be calculated from analytic models of the Milky Way and $\Phi_{\rm cloud}$ is the molecular cloud gravitational potential, one  can translate this bound to $k_\mu\sim1$. This early result was in contradiction with the local constraint~(\ref{eq:blatt}) if $\Delta\mu/\mu$ follows the gradient of the local gravitational potential. Indeed a logical possibility would be that a chameleon field is at work -- see Sect.~\ref{subsub2} -- which motivates the effort to obtain measurements in different environments.  \cite{molaro3} applied the ammonia method to an atlas of 193 dense protostellar and prestellar cores of low masses in the Perseus molecular cloud as well as he dark cloud L183 with observations of NH$_3$(1,1) and (2,2) inversion lines. They concluded that $\Delta\mu/\mu <10^{-7}$. Two similar analysis were performed first by \cite{levmusp2} using high resolution spectral observations of molecular core in lines of NH$_{3}$, HC$_{3}$N and N$_{2}$H$^{+}$ with 3 radio-telescopes to get $|\Delta\mu/\mu|<3\times10^{-8}$ between the cloud and the local laboratory environments. However, an offset was measured that could be interpreted as a variation of $\mu$ of amplitude $\Delta\bar\mu/\bar\mu=(2.2\pm0.4_{\text{stat}}\pm0.3_{\text{sys}})\times10^{-8}$. Then, a second analysis \citep{levmusp2b} mapped four molecular cores L1498, L1512, L1517, and L1400K selected from the previous sample in order to estimate systematic effects due to possible velocity gradients. The measured velocity offset, once 
expressed in terms of $\Delta\bar\mu$, gives $\Delta\bar\mu= (26 \pm 1_{\text{stat}} \pm 3_{\text{sys}})\times10^{-9}$. The same technique was followed by \cite{NEW_Levshakov:2013ufa,NEW_Levshakov:2013oja} with observation from the Medicina 32-m (towards the cores L1512 and L1498 are among the narrowest molecular lines known in the interstellar medium) and the Effelsberg 100-m radiotelescopes (to observe the two molecular cores L1512 and L1498). They revealed a systematic error in the radial velocities of an amplitude of about $0.02~ \unit{km.s^{-1}}$ that was conservatively assign to the possible systematic error of the Medicina dataset, hence concluding
\begin{equation}\label{E.L1498}
\Delta\bar\mu/\bar\mu=(0.1\pm2.2)\times10^{-8}\,\hbox{[L1521]}
\qquad
\Delta\bar\mu/\bar\mu=(-0.1\pm2.3)\times10^{-8}\,\hbox{[L1498]}
\end{equation}
 leading to  $\Delta\mu/\mu<2\times 10^{-8}$ at 3$\sigma$.\\
 
\paragraph{{\rm CO} vs {\rm C{\sc i}} and {\rm C{\sc ii}}}\ 

Following the same method as in  Sect.~\ref{subsec358}, \cite{levFsp}  measured the offsets between the radial velocities of the rotational transitions of carbon-13 monoxide and the fine structure transitions of neutral and singly ionized carbon  that probe the variation of $F=\aem^2\mu$ over the galaxy. It concluded that
\begin{equation}
 |\Delta F'/F'| < 3.7\times10^{-7}
\end{equation}
between high (terrestrial) and low (interstellar) densities of baryonic matter. Combined with the previous constraint on $\mu$, it implies that $ |\Delta\aem/\aem| < 2\times10^{-7}$. This was updated \citep{levrio} to $ |\Delta F'/F'| < 2.3\times10^{-7}$ so that $ |\Delta\aem/\aem| < 1.1\times10^{-7}$. A later study  \citep{NEW_Levshakov:2017ivg} with $^{12}$CO and $^{13}$CO in the dark cloud L1599B  concluded that $ |\Delta F'/F'| < 3\times10^{-7}$. 

An analysis of [C{\sc ii}] lines (Herschell observatory) and CO(2,1) (IRAM) from the regions BCLMP691 and BCLMP303  close to the dynamical center of the Triangulum galaxy (M33) led \cite{NEW_Levshakov:2017ivg} to set 
\begin{equation}
 \Delta F'/F' =(-0.3\pm4.6) \times10^{-7}\qquad \hbox{[M33]}.
\end{equation}
\cite{NEW_Levshakov:2019rqv}  applied this method  to the  Magellanic Clouds dwarf galaxies that are known to be dark matter dominated. Thanks to the Herschel Space Observatory they concluded that $\Delta F'/F'<2\times10^{-7}$ after averaging over 9 positions in the LMC while one position observed with a higher spectral resolution gave
\begin{equation}
\Delta F'/F' = (-1.7\pm 0.7)\times 10^{-7}.
\end{equation}
This offset is still not fully understood and further investigations are required to test whether it can be related to chemical segregation in the emitting gas or merely due to Doppler noise. Nevertheless this opens a window to track a specific coupling to dark matter.

\paragraph{CH}\ 
 
 \cite{NEW_Truppe:2013ypa} developed a spectroscopic method for pulsed beams of cold molecules, and use it to measure the frequencies of microwave transitions in CH with accuracy down to 3 Hz to compare this transition to the ground state $\Lambda$-doublet transition of OH. From 5 absorption systems toward 5 stellar sources,  each system having a density estimated 10$^{19}$ times smaller than on Earth, they conclude that $\Delta F'/F'=(-0.7\pm2.2)\times10^{-7}$ after averaging on their data.

\paragraph{CH$_3$OH}\ 

The detection of methanol in the Galaxy led \cite{NEW_Levshakov:2011su} to conclude that  it can reach constraint of $\Delta\mu/\mu<29\times10^{-9}$ at 1$\sigma$ so that this technique deserves attention. \cite{NEW_Dapra:2017umy} applied it to the dense dark cloud core L1498 in the Taurus-Auriga complex. They detected 5 methanol transitions towards  two different positions (IRAM) leading to the two independent constraints $\Delta\mu/\mu(1)=(-3.2\pm2.0)\times10^{-8}$ and $\Delta\mu/\mu(2)=(-3.8\pm6.6)\times10^{-8}$ with weighted mean
\begin{equation}
\Delta\mu/\mu=(-3.3\pm1.9)\times10^{-8}\quad\hbox{[L1498]}.
\end{equation}
Combined with (\ref{E.L1498}) for L1498, one obtains $\Delta\mu/\mu=(-3.2\pm1.5)\times10^{-8}$ and, if averaged with L1521, $\Delta\mu/\mu=(-1.2\pm0.9)\times10^{-8}$ at 1$\sigma$  in the Milky Way.

\cite{NEW_Levshakov:2021oco} measured $\Delta\mu/\mu$  in the Milky Way disk thanks to methanol masers from the northern Galactic hemisphere distributed over the galactocentric distance range of 4-12~kpc. Only 7 maser sources out of 229 could be  selected from  the MSX Source (RMS) catalogue and 11 objects from 144 sources  from the Bolocam Galactic Plane Survey. The analysis allowed them to set a constraint on the spatial variation of $\mu$ in the Galaxy and its absolute variation in the range 4-12~kpc,
\begin{equation}
|\Delta\mu/\mu_{\rm 4-12~kpc}|<2\times10^{-8}, \qquad
k_\mu<2\times10^{-9}.
\end{equation}

\subsubsection{Towards strong field environments}\label{secstrongs}

Measuring fundamental constants in stronger field regions opens a window to  models with environmental dependencies such as the chameleon theory and better tests on their dependencies on the gravitational potential. It was proposed by \cite{NEW_Berengut:2013dta} to measure $\aem$ in strong field environment using metal lines in the absorption spectra of white dwarfs since their local gravitational potential is higher than in the Solar system.

\paragraph{White Dwarf atmosphere.}\ 

White dwarfs are interesting systems to test for the variation of fundamental constants. First, they have a strong surface gravity,
\begin{equation}\label{e.logg}
\log g\equiv \log \left(\frac{GM}{r^2}/1\,\unit{cm.s^{-2}}\right)\,,
\end{equation} 
reaching 7--8. Second, \emph{DA white dwarfs} have a pure hydrogen photosphere but thanks to levitation and accretion from the interstellar medium, metallic elements can also be found in their atmosphere. Third, they are bright in UV, require pace observation and comprise about 85\% of all white dwarfs.

This idea was first applied by  \cite{NEW_Berengut:2013dta} to  the hot hydrogen rich white dwarf G191-B2B located at about 45~pc from the Earth observed by HST Imaging Spectrograph with absorption lines from Fe~{\sc v} and Ni~{\sc v} in its atmosphere. With a radius of $0.22 R_\odot$ and a mass of $0.51 M_\odot$, one estimates that $g\equiv GM/R^2\sim 10.5 g_\odot\sim 280 g_{\rm Earth}$ so that $\Phi_{\rm N}/c^2\sim 4.9\times10^{-5}$ much larger than $3\times10^{-10}$ on the Earth orbit. Comparing with laboratory spectrum, the 106 Fe~{\sc v} transitions led to, using Eq.~(\ref{defki}),
\begin{equation}
 \Delta\aem/\aem =(4.2\pm1.6)\times10^{-5}\quad\hbox{(Fe~{\sc v}, G191-B2B)}
\end{equation}
showing a 2.6$\sigma$ deviation from zero, that can be rephrased as $k_{\aem}=0.7\pm0.3$. The analysis of 32 Ni~V transitions gave
\begin{equation}
 \Delta\aem/\aem =(-6.1\pm5.8)\times10^{-5}\quad\hbox{(Ni~{\sc v}, G191-B2B)}
\end{equation}
showing a 1.05$\sigma$ deviation from zero. These two measurements on the same system seem inconsistent since they indicate variation with different signs.  \cite{NEW_Berengut:2013dta}  suggested that this inconsistency is due to a systematic effect in the laboratory wavelengths. This first analysis has the advantage to use typically hundreds of lines compare to tens for QSO observation. \cite{NEW_Bainbridge:2017lsj}  presented preliminary results on a similar analysis by considering new laboratory data and by including nine more hot, bright white dwarfs and sub-dwarfs and three bright white dwarfs known to have photospheric Fe and Ni absorption lines (see their Fig.1 depicting $\Delta\aem/\aem$  for various surface gravity ($\log g$ ranging from 3.9 to 7.9). 

\cite{NEW_Hu:2018lwv} estimated the effect of the stellar magnetic field that will alter absorption line profiles and introduce additional uncertainties in measurements  of $\aem$. In  the case of G191-B2B they obtained an upper limit on its magnetic field of $B<2300$~G (3$\sigma$) and concluded that the impact of quadratic Zeeman shifts on measurements of $\aem$ is 4 orders of magnitude below laboratory wavelength uncertainties. \cite{NEW_Hu:2020zeq} proposed a new analysis with two new and independent laboratory samples for Fe~{\sc v}, the re-reduction of HST data, new atomic calculation of the sensitivities for Fe~{\sc v}, and a better analysis of the blending of the absorption lines. This led to
\begin{eqnarray}\label{e.FeV}
&& \Delta\aem/\aem =[6.36\pm(0.33_{\rm stat}+1.94_{\rm syst})\times10^{-5}\nonumber\\
&& \Delta\aem/\aem =[4.21\pm(0.47_{\rm stat}+2.35_{\rm syst})\times10^{-5}
\end{eqnarray}
for each of the laboratory data, showing that the systematic uncertainties are dominated by laboratory wavelength precision, see e.g., \cite{NEW_Hu:2019bfy,NEW_Lee:2022krb} for an analysis of the systematic effects. \cite{Lee:2024uqw} considered the effects of continuum placement error that was concluded  to impact significantly the $\aem$ measurements. From the analysis of Ni~{\sc v}, they concluded for an inconsistency with  the Fe~{\sc v} measurements~(\ref{e.FeV}) -- the least negative of their measurements being $\Delta\aem/\aem= (-1.462\pm1.121)\times10^{-5}$ -- suspecting that this $3.2\sigma$ difference arises from unknown laboratory wavelength systematics.

\cite{NEW_Bagdonaite:2014mfa,NEW_Salumbides_2015} followed the same strategy while focusing on the Lyman transitions molecular H$_2$ in the atmosphere of the white dwarfs GD133  with a surface potential $\Phi_{\rm N}/c^2\sim 1.2\times 10^{-4}$ and G29-38 (with $\Phi_{\rm N}/c^2\sim 1.9\times 10^{-4}$) observed with the HST Cosmic Origin Spectrograph. They concluded 
\begin{eqnarray}
&& \Delta\mu/\mu =(-2.3\pm 4.7)\times10^{-5}   \qquad \hbox{GD133}\nonumber\\
&& \Delta\mu/\mu =(-5.8\pm 3.7)\times10^{-5}   \qquad \hbox{GD29-38}.
\end{eqnarray}

\begin{table}[t]
\caption[Constraints on $\mu$ and $\aem$ in strong field environments]{Summary of the constraints on the variations of $\mu$ and $\aem$ in strong field environments.  We recall that $\Phi_{\rm N}/c^2\sim 9.8\times 10^{-9}$ and $g\sim 9.8~{\rm m s}^{-2}$ at the Earth surface.The two results by \cite{NEW_Hu:2020zeq}  rely on the same data but on different laboratory wavelengths.}
\label{tab-g-alphamu}
\centering
{\small
\begin{tabular}{ccccccc}
 \toprule
  Object & $\Phi_{\rm N}/c^2$&  constraint &constant & method & Ref. \\
 \midrule
  G191-B2B & $4.9\times10^{-5} $   &$(4.2\pm1.6)\times10^{-5}$ & $\aem$ & Fe~{\sc v} & \cite{NEW_Berengut:2013dta} \\
 		  &&  $(-6.1\pm5.8)\times10^{-5}$& $\aem$ & Ni~{\sc v} & \cite{NEW_Berengut:2013dta} \\
		   && $ [4.21\pm(0.47_{\rm stat}+2.35_{\rm syst})$  & $\aem$ & Fe~{\sc v} & \cite{NEW_Hu:2020zeq} \\
 		  & & $ [6.36\pm(0.33_{\rm stat}+1.94_{\rm syst})\times10^{-5}$ & $\aem$ & Fe~{\sc v} & \cite{NEW_Hu:2020zeq} \\
GD133		   & $1.2\times 10^{-4}$  & $(-2.3\pm 4.7)\times10^{-5} $ & $\mu$ & H$_2$ &   \cite{NEW_Bagdonaite:2014mfa}  \\
 GD29-38		  & $ 1.9\times 10^{-4}$&$(-5.8\pm 3.7)\times10^{-5}$ & $\mu$ & H$_2$ &  \cite{NEW_Bagdonaite:2014mfa} \\	
S0-6  & $2.4\times10^{-6}$ &$ (1.0\pm1.2)\times10^{-4}$ & $\aem $ &MM& \cite{NEW_Hees:2020gda}\\	
S0-12 & $1.6\times10^{-6}$ &$(-0.3\pm1.4)\times10^{-4}$  & $\aem $  &MM& \cite{NEW_Hees:2020gda}\\	
S0-13   &$9.4\times10^{-7}$&  $(0.03\pm1.3.5)\times10^{-4}$ & $\aem$ &MM& \cite{NEW_Hees:2020gda}\\	
S1-5    &$6.5\times10^{-7}$ &$(-0.7\pm2.4)\times10^{-4},$& $\aem$ &MM& \cite{NEW_Hees:2020gda}\\	
S1-23,     &$4.6\times10^{-7}$ &$(0.9\pm5.8)\times10^{-4}$& $\aem$ &MM& \cite{NEW_Hees:2020gda}\\	
\bottomrule
\end{tabular}
}
\end{table}

\paragraph{Solar twins}\ 

The method can be applied to any stellar atmosphere. The {\emph{Solar twin method} was proposed by \cite{NEW_Berke:2022fmj} who analysed the $\aem$-dependence of the absorption spectra of 17 stars, selected to have almost identical atmospheric properties as the Sun in order to reduce systematic effects, observed by ESO/VLT HARPS spectrograph. A weighted mean of the individual stellar constraints \citep{NEW_Murphy:2022pzu} (see \cite{NEW_Berke:2022rjk} for discussion of systematic errors) gave
\begin{equation}
 \Delta\aem/\aem =[7\pm 5_{\rm stat}\pm11_{\rm syst})\times10^{-6}
 \end{equation}
on a local neighborhood of 50~pc around the Sun.

\paragraph{Compact objects}\ 

Stronger field environments exist, e.g., around black holes. \cite{NEW_Hees:2020gda} analyzed the spectra of 5 late-type evolved giant stars from the S-star cluster orbiting the central supermassive black hole SgrA$^*$ in a similar way as for the many-multiplet method. For each star, one got $(\Delta\aem/\aem,\Phi_{\rm N}/c^2,)$ (individual constraints are gathered in Table~\ref{tab-g-alphamu}) and the combined analysis imposes
\begin{equation}
 \Delta\aem/\aem =(1.0\pm5.8)\times10^{-9}
 \end{equation}
between the Galactic center and the Earth. Indeed, those stars are relatively far from the black hole so that $\Phi_{\rm N}$ is of the same order as for white dwarf stars but the eccentricity of their trajectory is an asset to control systematics.

\paragraph{Further ideas}\ 

To reach higher fields, it was proposed \citep{NEW_Bambi:2013mha} that the analysis of fluorescent emission lines in the reflection spectra of black holes can potentially test variations of fundamental constants in gravitational fields up to $\Phi_{\rm N}/c^2\sim1$ however the construction of a theoretical model to analyze a full reflection spectrum in the presence of variations of fundamental constants remains a challenge (see the example of \cite{NEW_Davis:2016avf} who considered a model in which the masses of elementary particles depend on a scalar field around a black hole).

From the theory side, let us mention that the effects of varying fundamental constants on the white dwarf mass-radius relation were recently investigated by \cite{NEW_Magano:2017mqk} for polytropic models.  It indicates that independent measurements of the mass and radius, together with direct spectroscopic measurements of $\aem$ in white dwarf atmospheres such as those discussed above, could constrain unification scenarios which interestingly are almost orthogonal to those coming from atomic clocks (see Sect.~\ref{secRS}).}

\subsection{Cosmological scales}\label{subsubb}

\subsubsection{Primordial fluctuations}

During inflation, any light scalar field develop super-Hubble fluctuations of quantum origin, with an almost scale invariant power spectrum (see Chapt.~8 of \citealt{peteruzanbook}). It follows that if the fundamental constants depend on such a field, their value must fluctuate and have a non-vanishing correlation function. More important these fluctuations can be correlated with the metric perturbations. More speculative is the effect that such fluctuations can have during preheating after inflation since the decay rate of the inflaton in particles may fluctuate on large
scales \citep{modfluc1,modfluc2}.

In such a case, the fine-structure constant will behave as $\aem=\aem(t) +\delta\aem(\bx,t)$, the fluctuations being a stochastic variable. As seen earlier, $\aem$ enters the dynamics of recombination, which would then become patchy. This has several consequences for the CMB anisotropies. In a very general way, assume a constant, $X$ say, depends on the local value of a dynamical scalar field $\phi$. The value of $X$ at the observation point is compared to its value here and today,
$$
\Delta X/X_0 \equiv X(\phi)/X(\phi_0) -1.
$$
Decomposing the scalar field as $\phi=\phi_0+\Delta\phi$, one gets that $\Delta X/X_0=s_X(\phi)\Delta\phi$, with $s_X$ defined in Eq.~(\ref{def_sphi}). Now the scalar field can be decomposed into a background and perturbations as $\phi = \bar\phi(t)+\delta\phi(\bx,t)$ where the background value depends only on $t$ because of the Copernican hypothesis. It follows that
\begin{eqnarray}
 \frac{\Delta X(\bx,t)}{X_0} &=& s_X(\bar\phi)[\bar\phi(t)-\phi_0] +
 \lbrace s_X(\bar\phi)+s'_X(\bar\phi)[\bar\phi(t)-\phi_0] \rbrace\delta\phi(\bx,t)\nonumber\\
 &\equiv& s_X(\bar\phi)\Delta\bar\phi + {\cal{S}}_X(\bar\phi)\delta\phi(\bx,t).
\end{eqnarray}
The first term of the r.h.s.\ depends only on time while the second is spacetime dependent. It is also expected that the second term in the curly brackets is negligible with respect to the first, i.e., ${\cal{S}}_X(\bar\phi)\sim s_X(\bar\phi)$. It follows that one needs $\delta\phi(\bx,t)$ not to be small compared to the background evolution term $\Delta\bar\phi$ for the spatial dependence to dominate over the large scale time dependence. This can be achieved for instance if $\phi$ is a seed field whose mean value is frozen. Because of statistical isotropy, and in the same way as for CMB anisotropies (see, e.g., \citealt{peteruzanbook}), one can express the equal-time angular  power spectrum of $\Delta X/X_0$ for two events on our past lightcone as
\begin{equation}
 \left\langle \frac{\Delta X(\bn_1,r,t)}{X_0}\frac{\Delta X(\bn_2,r,t)}{X_0}\right\rangle =
 \sum_\ell\frac{2\ell+1}{4\pi}C_\ell^{(XX)}(z) P_\ell(\bn_1\cdot \bn_2).
\end{equation}
If $\delta\phi$ is a stochastic field characterized by its power spectrum, $\langle \delta\phi(\bk_1,t)\delta\phi(\bk_2,t)\rangle=P_\phi(k,t)\delta(\bk_1+\bk_2)$ in Fourier space, then
\begin{equation}
 C_\ell^{(XX)}(z) = \frac{2}{\pi}{\cal{S}}^2_X[\bar\phi(z)] \int P_\phi(k,z)j_\ell[k(\eta_0-\eta)] k^2\dd k,
\end{equation}
$j_\ell$ being a spherical Bessel function. For instance, if $P_\phi\propto k^{n_s-1}$ where $n_s$ is a spectral index, $n_s=1$ corresponding to scale invariance, one gets that $\ell(\ell+1)C_\ell^{(XX)}\propto\ell^{n_s-1}$ on large angular scales.

Besides, similarly to weak gravitational lensing, these fluctuations of $\aem$ modify the mean power spectra (this is a negligible effect)  and induce a curl component (B mode) to the polarization \citep{sigur}. Such spatial fluctuations also induce non-Gaussian temperature and polarization correlations  in the CMB \citep{sigur,pub}. Such correlations have not allowed to set observational constraints yet but they need to be included for consistency, see e.g., the example of CMB computation in scalar-tensor theories \citep{cmb-G1}. The effect on large-scale structure was also studied in \cite{motablss,balss} and the Keck/HIRES QSO absorption spectra showed \citep{q-murphyAD} that the correlation function of the fine-structure constant is consistent on scales ranging between 0.2 and 13~Gpc.

From the 4-point correlation of  the CMB temperature anisotropies of the Planck data, \cite{NEW_OBryan:2013nip} constrained the fractional rms of the fluctuations of a stochastic $\aem$ to $(1.34\pm5.82)\times10^{-2}$ at 95\% C.L. on scales larger that 10~deg. \cite{NEW_Smith:2018rnu} put a 95\% C.L. limit on the amplitude of a scale-invariant angular power spectrum of $\aem$ fluctuation, $C_\ell(\aem)=A\ell(\ell+1)$,  of $A<1.6\times10^{-5}$ form Planck measurements of the temperature and polarization power spectrum.

\subsubsection{Dipolar modulation}\label{subsec-dipalpha}

The astronomical measurements of the fundamental constants described in Sect.~\ref{subsec33} have been used to question the homogeneity of $\aem$ and $\mu$ by constraining a dipole modulation. Such analysis may reveal a  spatial variation that could have consequences for the validity of the Copernican principle. Indeed if the putative scalar field $\varphi$ causing the variation had a dipole, it would reflect in the distribution of the fundamental constants. Such an assumption enlarges the space of free parameters by including the dipole direction ${\bf d}$ in the data analysis.\\

\noindent$\bullet$ {\bf QSO data.} In the analysis of the spectroscopic data from QSO, it has been claimed \citep{berenspa,webspace,NEW_King:2012id} that $\aem$ may have a dipolar variation that would explain consistently the data from the Southern and Northern hemispheres (see Sect.~\ref{MMQSO}). \cite{webspace} decomposed the value of the fine structure of each absorption system as
\begin{equation}\label{e.dippara}
\frac{\Delta\aem}{\aem}(z,{\bf n}) = A(z) + B(z)\cos\theta
\end{equation}
where $\theta$ is the angle between the dipole direction and the line of sight ${\bf n}$, i.e., $\cos\theta = {\bf n}.{\bf d}$. Three main models where considered for $B(z)$:
\begin{itemize}
\item (dip1) $B=B_0$, 
\item (dip2): $B=B_0r(z)$ with $r(z)/c$ the look-back time,
\item  (dip3): $B=B_0z^\beta$. 
\end{itemize}
\cite{webspace} concluded that in scenario (dip1) the dipole has an amplitude 
\begin{equation}\label{e.australian}
B=(1.02\pm0.21)\times10^{-5}
\end{equation}
 and direction  (RA,DEC)$=(17.4\pm0.9~{\rm hour}, -58^{\rm o}\pm 9^{\rm o})$. In the scenario (dip2)
 \begin{equation}\label{e.australian2}
B=(1.01\pm0.25)\times10^{-6}\unit{Gly^{-1}}
\end{equation}
 with almost the same direction. The analysis by \cite{NEW_Berengut:2012ep} concluded that $A=-0.19(8)\times10^{-5}$ and $B=0.106 (22)\times 10^{-5}$~Gly$^{-1}$ for the parameterisation (dip2). \cite{NEW_Mariano:2012wx} pointed out that the $\aem$ dipole is anomalously aligned with the dark energy dipole obtained through the Union2 sample at 2$\sigma$ level \citep{NEW_Antoniou:2010gw} and suggested a model (“extended topological quintessence'') to explain the correlation between the two dipoles -- see also \cite{NEW_Bronnikov:2013xh} for a model with dark energy and $\aem$ variations arising from extra-dimensions. We refer to \cite{NEW_Aluri:2022hzs} for a review on the different dipoles that appear in cosmology and to \cite{NEW_Mariano:2012ia} for their comparison. As discussed in Sect.~\ref{subsec33}, the QSO results can be performed with the archival or dedicated sample, which do not agree. Concerning a pure dipole (dip1), \cite{NEW_Aluri:2022hzs} got (see \cite{NEW_Pinho:2016mkm} for an earlier analysis)
\begin{equation}
B=(9.4\pm2.2)\times 10^{-6} \quad\hbox{(archival)}\qquad
|B|\leq2.9\times 10^{-6} \quad\hbox{(dedicated)}\,.
\end{equation}
Hence the Archival data have a statistical preference, at just over four standard deviations, for a dipole with an amplitude of about 9 ppm, while the Dedicated data show no preference for a dipole, and the amplitude is constrained to be less than about 3 ppm at the 95.4\% C.L.. To finish, from their analysis of VLT/X-SHOOTER, \cite{NEW_Wilczynska:2020rxx} (see Sect.~\ref{MMQSO}) constrained the amplitude of a constant dipole to  $A = (-0.70\pm0.16)\times10^{-5}$.

Several complementary tests have been proposed. First, one can consider the analogous high-resolution spectroscopic constraints on $\mu$ (see Table~\ref{tab02}). They tend to indicate that a pure dipole in $\mu$ is constrained to $|B_\mu|<2\times10^{-6}$ \citep{NEW_Aluri:2022hzs}. In any model where $\aem$ and $\mu$ variations are correlated one would expect the dipole for the two constants to be aligned. While \cite{berenspa} found a weak indication for a $\mu$-dipole aligned with the $\aem$-dipole  with $B_\mu=(2.6\pm1.3)\times10^{-6} {\rm Gly}^{-1}$ using parameterisation (dip2), this was not confirmed by the analysis of \cite{NEW_Aluri:2022hzs}. Hence, there is currently no robust evidence of a dipole in $\aem$ and $\mu$ at a level of a few $10^{-6}$.\\

\noindent$\bullet$ {\bf SNIa data.} \cite{NEW_Negrelli:2018toq} proposed to use the effect of the variation of the constant on SNIa and hence on the Hubble diagram to conclude from the SNIA compilations Union~2.1 that $B=(-3.11\pm1.16)\times10^{-2}$ and JLA  $B=(0.22\pm0.47)\times10^{-2}$. Indeed such an analysis deriving from the luminosity distance entangles any dipolar variations that can arise from the measurement or the cosmology itself \citep{NEW_Chang:2012fmz} so that they are not robust constraints. See also \cite{NEW_Karpikov:2015sra} for an analysis spacetime variations of the Ni-46 decay rate from SNIa light curves and a potential window on $G_{\rm F}$ or $v$.\\
 
\noindent$\bullet$ {\bf Galaxy clusters.} \cite{ NEW_Galli:2012bf} extended  her method based of galaxy clusters to investigated the signature of the spatial variation of $\aem$ using multi-frequency measurements of the thermal Sunyaev-Zeldovich (see Sect.~\ref{subsecSZ}) to conclude that $B = (-5.5\pm 7.9)\times 10^{-3}\,{\rm Gly}^{-1}$  for parameterisation (dip2) assuming the direction of dipole is fixed to the QSO dipole~(\ref{e.australian}).  A similar analysis by \cite{NEW_Bora:2020sws} got  $B = (-3\pm 3)\times 10^{-3}\,{\rm Gly}^{-1}$. \cite{NEW_deMartino:2016tbu,NEW_DeMartino:2016hxb} applied the same method to a catalog of 618 X-ray clusters including Planck data to measure the thermal SZ effect at the location of the clusters. They concluded that $A=(6\pm4)\times10^{-3}$ and $B=(8\pm9)\times10^{-3}$ for parameterisation (dip1) and  $A=(3\pm3)\times10^{-3}$ and $B=(6\pm5)\times10^{-3}~{\rm Gly}^{-1}$ for parameterisation (dip2) assuming the direction fixed to the QSO dipole~(\ref{e.australian}) while they got $A=(2.1\pm2.9)\times10^{-2}$ and $B=(-3\pm1.4)\times10^{-2}$  for parameterisation (dip1) and  $A=(1.9\pm1.1)\times10^{-2}$ and $B=(-2.7\pm5.1)\times10^{-3}~{\rm Gly}^{-1}$ for parameterisation (dip2) if the direction of the dipole is not fixed.\\

\noindent$\bullet$ {\bf [O\,{\sc iii}] emission spectra.}  \cite{NEW_Jiang:2024hiv} tested the spatial variation in  $\aem$ using the  [O\,{\sc iii}] emission  of a JWST galaxy sample with $3<z<7$ in four directions on the sky. The $\Delta\aem/\aem$ values in the different directions are also consistent with zero within $1\sigma$ error of $\sim 10^{-4}$, suggesting no spatial variation up to $z\sim9$.\\
 
\noindent$\bullet$ {\bf CMB.} To finish with the CMB, it was demonstrated that a spatial variation of $\aem$ would induce a dipolar modulation of CMB anisotropies \citep{Moss:2010qa,NEW_Aiola:2015rqa}, but at a level incompatible with existing constraints \citep{wall3}. The study by \cite{ NEW_Planck:2014ylh} concluded that the relative constant amplitude of a dipolar modulation of $\aem$ is constrained to 
 \begin{equation}
B=(-2.4\pm3.7)\times 10^{-2} 
\end{equation}
from the Planck data \citep{NEW_Planck:2013nga}.
 
\begin{table}[t]
\caption[Constraints on a cosmological dipolar modulation of $\aem$]{Summary of the constraints on a dipolar modulation of $\aem$ for the two ansatz (dip1) and (dip2) of the parameterisation~(\ref{e.dippara}). A star indicates that the direction of the dipole is fixed to the one of (\ref{e.australian}) in the analysis.}
\label{tab-dipole}
\centering
{\small
\begin{tabular}{ccccr}
 \toprule
  Method & Model & $A$ & $B$ & Ref. \\
 \midrule
 QSO & dip1 &           -                                   & $(1.02\pm0.21)\times10^{-5}$                     &   \cite{webspace}  \\
 QSO & dip2 &           -                                   & $(1.01\pm0.25)\times10^{-6}\unit{Gly^{-1}}$                     &   \cite{webspace}  \\ 
  QSO & dip2 & $-0.19(8)\times10^{-5}$  &  $0.106 (22)\times 10^{-5}$~Gly$^{-1}$   &   \cite{NEW_Berengut:2012ep}  \\
 QSO (Archival) & dip1  & -&  $(9.4\pm2.2)\times 10^{-6} $ &\cite{NEW_Aluri:2022hzs}  \\
 QSO  (Dedicated)  & dip1 & - & $2.9\times 10^{-6} $ & \cite{NEW_Aluri:2022hzs} \\
 \midrule
  SNIa (Union-2.1) & dip1 & - & $(-3.11\pm1.16)\times10^{-2}$ & \cite{NEW_Negrelli:2018toq} \\
  SNIa (JLA) & dip1 & - & $(0.22\pm0.47)\times10^{-2}$ & \cite{NEW_Negrelli:2018toq} \\
 \midrule
   Clusters &dip2* & - & $(-5.5\pm 7.9)\times 10^{-3}\,{\rm Gly}^{-1}$ &  \cite{ NEW_Galli:2012bf}\\
  Clusters & dip2* & - & $ (-3\pm 3)\times 10^{-3}\,{\rm Gly}^{-1}$&  \cite{ NEW_Bora:2020sws}\\
  Clusters & dip1* & $(6\pm4)\times10^{-3}$ & $(8\pm9)\times10^{-3}$  &  \cite{ NEW_DeMartino:2016hxb} \\
  Clusters & dip2 *& $(3\pm3)\times10^{-3}$  & $(6\pm5)\times10^{-3}~{\rm Gly}^{-1}$ &  \cite{ NEW_DeMartino:2016hxb}\\
  Clusters & dip1 & $(2.1\pm2.9)\times10^{-2}$ & $(-3\pm1.4)\times10^{-2}$ &  \cite{ NEW_DeMartino:2016hxb}\\
  Clusters & dip2 & $(1.9\pm1.1)\times10^{-2}$ & $(-2.7\pm5.1)\times10^{-3}~{\rm Gly}^{-1}$  &  \cite{ NEW_DeMartino:2016hxb}\\
 \midrule
   CMB & dip1 & - & $(-2.4\pm3.7)\times 10^{-2} $  &  \cite{NEW_Planck:2014ylh}\\
\bottomrule
\end{tabular}
}
\end{table}
 
\subsubsection{Wall of constants}\label{secwallalpha}

When the cosmological history of the evolution of fundamental constants is interpreted in terms of a time variation which, as we we have seen so far, it has to face the problem of strong local constraints. This has motivated \cite{wall1} to interpret the observations as a spatial variation on cosmological scales thanks to the hypothesis of the existence of a low energy domain wall produced in the spontaneous symmetry breaking involving a dilaton-like scalar field coupled to electromagnetism. Domains on either side of the wall exhibit slight differences in their respective values of $\aem$; see Fig.~\ref{fig-wall}. If such a domain wall is present within our Hubble volume, absorption spectra at large redshifts may or may not provide a variation in $\aem$ relative to the terrestrial value, depending on our relative position with respect to the wall and the direction observation. 

\begin{figure}[htbp]
  \vskip-.25cm
  \centerline{\includegraphics[scale=0.45]{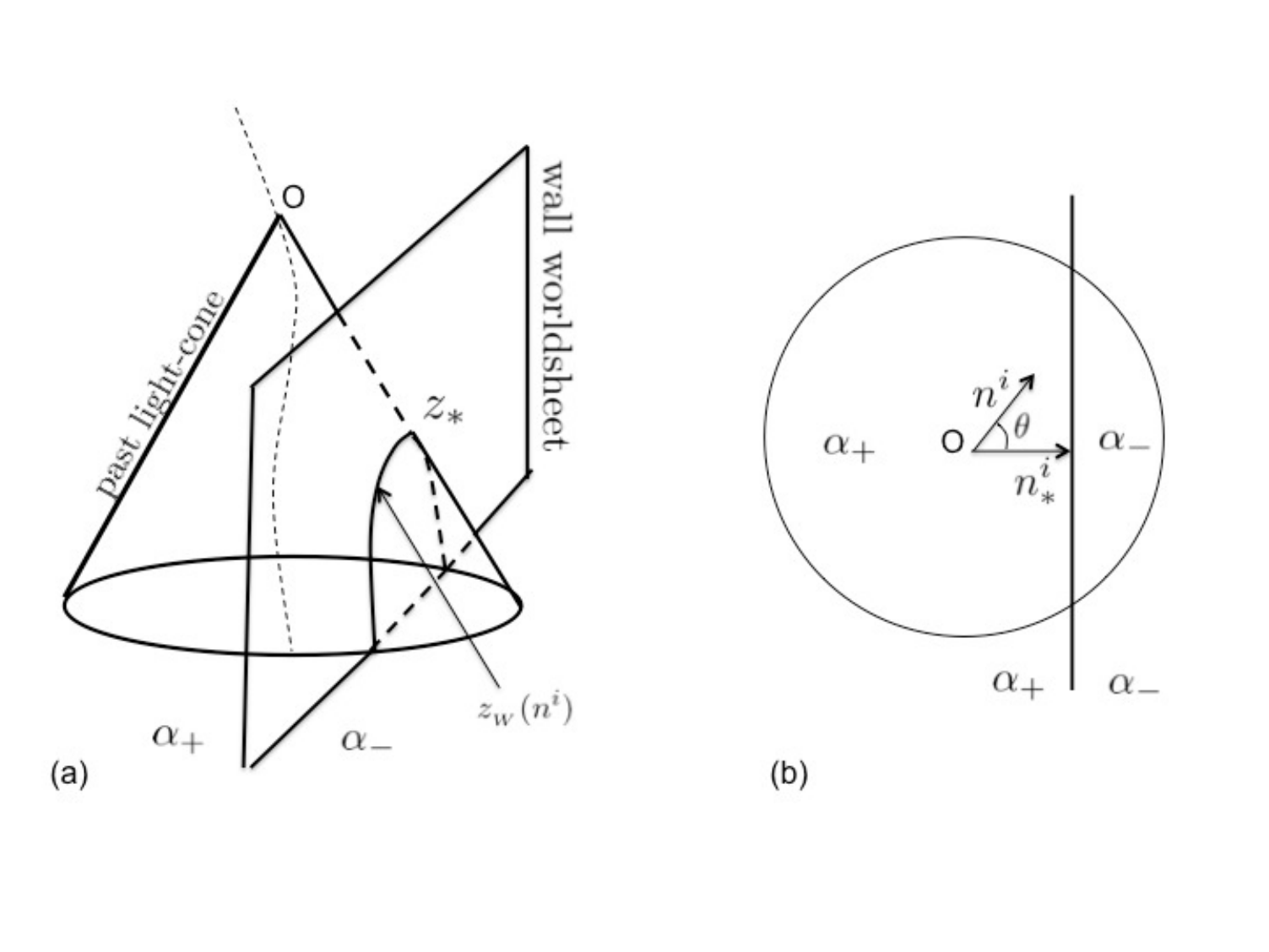}}
  \vskip-1.25cm
  \caption[Geometry of the wall of fundamental constants]{A domain wall is assumed to cross our Hubble volume, hence intersecting our past light-cone on a 2-dimensional spatial hypersurface characterized by the redshift of the wall in a direction ${\bf n}$. $z_*$ is the lowest redshift at which the wall can be observed (left). On a constant time hypersurface, this leads to a spatial distribution of domains in which $\aem$ is mostly constant (right). From \cite{wall1}.}
  \label{fig-wall}
\end{figure}

The first model \citep{wall1} relies on the generalisation of \cite{beken1},
\begin{eqnarray}
 S &=&\int \left[ \frac{R}{16\pi G} -\frac{1}{2}(\partial_\mu\phi)^2  +V(\phi)+\frac{1}{4}B_F(\phi) F^2 \right. \nonumber\\
    && \left.\qquad\qquad\qquad- \sum_j  \bar\psi_j\left( i \dslash- B_j(\phi) m_j \right)\psi_j
 \right]\sqrt{-g}\dd^4 x
\end{eqnarray}
with $\phi$ is assumed to have the quartic potential
\begin{equation} \label{quartic}
 V(\phi) = \frac{1}{4}\lambda (\phi^2 - \eta^2)^2
\end{equation}
and a coupling to the Faraday tensor of electromagnetism  as well as to the fermions $\psi$ with the coupling functions $B_i$  of the form
\begin{equation}
 B_i(\phi) = \exp\left({\xi_i \frac{\phi}{M_*}}\right)\simeq 1 + \xi_i \frac{\phi}{M_*}.
\end{equation}
The coefficients $\xi_i$ are constant and $M_*$ is a mass scale. It was assumed for simplicity that only $\xi_F$ is non-vanishing at tree level. Indeed, $\phi$ inevitably  couples to nucleons radiatively. To reproduce a change in $\aem$ through the domain wall matching the claimed spatial variation by \cite{webspace}, one needs
\begin{equation}
 \frac{\Delta \aem}{\aem} \simeq  2\xi_F  \,  \frac{\eta}{M_*}   \sim {\rm few} \, \times 10^{-6} \,.
\end{equation}
The contribution of the wall  to the energy budget of the universe is
\begin{equation}\label{omega-today}
 \Omega_{\rm wall}  \simeq\left(\frac{\eta}{100\,{\rm MeV}}\right)^{3}\,.
\end{equation}
his led to the idea that  $\eta = {\cal O} \left( {\rm MeV} \right)$ to construct a fully viable model.

Our local Hubble bubble can then be thought as a patchwork of domains each enjoying constants with different values as, what could be named, a ``\emph{microlandscape of constants}". Indeed, no variation of any constant will be detectable within each domain but they can be revealed on the past light cone of any observer. Such a model illustrates the complementarity between local and astrophysical constraints and give a counter example to claims that local constraints on the variation of $\aem$ from atomic clocks are sharper than those obtained from QSO absorption spectra. \cite{NEW_Olive:2012ck} showed that a single wall configuration  is statically comparable to a dipole fit (but is a big improvement over a weighted mean) and that a two-wall solution is a far better fit (despite adding 3 parameters over the single wall solution).  \cite{Nielsen:2023drs} actually argued that such domain walls can form a network with typical distance between walls of 100~millions light-year s,hence defining a large number of regions with different low-energy physics within our observable universe.

The model was extended to a runaway potential \citep{NEW_Chiba:2011en} in order to construct an toy model that could explain simultaneously time \citep{q-webprl99,q-murphy03a} and space \citep{webspace} variations that were thought to be indicated by QSO at the time. It was then reformulated as a $F(R)$ theory by \cite{NEW_Bamba:2011nm} and in the framework of the symmetron \citep{NEW_Silva:2013sla}.  \cite{NEW_Avelino:2014xsa} concluded that in order for such a wall to be responsible for a spatial variation of $\aem$ compatible with \citep{webspace,NEW_King:2012id} the domain wall network shall contribute to $\Omega_{\rm wall}\in[10^{-10},10^{-5}]$ to the matter budget of the universe. \cite{NEW_Stadnik:2020bfk} showed that for domain walls within the symmetron model with $B_F\propto\phi^2$, the current constraints imply $\Omega_{\rm wall}\ll10^{-10}$; see Sect.~\ref{secTopDef} for a discussion of the detection of topological defects in the laboratory.

\subsubsection{Non-Copernican field configuration}

Another possibility to explain a dipolar modulation of some fundamental constant would be that the Copernican principle is not fully satisfied, such as in various void models. Then the background value of $\phi$ would depend, e.g., on $r$ and $t$ for a spherically symmetric spacetime (such as the Lema\^{\i}tre--Tolman--Bondi spacetime). This would give rise to a dipolar modulation of the constant if the observer is not located at the center of the universe. Note, however, that such a cosmological dipole would also reflect itself, e.g., on CMB anisotropies; see \cite{NEW_Cusin:2016kqx}. Similar possibilities are also offered within the chameleon mechanism where the value of the scalar field depends on the local matter density; see Sect.~\ref{subsub2}.

\subsection{Universality of free fall} \label{subsec22}

The universality of free fall can also be used to constrain large scale spatial variation of the fundamental constant.

Indeed, the Solar system barycenter moves, on laboratory timescales, almost linearly with a constant speed, at about 250 km/s around the Milky Way with a period of about 240 million years and at about 370 km/s with respect to the comoving rest frame defined by the observed cosmic microwave background \citep{NEW_Planck:2018vyg}. Any spatial variation of the constants on these scales will translate as a time variation in the Earth restframe. Indeed the galactic modulation cannot be tested directly but would have to be taken into account to interpret Oklo data for which the timescales are identical to the Sun orbital period. A cosmological gradient, as e.g., discussed in Sect.~\ref{subsec-dipalpha}, will result in a linear temporal drift of the constants, $\Delta\alpha_i(t)/\alpha_i= {\bf v}_{\odot}.\nabla \ln\alpha_i$, hence linking a local measurement to a cosmological variation.

Estimating from  Sect.~\ref{subsec-dipalpha} that $\cos( {\bf v}_{\odot},\nabla \ln\alpha)\sim0.1$, the constraint by \cite{Stadnik:2020kcz} can be updated to take into account~(\ref{e.clock-alpha}) on $\aem$ by \cite{NEW_Filzinger:2023zrs} to deduce that
$$
\vert \nabla\ln\aem\vert \lesssim 3.5 \times 10^{-6} \unit{Glyr^{-1}},
$$
which is competitive with the bounds listed in Table~\ref{tab-g-alphamu}, highlighting the strong connection between local and cosmological constraints.

\cite{NEW_Damour:2011fa} investigated the perturbing influence of a differential force with a fixed direction acting on a gravitationally bound two-body system, such as the Earth--Moon system. This ``gravitational Stark effect'' \citep{Damour:1991rq} is an example of singular perturbation where a small perturbing force can have a large effect. In particular, elliptic orbits undergo a complicated secular evolution that can be constrained. It was estimated that LLR experiment can probe a gradient of
$$
\vert\nabla \ln X_{\rm q}\vert \lesssim 2.6 \times 10^{-6} \unit{Glyr^{-1}},
$$
again comparable to the bounds lof Table~\ref{tab-g-alphamu}.

\section{Why are the constants just so?}\label{section5}

The numerical values of the constants are not determined by the laws of nature in which they appear. One of the most outstanding questions ipertains to the values of the fundamental parameters of the Standard Model so that one may wonder why they have the values we observe.  In particular, as pointed out by many authors (see below), the constants of nature seem to be fine tuned \citep{leslie}. Many physicists take this fine-tuning to be an explanandum that cries for an explanans, hence following \cite{hoyle1} who wrote that ``one must at least have a modicum of curiosity about the strange dimensionless  numbers that appear in physics.'' The answer may lie in an as yet undefined future theory, such as a complete string theory, in which case there is hope of a deeper understanding. It is also possible that our Universe is part of a larger structure or a Multiverse, but we have no means to know. In this case, the observed values may be environmental and reflect an observational bias through the anthropic principle, which absolves us, the humans on Earth, from the duty of explaining the values of the governing constants, at least for the time being, until data at higher energies are available and eventually change our description of physics.

We refer to \cite{NEW_finetuning} for an extensive review of these questions, to \cite{NEW_Donoghue:2016tjk} for a review on the motivations for a multiverse and a discussion of the fine-tunings in particle physics, to \cite{NEW_Barnes:2011zh,NEW_Barnes:2017grv,NEW_Barnes:2017rpb,NEW_Adams:2019kby,NEW_Barnes:2021vbx,NEW_Sandora:2022ilr} for detailed discussions of the fine-tunings and discussions of the evaluation of the level of fine-tuning in Bayesian approach. 

\subsection{Universe and multiverse approaches}

Two possible lines of explanation are usually envisioned: a \emph{design or consistency hypothesis} and  an \emph{ensemble hypothesis}, that are indeed not incompatible together. The first hypothesis includes the possibility that all the dimensionless parameters in the ``final'' physical theory will be fixed by a condition of consistency or an external cause. In the ensemble hypothesis, the universe we observe is only a small part of the totality of physical existence, usually called the multiverse. This structure needs not be fine-tuned and shall be sufficiently large and variegated so that it can contain as a proper part a universe like the one we observe the fine-tuning of which is then explained by an \emph{observation selection effect} \citep{borstrom}.

These two possibilities send us back to the  large number hypothesis by \cite{dirac37} that has been used as an early motivation to investigate theories with varying constants. The main concern was the existence of some large ratios between some combinations of constants. As we have seen in Sect.~\ref{subsecGUT}, the running of coupling constants with energy, dimensional transmutation or relations such as Eq.~(\ref{QCDscale}) have opened a way to a rational explanation of very small (or very large) dimensional numbers.  This follows the ideas developed by \cite{eddi1,eddi2} aiming at deriving the values of the constants from consistency relations, e.g., he proposed to link the fine-structure constant to some algebraic structure of spacetime. \cite{dicke61} pointed out another possible explanation to the origin of Dirac  large numbers: the density of the universe is determined by its age, this age being related to the time needed to form galaxies, stars, heavy nuclei\dots. This led \cite{carter74} to argue that these numerical coincidences should not be a surprise and that conventional physics and cosmology could have been used to predict them,  at the expense of using the anthropic principle.

The idea of such a structure called the \emph{multiverse} has attracted a lot of attention in the past years and we refer to \cite{carrbook} for a more exhaustive account of this debate. While many versions of what such a multiverse could be, one of them finds its root in string theory. In 2000, it was realized \citep{boussoP} that vast numbers of discrete choices, called flux vacua, can be obtained in compactifying superstring theory. The number of possibilities is estimated to range between $10^{100}$ and $10^{500}$, or maybe more. No principle is yet known to fix which of these vacua is chosen. Eternal inflation offers a possibility to populate these vacua and to generate an infinite number of regions in which the parameters, initial conditions but also the laws of nature or the number of spacetime dimensions can vary from one universe to another, hence being completely contingent. It was later suggested by \cite{sus03} that the anthropic principle may actually constrain our possible locations in this vast string landscape. This is a shift from the early hopes \citep{kane} that M-theory may conceivably predict all the fundamental constants uniquely.

Indeed such a possibility radically changes the way we approach the question of the relation of these parameters to the underlying fundamental theory since we now expect them to be distributed randomly in some range. Among this range of parameters lies a subset, that we shall call the \emph{anthropic range}, which allow for universe to support the existence of observers. This range can be determined by asking ourselves how the world would change if the values of the constants were changed, hence doing \emph{counterfactual cosmology}. However, this is very restrictive since the mathematical form of the law of physics managed as well and we are restricting to a local analysis in the neighborhood of our observed universe. The determination of the anthropic region is not a prediction but just a characterization of the sensitivity of ``our'' universe to a change of the fundamental constants \emph{ceteris paribus}. Once this range is determined, one can ask the general question of quantifying the probability that we observe a universe as ours, hence providing a probabilistic prediction. This involves the use of the anthropic principle, which expresses the fact that we observe are not just observations but observations made by us, and requires us to state what an observer actually is \citep{neal}.

\subsection{Fine-tunings and determination of the anthropic range}

As we have discussed in the previous sections, the outcome of many physical processes are strongly dependent  on the value of the fundamental constants.  One can always ask the scientific question of what would change in the world around us if the values of some constants were changed, hence doing some counterfactual cosmology in order to determine the range within which the universe would have developed complex physics and chemistry, what is usually thought to be a prerequisite for the emergence of complexity and life (we emphasize the difficulty of this exercise when it goes beyond small and local deviations from our observed universe and physics, see, e.g., \cite{harnik} for a possibly life supporting universe without weak interaction). In doing so, one should consider the fundamental parameters entering our physical theory but also the cosmological parameters. We refer to \cite{NEW_Adams:2019kby} for an extensive and detailed discussion of the degree of fine-tuning of our description of the world.

\paragraph{Naturalness and the macroscopic world}\

Our description of the universe is plagued with incomprehensible instance  of anti-naturalness. As we have discussed in Sect.~\ref{secLCDM}, the first is related to the value of the cosmological constant, the observed value of which is some 120 orders of magnitude smaller than the Planck scale $M_{\mathrm{P}}^4$, believed to be the only fundamental scale; see \cite{Bianchi:2010uw,Gegelia:2019fjx} for opposite views on this belief. The second lies in the fact that the electroweak scale is 17 orders of magnitude lower than the Planck scale. The facy that the QCD vacuum angle $\theta$ was 10 orders of magnitude lower than its expected value $\theta\sim 1$ has been addressed by \cite{NEW_Lee:2020tmi}. These two anti-naturalness problems are related to the fact that ``\emph{we live in a big universe that contains big things.}'' 

Indeed, the suppression of the cosmological constant is vital for the existence of a large and old universe. Even if it were a few orders of magnitude larger, the Universe would have entered an inflationary phase before the onset of galaxy formation.  Indeed, the typical radius and typical age of the universe are given by $H_0^{-1}\sim\sqrt{3/\Lambda}\sim 10^{60}\ell_{\rm P}$. In other word the fact that we leave in a old and big universe is a consequence of the cosmological constant non-naturalness.

Then, concerning matter and the macroscopic structures we observe to exist, the smallness of the u and d quark masses compared to  $\Lambda_{\rm QCD}$ and the fact that $m_{\rm u} < m_{\rm d}$ are crucial for the genesis of heavier elements in stars. Then, following \cite{carr79} -- see also \cite{star-adam} -- we can estimate the typical mass of macroscopic objects such as stars and planets by roughly looking at the balance between the intensities of gravitation and the pressure of the force opposing the gravitational collapse for a given mass $M$.  The minimum mass necessary to sustain nuclear fusion is determined by the criterium that the star's central temperature is high enough leading to \citep{carr79}
$$
M_{\rm min,star} \sim \frac{1}{2}\mu^{3/4}\aem^{3/2} M_*
$$
with $M_*$ the characteristic stellar mass, $M_*= \ag^{-3/2} m_{\rm p}$.  Similarly the typical mass of a rocky planet is determined by the requirement that the electromagnetic interaction counterbalance gravity, leading to
$$
M_{\rm planet} \sim \left(\frac{\aem}{\alpha_{\rm G}}\right)^{3/2} m_{\rm p}=\aem^{3/2}M_*
$$
while the critical mass scale of life form \citep{Page_2009} is
$$
M_{\rm life} \sim \epsilon_c^{3/4}\left(\frac{\aem}{\alpha_{\rm G}}\right)^{3/2} m_{\rm p} =   (\epsilon_c\aem\alpha_{\rm G})^{3/4} M_*
$$
where $\epsilon_c\sim10^{-3}$ is the chemical conversion factor. As a consequence, there exist macroscopic objects because gravity is weak compared to the weak scale. For planets, it allows to pack $(\aem/\ag)^{3/2}$ particles without turning to a black hole. Such an object will have a typical size of $(\aem/\ag)^{1/2}a_0\sim 10^4$~km. The separation of scales then implies that there exist classes of objects with typically the same sizes and masses, hence fixing their dimension in units of the Bohr radius $a_0$ independently of its value. For instance, the minimal mass of a neutron star and its typical size are given by $\alpha_{\rm S}M_*$ where the fine structure constant has been replaced by the strong structure constant.

As a conclusion the anti-naturalness problems concerning the cosmological constant and the weak scale ensure that we live in a big universe containing big objects. This is the first step toward the anthropic principle.

\paragraph{Fine tuning on the particle physics parameters}\

Then, there are several constraints that the fundamental parameters listed in Table~\ref{tab-list} have to satisfy in order for the universe to allow for complex physics and chemistry. Let us stress, in a non-limiting way, some examples.

\begin{itemize}
\item It has been noted that the stability of the proton requires $m_{\mathrm{d}}-m_{\mathrm{u}}\gtrsim\aem^{3/2}m_{\mathrm{p}}$.  The anthropic bounds on $m_{\mathrm{d}}$, $m_{\mathrm{u}}$ and $m_{\mathrm{e}}$ (or on the Higgs vev) arising from the existence of nuclei, the di-neutron and the di-proton cannot form a bound state, the deuterium is stable have been investigated in many works \citep{ag1,ag2,hogan00,dentfair,dono2,hogan1,damourdono,dono3}, even allowing for nuclei made of more than 2 baryon species \citep{jaffeq}. Typically, the existence of nuclei imposes that $m_{\mathrm{d}}+m_{\mathrm{u}}$ and $v$ cannot vary by more that 60\% from their observed values. 
\item If the difference of the neutron and proton masses where less than about 1~MeV, the neutron  would become stable and hydrogen would be unstable \citep{rozental88,hogan2} so that helium would have been the most abundant at the end of BBN. Hence, the whole history of the formation and burning of stars would have been different. It can be deduced that \citep{hogan00} one needs $m_{\mathrm{d}}-m_{\mathrm{u}}-m_{\mathrm{e}}\gtrsim1.2 \unit{MeV}$ so that the universe does not become all neutrons; $m_{\mathrm{d}}-m_{\mathrm{u}}+m_{\mathrm{e}}\lesssim3.4 \unit{MeV}$ for the $pp$ reaction to be exothermic and $m_{\mathrm{e}}>0$ leading to a finite domain. Note that among the fine-tuning, one often considers the diproton arguing that if it were bound, stars would burn $10^{18}$ times brighter and faster via strong interactions, resulting in a universe that would fail to support life. \cite{NEW_Barnes:2015mac,NEW_Barnes:2017epx,NEW_Adams:2021wen} demonstrated  that stable, ``strong-burning'' stars adjust their central densities and temperatures to have familiar surface temperatures, luminosities and lifetimes so that there is no ``diproton disaster''. Similarly \cite{NEW_Coc:2012xk} showed that a stable Be-8 nuclei will not affect BBN.
\item A coincidence emerges from the existence of stars with convective and radiative envelopes, since it requires \citep{carr79} that $\ag\sim\aem^{20}$. It arises from the fact that the typical mass that separates these two behavior is roughly $\ag^{-2}\aem^{10}m_{\mathrm{p}}$ while the masses of star span a few decades around $\ag^{-3}m_{\mathrm{p}}$. Both stars seem to be needed since only radiative stars can lead to supernovae, required to disseminate heavy elements, while only convective stars may generate winds in their early phase, which may be associated with formation of rocky planets. This relation while being satisfied numerically in our universe cannot be explained from fundamental principles. 
\item Similarly, it seems that for neutrinos to eject the envelope of a star in a supernovae explosion, one requires \citep{carr79} $ \ag\sim\aw^{4}$.
\item As discussed in Sect.~\ref{secstellar}, the production of carbon seems to imply that the relative strength of the nuclear to electromagnetic interaction must be tuned typically at the 0.1\% level.
\item  \cite{NEW_Sandora:2022ilr} investigated the dependence of primordial abundances on physical constants and the implications  for the distribution of complex life for various proposed habitability criteria. Interestingly they exhibit a series of predictions (carbon-rich or carbon-poor planets are uninhabitable; slightly magnesium-rich planets are habitable; life does not depend on nitrogen abundance too sensitively; metal-rich planets and phosphorus-poor planets are habitable) that can be checked  by probing regions of our universe that closely resemble normal environments in other universes. Then, if any of these predictions are found to be wrong, the multiverse scenario would predict that the majority of observers are born in universes differing substantially from ours, changing our assumption about our typicality.
\end{itemize}

\paragraph{Fine tuning on the cosmological parameters}\

Other coincidences involve also the physical properties, not only of the physical theories, but also of our universe, i.e., the cosmological parameters summarized in Table~\ref{tab-cosmo}. Indeed those parameters are not on the same footing as fundamental constants since they may be though to vary on large scale if the Copernican principle does not hold on super-Hubble scales even if the physics remains unchanged. Let us remind some examples.

\begin{itemize}
 \item The total density parameter $\Omega$ must lie within an order of magnitude of unity. If it were
 much larger the universe will have re-collapsed rapidly, on a time scale much shorter that the main-sequence star
 lifetime. If it were to small, density fluctuations would have frozen before galaxies could form. Typically
 one expects $0.1<\Omega_0<10$. Indeed, most inflationary scenarios lead to $\Omega_0\sim1$ so that
 this may not be anthropically determined but in that case inflation should last sufficiently long
 so that this could lead to a fine tuning on the parameters of the inflationary potential.
 \item The cosmological constant was probably the first one to be questioned in an anthropical way \citep{weinberg87}.
 Weinberg noted that if $\Lambda$ is too large, the universe will start accelerating before structures had time
 to form. Assuming that it does not dominate the matter content of the universe before the redshift $z_*$
 at which earliest  are formed, one concludes that $\rho_V=\Lambda/8\pi G<(1+z_*)\rho_{\mat0}$.
 \cite{weinberg87} estimated $z_*\sim4.5$ and concluded that ``if it is the anthropic
 principle that accounts for the smallness of the cosmological constant, then we would expect the vacuum energy
 density $\rho_v\sim(10-100)\rho_{\mat0}$ because there is no anthropic reason for it
 to be smaller''. Indeed, the observations indicate  $\rho_v\sim2\rho_{\mat0}$
 \item \cite{tegrees} pointed out that the amplitude of the initial density
 perturbation, $Q$ enters into the calculation and determined the anthropic region in the
 plane $(\Lambda,Q)$. This demonstrates the importance of determining the parameters
 to include in the analysis.
 \item Different time scales of different origin seem to be comparable: the radiative cooling,
 galactic halo virialization, time of cosmological constant dominance, the age of the universe
 today. These coincidences were interpreted as an anthropic sign \citep{bousso}.
\end{itemize}

These are just a series of examples. For a multi-parameter study of the anthropic bound, we refer, e.g., to \cite{tegmark} and to \cite{hall} for a general anthropic investigation of the standard model parameters. Note that \cite{NEW_Landsman:2015cma} suggested that fine-tuning requires no special explanation at all, as an example of the difficulty to put the multiverse on solid theoretical grounds.

\subsection{Anthropic predictions}

The determination of the anthropic region for a set of parameters is in no way a prediction
but simply a characterization of our understanding of a physical phenomenon $P$ that
we think is important for the emergence of observers. It reflects that, the condition
$C$ stating that the constants are in some interval, $C\Longrightarrow P$, is equivalent
to $!P\Longrightarrow !C$.

The anthropic principle \citep{carter74} states that ``what we can expect to observe must be restricted
by the conditions necessary for our presence as observers''. It has received many interpretations
among which the \emph{weak anthropic principle} stating that ``we must be prepared to take account of the fact that our
location in the universe in necessarily privileged to the extent of being compatible with our existence
as observers'', which is a restriction of the Copernican principle often used in cosmology,
and the \emph{strong anthropic principle} according to which ``the universe (and hence the fundamental parameters
on which it depends) must be such as to admit the creation of observers within it at some stage.''
(see \citealt{barrowtip} for further discussions and a large bibliography on the subject).

One can then try to determine the probability that an observer measures the value $x$ of the constant $X$
(that is a random variable fluctuating in the multiverse and the density of observers depend
on the local value of $X$). According to Bayes theorem, 
$$
P(X=x|\text{obs})\propto P(\text{obs}|X=x) P(X=x).
$$
$P(X=x)$ is the prior distribution, which is related to the volume of those parts of the universe in which
$X=x$ at $\dd x$. $P(\text{obs}|X=x)$ is proportional to the density of observers that are going
to evolve when $X=x$. $P(X=x|\text{obs})$ then gives the probability that a randomly selected observer is located in a region
where $X=x\pm \dd x$. It is usually rewritten as \citep{vilenkin}
$$
 P(x)\dd x = n_{\text{obs}}(x)P_{\text{prior}}\dd x.
$$
This higlights the difficulty in making a prediction. First, one has no idea of how
to compute $n_{\text{obs}}(x)$. When restricting to the cosmological constant,
one can argue \citep{vilenkin} that $\Lambda$ does not affect microphysics
and chemistry and then estimate $n_{\text{obs}}(x)$ by the fraction of
matter clustered in giant galaxies and that can be computed from a model
of structure formation. This may not be a good approximation when other
constants are allowed to vary and it needs to be defined properly. Second, 
$P_{\text{prior}}$ requires an explicit model of multiverse that would generate
sub-universes with different values $x_i$ (continuous or discrete) for $x$.
A general argument \citep{weinberg89} states that if  the range over which $X$ 
varies in the multiverse is large compared to the anthropic region $X\in[X_{\min},X_{\max}]$ 
one can postulate  that $P_{\text{prior}}$ is flat on $[X_{\min},X_{\max}]$. Indeed, such
a statement requires a measure in the space of the constants (or of the theories) that are allowed to vary.
This is a strong hypothesis, which is difficult to control. In particular if  $P_{\text{prior}}$
peaks outside of the anthropic domain, it would predict that the constants should lie on the
boundary of the anthropic domain \citep{ruba}. It also requires that there are sufficiently enough
values of $x_i$ in the anthropic domain, i.e., $\delta x_i\ll X_{\max}-X_{\min}$.
\cite{garvi} stressed that the hypothesis of a flat $P_{\text{prior}}$ for the cosmological
constant may not hold in various Higgs models, and that the weight can lower the mean viable
value. To finish, one wants to consider $P(x)$ as the probability that a random observer measures
the value $x$. This relies on the fact that we are a typical observer and we are implicitly making
a self sampling hypothesis. It requires to state in which class of observers we are
supposed to be typical -- the final result may depend on this choice \citep{neal} -- and
this hypothesis leads to conclusions such as the doomsday argument that have been
debated actively \citep{borstrom,neal}.

This approach to the understanding of the observed values of the fundamental constants (but also of the initial conditions of our universe) by resorting to the actual existence of a multiverse populated by a different ``low-energy'' theory of some ``mother'' microscopic theory allows us to explain the observed fine-tuning by an observational selection effect. It also sets a limit to the Copernican principle stating that we do not live in a particular position in space since we have to live in a region of the multiverse where the constants are inside the anthropic bound. Such an approach is indeed not widely accepted and has been criticized in many ways \citep{ellismulti, ellismulti2,stoeger, stark,  dpage, aguirre,vaas}. Several criticisms have emerged concerning probabilistic measurements of life-permitting intervals and differing opinions inasmuch as the constants of nature that must be considered. When talking about life, consensus is illusive largely because there is not even a consistent definition of what life is, e.g., \cite{NEW_Adams:2019kby} did not restrict to carbon-based life, \cite{NEW_Sandora:2022ilr}  focused on complex intelligent life and \cite{NEW_Collins} considered ``about embodied moral agents". \cite{NEW_Diaz-Pachon:2021urr} adopted a Bayesian approach to assign an  upper bound for the probability of tuning, which is invariant with respect to change of physical constants.

Among the issues to be answered before such an approach becomes more
rigorous, let us note: (\textit{1}) what is the shape of the string landscape;
(\textit{2}) what constants should we scan. It is indeed important to
distinguish the parameters that are actually fine-tuned in order to
determine those that we should hope to explain in this
way \citep{wilc, wilczek07}. Here theoretical physics is indeed
important since it should determine which of the numerical
coincidences are coincidences and which are expected for some
unification or symmetry reasons; (\textit{3}) how is the landscape populated;
(\textit{4}) what is the measure to be used in order and what is the correct
way to compute anthropically conditioned probabilities.

While considered as not following the standard scientific approach, this is the only existing window on some understanding of the values of the fundamental constants.

\section{Conclusions}\label{sectionconcl}

Over the past three decades, the fundamental constants have attracted a larger interest. They have become key players in testing General Relativity in the Solar system but also on the whole cosmological history, in constraining its extensions and in investigating dark matter and dark energy scenarios, providing informations that cannot be accessed only through the study of the large-scale structure of our universe. This complementarity with the other tests of General Relativity on astrophysical scales and the complementarity of local and cosmological constraints have been emphasized for many years \citep{Uzan:2003zq,jpu-revu3,ugrg,Jain:2013wgs} and shall now be full part of the cosmological lore.

The study of fundamental constants has witnessed tremendous progresses. Within a decade, the constraints on their possible space and time variations have flourished in many ways. They have reached higher precision and new systems, involving different combinations of constants and located at different redshifts, have been proposed. This has improved our knowledge on the validity of the equivalence principle and allowed us to test it on astrophysical and cosmological scales, as reviewed them in Sect.~\ref{section3} and~\ref{section4}. Many experimental observational progresses are expected in the coming years thanks to quantum laboratory technologies, or the developments of the E-ELT, radio observations, atomic clocks in space, or gravitational waves. We have provided new combined analysis of the constraints on the variation of fundamental constants.

Fundamental constants have an important role to play in cosmology, in view of its many anomalies \citep{NEW_Peebles:2022akh} and the puzzle of the dark sector. As we have seen, they may be involved in the Hubble tension (Sect.~\ref{secosmocte}) and in the lithium problem (Sect.~\ref{secBBNLi}). Besides, they are a powerful tool to constrain and eventually detect ultralight dark matter -- Sect.~\ref{secULDM} and topological defects (Sect.~\ref{secTopDef}). Each of these effects has a different signature; see Fig.~\ref{fig-clocksum}. More important the local constraints from atomic clock set strong bounds -- see Eq.~(\ref{edotphi0}) that needs to be satisfied by any viable scalar field model -- on any light scalar field that couple to the standard matter fields and hence on any dark energy model (Sect.~\ref{subsecphico}). They also allow us to question the isotropy of space (Sect.~\ref{subsec-dipalpha}) and hence the Copernican principle. Fundamental constants bridge our local physics to cosmological scale in a unique way and they have become to study any new degree of freedom and constrain its coupling to the standard model fields (see Fig.~\ref{fig-newdof}).

From a theoretical point of view, we had to spend time defining the concept of constant (Sect.~\ref{subsec12fc}). Structurally, the fundamental constants are now the heart of the definition of the International System of units (Sect.~\ref{subsec12met}) which we have also detailed. We pointed out the conceptual difference between fundamental units and fundamental parameters and the crucial fact that fundamental constants cannot be considered without a theoretical framework that needs to be specified. The first strong connection with theory arises from General Relativity and the test of the equivalence principle on which it rests (Sect.~\ref{subsec12}). Then, we have described in Sect.~\ref{section-theories} the high-energy models that predict such variations, as well as the link with the origin of the acceleration of the universe. In all these cases, a spacetime varying fundamental constant reflects the existence of an almost massless field that couples to matter. This will be at the origin of a violation of the universality of free fall and thus of utmost importance for our understanding of gravity and of the domain of validity of General Relativity. Huge progresses have been made in the understanding of the coupled variation of different constants (Sect.~\ref{subsec5.3}). While more model-dependent, this allows one to set stronger constraints and eventually to open an observational window on unification mechanisms. Again, its is important to stress that such theories are necessary to interpret the constraints consistently. Section~\ref{subsec81} has detailed the main models used in the literature. 

Section~\ref{section5} discussed the ideas that try to understand the value of the fundamental constants. While considered as borderline with respect to the standard physical approach, it reveals the necessity of considering a universe larger than our own, and called the multiverse. It will also give us a hint on our location in this structure in the sense that the anthropic principle limits the Copernican principle at the basis of most cosmological models. We have stressed the limitations of this approach and the ongoing debate on the possibility to make it predictive.

Given the growth of the number of studies, both experimental and theoretical, it becomes  almost impossible to keep track of all the evolution. This cannot be the goal of a review, that would necessarily miss some results and eventually misinterpret some others. This shall be a collaborative work for the community. In particular, we shall find an easy way for anyone to include the constraints arising from the fundamental constants in her/his analysis, and more particularly the local constraints that set strong constraints on cosmological models. Epilogue~\ref{app4a} proposes a scheme that could allow us to keep track of the evolutions in almost real time. It proposes a form, that can be used to summarize experimental conclusions in articles or be used to create an open database. Indeed it needs to be thought and designed collectively The goal would be to compile these data in a standardized way. This would offer traceability, avoid mistakes (since complied by its authors), be validated by the community, always up-to-date  and eventually be accessible online.

To conclude, the puzzle about the large numbers pointed out by Dirac has led to a better understanding of the fundamental constants and of their roles in the laws of physics. They are now part of the general tests of General Relativity, as well as a breadcrumbs to understand the origin of the acceleration of the universe and to more speculative structures, such as a multiverse structure, and possibly a window on string theory.

\vfil
 \begin{tcolorbox}
\section{Epilogue: Towards the collective creation of a community data base}\label{app4a}

 In order to construct a comprehensive and useful data basis, it is proposed that every article stating a constraint on some variation of fundamental constant tries to provide the following element at least in an appendix. This will allow traceability and possibility to compare the analysis as well as avoid cut/paste mistakes and misinterpretation
\paragraph{Proposition of a general typology}\

\begin{enumerate}
\item \underline{Physical system}. This can be any of the systems we have been discussed or eventually a new physical system such as 
\begin{itemize}
\item {\tt Equivalence principle}
\item {\tt Clock}
\item {\tt Oklo}
\item {\tt Meteorite dating}
\item {\tt Absorption spectra}
\item {\tt Emission spectra}
\item {\tt Stellar physics}
\item {\tt 21cm}
\item {\tt Galaxy cluster and SZ}
\item {\tt Lensing}
\item {\tt CMB}
\item {\tt BBN}
\item{\tt Pulsar timing}
\item {\tt Gravitational waves}
\end{itemize}
It can have different level of entry such as {\tt clocks/Rb-Cs} or {\tt Absorption spectra/0H-18cm}

\item \underline{Redshift of the measurement}. Or more generally spacetime position and environment.

\item \underline{Data set}. What are the data used in taure (absorption spectra/reference, CMB maps/release etc...)

\item \underline{Hypothesis on the physical system}. This shall include all the external parameters that are either fitted or fixed for the analysis. This could be the list of free cosmological parameters, nuisance parameters, parameters describing the environment (temperature, density,....)

\item \underline{Hypothesis on the fundamental constants}. This shall include all the fundamental constants that are either fitted or fixed for the analysis.

\item \underline{Hypothesis on the physical theory}. Specify hypothesis on the link, if any, on the variations of the constants or if one has assumed an independent variation or a single constant.

\item \underline{Raw constraints}. If possible one shall give the constraints on the observable, e.g., relative frequency drift.

\item \underline{Constraints on the constant}. This shall include 1$\sigma$ error bar and systematics error bar and the list of hypothesis/models used to link the observable to the constants.

\item \underline{List of possible systematic effects}. Description of the most important systematics of the system.

\item \underline{Other articles that have studied the same dataset}. If any. It could be a reanalysis (in what sens), an extension, a cross-study with another dataset etc...

\item \underline{Contact authors}.

\end{enumerate}
\end{tcolorbox}

\begin{acknowledgements}
I would like to thank all my collaborators on this topic, Alain Coc, Pierre Descouvemont, Sylvia Ekstr\"om, George Ellis, Georges Meynet, Nelson Nunes, Keith Olive, Cyril Pitrou and Elisabeth Vangioni as well as
B\'en\'edicte Leclercq, Roland Lehoucq, and Ludivine Bantigny.

I also thank many colleagues for sharing their thoughts on the subject with me, first at the Institut d'Astrophysique de Paris, Luc Blanchet, Michel Cass\'e, Gilles Esposito-Far\`ese, Bernard Fort, Guillaume Faye, Jean-Pierre Lasota, Yannick Mellier, Patrick Petitjean; at BIPM, Richard Davies, C\'eline Fellag-Ariouet, Martin Milton, Terry Quinn in France, Jibril Ben Achour,Joel Berg\'e, Francis Bernardeau, S\'ebastien Bize, Philippe Brax, Fran\c{c}oise Combes, Thibault Damour, Nathalie Deruelle, Hugo L\'evy, Martin Pernot-Borr\' as, Christophe Salomon, Peter Wolf; and to finish worldwide, John Barrow, Thomas Dent, Victor Flambaum, Bala Iyer, Lev Kofman, Carlos Martins, Paolo Molaro, David Mota, Michael Murphy, Jeff Murugan, Anan Srianand, Gabriele Veneziano, John Webb, Amanda Weltman, Christof Wetterich.
\end{acknowledgements}
\pagebreak

\begin{appendix}\normalsize
\section{System of units and the international system of units (SI)}\label{App0}

\subsection{System of units}\label{app0-1}

Originally, the sizes of the human body were mostly used to measure the length of objects (e.g., the foot and the thumb gave \emph{feet} and \emph{inches}) and some of these units can seem surprising to us nowadays (e.g., the \emph{span} was the measure of hand with fingers fully splayed, from the tip of the thumb to the tip of the little finger). Similarly weights were related to what could be carried in the hand: the pound,  the ounce, the dram\dots. Needless to say, this system had a few disadvantages since each country, region has its own system (for instance in France there was more than 800 different units in use in 1789). The need to define a system of units based on natural standard led to several propositions to define a standard of length (e.g., the \emph{mille} by Gabriel Mouton in 1670 defined as the length of one angular minute of a great circle on the Earth or the length of the pendulum that oscillates once a second by Jean Picard and Christiaan Huygens). The  real change happened during the French Revolution during which the idea of a universal and non anthropocentric system of units arose. In particular, the Assembl\'ee Nationale Constituante adopted the principle of a uniform system of weights and measures on 8 May 1790 and, in March 1791 a decree (these texts are reprinted in \citealt{ul-book}) was voted, stating that a quarter of the terrestrial meridian would be the basis of the definition of the \emph{meter} (from the Greek metron, as proposed by Borda): a meter would henceforth be one ten millionth part of a quarter of the terrestrial meridian. Similarly the \emph{gram} was defined as the mass of one cubic centimeter of distilled water (at a precise temperature and pressure) and the second was defined from the property that a mean Solar day must last 24 hours.

To make a long story short, this led to the creation of the metric system and then of the signature of \textit{La convention du m\`etre} in 1875. Since then, the definition of the units has evolved significantly. First, the definition of the meter was related to more immutable systems than our planet, which, as pointed out by Maxwell in 1870, was  an arbitrary and inconstant reference. He then suggested that atoms may be such a universal reference\footnote{\label{foot1} In his 1870 address, J.C. Maxwell states ``If we wish to obtain standards of length, time and mass which shall be absolutely permanent, we must seek them not in the dimensions, or motion or the mass of our planet, but in the wavelength, the period of vibration, and absolute mass of these imperishable and unalterable and perfectly similar molecules.''  This is the reason why he proposed to use the properties of what he thought were the most fundamental objects of nature at his time, namely atoms (which he then called molecules). Indeed, and as he explained, if a structure is truly fundamental, it cannot change in time. It is imperishable and unalterable. It follow that these references are to be found, as already discussed, in the most fundamental theories and in particular in its constants.}. In 1960, the International Bureau of Weights and Measures (BIPM) established a new definition of the meter as the length equal to 1650763 wavelengths, in a vacuum, of the transition line between the levels $2p_{10}$ and $5d_{5}$ of krypton-86. Similarly the rotation of the Earth was not stable enough for a reference. It was proposed in 1927 by Andr\'e Danjon to use the tropical year as a reference, as adopted in 1952. In 1967, the second was then related to an atomic transition, defined as the duration of 9\,162\,631\,770 periods of the transition between the two hyperfine levels of the ground state of caesium-133. To finish, it was decided in 1983, that the meter shall be defined by fixing the value of the speed of light in vacuum to $c = 299\,792\,458\unit{m\ s}^{-1}$ and we refer to \cite{bipm} for an up to date description of the SI system. Note that this version of the SI definition assumes that the permittivity of vacuum $\mu_0$ is also fixed so that the permeability of vacuum $\epsilon_0$ is also fixed.

This summary illustrates that the system of units is a human product and all SI definitions are historically based on non-relativistic classical physics and slowly drifted to quantum defined units since quantum technologies allow for a much lower uncertainty and a better control. The changes in the definitions were driven by the will to use more stable and more fundamental quantities so that they closely follow the progress of physics. This system has been created for legal use and indeed the choice of units is not restricted to SI.

\subsection{SI systems and the number of basic units} \label{app0-2}

The International System (SI) of Units defines seven basic units (see \cite{bipm} for details): the meter (m), second (s) and kilogram (kg), the ampere (A), kelvin (K), mole (mol)  and candela (cd), from which one defines secondary units. While needed for pragmatic reasons, this system of units is unnecessarily complicated from the point of view of theoretical physics. In particular, the kelvin, mole and candela are derived from the four other units since temperature is actually a measure of energy, the candela is expressed in terms of energy flux so that both can be expressed in mechanical units of length [L], mass [M] and time [T]. The mole is merely a unit denoting numbers of particles and has no dimension associated.

The status of the Ampere is interesting in itself. The discovery of the electric charge [Q] led to the introduction of a new units, the Coulomb (C). The Coulomb law describes the force between two charges as being proportional to the product of the two charges and to the inverse of the distance squared. The dimension of the force being known as [MLT$^{-2}$], this requires the introduction of a new constant $\varepsilon_0$ (which is only a conversion factor), with dimensions [Q$^{2}$M$^{-1}$L$^{-3}$T$^{2}$] in the Coulomb law, and that needs to be measured.  Indeed, another route is possible, and was actually adopted by Maxwell following Gauss, since the Coulomb law tells us that no new constant is actually needed if one uses [M$^{1/2}$L$^{3/2}$T$^{-1}$] as the dimension of the charge. In such a system of units, known as Gaussian units, the numerical value of $\varepsilon_0$ is 1. Hence the coulomb can be expressed in terms of the mechanical units [L], [M] and [T], and so will the ampere. This reduces the number of conversion factors, that need to be experimentally determined, but note that both choices of units assume the validity of the Coulomb law, which has its own domain of validity.

\subsection{Natural units}\label{app0-3}

The previous discussion tends to show that all units can be expressed in terms of the three mechanical units. It follows, as realized by Johnstone Stoney in 1874\footnote{After studying electrolysis in 1874, Johnstone Stoney suggested the existence of a ``single definite quantity of electricity''. He was able to estimate  the value of this elementary charge by means of Faraday's laws of   electrolysis. He introduced the term ``electron'' in 1894 and it  was identified as a particle in 1897 by Thomson.}, that these three basic units can be defined in terms of 3 independent constants.  He proposed \citep{stoney,barrowston} to use three constants: the Newton constant $G$, the velocity of light $c$ and the basic units of electricity, i.e., the electron charge $e$, in order to define, from dimensional analysis a ``natural series of physical units'' as
\begin{align}
 t_{\mathrm{S}} &= \sqrt{\frac{Ge^2}{4\pi\varepsilon_0 c^6}}\sim 4.59\times10^{-45} \unit{s},
 \nonumber\\
  \ell_{\mathrm{S}} &= \sqrt{\frac{G e^2}{4\pi\varepsilon_0 c^4}}\sim 1.37\times10^{-36} \unit{m},
 \nonumber\\
  m_{\mathrm{S}} &= \sqrt{\frac{e^2}{4\pi\varepsilon_0 G}}\sim 1.85\times10^{-9} \unit{kg},
  \nonumber
\end{align}
where the $\varepsilon_0$ factor has been included because we are using the SI definition
of the electric charge. In such a system of units, by construction, the numerical value
of $G$, $e$ and $c$ is 1, i.e., $c=1\times  \ell_{\mathrm{S}}\cdot t_{\mathrm{S}}^{-1}$ etc.

A similar approach to the definition of the units was independently proposed
by \cite{planck1} on the basis of the two constants $a$ and
$b$ entering the Wien law and $G$, which he reformulated later \citep{planck2} in terms
of $c$, $G$ and $\hbar$ as
\begin{align}
 t_{\mathrm{P}}=&\sqrt{\frac{G \hbar}{c^5}}\sim 5.39056\times10^{-44} \unit{s},
 \nonumber\\
  \ell_{\mathrm{P}}=&\sqrt{\frac{G\hbar}{c^3}}\sim 1.61605\times10^{-35} \unit{m},
 \nonumber\\
  m_{\mathrm{P}}=&\sqrt{\frac{\hbar c}{G}}\sim 2.17671\times10^{-8} \unit{kg}.
  \nonumber
\end{align}
The two systems are clearly related by the fine-structure constant since $e^2/4\pi\varepsilon_0$ $=$ $\aem hc$.

Indeed, we can construct many such systems since the choice of the 3 constants is arbitrary. For instance, we can build a system based on ($e, m_{\mathrm{e}}, h)$, that we would call the \emph{Bohr units}, well-suited to the study of the atom. The choice may be dictated by the system being studied (e.g., it is indeed far fetched to introduce $G$ in the construction of the units when studying atomic physics) so that the system is well adjusted in the sense that the numerical values of the computations are expected to be of order unity in these units.

Such constructions are very useful for theoretical computations but not adapted to measurement so that one needs to switch back to SI units. More important, this shows that, from a theoretical point of view, one can define the system of units from the laws of nature, which are supposed to be universal and immutable.

\subsection{The new (2018) SI system}\label{app0-4}

The 1983 redefinition of the meter was a first step toward natural units. Indeed the requirement of continuity, which is central to any redefinition of the SI system, forbids to set the numerical value of $c$ to unity. Even though the ``natural units" are not convenient and practical enough for high precision metrology, the  idea to define units from constants was taken very seriously, mostly because they appear, using the words by J.C. Maxwell (see footnote~\ref{foot1}), as the {\emph{most fundamental objects in our theories} and hence maybe the most stable objects on which one can build a stable system of units driving to move from physical artefacts to atomic properties to the physical constants; see Table~\ref{tab-SI0} for the effects of such a change on the accuracy of the meter.

\begin{table}[t]
\caption[Definitions of the meter and evolution of its uncertainties] {Evolution  absolute and relative uncertainties of the realisation of the meter with its definition. The shift to atomic units occurred in 1960 and to a fundamental constant in 1983.}
\label{tab-SI0}
\centering
{\small
\begin{tabular}{lllc}
 \toprule
Date &  Reference of the definition &Absolute uncertainty & Relative uncertainty
 \\
 \midrule
1798 & Earth meridian &0.1 mm & $10^{-4}$\\
1799 & Platine prototype & 0.01 mm & $10^{-5}$\\
1889 & Platine prototype & 0.1 $\mu$m & $10^{-7}$\\
1960 & Kr-86 wavelength &0.01 $\mu$m & $10^{-8}$\\
1983 &Speed of light in vacuum &0.1 nm & $10^{-10}$\\
2018 & Speed of light in vacuum&0.1 nm & $10^{-10}$\\
  \bottomrule
\end{tabular}
}
\end{table}

Let us first remind that the 1983 definition of the SI system was calling for improvements for many reasons. 
\begin{itemize}
\item The first is related to the kilogram. Since 1889, it was defined on the basis of a material artefact, the International Prototype of the Kilogram (IPK). Since then, it was compared almost every 40~years to official copies. This shows a relative drift of some tenth of microgram since 1889, which limits the relative accuracy of the kilogram to 50~ppm. Also, the PIK was kept in Paris making comparison difficult. Then, the fact that the kilogram is a material artefact implies that one can reach a relative accuracy of some $10^{-6}$ of masses of some kg but that such an accuracy was difficult to assess for vey small masses, e.g., for particles, or very large masses, e.g., in astrophysics. This led for instance to the use of atomic units in particle physics  with which one had a better accuracy for the masses of particles than in kg. To finish,  the definitions other base units such as the ampere, candela, mole depend on the one of the kilogram.
\item Concerning electronic units, the definition of the ampere through the mechanical force between two infinitely thin and long conductors was illusory. During some decades, this definition had been swapped in practice to definitions based on quantum effects so that electric standards were based on definitions outside the SI to reach higher precisions. This was in particular the case of the quantum Hall effect, related to the von Klitzning constant $R_{\rm K} = h/e^2$ and the Josephson effect related to the constant $K_{\rm J}=2e/h$. It follows that quantum techniques spread largely in one branch of metrology.
\item The Kelvin was defined as a fraction of the thermodynamic temperature of the triple point of water, hence depending on the properties of water and of its composition (purity, isotopic composition, etc.) Besides, the temperature scale was not adapted to extreme temperatures, typically above 2~000~K and below 20~K.
\item The definition of the mole mixes the concept of quantity of matter and mass, which was conceptually misleading.
\item To finish, from a structural point of view, the speed of light was fixed since 1983 to define the meter and the vacuum permeability $\mu_0$ was fixed in the definition of the ampere. Assuming Maxwell theory of electromagnetism, this implies that the vacuum permittivity $\varepsilon_0$ was also fixed. Hence constants started to play a key role in the definitions.
\end{itemize}
These considerations led to the idea to unify the SI system to solve these problems of consistency, stability and precision that became crucial with the development of quantum technologies in particular. The idea to redefine the kilogram in terms of a fixed value of the Planck constant \citep{karshen-kilo,NEW_Blaum:2019bxf} through the development of e.g., Watt balance paved the way to a  fully quantum-based (coherent and consistent) system of units based on fundamental constants; see e.g., \cite{NEW_PEIK201018}.

\begin{table}[t]
\caption[Defining constants of the International System of units]{The seven defining constants of the SI and the seven units they define.}
\label{tab-SI2}
\centering
{\small
\begin{tabular}{llll}
 \toprule
Defining constant& Symbol & Numerical value & Unit \\
 \midrule
 Hyperfine transition frequency of Cs & $\Delta\nu_{\rm Cs}$ & 9~192~631~770 & Hz \\
 Speed of light in vacuum &$c$  & 299~792~458 & m~s$^{-1}$ \\
Planck constant & $h$ &6.626~070 15 $\times10^{-34}$ & J s\\
elementary  charge &  $e$& 1.602~176~634 $\times10^{-19}$&C\\
Boltzmann constant & $k$  & 1.380~649 $\times10^{-23}$ &J K$^{-1}$\\
Avogadro number & $N_{\rm A}$ & 6.022~140~76 $\times10^{23}$ &mol$^{-1}$\\
luminous efficacy &  $K_{\rm cd}$ & 683 &lm W$^{-1}$\\
  \bottomrule
\end{tabular}
}
\end{table}

\begin{table}[t]
\caption[Comparison of the uncertainties CODATA-2017 vs new SI]{Uncertainties of the quantities entering the new definition of the SI compared to the one of the former definition of the SI.}
\label{tab-SI1}
\centering
{\small
\begin{tabular}{p{4cm}ccc}
 \toprule
 Quantity & Uncertainty (CODATA-2017) & Uncertainty (new SI) \\
  & $\times 10^{9}$& $\times 10^{9}$ \\
 \midrule
Speed of light &  0& 0\\
Planck constant & 10& 0\\
Electron charge & 5.2& 0\\
Boltzmann constant& 370  &0 \\
Avogadro number &10 & 0 \\
IPK & 0& 10 \\
Vacuum permeability & 0& 0.23\\
Vacuum permittivity & 0& 0.23\\
Water triple point & 0& 370\\
C-12 molar mass & 0& 0.45\\
  \bottomrule
\end{tabular}
}
\end{table}

Structurally, the new SI relies on a single physical artefact (as in the previous system, a hyperfine transition of Cs-133, together with the (exact) value of six constants $(c, h, e, k, N_{\rm A}, K_{\rm cd})$; see Table~\ref{tab-SI2} (the definitions below specify the exact numerical value of each constant when expressed in the corresponding SI unit). This set of defining constants has been chosen to provide a fundamental, stable and universal reference that simultaneously allows for practical realizations with the smallest uncertainties. 

For theorists, this definition is particularly satisfying for at least 2 reasons: (\textit{1}) it shed some light of the constants, the nature of which is source of many debates and (\textit{2})  everything rely on a clock. This is indeed the most universal way to define units in General Relativity since all units derive from the second and that the second, as defined, is the unit of proper time in the sense of the General theory of Relativity.  This revision of the SI, probably the most significant since its establishment, was adopted by the 26th CGPM (2018) and is documented\footnote{We also refer to the following online resources for an in depth description  of the new SI {\tt https://www.bipm.org/en/measurement-units}.}  in the 9th edition of the SI Brochure \citep{NEW_SI}. It  has entered into force on May 20th 2019. 
 \begin{tcolorbox}
{\bf Resolution 1 of the 26th CGPM (2018) --

On the revision of the International System of Units (SI)}

The General Conference on Weights and Measures (CGPM), at its 26th meeting,

{\bf considering}

[\ldots]

{\bf decides} that, effective from 20 May 2019, the International System of Units, the SI, is the system of units in which:

$\bullet$ the unperturbed ground state hyperfine transition frequency of the caesium 133 atom $\Delta \nu_{\rm Cs}$ is 9 192 631 770 Hz,

$\bullet$ the speed of light in vacuum $c$ is 299 792 458 m/s,

$\bullet$ the Planck constant $h$ is 6.626 070 15 $\times10^{-34}$ J s,

$\bullet$ the elementary charge $e$ is 1.602 176 634 $\times10^{-19}$ C,

$\bullet$ the Boltzmann constant $k$ is 1.380 649 $\times10^{-23}$ J/K,

$\bullet$ the Avogadro constant $N_A$ is 6.022 140 76 $\times10^{23}$ mol$^{-1}$,

$\bullet$the luminous efficacy of monochromatic radiation of frequency 540 $\times10^{12}$ Hz, $K_{\rm cd}$, is 683 lm/W,

where the hertz, joule, coulomb, lumen, and watt, with unit symbols Hz, J, C, lm, and W, respectively, are related to the units second, metre, kilogram, ampere, kelvin, mole, and candela, with unit symbols s, m, kg, A, K, mol, and cd, respectively, according to Hz = s$^{-1}$, J = kg m$^{2}$ s$^{-2}$, C = A s, lm = cd m$^{2}$ m$^{-2}$ = cd sr, and W = kg m$^{2}$ s$^{-3}$

[\ldots]

The integral text of the resolution can be accessed at \\{\small\url{https://www.bipm.org/en/committees/cg/cgpm/26-2018/resolution-1}}
 \end{tcolorbox}

The definition of the new SI has many consequences that shall be mentioned.
\begin{itemize}
\item It is a consistent and homogeneous system which ensures a gain in accuracy and precision with a better fiability and stability of the mass unit. It brings back electrical units within the SI and ensures continuity with the previous system while approaching the ideal of theoreticians. It relies on the best-tested equations of physics to connect the value of the constants to measurable quantities in the best tested theories at hand so that the SI system reflects our knowledge of physics at his best.
\item It contains only one physical artefact, an atomic clock, to define the second. This implies that any progress in Time-Frequency metrology will propagate to the other units. The actual reflection thus concerns the definition of the second, see e.g., \cite{NEW_Bize:2019bxt} and \cite{NEW_Dimarcq:2023lpj} for the actual roadmap.  
\item Besides, the new system makes no reference to any particular technology. A variety of experimental methods, referred to as ``mises en pratique'',  described by the CIPM Consultative Committees may be used to realize the definitions; see \cite{NEW_SI}. Realizations may be revised whenever new experiments with higher precision are developed; for this reason advice on realizing the definitions is not included in the definition but is available on the BIPM website. The use of a constant to define a unit disconnects definition from realization. This offers the possibility that completely different or new and superior practical realizations can be developed, as technologies evolve, without the need to change the definition.
\item Concerning the nature of the defining constants, it ranges from fundamental constants to technical constants, see Table~\ref{tab-SI2}. The seven constants are chosen in such a way that any unit of the SI can be written either through a defining constant itself or through products or quotients of defining constants. According to our former discussions, both the Planck constant $h$ and the speed of light in vacuum $c$ are properly described as fundamental (class C). The elementary charge $e$ corresponds to a coupling strength of the electromagnetic force via the fine-structure constant $\aem$ (class B). As discussed in this review, variable $\aem$ theories do exist but the constraints discussed in Sect.~\ref{section3} is a strong argument to think  that any effect of a variation of $\aem$ on foreseeable practical measurements can be excluded. Anyhow, such a variation will not impact the new SI since it will then be interpreted as a variation of $\mu_0$ and $\varepsilon_0$ (remember that $\aem$ is dimensionless). Note also that this choice is not unique and was at the core of many debates, both theoretical and experimental. For instance, shall one choose $e$ or $\varepsilon_0$ or the  vacuum impedance $Z_0=\mu_0 c$ to define the ampere? The choice emphasizes on the physical interpretation of the fine structure constant, seen either as a property of the vacuum or of matter\footnote{Assuming $\aem$ is a property of vacuum originates in string theories in which the dilaton determine the string coupling and can source a variation of $\aem$ and other constants. Hence one would then favor the choice of $e$ to define the ampere. If $\aem$ can also be seen as property of the electron since QED in flat spacetime relates the $g$-factor of the electron to a series in $\aem/2\pi$. It would then favor the choice of the Planck charge $q_{\rm P}= \sqrt{2\varepsilon_0 hc}=\sqrt{2h/Z_0}=e/\sqrt{\aem}$ which would correspond to keep $Z_0$ constant. To conclude, it is for practical and educational reasons that the elementary charge was preferred.}; see \cite{NEW_edef,NEW_Jeckelmann:2019xfw}. This highlights the importance on our reading of the physical laws. $\Delta\nu_{\rm Cs}$ the unperturbed ground-state hyperfine transition frequency of the caesium-133 atom is an atomic parameter (class A), which may be affected by the environment, such as electromagnetic fields. However, the underlying transition is well understood, stable and a good choice as a reference transition under practical considerations. It specifies the reference for all other units. The Boltzmann constant $k$ and the Avogadro constant $N_{\rm A}$ are  proportionality constants, the first between  temperature (with unit kelvin) and energy (with unit joule), whereby the numerical value is obtained from historical specifications of the temperature scale and the second  between the quantity amount of substance (with unit mole) and the quantity for counting entities (with unit one, symbol 1). The luminous efficacy of monochromatic radiation of frequency $540\times 10^{12}$~Hz, $K_{\rm cd}$, is a technical constant (class A) that gives an exact numerical relationship between the purely physical characteristics of the radiant power stimulating the human eye (W) and its photobiological response defined by the luminous flux due to the spectral responsivity of a standard observer (lm) at that frequency.
\item The definitions of the units can be understood as follows; see \cite{NEW_SI}. The exact relation $\Delta\nu_{\rm Cs} = 9~192~631~770$~Hz implies that the unit second is expressed in terms of the defining constant as
\begin{equation}
1~{\rm s} = \frac{9~192~631~770}{\Delta\nu_{\rm Cs}}.
\end{equation}
The definition of the meter derived from the exact relation $c=299~792~458$~m.s$^{-1}$, i.e.,
\begin{equation}
1~{\rm m} = \frac{c}{ 299 792 458} {\rm s}\simeq 30.663~319\frac{c}{\Delta\nu_{\rm Cs} }.
\end{equation}
The definition of the kilogram derives from the fact that the exact value of the Planck constant is set to 6.626~070~15$\times10^{-34}$~J~s, that can be inverted to give
\begin{equation}
1~{\rm kg} = \frac{h}{6.626~070~15 \times10^{-34}} {\rm m}^{-2} {\rm s} \simeq 1.475~5214\times 10^{40} \frac{h \Delta\nu_{\rm Cs} }{c^2}.
\end{equation}
The ampere derives from the exact value of $e$ so that
\begin{equation}
1~{\rm A} = \frac{e}{1.602~176634 \times10^{-19}} {\rm s}^{-1}  \simeq 6.789~6868 \times 10^{8} \Delta\nu_{\rm Cs} e.
\end{equation}
Since only $e$ is fixed, $\mu_0$ shall now be determined experimentally, as well as $\varepsilon_0$ -- see Table~\ref{tab-SI2} -- but the relation $\mu_0\varepsilon_0=1/c^2$ remains exact. The Kelvin derives from the Boltzmann constant as
\begin{equation}
1~{\rm K} = \frac{1.380~649\times10^{-23}}{k} \,{\rm kg~m}^{-2}{\rm s}^{-2}  \simeq 2.266~6653 \times 10^{8} \frac{\Delta\nu_{\rm Cs} h}{k}.
\end{equation}
The mole is not related to a dimension since it is a pure number. The definition in terms of the Avogadro  number has the advantage to disentangle the amount of matter from the notion of mass. As a consequence, the molar mass of carbon-12, previously exactly fixed to $0.012$~ kg/mol, is no longer known exactly and must be determined experimentally and has a relative standard uncertainty of $4.5\times10^{-10}$. To finish, the candela is 
\begin{equation}
1~{\rm cd} = {K_{\rm cd}}{683}~{\rm kg~m}^{2}{\rm s}^{-3}{\rm sr}^{-1}  \simeq 2.614~8305 \times 10^{10} \Delta\nu_{\rm Cs}^2 h K_{\rm cd}.
\end{equation}
\item Note that with such definitions, primary realizations of the kilogram and the Kelvin can be established, in principle, at any point in the mass scale or  the temperature scale.
\item There is a big shift in the way to think about constants and units. For instance, the previous definition of the kilogram fixed the value of the mass of the IPK to be equal to one kilogram exactly and the value of the Planck constant $h$ had to be determined with a given uncertainty by experiment. The present definition fixes the
numerical value of $h$ exactly and the mass of the IPK has now to be determined by experiment and is thus known with an uncertainty. Thus, uncertainties are shifted from the constants to the physical realization of the units, which then propagates to all measurements (see Table~\ref{tab-SI2}). It follows that the temperature of the triple point of water, $\mu_0$, $\varepsilon_0$ and the molar mass of carbon-12 are no more exact and their values and uncertainties shall be determined experimentally. Hence, one can see the exact values of the defining constant as a thermostat which fixes the realization of the units. Indeed, none of the values of the dimensionless parameters are affected by the change of definition.
\item  All conversion factors between energy units (Joule, eV, kg.$c^2$, ... ) become exact since all conversions involve constants the values of which are now exact.
\item The whole system relies on atomic and quantum physics. Besides, the \emph{quantum metrology Triangle} experiments (see \citealt{NEW_Scherer:2012ai} for a review) combine three quantum electrical effects (the Josephson effect, the quantum Hall effect and the single-electron transport effect) used in metrology and allow for important fundamental consistency tests on the validity of commonly assumed relations between fundamental constants and the quantum electrical effects.
\end{itemize}

This new SI is probably the best system one can have constructed since it builds on our best understanding of the laws of nature and on the parameters which are thought to be the most fundamental so far, the fundamental constants.

\subsection{Do we actually need 3 natural units?}\label{app0-5}

This is an issue debated at length. For instance,  \cite{duff01} respectively argue for none, three and two (see also \citealt{wignall00,Matsas:2007zz,NEW_Duff:2014mva,Matsas:2023lmp}). Arguing for no fundamental constant leads to consider them simply as conversion parameters. Some of them are, like the Boltzmann constant, but some others play a deeper role in the sense that when a physical quantity becomes of the same order as this constant, new phenomena appear; this is the case, e.g., of $\hbar$ and $c$, which are associated respectively to quantum and relativistic effects. \cite{okun91} considered that only three fundamental constants are necessary, as indicated by the International System of units. In the framework of quantum field theory + General Relativity, it seems that this set of three constants has to be considered and it allows one  to classify the physical theories (with the famous \emph{cube of physical theories}). However, \cite{veneziano86} argued that in the framework of string theory one requires only two dimensionful fundamental constants, $c$ and the string length $\lambda_s$. The use of $\hbar$ seems unnecessary since it combines with the string tension to give $\lambda_s$. In the case of the Nambu--Goto action $S/\hbar=(T/\hbar)\int\dd(Area)\equiv \lambda_s^{-2}\int\dd(Area)$ and the Planck constant is just given by $\lambda_s^{-2}$. In this view, $\hbar$ has not disappeared but has been promoted to the role of a UV cut-off that removes both the infinities of quantum field theory and singularities of General Relativity. This situation is analogous to pure quantum gravity \citep{novikov82} where $\hbar$ and $G$ never appear separately but only in the combination $\ell_{\mathrm{Pl}}=\sqrt{G\hbar/c^{3}}$ so that only $c$ and $\ell_{\mathrm{Pl}}$ are needed. \cite{volovik02} made an analogy with quantum liquids to clarify this. There, an observer knows both the effective and microscopic physics so that he can judge whether the fundamental constants of the effective theory remain fundamental constants of the microscopic theory. The status of a constant depends on the considered theory (effective or microscopic) and, more interestingly, on the observer measuring them, i.e., on whether this observer belongs to the world of low-energy quasi-particles or to the microscopic world.

\section{Notations}

\subsection{Constants}\label{app1}

The notations and numerical values of the constants used in this review are summarized in Table~\ref{tab-list} and Table~\ref{tab-list2}.

\subsection{Sensitivity coefficients}\label{app2}

The text introduces several sensitivity coefficients. We recall their definitions here.
\begin{itemize}
 \item Given an observable $O$, the value of  which depends on a set of  primary parameters $G_k$, the sensitivity of the measured value of $O$ to these parameters is
\begin{equation}\label{e.ck}
  \frac{\dd \ln O}{\dd \ln G_k} = c_k.
 \end{equation}
The value of the quantities $c_k$ requires a physical description of the system.
\item The parameters $G_k$ can be related to a set of fundamental constants $\alpha_i$ and we define
\begin{equation}\label{e.dki}
  \frac{\dd \ln G_k}{\dd \ln\alpha_i} = d_{ki}.
 \end{equation}
 The computation of the coefficients $d_{ki}$ requires one to specify the
 theoretical framework and depends heavily on our knowledge of
 nuclear physics and the assumptions on unification.
\item A particular sets of parameters $d_{ki}$
 has been singled out for the sensitivity of the mass of a body $A$
 to a variation of the fundamental constants
\begin{equation}\label{e.fA}
  \frac{\dd \ln m_A}{\dd \alpha_i} = f_{Ai}.
 \end{equation}
 One also introduced
\begin{equation}\label{e.lambdaAi}
  \frac{\dd \ln m_A}{\dd \ln\alpha_i} = \lambda_{Ai}
 \end{equation}
 so that 
 \begin{equation}\label{e.lambdafi}
  \lambda_{Ai} = \alpha_if_{Ai}.
 \end{equation}
\item In models where the variation of the fundamental constants are induced by the variation of a scalar field we defined
\begin{equation} \label{def_sphi}
  \frac{\dd \ln\alpha_i}{\dd \varphi} = s_{i}(\varphi),
 \end{equation}
and 
\begin{equation} \label{def_betaA}
  \frac{\dd \ln m_A}{\dd \varphi} = \alpha_A(\varphi),
 \end{equation}
 which can be expressed in terms of the previous sensitivities as
 $$
 \alpha_A(\varphi) = \sum_i f_{Ai}\alpha_i s_i(\varphi).
 $$
 We also recall that the dimensionless scalar field $\varphi$ and the one with dimension of mass $\phi$ are related by
 \begin{equation}
 \phi = M_*\varphi
 \end{equation}
 where $M_*$ is a mass scale, often chosen as 
 $$
 M_*=\sqrt{8\pi G_*}.
 $$
 
\item In  the class of models the variation of the constants can be related
to the gravitational potential by
\begin{equation}\label{e.fi}
  \frac{\dd \ln\alpha_i}{\dd \Phi_{\rm N}} = k_{i}.
 \end{equation}  
 \end{itemize}

\section{Background cosmological spacetime}\label{app3}

We consider that the spacetime is described by a manifold ${\mathcal M}$ with metric $g_{\mu\nu}$ with signature $(-,+,+,+)$. In the case of a Minkowski spacetime $g_{\mu\nu}= \eta_{\mu\nu}$.

In the cosmological context, the universe is described by a Friedmann--Lema\^{\i}tre spacetime with metric
\begin{equation}
 \dd s^2 = - \dd t^2 + a^2(t)\gamma_{ij}\dd x^i\dd x^j
\end{equation}
where $t$ is the cosmic time, $a$ the scale factor and $\gamma_{ij}$ the metric on the constant time hypersurfaces. The Hubble function is defined as 
\begin{equation}
H\equiv \dot{a}/a
\end{equation}
with a dot referring to a derivative with respect to the cosmic time $t$. We also define the redshift by the relation $1+z=a_0/a$, with $a_0$ the scale factor evaluated today. It is usual to introduce the conformal time $\eta$ by $\dd t= a\dd\eta$ and use a prime for derivative with respect to $\eta$. The deceleration parameter $a$ is defined as
\begin{equation}\label{e.defq}
q=-\frac{\ddot a}{a H^2} = \frac{\dot H}{H^2}-1
\end{equation}
that the Universe accelerates if $q<0$ and decelerates otherwise.

The evolution of the scale factor is dictated by the Friedmann equation
\begin{equation}\label{e.FLeq}
 H^2= \frac{8\pi G}{3}\rho - \frac{K}{a^2} + \frac{\Lambda}{3},
\end{equation}
where $\rho= \sum_i\rho_i$ is the total energy density of the matter components in the universe. Assuming the species $i$ has a constant equation of state $w_i=P_i/\rho_i$, each component evolves as $\rho_i= \rho_{i0}(1+z)^{2(1+w_i)}$. The Friedmann equation can then be rewritten as
\begin{equation}
 \frac{H^2}{H_0^2}= \sum \Omega_i (1+z)^{3(1+w_i)} +  \Omega_K(1+z)^2 + \Omega_\Lambda,
\end{equation}
with the density parameters defined by
\begin{equation}
 \Omega_i\equiv \frac{8\pi G\rho_{i0}}{3H_0^2},\qquad
  \Omega_i\equiv -\frac{K}{3H_0^2},\qquad
 \Omega_\Lambda\equiv \frac{\Lambda}{3H_0^2}. 
\end{equation}
They clearly satisfy $\sum \Omega_i +  \Omega_K+ \Omega_\Lambda=1$.

The value of the Hubble parameter today is
\begin{equation}\label{e.hubble}
H_0= 100\, \hub\unit{km.s^{-1}Mpc^{-1}} =  \frac{\hub}{9.78\times 10^{9}\unit{yr}} =\frac{\hub}{3.09\times 10^{17}\unit{s}},
\end{equation}
where $\hub$ is a dimensionless number, usually denoted by $h$ in the literature but this notation avoids confusion with the Planck constant. For the sake of numerical computations, we shall use the values of the cosmological parameters as estimated by the latest Planck satellite data analysis  \citep{NEW_Planck:2018vyg}.

Concerning the properties of the cosmological spacetime, we follow the notations and results of \cite{peteruzanbook}.

\begin{table}[t]
\caption[Main cosmological parameters of the standard $\Lambda$CDM  model.]{Main cosmological parameters in the standard $\Lambda$-CDM model. The first part of the table gives the parameters of the background minimal standard cosmological parameters, that is the reduced Hubble parameters together with 7 density parameters -- only 6 of them being independent since $\sum\Omega_i=1$ -- and the baryon-to-photon ratio. The second part gives the parameters related to the density perturbations for the large-scale structures of the universe. Then, the last part gives several extensions such as a dark energy component and gravity waves. Note that often the spatial curvature is set to $\Omega_K=0$. (See \citealt{NEW_Planck:2018vyg}).}
\label{tab-cosmo}
\centering
{\small
\begin{tabular}{p{6cm}ll}
 \toprule
 Parameter & Symbol & Value \\
 \midrule
  Reduced Hubble constant & $h_0$ & $0.6766\pm0.0042$ \\
  Photon density  & $\Omega_\gamma h_0^2$ & $2.471 \times 10^{-5}$ \\
   Neutrino density  & $\Omega_\nu h_0^2$ & (0.0005\,--\,0.023) \\
  Dark matter density  & $\Omega_{\mathrm{CDM}}h_0^2$ & $0.11933 \pm 0.00091$ \\
  Baryon density  & $\Omega_{\mathrm{b}}h_0^2$ & $0.02242\pm0.00014$ \\
  Matter density  & $\Omega_{\mathrm{m}}h_0^2$ & $0.14240\pm0.00087$ \\
  Matter density  & $\Omega_{\mathrm{m}}$ & $0.3111\pm0.0056$ \\
  Cosmological constant & $\Omega_\Lambda$ & $0.6889\pm0.0056$\\
  Spatial curvature & $\Omega_K$ & 0.011(12)\\
  Baryon-to-photon ratio & $\eta = n_\baryon/n_\gamma$ & $6.12(19) \times 10^{-10}$ \\
  \midrule 
  Scalar modes amplitude & $Q$ & $(2.0 \pm 0.2) \times 10^{-5}$  \\
  Scalar spectral index & $n_S$ & 0.958(16) \\
 \midrule 
  Dark energy equation of state & $w$ & --0.97(7) \\
  Scalar running spectral index & $\alpha_S$ & --0.05\,$\pm$\,0.03\\
  Tensor-to-scalar ratio & T/S &  $<$\,0.36\\
  Tensor spectral index & $n_T$ & $<$\,0.001 \\
  Tensor running spectral index & $\alpha_T$ &  ?\\
  \bottomrule
\end{tabular}
}
\end{table}

\section{Equation of motion for a charged particle}\label{app5}

In a theory in which the matter fields couple to the metric $A^2(\varphi)g_{\mu\nu}$, the equation of a point particle of mass $m$ and charge $q$ is
\begin{equation}\label{e.action}
S_{\rm pp}=-c^2\int m A(\varphi) \sqrt{-g_{\mu\nu}u^\mu u^\nu}\dd\tau+ q \int {\cal A}_\mu u^\mu \dd\tau.
\end{equation}
The equation of motion follows as
\begin{eqnarray}\label{e.motion}
mc^2 \gamma^\mu = \frac{q}{A(\phi)}{F^\mu}_\nu u^\nu - mc^2\frac{\partial \ln A}{\partial\phi}\perp^{\mu\nu}\nabla_\nu\varphi
\end{eqnarray}
with  $\perp_{\mu\nu}\equiv g_{\mu\nu}+u_\mu u_\nu/c^2$ the projector on the 3-space normal to $u^\mu$, which indeed ensures that $u^\mu u_\mu=-c^2$; see e.g., \cite{NEW_Uzan:2010pm}. It follows that the fifth force,
\begin{equation}\label{e.5th}
F^\mu =  - mc^2\alpha(\varphi)\perp^{\mu\nu}\nabla_\nu\varphi,
\end{equation}
remains perpendicular to the 4-velocity, $u_\mu F^\mu=0$.

} 
\end{appendix}

\phantomsection
\addcontentsline{toc}{section}{References}
\bibliographystyle{spbasic-FS}      
\bibliography{refsNEW}   
\end{document}